\documentclass[11pt,a4paper]{book}
\pdfoutput=1
\usepackage{amsmath}
\usepackage{amssymb}
\usepackage{amsfonts}
\usepackage{appendix}
\usepackage{array}
\usepackage[english]{babel}
\usepackage{calc}
\usepackage{color}
\usepackage{colortbl}
\usepackage{epigraph} 
\usepackage{eso-pic}
\usepackage{fancyhdr}
\usepackage[pdftex]{graphicx}
\usepackage{makeidx}
\usepackage{moreverb}
\usepackage{multirow}
\usepackage{multicol}
\usepackage{verbatim}
\usepackage[pdftex,
        colorlinks=true,
        urlcolor=rltblue,
        anchorcolor=rltbrightblue,
        filecolor=rltgreen,
        linkcolor=rltred,
        menucolor=webblue,
        citecolor=webgreen,
        pdftitle={Strategies for precision measurements
          of the charge asymmetry of the W boson mass
          at the LHC within the ATLAS experiment},
        pdfauthor={Florent FAYETTE},
        pdfsubject={Ph.D. thesis dissertation},
        pdfkeywords={CERN LHC ATLAS 
          W-boson W-boson-mass 
          charge-asymmetry 
          W-mass-charge-asymmetry electroweak QCD 
          Drell-Yan charged-Drell-Yan
          Monte-Carlo WINHAC},
        pagebackref,
        pdfpagemode=None,
        bookmarksopen=true]{hyperref}

\makeindex

\renewcommand{\chaptermark}[1]{\markboth{#1}{}}

\renewcommand{\headrulewidth}{0.5pt}
\renewcommand{\footrulewidth}{0pt}
\fancypagestyle{plain}{
\fancyhead{}
\renewcommand{\headrulewidth}{0pt}
}

\makeatletter
\newcommand\AtUpperLeftCorner[3]{%
\begingroup
\@tempdima=0cm
\@tempdimb=\paperheight
\ifodd\c@page
\advance\@tempdima#1
\else
\advance\@tempdima#1
\fi
\advance\@tempdimb-#2
\put(\LenToUnit{\@tempdima},\LenToUnit{\@tempdimb}){#3}%
\endgroup
}
\newcommand\AtUpperRightCorner[3]{%
\begingroup
\@tempdima=\paperwidth
\@tempdimb=\paperheight
\ifodd\c@page
\advance\@tempdima-#1
\else
\advance\@tempdima-#1
\fi
\advance\@tempdimb-#2
\put(\LenToUnit{\@tempdima},\LenToUnit{\@tempdimb}){#3}%
\endgroup
}
\makeatother

\makeatletter
\newlength\twolinebox@linelength
\newlength\twolinebox@columnheight
\newcommand{\twolinebox}[2]{%
   \setlength{\twolinebox@linelength}%
             {\maxof{\widthof{#1}}{\widthof{#2}}}%
   \setlength{\twolinebox@columnheight}{\heightof{#1}+\depthof{#1}+0.2em+0.4em/2+\heightof{0}/2}%
   \raisebox{0pt}[\twolinebox@columnheight][\heightof{\vbox{\vskip0.2em\hbox to 
   \twolinebox@linelength {#1\hfil}\vskip0.4em\hbox to 
   \twolinebox@linelength {#2\hfil}}}+\depthof{\vbox{\vskip0.2em\hbox to 
   \twolinebox@linelength {#1\hfil}\vskip0.4em\hbox to 
   \twolinebox@linelength {#2\hfil}}}-\twolinebox@columnheight+0.2em]{\vbox to 
   \twolinebox@columnheight{\vskip0.2em\hbox to 
   \twolinebox@linelength {#1\hfil}\vskip0.4em\hbox to 
   \twolinebox@linelength {#2\hfil}}}%
}

\renewcommand{\thesubsubsection}{(\alph{subsubsection})}

\renewcommand\arraystretch{1.45}

\newcommand{\figtxt}[1]{\footnotesize{#1}}

\makeatletter
\renewcommand\frontmatter{%
    \cleardoublepage
  \@mainmatterfalse}
\renewcommand\mainmatter{%
    \cleardoublepage
  \@mainmattertrue}
\makeatother

\newcommand{\tw}      {\textwidth}

\newcommand{\ie}[0]   {\textit{i.e.}}
\newcommand{\eg}[0]   {\textit{e.g.}}
\newcommand{\etc}[0]  {\textit{etc}}
\newcommand{\mm}[1]   {\mathrm{#1}}
\newcommand{\mbf}[1]  {\mathbf{#1}}
\newcommand{\CHIEXP}[1]{^{\;(#1)}}
\newcommand{\CHIEXPPM}[2]{^{\;(#1)}_{\;(#2)}}

\newcommand\WINHAC[0] {\textsf{WINHAC}}
\newcommand\ZINHAC[0] {\textsf{ZINHAC}}
\newcommand\PHOTOS[0] {\textsf{PHOTOS}}
\newcommand\LHAPDF[0] {LHAPDF}
\newcommand\winhaci[0]{Winhac\_i}
\newcommand\Pythia[0] {\textsc{Pythia}}

\newcommand{\Cen}     {\mathrm{cen}}
\newcommand{\Min}     {\mathrm{min}}
\newcommand{\Max}     {\mathrm{max}}

\newcommand{\PD}      {\mathcal{PD}}
\newcommand{\MT}      {\mathcal{MT}}

\newcommand{\DfDx}[2]     {\frac{d\,{#1}}{d\,{#2}}}
\newcommand{\flatDfDx}[2] {d\,{#1}/d\,{#2}}
\newcommand{\cotan}       {\mathrm{cotan}\,}
\newcommand{\Mean}[1]     {\left<#1\right>}
\newcommand{\sg}[1]       {\mathrm{sg}_{#1}}

\newcommand{\percent}{\,\%}
\newcommand{\nsec}   {\,\mathrm{ns}}

\newcommand{\microm} {\,\mu\mathrm{m}}
\newcommand{\cm}     {\,\mathrm{cm}}
\newcommand{\m}      {\,\mathrm{m}}

\newcommand{\Tesla}  {\,\mathrm{T}}
\newcommand{\MeV}    {\,\mathrm{MeV}}
\newcommand{\GeV}    {\,\mathrm{GeV}}
\newcommand{\TeV}    {\,\mathrm{TeV}}

\newcommand{\cut}    {{\mathrm{cut}}}
\newcommand{\plus}   {{+}}
\newcommand{\minus}  {{-}}

\newcommand{\GF}     {{G_\mathrm{F}}}
\newcommand{\MW}     {{M_W}}

\newcommand{\GamW}   {{\Gamma_W}}

\newcommand{\Nc}     {{N_\mm{c}}}
\newcommand{\alphas} {{\alpha_\mm{s}}}
\newcommand{\dbar}   {\bar d}
\newcommand{\ubar}   {\bar u}
\newcommand{\sbar}   {\bar s}
\newcommand{\cbar}   {\bar c}
\newcommand{\bbar}   {\bar b}
\newcommand{\tbar}   {\bar t}

\newcommand{\smartssbar}{{s\hskip-7pt\hbox{$^{^{(\!-\!)}}$}}}
\newcommand{\smartccbar}{{c\hskip-7pt\hbox{$^{^{(\!-\!)}}$}}}
\newcommand{\smartbbbar}{{b\hskip-7pt\hbox{$^{^{(\!-\!)}}$}}}

\newcommand{\WRF}[0] {{$W$RF}}

\newcommand{\W}      {W}

\newcommand{\BFW}    {\mathbf{W}} 
\newcommand{\Wpm}    {{W^\pm}}

\newcommand{\Wp}     {{W^+}}
\newcommand{\BFWp}   {\mathbf{\Wp}}

\newcommand{\MWp}    {{M_\Wp}}
\newcommand{\GamWp}  {{\Gamma_\Wp}}

\newcommand{\Wm}     {{W^-}}   
\newcommand{\BFWm}   {\mathbf{\Wm}}

\newcommand{\MWm}    {{M_\Wm}}
\newcommand{\GamWm}  {{\Gamma_\Wm}}

\newcommand{\Wtolnu} {W\to l\,\nu_l}
\newcommand{\Wtoel}  {W\to e\,\nu_e}
\newcommand{\Wtomu}  {W\to \mu\,\nu_\mu}
\newcommand{\Wtotau} {W\to \tau\,\nu_\tau}

\newcommand{\lm}     {{l^-}}
\newcommand{\lp}     {{l^+}}
\newcommand{\lpm}    {{l^\pm}}
\newcommand{\nul}    {{\nu_l}}
\newcommand{\anul}   {{\bar\nu_l}}
\newcommand{\smartnuanu}
	             {{\nu_l\hskip-9pt\hbox{$^{^{(\!-\!)}}$}}}
\newcommand{\smartnuanul}{\smartnuanu}

\newcommand{\Em}     {{e^-}} 
\newcommand{\ep}     {{e^+}}
\newcommand{\epm}    {{e^\pm}}
\newcommand{\nue}    {{\nu_e}}

\newcommand{\mum}    {{\mu^-}}
\newcommand{\mup}    {{\mu^+}}

\newcommand{\numu}   {{\nu_\mu}}

\newcommand{\nutau}  {{\nu_\tau}}

\newcommand{\yW}     {{y_W}}
\newcommand{\pTW}    {{p_{T,W}}}

\newcommand{\pTlstar}{{p_{T,l}^\ast}}
\newcommand{\pTl}    {{p_{T,l}}}

\newcommand{\rhoTl}  {{\rho_{T,l}}}
\newcommand{\pTlp}   {{p_{T,\lp}}}
\newcommand{\pTlm}   {{p_{T,\lm}}}
\newcommand{\pTnu}   {{p_{T,\nul}}}
\newcommand{\mTlnu}  {{m_{T,\,l\,\nul}}}

\newcommand{\ETmiss} {{\slashii E_T}}
\newcommand{\etal}   {{\eta_l}}
\newcommand{\etalp}  {{\eta_\lp}}
\newcommand{\etalm}  {{\eta_\lm}}

\newcommand{\Aeta}   {{|\eta|}}

\newcommand{\costhetaWlwrf}{{\cos\theta_{W,l}^{\ast}}}
\newcommand{\thetaWlwrf}   {{\theta_{W,l}^{\ast}}}

\newcommand{\costhetaqlwrf}{{\cos\theta_{q,l}^{\ast}}}
\newcommand{\thetaqlwrf}   {{\theta_{q,l}^{\ast}}}
\newcommand{\val}     {{\mathrm{(v)}}}
\newcommand{\sea}     {{\mathrm{(s)}}}
\newcommand{\smartpap}{{p\hskip-7pt\hbox{$^{^{(\!-\!)}}$}}}
\newcommand{\ppbar}   {{p\,\pbar}}
\newcommand{\pbar}    {{\bar p}}
\newcommand{\pp}      {{p\,p}}
\newcommand{\dd}      {{d\,d}}

\newcommand{\qbar}    {{\bar q}}

\newcommand{\qbp}     {{\bar q'}}
\newcommand{\kT}      {{k_T}}
\newcommand{\pT}      {{p_T}}

\newcommand{\usbar}[1] {{\bar u({#1})}}
\newcommand{\vsbar}[1] {{\bar v({#1})}}
\newcommand{\us}[1]    {{u({#1})}}
\newcommand{\vs}[1]    {{v({#1})}}

\newcommand{\HVmAdn}[1]{{\gamma_{#1}\,{\tfrac{1}{2}}(1-\gamma^5)}}
\newcommand{\HVmAup}[1]{{\gamma^{#1}\,{\tfrac{1}{2}}(1-\gamma^5)}}

\newcommand{\shat}    {{\hat s}}

\newcommand{\sigmahat}       {{\hat \sigma}}
\newcommand{\DsigpartDobs}[1]{\frac{d\,\sigmahat}{d\,{#1}}}

\newcommand{\FlatDsigmaDobs}[1]
                             {{d\,\sigma}/{d\,{#1}}}

\newcommand{\Vckm}[2]   {{V_{#1#2}}}
\newcommand{\Vckmsqr}[2]{{\left|V_{#1#2}\right|}^2}
\newcommand{\PDF}[2]    {{f^{#1}_{#2}}}

\newcommand{\dof}       {{\mathrm{dof}}}
\newcommand{\chiD}      {{\chi^2}}
\newcommand{\chiDmin}   {{\chi_\mathrm{min}^2}}
\newcommand{\Asym}[1]   {\mathrm{Asym}^{(+,-)}\left(#1\right)}
\newcommand{\FlatAsym}[1]
                        {\mathrm{Asym}^{(+,-)}(#1)}
\newcommand{\DAsym}[1]  {\mathrm{DAsym}^{(+,-)}\left({#1}\right)}
\newcommand{\rec}       {{\mathrm{(rec.)}}}
\newcommand{\smear}     {{\mathrm{(smr.)}}}
\newcommand{\true}      {{\mathrm{(true)}}}

\newcommand{\DeltaPM}   {\Delta_{(+,-)}}
\newcommand{\lumi}      {{\mathcal{L}}}
\newcommand{\lumilow}   {{\mathcal{L}_\mathrm{low}}}
\newcommand{\lumihigh}  {{\mathcal{L}_\mathrm{high}}}
\newcommand{\evts}      {{\mathrm{events}}}

\newcommand{\es}        {\varepsilon}

\newcommand{\granularity}
                        {{\Delta\eta\times\Delta\phi}}

\def\slashii#1{\setbox0=\hbox{$#1$}            
  \dimen0=\wd0                                 
  \setbox1=\hbox{\sl/} \dimen1=\wd1            
  \ifdim\dimen0>\dimen1                        
     \rlap{\hbox to \dimen0{\hfil\sl/\hfil}}   
     #1                                        
  \else                                        
     \rlap{\hbox to \dimen1{\hfil$#1$\hfil}}   
     \hbox{\sl/}                               
  \fi}

\def\slashiv#1{#1\llap{\sl/}}

\definecolor{rltbrightred}{rgb}{1,0,0}
\definecolor{rltred}{rgb}{0.75,0,0}
\definecolor{rltdarkred}{rgb}{0.5,0,0}
\definecolor{rltbrightgreen}{rgb}{0,0.75,0}
\definecolor{rltgreen}{rgb}{0,0.5,0}
\definecolor{rltdarkgreen}{rgb}{0,0,0.25}
\definecolor{rltbrightblue}{rgb}{0,0,1}
\definecolor{rltblue}{rgb}{0,0,0.75}
\definecolor{rltdarkblue}{rgb}{0,0,0.5}

\definecolor{webred}{rgb}{0.5,.25,0}
\definecolor{webblue}{rgb}{0,0,0.75}
\definecolor{webgreen}{rgb}{0,0.5,0}

\definecolor{Black}{rgb}{0,0,0}
\definecolor{White}{rgb}{1,1,1}
\definecolor{VeryDarkRed}{rgb}{0.3,0,0}
\definecolor{VeryDarkGreen}{rgb}{0,0.3,0}
\definecolor{VeryDarkBlue}{rgb}{0,0,0.3}
\definecolor{DarkRed}{rgb}{0.45,0,0}
\definecolor{DarkGreen}{rgb}{0,0.45,0}
\definecolor{DarkBlue}{rgb}{0,0,0.45}
\definecolor{DarkCyan}{rgb}{0,0.45,0.45}
\definecolor{Red}{rgb}{1,0,0}
\definecolor{Green}{rgb}{0,1,0}
\definecolor{Blue}{rgb}{0,0,1}
\definecolor{vert2}{rgb}{0,0.75,0}
\definecolor{MyBlue}{rgb}{0,0.20,.70} %
\definecolor{Grey}{rgb}{0.45,0.45,0.45}
\definecolor{MyRed}{rgb}{0.70,0.,0.}%
\definecolor{hl}{rgb}{
                0.909803922,       
                0.82745098,               
                0.909803922}
\definecolor{tg}{rgb}{0.80,0.70,1.}
\definecolor{sg}{rgb}{0.90,0.85,1.}
\definecolor{fg}{rgb}{0.95,0.92,1.}
\definecolor{ex}{rgb}{0.89,0.71,0.33} 
\definecolor{Greymax}{rgb}{0.65,0.65,0.65}
\definecolor{Greycen}{rgb}{0.75,0.75,0.75}
\definecolor{Greymin}{rgb}{0.85,0.85,0.85}
\definecolor{DarkMagenta}{rgb}{0.75,0,0.75}
\definecolor{Gold}{rgb}{1.0,0.7,0.0}
\definecolor{White}{rgb}{1.0,1.0,1.0}
\definecolor{DarkOrange}{rgb}{1,0.3,0}
\definecolor{LightOrange}{rgb}{1,0.53,0}

\newcommand{\white}{\color{White}}

\begin{document}

\thispagestyle{empty}
\begin{center}

\Large{Laboratoire de Physique Nucl\'eaire et des Hautes \'Energies}\\

\vspace*{6.cm}

\huge{
\textbf{Strategies for precision measurements \\
of the charge asymmetry of the $\mathbf{W}$~boson mass \\
at the LHC within the ATLAS experiment}
}\\

\vspace*{3cm}

\Large{by}\\

\vspace*{0.5cm}

\Large{\textbf{Florent Fayette}}\\

\vspace*{1cm}

\Large{presented to the}\\
\vspace*{0.2cm}
\Large{Universit\'e de Paris VI --  Pierre~et~Marie~Curie}\\
\vspace*{0.2cm} 
\Large{for the degree of \emph{Docteur de l'Universit\'e Paris VI}} \\

\end{center}
\clearpage
\thispagestyle{empty}
{\white .}
\clearpage
\thispagestyle{empty}
This Ph.D. thesis was supervised by Dr.~Mieczys{\l}aw Witold~Krasny
at the LPNHE institute in Paris and it was eventually
submitted to the Universit\'e de Paris VI - Pierre et Marie Curie
to obtain the degree of \emph{Docteur de l'Universit\'e Paris VI} on January $16^\mm{th.}$ 2009
in front of a committee consisting of\:
Prof.~Michael Joyce (President of the committee),
Prof.~Stefan Tapprogge (Referee),
Prof.~Fabian Zomer (Referee),
Dr.~Thomas LeCompte,
Dr.~Gavin Salam and 
Dr.~Mieczys{\l}aw Witold~Krasny.

\vspace*{4.cm}
\begin{center}
\begin{minipage}[t]{0.8\tw}

\begin{center}
\Large{\textbf{Abstract}}\\
\end{center}
This thesis dissertation presents a prospect for a measurement of 
the charge asymmetry of the $W$~boson mass ($\MWp-\MWm$) at the LHC 
within the ATLAS experiment.
This measurement is of primordial importance for the LHC experimental program,
both as a direct test of the charge sign independent coupling of the $W$ bosons
to the fermions and as a mandatory preliminary step towards the precision measurement 
of the charge averaged $W$ boson mass. This last pragmatic point can be understood 
since the LHC specific collisions will provide unprecedented kinematics for 
the positive and negative channels while the SPS and Tevatron collider produced $\Wp$ and 
$\Wm$ on the same footing.
For that reason, the study of the asymmetries between $\Wp$ and $\Wm$ in Drell--Yan like processes 
(production of single $W$ decaying into leptons), studied to extract the properties of the $W$ boson, 
is described thoroughly in this document.

Then, the prospect for a measurement of $\MWp-\MWm$ at the LHC is addressed in a
perspective intending to decrease as much as possible the systematic errors that will inevitably
comes from the misunderstanding of both phenomenological and apparatus modeling.
For that matter strategies have been devised specifically for the present measurement 
to display robustness with respect to the main uncertainties.
These strategies consist of introducing new observables along with considering
specific LHC running modes and configurations of the ATLAS tracker.

Eventually we show that the present (2009) precision can be improved at the LHC by a factor of $20$
and argue that such a precision is beyond the reach of the standard measurement and 
calibration methods imported to the LHC from the Tevatron program.
\end{minipage}

\end{center}

\clearpage
\thispagestyle{empty}
{\white J'en ai plein le QCD.}
\clearpage

\thispagestyle{empty}
\begin{center}
{\Large \textit{Remerciements, podziekowania, acknowledgements, etc.}}
\vspace{0.5cm}\\
\end{center}

Mes premiers remerciements s'adressent \`a ceux qui ont {\oe}uvr\'e pour que j'obtienne 
une bourse pour r\'ealiser ma th\`ese au LPNHE, \`a savoir Anne-Marie Cazabat, 
Matteo Cacciari, Jean-Eudes Augustin, Philippe Schwemling et Mieczys{\l}aw Witold Krasny 
mon directeur de th\`ese.
En particulier je tiens \`a exprimer ma reconnaissance \`a Pascal Debu, Philippe Schwemling et 
Witold Krasny pour avoir dispens\'e avec parcimonie toutes injonctions pour rattraper
ce l\'eger retard que j'accusais ce qui m'aura permis de parachever ma th\`ese en toute qui\'etude.
Enfin je tiens \`a remercier Stefan Tapprogge et Fabian Zomer pour avoir accept\'e d'en \^etre les rapporteurs
et pour avoir studieusement pris en consid\'eration mon travail pendant leurs vacances
de No\"el. Cette gratitude s'adresse par ailleurs aux autres membres du jury, Michael Joyce,
Tom LeCompte, Gavin Salam et mon directeur.

Je remercie ce dernier, Mieczys{\l}aw Witold Krasny, pour m'avoir offert l'opportunit\'e 
de faire ma th\`ese avec lui et surtout pour avoir su composer, \`a la faveur de mes capacit\'es et 
de mes aspirations, un sujet de th\`ese idoine \`a ma d\'elicate personne. 
Par bien des aspects travailler sous sa direction sur le pr\'esent sujet fit largement honneur au 
travail de recherche tel que l'on peut se le figurer dans l'id\'eal. 
Ces remerciements ne sauraient rendre justice \`a ce que furent ces trois derni\`eres ann\'ees
en sa compagnie si je ne devais mentionner sa sympathie, sa bienveillance et pour ne m'avoir
jamais mis sous pression favorisant en cela le plaisir dans le travail \`a la 
production de r\'esultats \'equivoques.
Enfin l'originalit\'e du personnage, \`a mi-chemin entre le diable de Tasmanie et le ``road-runner'',
son g\'enie ainsi que son humour ont fait de cette collaboration une exp\'erience unique.

Just like to fire there is water, or to vodka there is mi\'od pitny to Witold Krasny 
there is Wies{\l}aw P{\l}aczek.
On many aspects I consider Wies{\l}aw P{\l}aczek as my second supervisor for all of his 
helps which improved the quality of my work over these last years.
I learned a lot from his rigor, his infallible skills in informatics and I am indebted
for all the multiple times he helped me using the \WINHAC{} Monte Carlo.
Besides I always appreciated his zen attitude and I also do not forget that most of my stays
in Krak\'ow would not have been possibly that smooth if it had not been for his attentions and 
his hospitality. Za wszystko dzi\c{e}kuj\c{e} Ci polski mistrzu Zen.

I am deeply grateful to Giorgos Stavropoulos whose help and support, while fighting with Athena,
allowed me to reach my first concrete results in my thesis.
His broad sympathy really made of this task an agreeable moment, which, in regard of what Athena
is, stands as a huge compliment.

Jestem wdzi\c{e}czny Andrzejowi Si\'odmokowi. Indeed, Andrzej brought equilibrium to tackle the
Krasny--P{\l}aczek doublet which would have been hard to tame all by myself 
(especially the M.W.K. component), besides I am grateful to him for his sympathy and his 
hospitality in Krak\'ow.
I also do not forget Katarzyna Rejzner whose recent arrival in our team brought additional fun  
and more importantly allowed to decrease the amount of krasnic activity Andrzej and I used
to deal with.

Je remercie les membres du groupe ATLAS du LPNHE pour leur accueil et plus pr\'ecis\'ement Fr\'ed\'eric Derue
pour s'\^etre montr\'e toujours disponible pour m'aider lors de mes confrontations
non-ASCII avec Athena ainsi qu'Emmanuel Busato dont les aides, m\^eme apr\`es son d\'epart pour 
Clermont--Ferrand, auront grandemement contribuer \`a am\'eliorer la qualit\'e de mon code.
Je tiens aussi \`a rendre hommage \`a Emmanuel ``Mini Manu'' Hornero, J\'er\^ome ``Jers'' Glisse et 
Rui Pereira pour avoir toujours pris le temps de s'int\'eresser aux probl\`emes
informatiques sur lesquels je butais. Je double mes louanges \`a l'endroit de Rui et Jers dont
les aides en Matplotlib auront contribuer \`a am\'eliorer la facture de ce document.
Je triple mes louanges \`a Mini Manu pour tout le temps qu'il m'aura accord\'e pour r\'esoudre les
obstacles techniques qui se sont dress\'es sur mon chemin.
Enfin remerciements particuliers \`a Pietro Cavalleri, J\'er\'emie Lellouch et Jers (une fois de plus).
Pietro, gentilshomme d'Italie dedans Paris, avec qui j'ai eu l'opportunit\'e de passer de
bons moments grace \`a son humour et un caract\`ere unique.
J\'er\'emie, du groupe des z\'eros du LPNHE, pour nos \'echanges d\'esinvoltes voire 
ouvertement oisifs qui, \`a l'aune de nos travaux, s'av\'er\`erent salutaires.
Jers enfin qui, outre toutes les aides en informatiques, m'aura prodigu\'e \`a maintes reprises 
l'occasion de rire de bon c{\oe}ur.

I thank the members of the ATLAS group of the Institute of Nuclear Physics in Krak\'ow
for their hospitality during my several stays amongst them, in particular Jolanta Olszowska and 
Janusz Chwastowski.
I am grateful to the nice fellows I met in Krak\'ow with whom I shared nice
moments especially Zofia Czyczula, Ewa Stanecka, Adam Matyja, Sebastian Sapeta, Hayk and Meri.
I have also a thought for all the poles from CERN especially Justyna and Janek.
Enfin de mani\`ere plus g\'en\'erale je rends hommage au peuple Polonais pour son accueil et
son savoir-vivre. 
Depuis les charmantes polonaises callipyges aux vieillards vout\'es par les ans
en passant par les vendeurs ambulants de pierogi et de vodka c'est toujours avec grande courtoisie, 
force d\'elicatesse et patience non feinte que l'on m'aura consid\'er\'e.
Ce faisant, tous mes s\'ejours \`a Cracovie, sans exception, furent autant de pauses ``civilisation'' 
qui auront \'emaill\'ees les vicissitudes inh\'erentes \`a la vie parisienne.

I would like to greet the people from the CERN Standard Model forum especially, 
Lucia DiCiacco, Nathalie Besson, Stefan Tapprogge, Thomas Nunneneman, Maarten Boonekamp and Troels Petersen.
Special thanks to Tom LeCompte for sharing his experience with a guenuine vivid and communicative enthusiasm.
I am also grateful to the following people with whom I had positive exchanges\,: David Rousseau,
Andrea Dell'Acqua, Wojciech Wojcik, Pawe{\l} Br\"uckman de Restrom, Mike Whalley, Borut Kersevan, 
Kristin Lohwasser, Fred Olness, Muge Karagoz Unel, Chris Hay, Ashutosh Kotwal, Mark Lancaster 
and Arkadiuz Weinstein.

Je suis reconnaissant \`a Benoit Loiseau pour son attention quant au bon d\'eroulement de ma th\`ese et 
je l'associe aussi \`a ma gratitude envers Bertrand Desplanques tous deux m'ayant permis de trouver asile 
scientifique \`a Grenoble lorsque je m'\'etais mis en d\'elicatesse avec une partie du milieu acad\'emique parisien.
Outre le sauvetage d'un avenir potentiel dans les sciences, les explications de Bertrand Desplanques 
au cours de mon stage sous sa direction m'auront apport\'ees beaucoup dans mon approche actuelle vis-\`a-vis
de la physique des particules.

Je remercie aussi mes coll\`egues du LPTHE\,:
Matteo Cacciari, gran maestro di quark pesanti, pour le stage sous sa direction et
pour m'avoir toujours fait profiter des ses explications limpides pendant le d\'ebut de ma th\`ese quand 
mon esprit penait \`a se calibrer sur le mode de pens\'ee polonais, 
Gavin Salam pour ses aides efficaces et pour sa relecture de mon manuscrit
enfin Bruno Machet pour sa grande bont\'e, son enthousiasme et toutes ces agr\'eables discussions qui, 
faisant \'echo aux discours de mes autres mentors, m'auront permis d'\'elargir mon ouverture d'esprit.
Enfin une mention particuli\`ere \`a R\'edamy Perez-Ramos pour ses aides et ses encouragements. 

Je m'incline devant Souad Rey, V\'eronique Joisin, Jocelyne Meurguey et Annick Guillotau pour leur
efficacit\'e dans l'organisation de mes d\'eplacements comme dans d'autres taches administratives.

Jestem wdzieczny mojej ma{\l}ej siostrze Ani ``Pani Ruda'' Kaczmarska i 
Mariusz Bucki whose sympathy brought me a lot during these last years especially when hustling
my way through krasnic mazes or when I was feeling hyper-weak rather than electroweak. 
I cherish all these nice moments I shared with them in Krak\'ow and thank them for the 
basics in polish they learned me, short, but yet enough to get myself into trouble.

Je rends un hommage particulier \`a Pascal Parneix dont la passion pour la physique et l'\'energie mise 
\`a son service auront marqu\'e une partie de mes \'etudes universitaires, pour ce stage bien sympathique 
pass\'e sous sa direction ainsi que pour ses conseils et ses aides dans les moments difficiles.

Je remercie Alain ``Docteur Folamour'' Mazeyrat pour son soutien dans les passages \`a vide et 
salue sa dext\'erit\'e qui sauva une carte m\`ere dont la perte aurait grandement fait d\'efaut \`a mon travail.
Merci aussi \`a Jean On\'esippe pour son aide, son \'ecoute et pour avoir distill\'e un peu de sa grande
sagesse antillaise dans mon esprit inquiet.
Je suis reconnaissant \`a Olivier Destin pour m'avoir aid\'e bien des fois en C++ et en \LaTeX{} avec 
une efficacit\'e hors du commun. Enfin encore plus important je le remercie pour tous nos \'echanges
passionants, divertissants ou hilarants qui m'auront permis de m'a\'erer l'esprit.

Pour conclure je remercie infiniment mes parents et ma s{\oe}ur pour leur soutien au cours de mes 
longues ann\'ees d'\'etudes sans lequel je n'aurais pas pu mat\'erialiser mes aspirations.
Chcia{\l}bym tak\.ze podzi\c{e}kowa\'c mojej ma{\l}ej i subtelnej Paulince za jej 
czu{\l}o\'s\'c i ciep{\l}o oraz za wszystkie sp\c{e}dzone razem chwile, kt\'ore tak 
bardzo pomog{\l}y mi przetrwa\'c trudny czas przed obron\c{a} pracy.

\thispagestyle{empty}

\cleardoublepage
\tableofcontents{}

\frontmatter
\chapter{Introduction}\label{chap_introduction}
\setlength{\epigraphwidth}{0.7\tw}
\epigraph{
L\`a trois cent mille personnes trouv\`erent place et brav\`erent pendant plusieurs heures une 
temp\'erature \'etouffante, en attendant l'arriv\'e du Fran\c{c}ais.
De cette foule de spectateurs, un premier tiers pouvait voir et entendre\,;\;un second tiers 
voyait mal et n'entendait pas\,;\;quant au troisi\`eme, il ne voyait rien et n'entendait pas 
davantage. Ce ne fut cependant pas le moins empress\'e \`a prodiguer ses applaudissements.
}%
{\textit{De la Terre \`a la Lune} \\\textsc{Jules Vernes}}

The actual description of the fundamental building blocks of matter and the interactions ruling
them is called the Standard Model. It is believed not to be the most fundamental description
but rather a phenomenological approximation for energies below the TeV scale.
In this model, two types of particles are to be distinguished. 
First are the quarks and leptons building up the matter and, second, are the bosons that mediate the 
interactions among them. Of all the four known fundamental interactions --the gravitation, 
electromagnetic, weak and strong interactions-- only the last three are now implemented within the 
mathematical formalism supporting the Standard Model. 
Amid these interactions the weak one, acting on both quarks and leptons, is mediated by the exchange
of the massive neutral $Z$ \index{Z boson@$Z$ boson} and two charged $\Wpm$ bosons.
The $\Wp$ and $\Wm$, which are antiparticle of each other, are the object of interest in this
dissertation.

The $Z$ boson has been observed in 1973 in the Gargamelle bubble chamber at CERN while the 
$\Wpm$ bosons were observed in single production in 1983 in the UA1 detector of the Super Proton 
Synchrotron (SPS) \index{SPS collider} $\ppbar$ collider, again at CERN.
This discovery confirmed the Glashow--Weinberg--Salam electroweak model.
Since then, the $\Wpm$ have been studied for the last decades at the Large Electron Positron (LEP) 
\index{LEP collider}\index{Tevatron collider|(}
$\ep\Em$ collider ($\Wp\Wm$ pair production) and at the Tevatron $\ppbar$ collider (single $W$ production).
These two experiments allowed to measure $W$ properties such as its mass $\MW$ or its width 
$\GamW$. These parameters are important since they provide, when combined to other Standard Model parameters, 
constraints on the Standard Model. For example the masses of the $W$ and of the top quark constrain
the mass of the hypothetical Higgs boson.
The specific $W$ study of this work aims at improving the experimental value of the 
charge asymmetry of the $W$ mass by studying single $W$ production at the Large Hadron Collider (LHC).
This measurement has so far not received much attention and, as a consequence, displays an accuracy
10 times larger than the one on the absolute mass $\MW$.
With the new possibilities that the LHC $\pp$ collider should offer for the next years we considered the 
prospect for a drastic decrease of the experimental error on the $\MWp-\MWm$ value using the
ATLAS multipurpose detector capabilities.
The first motivation for such a measurement is to refine the confirmation of the $CPT$ invariance 
principle \index{Symmetry!CPT@$CPT$} through the direct measurement of the $\Wpm$ masses, to complete 
the accurate $CPT$ test made by observing charged $\mup$ and $\mum$ life time decays. 
Other motivations will be given later.
Besides, as it will be shown, the $W$ bosons, despite the fact they will be produced with the same 
process as at the SPS \index{SPS collider} and the Tevatron colliders, will nonetheless display 
original kinematics due to the nature of the colliding beams. 
Indeed, while $\Wp$ and $\Wm$ are produced on the same footing in $\ppbar$ collision, this will not be 
the case anymore at the LHC.
The first step of this work is to understand the $\Wp$ and $\Wm$ kinematics in $\pp$ collisions.
After, this first stage providing all cards in our hands to understand the $W$ properties at the LHC, 
the rest of this work will focus on the improvement that could be provided to the $\MWp-\MWm$ measurement
using the ATLAS detector at the LHC.
Here, rather than reusing Tevatron tactics, we devised new strategies specific to this measurement
and to the LHC/ATLAS context.
\index{Tevatron collider|)}
The philosophy for these strategies --as it will be detailed-- aims at being as independent as 
possible of both phenomenological and experimental uncertainties, that cannot be fully controlled.
Eventually, we argue that the proposed strategies, could enhance the actual
accuracy on $\MWp-\MWm$ by a factor of $\approx 20$.

This work represents the second stage of a series of several publications aiming at providing
precision measurement strategies of the electroweak parameters for the upcoming LHC era. 
In the same logic, next steps will provide dedicated
strategies for the measurement of the absolute mass and width of the $W$ boson.

This work is the result of three years of collaboration in a team consisting of Mieczys{\l}aw Witold Krasny,
Wies{\l}aw P{\l}aczek, Andrzej Si\'odmok and the author.
Furthermore, the technical work that took place within the ATLAS software was made possible with 
substantial help from Giorgos Stavropoulos.

\mainmatter
\chapter{Phenomenological context and motivations}\label{chap_theo}
\setlength{\epigraphwidth}{0.7\tw}
\epigraph{
\quad- Depuis lors, continua Aramis, je vis agr\'eablement. 
J'ai commenc\'e un po\`eme en vers d'une syllabe\,;\;c'est assez difficile, 
mais le m\'erite en toutes choses est dans la difficult\'e. 
La mati\`ere est galante, je vous lirai le premier chant, 
il a quatre cents vers et dure une minute.\\
\quad- Ma foi, mon cher Aramis, dit d'Artagnan, qui d\'etestait presque autant les vers que le latin, 
ajoutez au m\'erite de la difficult\'e celui de la bri\`evet\'e, et vous \^etes s\^ur au moins que votre 
po\`eme aura deux m\'erites.}%
{\textit{Les Trois Mousquetaires}\\ \textsc{Alexandre Dumas}}

\index{QCD!Quarks|see{Quarks}}
\index{Valence Quarks|see{Quarks}}
\index{Sea Quarks|see{Quarks}}
\index{LHC!Luminosity|see{Luminosity}}
\index{Tevatron!Luminosity|see{Luminosity}}
\index{Drell--Yan processes for W@Drell--Yan processes for $W$|see{$W$ boson}}
\index{W boson@$W$ boson!Helicity|see{Helicity}}
\index{Monte Carlo!Pythia@\Pythia{}|see{\Pythia{}}}
\index{Monte Carlo!WINHAC@\WINHAC{}|see{\WINHAC{}}}
\index{Charged lepton@Charged lepton from $W$ decay!Helicity|see{Helicity}}
\index{Quarks!Helicity|see{Helicity}}

\index{Transverse momentum!Of the charged lepton@Of the charged lepton from $W$ decay|see{Charged lepton from $W$ decay}}
\index{Transverse momentum!Of the quarks|see{Quarks}}
\index{Transverse momentum!Of the W boson@Of the $W$ boson|see{$W$ boson}}
\index{Rapidity!Of the W boson@Of the $W$ boson|see{$W$ boson}}
\index{Pseudo-rapidity!Of the charged lepton@Of the charged lepton from $W$ decay|see{Charged lepton from $W$ decay}}

This Chapter introduces the background of the present work.
The first part reviews in a nutshell the Standard Model which is the present paradigm to describe
the elementary particles and their interactions.
Then, a parenthesis is made on the experimental setting to already provide to the reader the 
global vision necessary to understand the rest of the Chapter.
For this purpose, the hadronic production of $W$ bosons and how their properties are extracted from leptonic 
decays are reviewed.
Next, motivations for the measurement of the $W$ mass charge asymmetry are given.

The second part introduces the notations and conventions used throughout the document.

The third part describes the phenomenological formalism used to study $W$~bosons produced in 
hadronic collisions and decaying into leptons, phenomenon also known as production of $W$ in 
Drell--Yan-like processes. 
This presentation, rather than being exhaustive, emphasises the relevant 
kinematics needed to understand the gist of $W$ physics in Drell--Yan-like processes and how,
from such kinematics, the $W$ properties like its mass and its width are extracted. 

Finally the Chapter closes on a presentation of the rest of the document.

\section{The Standard Model in a nutshell}
\index{Standard Model|(}
\subsection{Overview}
Based on the experience that seemingly different phenomena can eventually be interpreted with 
the same laws, physicists came up with only four interactions to describe 
all known physics processes in our Universe. 
They are the gravitational, electromagnetic, weak and strong interactions.
The electromagnetic, weak and strong interactions are described at the subatomic level by 
Quantum Field Theory (QFT), the theoretical framework that emerges when encompassing the features 
of both Special Relativity and Quantum Mechanics. 
In particular the  weak and electromagnetic interactions are now unified in QFT into the 
Glashow--Weinberg--Salam electroweak theory.
The description of the three interactions in QFT is called the Standard Model of elementary particles and their 
interactions, or simply Standard Model (SM).
The ``Standard'' label means that it is the present day reference, which,
although not believed to be the ultimate truth, is not yet contradicted by data.
This amends for the term ``phenomenology'' used in some applications in the SM, where phenomena 
are described with non fundamental models and, hence, non fundamental parameters.
Gravity, on the other hand, is not yet implemented in QFT, it is described by General Relativity. 
It mostly concerns Cosmology, that rules the behaviour of space--time 
geometry under the influence of massive bodies and in non-inertial frames.
It explains the structure of the Universe and its components on large scales, and eventually leads to
the Big Bang theory.
In the Standard Model, the effects of gravity are negligible as long as the energies stay below the 
Planck scale ($<10^{19}\GeV$).

\begin{table}[]
\begin{center}
\renewcommand\arraystretch{1.45}
\begin{tabular}{|c|l@{\kern\tabcolsep}>{\kern-\tabcolsep}ll@{\kern\tabcolsep}>{\kern-\tabcolsep}ll@{\kern\tabcolsep}>{\kern-\tabcolsep}lc|}
  \hline
   Fermions                & \multicolumn{2}{c}{\cellcolor{fg}{$1^{\mm{st}}$ generation}} 
                           & \multicolumn{2}{c}{\cellcolor{sg}{$2^{\mm{nd}}$ generation}} 
                           & \multicolumn{2}{c}{\cellcolor{tg}{$3^{\mm{nd}}$ generation}} & \cellcolor{ex}{$Q$}    \\
\hline\hline
  \multirow{2}{*}{Quarks}  &\cellcolor{fg}{$u$}    &\cellcolor{fg}{(up)}                &\cellcolor{sg}{$c$}     &\cellcolor{sg}{(charm)}          &\cellcolor{tg}{$t$}      &\cellcolor{tg}{(top)}               & \cellcolor{ex}{$+2/3$} \\
                           &\cellcolor{fg}{$d$}    &\cellcolor{fg}{(down)}              &\cellcolor{sg}{$s$}     &\cellcolor{sg}{(strange)}        &\cellcolor{tg}{$b$}      &\cellcolor{tg}{(bottom)}            & \cellcolor{ex}{$-1/3$} \\
\hline
\hline
\renewcommand\arraystretch{1.45}
  \multirow{2}{*}{Leptons} &\cellcolor{fg}{$\Em$}  &\cellcolor{fg}{(electron)}         &\cellcolor{sg}{$\mum$}   &\cellcolor{sg}{(muon)}           &\cellcolor{tg}{$\tau^-$}   &\cellcolor{tg}{(tau)}                 & \cellcolor{ex}{$-1$}   \\
                           &\cellcolor{fg}{$\nu_e$}&\cellcolor{fg}{(electron neutrino)}
                                                                                       &\cellcolor{sg}{$\nu_\mu$} & \cellcolor{sg}{(muon neutrino)}
                                                                                                                                                   & \cellcolor{tg}{$\nu_\tau$}&\cellcolor{tg}{(tau neutrino)}      & \cellcolor{ex}{\;\;\,$0$}    \\
\hline
\end{tabular}
\caption[The three generations of quarks and leptons]
          {\figtxt{The three generations of quarks and leptons. $Q$ is the electrical charge.}}
          \label{tab_quarks_leptons}
\renewcommand\arraystretch{1.45}
\end{center}
\end{table}
Before presenting the Standard Model in more details we present an overview of the particles
properties and denominations. At this stage, we already adopt natural units where $c=1$ and $\hbar=1$.
Table~\ref{tab_quarks_leptons} displays the elementary (\ie{} structureless) particles building the matter.
They are fermions\footnote{Fermion is the generic term used to qualify all particles whose spin 
is a ``half-value'' of the Planck constant, \ie{} $n\,\tfrac{1}{2}$ in units of $\hbar$, 
$n$ being an integer. Boson on the other hand qualifies particles whose spin 
is an integer of the Planck constant.} of spin $S=1/2$ and
come in two types, the quarks and the leptons, both present in three generations of doublets\,:
this leads to six different flavours of quarks or leptons.
The only thing that differentiates each generation is the masses of the particles~\cite{Amsler:2008zz}.
Quarks have fractional electrical charge --with respect to the charge $e$ of the electron-- 
and are sensitive to all interactions. 
Charged leptons are sensitive to the weak and electromagnetic interactions, while neutral leptons
(neutrinos) are only sensitive to the weak interaction.
Interactions among all fermions are due to the exchange of elementary particles
of boson type. The electromagnetic interaction is mediated by the exchange of neutral massless photons $\gamma$ 
between all particles that have an electrical charge. 
The weak interaction is mediated by three massive vector bosons\,: the electrically charged $\Wp$ and $\Wm$
and the electrically neutral $Z$. 
All particles bearing a weak isospin charge are sensitive to the weak interaction.
The strong interaction between particles having a color charge  is mediated by massless 
colored gluons $g$ of eight different kinds.

The quarks and leptons were discovered on a time scale that span no less than a century.
The electron was observed at the end of the $19^\mm{th}$ century~\cite{Thomson,Kauf1,Kauf2} 
while the muon and tau were observed
respectively in the thirties~\cite{PhysRev.51.884} and the seventies~\cite{PhysRevLett.35.1489}. 
The electron neutrino $\nu_e$ was found in the fifties~\cite{Reines:1953pu},
the $\nu_\mu$ ten years later~\cite{Danby:1962nd} and the $\nu_\tau$ after another forty years 
later~\cite{Kodama:2002dk}.
The up, down and strange quarks were observed in hadrons in the deep inelastic electron--nucleon
and neutrino--nucleon scattering experiments~\cite{RevModPhys.63.597,Eichten:1973cs}. 
The discovery of the charmed quark occurred in the
seventies~\cite{PhysRevLett.33.1404,Augustin:1974xw}, the bottom was discovered a few years 
later~\cite{Herb:1977ek}, while the top, due to its very large mass, was discovered in the nineties
at the Tevatron $\ppbar$ collider~\cite{Abe:1995hr,Abachi:1995iq}.
Concerning the gauge bosons, the photon was discovered with the theoretical interpretation of
the photo-electric effect~\cite{Einstein:1905cc} while the $\Wpm$ and $Z$ bosons were isolated in $\ppbar$ 
collisions~\cite{Arnison:1983rp,Banner:1983jy,Arnison:1983mk,Bagnaia:1983zx}.
The existence of gluons was deduced from the observation of hadronic jets generated in 
$\ep\Em$ collisions at high energies~\cite{Brandelik:1979bd,Barber:1979yr,Berger:1979cj}.

Each particle mentioned so far has its own anti-particle with opposite quantum numbers. 
Anti-particles of electrically charged particles are noted with the opposite sign of the charge 
(\eg{} $\ep$ is the anti-particle of $e^-$)
while neutral particles are noted with a bar upon them (\eg{} $\anul$ is the anti-particle of $\nul$).
Some particles are their own anti-particles, \eg{} the photon and the $Z$ boson.
The rest of this section presents the main features of the Standard Model.
First, a glimpse at QFT is given, then we describe the electromagnetic, strong and weak 
interactions in QFT.

\subsection{The theoretical background of the Standard Model\,: Quantum Field Theory}
\index{Quantum Field Theory|(}
The beginning of the $20^\mm{th}$ century witnessed the emergence of two revolutions in physics\,:
Quantum Mechanics and Special Relativity.
Quantum Mechanics deals with phenomena below the atomic scale while 
Special Relativity, based on space--time homogeneity/isotropy,
describes laws of transformations between inertial frames.
Exploring deeper the subatomic world, the Heisenberg principle entails that,
as the length scale decreases, the momentum (energy) increases. 
This increase of the energy amends for the High Energy Physics (HEP) terminology used 
to refer to the elementary particles and their interactions.
In those conditions, taking into account Special Relativity is mandatory.

The Klein--Gordon and Dirac equations were the first relativistic generalisations of the 
Schr\"odinger equation. The problem with these equations is that they have negative energies
solutions which are difficult to interpret. This problem can be overcome if one considers that
a negative energy solution describes an anti-particle. 
An anti-particle can be interpreted as a positive energy particle travelling backward in time.
Besides, in this relativistic and quantum context, a solution describes a quantum field $\psi$ 
whose elementary excitations are the particles. 
Then, the number of particles is no longer fixed and the relativistic mass--energy equivalence 
accounts for the annihilation/creation of pairs of particle--anti-particle. 
This new framework is the Quantum Field Theory.

QFT describes local interactions of supposedly point-like particles. It bears its own 
difficulties, with the arising of infinities, which need to be consistently 
taken care of in the process of renormalisation.
Taking into account local Lorentz invariant QFT with the spin--statistic theorem leads to 
the $CPT$ symmetry, \ie{} the conservation the product of the charge conjugation $C$, 
parity $P$ and time inversion $T$ operators in any processes.
In particular, $CPT$ symmetry shows that $\MWp=\MWm$.

One method to quantify a field is to use the Feynman path integral. 
Path integral formulation of Quantum Mechanics relies on a generalisation of the least action 
principle of classical mechanics.
It is based on the Lagrangian density from which the equations of the dynamics can be deduced.
The Lagrangian density (called hereafter Lagrangian) of the field
under study contains its kinetic and potential energies.
Hence, the description of the dynamics of elementary particles under the influence of fundamental
forces consists to formulate the right Lagrangian with\,: 
(1) the kinetic energy of the free fields of the spin-$\tfrac{1}{2}$ quarks/leptons and of the 
spin-$1$ bosons mediators and (2) the potential term displaying their mutual interactions and 
--if any-- the bosonic field self-interaction terms.
A term of interaction is proportional to a coupling constant, say $g_\mm{int.}$, inherent to the 
interaction under study. 
Up to the fact that this constant is small enough, the calculus of the amplitude of probability 
can be developed into a perturbative expansion in powers of the coupling constant $g_\mm{int.}$.
All terms proportional to $g^n_\mm{int.}$ in the expansion involve processes with $n$ interaction
points (vertices) between quarks/leptons and bosons. We can associate to them drawings, 
called Feynman diagrams. Each one entangles all processes sharing the same topological representation 
in momentum space.

Finally let us mention that QFT is also extensively used in condensed matter physics, in particular in
many-body problems, and the interplay of ideas between this domain and HEP is very rich.
With the basics features of QFT presented above we now describe the electromagnetic, strong and 
weak interactions.
\index{Quantum Field Theory|)}

\subsection{Quantum Electrodynamics}
\index{QED!Basics|(}
The theory that describes the electromagnetic interactions in QFT is called Quantum Electrodynamics
(QED). The Lagrangian of QED includes the Dirac and Maxwell Lagrangians respectively to describe the 
kinetic energy of, say, the free electron field $\psi$ (its excitation are the electrons)
and of the free gauge field $A^\mu$ (its excitation are the photons). 
It also contains an interaction term between the photon and electron fields which is
proportional to $e$, the  charge of the electron. 
The charge $e$ is linked $\alpha$, the fine structure constant of QED, via $\alpha\equiv e^2/4\,\pi$.
Since $\alpha\approx 1/137$ the expansion in power of $\alpha$ is feasible. 
Still, when taking into account Feynman diagrams with loops, \ie{} quantum fluctuations involving 
particles with arbitrary four-momenta, the calculus diverges. 
This reflects the conflict between the locality of the interactions inherited from Special 
Relativity and Quantum Mechanics that allows virtual processes to have arbitrary high energies. 
Divergences come from terms like $\ln(\Lambda/m)$, where $\Lambda$ is a temporary unphysical 
cut-off parameter and $m$ the mass of the electron.

The procedure to get rid of those divergences is called renormalisation.
It consists in redefining the fundamental parameters of the theory by realising that
the charge $e$ used so far is already the one resulting from all quantum loops. 
These $\ep\Em$ loops affect the value of the bare charge $e_0$ that one would observe if there 
was no interaction. 
Expressing in a perturbative series $e$ as a function of $e_0$ and $\Lambda$ and, then,
writing in the expression of the amplitude $e_0$ as a function of $e$ allows to get rid 
of the divergences when $\Lambda\to+\infty$.
The effect of these loops depends on the energy, for that reason 
a renormalisation scale energy $\mu_r$  is chosen close to a characteristic energy scale 
in the process.
Since $\mu_r$ is arbitrary the charge must obeys an equation expressing the invariance of 
physics with respect to $\mu_r$. This leads to the formulation of the renormalisation equations.
Eventually the $\ep\Em$ loops screen the bare charge.
Then, the charge $e$ and mass of the electron, in our example, are no longer fundamental parameters 
but rather effective parameters (running coupling constants) which values depends of $\mu_r$.

Gauge invariance is an essential tool in proving the renormalisability of a gauge theory.
For QED, it appears that equations are invariant under a local transformation of the 
phase factor of the field $\psi$. 
The invariance of the Lagrangian under this particular transformation directly governs the 
properties of the electromagnetic interaction. Group theory allows to describe these properties, in the
case of QED this group is $U(1)_\mm{QED}$.
Consequent to those developments, physicists tried to describe the remaining interactions 
with the help of gauge invariant theories, too.
\index{QED!Basics|)}

\subsection{Quantum Chromodynamics}
\index{QCD!Basics|(}
The description of the strong interaction by a gauge theory became sensible when the
elementary structure of the hadrons was discovered in lepton--hadron deep inelastic 
scattering (DIS) experiments. Those experiments shed light on the partons, the hypothetical
constituents of the hadrons, which become quasi free at high energies.
The partons are of two types, the quarks and the gluons~\cite{GellMann:1964nj,Zweig:1964pd}
and, at first, the structure of the hadrons was found to satisfy some scaling properties.

The observation of hadrons made quarks of the same flavour and spin
hinted at the existence of a new quantum number attached to quarks that could eventually help to
anti-symmetrise the wave functions of such hadrons, and save the Fermi exclusion principle.
It was then postulated that the group describing the strong interaction is 
$SU(3)_\mm{c}$, where c stands for color.
Indeed, in analogy with the additive mixing of primary colors, quarks hold three primitive 
``colors'' (``red'', ``green'' and ``blue'') that, when combined, give ``white'' hadrons,
which justifies the term of chromo in Quantum Chromodynamics (QCD), 
the QFT model of the strong interaction~\cite{Gross:1973ju,Gross:1974cs,Politzer:1973fx}.
This group is non-abelian which pragmatically implies that the interactions carriers,
the gluons, are colored and can interact among themselves, in addition to interacting
with quarks. There are eight colored gluons to which we can associate the 
gauge field $G_a^\mu$ where $a=1,2,\dots 8$.

This last property implies striking differences with respect to QED.
In addition to gluon--quark vertices, equivalent to photon-lepton/quark vertices in QED,
there are now three and four gluons vertices.
During the renormalisation of the strong interaction coupling constant $g_\mm{s}$,
while the quarks loops screen the color charge, the gluons loops magnifies it and eventually
dominate.
As a consequence $g_\mm{s}$, or equivalently $\alphas\equiv g^2_\mm{s}/4\,\pi$, is a decreasing function
of the renormalisation energy $\mu_r$.
This justifies \textit{a posteriori} the asymptotic freedom hypothesis for the partons in DIS, 
and also, qualitatively, the confinement of quarks in hadrons.
Hence, at some point, when the energy involved in a QCD process becomes small, $\alphas$
gets large which prohibits the use of perturbative calculations. The energy scale marking this
frontier is of the order of $0.1\GeV$.
Below this limit, other techniques have to be employed to make calculations (\eg{} lattice QCD).
Note, finally, that the change of $\alphas$ with the energy implies that the formerly observed
scaling property must be violated, which was observed and formalised in particular with
DGLAP equations~\cite{Gribov:1972ri,Altarelli:1977zs,Dokshitzer:1977sg}.
Note also that the renormalisation constraints impose to have as many quarks as there are leptons
which twice allowed to predict new quarks\,: the charm to match, with the $u$, $d$ and $s$,
the two first leptons families and eventually the top and bottom quark doublet when the tau lepton 
was discovered.
\index{QCD!Basics|)}

\subsection{Electroweak interactions}

\subsubsection{Weak interactions and the path to the electroweak model}
The weak interaction was discovered in nuclear $\beta$--decay and
recognised to drive as well the muon decay.
The first model was a point-like interaction involving charged currents (to account for the
change of flavor) and proportional to the Fermi constant $\GF$.
The peculiarity of this interaction is that it violates parity, leading to the
vector minus axial-vector ($V-A$) nature of the weak currents.
Hence, weak interactions couple only left-handed particles and right-handed 
anti-particles.

The problem of the Fermi model is that it collapses above $\approx 300\GeV$ by giving inconsistent
predictions. Using unitarity constraints, one can shape the form of the more fundamental model
ruling the weak interactions.
First, based on the previous examples of QED and QCD, the weak interaction can be assumed to be
mediated by the exchange of heavy charged vector bosons. In this new context the propagator of
the $W$ boson damps the rise with energy. Assuming that the coupling strength
of the $W$ to fermions is comparable to the one of QED, one finds that the mass of the 
$W$ must be around $100\GeV$. Hence, due to the mass of the $W$ the range of the 
weak interaction is of the order of $1/\MW$.

In addition to charged currents the theory needs neutral currents as well.
Let us consider, for example, the process $\ep\,\Em\to\Wp\,\Wm$ that occurs with an exchange of a
neutrino. When the $W$ bosons are produced with longitudinal polarisation their
wave functions grows linearly with the energy while the exchange of a neutrino predicts
a growth quadratic with the energy.
To overpass this new breakdown of unitarity a neutral boson, the $Z$, must take part in the
process like $\ep\,\Em\to Z\to\Wp\,\Wm$ with a tri-linear coupling to the $\Wp$ and $\Wm$.

A consequence of this tri-linear coupling is that, now, scattering of vector 
bosons can be observed. In the case of $\Wp\,\Wm\to \Wp\,\Wm$ the amplitude of longitudinally polarised
$W$ bosons, built-up by the exchange of $Z$, grows as the fourth power of the energy in the center 
of mass of the collision.
This leading divergence is canceled by introducing a quadri-linear coupling among the weak 
bosons. Still, unitarity is not yet restored for asymptotic energies since the amplitude still grows
quadratically with the energy.

At this stage, two solutions can be envisaged. The scattering amplitude can be damped, either 
by introducing strong interactions between the $W$ bosons (technicolor model), or by introducing 
a new particle, the scalar Higgs boson $H$, which interferes destructively with the exchange of 
weak bosons.

We have seen that the coupling strength of the weak and electromagnetic interactions are
of the same order. Besides we have seen that the road to preserve unitarity is 
very much linked to the handling of longitudinally polarised $W$ bosons.
We now give a brief presentation of the electroweak model that
unifies electromagnetic and weak interactions.

\subsubsection{The electroweak model}
Since weak interactions couple only left particles and right anti-particles, finding 
a gauge theory to cope with this implies that all elementary particles have to be considered, 
in a first approximation, to be massless.
The path to the electroweak model can be presented in two steps. 
In the first step, the Glashow--Weinberg--Salam (GWS) model~\cite{Glashow:1961tr,Salam:1964ry,Weinberg:1967tq} 
unifies the weak and electromagnetic interactions 
up to the approximation that all quarks, leptons and vector bosons are massless, QCD can be 
added along to the GWS model, still with massless particles.
The second step describes the mechanism that allows massive particles 
without explicitly breaking the gauge invariance constructed earlier. This is the 
Brout--Engler--Higgs--Kibble mechanism~\cite{Englert:1964et,Higgs:1964ia,Guralnik:1964eu}.

The fact that elementary fermions were found in doublets and the desire to unify the weak
and electromagnetic interactions in a gauge invariant theory led to the gauge groups 
$SU(2)_L\otimes U(1)_Y$.

The $SU(2)_L$ takes into account the fact that left-handed fermions
are found in weak isospin doublet, \ie{} $I=1/2$ which, for the conventionally used third component, 
means $I_3=\pm 1/2$, while right-handed fermions are found in singlets.
Let us note also that the left-handed quarks eigenstates in weak interactions differ from the mass eigenstates\,;
the convention tends to write down the lower isospin states $q'$ as linear combinations of the eigenstate
masses $q$, like $q'_i=\sum_j V_{ij}\,q_j$, where $i$ and $j$ are the flavors of the quarks and the 
elements $V_{ij}$ are the elements of the Cabibbo--Kobayashi--Maskawa (CKM)~\cite{Cabibbo:1963yz,Kobayashi:1973fv}
\index{Electroweak!CKM matrix elements}
$3\times 3$ matrix.
The non-abelian properties of the $SU(2)$ group ensures that some mediators of the 
interaction will have a charge and then will be assimilable to the $W$ bosons. 
At this stage, there are 3 fields, $W^{1,\mu}$, $W^{2,\mu}$ and $W^{3,\mu}$ that couples to 
particles with a weak isospin with a coupling constant $g$.

The group $U(1)_Y$, although different from $U(1)_\mm{QED}$, is chosen so that, eventually, some of
its components will give back QED. It governs the interaction of weakly hypercharged particles $Y$ 
coupled to a gauge field $B^\mu$ via a coupling constant $g'$.

In this context one imposes the electric charge $Q$ of a particle to be linked to the
weak isospin and hypercharge via the Gell-Mann--Nishijima equation\,: $Q=\tfrac{1}{2}Y+I_3$.
Things can then be recast in terms of electric charge to purposely make the 
$W$ bosons appear\,; from the weak hypercharge and isospin terms we end up with three terms respectively
displaying positively, negatively and neutrally charged currents.
In this new basis the $W$ bosons fields are linear combinations of the $W^{1,\mu}$ and $W^{2,\mu}$.
We are left with a neutral term involving the gauge fields $B^\mu$ and $W^{3,\mu}$.
These fields, though, are not yet the photon and the $Z$.
In fact, the latter appear to be admixtures of $B^\mu$ and $W^{3,\mu}$.
The angle that governs this mixture is called the Weinberg angle $\Theta_\mm{W}$.
Weak and electromagnetic interactions are now unified and described by three parameters\,:
how they mix via $\Theta_\mm{W}$ and their individual coupling strength $\GF$ and $\alpha$.

Like mentioned previously QCD can be added such that three gauge theories finally account for
the three interactions, \ie{} $SU(3)_\mm{c}\otimes SU(2)_L\otimes U(1)_Y$.
Nonetheless this model presents a few flaws. 
The problem of the unitarity above $300\GeV$ is partially fixed by introducing massive vector
bosons but the unitarity still breaks down around $1\TeV$ in $\Wp\,\Wm\to \Wp\,\Wm$,
where the $W$ are polarised longitudinally. 
Last but not least, some elementary particles are massive but 
if one enters mass terms explicitly in the equations the gauge invariance is broken. 
To keep the gauge invariance properties of the GWS model and take into account the masses a
mechanism that spontaneously breaks the symmetry of the solutions had to be devised.

The solution was directly inspired by condensed matter physics, more precisely, from
supra-conductivity where photons can become massive due to the non-symmetric
fundamental state of the scalar field ($S=0$) describing electrons pairs.
Here, the electroweak symmetry breaking (ESB) is realised by adding to the previous model a Higgs 
scalar field which, to respect $SU(2)_L\otimes U(1)_Y$, comes as a weak isospin doublet 
with an hypercharge. 
A part of the potential for the Higgs field is chosen so that the vacuum energy is degenerate.
This complex Higgs doublet makes a total of four degrees of freedom.
Three of these degrees mix with the $\Wpm$ and $Z$ bosons and provide them with a third
longitudinal spin state which makes them massive while the remaining one becomes the massive Higgs boson. 
The fermions acquire their masses via Yukawa couplings with the Higgs field.
The Higgs boson is the last missing piece of the present day formulation of the Standard Model.
If it exists and if it perturbatively interacts, its experimental observation 
will validate the Standard Model.

\subsection{Summary}
We briefly presented the quantum and relativistic framework for the dynamics of the 
quarks and leptons sensitive to the electromagnetic, weak and strong interactions, as well as
how massive particles gain their masses dynamically by interacting with the Higgs scalar field.
The Standard Model has a highly predictive power in its actual form but
its main problem still lies in the understanding of the origin and nature of the masses of the
particles. Although the ESB mechanism accounts for them while respecting the gauge 
invariance, its addition is mainly \textit{ad hoc}. 

This summary of the Standard Model did not had the pretension to be exhaustive.
In particular, the wide variety of experiments, such as hadron--hadron, lepton--nucleon or 
electron--positron colliders, that provided essential results, were not credited to keep this 
presentation as short as possible.
As a consequence the reader is invited to refer to classical textbooks with more details and 
references to historical papers\,:
for QFT/SM in order of accessibility 
Refs.~\cite{CohenTannoudji,AitchisonAndHey1,AitchisonAndHey2,Ryder,Peskin:1995ev}, for details in 
QCD Ref.~\cite{Ellis:1991qj} and in the Electroweak Model Ref.~\cite{Spiesberger:2000ks} for example.

\index{Standard Model|)}

\section{The W mass charge asymmetry in the Standard Model}
\subsection{A first overview from the experimental point of view}
The present work takes place in the context of collider physics, more specifically
within the Large Hadron Collider (LHC) that should accelerate in a large ring counter-rotating 
hadrons --most of the time protons-- and make them collide at several interaction points with an
energy $\sqrt S$ in the center of mass of $\sqrt{S}=14\TeV$.\index{LHC} 
The observation of the particles emerging from these hadronic collisions is achieved by several 
detectors located in the vicinity of the interaction points. 
Among them is the ATLAS detector whose capabilities were used in this analysis.
The LHC and the ATLAS detector will be described in more details in the next Chapter.
Also, worth mentioning, is the Tevatron which, for the last decades before the LHC, has been the 
largest circular accelerator. The Tevatron collider accelerates \index{Tevatron collider} 
counter rotating protons and anti-protons at energies in the center of mass of $\sqrt{S}\approx 2\TeV$.

Amid all the difficulties entering such experimental analysis, two are to be noticed.
The first, inherent to high energy physics, is that most of the exotic particles 
cannot be observed directly due to their short life time but are rather detected indirectly 
from the observations of their decays displaying specific kinematics.
The second difficulty, specific to hadronic processes, is that from the theoretical point of view 
physicists speak in terms of quarks and gluons but from the experimental point of view only hadrons 
and their respective decays --if any-- can be observed. 
When colliding hadrons this last problem is unavoidable due to the nature of the initial state.

In hadronic collisions the extraction of the $W$ and $Z$ bosons properties are made
studying their leptonic decays which display a distinguishable signal due to the high energy leptons 
in the final state.
This, in particular, allows to get rid of the problems inherent to QCD in the final state.
Let us remark the decay into the tau channel is not considered as the short life time of the
latter makes it not ``directly'' observable in a detector.

\index{Z boson@$Z$ boson}
In the case of the  $Z$ boson, where $Z\to\lp\,\lm$, things are simple as the direct observation of the
two charged leptons gives access, via their invariant mass $m^2_{\lp\lm}$, to the invariant mass of the $Z$.
In the case of the $W$ the presence of a neutrino in the final state complicates things a lot more. 
Indeed, because the neutrino does not interact with any part of the detector its kinematics can
be only deduced from the overall missing energy for a given event, which will never be as precise
as a direct measurement.
Multipurpose detectors instruments are more precise in the transverse direction of the beam axis, 
since this is where interesting physics with high $\pT$ particles occur. 
The presence of the beam-pipe leaves the very forward region less hermetic in term of calorimetry, 
forbidding eventually to measure the longitudinal component of the neutrino.
This leaves then only the transverse momenta of the two leptons to extract the $W$~boson 
mass. As it will be shown in this Chapter, the shapes of these transverse momenta depends on
$W$ properties such as its mass $\MW$ and its width $\GamW$.
Nonetheless, because here we cannot access any Lorentz invariant quantities, the kinematics of the $W$ 
boson and of the leptons in the $W$ rest frame needs to be known with accuracy
to model precisely enough the observed kinematics of the leptons in the laboratory frame.
Hence the extraction of the $W$ properties proves to be a real challenge from both phenomenological 
and experimental point of views.

\subsection{Motivations for a measurement of the W mass charge asymmetry}
\index{Symmetry!CPT@$CPT$|(}
\index{W boson@$W$ boson!Mass charge asym@Mass charge asymmetry $\MWp-\MWm$|(}
As demonstrated by Gerhard L\"uders and Wolfgang Pauli~\cite{Pauli}, any Lorentz-invariant quantum 
field theory obeying the principle of locality must be $CPT$-invariant.
For theories with spontaneous symmetry breaking, the requirement of the Lorentz-invariance concerns
both the interactions of the fields and the vacuum properties. In $CPT$-invariant quantum field 
theories, the masses of the particles and their antiparticles are equal.

The Standard Model is $CPT$-invariant. In this model, the $\Wp$ and $\Wm$ bosons are constructed  
as each own antiparticle, which couple to leptons with precisely the same $SU(2)$ strength $g_{W}$.
The hypothesis of the exact equality of their masses is pivotal for the present understanding of 
the electroweak sector of the Standard Model.\index{Electroweak}
It is rarely put in doubt even by those who consider the $CPT$ invariant Standard Model as 
only a transient model of particle interactions.
However, from a purely experimental perspective, even such a basic assumption must be checked 
experimentally to the highest achievable precision. \index{Symmetry!CPT@$CPT$|)}

The most precise, indirect experimental constraint on equality of the $\Wp$ and $\Wm$ masses can be
derived from the measurements of the life time asymmetries of positively and negatively charged 
muons~\cite{Bailey:1978mn,Bennett:2004pv,Amsler:2008zz}. 
These measurements, if interpreted within the Standard Model framework, verify the equality of the 
masses of the $\Wp$ and $\Wm$ bosons to the precision of $1.6\MeV$. 
Such a precision cannot be reached with direct measurements of their mass difference.
The experimental uncertainty of the directly measured mass difference from the first CDF run and
reported by the Particle Data Group~\cite{Amsler:2008zz} is $\MWp-\MWm=-190 \pm 580\MeV$, 
\ie{} about $400$ times higher. 
More recently the CDF collaboration~\cite{Aaltonen:2007ps} measured $\MWp - \MWm$ to be 
$257 \pm 117\MeV$ in the electron decay channel and $286 \pm 136\MeV$ in the muon decay 
channel.
The Table~\ref{table_cdf_mw_mwp_mwm} sums up the measured values at CDF for the last decades
for both $\MW$ and $\MWp-\MWm$ at the time being.
\index{W boson@$W$ boson!Mass@Mass $\MW$!In CDF}
\index{W boson@$W$ boson!Mass charge asym@Mass charge asymmetry $\MWp-\MWm$!In CDF}
Note that in this thesis the experimental measurement will be focused on CDF results since up
to this day it is the only collaboration that published experimental values for $\MWp-\MWm$.
\begin{table}[]
\begin{center}
\renewcommand\arraystretch{1.45}
\begin{tabular}{lr@{$\,\pm\,$}lr@{$\,\pm\,$}lc}
  \hline
  Channel               & \multicolumn{2}{c}{$\MW$ [$\mm{GeV}$]} & 
  \multicolumn{2}{c}{$\MWp-\MWm$ [$\mm{GeV}$]} & Year\\
  \hline\hline
  $W\to l\,\nul$        & $79.910$&$0.390$           & $-0.190$&$0.580$ & 
  1990,1991 \cite{Abe:1990pp,Abe:1990tq}\\
  \hline
  $W\to \mu\,\numu$     & $80.310$&$0.243$             & $0.549$ & $0.416$ & 
  \multirow{3}{*}{1995 \cite{Abe:1995np,Abe:1995nm}} \\
  $W\to e\,\nue$        & $80.490$&$0.227$             & $0.700$&$0.290$ &\\
  $W\to l\,\nul$        & $80.410$&$0.180$           & $0.625$&$0.240$ &\\
  \hline
  $W\to \mu\,\nu_\mu$   & $80.352$&$0.060$         & $0.286$&$0.152$ &  
                                  \multirow{3}{*}{2007 \cite{Aaltonen:2007ypa,Aaltonen:2007ps}}\\
  $W\to e\,\nu_e$       & $80.477$&$0.062$         & $0.257$&$0.117$ &\\
  $W\to l\,\nul$        & $80.413$&$0.048$       
& \multicolumn{2}{l}{\qquad\quad\;$\huge{\times}$}\\
 \hline
\end{tabular}
\caption[Sum up of the measured values of $\MW$ and $\MWp-\MWm$ with the CDF detector for the last decades
(1990 $\to$ 2007)]
          {\figtxt{Sum up of the measured values of $\MW$ with the CDF detector for the last decades 
              (1990 $\to$ 2007).
              Each of result is obtained for the considered collected data in each
              publication, \ie{} with no combinations with previous results from CDF or other experiments.
              The two references next to the year indicate\,: (1) the results and (2) the detailed 
              $W$ mass analysis.
              The errors are the one obtained when adding up quadratically the statistical and 
              systematic errors.}}

\renewcommand\arraystretch{1.45}
        \label{table_cdf_mw_mwp_mwm}
        \index{W boson@$W$ boson!Mass charge asym@Mass charge asymmetry $\MWp-\MWm$!In CDF}
        \index{W boson@$W$ boson!Mass@Mass $\MW$!In CDF}
        \index{CDF detector!W properties@$W$ properties measurements|see{$W$ boson}}
\end{center}
\end{table}

\index{Tevatron collider|(}
These measurements provide to this date the best model independent verification of the equality of 
the masses of the two charge states of the $W$~boson. They are compatible with the charge symmetry 
hypothesis. It is worth stressing, that the present precision of the direct measurement of the 
charge averaged mass of the $W$~boson derived from the combination of LEP \index{LEP collider} 
and Tevatron results and under the assumption that $\MWp=\MWm$ is $\MW=80.398 \pm 0.025\GeV$.
It is better by a factor $10$ than the precision of the direct individual measurements of the 
masses of its charged states. 
 
In top of the obligatory precision test of the $CPT$-invariance \index{Symmetry!CPT@$CPT$}
of the spontaneously broken 
gauge theory with \textit{a priori} unknown vacuum properties, we are interested to measure $\MWp-\MWm$ 
at the LHC for the following three reasons.
Firstly, we wish to constrain the extensions of the Standard Model in which the effective coupling
of the Higgs particle(s) to the $W$~boson depends upon its charge. Secondly, contrary to the 
Tevatron case, the measurement of the charge averaged mass at the LHC cannot be dissociated from, 
and must be preceded by the  measurement of the masses of the $W$~boson charge states.
Therefore, any effort to improve the precision of the direct measurement of the charge averaged mass
of the $W$~boson and, as a consequence, the indirect constraint on the mass of the Standard Model
Higgs boson, must be, in our view, preceded by a precise understanding of the $W$~boson charge 
asymmetries. Thirdly, we would like to measure the $W$~boson polarisation asymmetries at the LHC. 
Within the Standard Model framework the charge asymmetries provide an important indirect access to 
the polarisation asymmetries. 
This is a direct consequence of both the $CP$ conservation \index{Symmetry!CP@$CP$} in the gauge 
boson sector and the purely ($V-A$)-type \index{Electroweak!VmA@$V-A$ coupling} 
of the conjugation ($C$) and parity ($P$) violating coupling of the $W$~bosons to fermions. 
Any new phenomena contributing to the $W$~boson polarisation asymmetries at the LHC must thus be 
reflected in the observed charge asymmetries.

The optimal strategies for measuring the charge averaged mass of the $W$~boson and for measuring 
directly the masses  of its charge eigenstates are bound to be different. Moreover, the optimal 
strategies are bound to be different at the LHC than at the Tevatron. 

At the Tevatron, the nature of the colliding beam makes it so the production of $\Wp$ is the same
than $\Wm$ up to a $CP$ \index{Symmetry!CP@$CP$} transformation. 
Then, producing equal numbers of the $\Wp$ and $\Wm$ bosons,
the measurement strategy was optimised to achieve the best precision for the charge averaged mass of
the $W$~boson. 
For example, the CDF collaboration~\cite{Aaltonen:2007ps} traded off the requirement of the 
precise relative control of the detector response to positive and negative particles over the full 
detector fiducial volume, for a  weaker requirement of a precise relative control of charge averaged
biases of the detector response in the left and right sides of the detector.
Such a strategy has provided the best up to date measurement of the charge averaged $W$~boson mass, 
but large measurement errors of the charge dependent $W$~boson masses as seen above in 
Table~\ref{table_cdf_mw_mwp_mwm}.
More detailed explanations on this experimental development are given in Chapter~\ref{chap_atlas_exp}
Appendix~\ref{cdf_tracker}.

If not constrained by the beam transfer systems, the best dedicated, bias-free strategy of 
measuring of $\MWp - \MWm$ in proton--anti-proton colliders would  be rather straightforward. 
It would boil down to collide, for a fraction of time, the direction interchanged beams of protons 
and anti-protons, associated with a simultaneous change of the sign of the solenoidal $B$-field in 
the detector fiducial volume. Such a measurement strategy cannot be realised at the Tevatron leaving
to the LHC collider the task of improving the measurement precision. 

The statistical precision of the future measurements of the $W$~boson properties at the LHC 
will be largely superior to the one reached at the Tevatron. Indeed, where this error was of 
$\delta_\MW^\mm{(stat.)}=34\MeV$ at the Tevatron for an integrated luminosity $L$ of 
$200\,\mm{pb}^{-1}$~\cite{Aaltonen:2007ps}\index{Luminosity!Integrated!Used in CDF II MW analysis@Used in CDF II $\MW$ analysis}
at the LHC, for the same measurement, in just one year 
of $\pp$ collisions at low luminosity ($L=10\,\mm{fb}^{-1}$) the statistical error should approximately reach 
$\delta_\MW^\mm{(stat.)}\approx 5\MeV$.
On the other hand, it will be difficult to reach comparable or smaller systematic errors. 
At the Tevatron they equalise to the statistical error, \ie{} $\delta_\MW^\mm{(sys.)}=34\MeV$ 
while at the LHC the primary goal is, to constrain the hypothesised Higgs mass, to reach 
$\delta_\MW^\mm{(sys.)}=15\MeV$.
The measurements of the $W$~boson mass and its charge asymmetry can no longer be factorised and 
optimised independently.
The flavour structure of the LHC beam particles will have to be controlled with a significantly 
better precision at the LHC than at the Tevatron. While being of limited importance for the  
$\MW$ measurement at the Tevatron, \index{W boson@$W$ boson!Mass@Mass $\MW$!At the Tevatron}
the present knowledge of the momentum distribution asymmetries 
of\,:\;(1) the up and down valence quarks and (2) of the strange  and charm  quarks in the proton 
will limit significantly the measurement precision.   
The `standard candles', indispensable for  precise experimental control of the reconstructed lepton
momentum scale -- the $Z$~bosons and other ``onia'' resonances -- will be less powerful for 
proton--proton collisions than for the net zero charge proton--anti-proton collisions. 
Last but not least, the extrapolation of the strong interaction effects measured in the $Z$~boson 
production processes to the processes of $W$~boson production will be more ambiguous due to an 
increased contribution of the bottom and charmed quarks. 

Earlier studies of the prospects of the charge averaged $W$~boson mass measurement by the 
CMS~\cite{Buge:2006dv} and by the ATLAS~\cite{Besson:2008zs} collaborations ignored the above 
LHC collider specific effects and arrived at rather optimistic estimates of the achievable 
measurement precision at the LHC.\index{W boson@$W$ boson!Mass@Mass $\MW$!At the LHC (prospects)}
In our view, in order to improve the precision of the Tevatron experiments, both for the average and
for the charge dependent masses of the $W$~boson, some novel, dedicated strategies, adapted to the 
LHC environment must be developed. \index{Tevatron collider|)}
Such strategies will  have to employ full capacities of the collider and of the detectors in the aim
to reduce the impact of the theoretical, phenomenological and measurement uncertainties on the 
precision of the $W$~boson mass measurement at the LHC.
\index{W boson@$W$ boson!Mass charge asym@Mass charge asymmetry $\MWp-\MWm$|)}

\section{Notations and conventions}\label{notations_conventions}
In this section notations and conventions used through out the rest of the document are
introduced.

Let us remind to the reader that to simplify analytic expressions the natural unit 
convention $c=1$ and $\hbar=1$ is adopted.
In this notation energies, masses and momenta are all expressed in electron-Volt (eV).
Nonetheless, although every MKSA units can be expressed in powers of eV,
cross sections --noted $\sigma$-- are expressed in powers of barns, where 
$1\,\mm{barn}\equiv 10^{-28}\,\mm{m}^2$.
Especially, unless stated otherwise, all differential cross sections $\flatDfDx{\sigma}{a}$ for a 
scalar observable $a$ produced in this work are all normalised to $\mm{nb}/[A]$, 
$A$ being the dimension of the observable $a$.

Both Cartesian and cylindrical coordinate basis are considered in the inertia laboratory frame.
They are defined already with respect to the experimental apparatus.
The interaction point, located in the center of the ATLAS detector, corresponds to the origin of
both coordinate systems.
In the Cartesian basis, colliding hadrons move along the $z$ axis, $+y$ points upward and $+x$ to 
the center of the LHC accelerating ring.
In the cylindrical basis, $r$ is the radius in the $x-y$ plane,
$\phi$ the azimuthal angle with respect to the $+x$~direction,
and $\theta$ the polar angle with respect to the $+z$~direction.
Unit vectors along these different directions are noted $\vec e_i$ where $i$ can stands for
$x$, $y$, $z$, $r$ or $\phi$. 
The components of a vector observable $\vec b$ along the axis $i$ is noted $b_i$.
All angles are expressed in radians unless stated otherwise.

To define the most relevant kinematics variables to collider physics we consider
the example of a particle which four-momentum, energy, momentum and absolute momentum are noted 
respectively $p$, $E$, $\vec p$ and $|\vec p|$.
The four-momentum of a particle in its co-variant representation and in the Cartesian 
basis writes
\begin{align}
p = \left( \begin{array}{c}
   E \\
\vec p
\end{array}\right)
     =
\left(\begin{array}{c}
   E    \\
   p_x  \\
   p_y  \\
   p_z  
\end{array} \right), \label{eq_four_mom}
\end{align}
where the energy $E$ and the momentum $\vec p$ of the particle are related to its invariant mass $m$
by the relation $E^2=\vec{p}^{\,2} +m^2$.
The form of Eq.~(\ref{eq_four_mom}) implies the same conventions for the time and space 
components of any other kind of four-vectors.
The helicity $\lambda$ of a particle is defined by the projection of its spin $\vec S$ \index{Helicity!Definition}
against the axis pointing in the same direction than the momentum of the particle
$\vec p$, that is analytically 
\begin{equation}
\lambda \equiv \vec S \cdot \frac{\vec p}{|\vec p|}.
\end{equation}
The transverse component $\pT$ in the $r-\phi$ plane of the vector $\vec p$ is defined by
\begin{equation}
p_T \equiv \sqrt{p_x^2+p_y^2}.\label{eq_def_pT}\index{Transverse momentum!Definition}
\end{equation}
In that notation transverses energies and missing transverse energies are written $E_T$ and $\ETmiss$.
In ATLAS, and other multipurpose detectors, the central tracking sub-detector bathes in a solenoidal 
magnetic field $\vec B= |\vec B|\,\vec e_z$. In this context, the transverse curvature $\rho_T$, 
defined as the projection of a particle track on the $r-\phi$ plane, is related to the particle's 
transverse momentum via
\begin{equation}
\rho_T \equiv 1/\pT.
\end{equation}
A particle kinematics can be unequivocally described by its azimuth $\phi$, its transverse momentum
$\pT$ and its rapidity $y$ defined by 
\begin{equation}
y \equiv \frac{1}{2}\,\ln\left(\frac{E+p_z}{E-p_z}\right)\index{Rapidity!Definition}.
\end{equation}
The rapidity is additive under Lorentz transformations along the $z$ axis.
For massless/ultra-relativistic particles the rapidity equals the pseudo-rapidity $\eta$
which is related to the particle polar coordinate by the following relation
\begin{equation}
\eta\equiv -\ln\left(\tan(\theta/2)\right)\label{eq_def_eta}\index{Pseudo-rapidity!Definition}.
\end{equation}
Finally to understand more deeply some important physics aspects it is better to consider them
in the $W$ Rest Frame (\WRF{}) rather than in the laboratory frame (LAB) where fundamental dynamical 
patterns are blurred by the add-up of the effects of the $W$ boosts.
Variables considered in the \WRF{} are labeled with a $\ast$ superscript while the one with no 
particular sign are to be considered in the laboratory frame.
More details on relativistic kinematics can be found in Ref.~\cite{Amsler:2008zz}.

Another useful variable that will be extensively used is the charge asymmetry,
which, for a scalar $a$, is noted $\Asym{a}$ and defined like
\begin{equation}
    \Asym{a} \;\equiv\; 
    \frac
        { \flatDfDx{\sigma^\plus}{a} - \flatDfDx{\sigma^\minus}{a} }
        { \flatDfDx{\sigma^\plus}{a} + \flatDfDx{\sigma^\minus}{a} },
        \label{eq_def_charge_asym}
        \index{Charge asymmetry!Definition} 
\end{equation}
where the $\plus$ and $\minus$ refers to the electrical charge of the particle under consideration.
Finally when the electrical charge is not made explicit it means we consider indifferently both positive and
negative particles, \ie{} hereafter $W\equiv \Wpm$ and $l\equiv\lpm$.

Different levels of understanding for the observables are considered, respectively 
the ``true'', the ``smeared'' and the ``reconstructed'' levels.
\index{True level|see{Monte Carlo}}
The true level, also called particle or generator level,\index{Particle level|see{Monte Carlo}}
\index{Generator level|see{Monte Carlo}} refers to the phenomenological prediction of a model or, 
to be more precise, to the best emulation a given Monte Carlo simulation can produce.
\index{Monte Carlo!Generator/particle/true level}
\index{Smeared level}
The smeared level refers to the true level convoluted with the finite resolution
of a detector. For an observable $a$, the link between the smeared distribution 
$\mm{Smear}(a)$ and the true one $\mm{True}(a)$ is
\begin{equation}
\mm{Smear}(a) \equiv \int d\,t\;\mm{True}(t-a)\;
                               \mm{Res}(t;\;a),
\end{equation}
where $\mm{Res}(t;\;a)$ is the function governing the response of the detector to an input of value
$t$ for the observable $a$.
Here, the general resolution performances of a detector will be usually given using rough Gaussian 
parametrisation
\begin{equation}
\mm{Res}(t;\;a)=\frac{1}{\sqrt{2\,\pi}\;\sigma_a}\mm{e}^{(-t^2/2\,\sigma_a^2)},
\end{equation}
where the variance $\sigma_a$ characterises the resolution of the detector for the observable $a$ 
of the considered particle.
\index{Reconstructed level}
Finally the collected data suffer from additional degradations coming from
misalignment, miscalibration, limited accuracy of algorithms, \etc{}. 
All those effects concur to give in the end a reconstructed observable deviating from the smeared 
value a perfect detector would provide. 
From the purely experimental point of view only the reconstructed level is relevant but
the intermediate levels are used for both pedagogical and pragmatic purposes.

\section{Generalities on the production of W boson in Drell--Yan like processes}
This section starts with a short presentation of the decay of both unpolarised and polarised spins states 
of a real $W$ boson which leads to the computation of its width $\GamW$.
These derivations will prove to be useful afterward and it allows 
as well to remind some basics related to the electroweak interaction.
After, the whole process $\mm{hadron}-\mm{hadron}\to W \to l\,\nul$ is presented.
The goal is to give an intuitive comprehension of the kinematics relevant to the $W$ production. 
Among these kinematics is the charged lepton transverse momentum that is used in the present 
document for extracting the mass of the $W$.

\subsection{W decay}\label{chap_phenoW_ssec_W_decay}
\subsubsection{Unpolarised $\BFW$~bosons}
\begin{figure}[!h] 
  \begin{center}
    \includegraphics[width=0.7\tw]{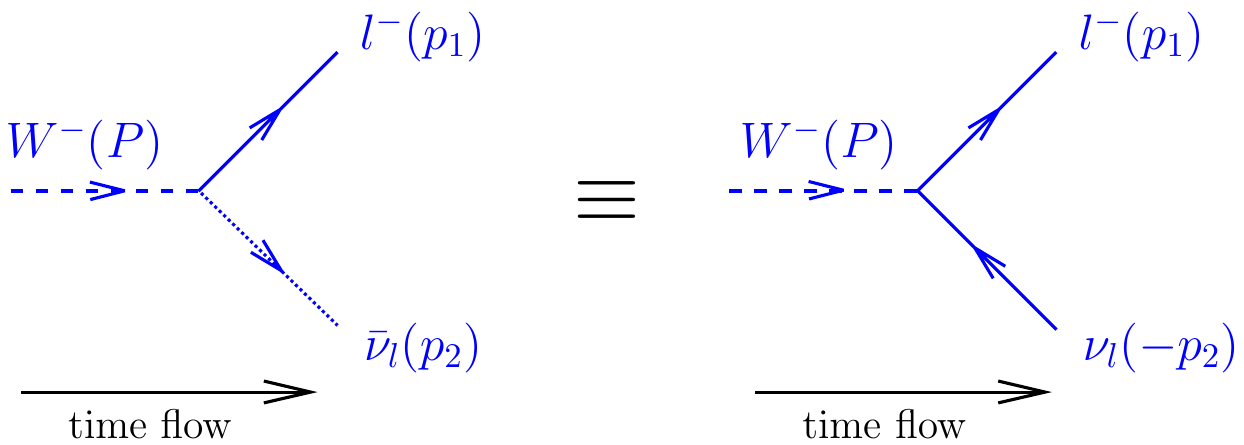}
    \caption[Feynman diagram of the leptonic decay of a $W$ boson at the Born level]
	    {\figtxt{Feynman diagram of the leptonic decay of a $W$ boson at the Born level.
            The diagram on the right represents the conventional way to handle anti-particles
            in perturbative calculation (see text for more details).}}
	    \label{fig_W_decay_born_feyn_diag}
  \end{center} 
\end{figure}

\index{W boson@$W$ boson!Width@Width $\GamW$|(}
The case of the leptonic decay of an unpolarised $W$~boson is considered through the
example of 
\begin{equation}
\Wm(P)\to\,\lm(p_1)\;\anul(p_2),
\end{equation}
where $P$, $p_1$ and $p_2$ are respectively the four-momenta of the $\Wm$ boson,
of the charged lepton $\lm$ and of the anti-neutrino $\anul$.
This process in the first perturbation order (Born level) is made of one Feynman diagram 
(Fig.~\ref{fig_W_decay_born_feyn_diag}). Here, and in all other Feynman diagrams, the
time flow goes from the left to the right.

The leptonic decay can be of any type (electronic, muonic or tau) since all leptons can be 
considered to be massless.
Indeed, in the $W$ Rest Frame ($W$RF) each charged lepton have an energy of the order of 
$E_l\sim \MW/2\approx 40\GeV$, hence the charged leptons have a Lorentz $\beta$ factor
of $\beta = 0.999\dots\sim 1$ making out of them ultra-relativistic 
particles\footnote{In fact for the case of the $\tau$, taking into account its mass
eventually leads after computation to affect the $W$ width at the level of $0.1\percent$, 
but we consider this to be negligible in the context of our discussion.}.
The amplitude of probability of this process is 
\begin{equation}
\mathcal{M}_\lambda = -i\,\tfrac{g}{\sqrt{2}}\;\es_\mu^\lambda(P)\, \usbar{p_1}\,
\HVmAup{\mu}\,\vs{p_2}.
\label{eq_w_lep_decay_prob_ampl}
\end{equation}
In this last equation, the weak force is embedded in the strength $g/\sqrt 2$ that \index{Electroweak!VmA@$V-A$ coupling|(}
couples particles sensible to the weak interaction and by the nature of this coupling represented
by the co-variant bi-linear term $\HVmAup{\mu}$ of Vector--Axial ($V-A$) form.
The coupling constant $g/\sqrt 2$ is usually expressed as a function of the Fermi constant $\GF$ 
that was formerly used when modeling weak interactions as contact interactions when occurring
at energies scales much lower than $\MW$. For historical reasons they are linked through the relation
\begin{equation}
\frac{\GF}{\sqrt 2} \equiv \frac{g^2}{8\,M_W^2}.\index{W boson@$W$ boson!Mass@Mass $\MW$}
\end{equation}
The $\Wm$ polarisation state of helicity $\lambda$ is represented by the four-vector 
$\es_\mu^\lambda(P)$, \index{Helicity!Of the W boson@Of the $W$ boson}
$\usbar{p_1}$ is the spinor of $\lm$ and $\vs{p_2}$ that of $\nul$.
We use here the convention which tends to consider that anti-fermions going forward
in time can be treated on the algebraic level as fermions going backward in time.
The spinors of backward traveling time fermions are noted $\vs{k}\equiv u(-k)$ for convenience.
Let us note in the $V-A$ term the operator $\tfrac{1}{2}(1-\gamma^5)$ which, 
applied to a spinor $u$, projects only its left-handed component $u_\mm{L}$
\begin{equation}
\tfrac{1}{2}(1-\gamma^5) \;u= u_\mm{L},
\end{equation}
while the operator $\tfrac{1}{2}(1+\gamma^5)$ projects the right component of the latter
\begin{equation}
\tfrac{1}{2}(1+\gamma^5)\;u = u_\mm{R}.
\end{equation}
In other words, the presence of the $V-A$ term in the current couples only left-handed fermions 
and right-handed anti-fermions in electroweak interactions.
\index{Electroweak!VmA@$V-A$ coupling|)}\index{Electroweak}

Usual ``diracologic'' calculus techniques allow to calculate 
$\overline{\left|\mathcal{M}_\lambda\right|^2}$, which corresponds to the Lorentz co-variant expression 
of the squared amplitude summed and averaged with respect to all 
possibles spins states for the $W$ and leptons. The differential decay rate writes in the \WRF{}
\begin{equation}
d\,\Gamma = \frac{1}{2\,\MW}\,\overline{\left|\mathcal{M}_\lambda\right|^2}\,
d\,\mm{Lips},
\end{equation}
where $d\,\mm{Lips}$, the Lorentz invariant phase space assuring energy-momentum conservation, 
can be reduced here to
\begin{equation}
d\,\mm{Lips} = \frac{1}{128\,\pi^2}\,\delta(E_l^\ast-{\MW}/{2})
\,d\,E_l^\ast\,d\,\Omega_l^\ast,
\end{equation}
where the energy $E_l^\ast$ and solid angle $\Omega_l^\ast$ are those of the charged lepton.
After the integration over the whole accessible phase space to the particles, the partial width 
$\Gamma\left(W\to\,l\,\nu_l\right)$ of this process reads
\begin{eqnarray}
\Gamma\left(W\to\,l\,\nu_l\right) &=& \frac{\GF}{\sqrt 2}\,\frac{M_W^3}{6\pi}, \\ 
                                  &\equiv& \Gamma_W^0,
\end{eqnarray}
which is a Lorentz scalar.
The result is the same when considering the case of the $\Wp$ decay.
The reader willing to have further details on the previous development or how to undergo such
perturbative calculation, can go to classic references in particle physics/quantum 
field theory such as Refs.~\cite{Halzen:1984mc,Peskin:1995ev}.

This first order expression for the partial width allows to calculate the total width $\GamW$
of the $\Wp$ or the $\Wm$.
The $W$ can decay into leptons or into a pair of quark--anti-quark that in turn decay to observable 
hadrons with a probability of one
\begin{equation}
\GamW = \Gamma(W\to\mm{leptons}) + \Gamma(W\to\mm{hadrons}).
\end{equation}
For the leptonic decay, in the ultra-relativistic approximation assumed so far, the width is simply
\begin{eqnarray}
\Gamma(W\to\mm{leptons}) &=& \Gamma(\Wtoel) + \Gamma(\Wtomu) + \Gamma(\Wtotau), \\
                         &=& 3\,\Gamma_W^0.
\end{eqnarray}
For the hadronic decay, the calculus is similar to the one for the leptons except that here\,:\;%
\index{Electroweak!CKM matrix elements|(}
(1) CKM matrix element intervene and mix the quarks flavors and (2) in top of summing/averaging on the 
quarks spin states the sum and average on their color charge states need to be done as well.
The $W$ can decay into all quarks flavors but the top which is forbidden by energy conservation
since $m_t>\MW$. 
The quarks possess masses such that they can be treated like the charged leptons as 
ultra-relativistic particles.
Then studying $W\to\,q\,\qbp\to\mm{hadrons}$ gives for the partial width
\begin{equation}
\GamW(W\to \mm{hadrons}) = \Nc\, \Vckmsqr{q}{q'}\,\Gamma_W^0,\label{eq_gamma_hadr}
\end{equation}
where $\Nc=3$ is the color number, $\Vckm{q}{q'}$ the element of the CKM matrix governing the mixing
angle between flavors $q$ and $\bar q '$. Summing over all quarks flavors in Eq.~(\ref{eq_gamma_hadr}) gives
\begin{equation}
\sum_{q\,\qbp} \Vckmsqr{q}{q'} = \sum_{q} \underbrace{\sum_{\qbp} \Vckmsqr{q}{q'}}_{=1},
\end{equation}
where the sum on $q'$ of the squared CKM elements translates the unitary of the CKM matrix 
and where $\sum_{q}$ is restricted to $u$ and $s$ flavors since the decay to the top is forbidden.
\index{Electroweak!CKM matrix elements|)}
Then in this context
\begin{equation}
\Gamma(W\to\mm{hadrons}) = 2\,\Nc\,\Gamma_W^0.
\end{equation}
Adding both leptonic and hadronic parts gives eventually in the Born level approximation
\begin{equation}
\Gamma_W = \left(3+2\,\Nc\right)\,\frac{\GF}{\sqrt 2}\,\frac{M_W^3}{6\pi},
\end{equation}
where $\Gamma_W^0$ has been replaced by its explicit form.
Adding QCD corrections in quarks decays transforms the latter expression to 
\index{QCD!Corrections in W decay into quarks@Corrections in $W$ decay into quarks}
\begin{equation}
\Gamma_W = \left(3+2\,\Nc\,\left[1+\tfrac{\alphas(\MW)}{\pi}\right]\right)
\,\frac{\GF}{\sqrt 2}\,\frac{M_W^3}{6\pi},
\end{equation}
where $\alphas(\equiv g^2_\mm{s}/4\,\pi)$ is evaluated at the energy scale imposed by the $W$ mass.
Following those developments it is assumed in the rest of this thesis that $\Gamma_\Wp=\Gamma_\Wm$.
\index{W boson@$W$ boson!Width@Width $\GamW$|)}

\subsubsection{Polarised $\BFW$~bosons}
\index{W boson@$W$ boson!Polarisation|(}
Considering the leptonic decay of a real $W$~boson with a specific polarisation state proves to be 
very helpful to understand the angular decay behaviour of the quarks or leptons.
The polarisation states of the two transverses ($\lambda=\pm 1$) 
and the one longitudinal mode $\lambda=0$ of a $W$ of momentum $\vec p_W=p_W\,\vec e_z$ can be 
expressed in the laboratory frame like
\begin{align}
\varepsilon^{(\lambda=\pm 1)} = \mp\frac{1}{\sqrt{2}}
\left(\begin{array}{c}
   0  \\
   1  \\
   \pm i \\
   0  
\end{array} \right),
\qquad\qquad
\varepsilon^{(\lambda=0)} = \frac{1}{\MW}
\left(\begin{array}{c}
   \left|\vec p_W\right|  \\
   0  \\
   0 \\
   E_W
\end{array} \right). \label{eq_w_pol_state}
\end{align}

\paragraph{Transverse polarisation states.}
\index{W boson@$W$ boson!Polarisation!Transverse states|(}
The example of the leptonic decay of a $\Wm$ is again considered, for a polarisation state of $\lambda=+1$. 
Substituting the expression of $\varepsilon^{(\lambda=+1)}$ in 
Eq.~(\ref{eq_w_lep_decay_prob_ampl}) gives the probability amplitude for 
$\Wm(\lambda=+1) \to \lm\,\anul$.
This amplitude squared and averaged on the leptons spins, gives the term 
$|\mathcal M_{\W^-(\lambda=+1)}|^2$ from which the differential decay width, in the \WRF{} is
\begin{equation}
\DfDx{\Gamma}{\costhetaWlwrf} 
\propto \left|\mathcal M_{\W^-(\lambda=+1)}\right|^2.
\end{equation}
Developing that expression gives, in the end, an angular dependency of
\begin{equation}
\DfDx{\Gamma}{\costhetaWlwrf} \propto (1-\cos \thetaWlwrf)^2, \label{eq_angle}
\end{equation}
where the angle $\thetaWlwrf$ is defined like 
\begin{equation}
\costhetaWlwrf \equiv \frac{\vec p_W \cdot \vec{p}_{l}^{\;\ast} }{|\vec p_W|\;|\vec{p}_{l}^\ast|}.
\label{eq_def_costhetaWlwrf}
\end{equation}

\index{Helicity|(}
\index{Helicity!Of the decaying leptons|(}
The angular dependency in Eq.~(\ref{eq_angle}) is an important key for understanding the 
leptons kinematics in the $W$ decay. 
It can actually be understood without going to refined calculations but
using only the helicity conservation rules in the high energy limit imposed on the leptons by 
the mass of the $W$.
As it was seen previously the electroweak interactions couples only left-handed fermions 
and right-handed anti-fermions.\index{Electroweak}
In the high energy limit chirality \index{Electroweak!Chirality} and helicity becomes the same, which means that the 
helicity is a conserved quantum number and only negative (positive) helicity fermions (anti-fermions)
are involved.\index{Helicity!In the high energy limit}
This explain why sometime in the literature, as a shortcut, but only in the high energy limit, 
negative helicities states are referred to as ``left'' and positive states as ``right''.

\begin{figure}[!h] 
  \begin{center}
    \includegraphics[width=0.9\tw]{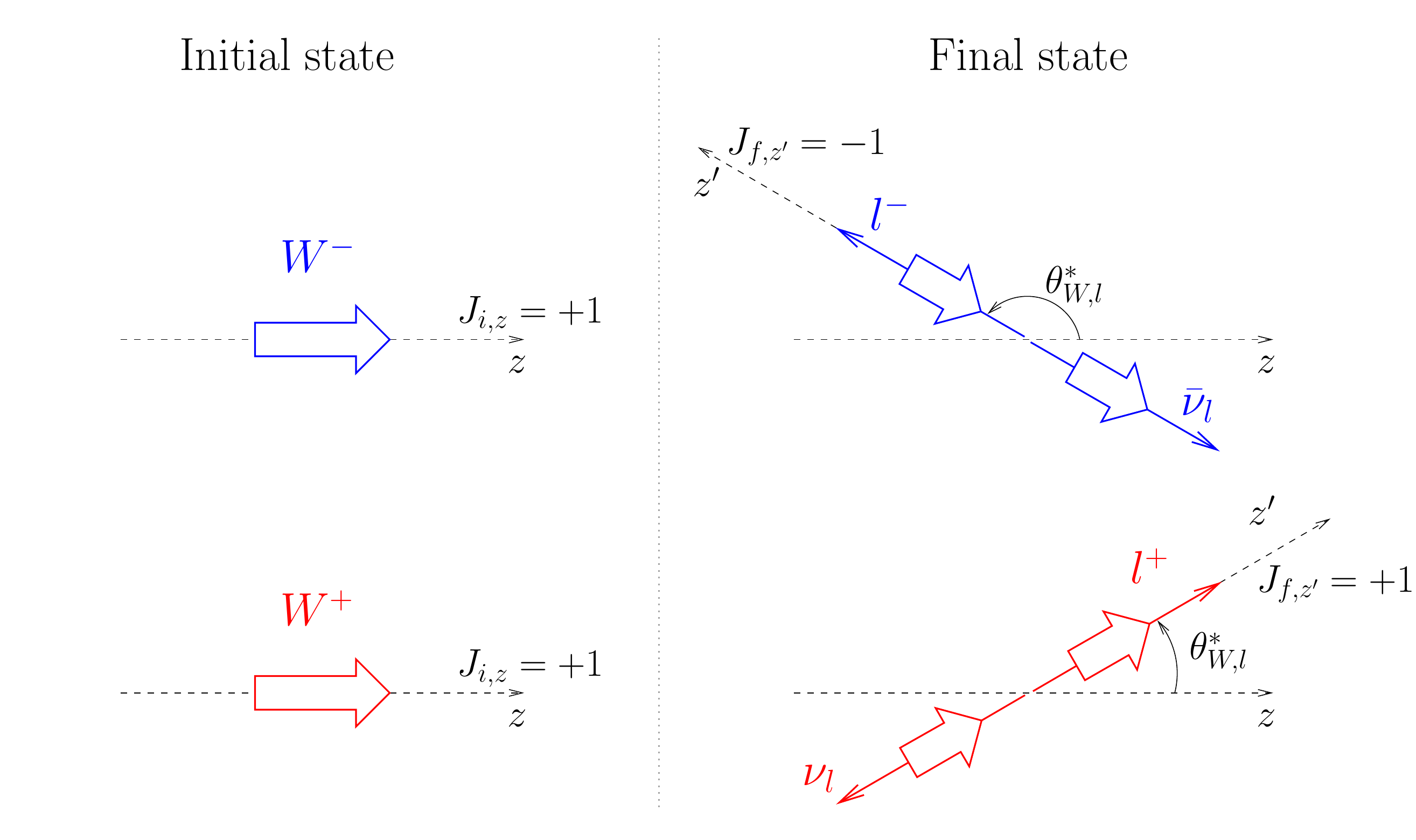}
    \caption[Representation in the \WRF{} of the leptonic decay of a positive helicity $W(\lambda=+1)$
              for both negative and positive]
	    {\figtxt{Representation in the \WRF{} of the leptonic decay of a positive helicity $\Wm$
               boson (up) and $\Wp$ boson (down).}}
	    \label{fig_Wm_decay}
  \end{center} 
\end{figure}
Now the previous example is treated from the helicity conservation point of view.
The decay $\Wm\to l\,\anul$ is depicted in the \WRF{} in the upper part of Fig.~\ref{fig_Wm_decay}.
The convention for the axis is the following.
In the initial state the $+z$ direction is parallel to the direction of the momentum of the $W$
in the laboratory frame while the $+z'$ direction points to the direction of the decaying charged
lepton in the final state.
In that context, the decay proceeds from an initial $W$ state with $J_{i,z}=+1$ to a final leptonic
state with $J_{f,z'}=-1$ and in both cases the system holds a total spin of $J=1$, thus the
amplitudes are proportional to the rotation matrices, it reads 
\begin{equation}
d_{J_{f,z'},J_{i,z}}^J(\thetaWlwrf) \equiv 
\left< J,\,J_{f,z'} \Big| e^{-i\thetaWlwrf\,J_y}\Big| J,\,J_{i,z} \right>,
\end{equation}
which expressions for commonly used spins are tabulated in many places, \eg{} 
Ref.~\cite{Amsler:2008zz}, then
\begin{eqnarray}
\DfDx{\Gamma}{\costhetaWlwrf} &\propto& \left|d^1_{-1,\,1}(\thetaWlwrf)\right|^2, \\
                              &\propto& (1-\cos \thetaWlwrf)^2.
\end{eqnarray}
Thus the most privileged configuration is the one where the initial and final spin projection 
$J_{i,z}$ and $J_{f,z'}$ are aligned, while on the other hand the configuration where the $\lm$ would
be emitted such that $z'=+z$ would be totally forbidden by helicity conservation.

The case of the decay of the $\Wp$ can be deduced using the same kind of argument except
this time the charged lepton is ``right'' and the neutrino is ``left'' 
as shown in the lower part of Fig.~\ref{fig_Wm_decay}, the angular dependency is then 
\begin{eqnarray}
\DfDx{\Gamma}{\costhetaWlwrf} &\propto& \left|d^1_{1,\,1}(\thetaWlwrf)\right|^2, \\
                                  &\propto& (1+\cos \thetaWlwrf)^2.
\end{eqnarray}

\paragraph{Longitudinal polarisations state.}
\index{W boson@$W$ boson!Polarisation!Longitudinal states|(}
For the decay of a $W$ possessing a longitudinal polarisation state a detailed calculus can be
carried out using the expression of $\es^{(\lambda=0)}$ with $p_W=0$ and $E_W=\MW$ in the \WRF{}.
Instead of doing so the angular dependency, that really matters, can be unraveled using again
rotations matrices
\begin{eqnarray}
\DfDx{\Gamma}{\costhetaWlwrf} &\propto& \left|d^1_{\pm1,\,0}(\thetaWlwrf)\right|^2 \\
                                  &\propto& \sin^2 \thetaWlwrf.
\end{eqnarray}
In this equation $J_{f,z'}=-1$ in the case of a $\Wm$ decay and $+1$ for the one of a $\Wp$.
Here there are no differences in the angular decay between the positive and negative channels.

\paragraph{Sum up for the decay of polarised W.}
The previous derivations can be generalised to include in the formulae both the $W$ boson charge $Q$ 
and helicity $\lambda$
\begin{eqnarray}
\DfDx{\Gamma_{W\to\, l\,\nu_l}}{\costhetaWlwrf}
  &\propto& \left(1 + \lambda\, Q\, \costhetaWlwrf\right)^2, \label{eq_WT_angle}\\
\DfDx{\Gamma_{W\to\, l\,\nu_l}}{\costhetaWlwrf}
  &\propto& \sin^2 \theta_{W,l}^{\ast}. \label{eq_WL_angle}
\end{eqnarray}
The sign in front of $\costhetaWlwrf$ can be deduced easily each time by deducing which direction
is privileged for the charged lepton from helicity conservation arguments point of view.
\index{Helicity|)}
\index{Helicity!Of the decaying leptons|)}
\index{W boson@$W$ boson!Polarisation|)}
\index{W boson@$W$ boson!Polarisation!Longitudinal states|)}
\index{W boson@$W$ boson!Polarisation!Transverse states|)}

\subsection{W in Drell--Yan-like processes at the LHC}\label{s_drell-yan}
\index{Drell--Yan processes for W@Drell--Yan processes for $W$!Generalities|(}
The Drell--Yan~\cite{Drell:1970yt} processes were originally defined, in hadron--hadron collisions
within the parton model, before identifying partons with quarks and later on, with gluons.
In these processes, a pair of partons, the hypothesised building blocks of nucleons, collide and 
annihilate giving in the final state a high invariant mass lepton pair $\lp\lm$.
Within QCD, this partonic reaction proved to be achieved at first order, by the 
quark--anti-quark annihilation via $q\,\qbp\to\gamma^\ast/Z\to\lp\,\lm$.
The main features of the formalism describing Drell--Yan survived to the rise of the QCD up to a few refinements.
Still, for historical reasons even though the whole present discussion takes place 
within QCD the term parton is still used and refers to quarks or gluons.

By extension the production of high invariant mass $l\nul$ pair through the production 
of an intermediate $W$ boson is referred to as ``Drell--Yan like processes'', or 
--for convenience-- Drell--Yan. 
Below reminders of the treatment of $W$ boson production in Drell--Yan is given.
Let us remark that the formalism described is applicable to the original Drell--Yan and to 
a wide variety of over hard scattering processes including jet and heavy flavour production.
Further general details on this topic can be found in 
Refs.~\cite{Barger,Ellis:1991qj,Nadolsky:2004vt}.

\subsubsection{Overview}
\begin{figure}[!h] 
  \begin{center}
    \includegraphics[width=0.7\tw]{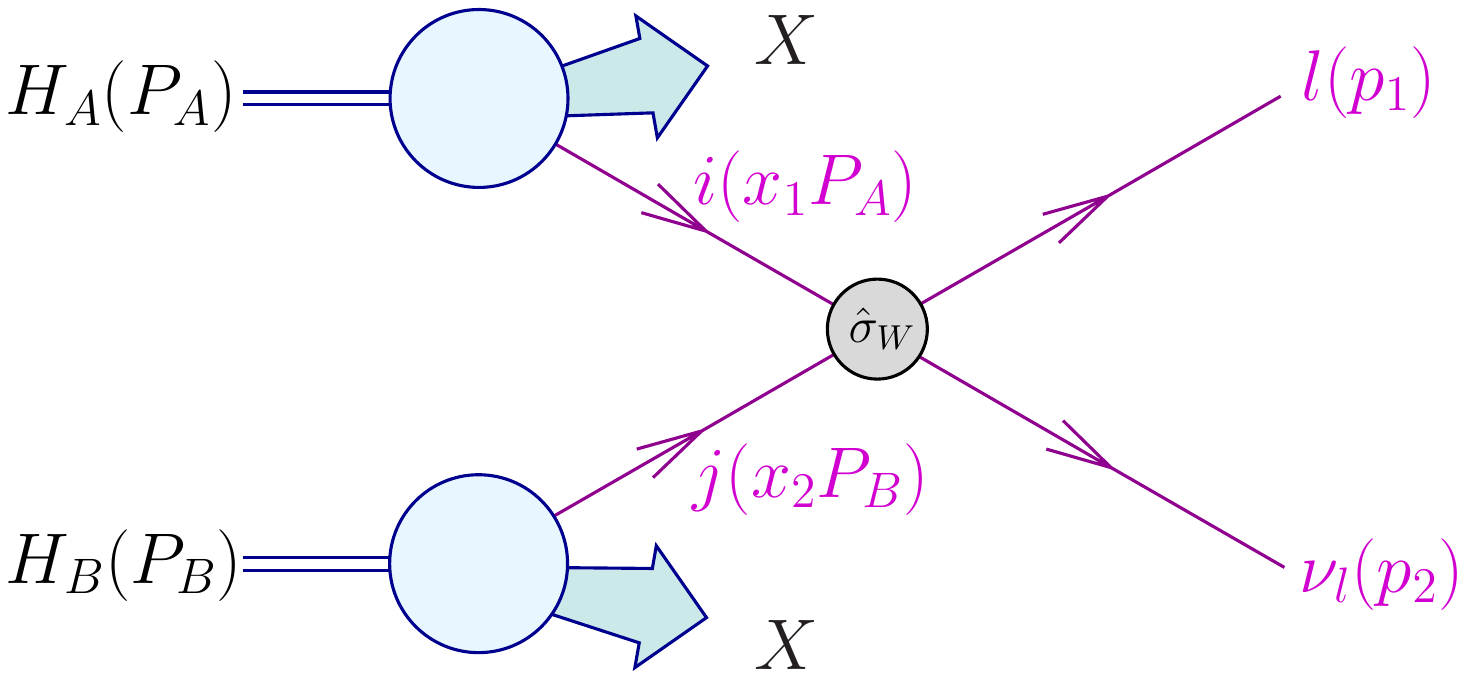}
    \caption[Representation of the process $\mm{hadron}\,\mm{hadron} \to W \to l\,\nul$]
	    {\figtxt{Feynman like diagram, \ie{} in momentum-space time-ordered fashion,
            of the process $H_A(P_A)\,H_B(P_B) \to W \to l(p_1)\,\nul(p_2)$.
            $X$ represents all the particles in the collision but the $W$ boson and its leptonic 
            decay.
            }}
	    \label{fig_hh_W_lnul}
  \end{center} 
\end{figure}

The production of a $W$ in Drell--Yan can be written
\begin{equation}
H_A(P_A)\,H_B(P_B) \to W \, X \to l(p_1)\,\nul(p_2)\,X.
\end{equation}
The associated Feynman like representation of this process is shown in Fig.~\ref{fig_hh_W_lnul},
where $H_{A}$ and $H_B$ are hadrons of four-momenta $P_A$ and $P_B$ accelerated by the 
collider bringing a total energy $\sqrt S$ in the center of mass.
The collision produces a pair of leptons $(l,\nul)\equiv \{(\lm,\anul),(\lp,\nul)\}$ 
of four-momenta $p_1$ and $p_2$.
The other particles produced in this collision noted $X$ are not considered at all.
The derivations that follow are applicable to the decay in the $\tau$ channel, nonetheless
the analysis being restricted to the electronic and muonic decays,
throughout the whole document $l\equiv\{e,\mu\}$ unless stated otherwise.

\index{Parton Distribution Functions (PDFs)!Definition|(}
\index{QCD!Hadronic cross section calculation|(}
The hadronic cross section $d^4\sigma_{ij}(S)$ for a given point of the quarks and leptons phase space,
$d^4\Phi\equiv d\,x_1\, d\,x_2\,d\,p_1\, d\,p_2$, and for a particular collision of partons
$i$ and $j$ can be expressed using the factorisation ``theorem'' like
\begin{equation}
\frac{d^4\sigma_{ij}(S)}{d\,x_1\, d\,x_2\,d\,p_1\, d\,p_2}=
\PDF{A}{i}(x_1,\mu_f)\, \PDF{B}{j}(x_2,\mu_f)\; d^2\hat\sigma_{ij}(x_1x_2S,p_1,p_2,\mu_r),
\label{eq_fact_theorem_d3}
\end{equation}
that is as the product of the partonic cross section $d^2\hat\sigma_{ij}$ and the parton 
distributions functions (PDFs) $\PDF{A}{i}$ and $\PDF{B}{j}$.

The PDFs are the density probabilities $\PDF{A}{i}(x_1)$ and $\PDF{B}{j}(x_2)$ for two partons 
$i$ and $j$, to carry before their hard scatter fractions $x_1$ and $x_2$ ($0<x_{1,2}<1$)
of the four-momenta of the hadrons they respectively income from. 
Purely theoretical derivation of PDFs are not computable as their description falls in the non 
perturbative regime of QCD where $\alphas$ is too large, therefore they are extracted from 
global fits to data from processes such deep inelastic scattering, Drell--Yan
and jet production at the available energy range fixed by colliders. 
The implementation of QED and QCD radiative corrections from quarks in the initial state are
universal, \ie{} independent from the process. 
These corrections contain mass singularities that can be factorised and absorbed in a 
redefinition (renormalisation) of the PDFs. 
The singularities are removed in the observable cross section while the PDFs becomes dependent
of a factorisation scale $\mu_f$ controlled by the DGLAP evolution 
equations~\cite{Gribov:1972ri,Altarelli:1977zs,Dokshitzer:1977sg}.
\index{Parton Distribution Functions (PDFs)!DGLAP}
This scale is to be identified --for example-- to a typical scale of the process, like the 
transverse momentum or in the present case by the mass of the resonance $\mu_f=M_W$.
\index{Parton Distribution Functions (PDFs)!Definition|)}

The partonic cross section $d^2\hat\sigma_{ij}$ corresponds to the probability that the partons 
$i$ and $j$, of four-momenta $x_1\,P_A$ and $x_2\,P_B$,
collide and create a $W$ resonance of mass $m_W=\sqrt{x_1x_2S}$ which in turn decays into leptons 
$l$ and $\nul$ respectively of four-momenta $p_1$ and $p_2$.
The hard scattering occurs at such energies that partons can be seen as free, \ie{} $\alphas$ is 
small. 
This allows to calculate the cross section using perturbation theory. 
Up to a given energy scale higher order Feynman diagrams are not directly computed but accounted by 
factorising their effect in the value of the constant $\alphas$.
This renormalisation scale is the one of the virtual resonance, which means here $\mu_r=\MW$.
The total partonic cross section is then developed in power of $\alphas(\MW)/(2\,\pi)$
\begin{equation}
\hat\sigma_{ij} = 
\hat\sigma_{ij}^{(0)} + 
\frac{\alphas(\MW)}{2\pi}\,\hat\sigma_{ij}^{(1)} + 
\left(\frac{\alphas(\MW)}{2\pi}\right)^2\,\hat\sigma_{ij}^{(2)} + 
\mathcal O(\alpha_\mm{s}^3).
\end{equation}
In this equation the first term of this series is the leading-order (LO), usually called the Born 
level, the second the next-to-leading order (NLO) correction of order $\alphas$ and so on the third
term adds up next-to-next-to-leading order (NNLO) corrections of order $\alpha_\mm{s}^2$.
The last term contains all corrections above $\alpha_\mm{s}^2$.
In practice only the first corrections are brought to a calculus as the number of Feynman diagrams
increase rapidly when going to higher order corrections.
Let us remark that a correction term of order $(k)$ is to be 
implicitly apprehended as $\hat\sigma_{ij}^{(k)}\equiv \hat\sigma_{ij}^{(k)}\delta^{(k)}_{ij}$
where
\begin{eqnarray*}
\delta^{(k)}_{ij} &=& 1 \quad\mbox{if the $ij$ collision contribute to $\hat\sigma_{ij}^{(k)}$ }, \\
\delta^{(k)}_{ij} &=& 0 \quad\mbox{if the $ij$ collision do not contribute to 
                            $\hat\sigma_{ij}^{(k)}$}.
\end{eqnarray*}

The total hadronic cross section is then deduced by embracing all available corrections to the
partonic cross section and by successively integrating Eq.~(\ref{eq_fact_theorem_d3})
over the accessible phase space to the leptons and partons. 
This means in the latter case integrating over all possibles $x_1$ and $x_2$, but also
by summing over all possible partonic collisions, which gives eventually
\begin{equation}
\sigma(S)= \sum_{i,j} \int_0^1 d\,x_1\int_0^1  d\,x_2\;
\PDF{A}{i}(x_1,\mu_f)\, \PDF{B}{j}(x_2,\mu_f)\; \hat\sigma_{ij}(x_1x_2S,\mu_r).
\label{eq_fact_theorem}
\end{equation}
\index{QCD!Hadronic cross section calculation|)}

From the experimentalist pragmatic point of view the matter of importance is to understand  
the kinematics of the leptons and how the $W$ properties can be extracted from them.
Hence, in what follows, basics on the partonic cross section and on the PDFs are
reminded with emphasis on the kinematics aspects rather than the dynamical issues.
After that, an overview of the relevant kinematics in $W$ production is given. 

\subsubsection{Partonic level}
\begin{figure}[!h] 
  \begin{center}
    \includegraphics[width=0.4\tw]{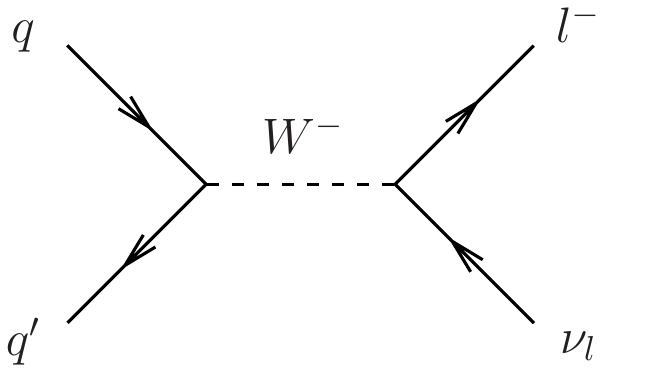}
    \caption[Feynman diagram for the production of a $W$ in $q\,\qbp\to \Wm \to \lm\,\anul$ at 
      the Born level]
	    {\figtxt{Feynman diagram for the production of a $W$ in 
                $q\,\qbp\to \Wm \to \lm\,\anul$ at the Born level.}}
	    \label{fig_feyndiag_wprod_bornlvl}
  \end{center} 
\end{figure}
The production of a $W$ at the Born level is made through quark--anti-quark annihilation,
which is illustrated here for the case of a $\Wm$ and noted
\begin{equation}
q(k_1)\,\qbp(k_2)\to \Wm \to \lm(p_1)\,\anul(p_2),
\end{equation}
where the four-momenta of the particles have been put into brackets.
This process, described by one Feynman diagram (Fig.~\ref{fig_feyndiag_wprod_bornlvl}),
has an amplitude of probability expressed as
 \begin{equation}
  \mathcal M = i\,
  \left[ \tfrac{g}{\sqrt 2}\Vckm{q}{\qbp}\,\vsbar{k_2}\,\HVmAup{\mu}\,\us{k_1}\right] \,
  \left\{ \frac{1}{\shat -M_W^2 +i\,\MW\GamW} \right\} \,
  \left[ \tfrac{g}{\sqrt 2}\,\usbar{p_1}\,\HVmAdn{\mu}\,\vs{p_2}\right].
\label{eq_qqbp_W_lnul}
\end{equation}
The pending expression for the process $q\,\qbp\to \Wp \to \lp\,\nul$ is obtained by
doing the substitution $k_1\leftrightarrow k_2$ and $p_1\leftrightarrow p_2$ in 
this last equation.
Let us note that in the case where $\shat \ll M_W^2$, that is for a low energy in the collision,
the propagator of order $\sim 1/M_W^2$ makes us rediscover the historical coupling 
constant $\GF/\sqrt 2$ from the Fermi model.

At Born level all partons have a purely longitudinal motion, which means that in the
present convention, the quark and anti-quark momenta are expressed in the Cartesian
coordinate basis like  
\begin{align}
k_1 = \frac{\sqrt{\shat}}{2}
\left(\begin{array}{c}
   1  \\
   0  \\
   0  \\
   1  
\end{array} \right),
\qquad\qquad
k_2 = \frac{\sqrt{\shat}}{2}
\left(\begin{array}{c}
   1  \\
   0  \\
   0  \\
  -1  
\end{array} \right). \label{eq_qqbp_4mom_LO}
\end{align}

\begin{figure}[!h] 
  \begin{center}
    \includegraphics[width=.6\tw]{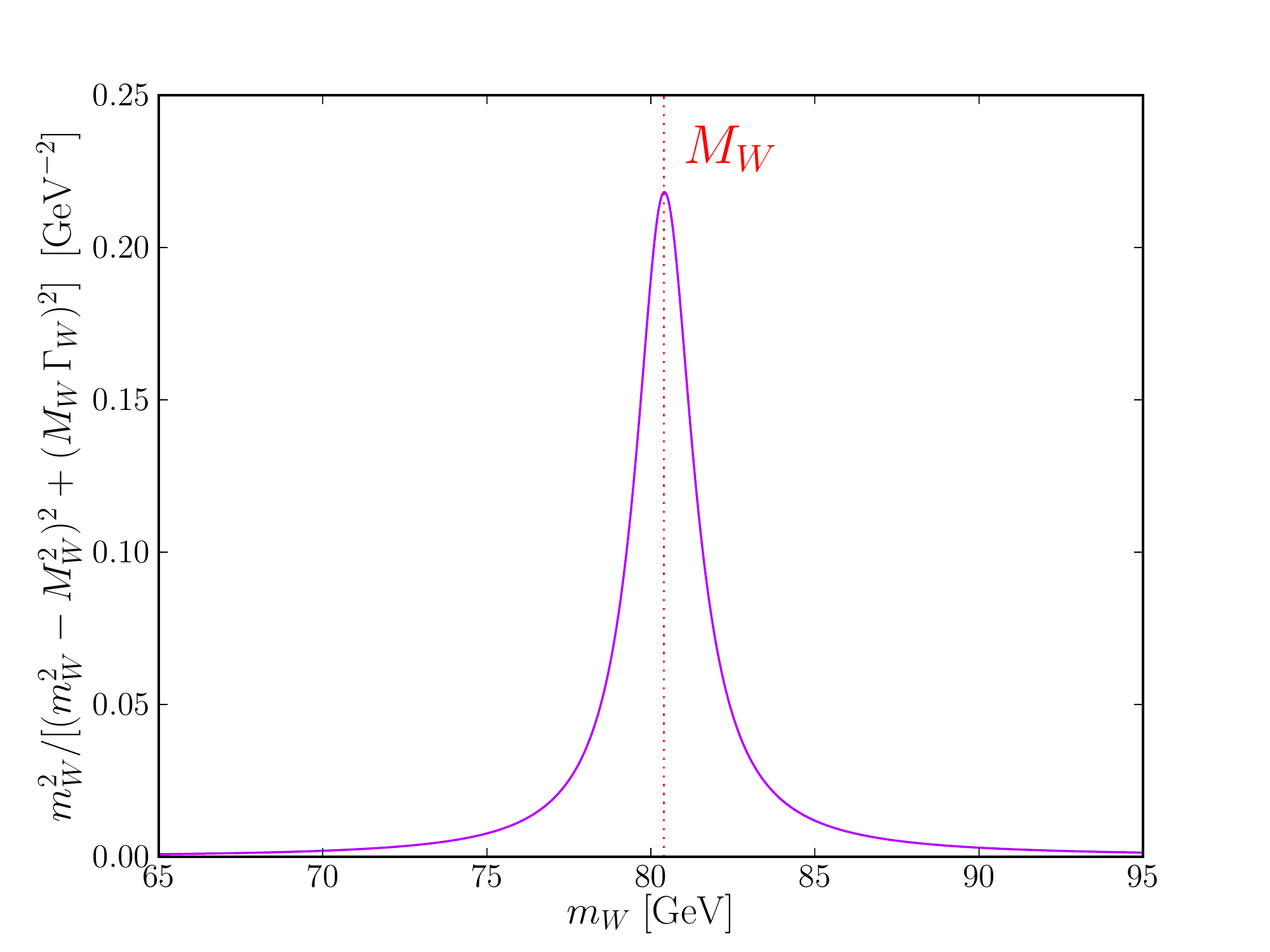}
    \caption[Breit-Wigner behaviour of the off-shell $W$ mass]
	    {\figtxt{Breit-Wigner behaviour of the off-shell $W$ mass.}}
	    \label{fig_mw_breit_wigner}
            \index{W boson@$W$ boson!Mass@Mass $\MW$}
  \end{center} 
\end{figure}

The total partonic cross section $\hat\sigma^{(0)}({q\,\qbp\to\Wtolnu})\equiv \sigmahat$
deduced from Eq.~\ref{eq_qqbp_W_lnul} reads
\begin{equation}
\sigmahat(\shat) = 
\frac{4}{3\pi} \,
\frac{\Vckmsqr{q}{\qbp}}{s_q s_{\qbp} \Nc} \,
\left( \frac{M_W^2\GF}{\sqrt 2} \right)^2 \,
\frac{\shat}{(\shat-M_W^2)^2+M_W^2\Gamma_W^2},  \label{sigmahat_tot_LO}
\end{equation}
where $s_q$ and $s_{\qbp}$ are the spins of the quark and of the anti-quark.
The previous formula is correct as well in the case of a $\Wp$ production.
The variable $\hat s$ belongs to a set of three others invariant Lorentz scalars,
known as the Mandelstam variables, and is defined by
\begin{eqnarray}
\hat s &\equiv& (k_1+k_2)^2=(p_1+p_2)^2,\\
       &\equiv& m_W^2,
\end{eqnarray}
where the last line makes the link with the invariant mass $m_W$ of the off-shell $W$.
The fraction in Eq.~(\ref{sigmahat_tot_LO}) depending of $\shat$ reveals the
Breit-Wigner resonant \index{W boson@$W$ boson!Breit-Wigner resonance} 
behaviour of the $W$, that is a peaked distribution centered on $\MW$ and 
whose width is controlled by $\GamW$.\index{W boson@$W$ boson!Width@Width $\GamW$}
It means a $W$ can be produced at any mass, as high or low as possible, but the probability this 
happens gets smaller and smaller as $m_W$ is far from the central value $\MW$ of the peak like as
shown on Fig.~\ref{fig_mw_breit_wigner}.
 
\begin{figure}[!h] 
  \begin{center}
    \includegraphics[width=.9\tw]{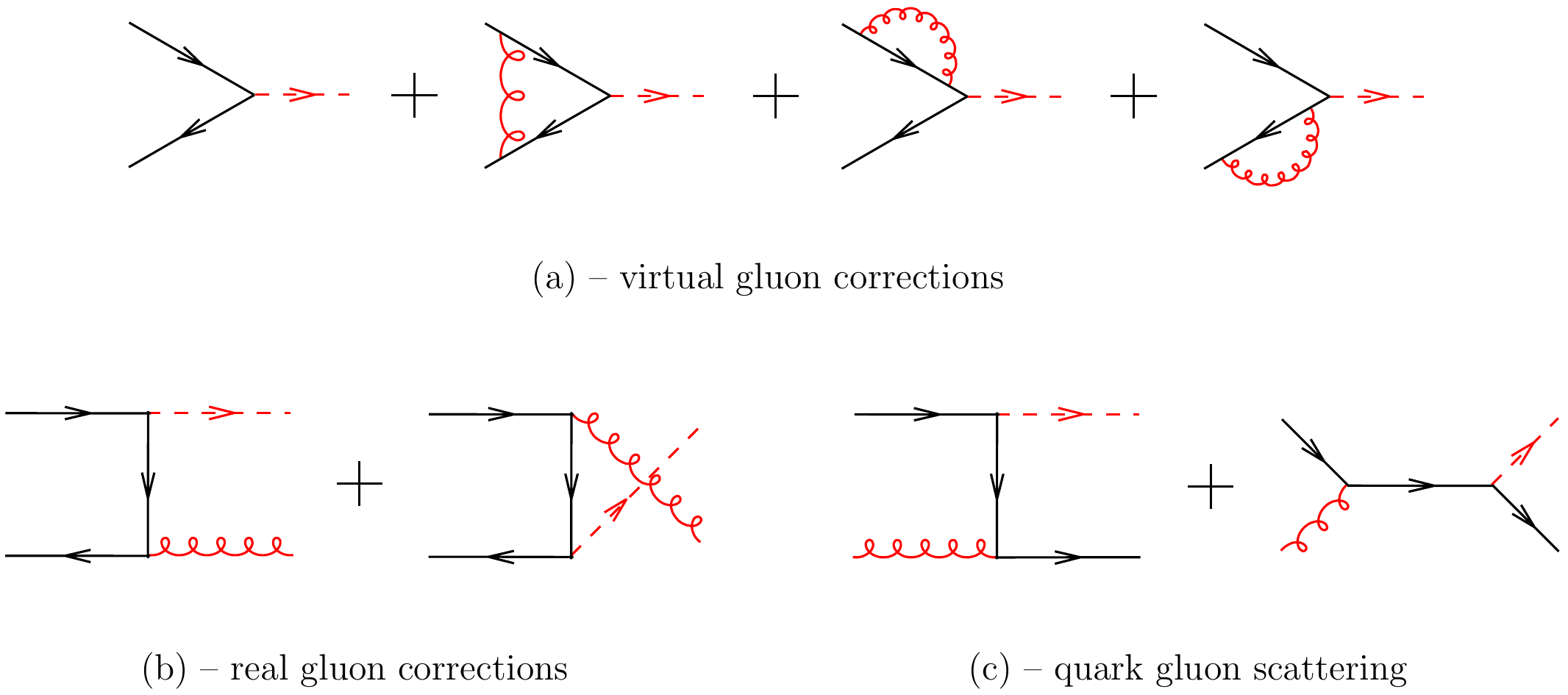}
    \caption[The leading- and next-to-leading-order Feynman diagrams for $W$ in Drell--Yan like]
	    {\figtxt{The leading- and next-to-leading-order Feynman diagrams for $W$ in Drell--Yan 
                like.}}
	    \label{fig_nlo}
  \end{center} 
\end{figure}

The NLO corrections of $\mathcal O(\alphas)$ comes from the three following contributions
in which $\gamma^\ast$ represents a virtual photon\,:\;%
(a) virtual gluon corrections to the LO $q\,\qbp \to \gamma^\ast\, g$,
(b) real gluon corrections $q\,g \to \gamma^\ast \, q$ and
(c) quark/anti-quark gluon scattering $q\,g \to \gamma^\ast\, q$, or $\qbp\,g \to \gamma^\ast\,\qbp$.
The Feynman diagrams associated to these corrections are displayed in Fig.~\ref{fig_nlo} 
(a), (b) and in (c) for the case of quark gluon scattering only.

\index{QED!Radiative corrections@Radiative correction in single $W$ production|(}
\index{Electroweak!Radiative corrections in W in Drell--Yan@Radiative corrections in $W$ in Drell--Yan}
In top of QCD corrections are electroweak radiative corrections briefly mentioned below.
The $\mathcal O(\alpha)$ EW radiative corrections to the $W$ in Drell--Yan can be divided in a gauge
invariant way into three parts\,:\;the initial state radiation (ISR), the initial-final state 
interferences (non factorisable corrections) and the final state radiation (FSR).
The leading ISR (mass singular) QED corrections can be absorbed as stated previously into the PDFs.
The non factorisable corrections are negligible in resonant $W$~boson production~\cite{Baur:1998kt}.
On the contrary, the FSR corrections affect considerably various $W$ observables among which is the
transverse momentum of the charged lepton whose shape is primordial for the extraction of the $\MW$.
For example in Ref.~\cite{Berends:1984qa}, the final state photonic correction was approximated and 
lead to a shift in the value of $\MW$ of $50-150\MeV$ for the Tevatron collisions \index{Tevatron collider}
at that time.
The development of these electroweak NLO corrections is beyond the scope of our discussion, the reader
interested to have more details on this particular topic can look at 
Refs.~\cite{Baur:1998kt,Dittmaier:2001ay}.
\index{QED!Radiative corrections@Radiative correction in single $W$ production|)}

Coming back to the LO, before the phase space integration the differential partonic cross section can
be written as a function of $\thetaWlwrf$,
\begin{equation}
\DsigpartDobs{\cos \theta_{W,l}^\ast} = 
\frac{3}{8} \,
\sigmahat(\shat) \,
\left( 1 + \lambda\, Q\,\cos\theta_{W,l}^\ast \right)^2,
\label{eq_dsigmahat_dcostheta0}
\end{equation}
the angular dependency can be traced back from the derivations made in Eq.~(\ref{eq_WT_angle}),
and where in the present context the polarisation $\lambda$ of the $W$ is ruled by 
the following inequality between the fractions of momenta $x_1$ and $x_2$ the partons bears
before the collision
\begin{eqnarray}
x_1 > x_2 &\Rightarrow& \lambda = +1, \\
x_1 < x_2 &\Rightarrow& \lambda = -1.
\end{eqnarray}
Note that at the LO there are no production of $W$ of longitudinal polarisation state due to the 
absence of a transverse momentum of the $W$ and in the massless quark approximation.
Nonetheless, from the algebraic point of view, rather than the $W$, the quarks and leptons are the 
objects we manipulate, hence in the continuity of the present calculus 
Eq.~(\ref{eq_dsigmahat_dcostheta0}) writes in function of $\theta_{q,l}^\ast$ --the angle between the 
charged lepton and the quark $q$ momenta-- like
\begin{equation}
\DsigpartDobs{\cos \theta_{q,l}^\ast} = 
\frac{3}{8} \,
\sigmahat(\shat) \,
\left( 1 - Q\,\cos\theta_{q,l}^\ast \right)^2.
\label{eq_dsigmahat_dcostheta}
\end{equation}

\begin{figure}[!h] 
  \begin{center}
    \includegraphics[width=1.\tw]{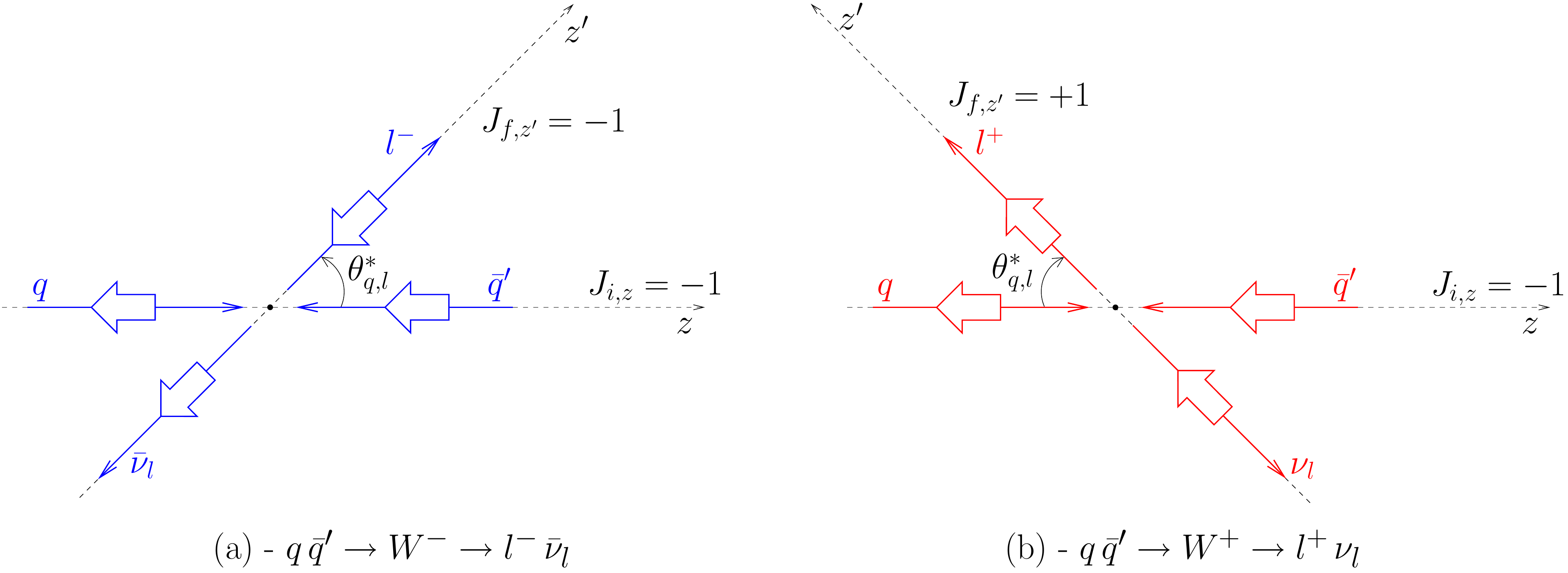}
    \caption[Opening angle between the quark and the charged lepton in the $W$ rest frame]
	    {\figtxt{Favored opening angle between the quark $q$ and the charged lepton $l$
                in the \WRF{} for the production of a $\Wm$ (a) and $\Wp$ (b).}}
	    \label{fig_quark_charged_lepton_angular_decay}
            \index{Helicity!Of the colliding quarks and the decaying leptons}
  \end{center} 
\end{figure}
The angular dependency can be found out quickly using the rotations matrices techniques.
Figure~\ref{fig_quark_charged_lepton_angular_decay} actually
gives the hint for the privileged angular decay configuration constrained by helicity conservation, 
for both $\Wm$ and $\Wp$ cases.
For example in the case of $q\,\qbp \to \Wm \to \lm\,\anul$, using rotation matrices techniques with 
the help Fig.~\ref{fig_quark_charged_lepton_angular_decay}.(a) gives 
\begin{eqnarray}
  \DfDx{\sigmahat}{\costhetaqlwrf} &\propto& \left|d^1_{-1,\,-1}(\thetaWlwrf)\right|^2, \\
                                     &\propto& (1+\cos \thetaWlwrf)^2.
\end{eqnarray}

\begin{figure}[!h] 
  \begin{center}
    \includegraphics[width=.7\tw]{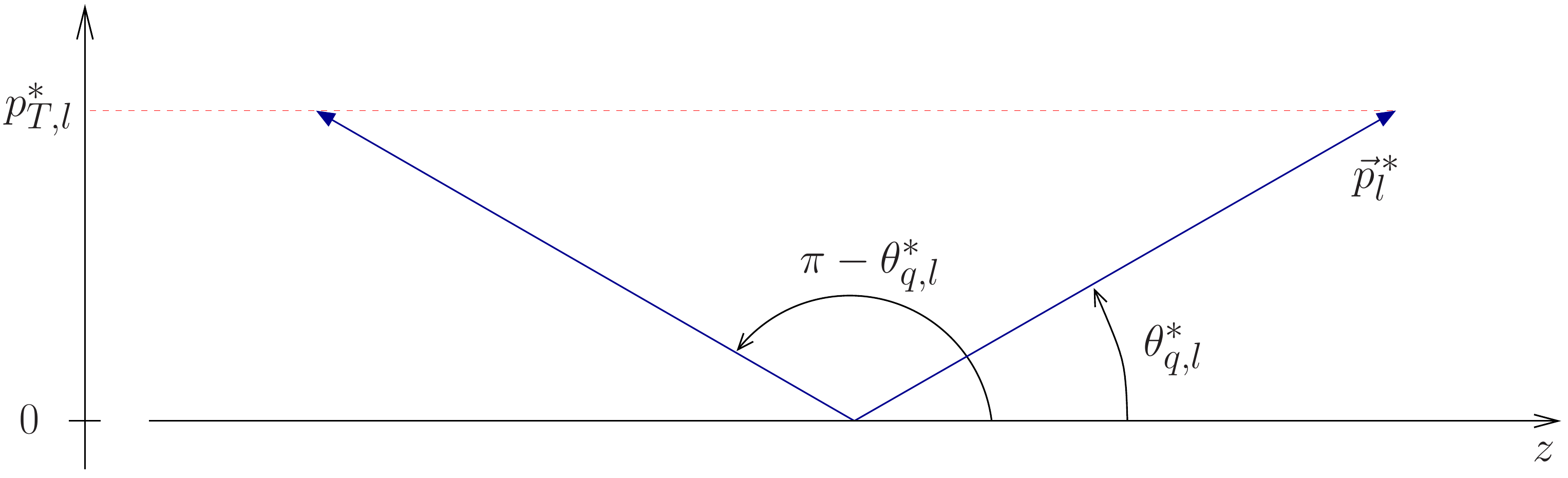}
    \caption[The two angular configurations contributing to the same value of $\pTlstar$]
	    {\figtxt{Representation of the two angular configurations that can 
                produce the same transverse momentum for the charged lepton.
            }}
	    \label{fig_costheta_to_pTl_WRF_jac}
  \end{center} 
\end{figure}
\begin{figure}[!h] 
  \begin{center}
    \includegraphics[width=.6\tw]{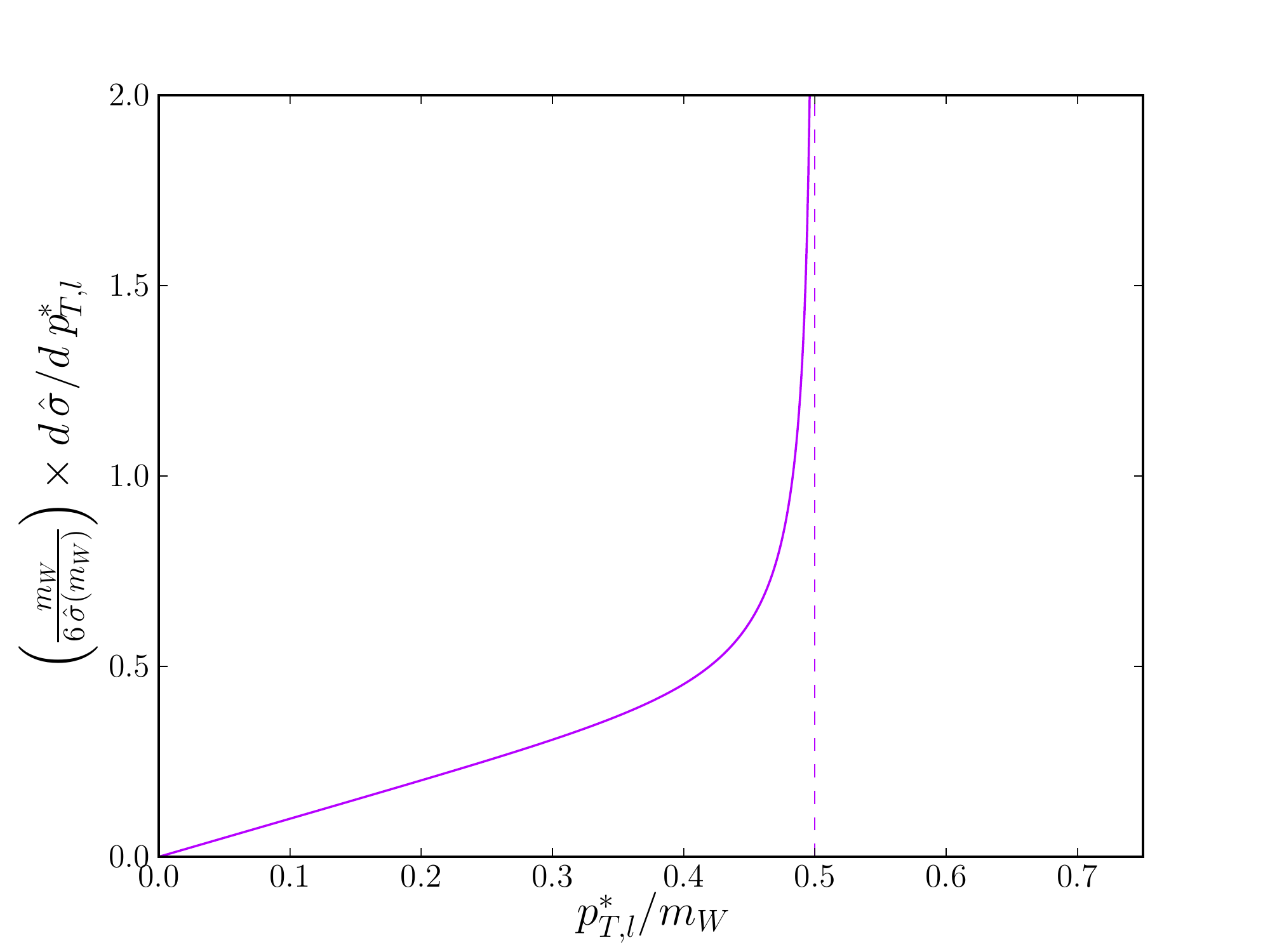}
    \caption[Transverse momentum of the charged lepton in the $W$ rest frame]
	    {\figtxt{Transverse momentum of the charged lepton in the $W$ rest frame.}}
	    \label{fig_pTl_wrf}
  \end{center} 
\end{figure}

Based on Eq.~(\ref{eq_dsigmahat_dcostheta}) the partonic cross section can be expressed as a 
function of the transverse momentum of the charged lepton in the \WRF{} $\pTlstar$.
This function will be be useful later to understand the behaviour of the charged lepton $\pT$ in the
laboratory frame. The change of variable is done by summing the two angular configurations
giving the same $\pTlstar$ as shown on Fig.~\ref{fig_costheta_to_pTl_WRF_jac}.
The analytical derivation writes
\begin{equation}
\DsigpartDobs{\pTlstar} = 
\left( \DsigpartDobs{\costhetaqlwrf}\Bigg|_{\thetaqlwrf} +
       \DsigpartDobs{\costhetaqlwrf}\Bigg|_{\pi-\thetaqlwrf}
\right)\times
\Bigg|\DfDx{\costhetaqlwrf}{\pTlstar}\Bigg|
\end{equation}
which gives finally
\begin{equation}
\DsigpartDobs{\pTlstar} = 
6 \,
\frac{\sigmahat(\shat)}{\shat} \,
\frac{\pTlstar\left(1-2\,{\pTlstar}^2/\shat\right)}{\sqrt{1-4\,{\pTlstar}^2/\shat}}
\label{eq_dsigma_dpTlstar}
\end{equation}
From the jacobian $|\flatDfDx{\costhetaqlwrf}{\pTlstar}|$ arise a term whose denominator 
shows a singularity when $\pTlstar=\sqrt{\shat}/2$. 
This singularity is visible in Fig.~\ref{fig_pTl_wrf} that represents the $\pTlstar$ distribution 
and for that reason it is called the jacobian peak.

This means this peak will eventually be visible at the hadronic level
but will be smoothed when integrating over all possible masses for the $W$.
Still, because the $\pTlstar$ distribution is weighted by $\sigmahat(\shat)$ only the
contributions for which the invariant mass is such that $\m_W\approx\MW$ will be preponderant
(cf. Fig.~\ref{fig_mw_breit_wigner}).
\index{W boson@$W$ boson!Mass@Mass $\MW$}
Therefore a peak is to be expected around $\pTlstar\approx \MW/2$ its smoothness being a direct consequence 
of the width $\GamW$.

\begin{figure}[!h] 
  \begin{center}
    \includegraphics[width=.9\tw]{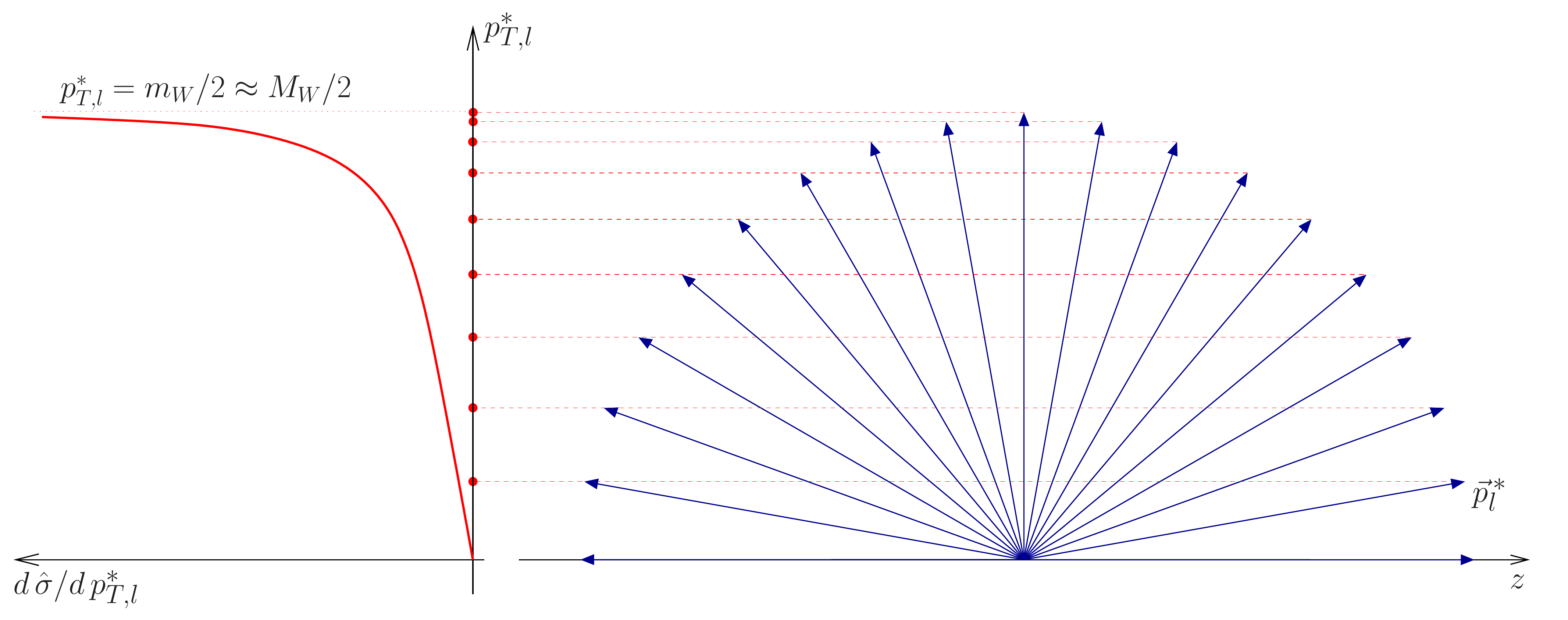}
    \caption[Transverse momentum of the charged lepton distribution : intuition]
	    {\figtxt{Schematic drawing giving a geometric intuition to apprehend the
                jacobian peak of the $\pTlstar$ distribution.}}
	    \label{fig_costheta_to_pTl_WRF}
  \end{center} 
\end{figure}

The jacobian peak in the $\pTlstar$ distribution can be understood from a more intuitive geometric 
point of view that consists to project the momenta of the charged lepton on the $r-\phi$ transverse 
plane like shown in Fig.~\ref{fig_costheta_to_pTl_WRF} where an isotropic decay of the charged lepton 
has been adopted for representation practicability.
As as can be seen the rise of the $\pTlstar$ is a direct consequence of the $\sin\thetaqlwrf$
function entering the projection $\pTlstar\equiv p^\ast_{l}\,\sin\thetaqlwrf$.

\subsubsection{Behaviour of the parton distribution functions}
\begin{figure}[!h] 
  \begin{center}
    \includegraphics[width=.9\tw]{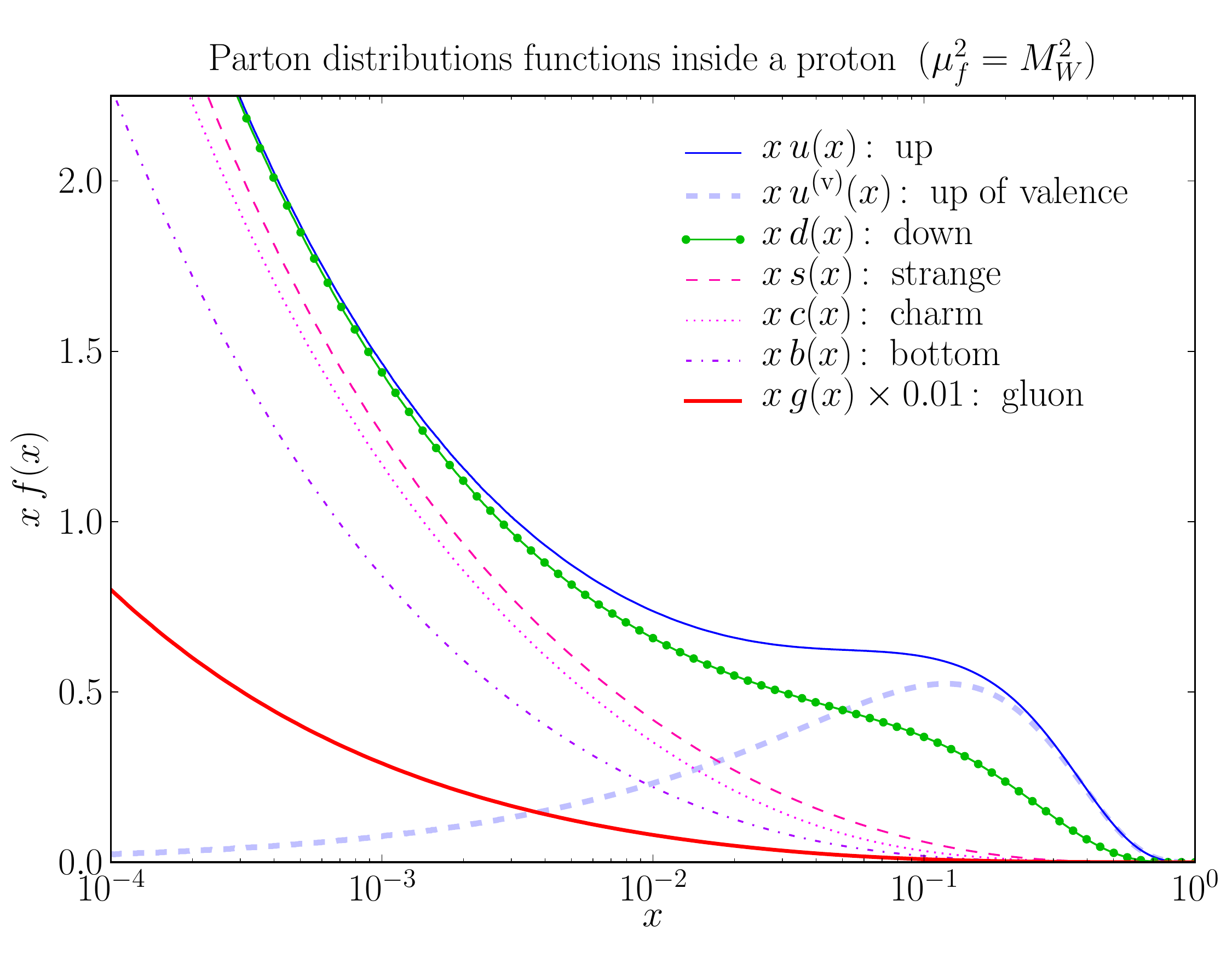}
    \caption[CTEQ6.1M parton distributions functions]
            {\figtxt{CTEQ6.1M parton distributions functions for the quarks and gluons.}
            }
            \label{fig_cteq61m}\index{Parton Distribution Functions (PDFs)!CTEQ}
  \end{center} 
\end{figure}

\index{QCD!Parton Distribution Functions (PDFs)|see{Parton Distribution Functions (PDFs)}}
\index{Parton Distribution Functions (PDFs)!Behaviour (CTEQ6)|(}
Here the PDFs behaviour in function of $x$ are reminded because of their strong influence on
the kinematics of the $W$ and in consequence on the leptons.
The quantity $\PDF{p}{a}(x)$ represents the probability to have a parton of type $a$ in a 
proton having a fraction of the four-momentum of the latter comprised between $x$ and $x+d\,x$.
Figure~\ref{fig_cteq61m} represents the PDFs extracted from the set 
CTEQ6.1M~\cite{Pumplin:2002vw}\index{Parton Distribution Functions (PDFs)!CTEQ} used in this analysis.

\index{Quarks!Valence quarks|(}
\index{Quarks!Sea quarks|(}
Two type of partons can be distinguished in a proton, the valence (v) quarks
and the sea (s) quarks.
The valence quarks --bound by the strong interaction-- are the elementary particles conferring 
to the proton its properties. 
The sea quarks arise as a consequence of the short-time fluctuation of the wave function of the 
proton.
As described in relativistic quantum physics particles can create and annihilate, which translates
in the present case that valence quarks/gluons radiate gluons which in turn can split again in 
gluon or pairs of quark--anti-quark.
A way to look at those distributions is to consider that a probing particle impinging on the proton 
can resolve in the transverse direction its smaller building blocks as better as it possess a high 
energy.
Indeed, using optics vocabulary, to a probe of momentum $p$ we can associate an intrinsic wave-length
$\lambda=1/p$ which is most likely to diffract with pattern of the same wave-length.
Bearing this fictitious probe in mind the behaviour of each flavours are recaptured.

The up and down quarks are predominant compared to any other flavors which means that as we would 
expect a probe will most likely hit a $u$ or a $d$ in proton.
Near $x\to 1$ the bumps reflect the high probability to hit a valence $u$ or $d$ quarks.
In Fig.~\ref{fig_cteq61m} the valence contribution for the case of the up quark is shown. 
Thus, the rise of the probability at low $x$ is due to the sea contributions starting from 0 at $x=1$
and increases as $x\to 0$.
The behaviour at small-$x$ is the consequence that probe starts to resolve more and more of 
de-localised sea quarks.
The $\ubar$ and $\dbar$ behaviour can be directly deduced from the removal of the valence bump
present for the $u$ and $d$ flavors. Let us stress than even though the masses of the $u$ and $d$
quarks are very close for $x<10^{-4}$ we observe that $\ubar<\dbar$.
\index{Quarks!Valence quarks|)}

The patterns in the strange, charm and bottom quarks can be understood from the previous explanations
for the up and down quarks of the sea. 
For a given $x$ the probability to observe a given flavour decreases as the quark mass increase. 
Indeed, it is harder to create virtual pairs of heavy quarks compared to lighter quarks given 
the same amount of energy explaining why $s(x)>c(x)>b(x)$.
Concerning the top flavour the energies scales we are working with are far too small to resolve 
any top quarks hence, $t(x)=0$.
The distributions of the $\sbar$, $\cbar$ and $\bbar$ are exactly the same in a first approximation.
For example, only very recently the strange quark asymmetry has been implemented in some PDF set.
Such corrections are not present in the CTEQ6.1M\index{Parton Distribution Functions (PDFs)!CTEQ}
set that was used in this work, then later on $q^\sea(x)=\bar q^\sea(x)$.

Finally there is the gluon contribution that needed to be divided to a factor of $100$ to be 
scaled to the frame.
\index{Quarks!Sea quarks|)}
\index{Parton Distribution Functions (PDFs)!Behaviour (CTEQ6)|)}

\subsubsection{Hadronic level at the improved leading order}\label{ss_hadr_lvl_iLO}
\index{Improved leading order|see{QCD}}
\index{QCD!Improved leading order|(}
Higher order corrections are necessary to get rid in the cross section as much as possible 
of the dependency from the unphysical renormalisation and factorisation scales due to the 
perturbation expansion truncation.
Nonetheless, the comprehension of the main features of the $W$ and leptons kinematics can be done
studying just an improved leading-order description of the phenomenon, and when necessary, use
higher order corrections to pin-point a particular effect in a given phase-space domain.

\begin{figure}[!h] 
  \begin{center}
    \includegraphics[width=.9\tw]{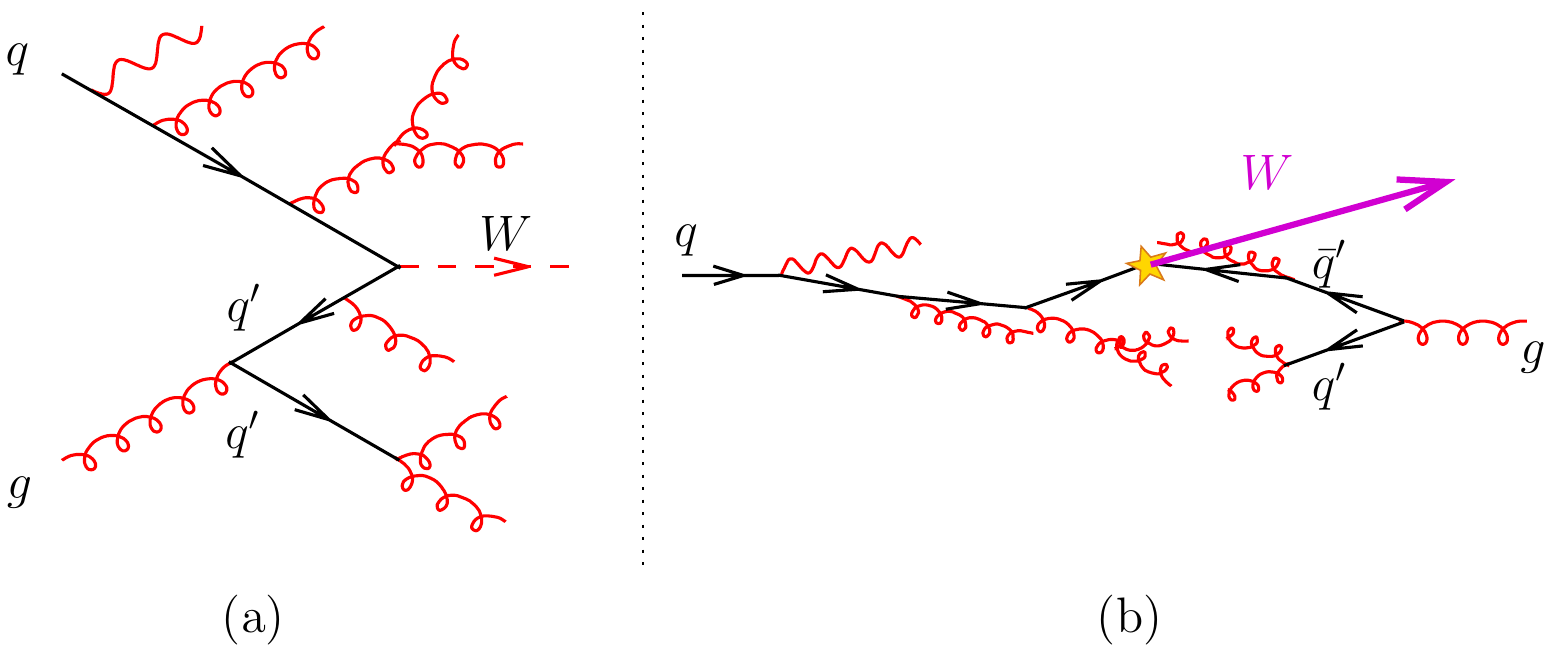}
    \caption[Radiation of real gluons and photons from the partons in the initial state]
	    {\figtxt{Example of the emission of real gluons and photons from the initial
                state partons in a Feynman-like representation (a) and in the associated
                $\vec x$-space representation (b).}}
	    \label{fig_qcd_qed_rad}
  \end{center} 
\end{figure}

This improved LO, considered in the rest of this Chapter, is defined by taking the LO expressions 
of the partonic cross section convoluted with the remormalised PDFs and by taking part
of the radiation of the partons in the initial state.
Due to momentum conservation, the radiation of gluons and photons off partons provide a realistic 
picture in which quarks produce a $W$ with a non zero transverse motion.
Note that by ``part of the radiation'' we mean that emission of real gluons and photons are taken into 
account but without embracing all other virtual loop corrections belonging to the same order in 
$\alphas$ or $\alpha$. In this scheme, Fig.~\ref{fig_qcd_qed_rad} represents an example of such
initial state radiation for a particular event.
\index{QCD!Improved leading order|)}

Before going any further common variables and conventions are reminded.
Usually, switching to a second set of variables $(\tau,\yW)$ equivalent to $(x_1,\,x_2)$ proves to 
be useful. This change of variables is defined by
\begin{eqnarray}
\tau &\equiv& x_1 x_2, \\
\yW  &\equiv& \frac{1}{2}\ln\left( \frac{E_W+p_{z,W}}{E_W-p_{z,W}} \right),
\label{eq_W_rapidity}
\end{eqnarray}
At leading order, using the expressions of the quark and anti-quark momenta from 
Eqs.~(\ref{eq_qqbp_4mom_LO}) gives for the rapidity the expression
\begin{equation}
\yW = \frac{1}{2}\,\ln\left(x_1/x_2\right), \label{eq_yW_LO}
\end{equation}
which for $x_1$ and $x_2$ when expressed in function of $\tau$ and $\yW$ give the following expressions
\begin{eqnarray}
x_1  &=& \sqrt{\tau}\,\mm{exp}(\yW), \label{eq_x1x2_in_func_tauyW0}\\
x_2  &=& \sqrt{\tau}\,\mm{exp}(-\yW).\label{eq_x1x2_in_func_tauyW}
\end{eqnarray}

\index{W boson@$W$ boson!Width@Width $\GamW$!Narrow width approximation|(}
To get rough estimations, in top of relying on LO expressions, the 
``narrow width approximation'' is commonly used, 
it is defined assuming $\GamW/\MW \ll 1 \Rightarrow\GamW=0$, the consequences for the 
mass and $\tau$ being then
\begin{eqnarray}
\GamW=0 &\Rightarrow& m_W=\MW, \\
\GamW=0 &\Rightarrow& \tau = \frac{\MW}{\sqrt{S}} \approx 6\times 10^{-3},
\label{eq_tau_narrow_width}
\end{eqnarray}
where for reminder $\sqrt S= 14\TeV$ in the case of $\pp$ collisions at the LHC.
The numerical value in the last equality gives an idea of the percentage of the total energy 
$\sqrt S$ needed to produce a $W$.
In this approximation the partonic cross section at LO can be expressed with a Dirac function.
The Breit--Wigner term in Eq.~(\ref{sigmahat_tot_LO}) 
becomes\footnote{For reminder the Cauchy-Lorentz function 
$\mathcal{L}(x;x_0,\es) =\tfrac{1}{\pi}\;\tfrac{\es}{(x-x_0)^2+\es^2}$
tends to a Dirac distribution $\delta(x-x_0)$ as $\es$ reaches zero with positive values.}
\begin{equation}
\lim_{\GamW\to 0}\frac{1}{\left(\shat-M_W^2\right)^2 + \left(\MW\GamW\right)^2} 
= \frac{\pi}{\MW\GamW}\;\delta\left(\shat - M_W^2\right),
\end{equation}
that is any dependency from $\shat$ vanishes, the partonic cross section is a constant which 
is non null only if the energy of the partons in the center of mass exactly equals $\MW$.
\index{W boson@$W$ boson!Width@Width $\GamW$!Narrow width approximation|)}

The gist of the $W$ and leptons kinematics are illustrated below with the process $W\to l\,\nul$, 
that is merging both positive and negative channels and looking at the $W$ and leptons kinematics.
The Monte Carlo used to produce these preliminaries pedagogical histograms is \WINHAC{}
which will be described with more details in Chapter~\ref{chap_winhac}.
All Monte Carlo predictions in this document are, unless stated otherwise, produced using \WINHAC{}.
The derivations presented below will serves as a base to understand later on
with more refinement the production of $W$ for each electrical charge channel separately.

\begin{figure}[!h] 
  \begin{center}
    \includegraphics[width=0.495\tw]{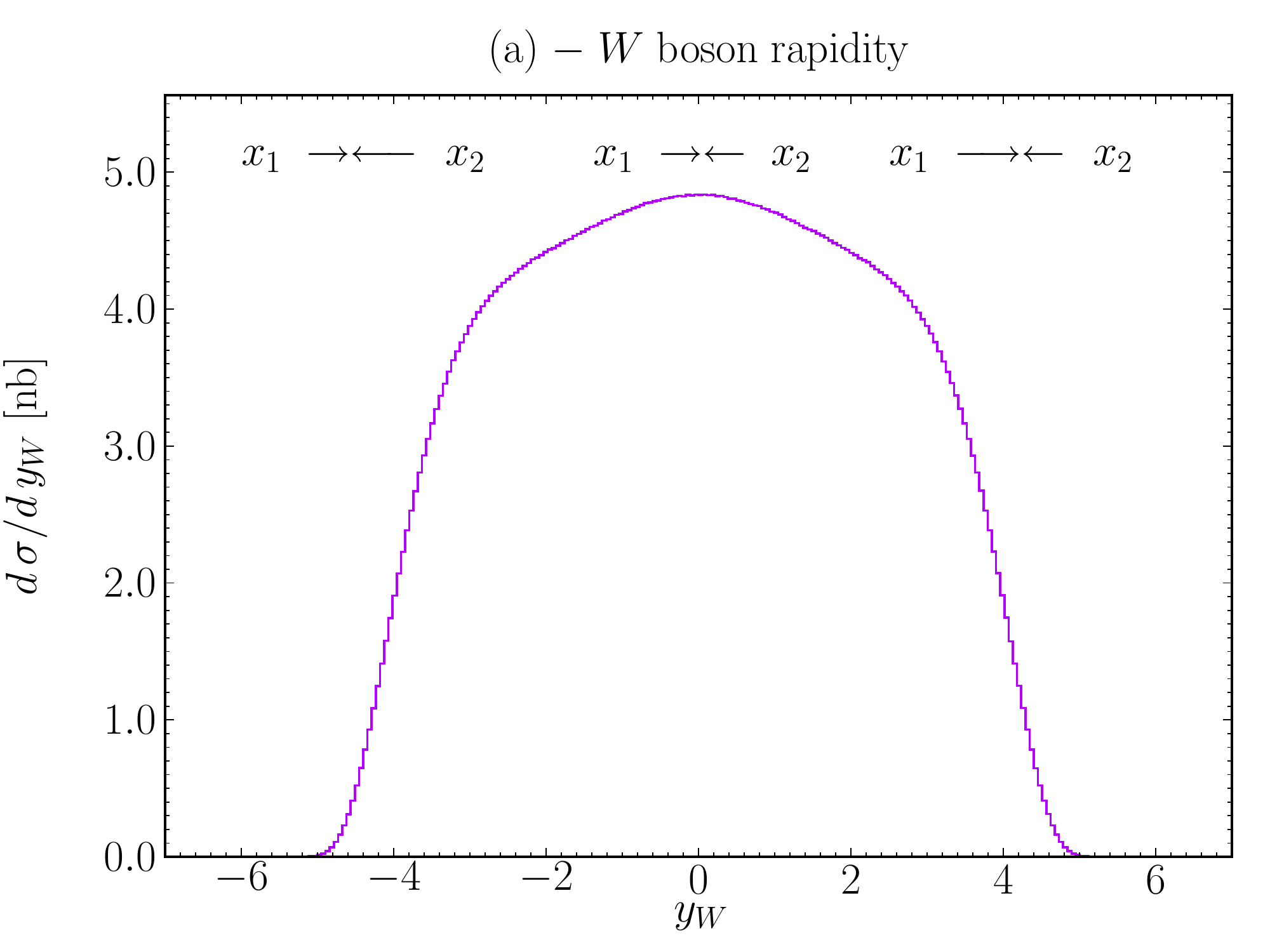}
    \hfill
    \includegraphics[width=0.495\tw]{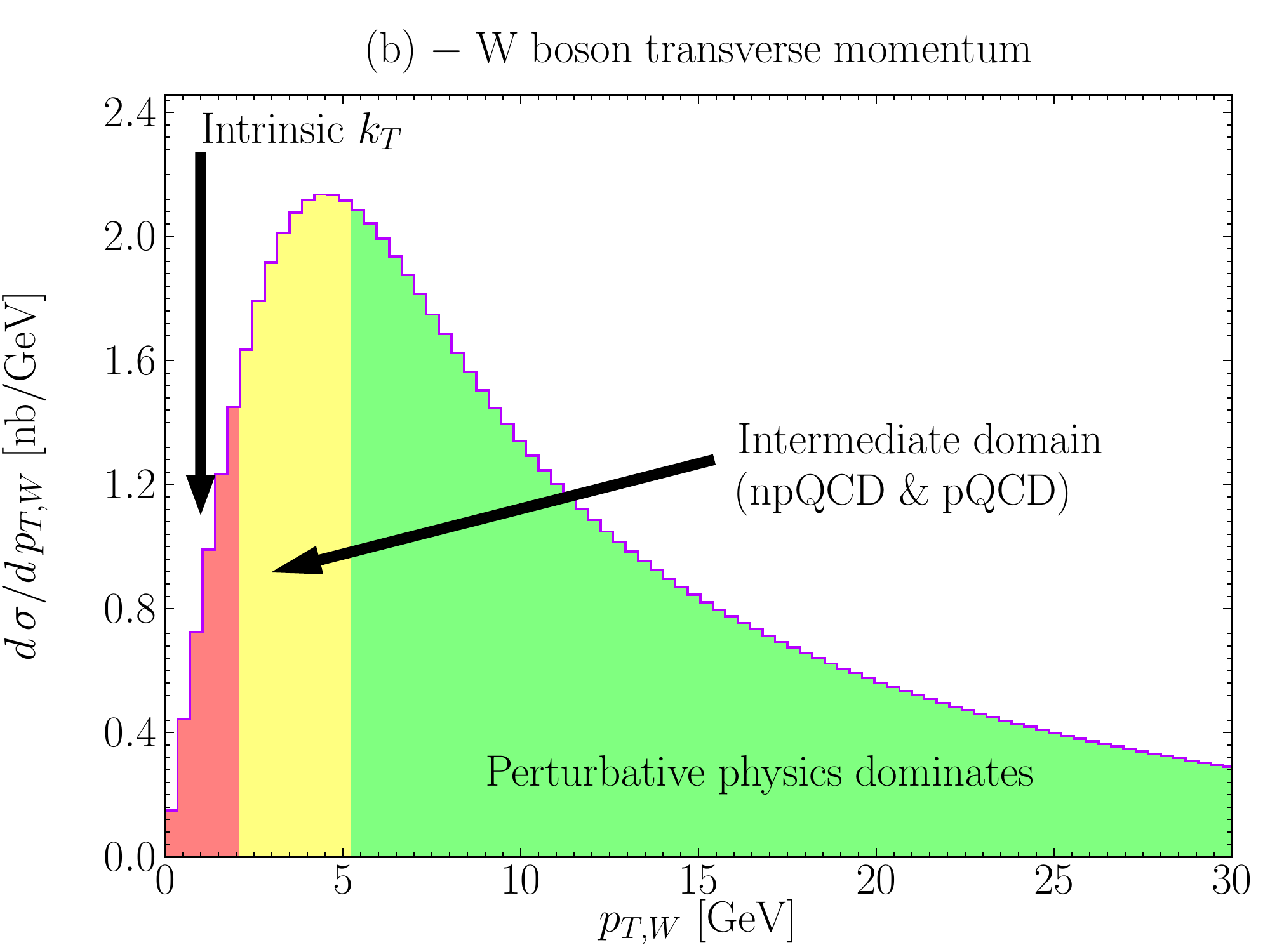}
    \caption[General features of a the $W$ kinematics in Drell--Yan]
	    {\figtxt{General features of a the $W$ rapidity (a) and 
                transverse momentum (b) in Drell--Yan at the improved LO order.
                The ``intermediate'' domain in $\pTW$ is ruled by both non-perturbative and
                perturbative QCD (npQCD \& pQCD).
                }}
	    \label{fig_yW_pTW}
  \end{center} 
\end{figure}
\paragraph{Rapidity of the W boson.}
\index{W boson@$W$ boson!Rapidity|(}
In the improved LO picture the rapidity approximately equals the LO expression Eq.~(\ref{eq_yW_LO}),
that is $\yW\approx \ln(x_1/x_2)$.
This form translates an unbalance between the fractions $x_1$ and $x_2$, and explains why the 
rapidity is strongly correlated to the PDFs.
To the leading order approximation we add the one of the narrow width, in that context
the $x_1$ and $x_2$ fractions entering the PDFs now reads
\begin{equation}
\sum_{i,j}\PDF{A}{i}(x_1)\,\PDF{B}{j}(x_2) \approx 
\sum_{i,j}\PDF{A}{i}\left(\tfrac{\MW}{\sqrt S}\,\mm{e}^{\yW}\right)\,
\PDF{B}{j}\left(\tfrac{\MW}{\sqrt S}\,\mm{e}^{-\yW}\right).
\end{equation}
The pattern of the rapidity distribution can be devised only from the previous expression, 
the weight from the partonic cross section being just a multiplicative factor.

The range for the rapidity can be deduced using the constraints on $x_1$ and $x_2$
\begin{equation}
0<x_{1,2}<1 \qquad\Rightarrow\qquad
|y_{W,\Max}| \approx \ln\left(\tfrac{\sqrt{S}}{\MW}\right),\label{eq_yWmax}
\end{equation}
which means a higher energy in the collision provides more kinematic energy to the $W$
while its mass $\MW$ reduces this kinematic energy.
Replacing in Eq.~(\ref{eq_yWmax}) $\MW$ and $\sqrt S$ by their values one find that at the LHC 
the rapidity range is approximately $|y_{W,\Max}|\approx 5$ as displayed in Fig.~\ref{fig_yW_pTW}.

The $W$ rapidity distribution tends to zero when $|\yW|\to 5$ because one of the fraction $x$ 
is necessarily tending to $1$ and as seen previously $f(x\to 1)\to 0$.
Considering, in a first approximation, that $x$ fractions are of the order of 
$x \sim \MW/\sqrt S$ one find that in average $x_\mm{LHC}\sim 6\times 10^{-3}$ while for the 
Tevatron energies $x_\mm{Teva.}\sim 4\times 10^{-2}$.
This shows how the energy in the LHC collisions will make small-$x$ contributions predominant with respect 
to the one of the Tevatron.\index{Tevatron collider}

In the region $\yW\approx 0$, the form of the rapidity seen in Eq.~(\ref{eq_yW_LO}) indicates that
$x_1\approx x_2$, which happens for sea quarks most of the time.
Indeed, substituting in Eqs.~(\ref{eq_x1x2_in_func_tauyW0}--\ref{eq_x1x2_in_func_tauyW})
$\MW/\sqrt S\sim 6\times 10^{-3}$ and $\yW=0$ gives $x_{1,2}\approx 6\times 10^{-3}$.
For these values of $x$ the probability of having a valence quark is small, most contributions comes
from sea quarks.

On the other hand, in the forward rapidity region, there must be an important unbalance between 
$x_1$ and $x_2$. In an extreme case scenario, this condition is fulfilled when a quark bears an 
important fraction $x_\mm{high}$ while the other one possesses a very small fraction $x_\mm{low}$, 
which is most probably occurring when a valence quark and a low energy sea quark collide.
This can be shown using rough values, substituting again in Eqs.~(\ref{eq_x1x2_in_func_tauyW0}--%
\ref{eq_x1x2_in_func_tauyW}) $\MW/\sqrt S\approx 6\times 10^{-3}$ and $|\yW|\approx 4$ gives, 
$x_{\mm{high}}\approx 0.3$ and $x_{\mm{low}}\approx 10^{-4}$.
The fall of the rapidity, occurring at $\yW\approx 3$, is then a consequence of the fall of the 
valence quarks PDFs densities as $x\to 1$.

Let us remark that the impossibility to measure the neutrino longitudinal component $p_{z,\nu}$
implies that while trying to resolve it using the other measurable kinematics one ends up with an 
equation of second degree in $p_{z,\nu}$ with two solutions leading in turn to two ambiguous solutions
for $\yW$. Nonetheless, even if the problem is not solvable on an event-per-event basis, some
solution based on the whole data event allows to partly overpass the problem
in the narrow-width approximation~\cite{Bodek:2007cz}.
\index{W boson@$W$ boson!Rapidity|)}

\paragraph{Transverse momentum of W boson.}
\index{W boson@$W$ boson!Transverse momentum|(}
\index{Quarks!Intrinsic transverse momenta|(}
The distribution of $\pTW$ (Fig.~\ref{fig_yW_pTW}.(b)) in this improved LO picture is the 
consequence of three main effects. 
First is the perturbative emission of real gluons and quark/anti-quark gluon scattering
(Fig.~\ref{fig_nlo}.(b,c)) which dominates at high $\pT$. 
As $\pT$ decreases the spectrum is governed by the re-summation of leading-logs, then finally
below $\pTW\approx 2\GeV$ the intrinsic transverse momentum of partons inside the proton dominates.

Since the intrinsic $\kT$ of partons modeling enters our analysis more details are given below.
This intrinsic $\kT$ can be initially apprehended as the consequence of the Heisenberg uncertainty 
principle applied to the confinement of the partons in a finite volume of the order of a Fermi. 
Still, a quick calculus shows this effect cannot totally account for the observed data
and this initial $\kT$ is nowadays depending on the energy involved in the collider.
Assuming a simple factorisation of this effect from the longitudinal motion ruled by the PDFs
the expression of the latter necessary to the cross section calculus can be accounted doing 
the following substitution
\begin{equation}
f(x) \;\to\; h(\vec{k}_T)\,f(x),
\end{equation}
where $h(\vec{k}_T)$ is the density of probability for a parton to have a transverse momentum
of $\vec{k}_T$. Modeled with a Gaussian distribution it can reads\,:
\begin{equation}
h(\vec{k}_T) =\frac{b}{\pi}\,\mm{exp}(-b\,k_T^2),
\end{equation}
where the average value of $\kT$ noted $\Mean{\kT}$ is $\Mean{\kT}=\sqrt{\pi/(4\,b)}$.
This values needs to be estimated for a particular collider to match to the data.
In the case of the Tevatron energies it is estimated to be of $\Mean{\kT}\sim 2.2\GeV$
and should be higher at the LHC.
\index{W boson@$W$ boson!Transverse momentum|)}
\index{Tevatron collider!Partons intrinsic transverse momentum}
\index{Quarks!Intrinsic transverse momenta!At the Tevatron energies}
\index{Quarks!Intrinsic transverse momenta|)}

\begin{figure}[!h] 
  \begin{center}
    \includegraphics[width=0.495\tw]{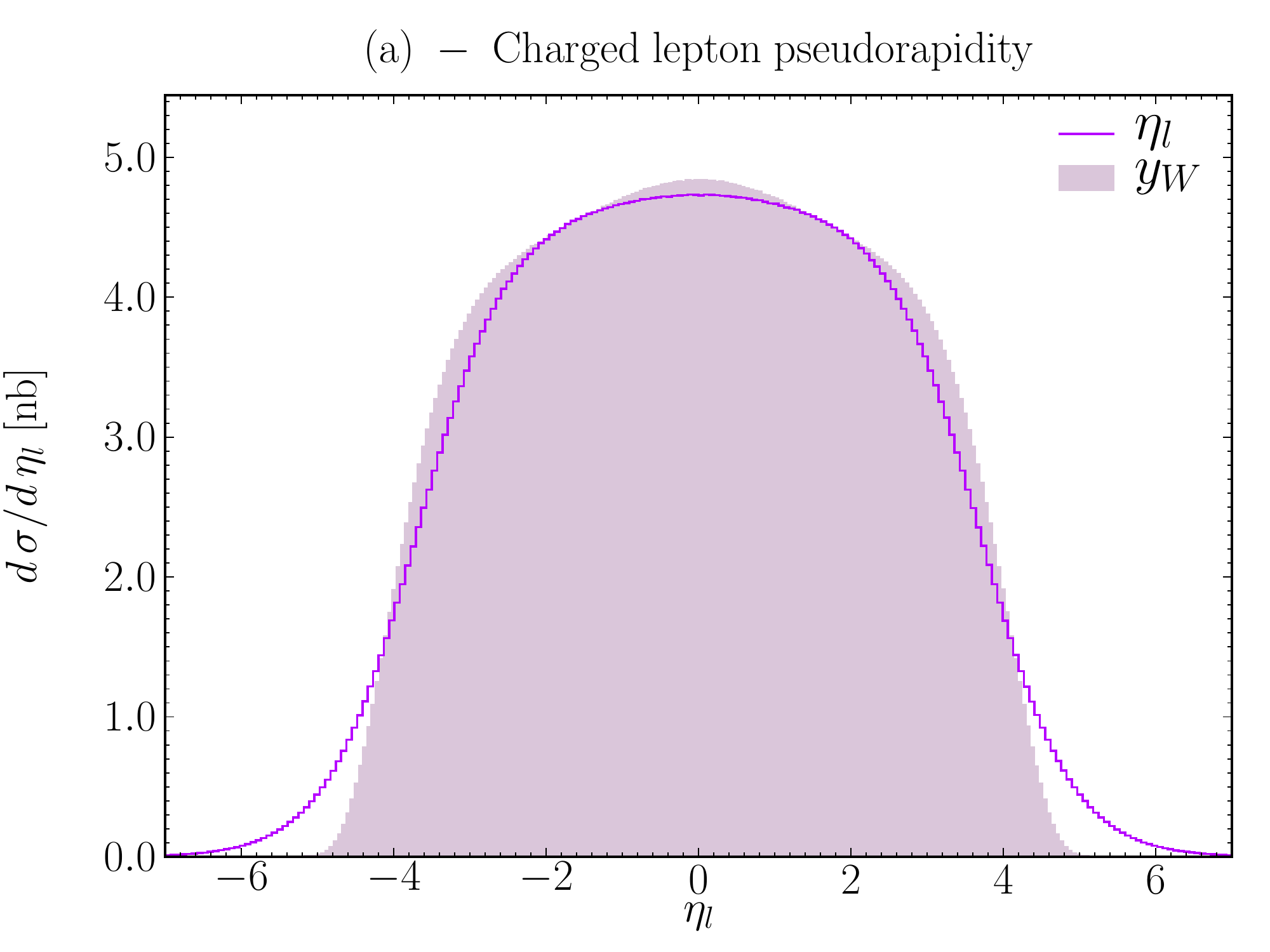}
    \hfill
    \includegraphics[width=0.495\tw]{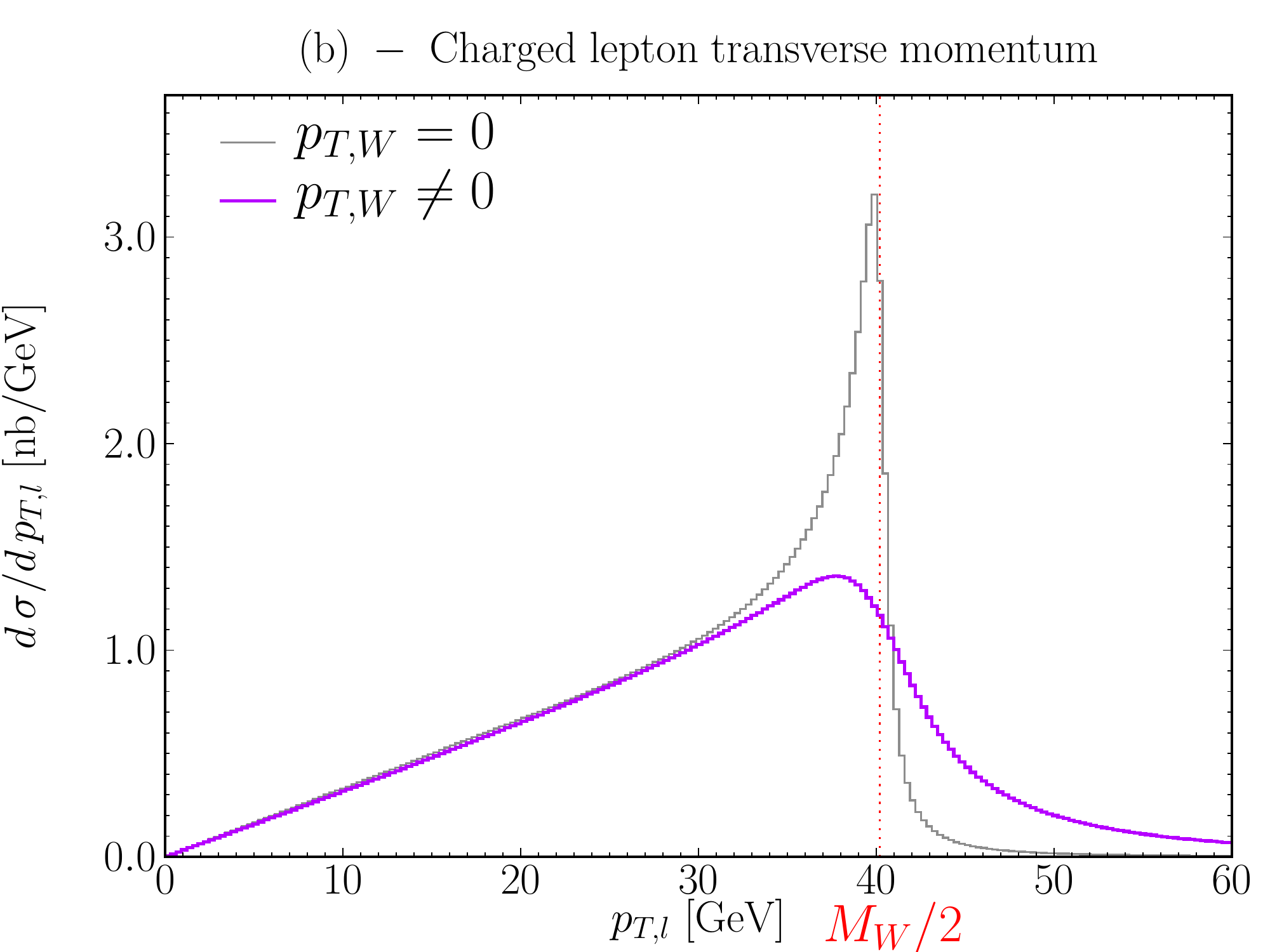}
    \caption[General features of the charged lepton kinematics decaying from $W$ in Drell--Yan]
	    {\figtxt{General features of the charged lepton pseudo-rapidity (c) 
                and transverse momentum (b) in Drell--Yan at the improved LO order.
                In the case of $\pTl$ both LO and improved LO are shown.}}
	    \label{fig_etal_pTl}
  \end{center} 
\end{figure}
\paragraph{Pseudo-rapidity of the charged lepton.}
\index{Charged lepton@Charged lepton from $W$ decay!Pseudo-rapidity|(}
The pseudo-rapidity of the charged lepton is obtained by adding to its intrinsic rapidity 
$ y_l^\ast$ in the \WRF{} the rapidity of the $W$, it reads at LO
\begin{equation}
\etal = \yW + y_l^\ast.\label{eq_etal_yW_ylast}
\end{equation}
The rapidity of the $W$, as shown previously, is related to the energy in the collision and 
the mass of the $W$ while the intrinsic rapidity $ y_l^\ast$ is a consequence of the angular
decay governed by the $V-A$ coupling.
Indeed, still at LO, $y_l^\ast$ takes the form
\begin{equation}
y_l^\ast = \frac{1}{2}\ln\left(\frac{1+\costhetaWlwrf}{1-\costhetaWlwrf}\right).
\label{eq_y_l_wrf_LO}
\end{equation}
The estimation of the contribution of the intrinsic rapidity of the lepton to the total 
pseudo-rapidity can be evaluated roughly.
We make the assumption that in average there is as much charged leptons spreading 
the fundamental $\yW$ distribution that narrowing it, which can be formalised stating there is 
as much $<\costhetaWlwrf>\approx 1/2$ than $<\costhetaWlwrf>\approx -1/2$.
Substituting these averaged values of $\costhetaWlwrf$ in Eq.~(\ref{eq_y_l_wrf_LO}) gives
$\Mean{y_l^\ast}=\pm 0.6$, \ie{} $|\Mean{y_l^\ast}|<|\Mean{\yW}|$.
Then, the main pattern of the $\etal$ distribution is mostly inherited from the one present
in the $\yW$ distribution.
This explains why $\etal$ distribution covers sensibly the same range than $\yW$ 
as can be seen in Fig.~\ref{fig_etal_pTl}.(c).
Because the $W$ rapidity cannot be measured the $\etal$ distribution is used as a substitute and
proves particularly useful in the PDFs study. More details on that topic in the context of
the ATLAS experiment can be found in Ref.~\cite{ATLASphdLohwasser}.

To be completely honest here, and tease the reader's curiosity, the assumption that
spreading and narrowing behaviour of the $\etal$ distribution occurs in the same proportions
is more or less verified at the Tevatron but it will not be the case at the LHC.
In the latter case the $y_\lp^\ast$ tends to narrow the inner $\yW$ spectrum while the 
$y_\lm^\ast$ spreads the latter.
Then, the previous calculus is justified since both charge being merged the narrowing and
spreading compensate each other and also because $\Mean{y_l^\ast}<\Mean{\yW}$.
\index{Charged lepton@Charged lepton from $W$ decay!Pseudo-rapidity|)}

\paragraph{Transverse momentum of the charged lepton.}
\index{Charged lepton@Charged lepton from $W$ decay!Transverse momentum|(}
\index{W boson@$W$ boson!Mass@Mass $\MW$!Link to the charged lepton@Link to the charged lepton $\pT$}
The distribution for the transverse momentum of the charged lepton at the hadronic level is
directly linked to the one at the partonic level. At leading order, $\pTW=0$ which implies
$\pTl=\pTlstar$. Then a jacobian peak around $\approx \MW/2$ is visible, its smearing  
coming from contributions in which $m_W\approx \MW$. 
Contributions displaying a value $m_W$ too far from the central value $\MW$ are, 
as it was shown in Eq.(\ref{eq_dsigma_dpTlstar}) highly improbable.
Then, the $\pTl$ distribution which is observable in a detector presents a jacobian peak which 
position allows to deduce the value of $\MW$.

Things gets more complicated when going to a more realistic scenario where the $W$ possesses a 
transverse motion. To understand that the relative angle between $\pTW$ and $\pTlstar$ in the
$r-\phi$ plane noted $\cos \phi_{W,l}^\ast$ is considered.
The sign of $\cos \phi_{W,l}^\ast$ give rise to ambivalent behaviour for $\pTl$.
In the first case, where $\cos \phi_{W,l}^\ast>0$, $\pTW$ is such that the total Lorentz boost 
based from $\vec p_W$ increase the value of $\pTl$.
In such events then $\pTl$ are shifted to higher values, to the right of the Jacobian peak. 
On the contrary, in $(\cos \phi_{W,l}^\ast<0)$-cases $\pTl$ are decreased, that is shifted to
the left of the Jacobian peak.

Figure~\ref{fig_etal_pTl}.(b) shows how much the transverse momentum of the $W$ smears the
sharpness of the Jacobian peak. 
That means that the extraction, from the bare $\pTl$ distribution, of $\MW$ with a Monte Carlo
imply a refined implementation of the $\pTW$.

Let us note also that the smearing of the jacobian peak being partially due to the $W$ width
the tail of the distribution beyond $\approx \MW/2$ can be used to measure $\GamW$. This is however
beyond the scope of the present document.
The reader interested in this topic can consult Refs.~\cite{Barger,Ellis:1991qj} for example.
\index{W boson@$W$ boson!Width@Width $\GamW$}
\index{Charged lepton@Charged lepton from $W$ decay!Transverse momentum|)}

\paragraph{Transverse momentum of the neutrino.}
The transverse momentum distribution of the neutrino displays the very same features than the one
of the charged lepton. 
Still from the experimental point of view, it is very different as $\pTnu$ can be deduced only
from the missing transverse energy given by the entire response of the calorimetry of a detector.
For that reason $\pTnu$ is preferentially noted $\slashiv{p}_{T,\nu}$ or $\ETmiss$.
The calibration of the missing transverse energy needs at least a few years, and even after that it cannot
compete with the study of the charged lepton, indeed as can be seen in 
Table~\ref{table_cdf_ii_MW_mT_pTl_pTnu} the CDF II results~\cite{Aaltonen:2007ps} 
shows large systematic errors on $\MW$ when using the $\slashiv{p}_{T,\nu}$ distribution.

\paragraph{Transverse mass of the lepton pair.}
\index{Transverse mass of the lepton pair!Definition}
\index{W boson@$W$ boson!Mass@Mass $\MW$!Link to the lepton pair transverse mass}
The transverse mass was suggested in Refs.~\cite{vanNeerven:1982mz,Smith:1983aa} to provide an alternative 
to measure the mass and width of the $W$ boson.
It consists of calculating the invariant mass $\mTlnu$ the lepton pair would have if 
$p_{z,l}$ and $p_{z,\nu}$ would be null.
For that reason it is called the 
transverse mass\footnote{This definition of the transverse mass should not be confused with the
one introduced sometimes in relativistic kinematics and defined as $m_T^2\equiv m^2 + p_T^2$,
like for example in the \textit{Kinematic} review in Ref.~\cite{Amsler:2008zz}.}
is defined like
\begin{eqnarray}
\mTlnu &\equiv& \sqrt{(p_{T,l} + p_{T,\nul})^2 - (\vec p_{T,l} + \vec p_{T,\nul})^2}, \\
       &=& \sqrt{2\,p_l\,p_\nul\,\left(1-\cos\phi_{l\nul}\right)}.
\end{eqnarray}
Now since the transverse momenta of both leptons enter as a product the influence of $\pTW$
upon $\pTl$ is counterbalanced by the opposite effect upon $\pTnu$.
At LO in the \WRF{} both leptons transverse momenta are equal, a purely longitudinal boost
does not affect these values, hence $\pTl=\pTnu$ and $\cos\phi_{l\nul}=0$ which gives for the
transverse mass the $\mTlnu=2\,\pTl$,
explaining why in this case the jacobian peak is located at $\MW$.

\begin{figure}[!h] 
  \begin{center}
    \includegraphics[width=1.\tw]{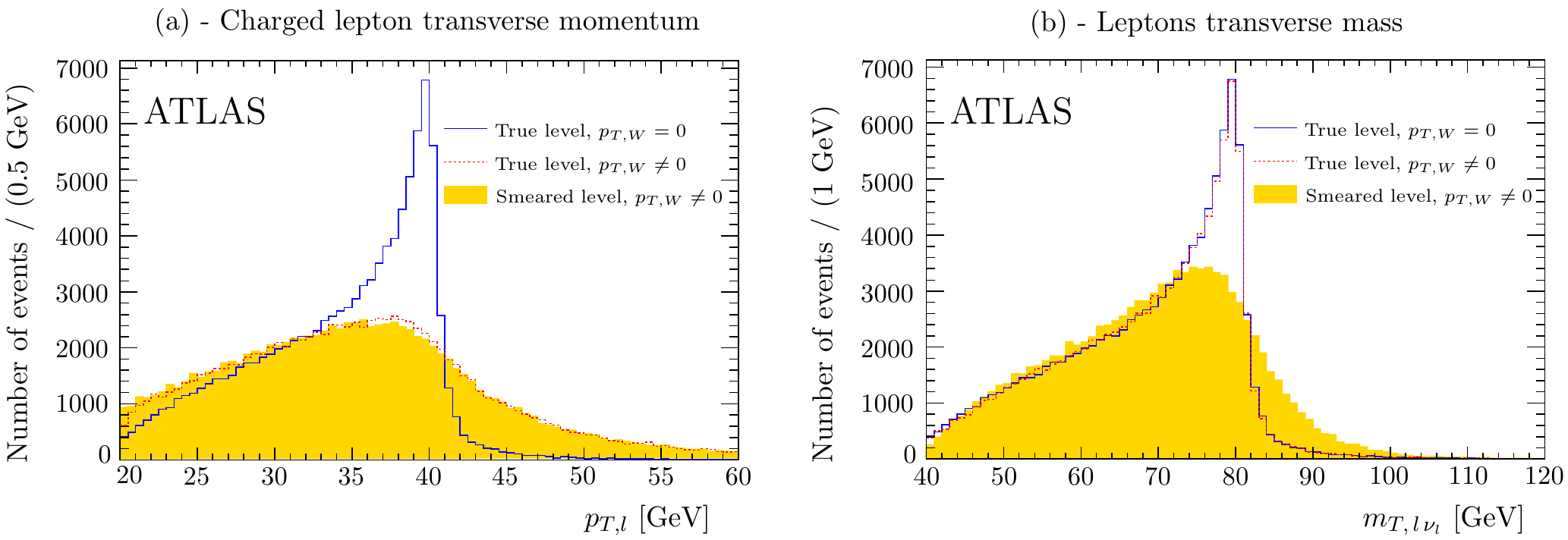}
    \caption[Transverse momentum of the electron and transverse mass of the 
      electron-neutrino pair at the particle level for $\pTW=0$, $\pTW\neq 0$ and
      for the ATLAS resolution.]
	    {\figtxt{Transverse momentum of the lepton (a) and of the transverse mass (b)
                at the true level for $\pTW=0$, $\pTW\neq 0$ and
                at the smeared level with $\pTW\neq 0$ based on the ATLAS detector resolution
                [Data extracted from~\cite{ATL-COM-PHYS-2008-243}].
            }}
	    \label{fig_pTl_hadr_look}
            \index{Transverse mass of the lepton pair}
            \index{Transverse momentum of the charged lepton}
  \end{center} 
\end{figure}

Figure~\ref{fig_pTl_hadr_look} represents the $\pTl$ and $\mTlnu$ distributions for three
different steps in the simulation of $W$ in Drell--Yan inside ATLAS~\cite{ATL-COM-PHYS-2008-243}.
Note that in each step trigger and acceptance cuts were made according to the ATLAS requirements,
in particular $\pTl$ and $\ETmiss$ cuts can be clearly distinguished looking at the low $\pT$
region of the histograms, for further details see Ref.~\cite{ATL-COM-PHYS-2008-243}.
The first case considered is the true level, that is the prediction from the Monte Carlo
at pure leading order where $\pTW=0$.
Both $\pTl$ and $\mTlnu$ jacobian peaks are sharp.
When adding up QCD corrections in the initial state the jacobian peak of $\pTl$ is very
smeared by $\pTW\neq 0$, while $\mTlnu$ still displays a sharp jacobian peak.
The last step consists to pass the latter event generations to a simulation of the
ATLAS detector. The good resolution for the charged lepton does not change a lot, 
while the bad resolution for $\ETmiss$ smears a lot the jacobian peak of $\mTlnu$.

\begin{table}[]
\begin{center}
\renewcommand\arraystretch{1.2}
\begin{tabular}{cc@{$\,\pm\,$}c@{$\,\pm\,$}l}
  \hline
  Distribution               & $\MW$&$\delta_\MW^\mm{\,(stat.)}$&$\delta_\MW^\mm{\,(sys.)}$ 
  [$\mm{GeV}$]\\
  \hline\hline
  $m_{T,e\nu_e}$               & $80.493 $ & $ 0.048 $ & $ 0.039$ \\
  $p_{T,e}$                   & $80.451 $ & $ 0.058 $ & $ 0.045$ \\
  $\slashiv{p}_{T,\nu_e}$      & $80.473 $ & $ 0.057 $ & $ 0.054$ \\
  \hline
  $m_{T,\mu\nu_\mu}$            & $80.349 $ & $ 0.054 $ & $ 0.027$ \\
  $p_{T,\mu}$                  & $80.321 $ & $ 0.066 $ & $ 0.040$ \\
  $\slashiv{p}_{T,\nu_\mu}$     & $80.396 $ & $ 0.066 $ & $ 0.046$ \\
  \hline
\end{tabular}
\renewcommand\arraystretch{1.45}
\caption[CDF II results for the mass of the $W$]
        {\figtxt{CDF II results for the mass of the $W$~\cite{Aaltonen:2007ps} 
            for the electronic and muonic channel
            using the transverse mass, charged lepton and neutrino transverse 
            distributions. The statistical error is noted $\delta_\MW^\mm{(stat.)}$
            and the systematic error $\delta_\MW^\mm{(sys.)}$.}}
\label{table_cdf_ii_MW_mT_pTl_pTnu}
\index{Transverse mass of the lepton pair}
\index{W boson@$W$ boson!Mass@Mass $\MW$!In CDF}
\end{center}
\end{table}
\index{Drell--Yan processes for W@Drell--Yan processes for $W$!Generalities|)}

\section{An invitation to the rest of the document}
Up to now, the motivation and the general context for a measurement of the $W$ charge asymmetry $\MWp-\MWm$
within the Standard Model paradigm have been described.
Now the outline of the rest of the document is presented through a short description of each 
Chapter aims and content.

Chapter~\ref{chap_atlas_exp} presents the experimental context in which this measurement can be 
achieved, that is at the LHC and more precisely with the ATLAS detector. 
Especially, the emphasis is made on the ATLAS tracker since the analysis was restricted using this
sub-detector capabilities.
Some highlights are made on the deformations of the tracker that should increase the systematic 
errors on $\MWp-\MWm$.

Chapter~\ref{chap_winhac} presents the tools that were used or implemented to carry out
our analysis to evaluate the ATLAS potential to measure $\MWp-\MWm$.
It almost exclusively treats about the Monte Carlo event generator \WINHAC{}.
The physics inside it is described along with the tools implemented downstream for 
the stand-alone analysis that was made.
The Chapter describes as well the work done to include this Monte Carlo event generator
inside the ATLAS software environment.

Chapter~\ref{chap_w_pheno_in_drell-yan} presents the detailed work that was done to understand the 
kinematics of the $\Wp$ and $\Wm$ at the LHC made necessary for the prospect of the measurement of
$\MWp-\MWm$ but as well as for our other ongoing effort made on $W$ properties extraction
such as $\MW$~\cite{Upcoming_MW,SiodmokPhD} or $\GamW$~\cite{Upcoming_GammaW}.
Actually we consider this Chapter to be of interest for every physicist working on the $W$ 
production in Drell--Yan at the LHC, for ATLAS, CMS and even LHC$b$ detectors.

Chapter~\ref{chap_W_mass_asym} presents dedicated strategies to measure $\MWp-\MWm$ and
their qualitative evaluation.
After a short introduction of the general experimental context 
(trigger, acceptance, background, \etc{}.) the analysis strategies devised to measure
the $W$ boson mass charge asymmetry. 
The strategies were specifically designed for this measurement
to get rid as much as possible of the dependency from both Monte Carlo and apparatus
imperfections to decrease the impact of systematic errors on $\MWp-\MWm$.

A Conclusion closes the document by summing up the results obtained with the proposed methods and,
by taking a step backward, localise the importance of the present work amid the ongoing effort
on precision measurements of the $W$ boson properties at the LHC.

\chapter{The ATLAS experiment}\label{chap_atlas_exp}
\setlength{\epigraphwidth}{0.7\tw}
\epigraph{
Atlas\,: ``Where is he ?
Where is your titan  ?!
I will show him sorrow! I will show him pain ! I will show him Atlas !
I am the champion this city, this land, this orb needs.''
}%
{\textit{Superman \#677 - In The SHADOW Of ATLAS (August 2008)}}

The Large Hadron Collider 
(LHC)~\cite{LHC:1987nr,LHC:CERN-2004-003-V-1,LHC:CERN-2004-003-V-2,LHCwebsite,Evans:2008zz} 
will be the largest circular accelerator of hadrons ever build providing unprecedented energy 
and luminosity.
It is buried $\approx 100-150\,\mm{m}$ underground within the CERN~\cite{CERNwebsite} facility near 
Geneva.
This accelerator was designed to fulfill the increasing needs to probe ever smaller lengths scales
to unravel the nature of the interactions of elementary particles.
For that purpose the LHC accelerates in its ring counter rotating bunches of hadrons 
at ultra relativistic speed and collide them in four distinct points.
The particles produced during these collisions are studied using one or several detectors built 
in the vicinity of each of the four interaction points. 
Among them is ATLAS~\cite{ATLAStdr,ATLASatCERN:2008zz} a general purpose detector.

In this chapter, two of the most fundamentals collider parameters from the physicist point of 
view, \ie{} the energy in the center of mass and the luminosity, are reminded in the LHC context.
Then, the ATLAS detector is presented by reviewing the geometry and technology implemented within 
each of its sub-detectors\,: the tracker, the calorimeters and the muon spectrometer.
Especially the tracker, which performances are used in this work, is described with more details.
Some emphasis are made on the aspects relevant to the measurement of $\MWp-\MWm$ through the observation 
of the process $\mm{hadron-hadron}\to W\to l\,\nul$ where $l$ for reminder stands for 
$l\equiv\{e,\mu\}$.

%
%
%
\section{The Large Hadron Collider}
\subsection{The collider}\label{ss_lhc}\index{LHC|(}

The need to produce physics resonances of ever higher masses at noticeable rates within colliders 
requires higher energy $\sqrt{S}$ \index{LHC!Energy of the colliding beams} 
in the center of mass of the collision as well as higher 
luminosity $\lumi$\index{Luminosity!At the LHC (expected)}.
The importance of synchrotron radiation in $\ep/\Em$ collisions is such that it proved to be 
prohibitive above LEP~\cite{LEP1} \index{LEP collider} energies to use circular collider.
Then, the idea of studying non~$\ep\Em$ collisions using the LEP tunnel began to be studied in 
1984~\cite{Asner:1984ys} and at the time the LEP collider shutdown in 2000, the project of
colliding hadrons was decided~\cite{LHC:1987nr}. 
So far hadron--hadron colliders were designed for $\ppbar$ collisions like at the SPS \index{SPS collider}
or the Tevatron colliders.
Above $3\TeV$ the cross sections for $\pp$ and $\ppbar$ collisions are comparable but for the 
latter case the luminosity would have been inferior to a factor $100$ at the LHC due to the 
difficulty to produce anti-protons.
On the other hand collisions of particles holding the same charge impose \textit{a priori} to have 
two independent rings having magnets of opposite polarity with separate cryostat which 
would have turned out to be more expensive.
For the LHC a special two-in-one dipole magnet was conceived in the aim to be more economic and compact.
Thus, $\pp$ collisions being more easily achievable and less costly it was the chosen option.
The LHC will as well make collisions of heavy positively charged ions.

From the physics point of view the most striking difference between hadron--hadron and $\ep\Em$ 
collisions is that in the primer case the colliding particles are composite and gives for each 
inelastic process of interest to the physicists an important number of other particles.
Hence, the LHC is primarily designed for discoveries rather than for precision 
measurements, still, its high center of mass energy and luminosity combined to the detectors 
performances should allow precision measurements.
Another consequence of the hadrons composition is that partons participating in an inelastic
scattering borrow only part of the whole disposable energy $\sqrt S$.
For the case of $W$ production in Drell--Yan we have seen in Eq.~(\ref{eq_tau_narrow_width}) that 
in average $\approx 0.6\percent$ of $\sqrt S$ is used.

\begin{figure}[!h] 
  \begin{center}
    \includegraphics[width=1.\tw]{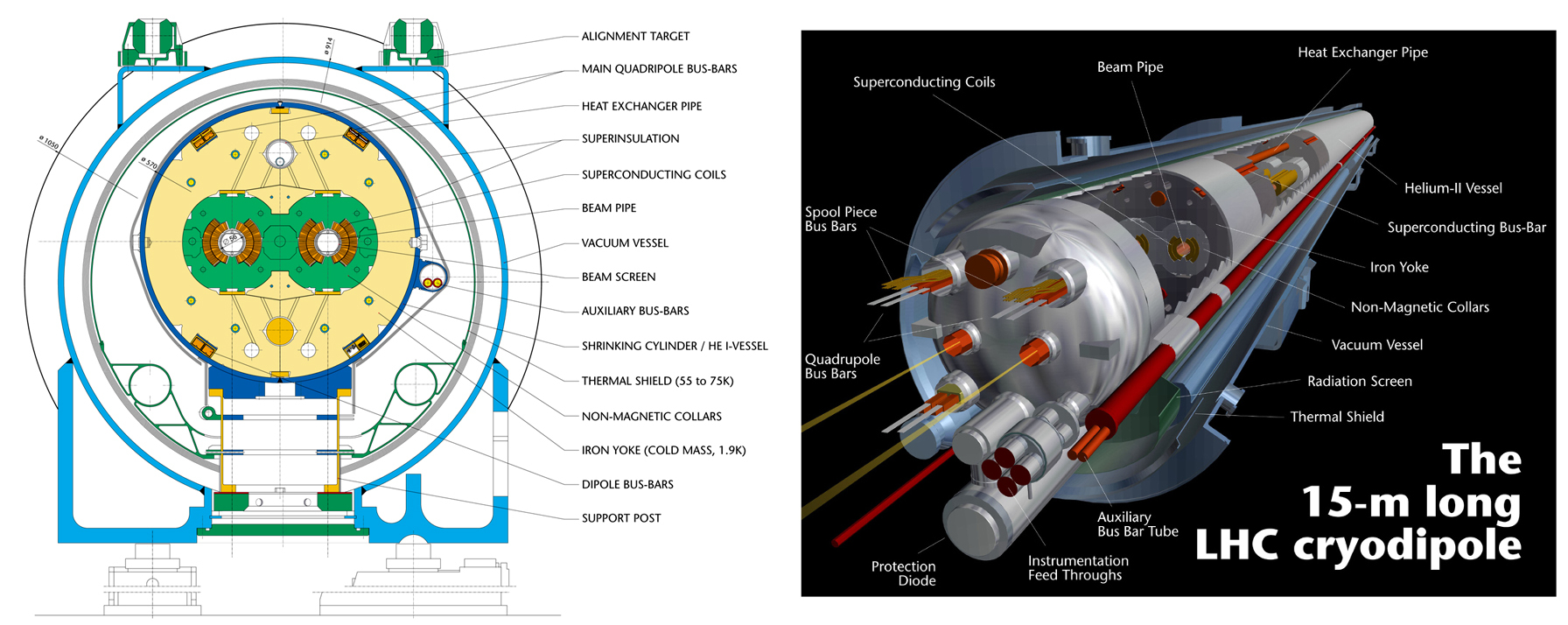}
    \vfill
    \includegraphics[width=1.\tw]{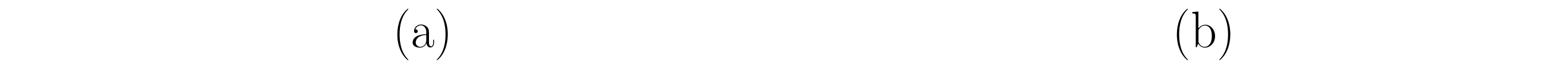}
    \caption[The LHC superconducting magnet]
            {\figtxt{The LHC superconducting magnet cross section (a) and in a 3D rendition (b)
              [\copyright CERN Geneva].}
            }
            \label{fig_dipole}
  \end{center} 
\end{figure}

Fig.~\ref{fig_dipole} shows the different parts entering each superconducting magnet dipole
that were assembled along the 27 km circumference of the LHC ring. 
Each proton beams are moving in opposite directions inside each of the two main dipole aperture 
beam-pipe.
The dipole generates a magnetic field of $8.33\Tesla$ to maintain the protons beams along the 
ring circumference, other magnets are present to focus and correct any deviations of the beams.

\begin{figure}[!h] 
  \begin{center}
    \includegraphics[width=.9\tw]{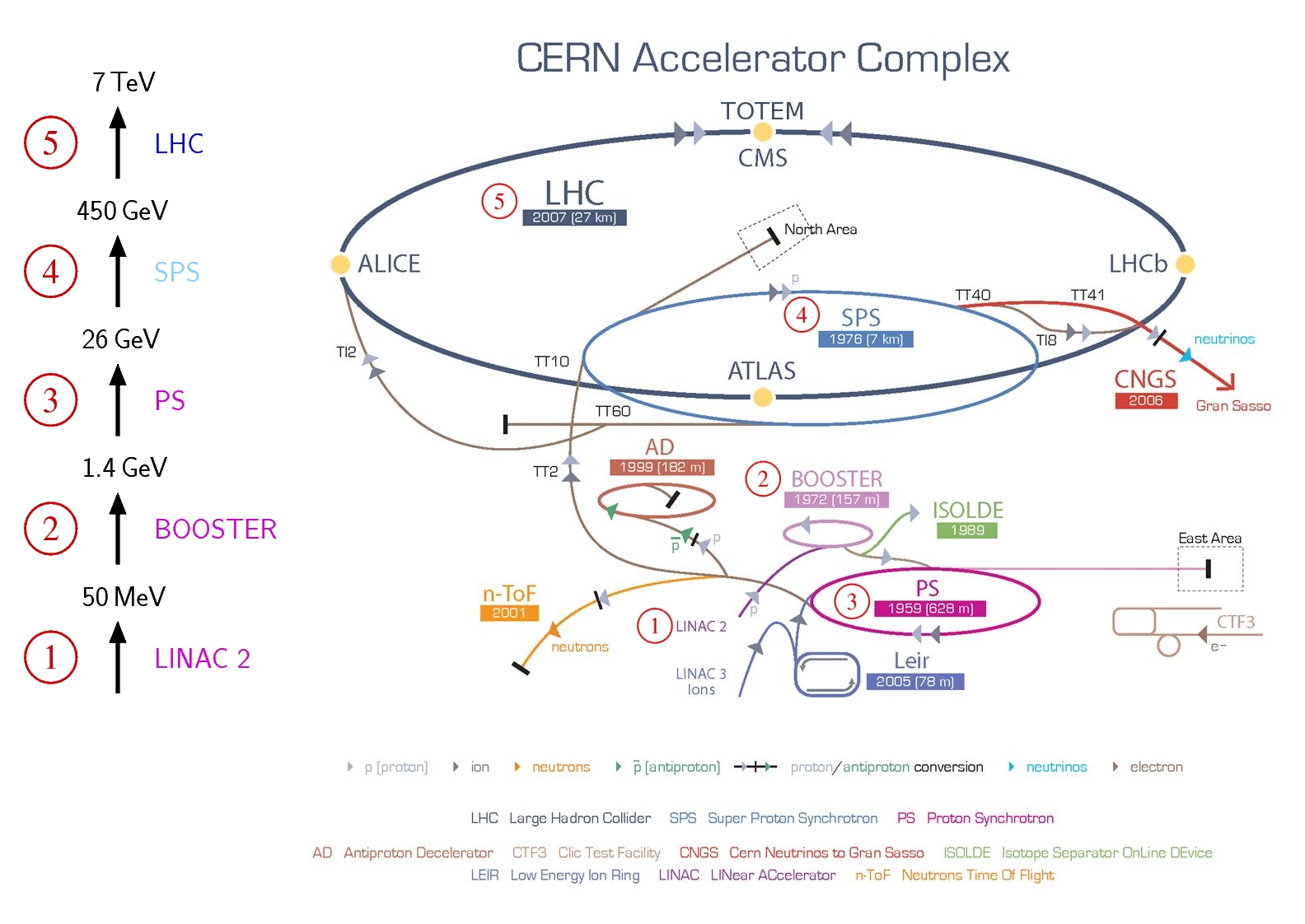}
    \caption[The CERN accelerator complex]
            {\figtxt{Representation of the different steps, within the CERN accelerator complex,
                necessary to inject protons in the LHC ring
                [Figure adapted from CERN-DI-0606052 \copyright CERN Geneva].}
            }
            \label{fig_lhc_accel_cplx}
  \end{center} 
\end{figure}

The different steps of the protons acceleration and injection inside the LHC ring are shown in 
Fig.~\ref{fig_lhc_accel_cplx}.
The proton beams, split in bunches of oblate volume, are successively accelerated 
by the LINAC2, the BOOSTER, the PS and the SPS \index{SPS collider} accelerators.
The SPS injects proton bunches of $450\GeV$ in the LHC superconducting magnet, 
half moving in one direction and the over half in the opposite direction.
Each proton in a bunch has an energy of $E_p=7\TeV$ providing a total energy in the center 
of mass of $\sqrt{S}=14\TeV$. The total $\pp$ cross section  $\sigma_{\pp}^\mm{tot.}$ and its inelastic
part $\sigma_{\pp,\mathrm{inel.}}^\mm{tot.}$ are approximately equals to
\begin{eqnarray}
\sigma_{\pp}^\mm{tot.}              &\approx& 100\,\mathrm{mb}, \label{eq_sigma_pp_tot}\\
\sigma_{\pp,\mathrm{inel.}}^\mm{tot.} &\approx& \;\,80\,\mathrm{mb}. \label{eq_sigma_pp_tot_inel}
\end{eqnarray}
The volume of a proton bunch can be defined using Gaussian functions which full widths at half 
maximum are, with respect to the $z$-axis along which protons move, of $\sigma_T=15\microm$ 
in the transverse direction and $\sigma_z=5.6\cm$ in the longitudinal direction.
Each bunch, composed of an order of $\sim 10^{11}$ protons, are separated by a time lapse of 
$25\nsec$ ($7.5\m$).

The luminosity $\lumi$\index{Luminosity!At the LHC (expected)|(} is defined at the LHC like
\begin{equation}
\lumi = \frac{N_p^2\,k_b\,f\,\gamma}{4\pi\,\varepsilon_n\,\beta^\ast}\;F,
\end{equation}
where $N_p$ is the number of protons per bunch,
$k_b$ the number of bunches,
$f$ the revolution frequency around the ring,
$\gamma$ the relativistic Lorentz factor related to the proton velocity ($\gamma\approx 7500$),
$\varepsilon_\mm{n}$ the normalised emittance
and $\beta^\ast$ the beta function at the interaction point.
The factor $F\approx 0.9$ accounts for the reduction of interaction numbers due to the full 
crossing angle $\theta_{\mm{c}}$ and other characteristics of the beam configuration, it reads\,:
\begin{equation}
F={1}\Big/{\sqrt{1+\left(\frac{\theta_\mm{c}\,\sigma_z}
{2\,\sqrt{\varepsilon_\mm{n}\,\beta^\ast}}\right)^2}}.
\end{equation}

The number of produced inclusive events $N^\mm{(incl.)}_\evts$ for a process with an inclusive
cross section $\sigma_\mm{process}$ over a time lapse $\Delta t=t_2-t_1$ 
is related to the integrated luminosity $L$ over $\Delta t$ via
\begin{equation}
N^\mm{(incl.)}_\evts = \sigma_\mm{process}\;L \qquad\mbox{with }\qquad L\equiv\int_{t_1}^{t_2}\lumi\;dt,
\label{eq_N_evt_sigma_lumi}\index{Luminosity!At the LHC (expected)!Integrated}
\end{equation}
The nominal luminosity of the LHC\index{Luminosity!At the LHC (expected)!Nominal/High}, 
expressed in several convenient unit should reach\,:
\begin{equation}
\lumihigh = 10^{34} \,\mm{cm}^{-2}\,\mm{s}^{-1}
          = 10^7    \,\mm{mb^{-1}}\,\mm{s}^{-1}
          = 10^{-5} \,\mm{fb}^{-1}\,\mm{s}^{-1}
          = 10      \,\mm{nb}^{-1}\,\mm{s}^{-1}.
\end{equation}
Based on Eqs~(\ref{eq_sigma_pp_tot},\ref{eq_sigma_pp_tot_inel}), this gives a 
frequency of $\sim 1\,\mm{GHz}$ of produced events among which inelastic events occur
at the rate of $0.8\,\mm{GHz}$.
This will eventually lead each event of interest to be accompanied by an average of $20$ 
inelastic events also referred to as the ``pile-up''\index{Pile-up}. 
For this reason the LHC will use during the first three years a lower 
luminosity\index{Luminosity!At the LHC (expected)!Low} 
$\lumilow=\lumihigh/10$ for which the smaller pile-up ($\approx 2$ events) will allow physicists 
to understand first the detectors before exploiting them at the design luminosity.

Most of the analysis in this work are considered for one year of LHC data taking
at low luminosity, which, due to technical reasons, was assumed to be of the order of 
$10^7\mm{s}$.
Thus, one year of data taking in ATLAS at low luminosity implies an integrated low luminosity of
\begin{equation}
L_\mm{low}^{\mm{1\,year}}=10\,\mm{fb}^{-1}=10^7\,\mm{nb}^{-1}. \label{eq_L_1_yr_low_lumi}
\index{Luminosity!At the LHC (expected)!Integrated}
\end{equation}
the choice for unit being cast on nb as the $W$ inclusive/cut cross sections are of this order.

The LHC ring is planed to accelerate and collide heavy ions as well. For the purpose of our 
measurement strategy the possibility of having run programs with light ions have been considered.
In such a context the total energy of $7\TeV$ for the accelerated hadron is shared between the 
different nucleons building it.
The energy $E_\mm{nucleon}$ of a nucleon belonging to an hadron of mass $A$ and charge $Z$ is
\begin{equation}
E_\mm{nucleon} = \frac{Z}{A}\times\,7\;[\mm{TeV}]. \label{eq_e_a1a2}
\end{equation}
Considering the collision of two hadrons $1$ and $2$ respectively of masses $A_1$ and $A_2$
the nucleon--nucleon ($n_1\,n_2$) center of mass energy is $\sqrt{S_{n_1\,n_2}}=E_{n_1}+E_{n_2}$ 
where the energies $E_{n_1}$ and $E_{n_2}$ of each nucleon are computed using Eq.~(\ref{eq_e_a1a2}).
Concerning the luminosity, it should be possible in the most optimistic case scenario to reach
$L_{A_1A_2} = {L_{pp}}/{A_1\,A_2}$, $L_{pp}$ corresponding to the standard integrated luminosity for 
$\pp$ collisions. In the present study, using these rules for the case of $\dd$ collisions that 
were occasionally considered, gives
\begin{eqnarray}
E_\mm{nucleon} &=& 3.5\TeV, \\
L_{dd}         &=& L_{pp}/4.\index{Luminosity!At the LHC (expected)!Integrated}
\end{eqnarray}
To sum up for $\pp$ and $\dd$ collisions, the nucleons energies and the integrated luminosity 
for one year at low luminosity are
\begin{eqnarray}
\pp &:& L_{pp}=10.0 \,\mm{fb}^{-1},\index{Luminosity!At the LHC (expected)!Integrated}
\qquad E_\mm{nucleon}=7.0\TeV \quad(\Rightarrow E_p=7\TeV), \label{eq_pp_Ecm_lumi} \\ 
\dd &:& L_{dd}=2.50 \,\mm{fb}^{-1},\index{Luminosity!At the LHC (expected)!Integrated}
\qquad E_\mm{nucleon}=3.5\TeV \quad(\Rightarrow E_d=7\TeV). \label{eq_dd_Ecm_lumi}
\end{eqnarray}
A number of produced events related to a process is computed using Eq.~(\ref{eq_N_evt_sigma_lumi}). 
\index{Luminosity!At the LHC (expected)|)}

\subsection{The LHC experiments}
\index{LHC!Experiments|(} 
The detectors layout around the LHC ring is shown in Fig.~\ref{fig_lhc_accel_cplx}.
The actual approved experiments are\,:\;Alice, ATLAS, CMS, LHC$b$, LHCf and TOTEM.

ATLAS (A Toroidal LHC ApparatuS) and 
CMS~\cite{:2008zzk}\index{LHC!Experiments!CMS} (Compact Muon Solenoid) 
are two detectors made for general studies of the physics at the LHC, 
\ie{} refine the Standard Model parameters, confirm/infirm the existence of the Higgs boson and 
study new physics signatures (hypothesised BSM models, new signatures).
Both can measure the signatures of high $\pT$ objects such as $e$, $\gamma$, $\mu$, $\tau$,
jets, $b$-jets, missing transverse energy $\ETmiss$, etc.

Alice~\cite{Aamodt:2008zz}\index{LHC!Experiments!Alice} 
(A Large Ion Collider Experiment) has been created to study lead ions 
collisions to possibly materialise a state of matter known as quark--gluon plasma, which may have 
existed soon after the Big Bang.

LHC$b$~\cite{Alves:2008zz}\index{LHC!Experiments!LHCb} 
(Large Hadron Collider beauty) is dedicated to the study of $b$~quarks 
decays in hadrons to understand the mechanism of $CP$ violation that could explain 
the matter/anti-matter asymmetry in the Universe.

The LHCf~\cite{Adriani:2008zz}\index{LHC!Experiments!LHC-f} 
(Large Hadron Collider forward) experiment uses forward particles created 
in collisions as a source to simulate cosmic rays in laboratory conditions.

TOTEM~\cite{Anelli:2008zza}\index{LHC!Experiments!TOTEM} (TOTal Elastic and diffractive cross 
section Measurement) will measure the general properties of $\pp$ collisions such as the total 
cross section and the luminosity.
\index{LHC!Experiments|)} 
\index{LHC|)}

%
%
%
%
\section{The ATLAS detector}\index{ATLAS detector|(}

\subsection{Detector requirements}
The ATLAS detector was designed in function of the expected physics signatures,
the high energies involved in the hadrons collisions and the high luminosity context 
present at the LHC.
This imposed high constraints over the detector performances, size and trigger system.
Assuming an operation life of ten years or so the detecting devices and their associated
electronics must stand high radiation due to the important particles fluxes.
The other problem is that in average each inelastic scattering of interest that triggers the 
apparatus is accompanied by usually $\approx 20$ other non-interesting inelastic events.
To decrease as much as possible this pile-up impact the detector needs, using a highly 
efficient trigger, to provide a precise and fast detector response.\index{Pile-up}
Also the detector needs a high granularity to lift as much as possible overlapping ambiguities 
between the processes of interest and the pile-up. 

The size of the ATLAS detector is directly related to the $\sim\mm{TeV}$ energy scale of the 
produced particles which needs to be contained.
For example, electrons of $1\TeV$ are absorbed by $30$ radiation length ($X_0$), 
pions of $1\TeV$ by $11$ absorption length ($\lambda$)
and measuring momenta of muons of $1\TeV$ needs bending power of several $\Tesla\,\mm{m}$.
Another reason for the size of the ATLAS detector is the choice made for the magnetic field
which, contrary to CMS, is separated in two pieces.
The first one (solenoid) used by the inner tracking detector allows a good charged particle 
momentum resolution and reconstruction efficiency while the muons 
--not stopped by the calorimetry-- have their momenta resolved by large magnets generating
a toroidal over a large range of momenta.

Turning now to the events signatures, ATLAS needs to identify extremely rare events, 
some as low as representing $10^{-14}$ of the total $\pp$ cross section.
In the LHC context, lepton identification is challenging due to the high QCD background 
(\eg{} the electrons jet ratio is $\epm/\mm{jet}\sim 10^{-5}$) and 
Higgs/BSM hypothesised signatures constrained strongly each sub-detectors performances.

These general requirements reminded, we give an overview of the ATLAS detector.

\subsection{Overview of the ATLAS detector}
\index{ATLAS detector!Overview of the detector|(}
\begin{figure}[!h] 
  \begin{center}
\includegraphics[width=.8\tw]{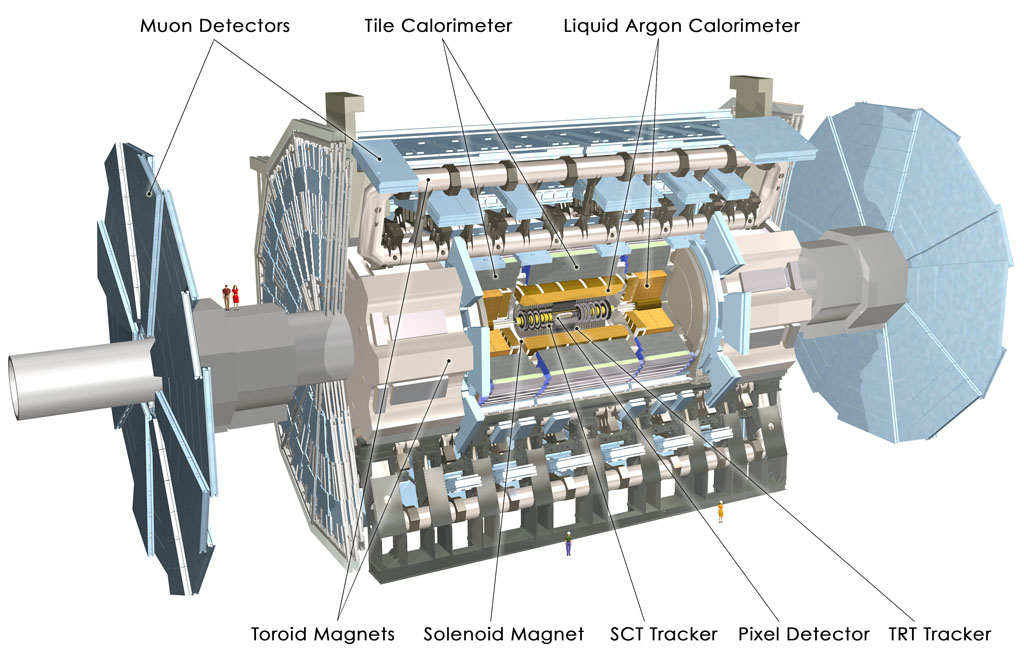}
    \caption[Overview of the ATLAS detector]
            {\figtxt{Overview of the ATLAS detector [CERN-GE-0803012 \copyright~CERN Geneva].}
            }
            \label{fig_atlas_overview}
  \end{center} 
\end{figure}
\paragraph{Detector.}
The ATLAS detector is nominally forward-backward symmetric with respect to the interaction point.
The sub-detectors layout follows basics rules common to all hadronic colliders.
Starting from the beam-pipe one can find a tracking device, then electromagnetic and hadronic 
calorimeters and finally a detector for muons.
The inner tracker and the muon spectrometer are immersed in magnetic fields generated respectively
by a thin superconducting solenoid and three large superconducting toroidal magnets.
This general setup can be seen in Fig.~\ref{fig_atlas_overview}.

Each sub-detector is composed mainly of three pieces, a barrel for high $\pT$ particles and 
two end-caps located symmetrically from the interaction point to optimise the solid angle coverage.
Usually, the transition regions between the barrels and the end-caps, also called the transition region,
are used to pass the power cables and the readout cables of the devices.

\index{ATLAS detector!Tracker}
First, the tracker, also referred to as the Inner Detector (ID), bathes in a $2\Tesla$ magnetic 
field generated by a surrounding solenoidal magnet.
It is made of semi-conductor silicon vertex detectors in the inner layers (pixel and semi-conductor 
tracker) and of straws trackers in its outer part.
Charged particles, as they pass each layer, leave measurable hits that allow to reconstruct
their trajectories and creation points (vertex) while their transverse momenta are deduced 
from their tracks deflection by the magnetic field.

After comes the sampling calorimeters made of alternate layers of absorbers and active detector
medium.
First is the electromagnetic (EM) calorimeter made of lead (absorber) and liquid argon
(LAr) for the active medium. It measures with a high granularity and excellent performances 
in energy and position resolution the electrons and photons.
The hadronic calorimeter measures the energy of hadrons and QCD jets.
It is made in the barrel of steel plates (absorber) and scintillator-tiles (active detector medium),
in the end-cap both tile and LAr calorimeters are used.

Around the hadronic calorimeter is the muon spectrometer bathing in the air-core toroid 
magnetic field. The high bending power of the magnetic field provides an excellent muon momentum 
resolution using three layers of high precision tracking chambers to detect the muon passage.
Muons properties are measured using drift tubes and multiwire chambers.
The spectrometer uses as well other chambers relying on cathode-plate and multi-wire chambers
to trigger for high $\pT$ muons events.
The muon detector defines the overall size of ATLAS which is $\approx 20\,\mm{m}$ of diameter and 
$\approx 45\,\mm{m}$ long. 
The total weight of the detector reaches $\approx 7000$ tons.

\index{ATLAS detector!Trigger}\index{ATLAS detector!TDAQ}
\paragraph{Trigger.}
The harvest of the data measured by those sub-detectors is achieved using a fast trigger.
The trigger system selects events displaying interesting signatures among the plethora
of events produced at the rate of $\sim 1\,\mm{GHz}$ at nominal luminosity.
For the first time the collected statistic for large scale processes such as $W$ or $Z$ production 
will be limited by the bandwidth and read out systems rather than by the produced events.
The trigger is split into three Levels (L), L1, L2 and the event filter, the two last one being
referred to as the high-level trigger.
Each level refines the decision outgoing from the precedent level and, if necessary, apply
additional selection criteria.

The L1 trigger treats subsets from each sub-detector information, it scans for high transverse 
momentum muons, electrons, photons, jets, $\tau$~leptons decaying into hadrons as well as large 
missing and total transverse energies. It also spots in each event to regions of interests defined
as regions within the detector displaying interesting features.
The decision to accept/reject an event is made within $2.5\,\mu\mm{s}$ and decreases the rate of 
incoming events to $\sim 75\,\mm{kHz}$.
The L2 trigger reduces within $\sim 40\,\mm{ms}$ the events rate to $\sim 3.5\,\mm{kHz}$ 
based on full data information within the regions of interested input by L1.
The last stage of the event selection, made by the event filter, reduces the rate of events to 
$\sim 200\,\mm{Hz}$. This step, lasting $\sim 4\,\mm{s}$, is carried out offline using analysis 
procedure and supply an event of size $1.3$ Megabyte.

Already at this point we mention few trigger information relevant for both selection and analysis
of $W$ properties from Drell--Yan production. More details will be given on these topics later on.
Among few other requirements, the trigger should be activated for isolated electrons/muons with high
$\pT$ threshold of the order of $p_{T,l}>20-25\GeV$ while the data from such events should be 
studied in the range $30\GeV<p_{T,l}<50\GeV$.

\paragraph{Forward detectors.}\index{ATLAS detector!Forward detectors|(}
Also worth mentioning are three smaller detectors in the forward region associated to ATLAS.
From the interaction point, at $\pm 17\,\mm{m}$ is
LUCID (LUminosity measurement using {\v C}erenkov Integrating Detector),
at $\pm 140\,\mm{m}$ is ZDC (Zero-Degree Calorimeter)
and at $\pm 240\,\mm{m}$ is ALFA (Absolute Luminosity For ATLAS)
LUCID and ALFA roles are to determine the luminosity delivered to ATLAS \index{ATLAS detector!Luminosity} and
ZDC to have a key role to determine the centrality of heavy ions collisions.\\

\begin{figure}[!h] 
  \begin{center}
    \includegraphics[width=.65\tw]{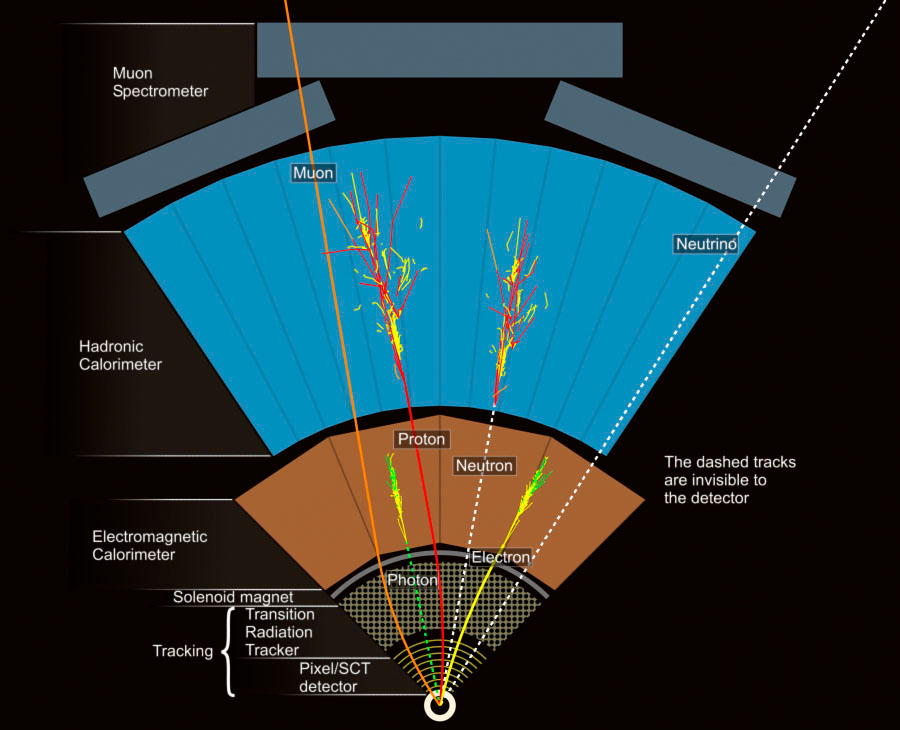}
    \caption[Overview of particles passage through the ATLAS sub-detectors]
            {\figtxt{Overview of some particles passage through the ATLAS tracker,
                electromagnetic calorimeter, hadronic calorimeter and eventually through 
                the first layers of the muon spectrometer. 
                Toroidal magnets located at higher radii are not visible in this picture
                [CERN-GE-0803022 \copyright CERN Geneva].}
            }
            \label{fig_atlas_particles_overview}
  \end{center} 
\end{figure}

Figure~\ref{fig_atlas_particles_overview} concludes this overview by recapturing the interactions 
of some particles with the different sub-detectors.
\index{ATLAS detector!Forward detectors|)}

The rest of the Chapter presents the technology and geometry of each ATLAS sub-detectors.
First the calorimeters and muon spectrometer are presented.
In top of a general description, the link of their role on the measurement of $W$ in Drell--Yan is 
mentioned mostly from the trigger point of view since the tracker, so far, does not have any trigger
on its own. 
Hence the electromagnetic calorimeter and muon spectrometer trigger efficiencies 
for events displaying respectively isolated high $\pT$ electrons and muons are cited.
These kind of studies associates as well $\ETmiss$ due to the final state neutrino. 
However, since in the present analysis the latter is not taken into account,
the $\ETmiss$ trigger is not considered to be relevant although its influence on the results 
was controlled. More details will be given in Chapter~\ref{chap_W_mass_asym}.
In the second and final part, a detailed description of the inner detector is given.
In what follows, dimensions of devices are given by their radial(longitudinal) extensions 
$r$($L$) and their absolute pseudo-rapidity acceptance $\Aeta$. 
\index{ATLAS detector!Overview of the detector|)}

\subsection{Calorimetry}
\begin{figure}[!h] 
  \begin{center}
    \includegraphics[width=.8\tw]{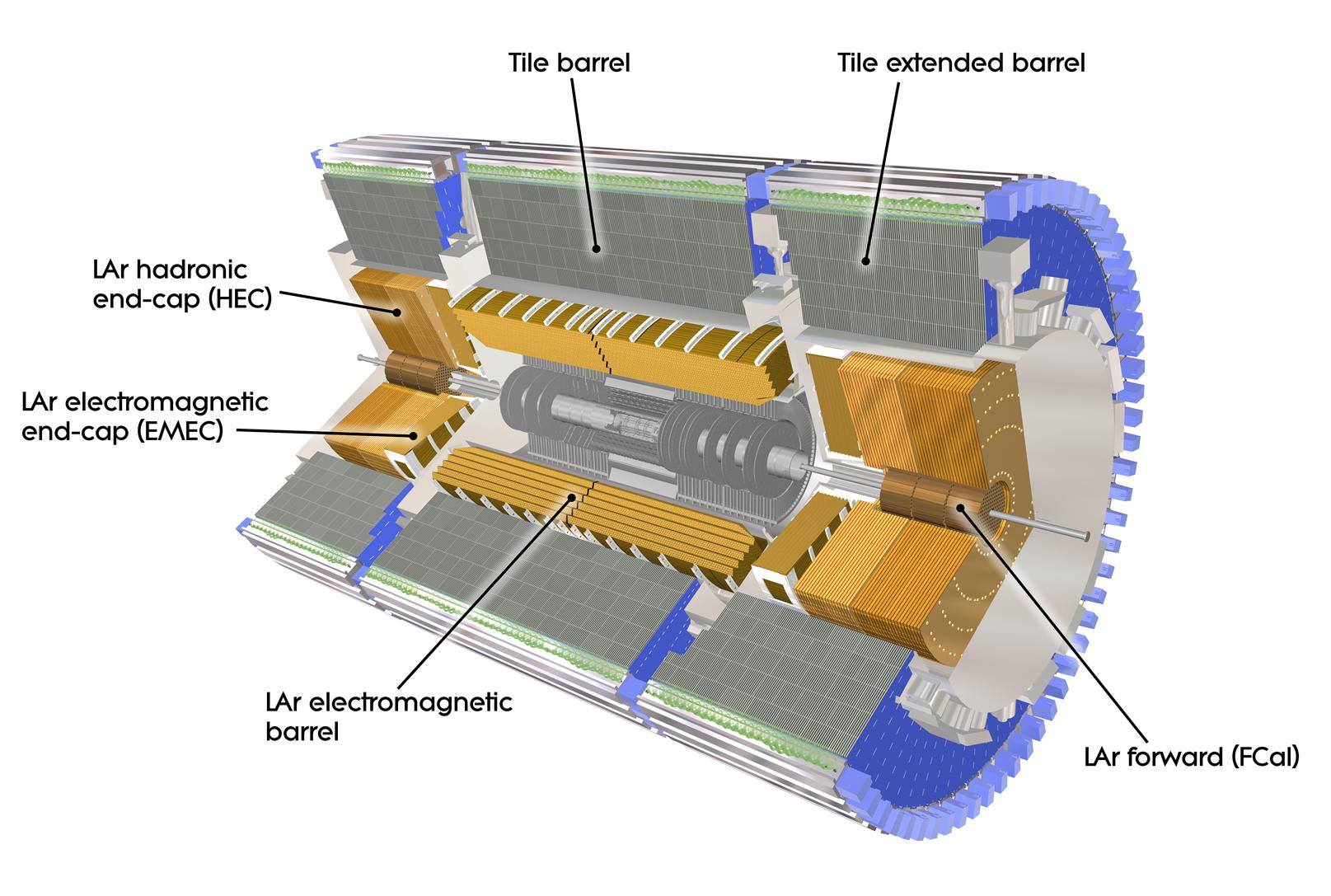}
    \caption[ATLAS calorimetry overview]
            {\figtxt{Overview of the ATLAS calorimetry. 
                Near the beam-pipe the tracker is visible, 
                surrounding it is the EM calorimeter 
                and beyond the hadronic calorimeter.
                Both barrel and end-caps elements are displayed.
                [CERN-GE-0803015 \copyright~CERN Geneva].}
            }
            \label{fig_atlas_calo_overview}
  \end{center} 
\end{figure}

\index{ATLAS detector!Calorimeters|(}
Both electromagnetic calorimeter~(EMC) and hadronic calorimeter~(HC) geometry, technology and 
performances are reviewed. 
Their set up can be seen in Figs.~\ref{fig_atlas_calo_overview} and ~\ref{fig_atlas_tile_calo}.

\subsubsection{Electromagnetic calorimeter}
\begin{figure}[!h] 
  \begin{center}
    \includegraphics[width=.5\tw] {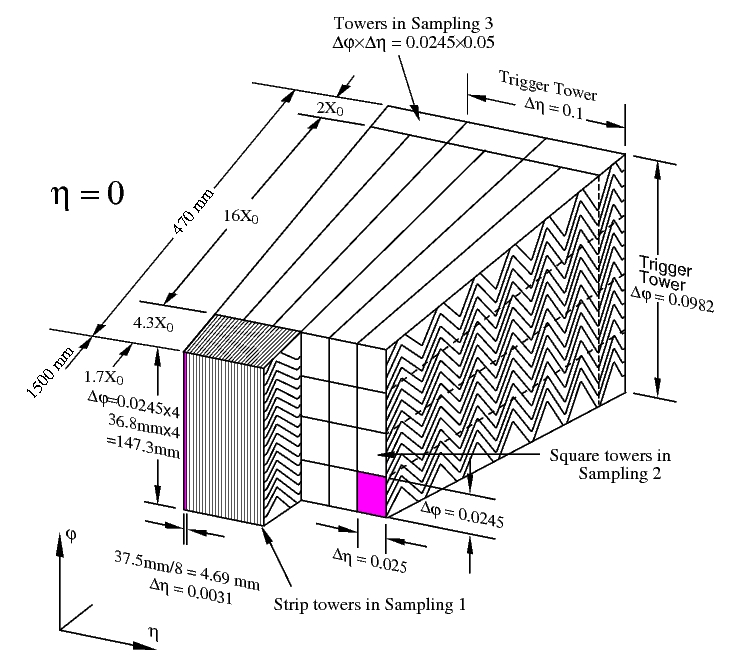}
    \hfill
    \includegraphics[width=.45\tw]{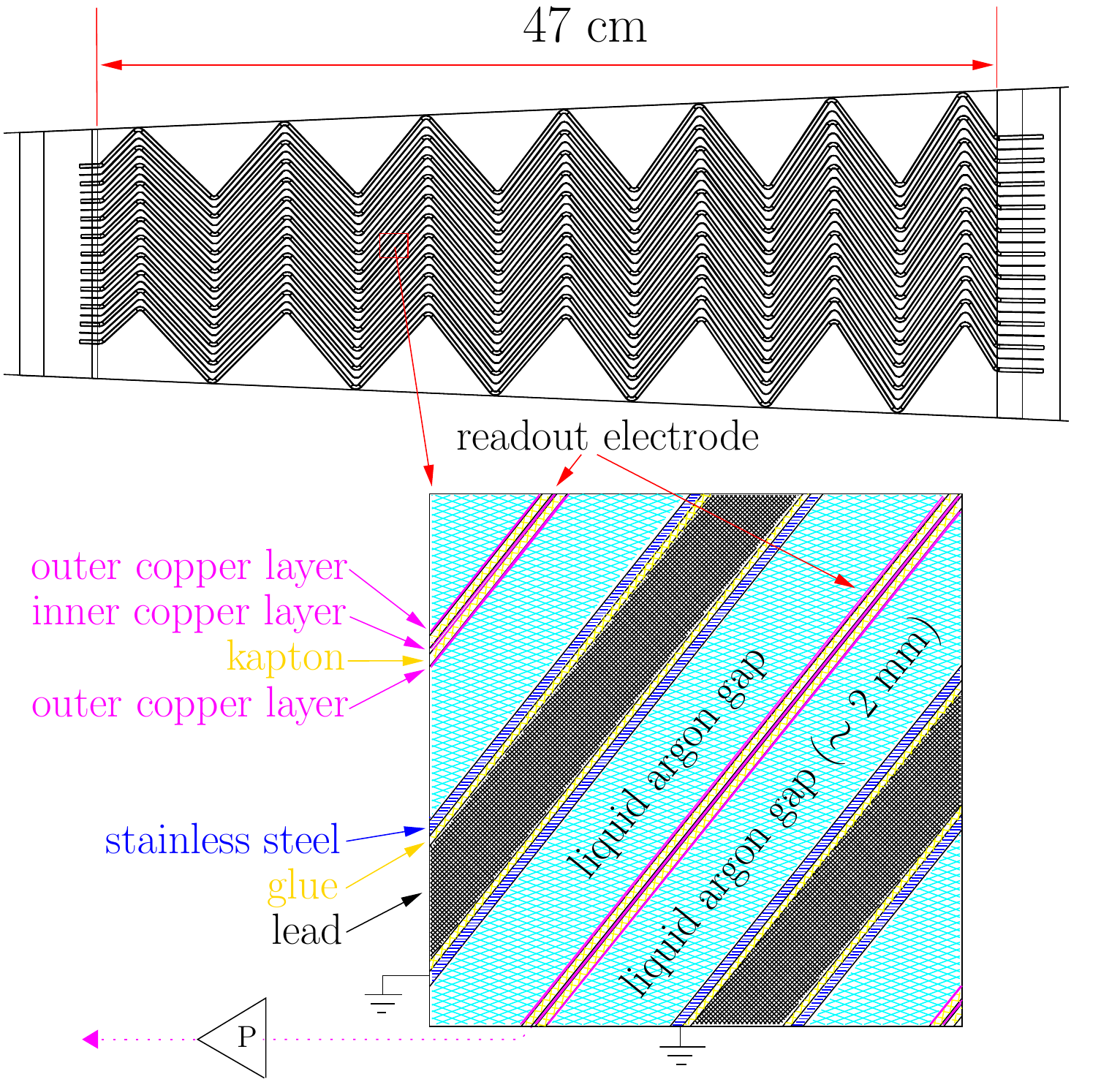}
    \vfill
    \includegraphics[width=1.\tw]{./figures/chapter_2/fig2_00-ab_lbls.pdf}
    \caption[ATLAS electromagnetic calorimeter structure]
            {\figtxt{Schematic view of an EMC barrel module (a)~\cite{AtlasLAr:1996fq}
                and detailed view of the accordion structure (b)
                [Figures adapted from~\cite{Airapetian:1996iv}].}
            }
            \label{fig_atlas_emc_calo_barrel_endcap}
  \end{center} 
\end{figure}

\index{ATLAS detector!Calorimeters!Electromagnetic|(}
\paragraph{Technology.}
The EMC~\cite{AtlasLAr:1996fq,Airapetian:1996iv} is made of accordion shaped layers of lead 
plates-kapton electrodes bathing in liquid argon as depicted in 
Fig.~\ref{fig_atlas_emc_calo_barrel_endcap}.
The accordion geometry, symmetric in $\phi$, presents a full and crack-free azimuthal coverage.
The readout of the high voltage electrodes are maintained at equal distance from two lead sheets 
using honeycomb spacers.
A high energy electron or photon as it passes through an absorber looses energy respectively via
bremsstrahlung $\epm\to\epm\,\gamma$ or pair production $\gamma\to\ep\Em$.
These produced particles in their turn interact with the other absorbers creating a shower of 
particles.
Particles from this shower excite liquid argon from which ionised electrons --as they drift to the
electrode-- allow to find eventually the shape and the total energy yielded by the incoming 
electron/photon.

\paragraph{Geometry.}
The barrel ($1.25\,\mm{m}<r<2.25\,\mm{m}$, $L=6.4\,\mm{m}$, $\Aeta<1.475$, $\ge 22\,\mm{X_0}$) 
is made of two identical half-barrels with the accordion waves running in the $r$~axis.
Both are stored with the solenoid in the same cryostat.

The two end-caps ($330\,\mm{mm}<r<2098\,\mm{mm}$, $L=63\,\mm{cm}$, $1.375<\Aeta<3.2$, $\ge 24\,
\mm{X_0}$) are stored with the hadronic end-caps and forward calorimeters in the same cryostat.
Here the waves of the accordion are parallel to the $z$~axis.
Each end-caps are mechanically split into two coaxial wheels in an approximately projective 
geometry at $\Aeta=2.5$.
The external wheel end-cap ($1.375<\Aeta<2.5$) together with the barrel provide precision 
measurements with a granularity of the cells of the order of $\granularity\sim 0.025\times0.025$.
Liquid argon presamplers are implemented upstream of both the barrel and external wheel end-cap
in the aim to  correct for the loss of energy of the electrons and photons before they enter the
calorimetry.

\paragraph{Performances.}
For tracker based studies the energy $E$ of electrons measured in the EMC enters as 
references to tracker measurement through the ratio $E_\epm^\mm{(EMC)}/p_\epm^{(\mm{ID})}\equiv E/p$.
The intrinsic energy resolution in the barrel was found, using test-beams~\cite{Aharrouche:2006nf},
to be for a Gaussian fit resolution
\begin{equation}
\frac{\sigma_E}{E^\true}=\frac{0.1}{\sqrt{E^\true}} \oplus  0.0017.
\end{equation}
In the previous equation $E$ is expressed in GeV, the (true) labels refers to the true level 
and the $\oplus$ symbol means that the two terms are added in quadrature.

\begin{figure}[!h] 
  \begin{center}
    \includegraphics[width=.495\tw]{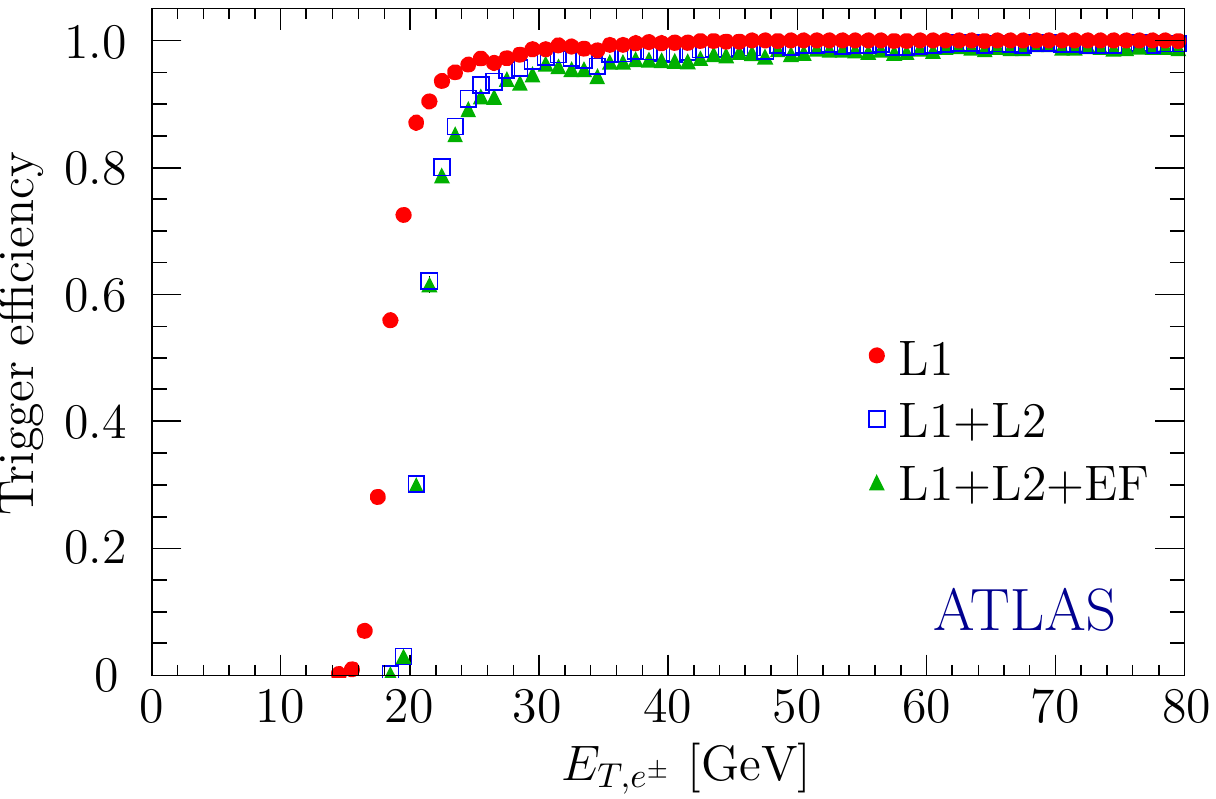}
    \caption[Simulation on ATLAS electron trigger efficiency for $p_{T,e}=20\GeV$]
            {\figtxt{Simulation on ATLAS electron trigger efficiency for $p_{T,e}=20\GeV$ 
                [Taken in~\cite{AtlasEgammaCSC}].}
            }
            \label{fig_atlas_elec_trigger_eff}
  \end{center} 
\end{figure}

\paragraph{Trigger on electrons.}
The trigger on electrons should not affect the data used in the analysis.
Indeed, simulation (Fig.~\ref{fig_atlas_elec_trigger_eff}) shows that the efficiency for electrons
with a threshold of $p_{T,e}=20\GeV$ reach the Efficiency=1 plateau already for $40\GeV$
where the data enters our analysis.
This behaviour is expected to be the same for separated positrons and electrons selection.
\index{ATLAS detector!Calorimeters!Electromagnetic|)}

\subsubsection{Hadronic calorimeter}
\index{ATLAS detector!Calorimeters!Hadronic|(}
\begin{figure}[!h] 
  \begin{center}
    \includegraphics[width=.4\tw]{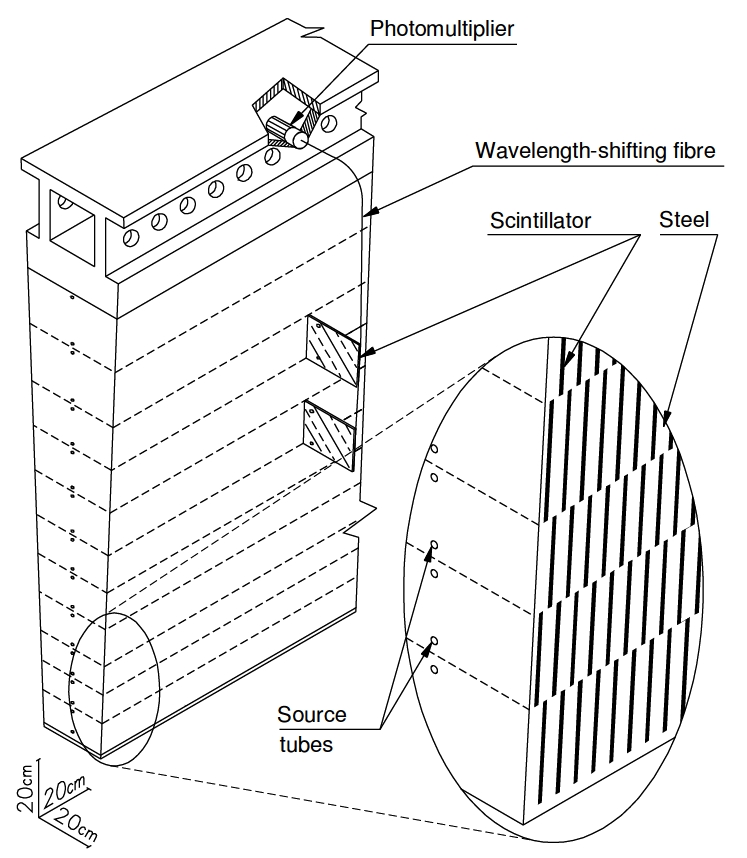}
    \hfill
    \includegraphics[width=.45\tw]{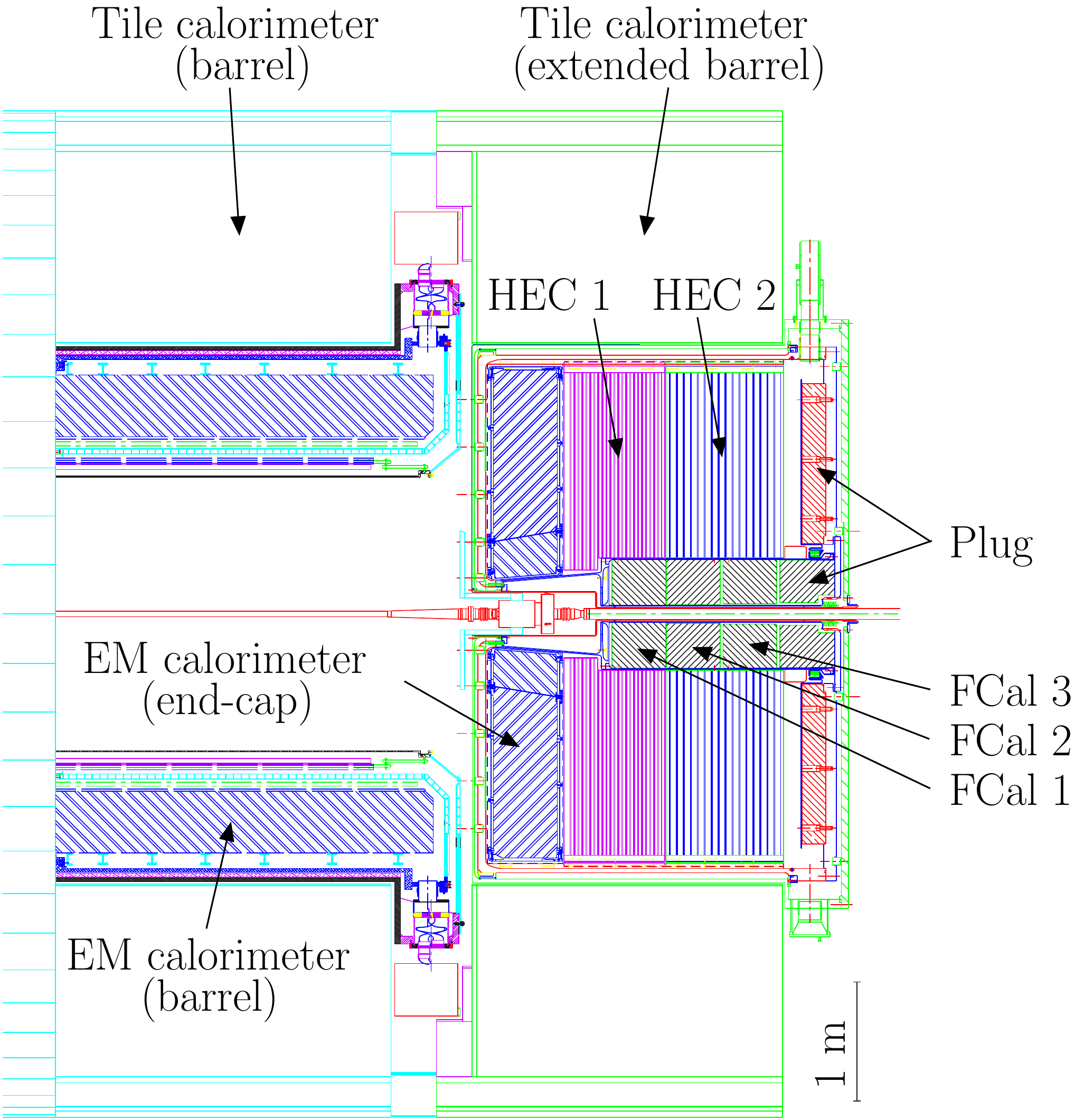}
    \vfill
    \includegraphics[width=1.\tw]{./figures/chapter_2/fig2_00-ab_lbls.pdf}
    \caption[ATLAS hadronic tile calorimeter and end-cap ]
            {\figtxt{Schematic representations of the tile calorimeter geometry with its structure 
                with its optical readout in (a) 
                and of the electromagnetic/hadronic end-caps and forward calorimeters
                in (b) [Fig.(a) taken in~\cite{ATLASatCERN:2008zz}, Fig.(b) adapted 
                  from~\cite{AtlasLAr:1996fq}].}
            }
            \label{fig_atlas_tile_calo}
  \end{center} 
\end{figure}
\paragraph{Technology.}
The hadronic calorimeter ~\cite{AtlasLAr:1996fq,AtlasTileCalo:1996fr} uses both LAr and 
scintillators tiles technology.

The tile calorimeter is used in the barrel. 
It is a sampling calorimeter made of alternate layers of steel plates (absorbers) and scintillating
tiles (active medium) orthogonal to the $z$~axis (Fig.~\ref{fig_atlas_tile_calo}.(a)).
Ionising objects (hadrons or jets) as they pass through the tiles induce the production of 
ultra-violet scintillation light which is converted to visible light by wavelength-shifting fibres. 
The fibres are grouped together and coupled to photo-multipliers.
Just like for the EM calorimeter the shape and energy of the object is measured as the object
yields all its energy in form a shower.

Sampling calorimeter made of flat copper plates (absorber) and LAr (active medium) 
are used for the end-cap calorimeter while the forward calorimeter uses both LAr in association with
copper and tungsten (absorbers). 

\paragraph{Geometry.}
The tile calorimeter ($2.28\,\mm{m}<r<4.25\,\mm{m}$, $\Aeta<1.7$, $\ge 7.4\,\lambda$) 
is made of one barrel ($L=5.8\,\mm{m}$, $\Aeta<1.0$) and two extended barrels ($L=2.6\,\mm{m}$, 
$0.8 < \Aeta < 1.7$).

In the forward pseudo-rapidity the hadronic end-cap (HEC) and forward calorimeter (FCal) are
implemented to enhance the hermetic confinement of the produced particles to refine the measurement of 
$\ETmiss$ (Fig.~\ref{fig_atlas_tile_calo}.b).
The end-cap calorimeter ($r<2030\,\mm{mm}$, $L=1818\,\mm{mm}$, $1.5<\Aeta<3.2$) 
is split in a front wheel (HEC~1\,:\;$372-475\,\mm{mm}<r<2030\,\mm{mm}$) 
and a rear wheel (HEC~2\,:\;$475\,\mm{mm}<r<2030\,\mm{mm}$).

The forward calorimeter ($3.1<\Aeta<4.9$) measures the energy of the intense particles flux in this 
forward region, it is divided in three layers of equal length, first an electromagnetic and then 
two hadronic calorimeters.
It relies on copper--LAr in the first layer (FCal~1) for electromagnetic calorimetry and 
on tungsten-LAr in the two last hadronic calorimeters (FCal~2~\&~3).

The granularity in the barrel and extended barrels is of the order of $\granularity\sim 0.1\times 
0.1$ and of $\granularity\sim 0.2\times 0.2$ in the forward calorimeters.
\index{ATLAS detector!Calorimeters!Hadronic|)}
\index{ATLAS detector!Calorimeters|)}

\subsection{Muon system}
\index{ATLAS detector!Muon spectrometer|(}
\begin{figure}[!h] 
  \begin{center}
    \includegraphics[width=.65\tw]{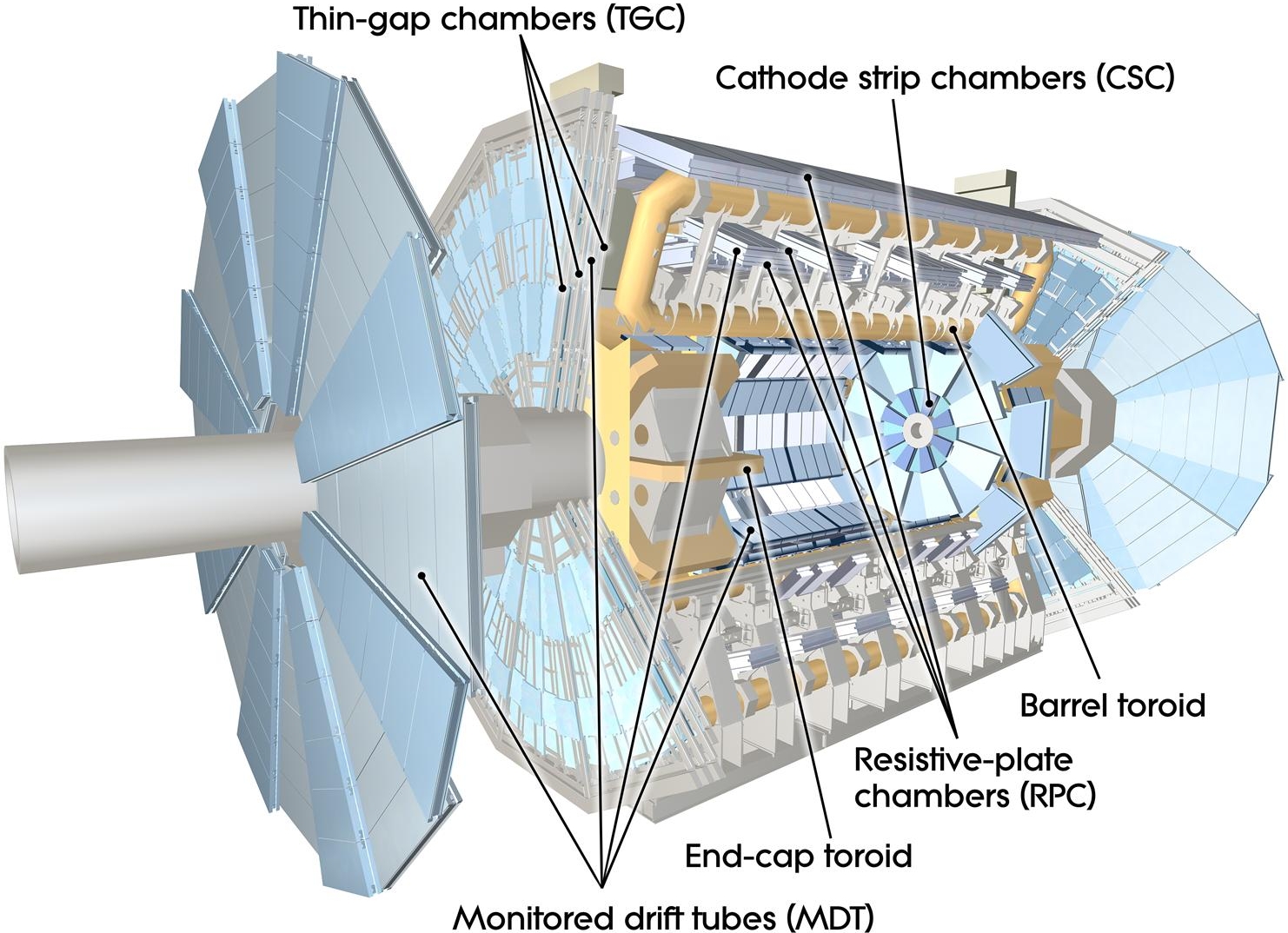}
    \caption[ATLAS muon spectrometer overview]
            {\figtxt{Overview of the ATLAS muon spectrometer
                [CERN-GE-0803017 \copyright~CERN Geneva].}
            }
            \label{fig_atlas_muon_overview}
  \end{center} 
\end{figure}
The muon precision tracking instruments and the toroidal magnet are reviewed.
Information on the muon triggering relevant to the present analysis is given as well.
An overview of the muon spectrometer can be seen in Fig.~\ref{fig_atlas_muon_overview}.

\paragraph{Technology.}
The bending of the muons is done in the $r-z$ plan by a toroidal field which strength can reach
$3\Tesla$ in the barrel and $6\Tesla$ in the end-cap.
Nonetheless, the magnetic field bending power is characterised rather by its integral over the
track length inside the tracking volume $BL\equiv\int \vec B\cdot d\,\vec l$ $[\Tesla\,\mm{m}]$, 
where $d\,\vec l$ is an infinitesimal track portion.
Large values of $BL$ are necessary to make precise measurements of $\sim 1\TeV$ muon tracks.

\begin{figure}[!h] 
  \begin{center}
    \includegraphics[width=.4\tw]{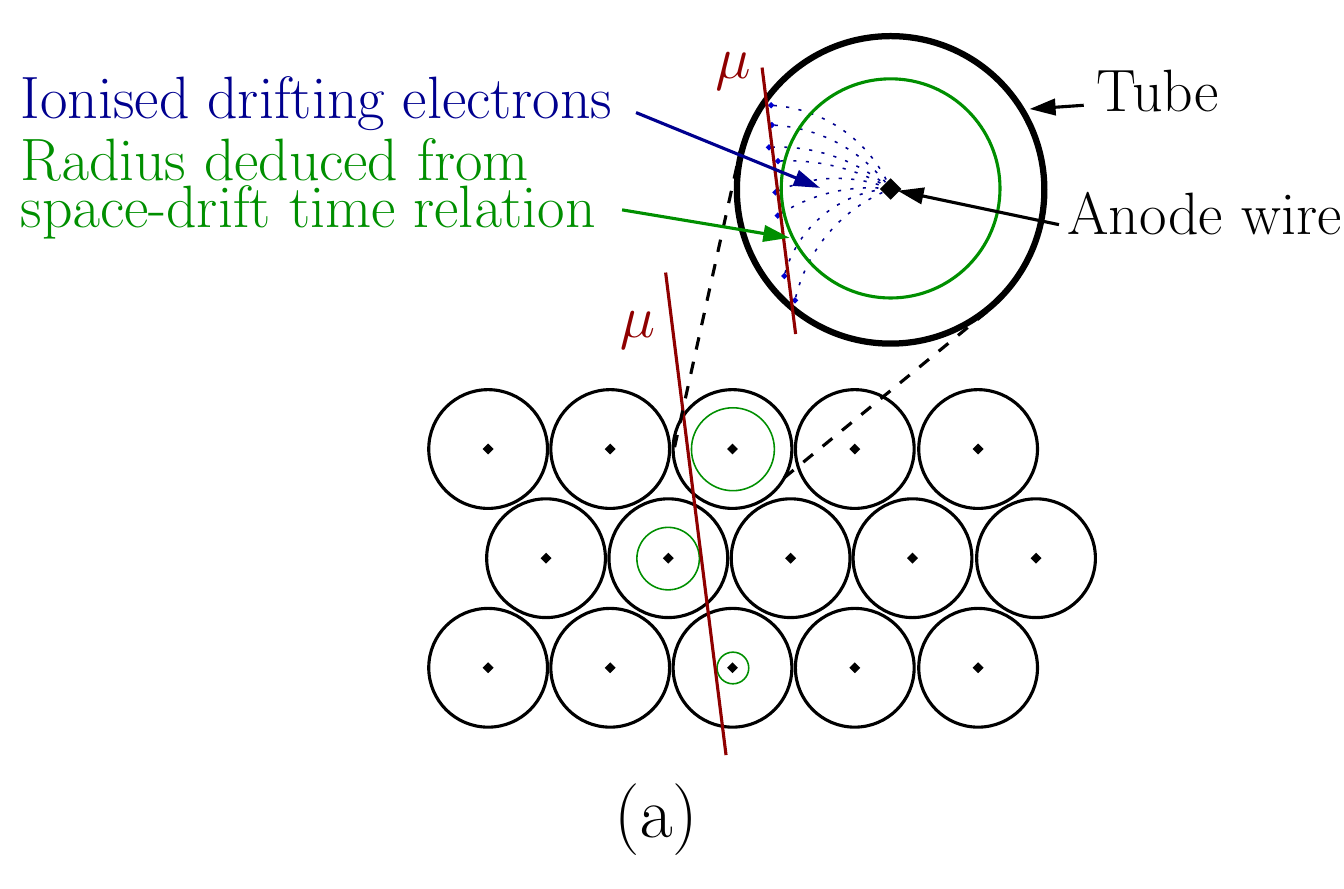}
    \hfill
    \includegraphics[width=.45\tw]{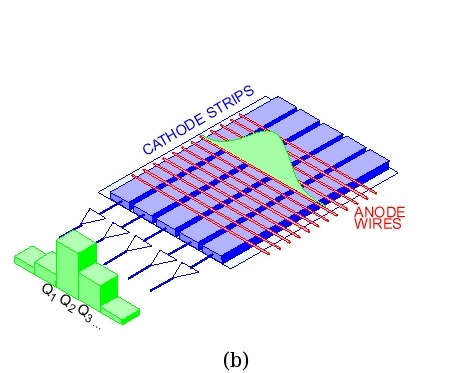}
    \caption[Muon spectrometer MDT and CSC functioning principle]
            {\figtxt{Principle of operation for MDT (a) and CSC devices
                (b)~[Figure (b) taken in~\cite{ATLASatCERN:2008zz}].}
            }
            \label{fig_atlas_muon_mdt_csc}
  \end{center} 
\end{figure}

\index{ATLAS detector!Muon spectrometer!Monitored Drift Tubes (MDT)}
Over most of the pseudo-rapidity range, precise measurement of the track coordinates in the 
principal bending direction is provided by Monitored Drift Tubes (MDT).
As muons pass through a pressurized gas mixture filling the tubes, 
ionised electrons drift to the anode wire. The radius of the particle passage in the tube is
deduced using space drift--time relation.
Eventually, all collected radii give the muon track as shown in Fig.~\ref{fig_atlas_muon_mdt_csc}.(a).

\index{ATLAS detector!Muon spectrometer!Cathode Strip Chambers (CSC)}
For larger pseudo-rapidity, Cathode Strip Chambers (CSC), which are multiwire proportional 
chambers with cathodes segmented into strips, are used due to their higher rate capability and 
time resolution. Both cathodes are segmented. 
The one with strips orthogonal to the direction of the wires measure the precision coordinate,
while the other with strips parallels to the wires provides the transverse coordinate.
The position of the track is deduced by interpolating the charged induced among adjacent strips 
(Fig.~\ref{fig_atlas_muon_mdt_csc}.(b)).

The trigger chambers use Resistive Plate Chambers (RPC) in the barrel and Thin Gap Chambers 
(TGC) in the end-cap. 
In top of their triggering primer role, these chambers provides second coordinate measures in
the bending plane $r-z$ to cross check the ones from the MDTs.

\paragraph{Geometry.}
The magnetic field is generated in the barrel using three large air-core toroidal magnet
($9.4\,\mm{m}<r<20.1\,\mm{m}$, $L=25.3\,\mm{m}$, $\Aeta<1.4$, $BL=1.5-5.5\Tesla\,\mm{m}$),
each of them composed by eight coils arranged radially and symmetrically around the $z$ axis. 
In the end-cap two smaller magnets ($1.65\,\mm{m}<r<10.7\,\mm{m}$, $L=5\,\mm{m}$, $1.6<\Aeta<2.7$, 
$BL=1.0-7.5\Tesla\,\mm{m}$) are inserted into both ends of the barrel toroid.
In the transition region ($1.4<\Aeta<1.6$), tracks deflection are done by a combination of both
barrel and end-caps magnet field.

MDTs configuration follow a projective geometry and display a $\phi$ orientation of the wires
in both barrel and end-cap.
In the barrel, MDT chambers are arranged in three concentric cylindrical shells ($\Aeta<2.0$) 
around the beam axis at radii of approximately $5\,\mm{m}$, $7.5\,\mm{m}$ and $10\,\mm{m}$.
In the end-cap they are confined inside large wheels ($\Aeta<2.7$) perpendicular to the $z$ axis at 
$z\approx 7.4\,\mm{m},10.8\,\mm{m},14\,\mm{m}$ and $21\,\mm{m}$.
CSCs ($2.0<\Aeta<2.7$) are arranged in wheels with approximate radial orientation of the wires.
RPCs and TGCs are respectively used in the barrel ($\Aeta<1.05$) and in the end-caps 
($1.05<\Aeta<2.7$), their implementation based on the one of the MDT and CSC modules.
There are in total three layers of RPCs in the barrel and three layers of TGCs in the end-cap.

\begin{figure}[!h] 
  \begin{center}
    \includegraphics[width=.9\tw]{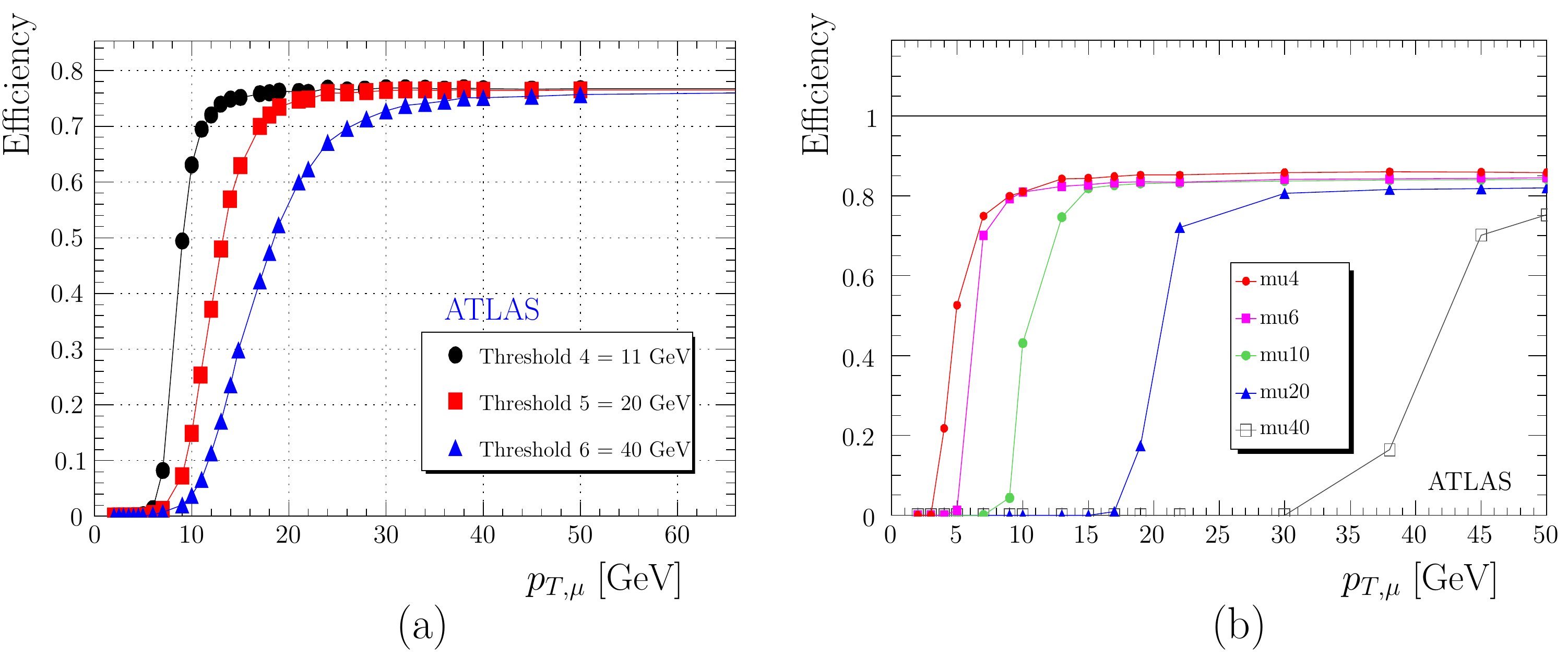}
    \caption[Muon trigger efficiency for different $pT$ thresholds for the Level 1 and
      Event Filter]
            {\figtxt{
                L1 barrel efficiency (a) and event filter efficiencies with the MuId algorithm (b) 
                for several $p_{T,\mu}$ thresholds 
                [Histograms extracted from~\cite{ATLASMuonTriggerCSCnote}].}
            }
            \label{fig_atlas_muon_trigger}
  \end{center} 
\end{figure}

\paragraph{Trigger on muons.}
Again, like for the electron the muon trigger efficiencies has already reached a plateau 
at $40\GeV$ where the data starts to enter the analysis (Fig.~\ref{fig_atlas_muon_trigger}).
Still here, contrary to the EMC, some asymmetries can arise between $\mup$ and $\mum$ due to the
toroidal topology of the magnetic field which, for a given side of the detector, is in-bending 
for a charge and out-bending for the opposite charge.
Nonetheless such effect should be mostly insignificant with respect to other apparatus limitations.
\index{ATLAS detector!Muon spectrometer|)}

\section{The ATLAS inner detector}\label{s_tracker}
\index{ATLAS detector!Tracker|(}
\subsection{Description of the inner detector}
\begin{figure}[!h] 
  \begin{center}
\includegraphics[width=.75\tw]{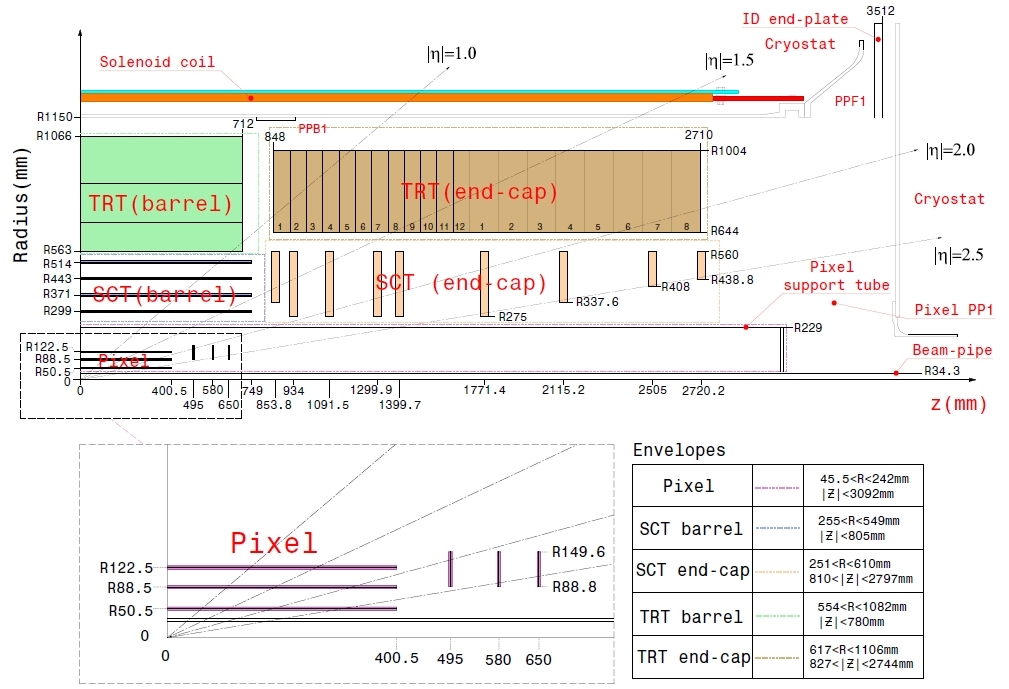}
    \caption[Longitudinal cross section of the ATLAS tracker]
            {\figtxt{Longitudinal cross section of the ATLAS tracker
                [Figure taken from~\cite{ATLASatCERN:2008zz}].
              }
            }
            \label{fig_atlas_tracker_z_xsection}
  \end{center} 
\end{figure}
\begin{figure}[!h] 
  \begin{center}
\includegraphics[width=.75\tw]{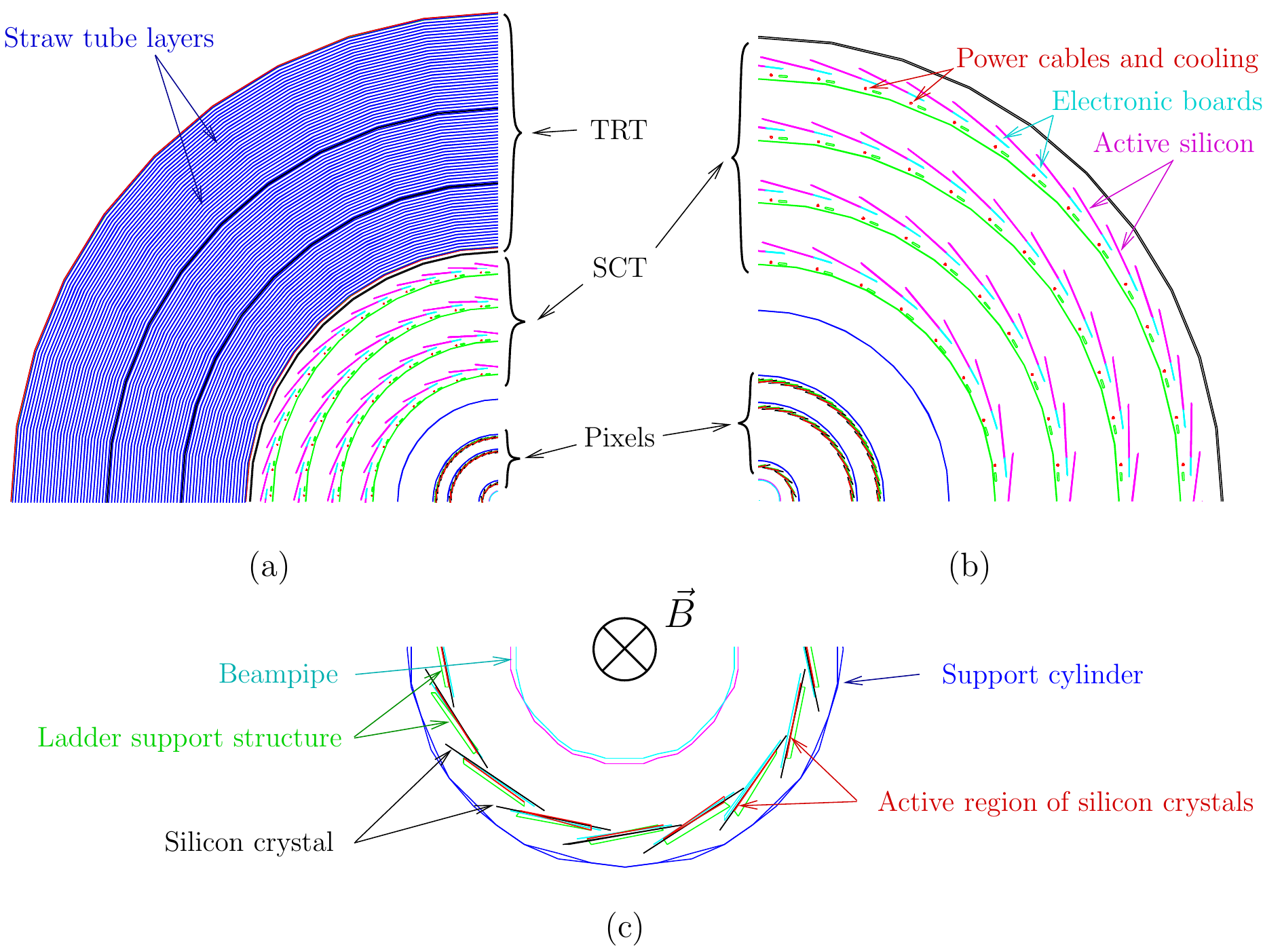}
    \caption[Radial cross section of the ATLAS tracker precision region ]
            {\figtxt{Radial cross section of the ATLAS tracker precision region,
                showing the three detectors (a), a zoom on the SCT and pixel detector (b)
                and finally a zoom on the first layer of the pixel detector with the 
                the direction of the magnetic field $\vec B$ in (c)
                [Figures adapted from~\cite{AtlasInnerDetector:1997fs}].
              }
            }
            \label{fig_atlas_tracker_r_xsection}
  \end{center} 
\end{figure}
The ATLAS inner 
detector~\cite{AtlasInnerDetector:1997fs,AtlasInnerDetector:1997ft,AtlasInnerDetectorCSC}
pixel, SCT and TRT elements along with the central solenoid are described below.
Their implementation is displayed with details on Figs.~\ref{fig_atlas_tracker_z_xsection} 
and \ref{fig_atlas_tracker_r_xsection}.
Since the analysis strategy to decrease systematic errors relies on the possibility to invert 
the magnetic field of the solenoid, mentions on the drift of charge carriers in the modules
were found worth to be noticed.\index{ATLAS detector!Tracker}

The Inner Detector is designed to provide a fine pattern recognition,
an excellent momentum resolution and both primary and secondary vertex measurements for
charged particles tracks displaying  $\pT>0.5\GeV$ and within the pseudo-rapidity range $\Aeta<2.5$.
This is achieved by using three independent but complementary sub-detectors. 
At inner radii, high-resolution pattern recognition performances are done by discrete 
space-points from silicon pixel layers and stereo pairs of silicon micro-strip (SCT) layers. 
At larger radii the Transition Radiation Tracker (TRT), made of gaseous proportional counters
embedded in radiator material, allows continuous track following.
Each track leaves at least 7 hits in the precision tracking (pixel and SCT) and
an average of $30$ hits in the TRT.\index{ATLAS detector!Tracker}

The central solenoid with the pixel, SCT and TRT detectors are now described.\index{ATLAS detector!Tracker}

\subsubsection{Central solenoid}\index{ATLAS detector!Tracker!Solenoid|(}
\begin{figure}[!h] 
  \begin{center}
    \includegraphics[width=.9\tw]{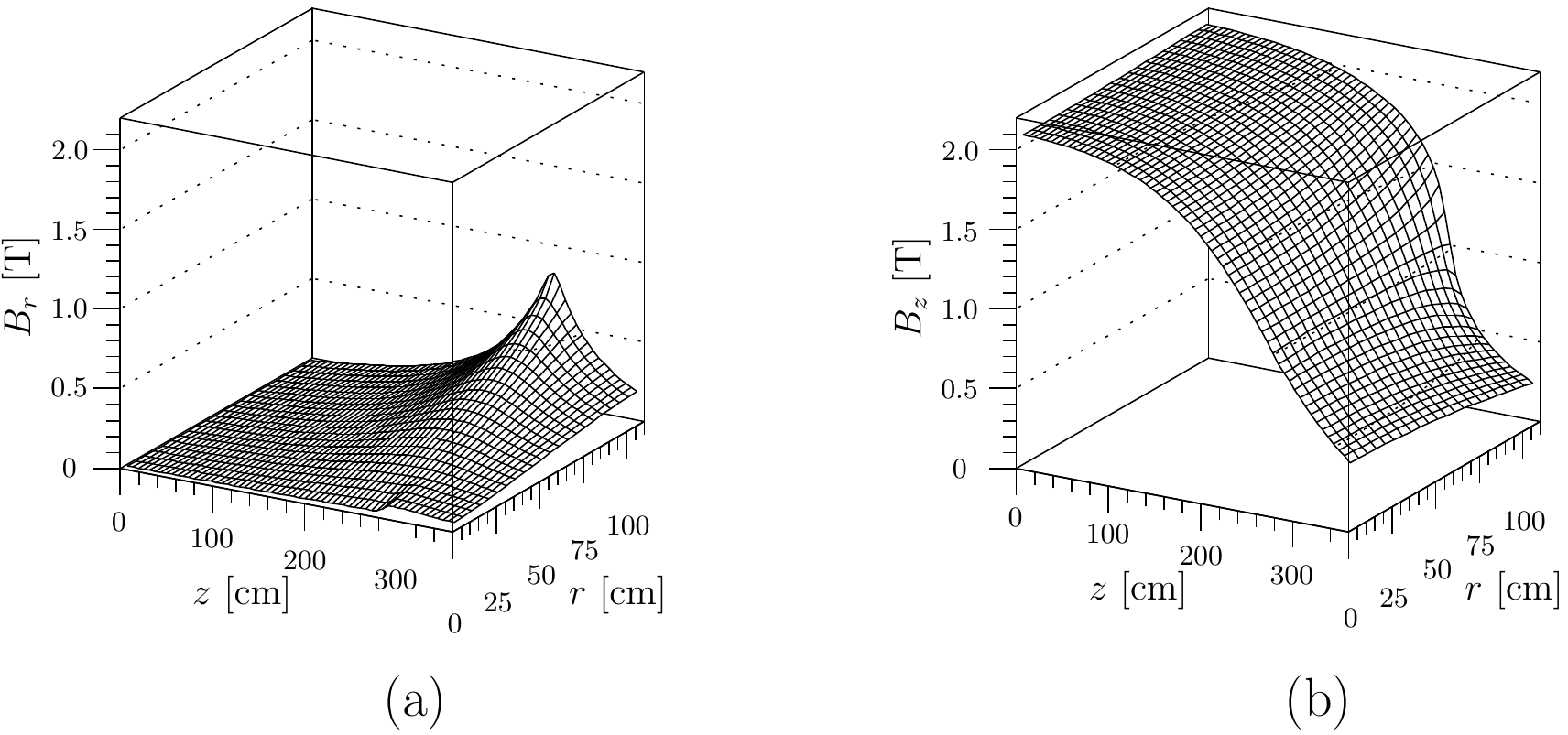}
    \caption[Magnetic field lines of the ATLAS central solenoid]
            {\figtxt{Mapping of the radial (a) and longitudinal (b) components of the magnetic 
                field generated by the central solenoid [Taken in~\cite{AtlasTDRvol1:1999fq}].
              }
            }
            \label{fig_atlas_solenoid}
  \end{center} 
\end{figure}
The solenoid coil, surrounding the tracker, generates a magnetic field to bend the tracks of 
charged particles emerging from the interaction point allowing, with the tracker instruments,
to identify and measure their transverse momenta.

The solenoid coil~\cite{Yamamoto:2008zze} ($1.23\,\mm{m}<r< 1.28\,\mm{m}$, $L=5.8\,\mm{m}$) 
provides a magnetic field
\begin{equation}
\vec B = B_r\,\vec e_r + B_z \,\vec e_z.
\end{equation}
In that equation, $B_z$ is the main component of the field and the radial component $B_r$,
optimally null at $z=0$, grows with $|z|$ due to border effects and the influence of iron in the 
tile calorimeter (Fig.~\ref{fig_atlas_solenoid}).
Charged particles are bent predominantly in the $r-\phi$ plane with bending powers of 
$2\Tesla\,\mm{m}$ at $\eta=0$ decreasing to $0.5\Tesla\,\mm{m}$ at $\Aeta<2.5$.
Besides, since the solenoid length is shorter than the tracker the field inhomogeneities in the 
forward region need to be accounted using a field map in both simulation and reconstruction.

One aspect in the present study relies on the capability of inverting the magnetic field of the 
solenoid which, although not programmed so far, is possible as stated in 
Ref.~\cite{Yamamoto:2008zze}.
The consequences of this operation, from the physics analysis point of view, will be discussed in
the analysis in Chapter~\ref{chap_W_mass_asym}.\index{ATLAS detector!Tracker!Solenoid|)}

\subsubsection{Pixel detector}\index{ATLAS detector!Tracker!Pixel detector|(}
\index{Pixel detector|see {ATLAS!Tracker!Pixel detector}}
\begin{figure}[!h] 
  \begin{center}
    \includegraphics[width=.4\tw]{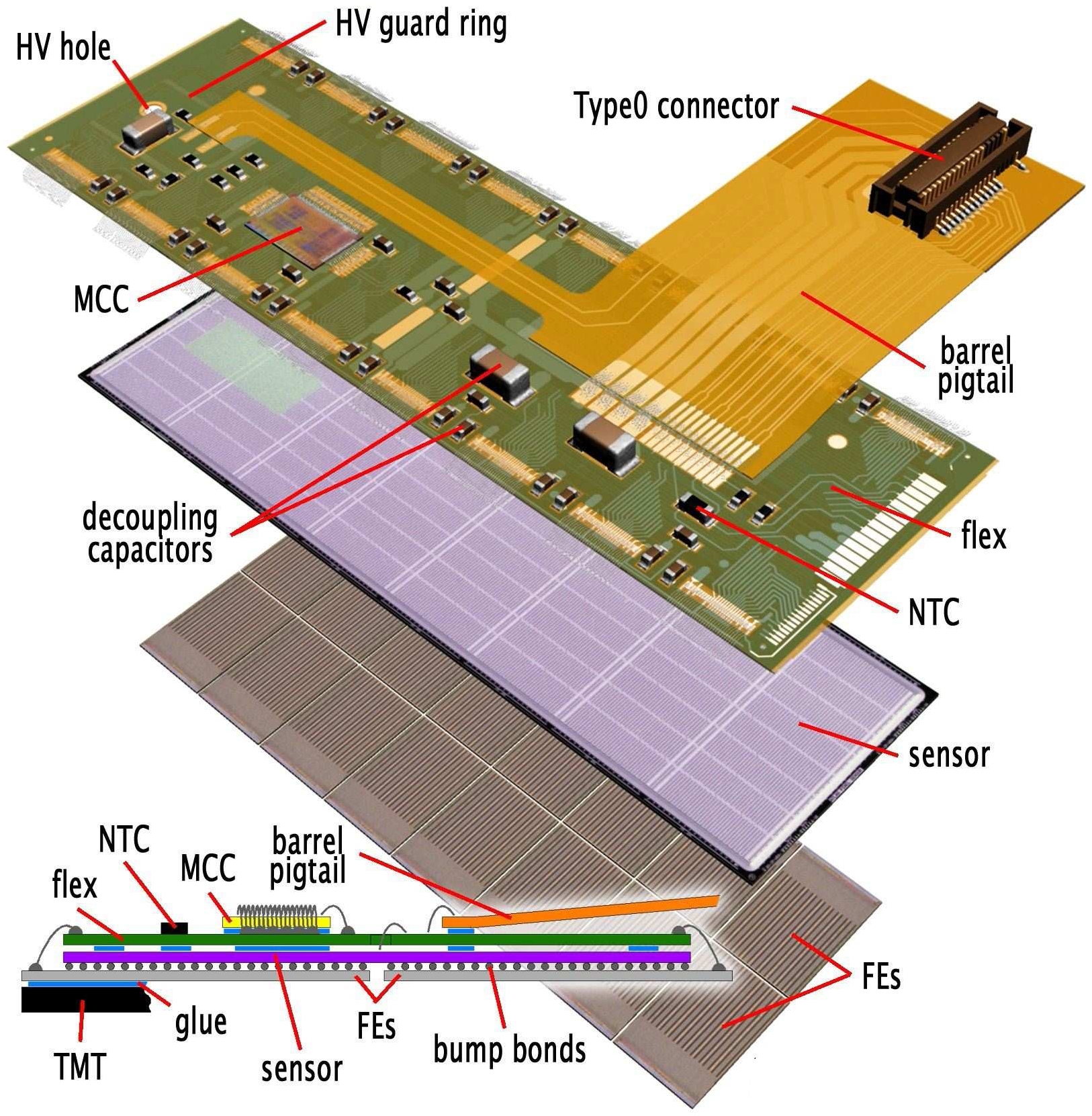}
    \hfill
    \includegraphics[width=.5\tw]{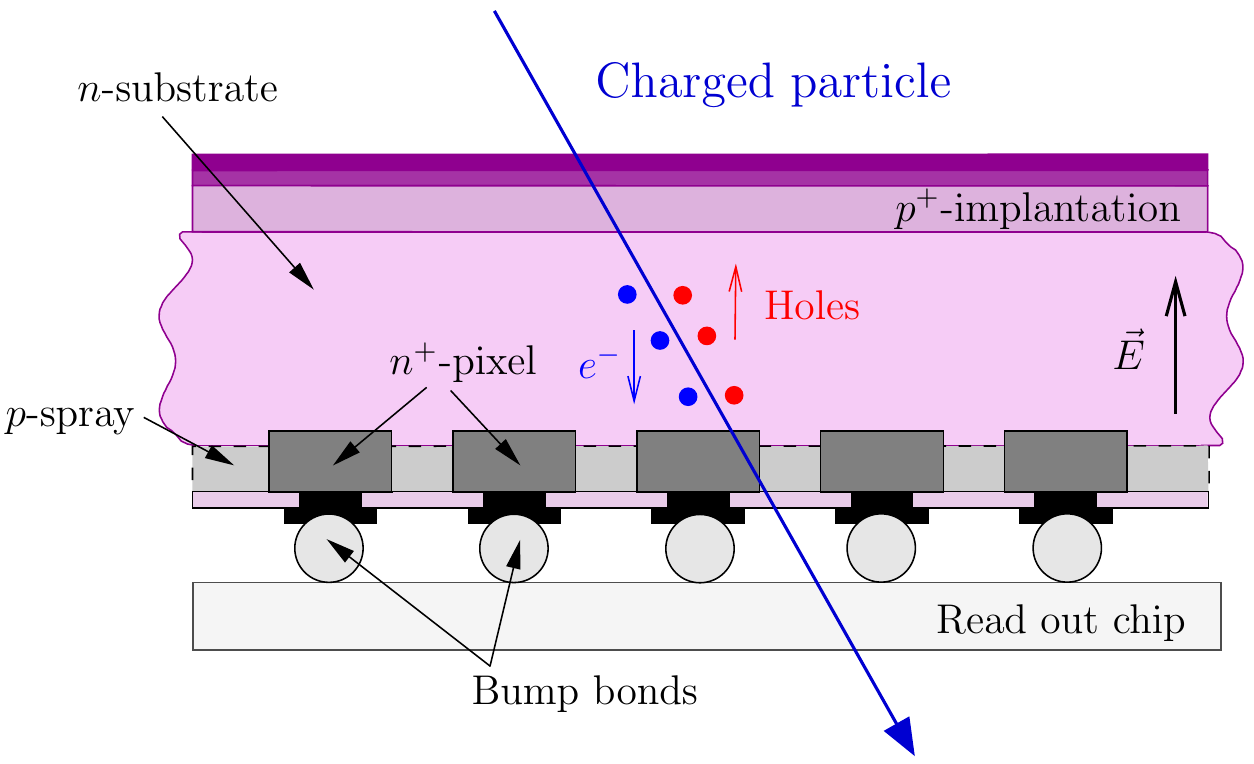}
    \vfill
    \includegraphics[width=1.\tw]{./figures/chapter_2/fig2_00-ab_lbls.pdf}
    \caption[Pixel module structure from the ATLAS tracker and principle of operation]
            {\figtxt{Pixel module structure of the ATLAS tracker (a)
                and principle of operation (b) [(a) taken in ~\cite{ATLASatCERN:2008zz}].
                }
            }
            \label{fig_atlas_pixel}
  \end{center} 
\end{figure}
The measurements from the pixel detector~\cite{AtlasInnerDetector:1997ft} 
are fundamental as they provide information on charged particles before they yield energy to the 
apparatus. 
Silicon detectors are used to provide fine vertexes information with high 
granularity~\cite{Damerell:1995fj}.
The cells are made, as shown on Fig.~\ref{fig_atlas_pixel}.(a), using silicon sensor layers 
of size $\mm{width}\times\mm{length}=50\times 400\,\mu\mm{m}^2$ segmented in both width and 
length to provide the pixel information.
Incoming charged particles ionise pairs of electrons/holes in the silicon.
The bias voltage applied in the silicon makes electrons drift to the $n^+$-side readouts.
Bump bonds transmit the collected charge to the front end electronic allowing to decipher which 
pixel was hit (Fig.~\ref{fig_atlas_pixel}.(b)).

Pixels modules --all identical in design-- are dispatched between a central and two forwards 
parts. The central part is made of three concentric cylinders and the forward one are made each
of three disks orthogonal to the $z$ axis. This provide for each particle track 3 hits.

\begin{figure}[!h] 
  \begin{center}
    \includegraphics[width=.8\tw]{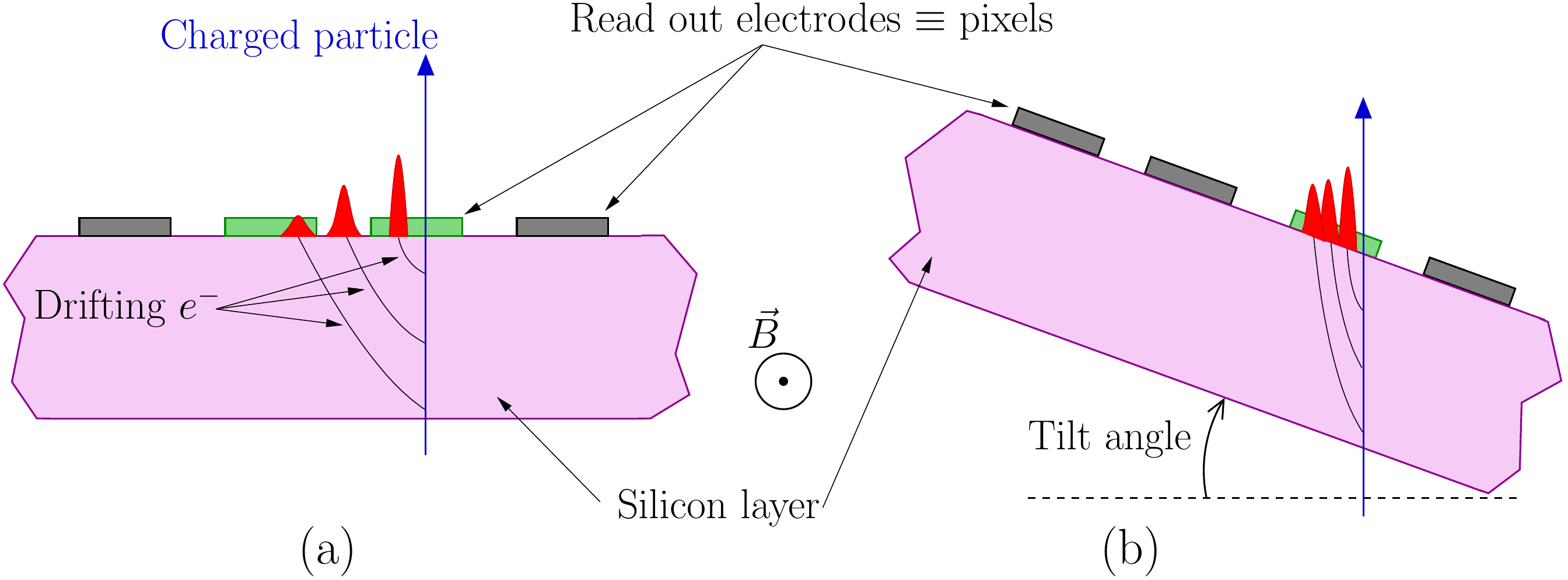}
    \caption[Alignment of the effective charge drift direction with the particle trajectory
    in a silicon pixel module]
            {\figtxt{Schematic representation --simplified with respect to Fig.~\ref{fig_atlas_pixel}.(b)--
                of the electrons drift in a barrel pixel module
                with no tilt (a) and with a tilt (b) that aligns
                the charge drift carriers direction with the one of the charged particle.
                The drift trajectories are not straight lines since the electric field
                is not constant in the depleted region, for more details see for 
                example Ref.~\cite{Gorelov:2001ca}.}
            }
            \label{fig_atlas_pixel_la}
  \end{center} 
\end{figure}

In the central barrel, as can be seen in Fig.~\ref{fig_atlas_tracker_r_xsection}.(b), 
modules are tilted in $\phi$ with respect to the tangent position. 
This tilt provides an overlap of the active area of the modules in the $\phi$-direction which 
enhance the hermetic confinement of particles.
It also ensures a better spatial resolution via the alignment of the effective charge drift 
--induced by the $\vec B$ field via Lorentz force-- direction with the particle 
trajectory~\cite{Gorelov:2001ca} as seen in Fig.~\ref{fig_atlas_pixel_la}.

Layers are segmented like $r\Delta\phi\times\Delta z = 10\,\mu\mm{m}\times 115\,\mu\mm{m}$.
In the forward barrel, that same segmentation correspondence is $r\Delta\phi\times\Delta r = 10
\,\mu\mm{m}\times 115\,\mu\mm{m}$.\index{ATLAS detector!Tracker!Pixel detector|)}

\subsubsection{Semi-Conductor tracker}
\index{ATLAS detector!Tracker!Semi-Conductor Tracker (SCT)|(}
\begin{figure}[!h] 
  \begin{center}
    \includegraphics[width=.45\tw]{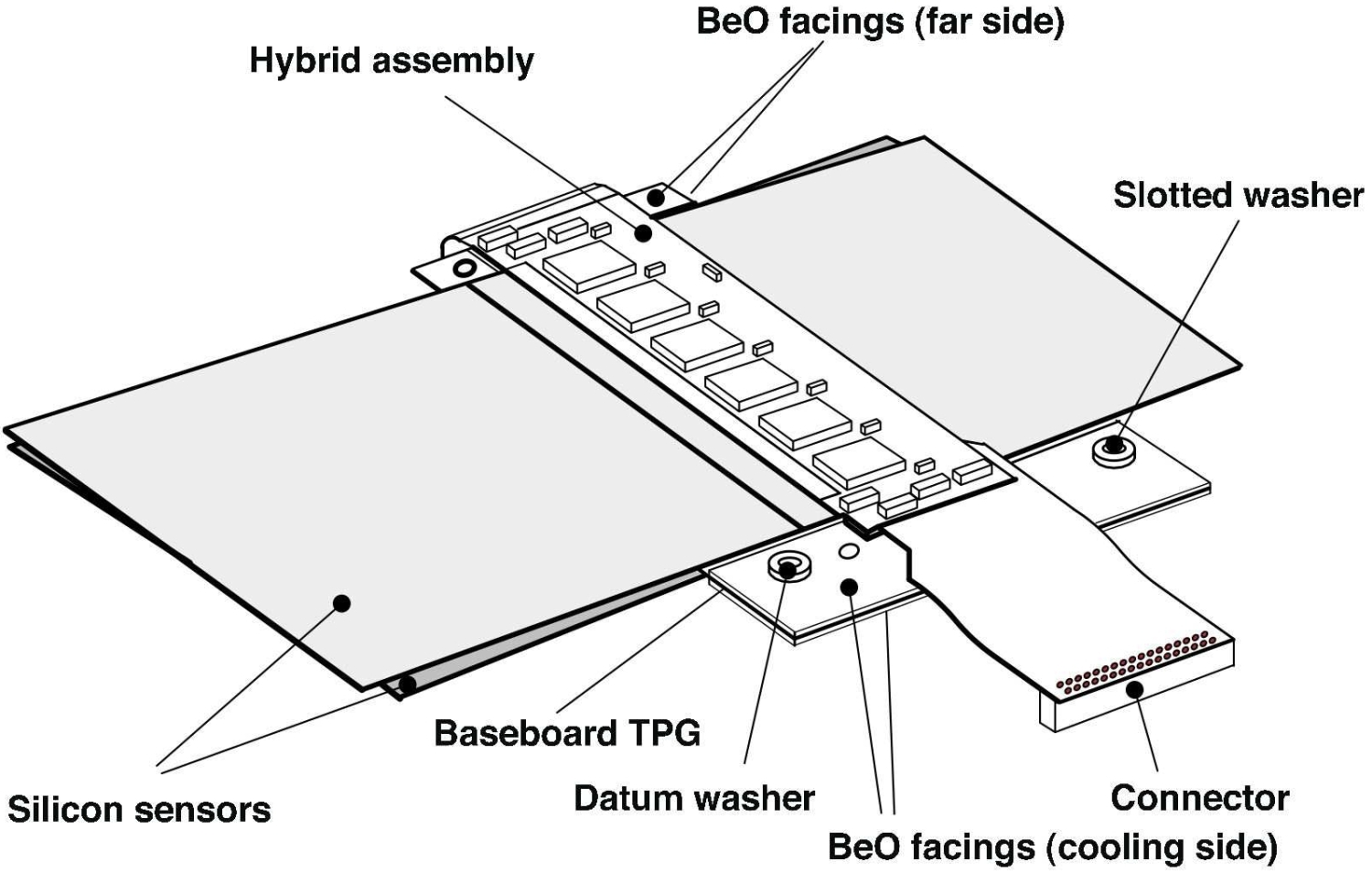}
    \hfill
    \includegraphics[width=.45\tw]{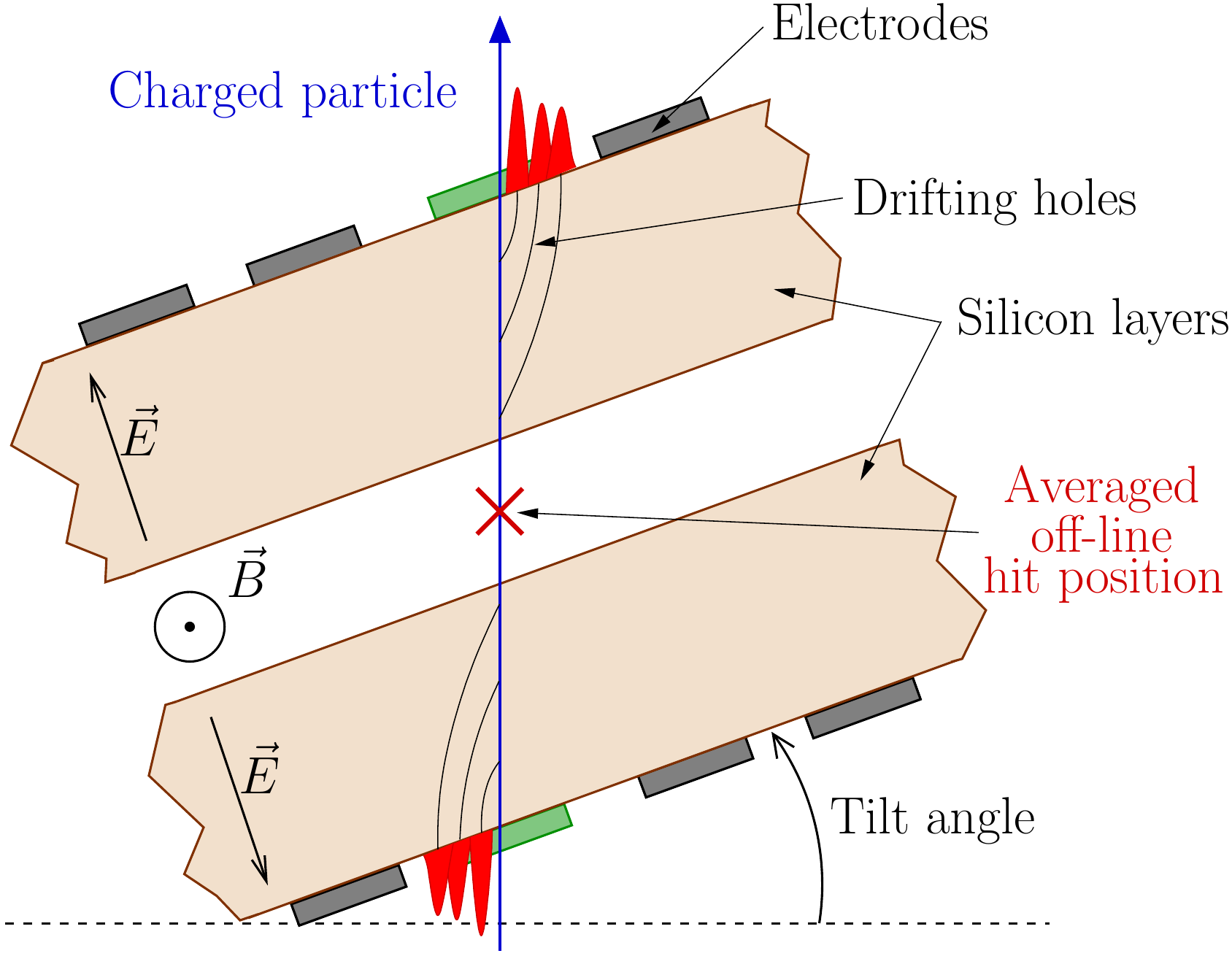}
    \vfill
    \includegraphics[width=.9\tw]{./figures/chapter_2/fig2_00-ab_lbls.pdf}
    \caption[SCT module structure and principle of operation]
            {\figtxt{SCT module structure (a) and principle of operation (b) 
                [(a) taken in ~\cite{ATLASatCERN:2008zz}].
                }
            }
            \label{fig_atlas_sct}
  \end{center} 
\end{figure}
The SCT~\cite{AtlasInnerDetector:1997ft,ATLASphdCornelissen} 
is based upon silicon micro-strip detector. 
Each module is made of two silicon sensor layers segmented in strips put back to back and 
rotated by 40 mrad to enhance the $z$ measurement (Fig.~\ref{fig_atlas_sct}.(a)).
The principle of operation is similar to pixel detector modules. 
Here the holes drift to the strips while the electrons drift to the back of the sensor. 
The strips are read out by a front-end chip, which measures the induction signal of the drifting
holes/electrons pairs.
Especially as shown on Fig.~\ref{fig_atlas_sct}.(b) the Lorentz drift in each layer are 
working in opposite directions. The position of the hit is averaged offline from the two sides
hit positions.
The active area is $\mm{width}\times\mm{length}=61.6\times 62\,\mm{mm}^2$ and the modules are 
segmented in both width and length.

The layout is made of four layers in the barrel and of nine disks in the end-cap orthogonal to the 
$z$~direction.
In the barrel modules are slightly tilted from the tangent position 
(Fig.~\ref{fig_atlas_tracker_r_xsection}.(b)) for the same reasons than for the pixel detector.
The segmentation in the barrel is $r\Delta\phi\times\Delta z = 580\,\mu\mm{m}\times 17\,\mu\mm{m}$.
In the end-cap, modules are also mounted to display some overlap and the corresponding 
segmentation is of $r\Delta\phi\times\Delta r = 580\,\mu\mm{m}\times 17\,\mu\mm{m}$.
\index{ATLAS detector!Tracker!Semi-Conductor Tracker (SCT)|)}

\subsubsection{Transition radiation tracker}
\index{ATLAS detector!Tracker!Transition Radiation Tracker (TRT)|(}
\begin{figure}[!h] 
  \begin{center}
    \includegraphics[width=.4\tw]{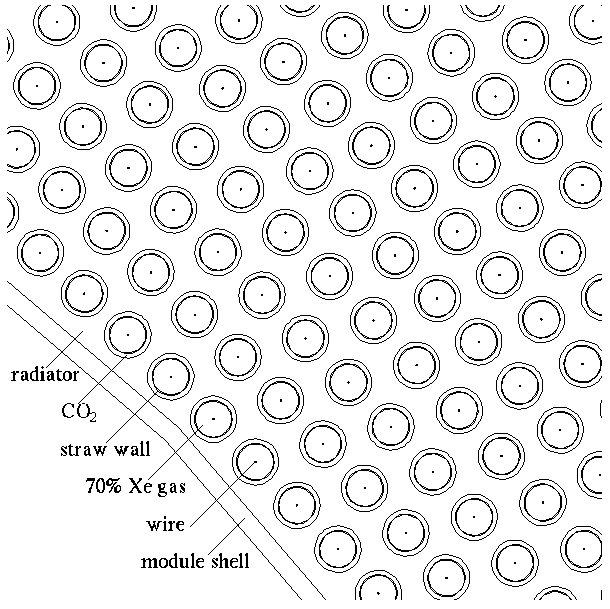}
    \hfill
    \includegraphics[width=.5\tw]{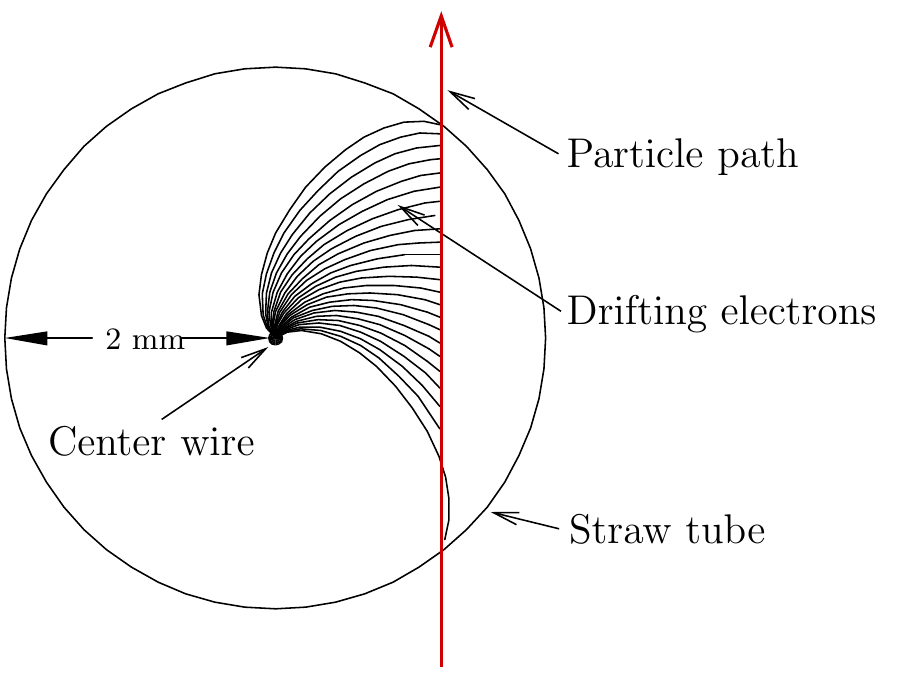}
    \vfill
    \includegraphics[width=1.\tw]{./figures/chapter_2/fig2_00-ab_lbls.pdf}
    \caption[TRT tubes assembly and principle of operation]
            {\figtxt{TRT tubes assembly (a) and principle of operation of one tube (b)
                [(a) taken and (b) adapted in/from~\cite{AtlasInnerDetector:1997fs}].
                }
            }
            \label{fig_atlas_trt}
  \end{center} 
\end{figure}
The TRT ~\cite{AtlasInnerDetector:1997ft} assures continuous tracking as well as electrons 
identification. 
Even if it is not as precise as the silicon trackers its long lever arm plays an important role in 
the momentum resolution.

The TRT is made of layers of gas filled straws interleaved with transition radiation material.
A charged particle passing through the straw ionise the electrons of the gas mixture 
(Fig.~\ref{fig_atlas_trt}.(b)).
A large potential difference is applied between the straw wall and the wire. 
The anode wire collects the energy of the ionised electrons (Fig.~\ref{fig_atlas_trt}.(b)).
Also, charged particles as they pass trough materials of different dielectric constant
(radiator $\to$ straw) radiate photons proportionally to their Lorentz $\gamma$~factor.
Part of the gas mixture in the straw is sensitive to photons and in consequence this ionisation 
energy adds up to the collection of the one induced by the incoming particle.
The threshold for a significant radiation ($\gamma\sim 1000$), depends then on the 
mass of the particle and help to identify electrons amid other heavier charged particles.
The TRT only provides information in $r\Delta\phi$ with an accuracy of $130\,\mu\mm{m}$.

In the barrel the straws are parallel to the $z$ axis and arranged in three cylindrical rings.
In each end-cap the wires are aligned to the radial direction and arranged in three sets of 
identical and independent wheels.
Again the precise dimensions and pseudo-rapidity coverage can be seen in 
Figs.~\ref{fig_atlas_tracker_z_xsection}.
\index{ATLAS detector!Tracker!Transition Radiation Tracker (TRT)|)}

\subsection{Track fitting and general performances}\label{ss_id_trck_fit_perf}
The track fitting is realised in three stages.
First the raw data from the pixel/SCT and TRT detectors are respectively converted into clusters
and calibrated circles. The SCT clusters are converted to space-points using a combination of the 
cluster from both sides of SCT modules (cf. Fig.~\ref{fig_atlas_sct}.(b)).
Then comes the track finding stage where algorithms~\cite{ATLASNEWT,AtlasInnerDetector:1997fs} 
essentially follow pattern recognition starting from the innermost pixel layers and goes outwards 
to the TRT. 
These algorithms, based on Kalman filter techniques~\cite{Fruhwirth:1987fm} 
and Global-$\chiD$~\cite{BruckmandeRenstrom:2005ha}, perform recognition of helices among the hits 
in the tracker.
In the final stage primary vertexes are resolved.
Also tracking information from the muon spectrometer is used as well to enhance the data on muons.
In the present work the expected stand-alone performances of the tracker are used exclusively.

The ID performances in $\pT$, $\theta$ and $\phi$ can be parametrised using 
Gaussian functions (cf. Ref.~\cite{AtlasTDRvol1:1999fq} \S\,3.3.1.6) where, 
up to the approximation the material and the solenoid field are uniform in $r$ writes
\begin{eqnarray}
%
\sigma_{1/\pT}       &=&  3.6\times 10^{-4} 
                         \oplus 
                         \frac{1.3\times 10^{-2}}{p_T^\true\,\sqrt{\sin\theta^\true}}
                         \quad[\mathrm{GeV}^{-1}],
                         \label{eq_rho_smearing}\\
%
\sigma_{\cotan\theta} &=& 0.7\times 10^{-3}  
                        \oplus 
                        \frac{2.0\times 10^{-3}}{p_T^\true\,\sin^{3/2}\theta^\true},
                        \label{eq_theta_smearing}  \\
%
\sigma_\phi          &=& 0.075\times 10^{-3} \oplus 
                         \frac{1.8\times 10^{-3}}{p_T^\true\sqrt{\sin\theta^\true}}
                        \quad[\mm{rad}].
\end{eqnarray}
where $p_T^\true$ are in GeV and (true) superscript means a kinematic is considered 
at the generator level.

\section{The weak modes affecting the inner detector}
\label{s_weak_modes}

\index{ATLAS detector!Tracker!Misalignment/weak modes|(}
\index{Tracker misalignment|see{ATLAS and CDF}}
\subsection{Misalignment and definition of the weak modes}
The tracker is build into an \textit{a priori} perfect ``blue-print'' configuration, 
but in reality modules assembly differs from such an ideal picture due to mounting limitations, 
mechanical stress, temperature variations, sagging due to gravity, \etc{}.
Since the tracker relies on the relative hits positions to measure the particles momenta and 
vertexes these small misalignment spoil the relevancy of reconstructed tracks.
Since the module positions cannot be touched after their assembly modules misalignment are 
accounted by performing a mapping of the detector shape and module positions to correct
the collected data. This procedure is called the alignment.
Since some of the constraints on the tracker are time dependent this alignment survey must be 
continually updated.

The alignment of the tracker is realised using two kind of methods\,: hardware-based and 
track-based methods.
Hardware methods make \textit{in situ} measurement of the shape of the support structure and
its change over time.
Track based alignment requires, using the least squares principle, that the measurements in the detector
are consistent with the assumed track model, \ie{} follows the expected track trajectory in the given 
$\vec B$ field and the scattering is consistent with the known amount of material.

\begin{figure}[!h] 
  \begin{center}
\includegraphics[width=.65\tw]{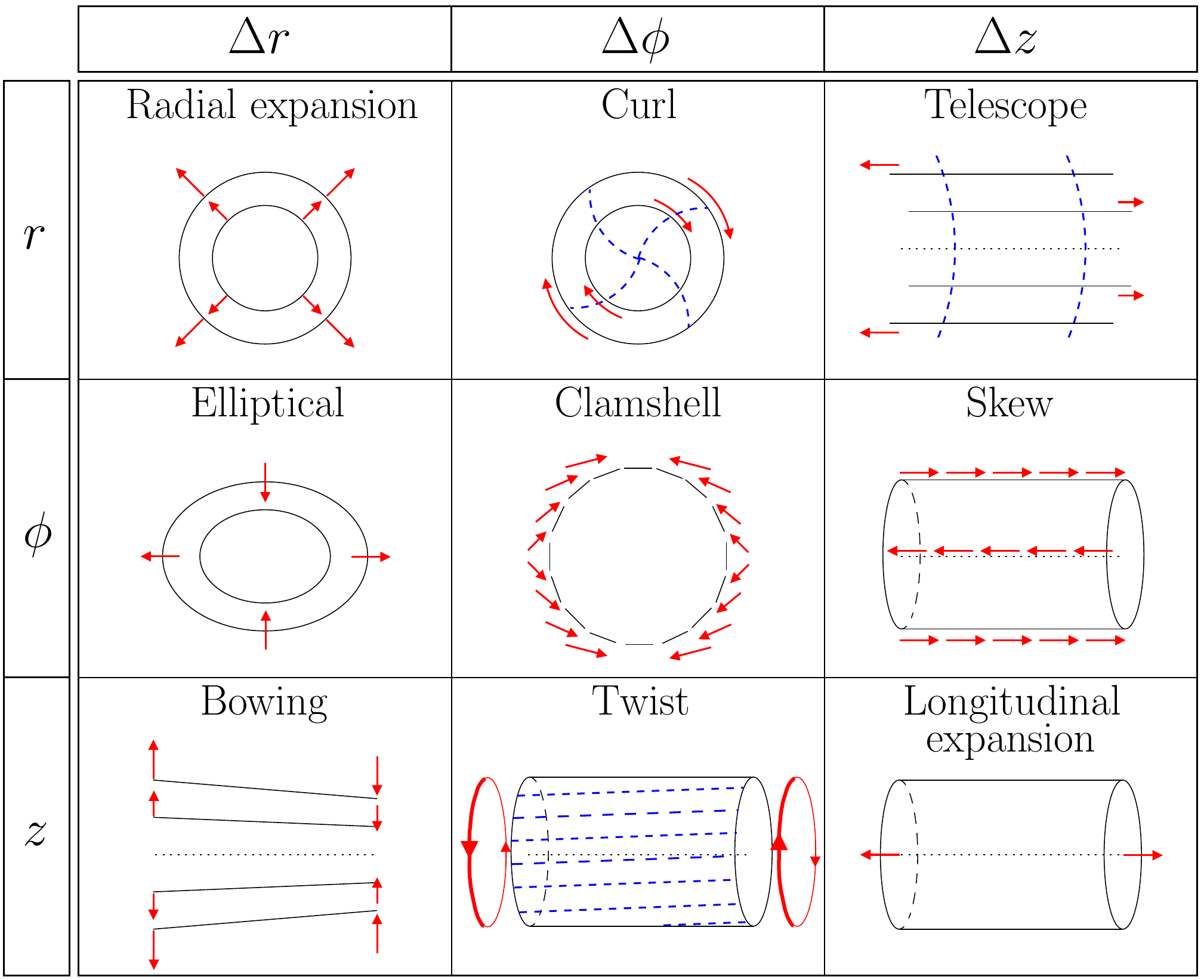}
    \caption[Weak modes affecting at the tracker at first order]
            {\figtxt{Weak modes affecting at the first order the alignment of the tracker.
                The arrows represent the constraints modules are submitted to and the dashed
                lines emphasise the consequences when not obvious to visualise.
                }
            }
            \label{fig_weak_modes}
  \end{center} 
\end{figure}

\index{Weak modes!In ATLAS tracker|(}
Still, despite these surveys some deformations are such that the fitted track even if 
being relevant from the tracking algorithm point of view --\ie{} an helix is recognised-- nonetheless 
deviates from the real track. 
These deformations can be --in a first approximation-- represented by a set of 9 simple
and independent distortions called the weak modes~\cite{Brown:2006zz},
parametrised using global deformations on the tracker.
Due to the symmetry of the problem, the parametrisation is made in the cylindrical coordinate 
system, hence the combination of deformations in $r$, $\phi$ and $z$ directions folded with 
$\Delta r$, $\Delta \phi$ and $\Delta z$ variations gives 9 weak modes.
The total misalignment of the tracker is a combination of these 9 modes gathered in 
Fig.~\ref{fig_weak_modes}.

More details on the alignment and weak modes can be found in 
Refs.~\cite{Brown:2006zz,ATLASphdStorig,ATL-COM-PHYS-2008-243}.

\subsection{Consequences of the weak modes on the error on the W boson mass measurement}
Curl and twists modes are responsible for limitations on the precision on the $W$~mass 
determination at CDF~II~\cite{Lukens:2003aq}.
In CDF~II systematic errors on $\MW$ are of the same order than the statistical 
errors, \ie{} $\delta_\MW^{\mm{(stat.)}}=\delta_\MW^{\mm{(sys.)}}= 34\MeV$ while at the LHC 
statistical errors will be of the order of $5\MeV$ already for one year at low luminosity.
For that reason the weak modes have to be addressed more thoroughly within ATLAS in regard of the difficulties
encountered in the CDF~II tracker (cf.~Appendix~\ref{cdf_tracker}).
For example, considering again our interest in $W$ production, a requested precision of $25\MeV$ 
for the measurement of the $\MW$ constrains track parameters and momentum uncertainties. 
For that purpose the degradation of the high $\pT$ tracks parameters due to misalignment 
have to be smaller than $20\percent$ while the systematic uncertainty on the 
momentum resolution needs to be smaller than $0.1\percent$~\cite{AtlasTDRvol1:1999fq,AtlasTDRvol2:1999fr}.

\index{W boson@$W$ boson!Mass charge asym@Mass charge asymmetry $\MWp-\MWm$}
Here the focus is made on the predominant modes increasing the error of $\MWp-\MWm$.
Like it will be explained thoroughly in Chapter~\ref{chap_w_pheno_in_drell-yan} the present work is
based on the knowledge of the transverse momenta $\pTl$ of charged leptons decaying from single 
$W$~bosons. 
Hence the attention is cast on modes affecting the reconstruction of positive and negative
transverse tracks.
Modes involving $\Delta z$ deformations are not considered as they do not degrade the resolution 
of $\pTl$, this leaves $\Delta r$ and $\Delta \phi$ modes which are discussed below.
Two kind of biases are considered, the ones biasing the positive and negative charged particles
tracks curvatures in the same direction (coherent biases) and the one affecting them in 
opposite directions (incoherent biases) the latter being the most important source of errors
for $\MWp-\MWm$.
In what follows no values are estimated, only the relative qualitative amplitudes between coherent and
incoherent biases.

The modes amplitudes biases are noted $\es$ and tagged using the $2\times 2$ matrix form 
displayed in Fig.~\ref{fig_weak_modes}, \ie{}
using the notation $r$ or $\Delta r\to 1$, $\phi$ or $\Delta\phi\to 2$ and $z$ or $\Delta z\to 2$.
The sign of a scalar $a$ is noted $\sg{a}\equiv a/|a|$.
\begin{figure}[!h] 
  \begin{center}
\includegraphics[width=1.\tw]{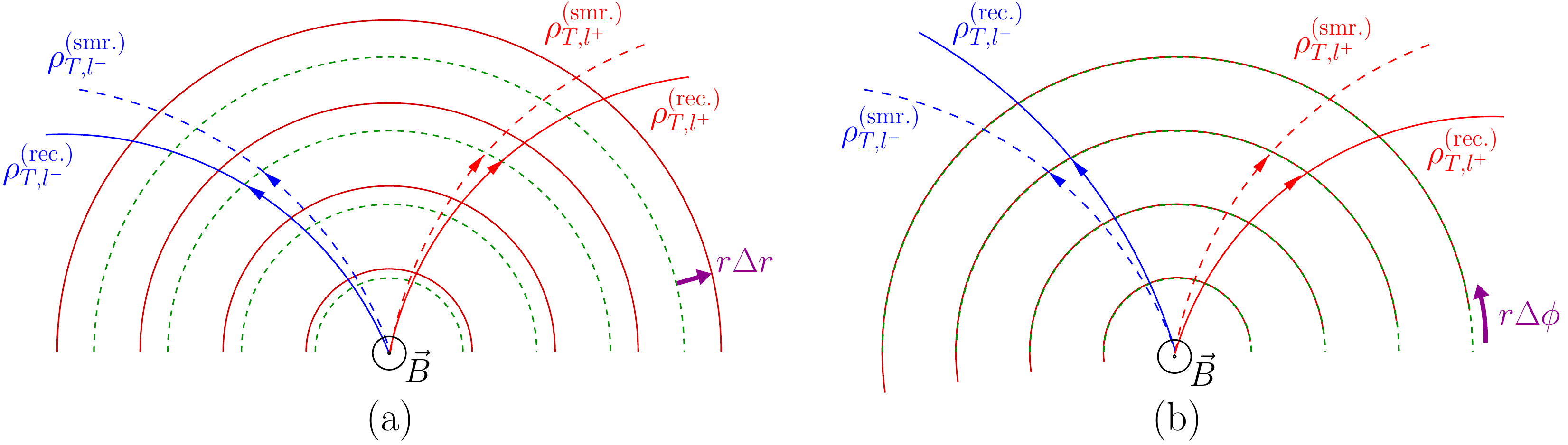}
    \caption[Radial expansion and curling modes consequences on positively and negatively
    charged particles reconstructed tracks]
            {\figtxt{Consequences of the radial expansion (a) and curling (b) distortions 
                on the positively and negatively charged particles reconstructed tracks. 
                In both cases, tracks of the same curvature, \ie{} 
                $\rho^\smear_{T,\lp}=\rho^\smear_{T,\lm}$, are considered.
                }
            }
            \label{fig_e11_e12_modes}
  \end{center} 
\end{figure}

\subsubsection{$\mbf{\Delta r}$ modes}

\paragraph{$\mbf{r\Delta r}$ : radial.}
The radial expansion ($r\Delta r>0$) and contraction ($r\Delta r<0$) infer coherent biases. 
This is shown in Fig.~\ref{fig_e11_e12_modes}.(a) in the case of a radial expansion while 
starting with smeared curvatures of the same values $\rho^\smear_{T,\lp}=\rho^\smear_{T,\lm}$.
The reconstructed curvatures can be written in the first order of a perturbation expansion in the 
parameter $\es_{11}(\propto\sg{\Delta r})$ that governs the radial expansion/contraction amplitude\,:
\begin{eqnarray}
\rho^\rec_{T,\lp} &=& \rho^\smear_{T,\lp}\,(1+\es_{11}), \\
\rho^\rec_{T,\lm} &=& \rho^\smear_{T,\lm}\,(1+\es_{11}).
\end{eqnarray}

\paragraph{$\mbf{\phi\Delta r}$ : elliptical.}
For this mode, the effect of the gravity will most likely flatten the tracker as depicted in 
Fig.~\ref{fig_weak_modes}. 
In the region where the nominal radius of the tracker is larger ($\Delta r>0$) positive and 
negative particle tracks curvature are overestimated.
On the contrary tracks are getting more straight in regions where the tracker's nominal radius
decreases ($\Delta r<0$).
Eventually these two effects should compensate each other and leave a residual coherent bias
such that $\es_{21}<\es_{11}$.

\paragraph{$\mbf{z\Delta r}$ : bowing.}
Here, for a given $z$ coordinate positive and negative reconstructed tracks
will be more straight (bent) if the tracker is expanding (contracting) with 
respect to its nominal radius. 
Biases coming from expansions are counter balanced by the one arising from contractions.
The final bias from the bowing should be a residue from the averaging of these two opposite 
constraints. In any of those cases, again, the coherent bias should verify $\es_{31}<\es_{11}$.

\subsubsection{$\mbf{\Delta \phi}$ modes}

\paragraph{$\mbf{r\Delta \phi}$ : curling.}
The curling of the tracker gives incoherent biases.
Figure~\ref{fig_e11_e12_modes}.(b) shows it in the case where $\sg{\Delta\phi}>0$ and with smeared 
curvatures are of the same values $\rho^\smear_{T,\lp}=\rho^\smear_{T,\lm}$.
The curling acts on the curvature like $\rho^\smear_{T,l}\to\rho^\smear_{T,l}+\delta_\mm{curl}$ where 
$\delta_\mm{curl}$ is the bias induced by the curl.
For small enough values of $\delta_\mm{curl}$ we assume --to keep the same parametrisation used up to now--
that the reconstructed curvatures can be written in the first 
order of a perturbation expansion in the parameter governing the curl amplitude 
$\es_{12}(\propto\sg{\Delta \phi})$\,:
\begin{eqnarray}
\rho^\rec_{T,\lp} &=& \rho^\smear_{T,\lp}\,(1+\es_{12}), \\
\rho^\rec_{T,\lm} &=& \rho^\smear_{T,\lm}\,(1-\es_{12}).
\end{eqnarray}

\paragraph{$\mbf{\phi\Delta \phi}$ : clam-shell.}
This mode is unlikely to affect any component of the inner 
tracker~\cite{DiscussionPawelBruckmanDeRestrom}.

\paragraph{$\mbf{z\Delta \phi}$ : twist.}
To illustrate the twist, a rotation of the tracker $\Delta\phi<0$ of the left side and of 
$\Delta\phi>0$ on the right side are considered.
Based on the understanding of the curl mode, on the left side positive tracks are believed
to be more bent while negative tracks are believed to be more straight. On the right side this effect 
is reversed. 
Just like for the bowing, antagonists modes of the same amplitudes cancel each other.
Eventually the most important twist should contribute with a term like $\es_{32}<\es_{12}$.

\subsubsection{Global effect of the 6 previous modes}
In conclusion, the reconstructed positively and negatively charged tracks should be biased like
\begin{eqnarray}
\rho^\rec_{T,\lp} &=& \rho^\smear_{T,\lp}\,(1+\es_{\Delta r}+\es_{\Delta \phi}), \label{eq_rhoTlprec}\\
\rho^\rec_{T,\lm} &=& \rho^\smear_{T,\lm}\,(1+\es_{\Delta r}-\es_{\Delta \phi}), \label{eq_rhoTlmrec}
\end{eqnarray}
where $\es_{\Delta r}$ and $\es_{\Delta \phi}$ are respectively the global coherent bias 
from ${\Delta r}$ and ${\Delta \phi}$ modes, \ie{}
\begin{eqnarray}
\es_{\Delta r}    &=& \es_{11}+\es_{21}+\es_{31}, \\
\es_{\Delta \phi}  &=& \es_{22}+\es_{32}.
\end{eqnarray}

Again, let us repeat that only qualitative estimations are made here.
This is justified as in our work these biases will be implemented with large worst case scenario
values to improve the robustness of our proposed analysis scheme.
\index{Weak modes!In ATLAS tracker|)}
\index{ATLAS detector!Tracker!Misalignment/weak modes|)}
\index{ATLAS detector!Tracker|)}
\index{ATLAS detector|)}
%
%
%
%
\cleardoublepage
\begin{subappendices}
%
%
%
%

\makeatletter\AddToShipoutPicture{%
\AtUpperLeftCorner{2cm}{2cm}{\ifodd\c@page\else\makebox[0pt]{\Huge$\bullet$}\fi}%
\AtUpperRightCorner{0cm}{2cm}{\ifodd\c@page\makebox[0pt]{\Huge$\bullet$}\else\fi}%
}\makeatother
\section{W mass charge asymmetry and tracker misalignment in CDF~II}\label{cdf_tracker}
\setlength{\epigraphwidth}{0.7\tw}
\epigraph{
(About the difficulties to measure $\MW$ because of tracker misalignment)\\
``You all remember how during your studies you were taught a solid could be described by only six 
degrees of freedom. Well. Forget about it. It's crap.''
}%
{\textit{CTEQ-MCnet Summer School 2008 - Standard Model lectures} \\\textsc{Tom LeCompte}}

\subsection{Context of the measurement of the W mass at CDF}
\index{Tevatron collider}
\index{CDF detector|(}
The CDF detector used in Tevatron Run II, labeled CDF~II~\cite{Lukens:2003aq}, 
is a multipurpose detector nominally forward-backward symmetric with respect to the interaction
point where protons and anti-protons collide at a center of mass energy of $\sqrt{S}=1.96\TeV$.
It is made, starting from the beam-pipe, of an inner tracker bathing in a $1.4\Tesla$ solenoidal 
magnetic field, an electromagnetic calorimeter followed by a hadronic calorimeter to contain and 
measure respectively the energies of electrons/photons and hadrons. Finally a muon spectrometer
surrounds the previous apparatus to measure the properties of muons.
The data is read-out on-line using the decisions of a three level trigger system.

The measurement of $\MW$~\cite{Aaltonen:2007ps,Aaltonen:2007ypa} in CDF~II is achieved using 
the tracker data for the muons and both tracker and electromagnetic calorimeter data for electrons.
The acceptance and resolutions for central electrons and muons are the same which means both 
channels enter with the same weight in the analysis.

\index{W boson@$W$ boson!Mass charge asym@Mass charge asymmetry $\MWp-\MWm$!In CDF|(}
\index{W boson@$W$ boson!Mass@Mass $\MW$!In CDF|(}
The extraction of $\MW$ is addressed via the usual observables such as $\pTl$,
$\mTlnu$ and $\slashiv{p}_{T,l}$ and via muon and electronic decays of the $W$.
Along all these information the difference between the masses of the positive and negative 
$W$~bosons is estimated as a mean of consistency check. 
Actually this last measurement is not to be apprehended as a real attempt to measure $\MWp-\MWm$.
Rather than that is has to be understood that the measurement is entirely focused on $\MW$ and
that the effects responsible for the low precision on $\MWp-\MWm$ were at no time 
addressed by the authors as long at it does not have a major role for the determination of $\MW$. 
Both $\MW$ and $\MWp-\MWm$ CDF results for these last years, as seen in Table~\ref{table_cdf_mw_mwp_mwm},
are recaptured here\,:
\begin{center}
\begin{tabular}{lr@{$\,\pm\,$}lr@{$\,\pm\,$}lc}
  \hline
  Channel               & \multicolumn{2}{c}{$\MW$ [$\mm{GeV}$]} & 
  \multicolumn{2}{c}{$\MWp-\MWm$ [$\mm{GeV}$]} & Year\\
  \hline\hline
  $W\to l\,\nul$        & $79.910$&$0.390$           & $-0.190$&$0.580$ & 
  1990,1991 \cite{Abe:1990pp,Abe:1990tq}\\
  \hline
  $W\to \mu\,\numu$     & $80.310$&$0.243$             & $0.549$ & $0.416$ & 
  \multirow{3}{*}{1995 \cite{Abe:1995np,Abe:1995nm}} \\
  $W\to e\,\nue$        & $80.490$&$0.227$             & $0.700$&$0.290$ &\\
  $W\to l\,\nul$        & $80.410$&$0.180$           & $0.625$&$0.240$ &\\
  \hline
  $W\to \mu\,\nu_\mu$   & $80.352$&$0.060$         & $0.286$&$0.152$ &  
                                  \multirow{3}{*}{2007 \cite{Aaltonen:2007ypa,Aaltonen:2007ps}}\\
  $W\to e\,\nu_e$       & $80.477$&$0.062$         & $0.257$&$0.117$ &\\
  $W\to l\,\nul$        & $80.413$&$0.048$       
& \multicolumn{2}{l}{\qquad\quad\;$\huge{\times}$}\\
 \hline
\end{tabular}
\end{center}
\medskip
The question one might ask is how come the absolute mass is measured with an error of 
$\approx 40\MeV$ 
then ? The trick is that incoherent biases are at work between the positively and negatively 
charged particle tracks and they get drastically reduced when both charges are merged.

The next subsection describes briefly the CDF~II central outer tracker used for the $W$ measurement.
After that a recapitulation of how the tracker misalignment affects charged particles curvature is made.
This shows how the experience from Tevatron physicists guided us to address, aware of the LHC/ATLAS
original features, the relevant weak modes for a future dedicated measurement of $\MWp-\MWm$ in ATLAS. 
It also explains why so far all experimental measurements of $\MWp-\MWm$ display such a low accuracy.

\subsection{Description of the CDF Central Outer Tracker}
\index{CDF detector!Tracker|(}
The CDF configuration is such that we borrow for its description the same conventions 
adopted for ATLAS (cf.~\S\,\ref{notations_conventions}).
The CDF~II uses silicon at lower radii and drift tubes technologies afterward.
The silicon detector is not detailed since its data was not used for the determination of $\MW$.

Around the silicon tracker is an open-cell drift chamber, the COT~\cite{Affolder:2003ep} 
which span the radial range $40\,\mm{cm}<r<137\,\mm{cm}$ and extend longitudinally 
for $|z|<155\,\mm{cm}$ ($|\eta|\approx 0.1$).

\begin{figure}[!h] 
  \begin{center}
    \includegraphics[width=.45\tw]{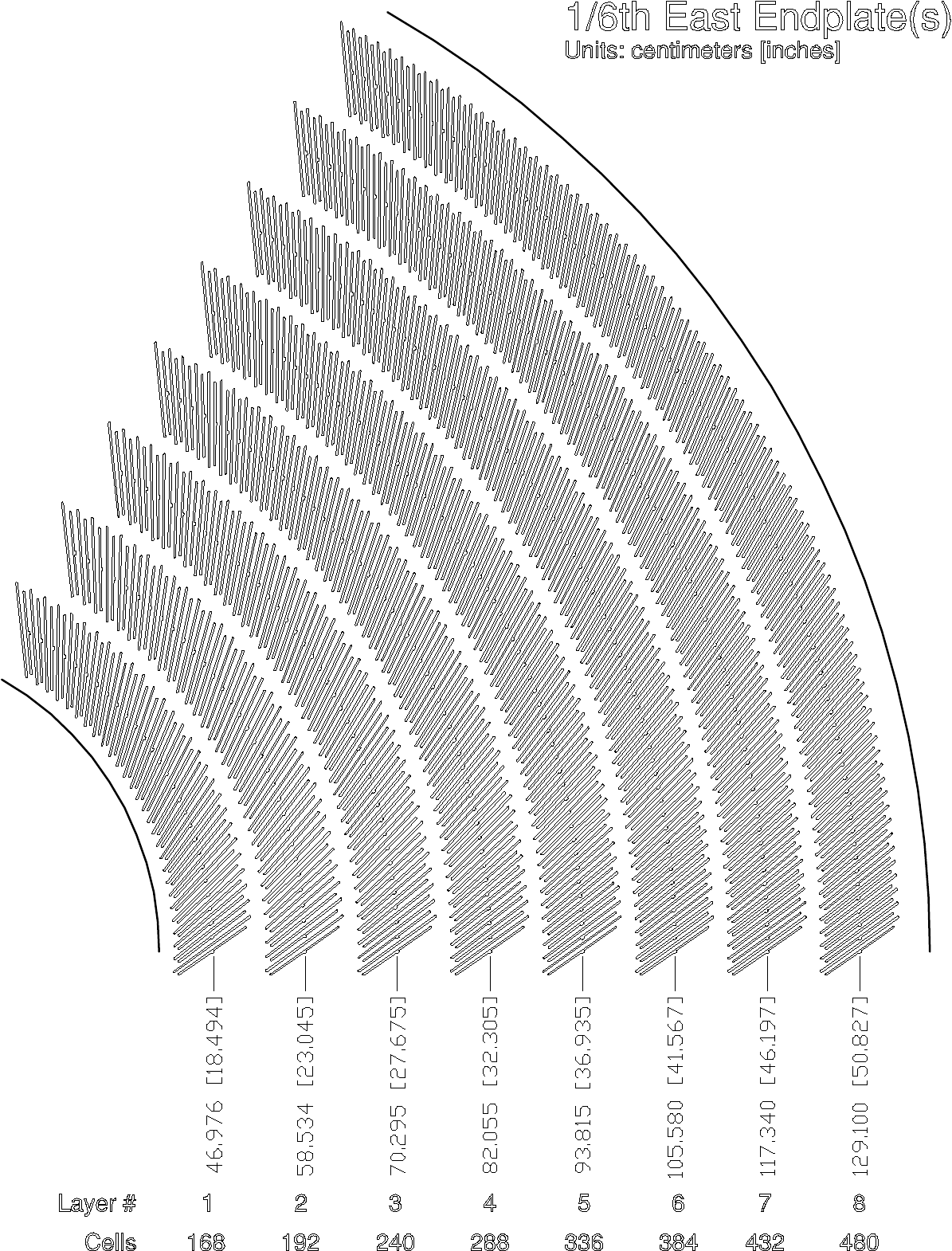}
    \hfill
    \includegraphics[width=.45\tw]{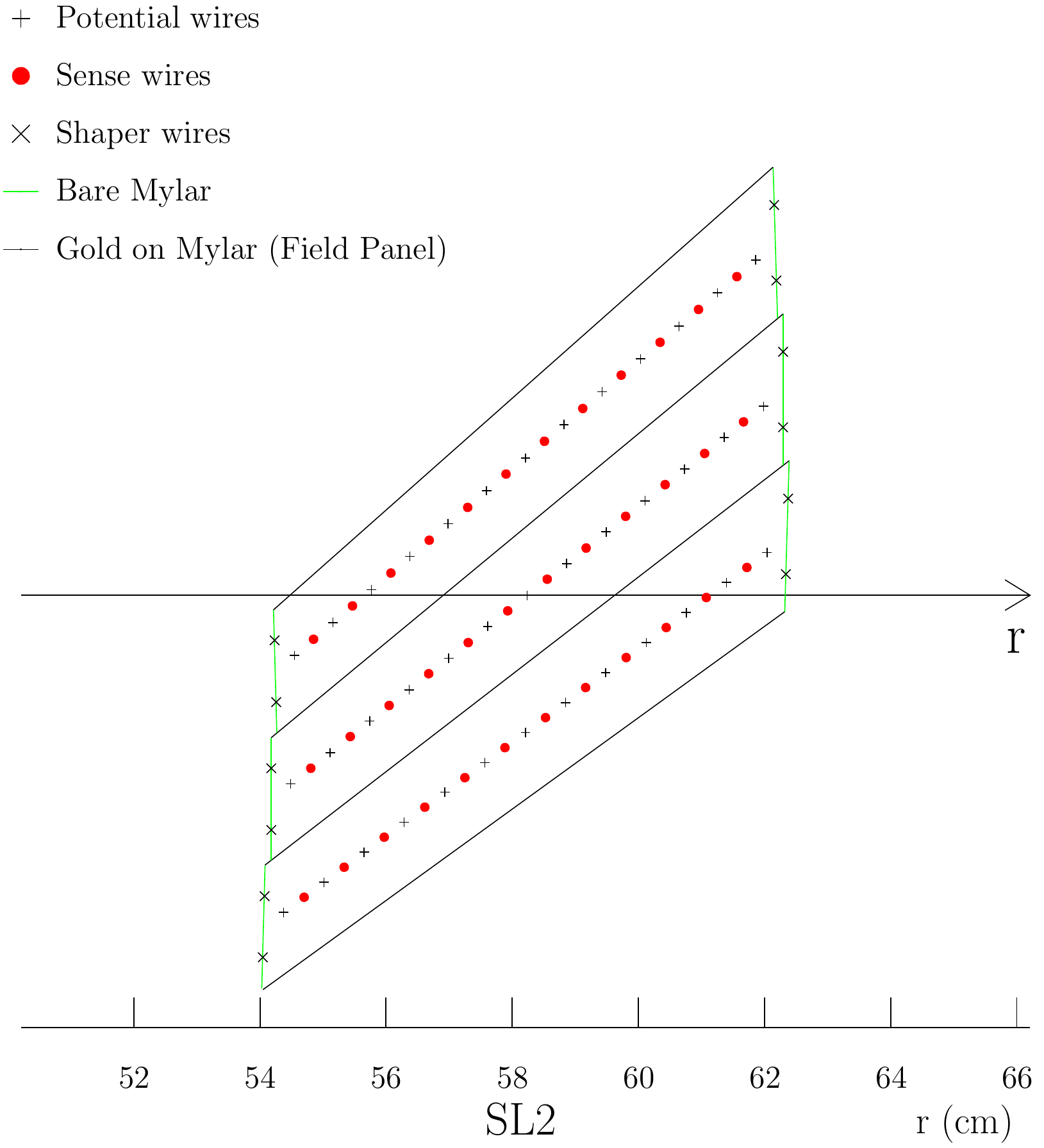}
    \vfill
    \includegraphics[width=1.\tw]{./figures/chapter_2/fig2_00-ab_lbls.pdf}
    \caption[CDF central outer tracker structure]
            {\figtxt{Structure of the CDF~II central outer tracker [Taken in~\cite{Lukens:2003aq}].
              }}
            \label{fig_cdf_tracker_overview}
  \end{center} 
\end{figure}

The COT, as displayed in Fig.~\ref{fig_cdf_tracker_overview}.(a) is made of eight concentric
``super-layers'' separated in azimuth into cells.
Each cell, as shown in Fig.~\ref{fig_cdf_tracker_overview}.(b), is made of sense wires and 
potential wires immersed in an ambient gas mixture. Ionised electrons from the passage of high 
energy charged particles drift under the influence of the electrostatic field to the sense wires
and yield their energy which allow to decipher the particle hit position.
The sense wires are attached at each extremities to end-plates which hold them into a string 
position. The tilt angle of the cells aims to make it so the ionised electrons travels 
approximately in azimuth to the sense wires under the combined influence of the solenoid magnetic
field and of the local electrostatic field.
Let us note that the cells move from their nominal geometry under the influence of gravity which
makes field sheets and wires sag. This eventually implies that the sense wires deflect toward a 
particular field sheet. To decrease this effect a support rod at $z=0$ connects the sense wires at
the center of the detector.
\index{CDF detector!Tracker|)}

\subsection{Influence of tracker misalignment on the W mass charge asymmetry}
\index{CDF detector!Tracker misalignment/weak modes|(}
\index{Weak modes!In CDF tracker|(}
We adopt in the rest of this Appendix the conventions of Ref.~\cite{Aaltonen:2007ps} to address the
curvature $c$ of a charged particle track in the $r-\phi$ plane. It is defined like
\begin{equation}
c\equiv q/(2\,R),
\end{equation}
where $q$ is the charge of the particle and $R$ is the radius of the track in the $r-\phi$ plan.
Charged leptons have their transverse momenta deduced from $c$ which is measured by the hits left
in the COT. For that purpose the position of the cells needs to be determined precisely to have a 
good accuracy for the momentum scale.
The calibration of this momentum scale is determined starting with the \textit{a priori} position
of the cells. Then \textit{in situ} measures correct for the electrostatic and gravity sag biases,
cosmic ray muon data allow to add up \textit{in situ} corrections and finally track based 
corrections are applied using data calibration.
For these last corrections first comes the study $W\to e\,\nu$ to reduce relative curvatures
biases between positive and negative particles and finally absolute scale data calibration using
$J/\psi$, $\Upsilon$ and $Z$~bosons decays to $\mup\mum$. 

The reconstructed track curvature $c_\mm{r}$ can be expressed as a function of the true curvature 
$c_\mm{t}$ in a Taylor expansion around zero which is justified by the high $\pT$ of the charged 
leptons of interest ($c_\mm{t}\approx 0.02-0.03$). It reads
\begin{equation}
c_\mm{r} = \es_1 + (1+\es_2)\,c_\mm{t} + \es_3\,c_\mm{t}^2 + \es_4\,c_\mm{t}^3 + \dots,
\label{eq_c_rec}
\end{equation}
which now allows us to consider the case of two positively and negatively charged leptons 
respectively of curvatures $c_{\mm{r},+}$ and $c_{\mm{r},-}$, having the same true transverse 
momentum (\ie{} $c_{\mm{t},-} = - c_{\mm{t},+}$). Their reconstructed tracks are written
\begin{eqnarray}
c_{\mm{r},+} &=&
\es_1 + (1+\es_2)\,c_{\mm{t},+} + \es_3\,c_{\mm{t},+}^2 + \es_4\,c_{\mm{t},+}^3 + \dots,\label{eq_crp}\\
c_{\mm{r},-} &=&
\es_1 - (1+\es_2)\,c_{\mm{t},+} + \es_3\,c_{\mm{t},+}^2 - \es_4\,c_{\mm{t},+}^3 + \dots,\label{eq_crm}
\end{eqnarray}
Now, remembering the goal is to eventually merge positive and negative channels for the extraction of
$\MW$, flipping the sign of the curvature in Eq.~(\ref{eq_crm}) 
(\ie{} $c_{\mm{r},-}\to -c_{\mm{r},-}$) to get rid of the charge sign and averaging this new 
expression with Eq.~(\ref{eq_crp}) gives an average track curvature
\begin{equation}
\tfrac{1}{2}(c_{\mm{r},+} - c_{\mm{r},-})  = (1+\es_2)\,c_{\mm{t},+} + \es_4\,c_{\mm{t},+}^3.
\end{equation}
With that development we see all terms of even power of $c_\mm{t}$ are cancelled when averaging,
the term linear in $c_\mm{t}$ scales the true curvature and is deduced from momentum calibration.
The term $\es_4\,c_{\mm{t},+}^3$ is the first one to affect the determination of $\MW$ but it can be 
neglected because of the high transverse momentum of the leptons. 
On the other hand, the direct average of the signed curvatures (Eqs.~(\ref{eq_crp}--\ref{eq_crm})) 
that should ideally be equal to zero, leads to constraints on $\es_1$, indeed
\begin{eqnarray}
\tfrac{1}{2}(c_{\mm{r},+} + c_{\mm{r},-})  &\equiv& \es_1 + \es_3\,c_{\mm{t},+}^2 + \dots,\\
                                  &\approx& \es_1,\label{eq_av_sg_cr}
\end{eqnarray}
where in the last line higher order terms are neglected compared to $\es_1$ and because as we saw
they are not worth to be considered for a measurement of $\MW$. 
Then, $\es_1$ is constrained using data calibration since the relative difference between 
$(E/p)_\ep-(E/p)_\Em$ should be zero in the absence of misalignment. 
The parametrisation of Eq.~(\ref{eq_av_sg_cr}) can be written like
\begin{equation}
\tfrac{1}{2}(c_{\mm{r},+} + c_{\mm{r},-})
= a_0 + a_1\,\cotan\theta + a_2\,\cotan^2\theta + b_1\,\sin(\phi+0.1) + b_3\,\sin(3\,\phi+0.5),
\end{equation}
where the terms $a_0$, $a_1$ and $a_2$ can be interpreted as distortions of the COT.

\begin{figure}[!h] 
  \begin{center}
    \includegraphics[width=.6\tw]{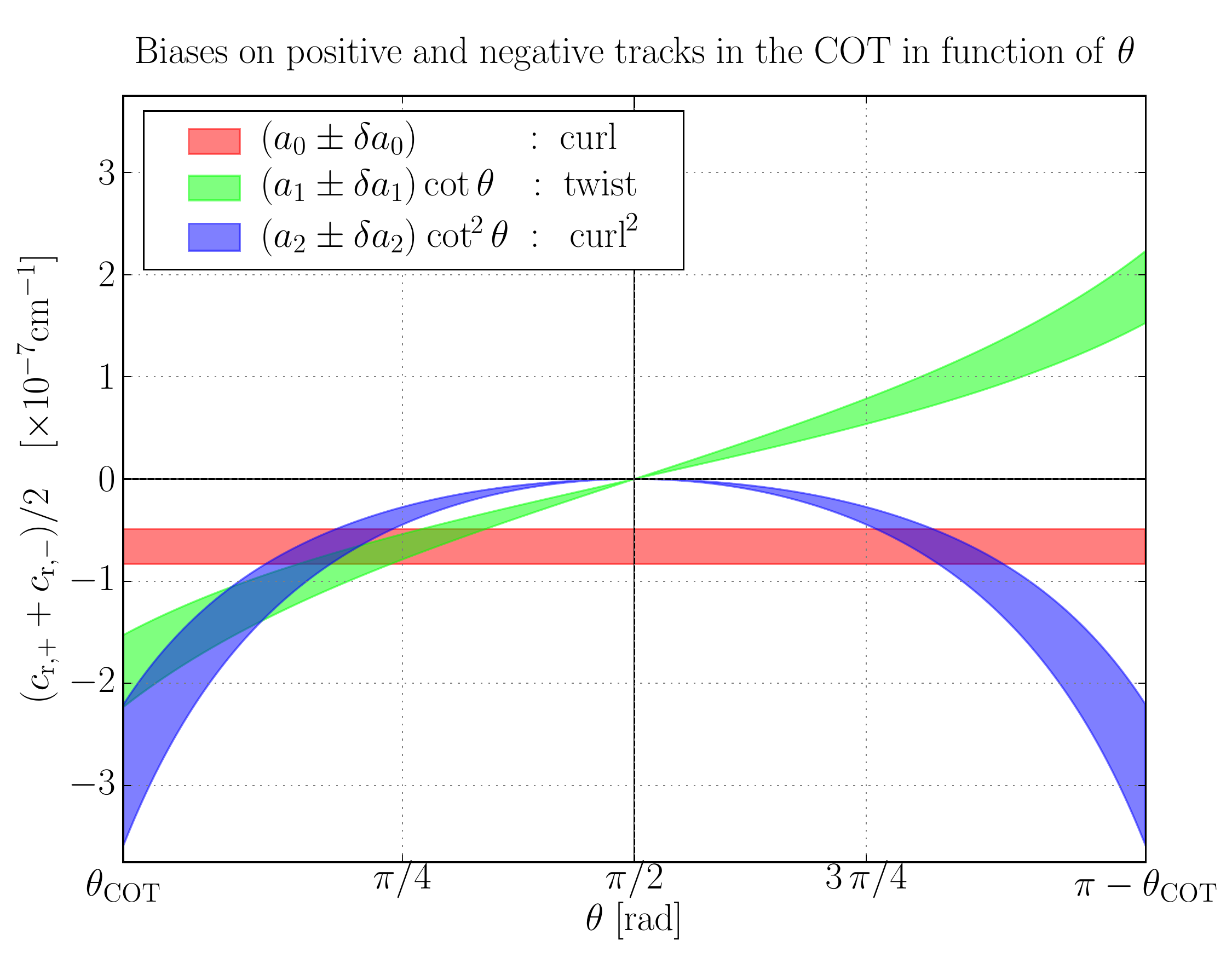}
    \caption[CDF~II central outer tracker distortions between positive and negative tracks]
            {\figtxt{Representation of the spatial distortion of the CDF~II COT 
                in function of $\theta$.}
            }
            \label{fig_cdfii_cot_weak_modes_1}
  \end{center} 
\end{figure}

Figure~\ref{fig_cdfii_cot_weak_modes_1} illustrates these distortions as a function of the more 
intuitive $\theta$ observable. Using the vocabulary of weak modes (cf.~\S\,\ref{s_weak_modes}),
the term in $a_0$ corresponds to a curl of the tracker while the term $a_1$ corresponds to a twist 
between the left and right end-plates.
The term proportional to $a_2$ goes beyond the first order approximation of the weak modes and
is characteristic of the COT construction, it corresponds to a curl of second order where the 
left and right end-plates are being rotated in the same $\phi$ direction but the center of the 
tracker is maintained to its original position by the support rod.
The terms in $b_1$ and $b_3$ corresponds to mis-measurements of the beam position.

\begin{figure}[!h] 
  \begin{center}
\hspace*{-1.3cm}
\includegraphics[width=.6\tw]{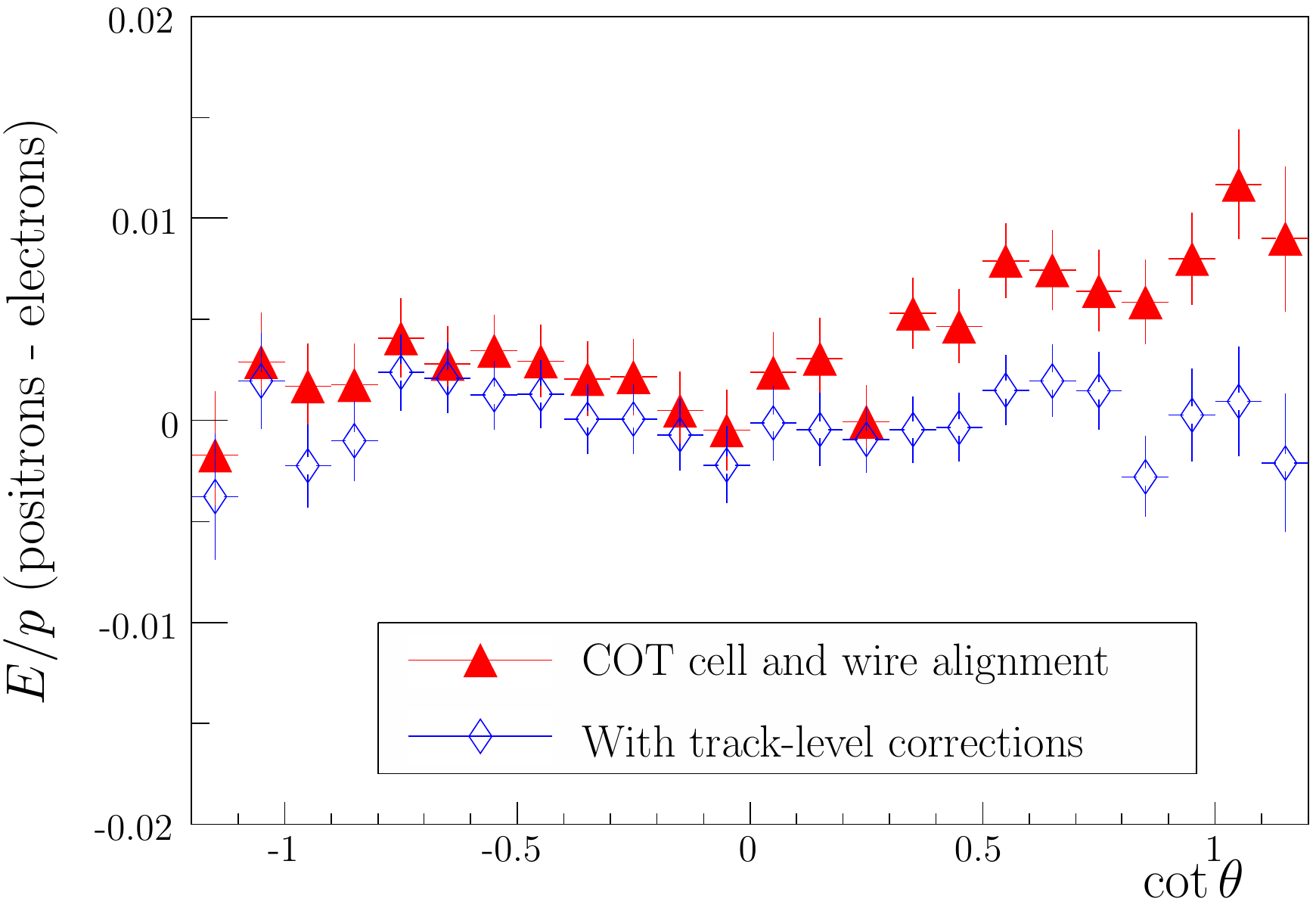}
    \caption[CDF~II $(E/p)_\ep-(E/p)_\Em$ as a function of $\cotan\theta$ before and after 
      corrections]
            {\figtxt{$(E/p)_\ep-(E/p)_\Em$ in function of $\cotan\theta$
                before and after corrections [Taken in~\cite{Lukens:2003aq}].}
            }
            \label{fig_cdfii_cot_weak_modes_2}
  \end{center} 
\end{figure}

Figure~\ref{fig_cdfii_cot_weak_modes_2} shows now the tracker misalignment consequences on 
$(E/p)_\ep-(E/p)_\Em$ before and after corrections in function of $\cotan\theta$.

In conclusion, this measurement dedicated to extract $\MW$ merges both $\Wp$ and $\Wm$ measurement
which gives data where half of the tracks are biased in one direction and the over half of
the same amount but in the opposite direction.
Eventually all those corrections lead to a precision on tracker momentum resolution of
$\Delta p/p \approx -1.50 \times 10^{-3}$ which in turn leads for all $\pTl$, $\mTlnu$ and 
$\slashiv{p}_{T,\nul}$ fits to a systematic error of $17\MeV$ for the measurement of $\MW$.
But, when performing fit analysis for separated $\Wp$ and $\Wm$ the curl distortion 
($a_0$ and $a_2$) give rise as it was seen in the core the Chapter to incoherent biases between
the positively and negatively charged tracks curvature responsible for the important difference of
the order of $\approx 2\,\sigma$.
From that experience we can state that the LHC capabilities and a dedicated measurement strategy should
considerably improve the accuracy on the value $\MWp-\MWm$.
\index{W boson@$W$ boson!Mass@Mass $\MW$!In CDF|)}
\index{W boson@$W$ boson!Mass charge asym@Mass charge asymmetry $\MWp-\MWm$!In CDF|)}
\index{CDF detector!Tracker misalignment/weak modes|)}
\index{Weak modes!In CDF tracker|)}
\index{CDF detector|)}
\end{subappendices}
\cleardoublepage
\ClearShipoutPicture

\chapter{The Monte Carlo event generator \WINHAC{}} \label{chap_winhac}
\setlength{\epigraphwidth}{0.56\tw}
\epigraph{veillant\\
\hspace*{1.25cm}doutant\\
\hspace*{2.5cm}roulant\\
\hspace*{3.75cm}brillant et m\'editant\\
\hspace*{5cm}avant de s'arr\^eter\\
\hspace*{3.75cm}\`a quelque point dernier qui le sacre\\
\bigskip
\hspace*{3.75cm}Toute Pens\'ee emets un Coup de D\'es}%
{\textit{Un coup de d\'es jamais n'abolira le hasard}\\ \textsc{St\'ephane Mallarm\'e}}

So far a description of the theoretical and experimental context in which the prospect for a 
measurement of $\MWp-\MWm$ have been given.
This Chapter describes the tools that were used for that purpose, which consist mostly of a 
description of the Monte Carlo event generator \WINHAC{} used to simulate the production of 
$W$~bosons in Drell--Yan. The other tools entering in the generation and the analysis steps are also
described briefly.

This Chapter is divided into three parts.
The first one, after reminding the gist of Monte Carlo methods, describes the physics 
implemented inside \WINHAC{}.
The second part presents the work done within this thesis to implement \WINHAC{} inside the ATLAS 
software. 
The reader interested only in the physics thread can skip this technical part in a first reading.
Indeed, to the use of the full and refine ATLAS software, a lighter framework was preferred
to make the studies presented in Chapter~\ref{chap_w_pheno_in_drell-yan} and~\ref{chap_W_mass_asym}.
This personal framework is eventually presented in the third part of the Chapter.

\section{The Monte Carlo event generator \WINHAC{}}

\subsection{Monte Carlo methods}\index{Monte Carlo|(}
Monte Carlo (MC) methods~\cite{UlamMetropolis} provide in high energy physics efficient solutions 
to deal with numerical 
integration and, due to their very stochastic nature, to simulate inelastic scatterings for both
phenomenological and experimental specific needs. 
The principles of MC procedure is reviewed in a nutshell.

\subsubsection{Principle of Monte Carlo methods.}
Starting with one dimension, the numerical integration of a function $f$ between $a$ and $b$
can be recast as the product of the integration range $b-a$ by the average of the integrand over it,
that is
\begin{eqnarray}
  I &=& \int_{a}^{b}f(x)\,dx,\\ 
    &=& (b-a)\,\left<f\right>_{a,b}.
\end{eqnarray}
This last expression can be approximated using a discrete number of points $N$ within the range 
$[a,b]$, giving then
\begin{equation}
I \approx I_N\equiv (b-a)\times \frac{1}{N}\sum_{i=1}^N f(x_i).
\end{equation}
Two types of methods can be distinguished to evaluate $I_N$.
The first one consists essentially to slice regularly the integration range, these are called 
numerical quadratures. A broad range of refinements exist to increase the accuracy and convergence
speed of the result (see \textit{e.g.}~\cite{NR}).
Among them let us mention the trapezium and the Simpson rules which accuracies in one dimension 
are respectively converging like $\propto 1/N^2$ and $\propto 1/N^4$.
The second approach consists to pick up $N$ random points in $[a,b]$ to approximate 
$\Mean{f}_{a,b}$.
This is the principle of Monte Carlo procedure, which in the one dimension case display a 
convergence $\propto 1/\sqrt{N}$.

When the dimension of the integral increases to a higher dimension, say $d$, quadrature 
integration methods become very cumbersome as the integration volume, or phase space to already
adopt physics vocabulary, needs to be split into $N^d$ bits.
Thus, the trapezium and Simpson methods converge now like $\propto 1/N^{2/d}$ and $\propto 1/N^{4/d}$
whereas for the MC integration by still picking up $N$ random points in the phase space keeps 
a convergence $\propto 1/\sqrt{N}$. 
Then, Monte Carlo may be slow but their speed convergence is independent of the dimension of the 
integral. 

Since at the LHC the typical number of produced particles in an inelastic scattering is of the
order of $n\sim 1,000$ it implies a dimension $d=3^n-4$ where the subtraction of $4$ to the degrees
of freedom is a consequence of the energy/momentum conservation.

\subsubsection{Monte Carlo in High Energy Physics.}
Coming back to high energy physics, estimating the probability of occurrence of a LHC reaction 
$A+B\to C+D$ demands the computation of its cross section which can be roughly written
\begin{equation}
\sigma = \int_{d=3^n-4}\left(\DfDx{\sigma}{\Phi}\right)\,d\,\Phi,
\end{equation}
where $d\,\Phi$ is an infinitesimal element of the phase space of the process
and $\flatDfDx{\sigma}{\Phi}$ is calculated with perturbative QCD/EW/BSM along with the relevant 
and available higher corrections depending on the nature of the process.
Let us emphasise that here, picking up a random element $d\,\Phi$ corresponds to a particular physical
configuration for the event, that is four-momenta of the colliding partons, of the decaying 
particles as well as other partons and photons which have been radiated.
Hence, the record of these random $d\,\Phi$ elements, in top of helping to compute $\sigma$,
represents a collection of events which can be kept and used to plot afterward any desired 
distribution for an observable $a$ in the form $\flatDfDx{\sigma}{a}$.
Note this would not be possible with quadrature procedures where from the start the calculus 
should be aimed to resolve a particular distribution.
This collected events are referred to as ``weighted events'' \index{Monte Carlo!Weighted events} 
as they are indeed weighted by the theoretical prediction $\flatDfDx{\sigma}{\Phi}$ associated to 
their production.

Nonetheless, these weighted events are not a simulation of real physics processes. 
Indeed in nature a physical event do not come with the information whether it is frequent or rare, 
only harvesting large enough data allows to see it. 
Still, from the simulation of weighted events a simple trick can be used to emulate the outcome
of unweighted events\index{Monte Carlo!Unweighted events}, that is events with a weight of $1$.
The principle is the following. In a local region of phase space a maximum $\mm{MAX}$ is determined 
for the integrand, \ie{} $\mm{MAX}>\flatDfDx{\sigma}{\Phi}$.
When a weighted event is generated a random number $r$ is draw in the range $[0,1]$ then
\begin{itemize}
\baselineskip 1pt
\item[-] If $r<(\flatDfDx{\sigma}{\Phi})/\mm{MAX}$ the event is accepted with a weight of $1$.
\item[-] If $r>(\flatDfDx{\sigma}{\Phi})/\mm{MAX}$ the event is simply rejected.
\end{itemize}
This rejection method allows to produce events with the same frequency of occurrence than in reality. 
Histograms produced with weighted and unweighted events are the same in the asymptotic limit
$N\to +\infty$ (in practice when $N$ is large enough so the Central Limit Theorem can be applied). 
Let us note this trick is possible because the integrand is positive definite which is true only at 
LO in QCD, the QCD NLO treatment becomes then very delicate 
, \eg{} Ref.~\cite{Sjostrand:2005CERNlectures}.

Monte Carlo event generators opens, from the experimentalist point of view, a wide range of 
possibilities. During the generation, acceptance cuts as well as detector smearing on the generator 
predictions can be applied which justify their extensive use in both R\&D prospects and real data 
analysis.

Also, worth mentioning is the generation of random numbers.
\index{Monte Carlo!Random/pseudo-random numbers}
It is impossible for an algorithm to generate random numbers since, by definition, its behaviour is
deterministic by essence. 
Hence random numbers are in fact sequence of pseudo-random numbers outgoing small algorithms. 
The quality of these sequence of number comes from their periodicity --the time it takes to repeat 
the series-- and some short-range correlations between the generated numbers (see \eg{} 
Ref.~\cite{Knuth}).
To better emulate the non correlation between several simulated data, different long sequences can be
generated using different seeds in the algorithm initialisation.
In the rest, for convenience, the term random is used instead of pseudo-random.

We can distinguish two types of Monte Carlo, the one using the MC technique only for integration 
and the other which profit of MC properties to simulate physics processes. 
The latter are referred to as ``event generator'' or simply ``generators''. 
\index{Monte Carlo!Event generator}

Another way to classify Monte Carlo, from the physics point of view this times, is to look at their 
domain of applications. 
On the one hand stands a few general purpose MC that possess a wide range of inelastic scattering 
($2\to 2$ or $2\to 3$) implemented for both SM and some BSM processes. They also provide the 
radiation of photons from leptons and photons/gluons from quarks 
(called in that context QCD ``parton shower''\index{Monte Carlo!Parton Shower}),
the hadronisation of jets, decays of unstable particles and the underlying event.
The most frequently used general purpose MC are \Pythia%
~\cite{Sjostrand:2006za,Sjostrand:2008vc,PythiaHomepage},\index{Pythia@\Pythia{} Monte Carlo event generator} 
Herwig++~\cite{Bahr:2008pv,HerwigHomepage} and Sherpa~\cite{Gleisberg:2003xi,SherpaHomepage}.
The other category are MC focused on specific processes but that embrace more corrections at the 
level of the hard process ($2\to n$) compared to the general purpose MC. 
To obtain eventually physical events just like the one we observe in experiments they are interfaced 
to general purpose MC.
These tools are quite numerous and span large domains of physics, this can go from standard
processes with higher QCD/EW corrections to the supply of corrections to existing MC events such
as higher QED radiation \index{QED!Radiative corrections@Radiative correction in single $W$ production}
in the final state like PHOTOS~\cite{Barberio:1990ms,Barberio:1993qi,Golonka:2006tw} 
or management of $\tau$ or Higgs decays done respectively by 
TAUOLA~\cite{Jadach:1990mz,Jezabek:1991qp,Jadach:1993hs} and HDECAY~\cite{Djouadi:1997yw}.

The refinements brought to the art of MC integration and event generation in High Energy Physics
goes far beyond the short overview made here.
Among them is the variance reduction to improve convergence speed, the handling of singularities
in the integrand, and the delicate issue to generate events at NLO in QCD.
All those points and many others are addressed in dedicated documents.

There is, so far, no classic textbooks on the use of MC techniques in high energy physics. 
Nonetheless, the reader eager to learn more on the subject is invited to look at the thesis of 
Michael Seymour~\cite{SeymourPhD} where one chapter provides a short yet thorough description of 
the matter. 
Other relevant sources are the presentations given by Monte Carlo experts in conferences 
(see \eg{}~\cite{Sjostrand:2006su,Sjostrandtalks,Seymourtalks,Placzektalks,CTEQtalks}).

\subsubsection{Weakness of Monte Carlo methods.}
So far Monte Carlo methods have been promoted as the best tools one can work with in experimental
high energy physics. Nonetheless, to be completely objective some light is cast on a few of their
weaknesses.

The use of Monte Carlo simulation are safe when it comes to pragmatic application. 
To illustrate this let us consider the example of nuclear safety where the control of the behaviour 
of a nuclear facility relies on Monte Carlo simulations. 
In no case the underlying model implemented have to be perfect, as long as it sticks to physical 
measurements. Then, using phenomenological models altogether with empirical laws are correct as long as 
the intended goal --the control of the facility-- is assured.

In high energy physics the aim is different. 
MC generators are introduced to simulate known processes but as well to help unravel possible 
deviations in paradigm models or even discover brand new processes.
Now the actual models are far from being perfect, large parts describing inelastic scatterings are 
modeled by non perturbative QCD, which relies on complex --and sometime empirical-- parametrisation 
(\eg{} hadronisation).
This poses a problem for the relevancy of an experimental analysis.

The awareness of these imperfection's acted as an incentive in the present work to come up with 
analysis strategies which had to adapt to a given measurement to get as much as independent from 
MC imperfections. 
Also, in a first step, rather than using several Monte Carlo the choice for a deeper understanding 
of the used tool, \WINHAC{} was adopted this being justified by the latter strategy to make an 
analysis.

To conclude with this parenthesis, some quotations on this particular topic.
The first one is extracted from a talk by J.D. Bjorken~\cite{JDBjorken} as noted by 
Torbj{\"o}rn Sj{\"o}strand in~\cite{Sjostrand:2005CERNlectures}
\begin{quote}
\textit{``The Monte Carlo simulation has become the major means of visualization of not only 
detector performance but also of physics phenomena. So far so good. But it often happens that the 
physics simulations provided by the Monte Carlo generators carry the authority of data itself. 
They look like data and feel like data, and if one is not careful they are accepted as if they were
data.''}
\end{quote}
The second one is from the authors of \Pythia{} who warn the users about the traps one can fall into
\begin{quote}
You must be very careful when you formulate the questions\,: any ambiguities will corrupt the reply 
you get. And you must be even more careful not to misinterpret the answers\,; in particular not to 
pick the interpretation that suits you before considering the alternatives. 
Finally [\dots] the current authors might unwittingly let a bug free in the program \Pythia{}.
\end{quote}
In conclusion a warning from the authors of Ref.~\cite{QCDHEET}
\begin{quote}
Monte Carlo event generators are complicated programs that will almost inevitably contain bugs,
incorrect assumptions and ill-chosen parameters. It is therefore vital that a user does not take
any results at face value. As a minimum at least two completely independent programs should be
used in any physics studies.
\end{quote}

\subsection{The Monte Carlo event generator \WINHAC{}}\label{ss_tools_WINHAC}
\index{WINHAC@\WINHAC{} Monte Carlo event generator|(}

The main tool that used in the present study is the Monte Carlo event generator 
\WINHAC{}~\cite{Placzek2003zg,CarloniCalame:2004qw,Gerber:2007xk,Bardin:2008fn}.
It has been developed in FORTRAN~77 (F77) by Wies{\l}aw P{\l}aczek and Stanis{\l}aw Jadach from 
the Cracow theoretical group which holds a leading role in term of electroweak radiative 
corrections in Monte Carlo.

\WINHAC{} is dedicated to precision description of the charged-current Drell--Yan process.
It has been thoroughly tested and cross-checked with independent calculations%
~\cite{Placzek2003zg,CarloniCalame:2004qw,Golonka:2005pn,Bardin:2008fn}.
This MC program has already been used in previous studies of experimental prospects 
for exploring the electroweak symmetry breaking mechanism~\cite{Krasny:2005cb},
in our ongoing effort for precision measurement of the Standard Model parameters 
at the LHC within the ATLAS experiment~\cite{Krasny:2007cy,Fayette:2008wt,Upcoming_MW,Upcoming_GammaW}.

At the time of the redaction of this dissertation, the most recent version of \WINHAC{} is
release 1.30~\cite{WINHAC:MC}. It features the exclusive Yennie--Frautschi--Suura 
exponentiation~\cite{yfs:1961} of QED effects, \ie{} the radiation of $n$ photons in the final state
\index{QED!Radiative corrections@Radiative correction in single $W$ production}
\begin{equation}
q + \qbp \to \Wpm \to \lpm + \smartnuanul + \gamma_1 + \gamma_2 + \dots + \gamma_n,
\end{equation}
with $n=\{0,1,\dots\}$, also referred to as multi-photon radiation.
It also includes ${\cal O}(\alpha)$ electroweak corrections for the full charged-current Drell--Yan 
process at the parton level, for more details see Ref.~\cite{Bardin:2008fn}.
\index{Parton Distribution Functions (PDFs)!LHAPDF|(}
This parton-level process is convoluted with the parton distribution functions (PDFs) provided by 
the \LHAPDF{} package~\cite{Whalley:2005nh} which includes a large set of recent PDF parametrisation
by several groups. \WINHAC{} is also interfaced with the \Pythia{}~6.4~\cite{Sjostrand:2006za} MC 
event generator for the QCD/QED initial-state parton shower as well as for the hadronisation.
\index{Pythia@\Pythia{} Monte Carlo event generator!Interface to WINHAC@Interface to \WINHAC{}}
Technical detail, \Pythia{} and \WINHAC{} are accessing to LHAPDF density functions through the 
LHAGLUE interface --present in LHAPDF-- which mimics the procedure that was used formerly to use 
PDFLIB~\cite{PlothowBesch:1992qj} the ancestor of LHAPDF.
\index{Parton Distribution Functions (PDFs)!LHAPDF|)}
Several type of collisions are made available\,:\;proton--proton, proton--anti-proton, proton--ion, 
ion--ion, where each ion beam is defined by its charge number $Z$, atomic number $A$ and energy in 
the center of mass. For $Z>2$, the nuclear shadowing effects from 
Refs.~\cite{Eskola:1998df,Eskola:1998iy} can be optionally switched on.

The parton-level matrix elements are calculated numerically from spin amplitudes%
~\cite{Placzek2003zg}. 
This allows for studies of any spin effects in the charged-current Drell--Yan process. In fact,
\WINHAC{} provides options for generation of processes with pure transversely or 
pure longitudinally polarised $W$~bosons at the Born level.

In addition to the charged-current Drell--Yan process, \WINHAC{} includes the neutral-current 
Drell--Yan process (with $\gamma^\ast+Z$ bosons in the intermediate state), however at the Born level 
only. For precision description of this latter process, similar to the former one, a dedicated MC 
event generator called \ZINHAC{}~\cite{ZINHAC:MC} is being developed in C++ by Wies{\l}aw P{\l}aczek
and Andrzej Si\'odmok. In the future, these twin MC generators can be used for precision 
studies/analyses of the Drell--Yan processes including the QED/EW corrections.

The generation of random numbers in \WINHAC{} is achieved using different classic algorithms.
The one used for our work is RANMAR~\cite{James:1988vf,Ranmar:1990} which displays a 
periodicity of $2^{144}$.

For this study version 1.23 of \WINHAC{} has been used which for all aspects investigated and 
adopted strategies is equivalent to the latest version. They differ in description of QED/EW 
corrections but these have not been included in the present work.
In Appendix~\ref{app_winhac_evt} an example of a \WINHAC{} summary event is given.

As conclusion, in Table~\ref{table_wz_in_mc} we stress the place of \WINHAC{}
with respect to multi-purpose and a few specialised Monte Carlo that can produce $W$ or $Z$
in Drell--Yan.
In this table MC event generator (Event Gen.) are distinguished from the one using only MC as a mean 
of integration (Histograms). 
As it can be seen there is up to this date (2009) no Monte Carlo which hold QCD and EW corrections
at the same level of detail. 

\index{QED!Radiative corrections@Radiative correction in single $W$ production|(}
\begin{table}
\begin{center}
\renewcommand\arraystretch{1.2}
\begin{tabular}{l c c c c c}
  \hline
  \textbf{Monte Carlo} & \textbf{Refs.} & \textbf{Process} & \textbf{QCD} & \textbf{EW} & \textbf{Type}  \\
  \hline\hline
  \WINHAC{}&~\cite{Placzek2003zg,WINHAC:MC} 
  & $W$      & $\mm{PDF}(x)$, impr. LO & $\mathcal O(\alpha)$+QED FSR & Event Gen. \\
  \hline
  HORACE&~\cite{CarloniCalame:2003ux,HORACEhomepage}
  & $W$, $Z$ & $\mm{PDF}(x)$, impr. LO & $\mathcal O(\alpha)$+QED PS  & Event Gen. \\
  \Pythia{}&~\cite{Sjostrand:2006za,PythiaHomepage} 
  & $W$, $Z$ & $\mm{PDF}(x)$, impr. LO & LO                           & Event Gen. \\
  HERWIG&~\cite{Corcella:2000bw,HerwigF77Homepage}  
  & $W$, $Z$ & $\mm{PDF}(x,p_T)$, impr. LO     & LO                           & Event Gen. \\
  Herwig++&~\cite{Bahr:2008pv,HerwigHomepage}  
  & $W$, $Z$ & $\mm{PDF}(x,p_T)$, NLO     & LO                           & Event Gen. \\
  Sherpa&~\cite{Gleisberg:2003xi,SherpaHomepage}    
  & $W$, $Z$ &  $\mm{PDF}(x,p_T)$, impr. LO    & LO                          & Event Gen. \\
  MC@NLO&~\cite{Frixione:2002ik,MCATNLOHomepage}     
  & $W$, $Z$ &  parton shower, NLO        & LO                           & Event Gen. \\
  AcerMC&~\cite{Kersevan:2004yg,AcerMCHomepage}    
  & $W$, $Z$ &  $\mm{PDF}(x)$, LO         & LO                           & Event Gen. \\
  ResBos-A&~\cite{Cao:2004yy,ResBosAHomepage} 
  & $W$, $Z$ &  $\mm{PDF}(x,p_T)$, NLO    & FS $\mathcal O(\alpha)$      & Histograms \\
  ResBos&~\cite{Balazs:1997xd,ResBosHomepage}
  & $W$, $Z$ &  $\mm{PDF}(x,p_T)$, NLO    & LO                           & Histograms \\
  \hline
  WGRAD&~\cite{Baur:1998kt,WGRADhomepage}     
  & $W$      & $\mm{PDF}(x)$, LO          & $\mathcal O(\alpha)$         & Histograms \\
  ZGRAD2&~\cite{Baur:2001ze,ZGRAD2homepage}   
  & $Z$      & $\mm{PDF}(x)$, LO          & $\mathcal O(\alpha)$         & Histograms \\
  \texttt{SANC}&~\cite{Arbuzov:2005dd,SANChomepage}    
  & $W$, $Z$ & $\mm{PDF}(x)$, LO          & $\mathcal O(\alpha)$         & Histograms \\
  \hline
\end{tabular}
\renewcommand\arraystretch{1.45}
\caption[Overview of some Monte Carlo capable of simulating single $W$ or $Z$ production in hadronic
  colliders]
        {\figtxt{Overview of some Monte Carlo capable of simulating single $W$ or $Z$ production 
            in hadronic colliders. The quoted references corresponds to, first, the main reference
            and second to the software homepage for further references and details on the Monte Carlo.}}
\label{table_wz_in_mc}
\index{Pythia@\Pythia{} Monte Carlo event generator}
\end{center}
\end{table}
\index{QED!Radiative corrections@Radiative correction in single $W$ production|)}

Hence so far, Monte Carlo are usually combined to simulate with the closest possible accuracy 
the physics observables needed to be confronted to the data.
For example the extraction of $\MW$ for the CDF II run~\cite{Aaltonen:2007ps} is made using the 
following Monte Carlo
\begin{itemize}
\baselineskip 1pt
\item[-] $\pTW$ is simulated with ResBos~\cite{Balazs:1997xd} and making cross checks with 
DYRAD~\cite{Giele:1993dj} $W$+jet simulation.
\item[-] The photon radiation corrections in the final state are made by WGRAD~\cite{Baur:1998kt} 
and correct the $\pTl$, $\mTlnu$ and $\slashiv{p}_{T,\nul}$ distributions from ResBos.
\item[-] The background to $W\to e\,\nue$ ($W\to\tau\,\nutau$ and $Z/\gamma^\ast\to\ep\Em$) and
to $W\to \mu\,\numu$ ($W\to\tau\,\nutau$ and $Z/\gamma^\ast\to\ep\Em$) are simulated using \Pythia{}
v~6.129 and passed to a GEANT-based full simulation of the detector.
\index{Pythia@\Pythia{} Monte Carlo event generator}
\end{itemize}
\index{WINHAC@\WINHAC{} Monte Carlo event generator|)}
\index{Monte Carlo|)}

\section{Implementation of \WINHAC{} in the ATLAS software}
\subsection{Introduction}
This part of the Chapter covers the work achieved in the context of this thesis to implement 
\WINHAC{} in the ATLAS software environment.

As seen previously each detector needs, for both R\&D and data analysis, to have at its disposition
different Monte Carlo event generators to simulate the physics to be studied.
Their implementation needs to follow codified rules, just like any other tools in the experiment
software, for clarity's sake.
On the one hand most event generators obey to custom rules and conventions and can be implemented 
in  any programming language like F77, C/C++, \etc{}. On the other, for convenience reasons, 
experiments software environment relies on automated skeletons that treat all the present generators
with the same manner.
For instance, Monte Carlo events needs, among many other things, to be smeared to simulate the
particle interactions with the material of the detector. This simulation takes as input
a standard data format which is completely blind to the generator that produced it.
This example gives an idea of how a generator needs to fulfill a few requirements imposed by the
experiment.

This section is divided as follow.
After an overview of the main features of Athena~\cite{Duckeck:2005rb}, the software environment 
of ATLAS, follows a description of the context in which \WINHAC{} was introduced, that is in the 
simulation and reconstruction chain of the Monte Carlo events inside ATLAS.
The section ends on the implementation of \WINHAC{} and the validation of this work.

\subsection{Software environment of the ATLAS experiment} 
To better visualise the context of our discussion a brief overview of the software environment 
ATLAS relies on is given.

First of all, let us note the quantity of events selected by ATLAS together with the data outgoing
their analysis should represents each year a volume of information of the order of 
$10$~Peta-byte\footnote{$1\,\mm{Peta}\equiv 10^{15}$.}. 
Hence to overcome this challenge ATLAS had to aim for a highly decentralised storage and data 
management. This system obeys to a certain hierarchy where tasks are split into facilities called 
Tiers which principal activities are recaptured below. 

\index{ATLAS software!Tiers}
Although it will not be emphasised later on let us note that one of the philosophy in the use of
Tiers prescribes, for obvious safety reasons, to make a least one back up of each data batch.
The first data processing occurs at CERN in the unique Tier-0 facility where raw data are saved and
reconstructed into ESD\index{ATLAS software!Data format!Event Summary Data (ESD)}, 
AOD\index{ATLAS software!Data format!Analysis Object Data (AOD)} and 
TAG\index{ATLAS software!Data format!TAG} formats which description are gathered at the end of 
\S\,\ref{s_simul}.
This first data batch are shared among Tiers-1, of the order of ten all other the world.
The tenth of these raw data are stored to give new ESD, AOD and TAG.
Tiers-1 need to provide good accessibility for the data they store as well as the necessary capacity
to analyse them.
Tiers-2 provide work related to the calibration, the simulation and the analysis.
Finally Tiers-3 are made of local sources in each institute necessary to store custom data 
(\eg{} ntuple) and act as well as access points to upstream Tiers.
More details on the precise role of each Tier are given in Ref.~\cite{Duckeck:2005rb}.
This decentralisation of data processing and storage is provided by the Grid which allows to make
out of the ATLAS software a virtual facility split between several calculators spread throughout 
the world.

\index{ATLAS software!Athena|(}\index{Athena|see{ATLAS software}}
The informatics environment of ATLAS is called Athena~\cite{Duckeck:2005rb,ALTASwiki}.
Athena is an evolved version of the Gaudi~\cite{Barrand:2000qk,Barrand:2001ny} framework developed 
initially by the LHC$b$~\cite{Alves:2008zz} experiment and is now common to both ATLAS and LHC$b$
projects.
Amid the important features of Athena is the clear separation between the data and the algorithms
as well as between transient data (in memory) and persistent (in file) data.
The processing of the data for data selection, event simulation, reconstruction and analysis 
is governed by Athena.
Athena is object oriented (OO). Its structure is build mainly in C++, uses extensively the ROOT System%
\cite{Brun:1997pa,ROOThomepage} with some tools written in FORTRAN~77 or Java while the user 
interacts with Athena via Python.
\index{ATLAS software!Athena|)}
\subsection{Simulation and reconstruction of Monte Carlo events within ATLAS} 
\label{s_simul}

\begin{figure}[!h] 
  \begin{center}
    \includegraphics[width=.9\tw]{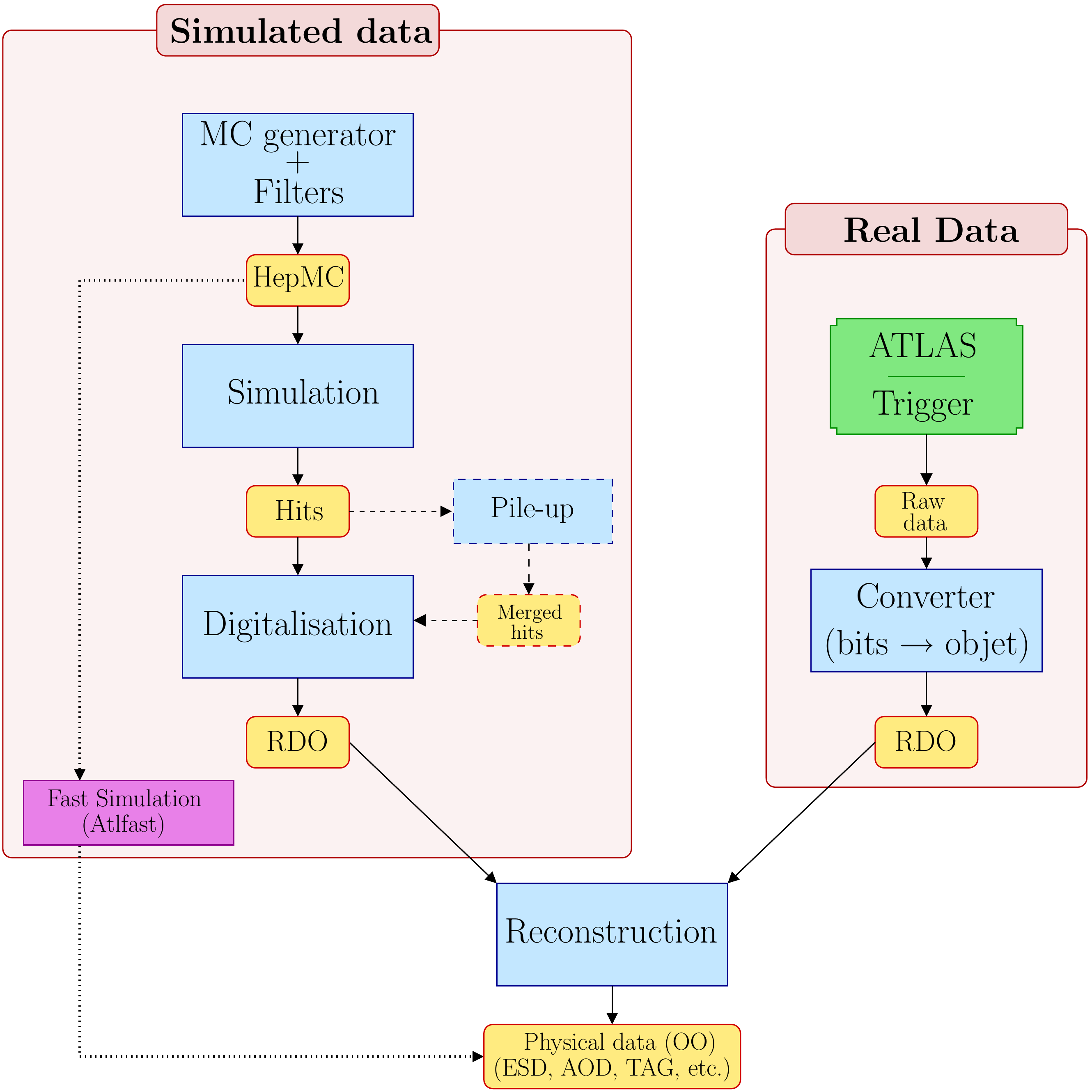}
    \caption[Schematic representation of the simulated/real data and reconstruction 
      processing chains]
	    {\figtxt{Schematic representation of the simulation (Monte Carlo data) and reconstruction
              of both Monte Carlo and real data.
              The add-up of the pile-up \index{Pile-up}
              (box with dashed borders) is an optional step.}}
	    \label{fig_simul}
  \end{center} 
\end{figure}

\index{ATLAS software!Athena!Simulation and reconstruction|(}
Here we give a description of the different steps that allow, starting from Monte Carlo events or 
from real data, to reach reconstructed data.
The first overview and the details that follow can be grasped looking at Fig.~\ref{fig_simul}. 
Let us remind more details on the data type mentioned are compiled at the end of this section.

\subsubsection{Overview of the simulation and of the reconstruction}
Starting with the real data. The event filter --the last process of the ATLAS trigger-- provides
as outputs raw data in byte stream format, \ie{} sequence of $1$/$0$.
These information are then converted to objects, the Raw Data Object (RDO) 
\index{ATLAS software!Data format!Raw Data Object (RDO)}transmitted in turn 
to reconstruction algorithms. Before describing the reconstruction chain, let us come back to the 
stages the simulated data must follow.
In the first step physical events are produced by Monte Carlo event generators, at this stage
acceptance cuts can be applied on some observables such as $p_T$, $\eta$, $\ETmiss$, \etc{}.
These events, as seen previously in \S\,\ref{notations_conventions}, are labelled ``truth events''
or ``generator level'' data.
They are transmitted to algorithms that simulate their passage in the ATLAS detector.
This step is by far the most CPU time consuming. The output are hits, that can be merged with 
pile-up \index{Pile-up} events, the latter receiving a special treatment.
After comes the digitisation which goal is to emulate the electronic read-out chain of the 
several sub-detectors of ATLAS. This gives eventually simulated RDO.
The simulation step finished the reconstruction chain is over-viewed.

Even though RDO are oriented object they are nonetheless raw data.
Also the goal of the reconstruction, as its name indicates, consists to reconstruct objects 
containing only relevant information for a physic analysis.

Finally let us mention the fast simulation of the detector.
Since the simulation part is very time consuming most experiments have tools with an approximated 
but faster simulation of the detector. In ATLAS, this fast simulation is called 
Atlfast~\cite{ATL-PHYS-98-131,ATL-PHYS-INT-2007-005:ATL-COM-PHYS-2007-012,AtlfastHomepage}. 
It shortcuts all simulation chain steps, that is from the particles four-momentum outgoing the 
generator it directly gives reconstructed data.

In the rest complementary information on the simulation and reconstruction chains are given.
For each step the emphasis is made first on physics aspects and then on technical aspects.
For further details on all technical aspects related to the Athena framework the reader is sent
back to the ATLAS Computing TDR~\cite{Duckeck:2005rb} and to the on-line workbook~\cite{ALTASwiki}.

\subsubsection{The simulation chain} \label{ss_simul}
\paragraph{Generation.}
The physics motivation being already known we move on directly to technical aspects with 
more emphasis with respect to other steps since the work for implementing \WINHAC{} 
occurred at this stage.

Inserting a MC within ATLAS consists to write inside Athena a C++ interface calling the algorithms
of the generator within its original libraries. Those instances are roughly\,:
\begin{itemize}
\baselineskip 1pt
\item[-] The initialisation, which essentially switches on the MC generator algorithms, read the input
parameters, \etc{}.
\item[-] The event loop in which the MC generate events.
\item[-] The finalisation where algorithms are shut down and the data are saved.
\end{itemize}
These three steps are realised for all the generators within Athena via the inheritance of the 
methods of the class GenzModule.

Concerning the libraries of the stand-alone code they are stored outside of the Athena framework
in AFS or in the GENSER (GENerator SERvice) repository.
The goal would be eventually to maintain all generators libraries by the GENSER LCG project%
~\cite{GENSERLCG} which would provide validated libraries for the needs of both theoreticians and 
experiments associated to the LHC.

The random numbers in Athena are provided by the use of the Athena Random Generator Service
which uses RanecuEngine~\cite{James:1988vf} maintained by CLHEP~\cite{CLHEP}.

The parton distributions functions are provided by the LHAPDF package~\cite{Whalley:2005nh}.
\index{Parton Distribution Functions (PDFs)!LHAPDF}\index{LHAPDF|see{Parton Distribution Functions}}

The output for the events is in the HepMC~\cite{Dobbs:2001ck} format that records information related
to each event. The HepMC format possess an infinite number of entries, the storage of the matrix
density in each vertexes, the flow pattern (\eg{} color) and their follow up, the record of the
used random numbers along with an arbitrary number of statistical weight that can be associated
to each event. 
This tool developed in C++ by ATLAS members has become the standard for the record of high energy
event and is now maintained by CLHEP.

\paragraph{Simulation.}
The goal of the simulation stage is to simulate the passage of particles generated by the MC within
the sub-detectors of ATLAS, \ie{} energy deposit in the calorimetry and tracks left in both
inner tracker and muon spectrometer.

This task is made by GEANT~4~\cite{Agostinelli:2002hh,Allison:2006ve}.
GEANT model the geometry and composition of the ATLAS detector, and simulate the physical process
occurring as particles pass through each cells. Hits record information related to the position,
yielded energies, identifications of activated elements, \etc{}.
For the pile-up \index{Pile-up} Athena provides a stock of simulated pile-up and select for each event
one random set and optionally merges it to the process simulated upstream.

\paragraph{Digitisation.}
Hits need to be converted into an output of the same format with the one provided by the ATLAS 
detector for real data.
For that purpose matters such as the propagation of charges (\eg{} in the tracker or in the LAr)
and of light (\eg{} in the tiles of the hadronic calorimeter) as well as the response of the
read out electronic needs to be emulated.

Contrary to the previous steps these tasks are very specific to the detector and cannot be 
accomplished without the physicist directly involved with the assembly and testing of each ATLAS
specific sub-detectors. This step provides output in RDO format.
\index{ATLAS software!Data format!Raw Data Object (RDO)}
Let us note all Monte Carlo truth information, kept so far for cross check, are removed from 
simulated RDO so that they exactly look like the format of the real data selected by the ATLAS 
trigger.

\subsubsection{Reconstruction of physic events} \label{ss_recon}
The goal of the reconstruction is to devise from raw data the vital information necessary to perform
an analysis.
To be more specific information related to photons, electrons, muons, taus, $K^0$, jets,
$\ETmiss$, primary vertexes, \etc{}. These collected data from each sub-detector are combined to
optimally reconstruct four-momenta for the widest range of momentum, pseudo-rapidity and whatever
value of the luminosity, all of that with the less background for particle identification.

\begin{figure}[!h] 
  \begin{center}
    \includegraphics[width=0.9\tw]{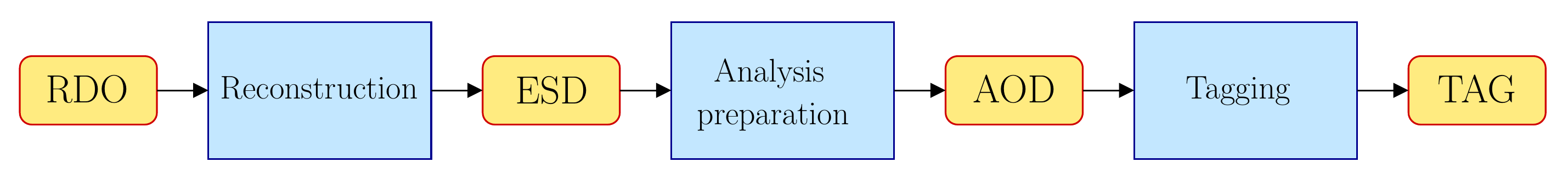}
    \caption[Schematic representation of the steps of the data reconstruction]
	    {\figtxt{Schematic representation of the steps of the data reconstruction.
              Boxes with sharp edges represent algorithms while the rounded boxes 
              represents data.}}
	    \label{fig_recon}
  \end{center} 
\end{figure}

The reconstruction is split in several steps depicted in Fig.~\ref{fig_recon}.
In a first stage is the individual reconstruction of the data outgoing the sub-detectors (tracking
and calorimetry) then comes the combined reconstruction which corresponds to the beginning of 
particle identification. An example of this step has been given for the case of the inner detector
tracks reconstruction in Chapter~\ref{chap_atlas_exp} \S\,\ref{ss_id_trck_fit_perf}. 
The output of this procedure are in ESD format.\index{ATLAS software!ESD}

In the second stage the preparation to the analysis starts with more complex reconstructed objects
(\eg{} $b$-tagging) and with a reduction of the data to the AOD format.

Finally data are tagged with the use of TAG files created from the AOD.

\subsubsection{Sum up of the different data format} \label{ss_datatype}
\index{ATLAS software!Data format|(}
As seen previously different data format exist at different stages of the simulation and 
reconstruction chain. The list below makes a summary of the one mentioned earlier.

\index{ATLAS software!Data format!Raw data}
\paragraph{Raw data.} These are the data provided by the event filter, the last stage of the
ATLAS trigger, and are written in byte stream format.
Each file contains the information related to a run and each event, delivered at a frequency
of $200\,\mathrm{Hz}$, should weight $1.6\,\mm{Mb}$.
This data format, contrary to the one that follow, is not object oriented.

\index{ATLAS software!Data format!Raw Data Object (RDO)}
\paragraph{Raw Data Object (RDO).}
The RDO is essentially an oriented object (C++) representation of the byte stream information
from the raw data.

\index{ATLAS software!Data format!HepMC}
\paragraph{HepMC.}
Format holding generator level information coming from MC generators, that is a purely 
theoretical information.

\index{ATLAS software!Data format!Hits}
\paragraph{Hits.}
Format carrying recording real interactions of simulated particles with the detector.
Hits carry information like position, energy deposit, identifier of the active element,
\etc{}. after simulating the theoretical events in the ATLAS detector.

\index{ATLAS software!Data format!Event Summary Data (ESD)}
\paragraph{Event Summary Data (ESD).} 
These files contain reconstructed information necessary to deduce the identification of particles,
the track refitting, the jet calibration, \etc{}.
The goal of this format is to avoid to access to raw data for every study out of the calibration
context or of the re-reconstruction. Those files are written in POOL~ROOT 
format~\cite{Brun:1997pa,ROOThomepage} with an average size of $500\,\mathrm{Kb}$ per event.

\index{ATLAS software!Data format!Analysis Object Data (AOD)}
\paragraph{Analysis Object Data (AOD).}
These POOL ROOT files are derived from ESD. 
They contain physical objects with information relevant to an analysis. 
It is on this format that in the long run most physical studies should be based upon.
Each even takes a size on disk of the order of $100\,\mathrm{Kb}$.

\index{ATLAS software!Data format!TAG}
\paragraph{TAG.}
These files are created in the goal to identify and select produced data. They contain information
on the data batches, each event occupy a size of $100\,\mathrm{Kb}$.

\index{ATLAS software!Data format!CBNT}
For conclusion several tests that were carried in this work to validate the implementation 
of the \WINHAC{} interface are based on the CBNT (ComBined NTuple) format which hold data in Trees
and leaves ROOT format.
\index{ATLAS software!Data format|)}
\index{ATLAS software!Athena!Simulation and reconstruction|)}
\subsection{Implementation of \WINHAC{} inside Athena} \label{s_impl_winhac}
\index{WINHAC@\WINHAC{} Monte Carlo event generator!Implementation inside Athena|(}
\subsubsection{The \winhaci{} interface}
In this part we present \winhaci{}, the interface between \WINHAC{} and Athena.
This work has been made in collaboration with Giorgos Stavropoulos to make of \WINHAC{} an ATLAS
approved Monte Carlo event generator used for the production of event samples as reported in 
Ref.~\cite{ATL-COM-PHYS-2008-243}.

\index{Parton Distribution Functions (PDFs)!LHAPDF|(}
The requirements from the Athena framework imposed to switch to its own random generator and to
the use of the LHAPDF libraries since in the version of \WINHAC{} used at that time (v~1.21)
PDFLIB was still use. The feedback to the authors allowed the code to evolve in later releases to 
use of LHAPDF as the default option (v~1.22 and upper).
Finally always at the time of this implementation the LHAPDF package did not provided
nuclear corrections necessary when studying ion-ion collisions.
Again to fulfill our needs the maintainer of the LHAPDF code included nuclear corrections 
(LHAPDF v5r2 and above) in the same way they were called so far via the LHAGLUE interface.
\index{Parton Distribution Functions (PDFs)!LHAPDF|)}

\WINHAC{} is an event generator dedicated to a particular process. 
Nonetheless it was not implemented like many others MC using the ``Les Houches'' format%
~\cite{Dobbs:2004qw} whose procedure consists to provide to a general purpose Monte Carlo a hard 
process to dress-up its kinematics with the QED/parton showers, hadronisation, decays, \etc{}.
\index{Pythia@\Pythia{} Monte Carlo event generator!Interface to WINHAC@Interface to \WINHAC{}|(}
As specified before, \WINHAC{} uses \Pythia{} parton shower and hadronisation scheme for its purpose
and this specificity was kept in the interface. This is justified for example by the fact that the 
parton shower in the initial state should eventually be replaced by one made by the authors.
To stick with the ATLAS standards an additional interface was written to overwrite the hard-coded
\Pythia{} options in \WINHAC{} by the default values imposed in Athena.
\index{Pythia@\Pythia{} Monte Carlo event generator!Interface to WINHAC@Interface to \WINHAC{}|)}

\subsubsection{Validation of the implementation}
This section describes essential results obtained to validate \winhaci{}.
In what follows all Monte Carlo which are referred to are used within the Athena framework.
First, tuned comparisons at the generator level were achieved with well validated MC.
The goal of these tests were to improve the interfacing of \WINHAC{} from both technical and 
physical point of view.
Complements on this work are available through the reports brought to the CERN forum%
~\cite{CERN_wh_talk1,CERN_wh_talk2}.

\paragraph{Tuned comparisons at the generator level.}
\index{WINHAC@\WINHAC{} Monte Carlo event generator!Event generation in Athena|(}
Tuned comparisons were achieved, \ie{} using the same input parameters in the Monte Carlo,
successively to confront \WINHAC{} to \Pythia{} and then to \Pythia{}+PHOTOS 
for $W\to \mu\,\nu_\mu$ with a statistic of $100,000$ events.
\index{Pythia@\Pythia{} Monte Carlo event generator}
\Pythia{} parameters were tuned with the one used to the \WINHAC{} default this being justified 
since at that time overwriting the hard-coded \Pythia{} parameters used in \WINHAC{} by the Athena
defaults had not yet been though through.

\begin{figure}[!h]
\begin{center}
 \includegraphics[width=0.495\tw]{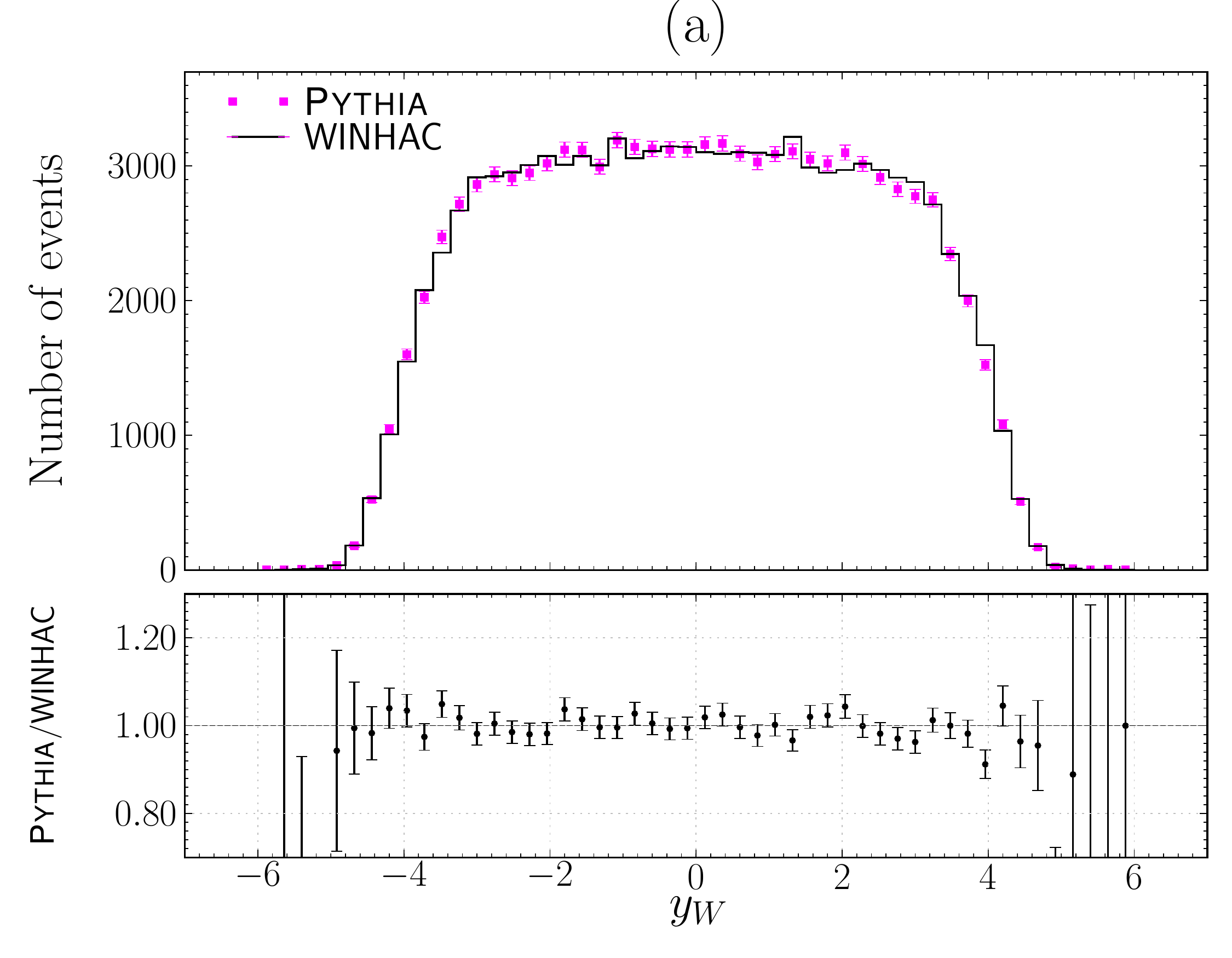}
   \hfill
 \includegraphics[width=0.495\tw]{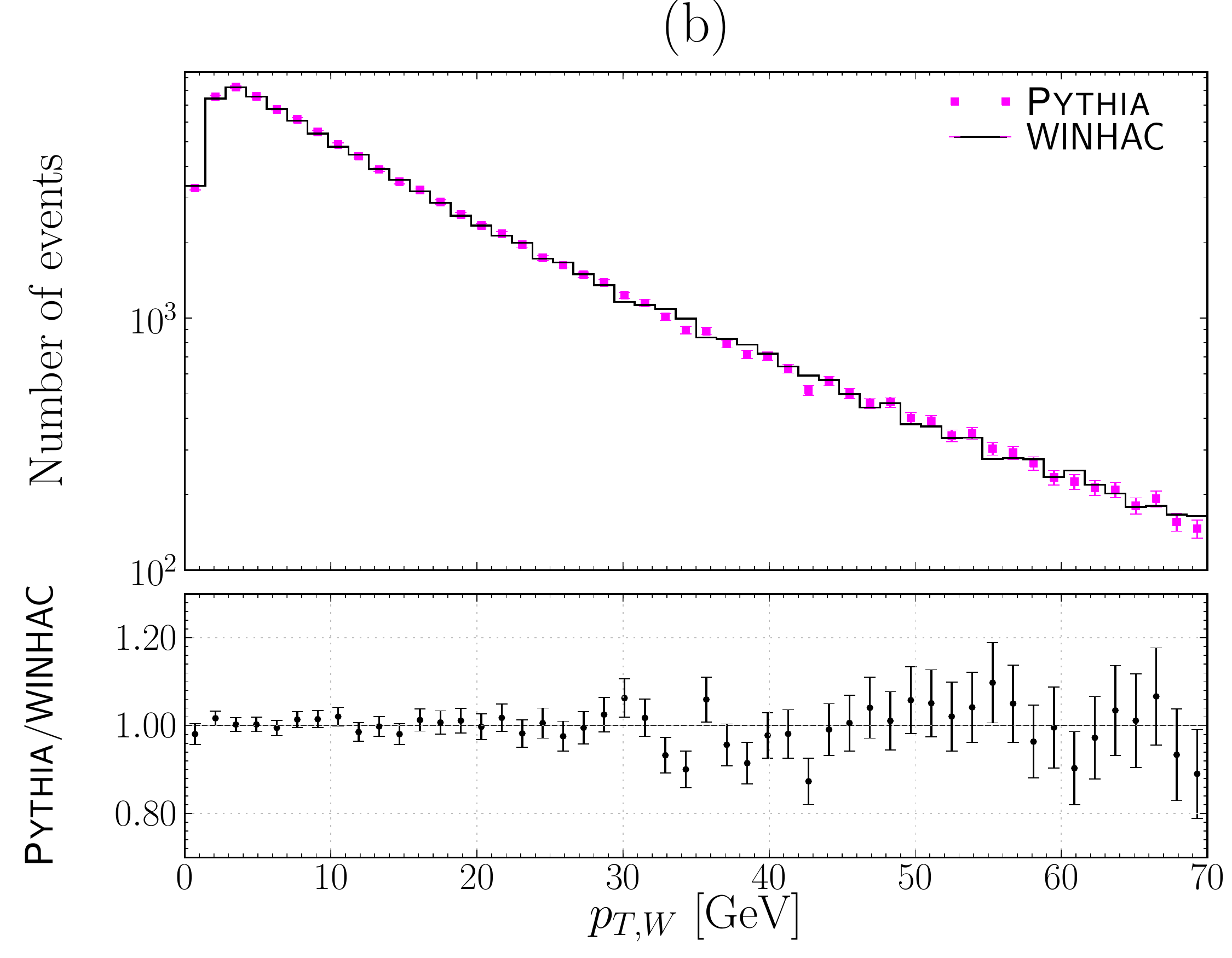}
 \vfill
 \includegraphics[width=0.495\tw]{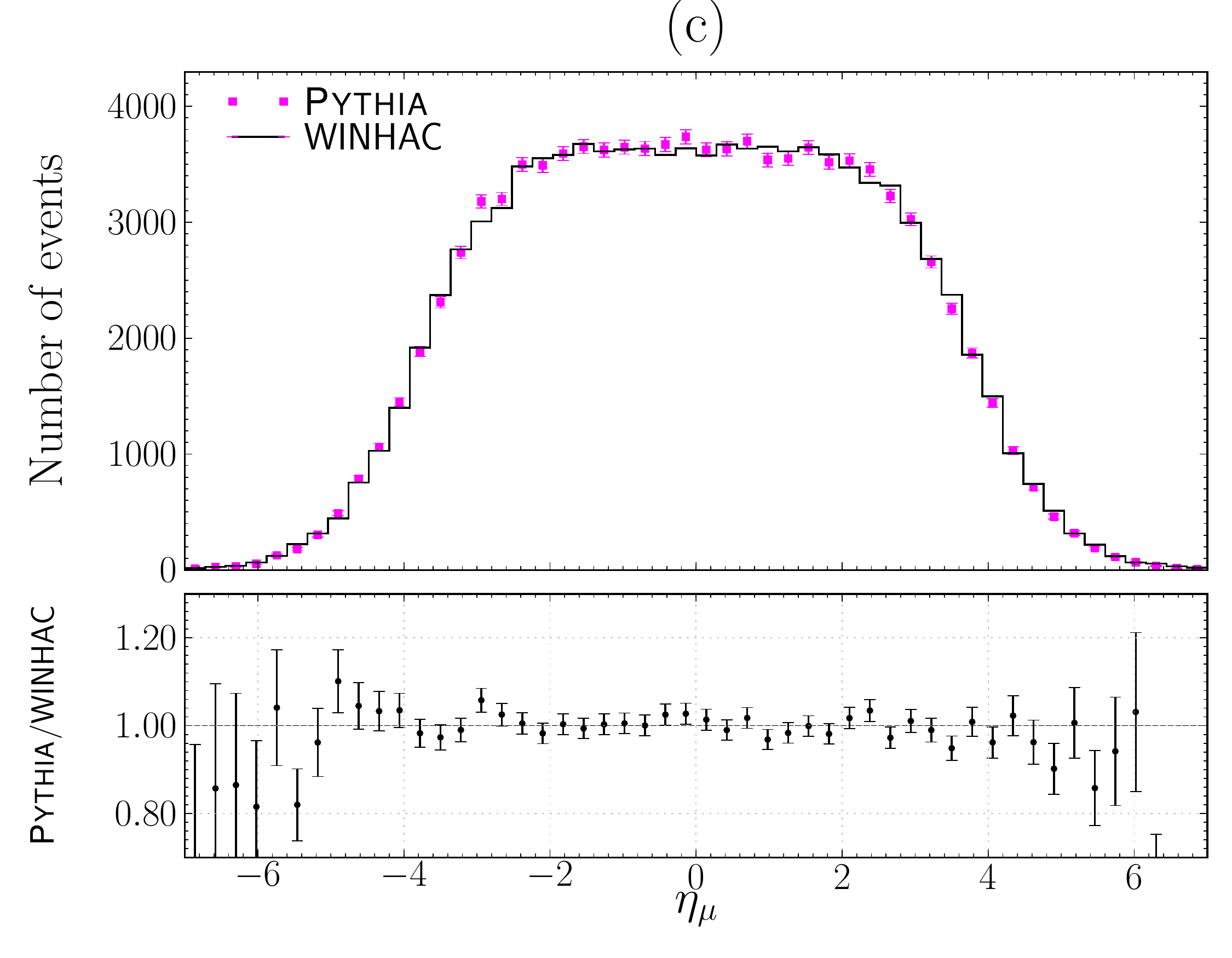}
 \hfill
 \includegraphics[width=0.495\tw]{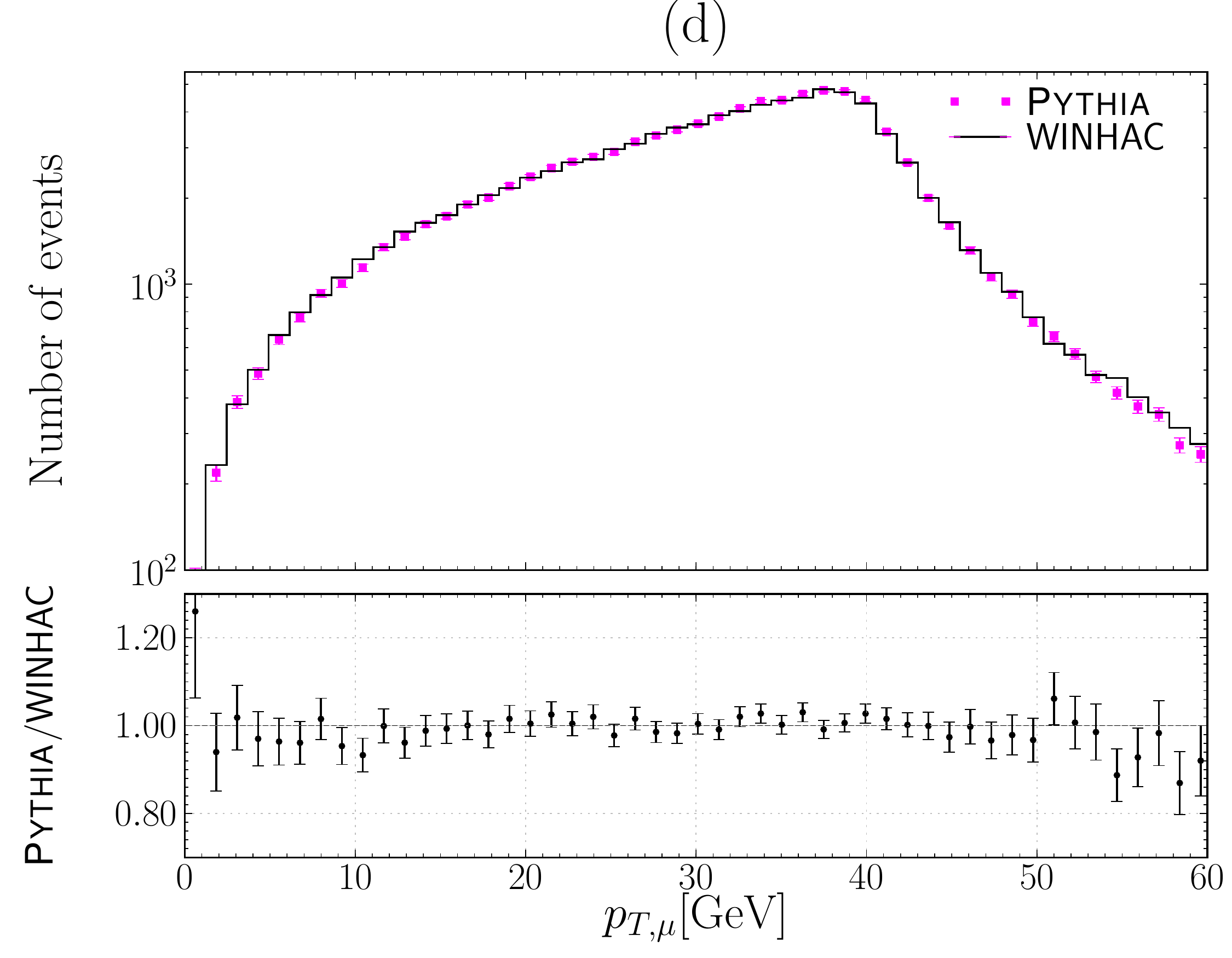}
    \caption[Tuned comparison between \Pythia{} and \WINHAC{} within the ATLAS software]
	    {\figtxt{Tuned comparison between \Pythia{} and \WINHAC{} within the Athena
              software, displaying $\yW$ (a), $\pTW$ (b), $\etal$ (c) and $\pTl$ (d) distributions.
              The lower frames in each plot displays refined comparison in the ratio 
              \Pythia{}/\WINHAC{}.}}
	    \label{fig_wh_vs_pyt}
  \end{center} 
\end{figure}
The first test, versus \Pythia{}, is set up at the Born level with, for only higher corrections,
the QCD and QED parton shower in the initial state.
This imply that both data due to the structure of \WINHAC{} are perfectly similar from both 
physical/technical point of view, indeed in both cases the parton shower was done by 
\Pythia{}~6.403.

In Fig.~\ref{fig_wh_vs_pyt} are respectively represented the rapidity and transverse momentum
of the $W$ and the pseudo-rapidity and transverse momentum of the charged leptons. The validation
using $\pTW$ is important as its distribution turns out to be very sensible from the parton shower
and from the intrinsic $\kT$ of partons (cf. Chapter~\ref{chap_theo} and
Fig.~\ref{fig_yW_pTW}.(b)).
In \Pythia{} this meant to take for these runs a Gaussian distribution for $\kT$ which corresponds
to use the value \texttt{MSTP(91)=1} with $\Mean{\kT}=1\GeV$ that latter corresponding to the switch
\texttt{PARP(91)=1}.
Another point, no multiple interactions were used (\texttt{MSTP(81)=0}).
For further details on the physical and technical aspects of these parameters see 
Ref.~\cite{Sjostrand:2006za}. As expected the data from both generator agree within the statistical
limit.

\begin{figure}[!h]
\begin{center}
  \includegraphics[width=0.495\tw]{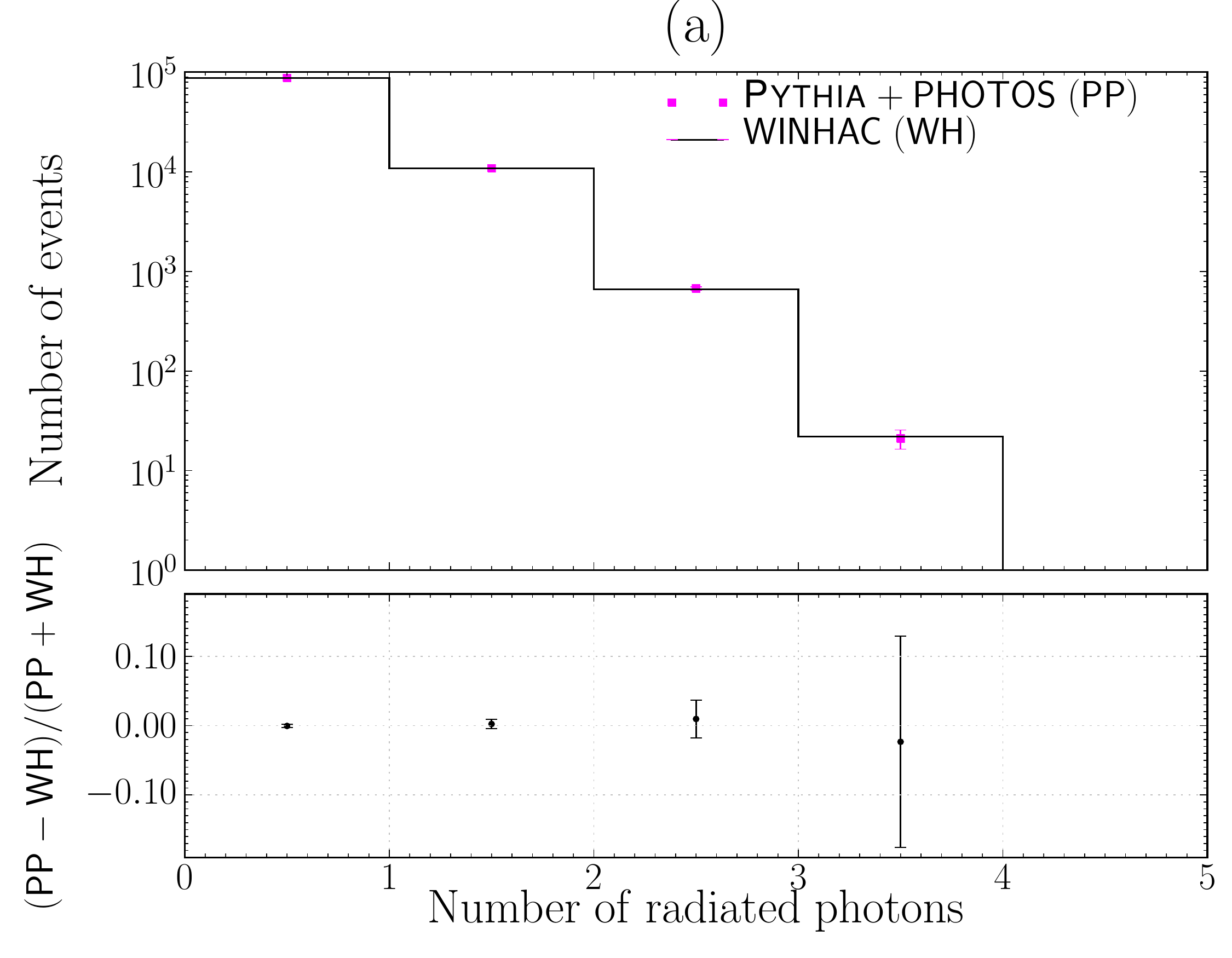}
  \hfill
  \includegraphics[width=0.495\tw]{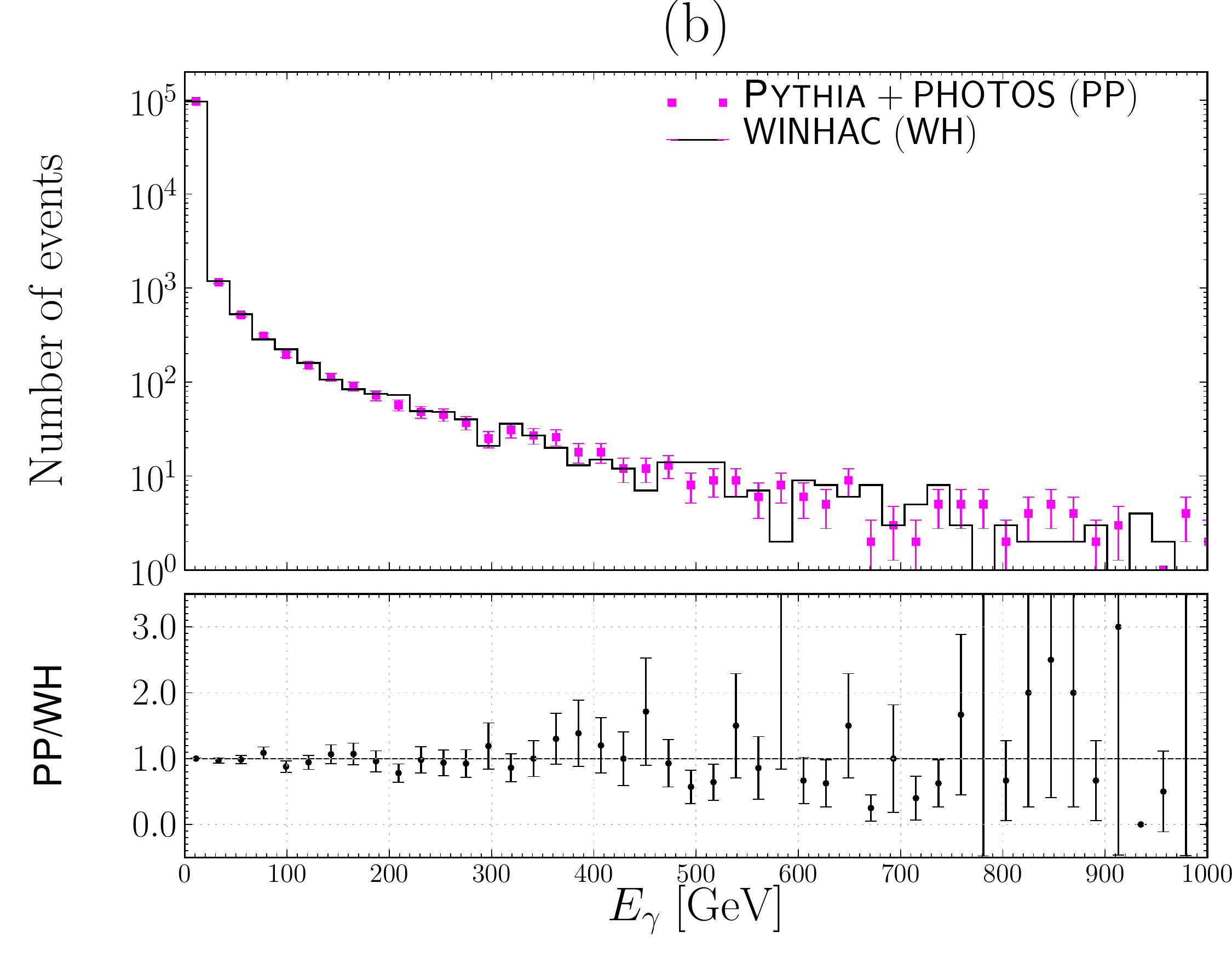}
  \vfill
  \includegraphics[width=0.495\tw]{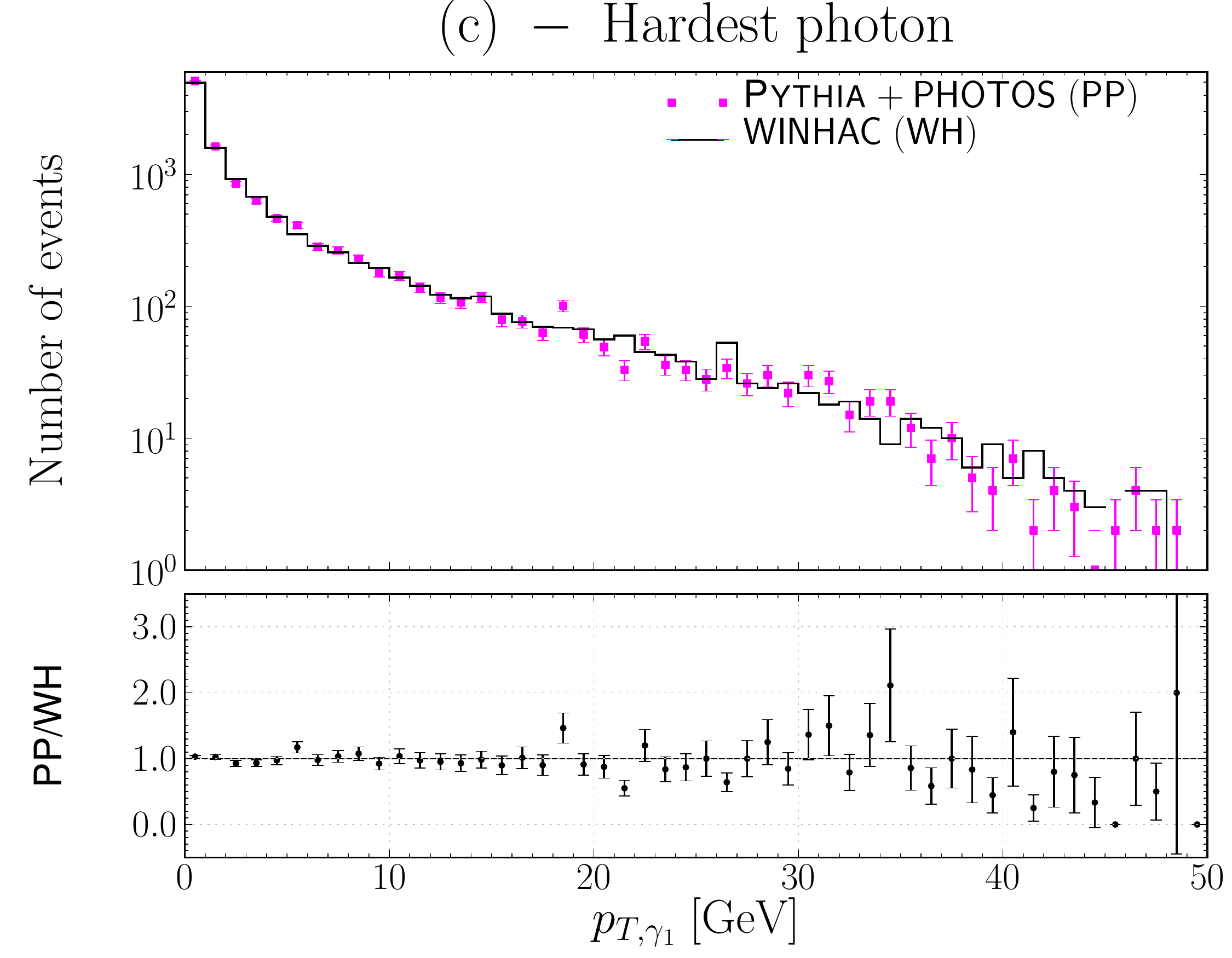}
  \hfill
  \includegraphics[width=0.495\tw]{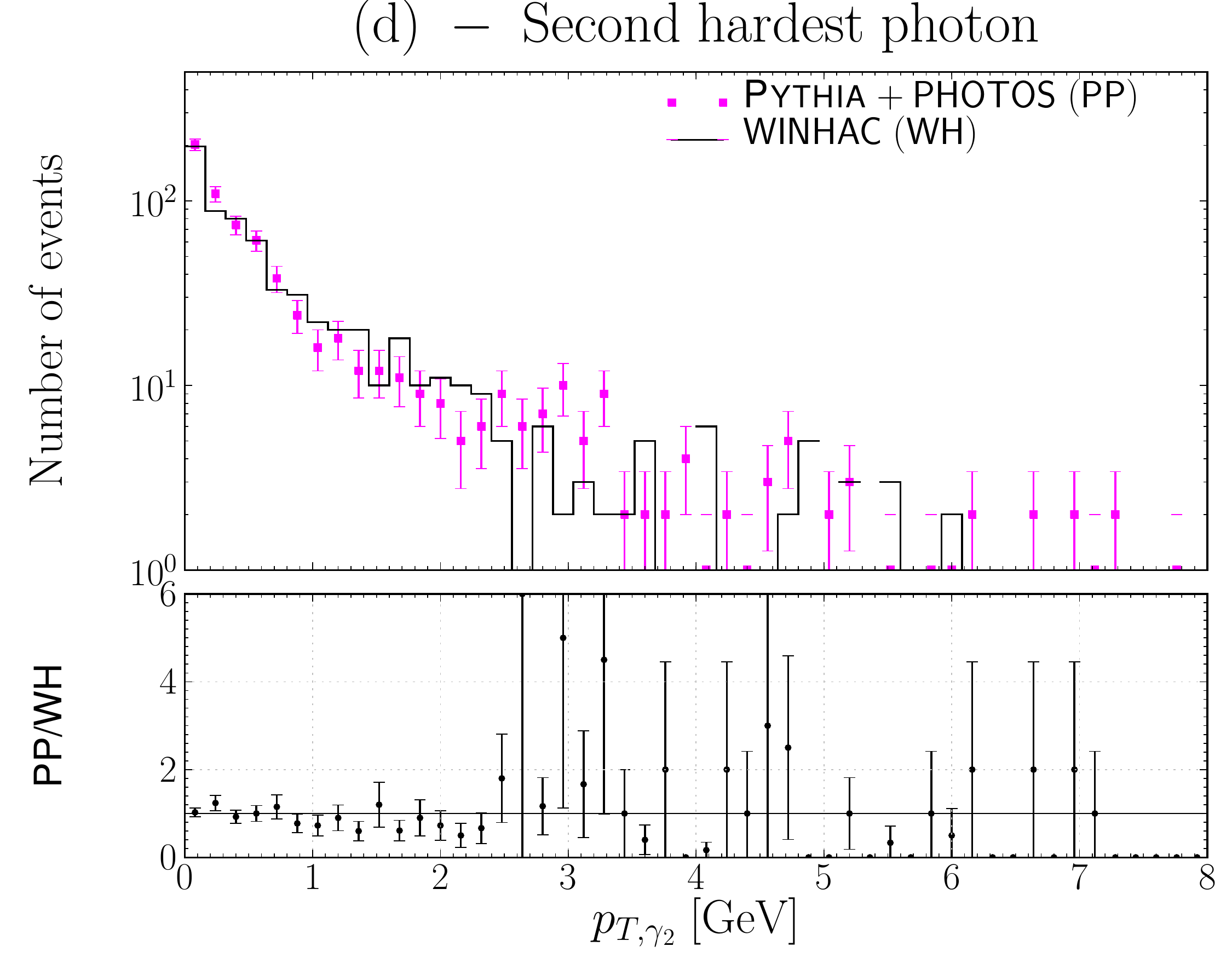}
    \caption[Tuned comparison between \Pythia{}+\PHOTOS{} and \WINHAC{}]
	    {\figtxt{Tuned comparison between \Pythia{}+\PHOTOS{} and \WINHAC{}
                showing the number of radiated photons (a), the total energy of the
                radiated photons (b), 
                the transverse momentum of the first hardest photons (c) and
                second hardest photons (d).}}
	    \label{fig_wh_vs_pytphot_1}
  \end{center} 
\end{figure}

\index{PHOTOS Monte Carlo!Interfaced to Pythia@Interfaced to \Pythia{}|(}
\index{Pythia@\Pythia{} Monte Carlo event generator!Interfaced with PHOTOS|(}
For the second test \WINHAC{} was used by adding up to the previous setting QED multi-photon 
radiation\index{QED!Radiative corrections@Radiative correction in single $W$ production}. 
On the other side to emulate higher QED radiation in \Pythia{} the former set up
was plugged to the PHOTOS~\cite{Barberio:1990ms,Barberio:1993qi,Golonka:2006tw}
generator which added QED corrections in the final state of the events.
In both cases, \WINHAC{} and \Pythia{}+PHOTOS, a common value was used to cut on the energies
of soft photons at the level of the analysis, that is $E_\gamma > 500\MeV$.
In Fig.~\ref{fig_wh_vs_pytphot_1}, the number of radiated photons are represented with the total 
energy of these photons. Below are represented the transverse momenta of the hardest and second
hardest photons defined respectively as, for a given event, the photon with the highest $\pT$ and 
the second highest $\pT$. Although the statistic is quite small for the histograms related to 
photons one can see there is a good agreement within the statistical error.

\begin{figure}[!h]
\begin{center}
 \includegraphics[width=0.495\tw]{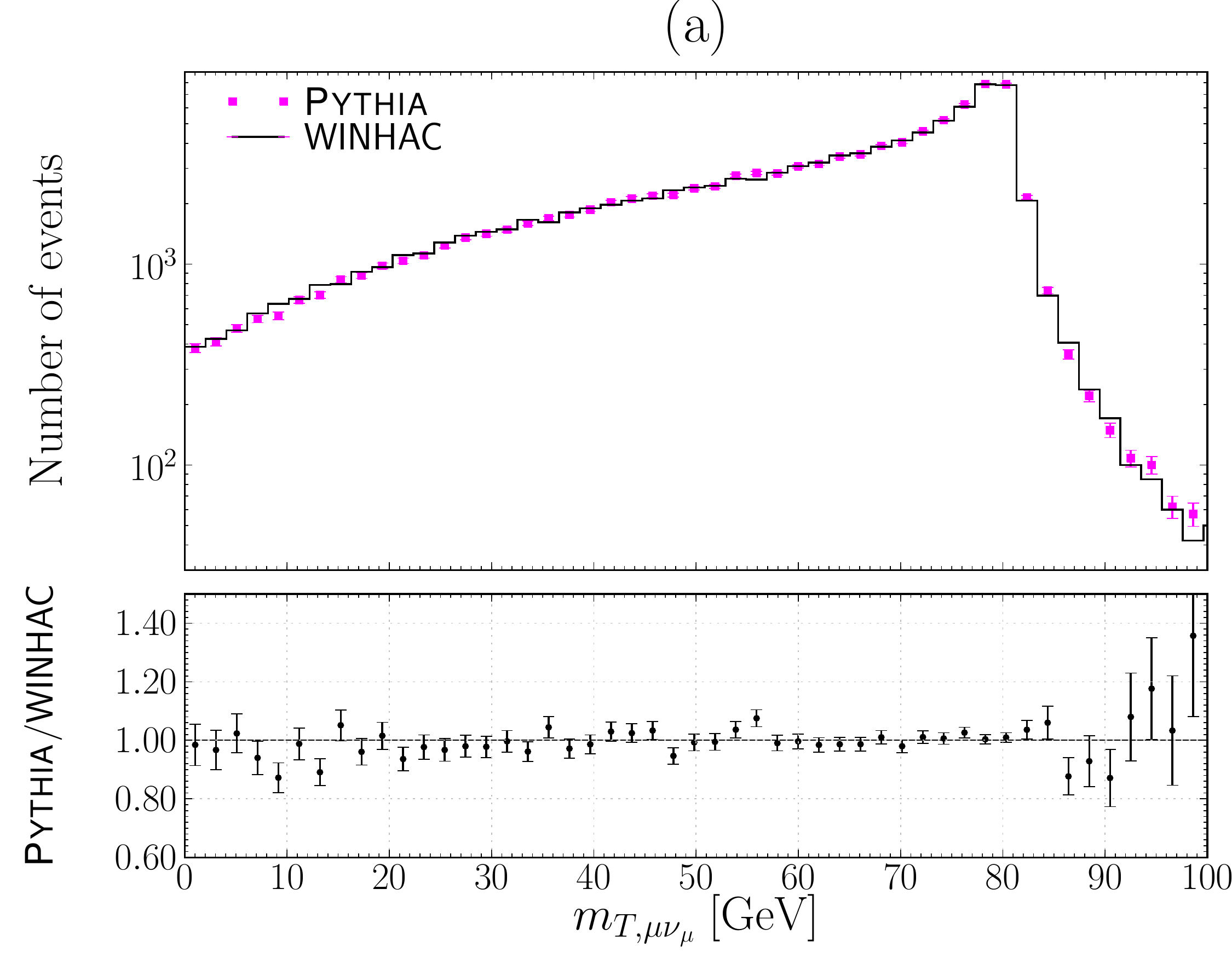}
   \hfill
 \includegraphics[width=0.495\tw]{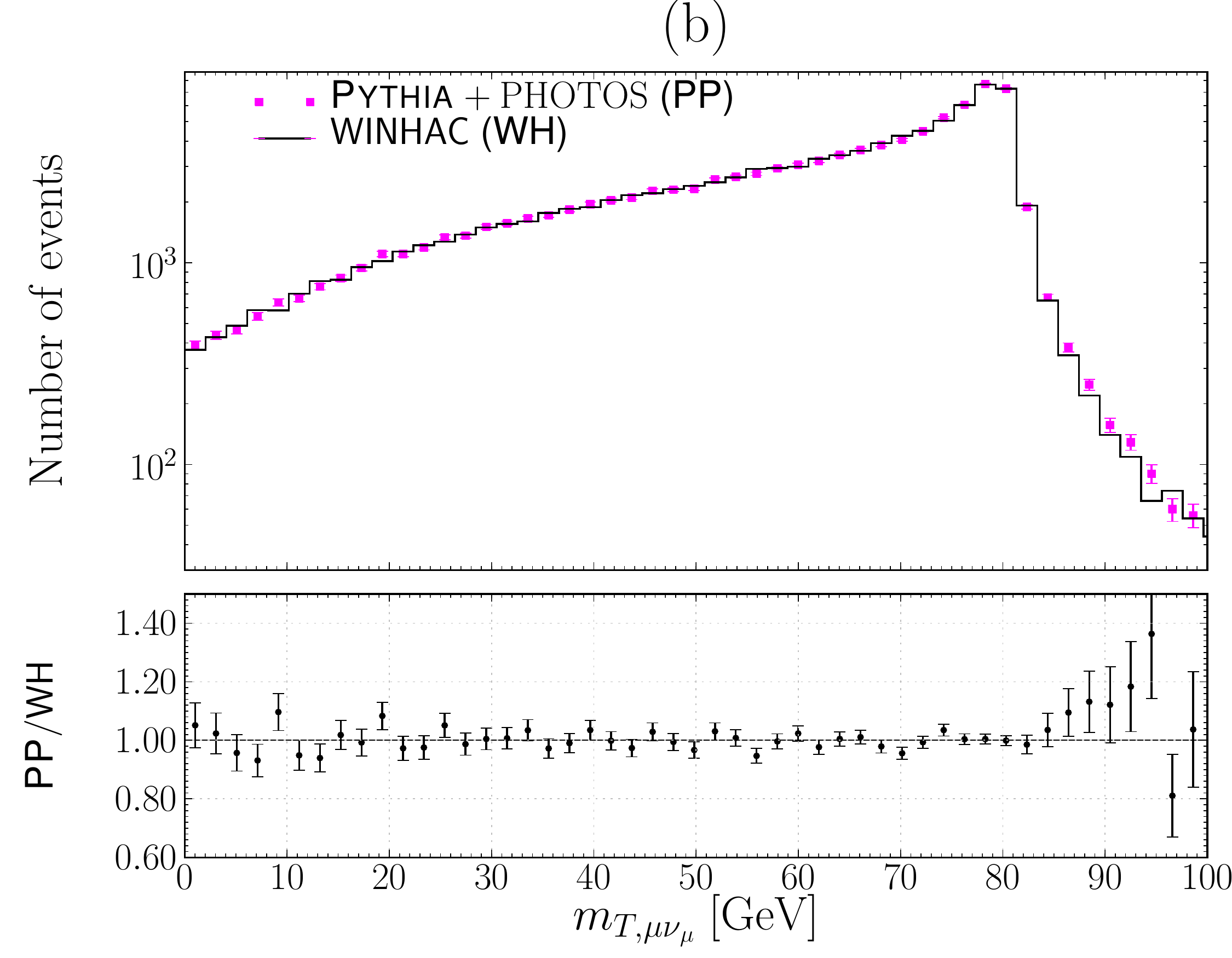}
    \caption[Tuned comparison within the ATLAS software between \Pythia{}/\Pythia{}+\PHOTOS{} 
      and \WINHAC{} for the lepton transverse mass]
	    {\figtxt{Tuned comparison based on the leptons transverse mass for
                \Pythia{} versus \WINHAC{} at the Born level (a) 
                and \Pythia{}+\PHOTOS{} versus \WINHAC{} at the Born+multi-photon QED radiation 
                (b).}}
	    \label{fig_py_pp_wh_mT}
  \end{center} 
\end{figure}

Some discrepancies were observed though while looking at the transverse mass.
In Fig.~\ref{fig_py_pp_wh_mT} the $m_{T,\mu\nu_\mu}$ distribution is represented for both the 
first validation set up and the present one. One can see in the latter case from the ratio frame
that after the jacobian peak \WINHAC{} displays a slight shift to higher $\pT$ with respect
to \Pythia{}(+\PHOTOS{}).
These discrepancies can be amended by the fact that here \WINHAC{} and PHOTOS do not treat QED radiation
on the same footing.\index{QED!Radiative corrections@Radiative correction in single $W$ production}
Indeed, PHOTOS do not take into account cases where the one or several photon
are emitted by the $W$ boson as well as the interference this creates with other diagrams where the 
charged lepton radiates photon(s).

\begin{table}[]
\begin{center}
\renewcommand\arraystretch{1.1}
\begin{tabular}{rrr}
  \hline
  $N_\gamma$    & \Pythia{}+PHOTOS & \WINHAC{} \\
  \hline\hline
  0            &  88374           & 88440 \\
  1            &  10927           & 10874 \\
  2            &    677           &   664 \\
  3            &     21           &    22 \\
  \hline
\end{tabular}
\renewcommand\arraystretch{1.45}
\caption[Radiated photons from \WINHAC{} and \Pythia{}+PHOTOS simulation]
        {\figtxt{Number of radiated photons $N_\gamma$ from \WINHAC{} 
            and \Pythia{}+PHOTOS simulations.}}
\label{table_wh_pp_photrad}
\end{center}
\end{table}
\index{PHOTOS Monte Carlo!Interfaced to Pythia@Interfaced to \Pythia{}|)}
\index{Pythia@\Pythia{} Monte Carlo event generator!Interfaced with PHOTOS|)}
\index{WINHAC@\WINHAC{} Monte Carlo event generator!Event generation in Athena|)}

\paragraph{Simulation and reconstruction tests.}
The last test consisted to pass \WINHAC{} events through the whole simulation and reconstruction 
chain. Tests were passed with success respectively for $W\to~e\,\nu_e,\;\mu\,\nu_\mu,\;l\,\nu_l$  
each time with a statistic of $1,000$ events using acceptance cuts of $\pTl>20\GeV$ and $|\etal|<2.5$.
The low statistic is justified because of the long time it takes to simulate events.
In each case the data was obtained in ESD, AOD and CBNT format.
\index{ATLAS software!CBNT}\index{ATLAS software!ESD}\index{ATLAS software!AOD}

\begin{figure}[!h]
\begin{center}
 \includegraphics[width=0.495\tw]{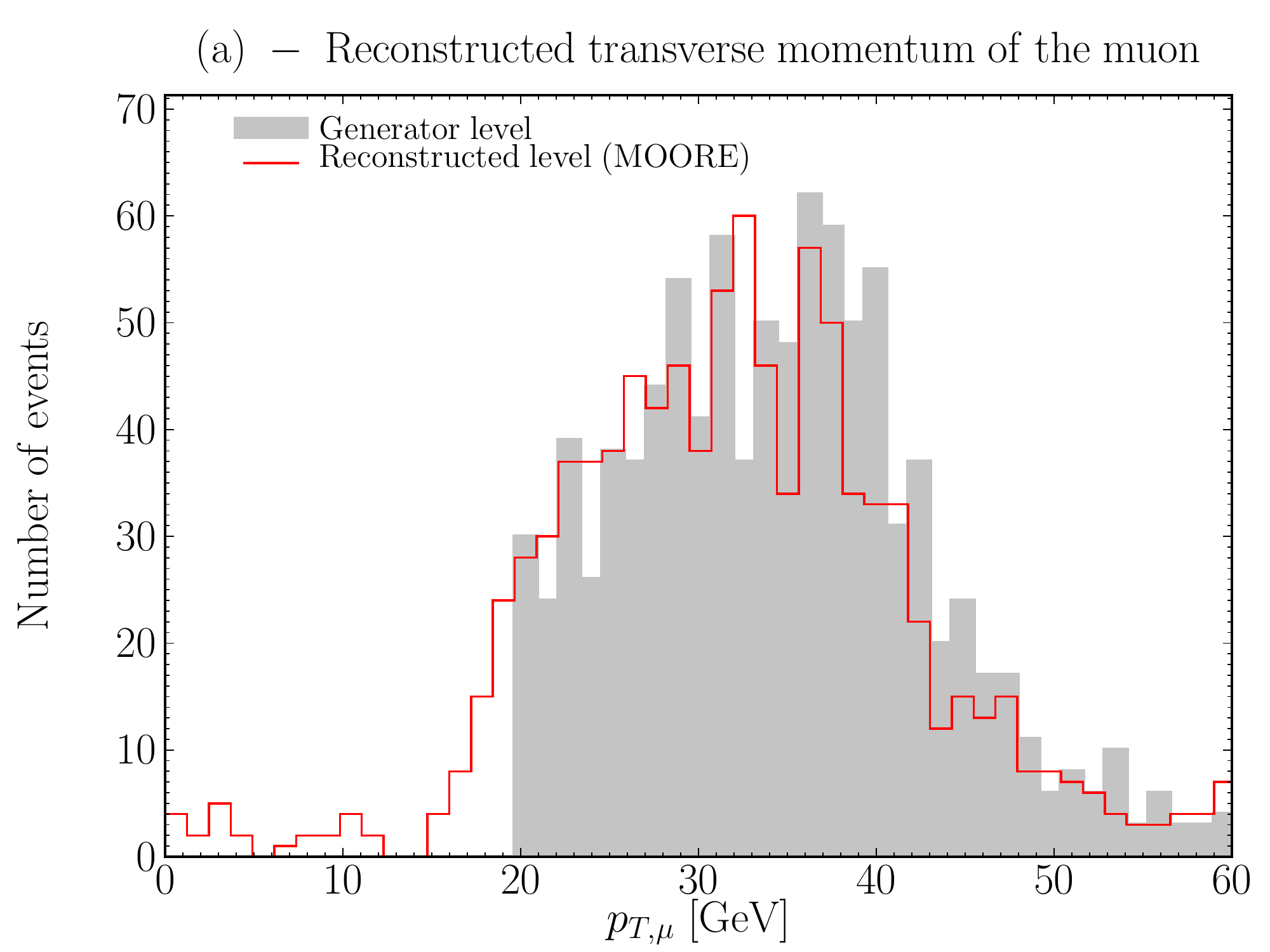}
 \hfill
 \includegraphics[width=0.495\tw]{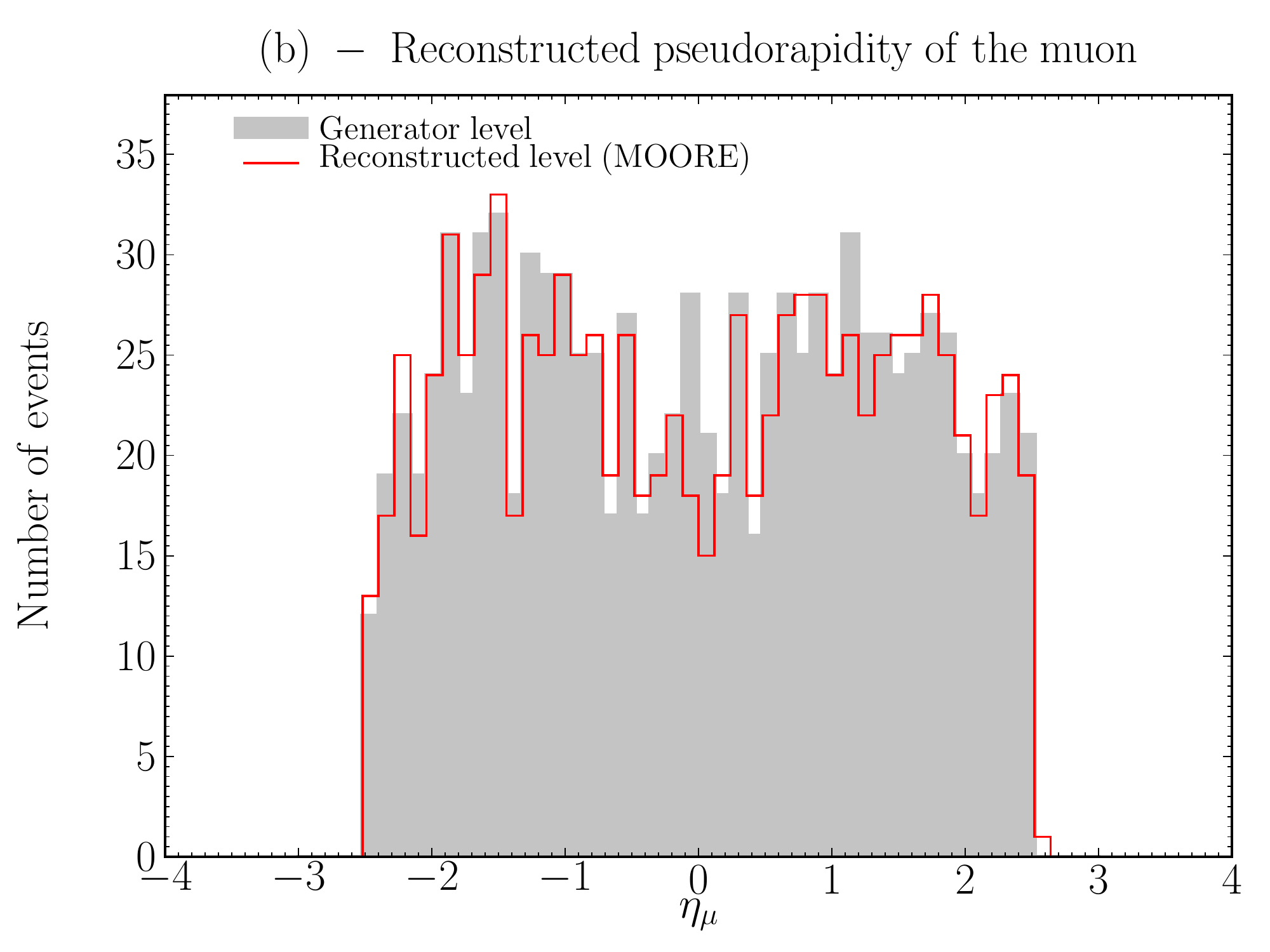}
    \caption[\WINHAC{} muon kinematics at the generator and reconstructed level]
	    {\figtxt{\WINHAC{} muon kinematics at the generator and reconstructed level
              for the transverse momentum (a) and the pseudo-rapidity (b).
              The acceptance cuts applied at the generation level are\,: $p_{T,\mu}>20\GeV$ and 
              $|\eta_\mu|<2.5$.}
	    }
	    \label{fig_mu_recon}
  \end{center} 
\end{figure}

Fig.~\ref{fig_mu_recon} presents respectively comparisons between the generator and reconstructed
data for the transverse momentum and pseudo-rapidity of the muon.
The reconstruction was made using the MOORE algorithm \cite{ATL-SOFT-2003-007}
that reconstructs tracks in the muon spectrometer.

\section{Framework of the analysis}\label{s_framework}
Like explained in the introduction of this Chapter,
rather than going to the refined --but heavy-- framework described above a simpler and more pragmatic 
approach was adopted. 
In this approach, the simulation of the detector for the analysis is achieved using Gaussian 
resolution to emulate the inner detector resolution.
The reason for that is that the statistic that needed to be generated is too huge to be passed 
through the whole simulation and reconstruction chain of the ATLAS detector.
Actually, analysis with real data are based on the same principle where instead of taking the whole
simulation for the detector a tuned fast simulation is used for all Monte Carlo simulation.
This is for example the strategy used by CDF~II for the measurement of $\MW$~\cite{Aaltonen:2007ps}.

The last argument becomes more tangible when looking quantitatively at the present analysis 
requirements. Studies were made for an integrated luminosity of $10\,\mathrm{fb}^{-1}$ which means
an order of $\sim 1.1\times 10^8~\Wp$ and $\sim 0.8\times 10^8~\Wm$.
In order to optimise the strategy to measure their masses, generation and simulation of 
${\cal O}(100)$ event samples was required. These samples correspond to specific biases in the 
detector response, or in the theoretical (phenomenological) modeling of the $W$-boson production 
processes.
In addition, a large number of unbiased event samples, for variable values of the masses of the 
$\Wp$ and $\Wm$ bosons, was simulated. For an assessment of the impact of the systematic biases on 
the overall measurement precision each of the above event sample must contains at least $10^8$ 
events in order to match the systematic and the statistical measurement precision. 

The presented analysis is then based on a total sample of ${\cal O}(10^{10})$ $W$ boson events. 
Generating, simulating and handling such a large event sample within reasonable limits of
the storage space and computing power is challenging.
Indeed, if we consider in the simplest case scenario where the gist of the event\footnote{ 
That is the four-momenta, types and genealogy of each particle entering in the hard process
($\sim 10$ particles per event).} would be recorded in double precision, this should require 
$\approx 500$ bytes for each event summaries, which leads eventually to store a Terabyte of data on 
disk.
Besides, even though one can compute an accurate experimental simulation using GEANT, 
the treatment of hundreds of millions of events would be prohibitive.
The computing of the desired histograms on a dedicated farm, using a fast simulation,
takes 1--2 days to get all the necessary event samples.

A short descriptions of both generation and analysis framework are given below.

\subsection{Generation framework}\label{ss_gen_framework}
\begin{figure}[!h]
\begin{center}
 \includegraphics[width=0.95\tw]{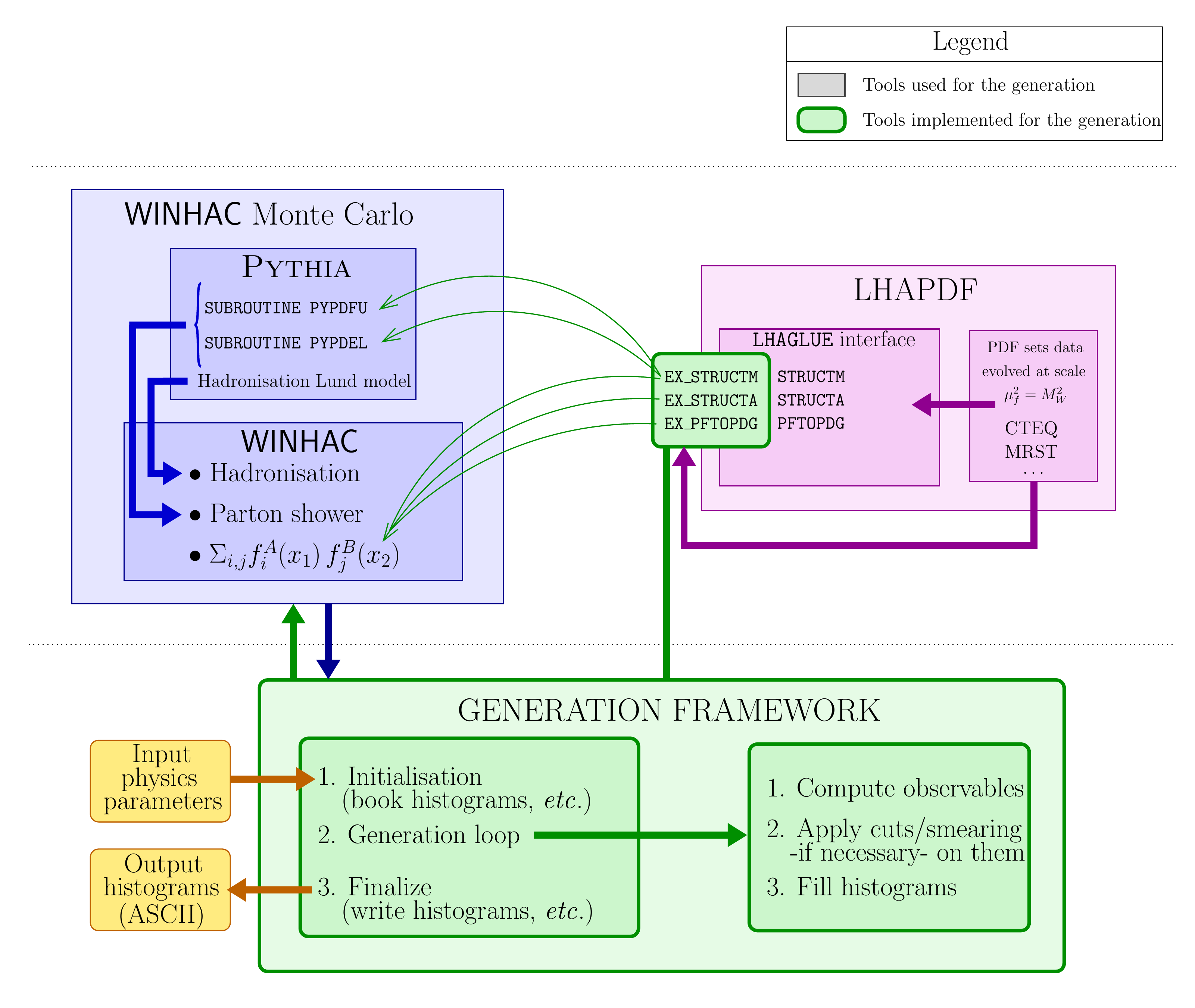}
    \caption[Generation framework schematic representation]
	    {\figtxt{Generation framework schematic representation.
                The round boxes are the code implemented to carry the analysis.}}
	    \label{fig_framework_generetion}
  \end{center} 
\end{figure}

\index{WINHAC@\WINHAC{} Monte Carlo event generator!Generation framework for the analysis|(}
In that state of mind, the generation is made using the \WINHAC{} event generator in stand-alone
mode. An overview of the main features of the generation framework is represented in 
Fig.~\ref{fig_framework_generetion}. For convenience the generation framework interfaced to 
\WINHAC{} was also written in FORTRAN~77.
This framework takes as input the physics parameters necessary for the initialisation and the 
running of \WINHAC{}. During this initialisation the observables of interest are booked.
Then in the generation loop, for each event, standard calling instances to \WINHAC{} implemented 
by its authors allow to access basics quantities such as four-momentum, particles type, 
\etc{}. from which observables of interest can be calculated. For practical reasons all analysis
were made using weighted events since it is faster.
At this stage, cuts or smearing to emulate the ATLAS detector resolution can be optionally applied. 
Eventually each observable fills the corresponding histogram and at the end of the generation the 
histograms are written in ASCII files along with the cross section of the process.
Already at this stage histograms are normalised to $\mm{nb}$ according to the conventions stated
in Chapter~\ref{chap_theo}\,\S\,\ref{notations_conventions}.

On our road to an experimental investigation a preliminary step consisted to look only at the 
generator level predictions with only optional phase-space cuts to put the spot on a particular 
behaviour. This is presented in the next Chapter.
Then the more realistic experimental analysis for $\MWp-\MWm$ was made using some acceptance cuts 
related to the ATLAS performances and the smearing of the data was computed according to 
Eqs.~(\ref{eq_rho_smearing}--\ref{eq_theta_smearing}) from \S\,\ref{ss_id_trck_fit_perf}.
This step us presented in Chapter~\ref{chap_W_mass_asym}.

\index{Parton Distribution Functions (PDFs)!LHAPDF|(}
\index{Pythia@\Pythia{} Monte Carlo event generator!Use of LHAPDF routines|(}
Let us note that in both cases, sometimes, changes of the original PDFs predictions needed to be
done. Still, for both security and practical reasons the LHAPDF package was not touched.
Instead an interface was written which lures \WINHAC{} and \Pythia{} to use custom values
instead of the original ones.
The subroutines that needed to be substituted to the use of the original ones are of the number of three
and belong to the LHAGLUE interface written to mimic the old-fashion way to call PDFs from the PDFLIB
package.
These three subroutines which \Pythia{} and \WINHAC{} depend on are briefly described.
The first one --the most fundamental-- is \texttt{STRUCTM} which gives for a given PDF set at
fraction $x$ and scale $\mu_f^2$ the parton density functions in the output form 
$x\,f_i(x,\mu_f^2)$, where $i$ is such that
\begin{equation*}
i = \{g,\,d,\,\dbar,u,\,\ubar,\,\smartssbar,\,\smartccbar,\,\smartbbbar\},
\end{equation*}
with, for reminder, $q^\sea=\qbar^\sea$, where (s) means sea for $s$, $c$ and $b$ flavors.
The two other subroutines, used exclusively by \WINHAC{}, are \texttt{PFTOPDG} and \texttt{STRUCTA} 
and both rely on the output from \texttt{STRUCTM}.
\texttt{PFTOPDG} provides the flavors in a one dimensional array which indexes, ranging from 
-6 to 6, are associated to the conventional ID number given to partons (see Ref.~\cite{Amsler:2008zz}, 
Monte Carlo particle numbering scheme section).
\texttt{STRUCTA} adds to the predictions of \texttt{STRUCTM} nuclear shadowing effects in the aim
to compute the PDFs for a hadron of mass $A$.

To these subroutines are substituted our custom copies 
\begin{itemize}
\baselineskip 1pt
\item[-] \texttt{EX\_STRUCTM} 
\item[-] \texttt{EX\_PFTOPDG}
\item[-] \texttt{EX\_STRUCTA}
\end{itemize}
the structure of the two last one are unchanged, they are just relying on \texttt{EX\_STRUCTM}
instead of the original subroutine. Then, any changes in the density functions are made 
by playing with the output values of \texttt{EX\_STRUCTM}.

Let us remark that in Fig.~\ref{fig_framework_generetion} only the main features are given and
in no case the whole chain of subroutine calling concatenation as it will eventually become obsolete
as both F77 versions of \WINHAC{} and \Pythia{} will be replaced to C++ versions.
Slight details where given for only \Pythia{} since this Monte Carlo being extensively used its 
subroutines are more familiar in the high energy physics community.
\index{Parton Distribution Functions (PDFs)!LHAPDF|)}
\index{WINHAC@\WINHAC{} Monte Carlo event generator!Generation framework for the analysis|)}
\index{Pythia@\Pythia{} Monte Carlo event generator!Use of LHAPDF routines|)}

\subsection{Analysis framework}
\begin{figure}[!h]
\begin{center}
 \includegraphics[width=0.65\tw]{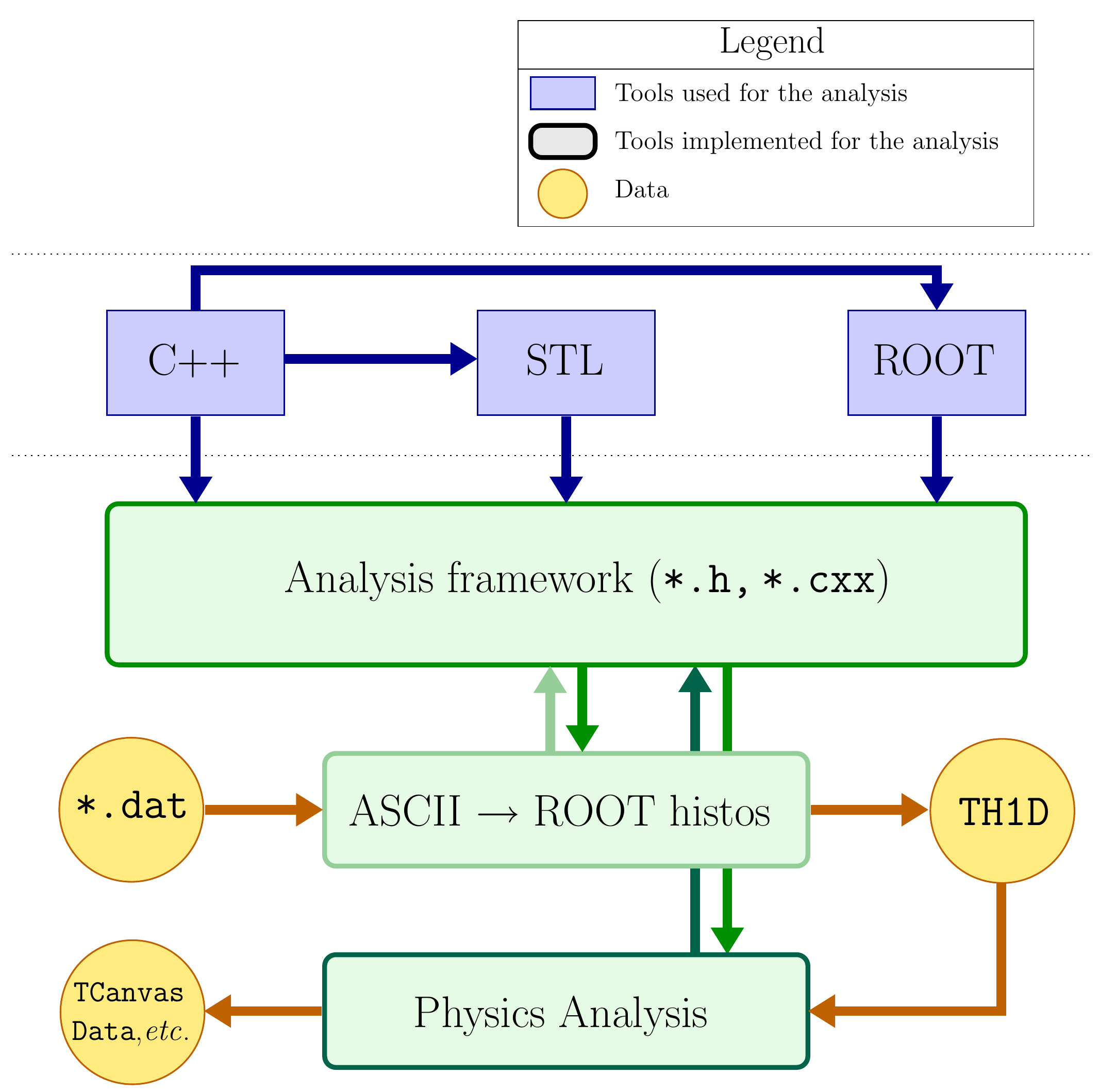}
    \caption[Analysis framework schematic representation]
	    {\figtxt{Analysis framework schematic representation.}}
	    \label{fig_framework_analysis}
  \end{center} 
\end{figure}

Here we describe the methods that were used to perform all kind of histograms and/or analysis 
from the \WINHAC{} event generations present in this document and in our ongoing studies%
~\cite{Upcoming_MW,Upcoming_GammaW}. This histograms/analysis framework is represented 
schematically on Fig.~\ref{fig_framework_analysis}.

The framework is made of a central set of functions written in C++ and gathered in one 
\texttt{*.cxx} file and its associated header file \texttt{*.h}.
This framework relies on C++ and STL (Standard Template Libraries) classes and also on ROOT 
classes~\cite{Brun:1997pa,ROOThomepage} from version 5.14.
The ROOT classes and methods that are used essentially to handle one-dimensional double precision 
histograms containers \texttt{TH1D} and other methods inherent to this class which
allow the user to perform basic operations on and between histograms. 
ROOT is also called to display histograms and related analysis results using the class 
\texttt{TApp}. Histograms are then displayed using the canvas class \texttt{TCanvas}, stored
inside \texttt{*.root} files (using the \texttt{TFile} class) and postscript files for later browsing.
Each step of the histograms/analysis consists always to the same principle which is to write a
C++ program to reach a given goal with the help of calls of functions from our framework.

Now, further details are given on the treatment of data in a chronological order.
In a first step, all the ASCII histograms obtained from the generation and sharing the same input 
parameters are converted to \texttt{TH1D} objects and stored in \texttt{*.root} files whose names 
are based on the main physical input parameters.

In a second intermediate step --not represented in Fig.~\ref{fig_framework_analysis}-- 
the newly created histograms are read from the \texttt{*.root} files they are stored in and 
displayed as \texttt{TCanvas} for visual cross-check.
Those canvas are saved in \texttt{*.root} and postscript files to allow visual control for safety.

Finally the last step consists to produce the desired results that is fine histograms or an
analysis. The first procedure was used extensively to understand the asymmetries in the mechanisms
of production between $\Wp$ and $\Wm$. For that purpose in top of drawing the two 
$(\flatDfDx{\sigma^\plus}{a})$ and $(\flatDfDx{\sigma^\minus}{a})$ distributions related to the 
observable $a$, another frame was drawn in the canvas to display discrepancies between them.
Optionally the difference, the ratio, or the charge asymmetry 
(Eq.~(\ref{eq_def_charge_asym})) of the two distributions was computed the latter being preferred 
due to the nature of the present work. This kind of analysis are presented in the next Chapter.

Concerning the physics analysis it involved to perform likelihood studies between a main 
($\mm{M}$) and $N$-templates event samples distributions which $n^\mm{th.}$ template is labeled $\mm{T}^{(n)}$.
This will be explained thoroughly in Chapter~\ref{chap_W_mass_asym} here only the technical 
principle is described.
It is very basic though, the main/templates distributions are read out from their respective 
\texttt{*.root} files and using dedicated functions of the framework a likelihood analysis is 
performed between each couple $\{\mm{M},\mm{T}^{(n)}\}$ couple. 
The main histograms and all 
templates are displayed along with the likelihood results in canvas and saved in \texttt{*.root}.
\index{WINHAC@\WINHAC{} Monte Carlo event generator!Implementation inside Athena|)}

\cleardoublepage
\begin{subappendices}
\makeatletter\AddToShipoutPicture{%
\AtUpperLeftCorner{2cm}{2cm}{\ifodd\c@page\else\makebox[0pt]{\Huge$\bullet$}\fi}%
\AtUpperRightCorner{0cm}{2cm}{\ifodd\c@page\makebox[0pt]{\Huge$\bullet$}\else\fi}%
}\makeatother
\section{Example of a \WINHAC{} summary event}\label{app_winhac_evt}
\setlength{\epigraphwidth}{0.5\tw}
\epigraph{Et toi mon code pourquoi bogues tu\\
\smallskip
Comme un prompteur stroboscopique\\
Je scanne les nigtlies et les logs}%
{\textit{Le prompteur stroboscopique}\\ \textsc{Guillaume Apollinaire$++$}}
\index{WINHAC@\WINHAC{} Monte Carlo event generator!WINHACevt@\WINHAC{} event example}
The two columns below represent a \WINHAC{} event displayed using a \Pythia{} routine.
This allows the reader to familiarise, if necessary, to a summary event of a high energy particle 
collisions from a Monte Carlo point of view. In each line are shown\,:
\begin{itemize}
\baselineskip 1pt
\item \texttt{I}\,: index of particle.
\item \texttt{particle/jet} : particle name or string (see Ref.~\cite{Sjostrand:2006za} for a 
definition of a string in the \Pythia{} Lund fragmentation model).
\item \texttt{KS}\,: gives the current status of the parton/particle. 
Codes from 1 to 10 correspond to currently existing partons/particles, while larger codes contain 
partons/particles which no longer exist, or other kinds of event information.
For further details see Ref.~\cite{Sjostrand:2006za}, \S\,5.2 \textit{Event Record} section.
\item \texttt{KF}\,: particle ID according to the convention used for the Monte Carlo numbering 
scheme~\cite{Amsler:2008zz}.
\item \texttt{orig}\,: \texttt{I}-wise origin of the particle
\item \texttt{p\_x}, \texttt{p\_y}, \texttt{p\_z}\,: Respectively the momentum components
$p_x$, $p_y$ and $p_z$ of the particle in the Cartesian basis attached to the 
laboratory inertia frame.
\item \texttt{E}, \texttt{m}\,: Energy $E$ of the particle in the Cartesian attached to 
the laboratory inertia frame and its invariant mass $m$.
\end{itemize}

\begin{SizeSPE}
\begin{multicols}{2}
\begin{verbatim}
                            Event listing (summary)

    I particle/jet KS     KF  orig    p_x      p_y      p_z       E        m

    1 !p+!         21    2212    0    0.000    0.000 7000.000 7000.000    0.938
    2 !p+!         21    2212    0    0.000    0.000-7000.000 7000.000    0.938
 ==============================================================================
    3 !dbar!       21      -1    1   -0.524    0.462  163.022  163.024    0.000
    4 !g!          21      21    2   -0.434   -0.604 -341.373  341.374    0.000
    5 !dbar!       21      -1    3   -1.249   -1.638  145.551  145.566    0.000
    6 !u!          21       2    4   -2.833   -4.851  -10.211   11.654    0.000
    7 !W+!         21      24    0   -4.082   -6.489  135.340  157.219   79.638
    8 !mu+!        21     -13    7  -27.004  -12.899  129.225  132.645    0.106
    9 !nu_mu!      21      14    7   22.922    6.410    6.115   24.575    0.000
 ==============================================================================
   10 (W+)         11      24    7   -4.082   -6.489  135.340  157.219   79.638
   11 mu+           1     -13    8  -27.004  -12.899  129.225  132.645    0.106
   12 nu_mu         1      14    9   22.922    6.410    6.115   24.575    0.000
   13 n0            1    2112    1    0.612   -0.125 5009.478 5009.478    0.940
   14 (ubar)    A  12      -2    4    1.858    0.212  -15.901   16.014    0.330
   15 (g)       I  12      21    4   -0.881    0.443    0.500    1.106    0.000
   16 (d)       V  11       1    2    0.285    0.502-1095.905 1095.905    0.330
   17 (sbar)    A  12      -3    4    6.255    3.150 -145.886  146.054    0.500
   18 (g)       I  12      21    4   -0.621   -0.442   -9.154    9.186    0.000
   19 (g)       I  12      21    4   -1.436    0.954   -1.378    2.207    0.000
   20 (g)       I  12      21    3    0.721    2.086   15.208   15.367    0.000
   21 (u)       V  11       2    1   -0.088   -0.336 1827.499 1827.499    0.330
   22 (s)       A  12       3    4   -0.651   -0.457 -101.017  101.021    0.500
   23 (g)       I  12      21    4   -1.165    1.421  -36.632   36.678    0.000
   24 (g)       I  12      21    4   -1.146   -0.715  -18.439   18.488    0.000
   25 (g)       I  12      21    4    0.189   -0.306   -0.994    1.057    0.000
   26 (uu_1)    V  11    2203    2    0.149    0.102-5562.721 5562.721    0.771
 ==============================================================================
   27 (string)     11      92   14    1.262    1.157-1111.306 1113.025   61.825
   28 pbar-         1   -2212   27    1.512    0.037  -11.182   11.323    0.938
   29 n0            1    2112   27   -0.200    0.205   -2.283    2.485    0.940
   30 (pi0)        11     111   27   -0.080   -0.029   -1.056    1.068    0.135
   31 pi+           1     211   27    0.146    0.009   -1.183    1.200    0.140
   32 (K*-)        11    -323   27   -0.072    0.091   -1.373    1.620    0.852
   33 (Sigma*bar-) 11   -3224   27    0.007    0.183   -0.680    1.553    1.384
   34 (eta)        11     221   27   -0.145    0.187   -1.360    1.485    0.547
   35 (Delta++)    11    2224   27   -0.652    0.380  -26.507   26.547    1.233
   36 (pi0)        11     111   27    0.501   -0.255   -7.725    7.747    0.135
   37 (eta)        11     221   27   -0.640   -0.258  -43.671   43.680    0.547
   38 pbar-         1   -2212   27    0.872    0.582  -50.855   50.875    0.938
   39 pi-           1    -211   27    0.019   -0.425  -24.915   24.919    0.140
   40 (Delta+)     11    2214   27   -0.079   -0.222 -124.802  124.808    1.287
   41 (eta)        11     221   27   -0.090   -0.070 -228.547  228.548    0.547
   42 (eta')       11     331   27    0.162    0.741 -585.166  585.167    0.958
   43 (string)     11      92   17    4.831    5.411 1686.290 2000.313 1075.931
   44 (K0)         11     311   43    1.405    0.574  -31.106   31.147    0.498
   45 pi+           1     211   43    3.304    2.070  -95.006   95.086    0.140
   46 (rho-)       11    -213   43    1.103    0.392  -20.498   20.543    0.672
   47 (pi0)        11     111   43    0.057   -0.097   -0.566    0.593    0.135
   48 (K*bar0)     11    -313   43    0.055   -0.031   -4.070    4.170    0.902
   49 (K*+)        11     323   43   -1.041    0.270   -4.085    4.325    0.930
   50 pi-           1    -211   43    0.280    0.659   -0.408    0.836    0.140
   51 (omega)      11     223   43   -0.670    0.038   -0.036    1.032    0.783
   52 (rho+)       11     213   43    0.102    0.550    5.466    5.530    0.623
   53 pbar-         1   -2212   43   -0.106    0.362    2.722    2.904    0.938
   54 p+            1    2212   43    0.158   -0.143    6.122    6.197    0.938
   55 (pi0)        11     111   43   -0.156   -0.089    3.386    3.393    0.135
   56 (rho0)       11     113   43   -0.054    0.502    3.075    3.205    0.747
   57 (omega)      11     223   43    0.571   -0.169   12.120   12.158    0.766
   58 (pi0)        11     111   43    0.254    0.584   16.962   16.974    0.135
   59 (rho0)       11     113   43    0.125    0.010   82.982   82.986    0.796
   60 pbar-         1   -2212   43   -0.852    0.086  544.959  544.961    0.938
   61 pi-           1    -211   43    0.370   -0.202   43.855   43.858    0.140
   62 (Delta++)    11    2224   43   -0.075    0.044 1120.417 1120.417    1.172
   63 (string)     11      92   22   -2.623    0.046-5719.802 5719.965   43.080
   64 (Sigma*0)    11    3214   63   -0.973    0.427  -81.646   81.665    1.393
   65 pi+           1     211   63    0.002    0.040   -1.701    1.707    0.140
   66 pbar-         1   -2212   63   -1.037   -0.226  -30.461   30.494    0.938
   67 (K0)         11     311   63    0.026   -0.234  -25.968   25.974    0.498
   68 (phi)        11     333   63   -0.227    0.399  -13.698   13.744    1.023
   69 (phi)        11     333   63   -0.725    0.035  -18.106   18.149    1.019
   70 (eta)        11     221   63    0.478   -0.677  -76.891   76.898    0.547
   71 K-            1    -321   63   -0.332    0.361 -119.187  119.189    0.494
   72 (eta')       11     331   63    0.028   -0.076 -324.691  324.692    0.958
   73 (Delta+)     11    2214   63   -0.042   -0.024-4353.561 4353.561    1.255
   74 pi+           1     211   63    0.180    0.020 -673.892  673.892    0.140
   75 gamma         1      22   30   -0.016    0.026   -0.050    0.059    0.000
   76 gamma         1      22   30   -0.063   -0.054   -1.005    1.009    0.000
   77 K-            1    -321   32    0.144   -0.102   -0.695    0.871    0.494
   78 (pi0)        11     111   32   -0.216    0.194   -0.677    0.749    0.135
   79 (Lambdabar0) 11   -3122   33   -0.107    0.325   -0.576    1.301    1.116
   80 pi-           1    -211   33    0.114   -0.142   -0.104    0.252    0.140
   81 (pi0)        11     111   34   -0.027   -0.021   -0.422    0.444    0.135
   82 (pi0)        11     111   34   -0.095   -0.013   -0.208    0.266    0.135
   83 (pi0)        11     111   34   -0.023    0.221   -0.730    0.775    0.135
   84 p+            1    2212   35   -0.728    0.292  -24.709   24.740    0.938
   85 pi+           1     211   35    0.076    0.088   -1.798    1.807    0.140
   86 gamma         1      22   36    0.014   -0.035   -0.288    0.290    0.000
   87 gamma         1      22   36    0.487   -0.220   -7.437    7.457    0.000
   88 (pi0)        11     111   37   -0.072    0.017   -4.828    4.830    0.135
   89 (pi0)        11     111   37   -0.226   -0.159  -13.516   13.520    0.135
   90 (pi0)        11     111   37   -0.342   -0.117  -25.327   25.330    0.135
   91 p+            1    2212   40   -0.286   -0.151 -111.230  111.235    0.938
   92 (pi0)        11     111   40    0.207   -0.071  -13.571   13.574    0.135
   93 (pi0)        11     111   41   -0.113   -0.013  -43.560   43.560    0.135
   94 (pi0)        11     111   41   -0.082    0.017  -94.359   94.359    0.135
   95 (pi0)        11     111   41    0.105   -0.074  -90.628   90.629    0.135
   96 pi+           1     211   42   -0.059    0.227  -99.601   99.602    0.140
   97 pi-           1    -211   42    0.179    0.111 -118.078  118.078    0.140
   98 (eta)        11     221   42    0.043    0.403 -367.487  367.488    0.547
   99 K_L0          1     130   44    1.405    0.574  -31.106   31.147    0.498
  100 pi-           1    -211   46    0.893    0.545  -15.636   15.672    0.140
  101 (pi0)        11     111   46    0.210   -0.153   -4.862    4.871    0.135
  102 gamma         1      22   47   -0.016    0.018   -0.027    0.037    0.000
  103 gamma         1      22   47    0.074   -0.115   -0.539    0.556    0.000
  104 (Kbar0)      11    -311   48   -0.240   -0.103   -2.248    2.317    0.498
  105 (pi0)        11     111   48    0.295    0.072   -1.823    1.853    0.135
  106 K+            1     321   49   -0.294    0.117   -1.161    1.301    0.494
  107 (pi0)        11     111   49   -0.747    0.153   -2.924    3.025    0.135
  108 pi-           1    -211   51   -0.155   -0.163   -0.109    0.286    0.140
  109 pi+           1     211   51   -0.448   -0.020    0.006    0.470    0.140
  110 (pi0)        11     111   51   -0.067    0.220    0.067    0.275    0.135
  111 pi+           1     211   52   -0.135   -0.024    1.032    1.050    0.140
  112 (pi0)        11     111   52    0.237    0.574    4.434    4.479    0.135
  113 gamma         1      22   55   -0.166   -0.084    3.357    3.362    0.000
  114 gamma         1      22   55    0.010   -0.005    0.028    0.031    0.000
  115 pi+           1     211   56   -0.192    0.081    2.198    2.212    0.140
  116 pi-           1    -211   56    0.137    0.421    0.878    0.993    0.140
  117 pi+           1     211   57   -0.085   -0.066    1.508    1.519    0.140
  118 pi-           1    -211   57    0.333    0.078    6.207    6.218    0.140
  119 (pi0)        11     111   57    0.323   -0.181    4.405    4.422    0.135
  120 gamma         1      22   58    0.046    0.033    2.242    2.243    0.000
  121 gamma         1      22   58    0.208    0.551   14.720   14.732    0.000
  122 pi+           1     211   59   -0.118   -0.322   36.308   36.310    0.140
  123 pi-           1    -211   59    0.243    0.333   46.674   46.676    0.140
  124 p+            1    2212   62    0.085    0.068  828.515  828.516    0.938
  125 pi+           1     211   62   -0.160   -0.023  291.902  291.902    0.140
  126 (Lambda0)    11    3122   64   -0.739    0.123  -61.419   61.434    1.116
  127 (pi0)        11     111   64   -0.234    0.304  -20.227   20.231    0.135
  128 (K_S0)       11     310   67    0.026   -0.234  -25.968   25.974    0.498
  129 K_L0          1     130   68   -0.116    0.221   -8.379    8.397    0.498
  130 (K_S0)       11     310   68   -0.111    0.178   -5.319    5.346    0.498
  131 K-            1    -321   69   -0.226   -0.026   -7.284    7.305    0.494
  132 K+            1     321   69   -0.500    0.061  -10.821   10.844    0.494
  133 (pi0)        11     111   70    0.080   -0.021  -16.763   16.764    0.135
  134 (pi0)        11     111   70    0.127   -0.169  -19.715   19.717    0.135
  135 (pi0)        11     111   70    0.270   -0.488  -40.412   40.417    0.135
  136 pi+           1     211   72    0.058   -0.015  -25.415   25.416    0.140
  137 pi-           1    -211   72    0.046    0.109  -69.547   69.547    0.140
  138 (eta)        11     221   72   -0.077   -0.169 -229.728  229.729    0.547
  139 p+            1    2212   73    0.154    0.148-3338.310 3338.310    0.938
  140 (pi0)        11     111   73   -0.197   -0.171-1015.251 1015.251    0.135
  141 gamma         1      22   78   -0.045    0.117   -0.198    0.235    0.000
  142 gamma         1      22   78   -0.171    0.076   -0.479    0.514    0.000
  143 nbar0         1   -2112   79   -0.145    0.351   -0.561    1.158    0.940
  144 (pi0)        11     111   79    0.038   -0.026   -0.015    0.143    0.135
  145 gamma         1      22   81   -0.025   -0.060   -0.386    0.392    0.000
  146 gamma         1      22   81   -0.002    0.039   -0.035    0.052    0.000
  147 gamma         1      22   82   -0.054   -0.071   -0.145    0.170    0.000
  148 gamma         1      22   82   -0.041    0.059   -0.063    0.096    0.000
  149 gamma         1      22   83   -0.069    0.149   -0.368    0.403    0.000
  150 gamma         1      22   83    0.046    0.072   -0.362    0.372    0.000
  151 gamma         1      22   88    0.003    0.064   -2.343    2.344    0.000
  152 gamma         1      22   88   -0.075   -0.047   -2.485    2.486    0.000
  153 gamma         1      22   89   -0.100   -0.104   -9.627    9.628    0.000
  154 gamma         1      22   89   -0.125   -0.055   -3.889    3.892    0.000
  155 gamma         1      22   90   -0.035   -0.055   -2.797    2.798    0.000
  156 gamma         1      22   90   -0.307   -0.061  -22.530   22.532    0.000
  157 gamma         1      22   92    0.145   -0.059  -12.182   12.183    0.000
  158 gamma         1      22   92    0.062   -0.012   -1.390    1.391    0.000
  159 gamma         1      22   93    0.008   -0.042   -5.984    5.985    0.000
  160 gamma         1      22   93   -0.121    0.029  -37.575   37.576    0.000
  161 gamma         1      22   94   -0.117    0.030  -75.044   75.044    0.000
  162 gamma         1      22   94    0.035   -0.013  -19.315   19.315    0.000
  163 gamma         1      22   95    0.007   -0.034   -4.705    4.705    0.000
  164 gamma         1      22   95    0.098   -0.040  -85.923   85.923    0.000
  165 gamma         1      22   98   -0.038    0.501 -336.480  336.481    0.000
  166 gamma         1      22   98    0.080   -0.097  -31.007   31.007    0.000
  167 gamma         1      22  101    0.022   -0.008   -1.605    1.605    0.000
  168 gamma         1      22  101    0.187   -0.145   -3.257    3.266    0.000
  169 K_L0          1     130  104   -0.240   -0.103   -2.248    2.317    0.498
  170 gamma         1      22  105   -0.001    0.023   -0.292    0.293    0.000
  171 gamma         1      22  105    0.296    0.049   -1.531    1.560    0.000
  172 gamma         1      22  107   -0.045    0.045   -0.209    0.218    0.000
  173 gamma         1      22  107   -0.702    0.109   -2.715    2.806    0.000
  174 gamma         1      22  110   -0.019   -0.015   -0.006    0.025    0.000
  175 gamma         1      22  110   -0.048    0.235    0.073    0.251    0.000
  176 gamma         1      22  112    0.088    0.092    1.106    1.113    0.000
  177 gamma         1      22  112    0.149    0.482    3.328    3.366    0.000
  178 gamma         1      22  119    0.278   -0.171    4.193    4.205    0.000
  179 gamma         1      22  119    0.045   -0.009    0.212    0.217    0.000
  180 n0            1    2112  126   -0.554    0.177  -48.857   48.870    0.940
  181 (pi0)        11     111  126   -0.185   -0.054  -12.562   12.564    0.135
  182 gamma         1      22  127   -0.081    0.191   -8.389    8.392    0.000
  183 gamma         1      22  127   -0.153    0.114  -11.838   11.839    0.000
  184 (pi0)        11     111  128   -0.086    0.005   -3.284    3.288    0.135
  185 (pi0)        11     111  128    0.112   -0.239  -22.684   22.686    0.135
  186 (pi0)        11     111  130    0.149    0.008   -1.913    1.923    0.135
  187 (pi0)        11     111  130   -0.260    0.171   -3.406    3.423    0.135
  188 gamma         1      22  133    0.049   -0.066   -5.428    5.429    0.000
  189 gamma         1      22  133    0.031    0.045  -11.335   11.335    0.000
  190 gamma         1      22  134    0.118   -0.192  -17.526   17.528    0.000
  191 gamma         1      22  134    0.009    0.023   -2.189    2.189    0.000
  192 gamma         1      22  135    0.137   -0.370  -29.014   29.017    0.000
  193 gamma         1      22  135    0.134   -0.118  -11.398   11.400    0.000
  194 gamma         1      22  138    0.000   -0.042 -214.479  214.479    0.000
  195 gamma         1      22  138   -0.077   -0.127  -15.250   15.250    0.000
  196 gamma         1      22  140   -0.142   -0.183 -725.604  725.604    0.000
  197 gamma         1      22  140   -0.054    0.012 -289.647  289.647    0.000
  198 gamma         1      22  144    0.079    0.006   -0.036    0.087    0.000
  199 gamma         1      22  144   -0.042   -0.032    0.021    0.056    0.000
  200 gamma         1      22  181   -0.137    0.004  -10.352   10.353    0.000
  201 gamma         1      22  181   -0.048   -0.059   -2.211    2.212    0.000
  202 gamma         1      22  184   -0.004   -0.046   -1.886    1.887    0.000
  203 gamma         1      22  184   -0.082    0.051   -1.397    1.401    0.000
  204 gamma         1      22  185    0.019    0.010   -0.734    0.734    0.000
  205 gamma         1      22  185    0.092   -0.249  -21.950   21.952    0.000
  206 gamma         1      22  186    0.078    0.070   -1.148    1.153    0.000
  207 gamma         1      22  186    0.071   -0.062   -0.765    0.771    0.000
  208 gamma         1      22  187   -0.067    0.060   -1.702    1.705    0.000
  209 gamma         1      22  187   -0.193    0.111   -1.704    1.718    0.000
                   sum:  2.00          0.00     0.00     0.00 14000.00 14000.00
\end{verbatim}
\end{multicols}
\end{SizeSPE}

\end{subappendices}
\cleardoublepage
\ClearShipoutPicture

\chapter{Phenomenology of $\BFWp$ and $\BFWm$ in Drell--Yan like processes at the LHC}
\label{chap_w_pheno_in_drell-yan}
\setlength{\epigraphwidth}{0.7\tw}
\epigraph{
Super Skrull Warrior\,:\;%
``I have trained my entire life to face you.'' \\
Black Panther\,:\;%
``Then you  have already lost. For I have trained my entire life to face the unknown.''
}%
{\textit{Black Panther \#39 - See Wakanda and Die (September 2008)} }

This Chapter presents the phenomenology of the $\Wp$ and $\Wm$~bosons production in Drell--Yan 
processes at the LHC. 
This preliminary work is mandatory before addressing any kind of study related to the extraction 
of the $W$ boson properties. 
Indeed, the LHC in top of supplying unprecedented luminosity and energy in collisions 
will provide --due to the nature of the colliding beams-- original kinematics with respect to the 
one inherent to $\ppbar$ collisions studied exhaustively for these last decades at the SPS 
\index{SPS collider} and Tevatron colliders.
These original LHC features demand to start with basics understanding before addressing a more
complete analysis prospect.

The Chapter is divided as follow. In a short introduction the context of the studies is described
from both physics and technical point of view.
Then comes the phenomenological understanding of the $\Wp$ and $\Wm$~bosons production in Drell--Yan, 
first by looking at the production of $W$ and then at the whole process by studying the properties
of the decaying charged leptons.

Let us stress that before reaching the refined understanding presented here more thorough analysis 
were done using complementary distributions, different physical input parameters,
for specific domains of the phase-space, \etc{}.
Nonetheless, to keep the discussion as clear as possible in the core of the Chapter all these 
exhaustive studies are compiled in Appendix~\ref{app_ppbar_pp_dd}.

\section{Context of these studies}
\label{s_W_prod_decay_study}
The most relevant kinematics to the $W$ in Drell--Yan are reviewed at the generator level, that is
looking only at the purely phenomenological level.

Among all the observables/pseudo-observables that entered in our study only a few were kept in the
core of the Chapter. They are of two categories. 
First are the pseudo-observables characterising the $W$ boson properties, its rapidity $\yW$ and 
transverse momentum $\pTW$. Then are the observables characterising the charged lepton properties, 
its pseudo-rapidity $\etal$ and transverse momentum $\pTl$ that allows to study the whole process.
The definitions of these quantities and the information they hold for $W$ in Drell--Yan have been 
over-viewed in Chapter~\ref{chap_theo} \S\,\ref{s_drell-yan}.\ref{ss_hadr_lvl_iLO} for the case 
of a charge blind study. Here, based on this previous overview a fine study of the positive and 
negative channel is considered.

The type of collisions considered are\,:\;$\ppbar$, $\pp$ and $\dd$.
The $\ppbar$ collision scheme is considered as a reference since in that case the production of 
$\Wp$ and $\Wm$ are on the same footing and the characteristics of the kinematics are well known
from the experience gathered from SPS \index{SPS collider} and Tevatron colliders.
Standard LHC $\pp$ collisions are analysed as well with isoscalar beams collisions on the example
of deuteron--deuteron ($\dd$) targets which is justified for both pedagogical means and because the 
strategy prospect in the next Chapter relies on such a scheme.
The nucleon--nucleon center of mass energy $\sqrt S$ is of $14\TeV$ for $\ppbar$ and $\pp$ collisions
and of $7\TeV$ for $\dd$ collisions according to the expected LHC capabilities (cf. 
Eqs.~(\ref{eq_pp_Ecm_lumi}--\ref{eq_dd_Ecm_lumi})). The choice of $14\TeV$ in the center of mass 
energy for $\ppbar$ is justified by the wish to put the energy scales and accessible phase space to 
the particles on the same footing as the one for $\pp$ collisions to stress the differences between 
these two collision schemes.
The total hadronic cross sections corresponding to these collisions and computed by \WINHAC{}  
are gathered in Table~\ref{table_xtot}.\index{W boson@$W$ boson!Hadronic cross sections in Drell--Yan}
\begin{table}[]
\begin{center}
\renewcommand\arraystretch{1.1}
\begin{tabular}{cccc}
  \hline
  Collider               & $\sigma_\mm{(incl.)}^+$ [nb] & $\sigma_\mm{(incl.)}^-$ [nb] 
                                                        & $\sigma_\mm{(incl.)}^\pm$ [nb]\\
  \hline\hline
  $\ppbar$               & $17.4$        & $17.4$  & $34.4$       \\
  $\pp$                  & $19.8$        & $14.7$  & $34.5$       \\
  $\dd$                  & $32.5$        & $32.4$  & $64.9$       \\
  \hline
\end{tabular}
\renewcommand\arraystretch{1.45}
\caption[Inclusive hadronic cross sections $\sigma_\mm{(incl.)}^+$, $\sigma_\mm{(incl.)}^-$ and 
            $\sigma_\mm{(incl.)}^\pm$ respectively for the $\Wp$, $\Wm$ and $W$ in Drell--Yan 
            with $\sqrt S=14\TeV$ for $\ppbar$ and $\pp$ collisions 
            and $\sqrt S_{n_1n_2}=7\TeV$ in the $n_1\,n_2$ nucleon--nucleon 
            center of mass energy for $\dd$ collisions (computed by \WINHAC{}).]
        {\figtxt{Inclusive hadronic cross sections $\sigma_\mm{(incl.)}^+$, $\sigma_\mm{(incl.)}^-$ and 
            $\sigma_\mm{(incl.)}^\pm$ respectively for the $\Wp$, $\Wm$ and $W$ in Drell--Yan 
            with $\sqrt S=14\TeV$ for $\ppbar$ and $\pp$ collisions 
            and $\sqrt S_{n_1n_2}=7\TeV$ in the $n_1\,n_2$ nucleon--nucleon 
            center of mass energy for $\dd$ collisions (computed by \WINHAC{}).}}
\label{table_xtot}
\index{W boson@$W$ boson!Hadronic cross sections in Drell--Yan}
\end{center}
\end{table}

Now details are given on the material presented below from a more technical point of view.
All the histograms were produced using the \WINHAC{} Monte Carlo event generator at the truth level
within improved leading order defined from the physics point of view in Chapter~\ref{chap_theo}
\S\,\ref{s_drell-yan}.\ref{ss_hadr_lvl_iLO} and from a Monte Carlo point of view in 
Chapter~\ref{chap_winhac} \S\,\ref{ss_tools_WINHAC}.
A common statistic of $200$ millions weighted events were used for each $\Wp$ and $\Wm$ channel and 
for all type of collisions. The decay of the $W$~bosons was opened to both electronic 
and muonic channels. Every histograms were generated with 200 bins and for visual convenience this
former binning was reduced later on at the analysis level when necessarily.
The remaining input parameters of importance for the generation are\,:
\begin{itemize}
\baselineskip 1pt
\item[-] $\MW = 80.403\GeV$, $\GamW = 2.141\GeV$
with the fixed-width scheme in the $W$~boson propagator.
\item[-] The partons intrinsic $\kT$ are modeled by the \Pythia{} Gaussian scheme 
\index{Quarks!Intrinsic transverse momenta!By Pythia@Generated by \Pythia{}}
(cf. Ref.~\cite{Sjostrand:2006za}) with $\Mean{\kT}=4\GeV$.\index{Pythia@\Pythia{} Monte Carlo event generator}
\item[-] PDF set\,:\; CTEQ6.1M \cite{Pumplin:2002vw}\index{Parton Distribution Functions (PDFs)!CTEQ}.
\end{itemize}

Let us remind the conventions used for the coordinate systems and the definitions of the common 
kinematics are gathered in Chapter~\ref{chap_theo} \S\,\ref{notations_conventions}. 
In this context, $\ppbar$ collisions are defined in the Cartesian basis, with the proton moving in the
$+z$ direction while the anti-proton moves in the $-z$ direction. 
In what follows the discrepancies between the positive and negative channels predictions are 
scrutinised using the charge asymmetry (Eq.~(\ref{eq_def_charge_asym})).

\section{Production of $\BFWp$ and $\BFWm$ bosons}\label{ss_W_prod}
\begin{figure}[!h] 
  \begin{center}
    \includegraphics[width=0.495\tw]{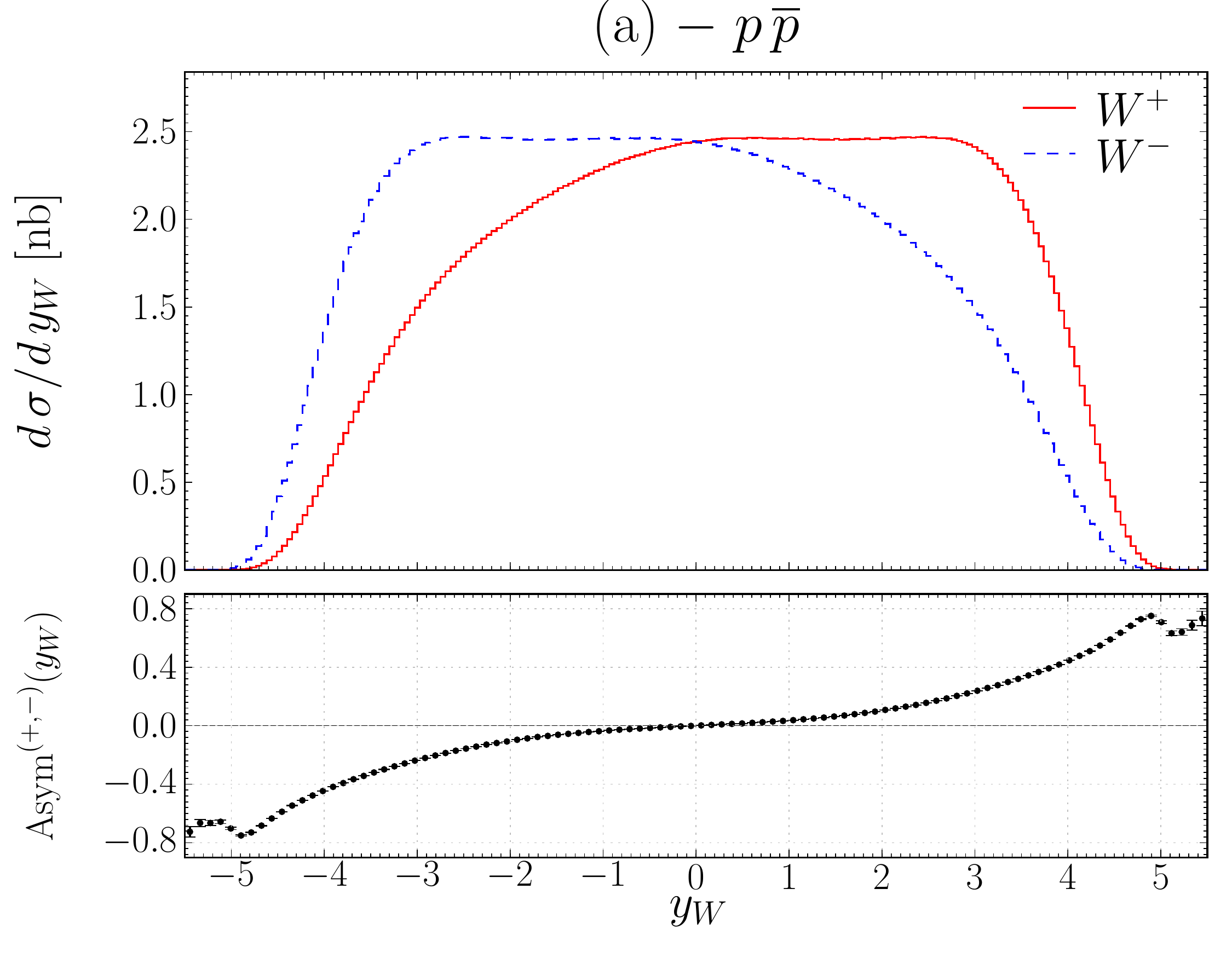}
    \hfill
    \includegraphics[width=0.495\tw]{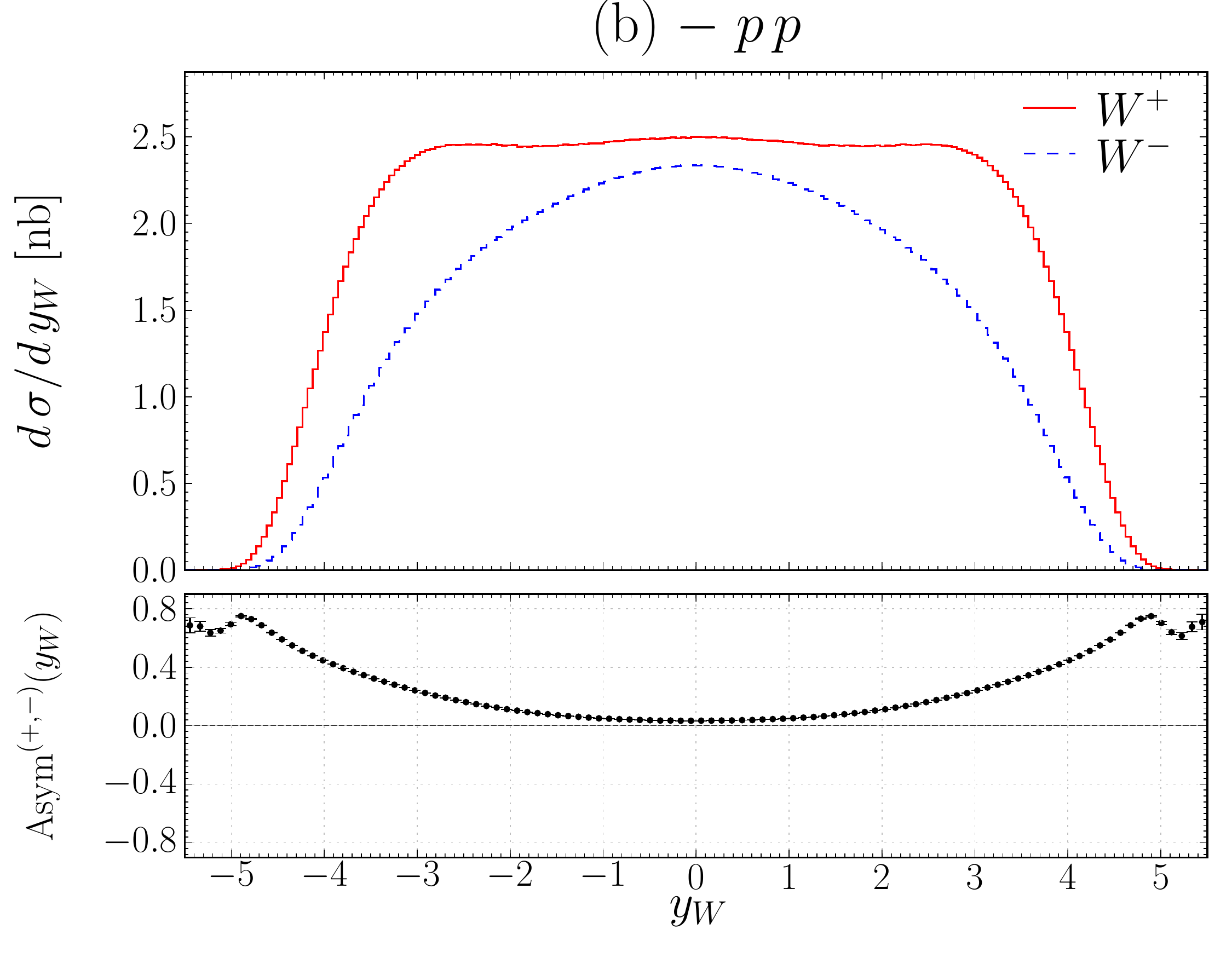}
    \vfill
    \includegraphics[width=0.495\tw]{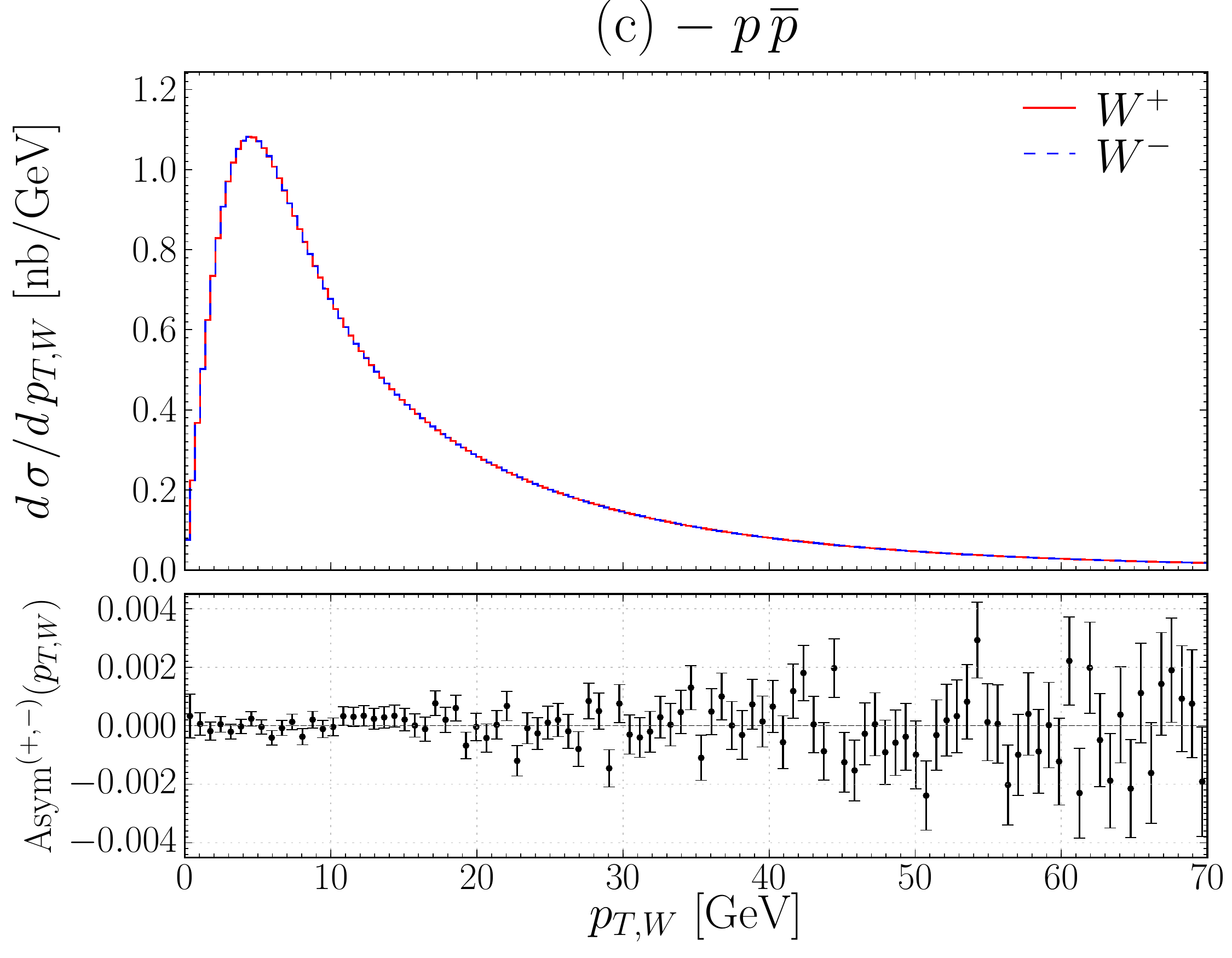}
    \hfill
    \includegraphics[width=0.495\tw]{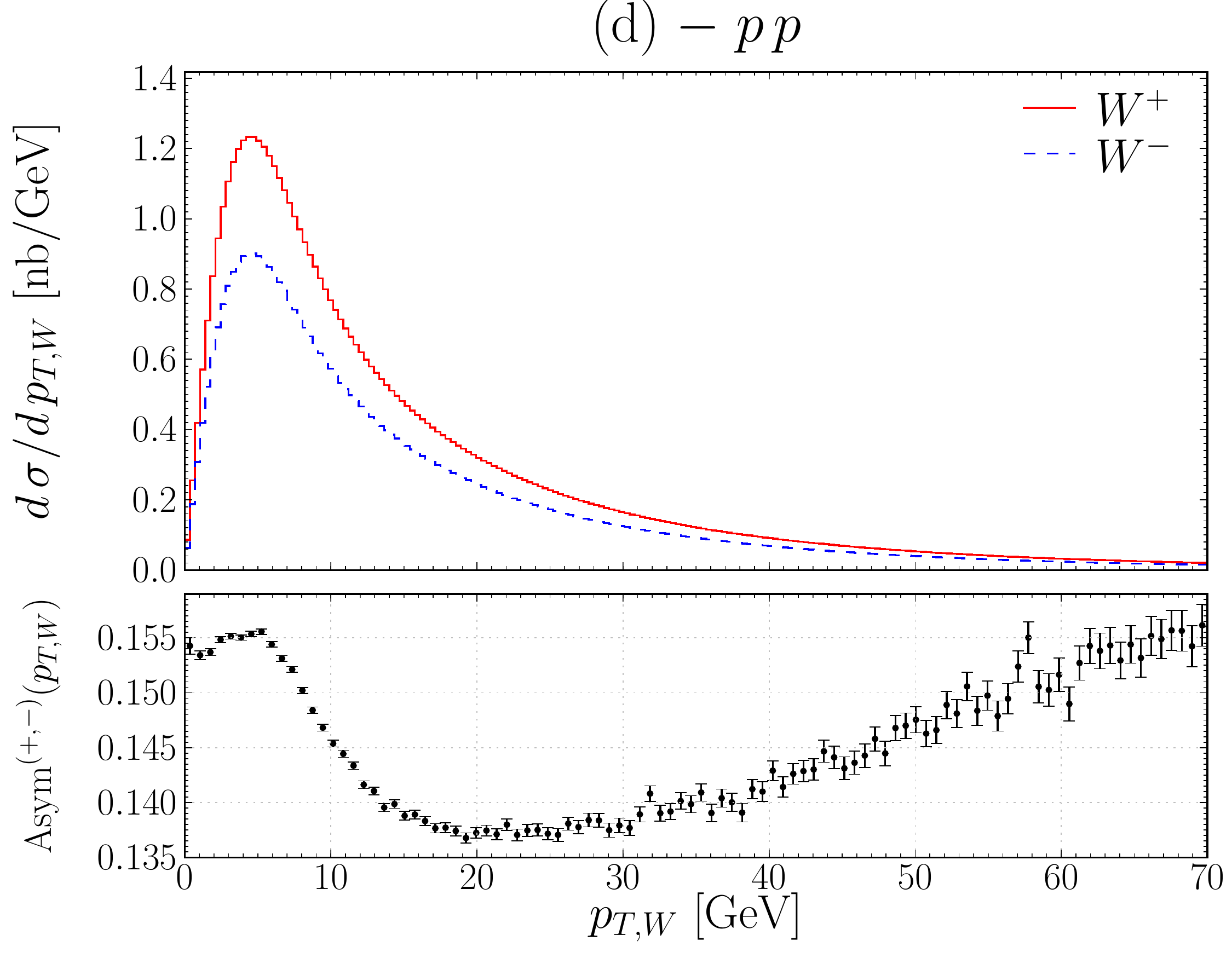}
    \caption[Rapidity and transverse momentum distributions of the $W$~bosons with their 
            charge asymmetries for $\ppbar$ and $\pp$ collisions]
            {\figtxt{The rapidity and transverse momentum distributions of the $W$~bosons with their
                associated charge asymmetries for $\ppbar$ (a, c) and $\pp$ (b, d) collisions.}}
            \label{fig_yW_pTW_ppb_pp}
  \end{center} 
\end{figure}

\index{W boson@$W$ boson!Transverse momentum|(}
\index{W boson@$W$ boson!Rapidity|(}
In this Section the $W$~boson production mechanism is discussed via the analysis of 
the $W$~boson rapidity $\yW$ and transverse momentum $\pTW$.
Figure~\ref{fig_yW_pTW_ppb_pp} presents the latter distributions for $\Wp$ and $\Wm$~bosons, 
as well as their associated charge asymmetries in $\ppbar$ and $\pp$ collisions.

\index{W boson@$W$ boson!Production in ppbar@Production in $\ppbar$ collisions|(}
Our observation starts with the well known $\ppbar$ kinematics.
Let us remind the $\Wp$ and $\Wm$ bosons are produced with the same dynamics due to the nature 
of the colliding beams. 
Indeed, the same amount of matter and antimatter are available to produce a $\Wp$ or a $\Wm$.
The most visible difference between the two processes occurs at the spatial level because
the proton and anti-protons are always impinging respectively from the $+z$ and $-z$ directions.
Thus the kinematics between the $\Wp$ and $\Wm$ event samples differ only from a mirror
reflection with respect to the $r-\phi$ plane passing by the origin.
More explicitly, to each phase space configuration of amplitude of probability $\mathcal M$ 
for the $\Wp$~boson 
corresponds another phase space configuration for the $\Wm$ in which $\theta_\Wm=\pi-\theta_\Wp$ and
that has the exact same amplitude.
The evidence for this behaviour can be seen already for the $W$ bosons longitudinal component when looking 
at the rapidity (Fig.~\ref{fig_yW_pTW_ppb_pp}.(a)) where one can see the $\Wp$ and $\Wm$ 
distributions are the same up to a vertical flip with respect to the $\yW=0$ origin.
Concerning the $\Wp$ and $\Wm$ transverse momenta the symmetry is such that it makes no differences
between the two channels when projecting on the $r-\phi$ plane as shown in 
Fig.~\ref{fig_yW_pTW_ppb_pp}.(c).
From the point of view of a $\MW$ extraction the important thing is that $\pTW$ distributions are independent
of the $W$~boson charge which means the leptons, as it will be confirmed, undergo the same boosts.
\index{W boson@$W$ boson!Production in ppbar@Production in $\ppbar$ collisions|)}

\index{W boson@$W$ boson!Production in pp@Production in $\pp$ collisions|(}
Now Figures~\ref{fig_yW_pTW_ppb_pp}.(b) and (d) show the corresponding distributions for $\pp$ 
collisions.
Here both rapidity distributions are symmetric with respect to $\yW=0$ which is the consequence of
the identical nature of the colliding targets.
The shapes of the spectra are different for the $\Wp$ and $\Wm$ bosons which reflects the
difference in the valence $u$ and $d$ quark content and properties in a proton. 
The $\yW$ distribution for the $\Wp$ boson is higher and wider with respect to the one for the $\Wm$ 
boson. 
First let us note there are twice as many valence $u$ quarks as valence $d$ quarks explaining
the higher differential cross section for $\Wp$.
Besides, $u$ quarks carry, in average, a higher fraction $x$ of the parent proton momentum
(cf.~Fig.~\ref{fig_cteq61m}), which, assuming $\yW$ respects to a good approximation the LO
expression of Eq.~(\ref{eq_yW_LO}), implies that $\Wp$ are more likely to have a higher absolute 
values rapidity. 
This explains the wider behaviour of $\flatDfDx{\sigma^+}{\yW}$ compared to
$\flatDfDx{\sigma^-}{\yW}$.

The charge asymmetry in the $\pTW$ distribution reflects both the differences in the relative 
cross sections but also, what will be crucial for the studies presented in this Chapter, in the 
shape of their distributions.  
The nontrivial shape of $\Asym{\pTW}$ is due to the flavour 
asymmetries in the distributions of quarks producing  $\Wp$ and $\Wm$ bosons.
The latter are predominantly driven by the $u$--$d$ quark asymmetries. 
\index{Electroweak!CKM matrix elements|(}
Then the CKM mixing involves
the other $s$, $c$ and $b$ flavours among which --as it will be shown-- the non equality of the $s$ 
and $c$ quark masses add up to the main $u$--$d$ charge asymmetries.
This can be analysed with the flavour structure of the $W$ bosons charge asymmetries by writing 
explicitly the simplified Born level formul\ae of the total cross section charge asymmetries for 
$\ppbar$ and $\pp$ collisions
\begin{eqnarray}
  \left(\sigma^{\plus} - \sigma^{\minus}\right)_{\ppbar}(S) 
  &=& 0, \label{eq_sWp_sWm_ppar}\\
  \left(\sigma^{\plus} - \sigma^{\minus}\right)_{\pp}(S) 
  &\propto& \iint dx_q\, dx_{\bar{q}}\, \Big\{
  \Vckmsqr{u}{d}\left[ u^{\val}(x_q)\,\bar d(x_{\bar{q}}) - d^{\val}(x_q)\,\bar u(x_{\bar{q}}) \right] 
  \nonumber\\
  & & \hspace{20mm}
  +\, u^{\val}(x_q) \left[ \Vckmsqr{u}{s}\,\bar s(x_{\bar{q}}) + \Vckmsqr{u}{b}\,\bar b(x_{\bar{q}}) 
  \right]
  \nonumber\\
  & & \hspace{20mm}
  - \,\Vckmsqr{c}{d}\,d^{\val}(x_q)\,\bar c(x_{\bar{q}})\Big\}\,
   \tilde\sigma_{q \bar{q}}(\hat{s}), 
  \label{eq_sWp_sWm_pp}
\index{Quarks!Valence quarks}
\index{W boson@$W$ boson!Hadronic cross sections in Drell--Yan!LO expressions}
\end{eqnarray}
where in these expressions $d,u,s,c$ and $b$ on the right hand side denote the PDFs of the 
corresponding quark flavours and the $\val$ superscript stands for valence quarks, 
$V_{ij}$ is the CKM matrix element for
the $i$ and $j$ flavours, while $\tilde\sigma_{q \bar{q}}(\hat{s})$ is the 
``CKM matrix element stripped'' partonic cross section for the $W$~boson production with 
$\hat{s} =  x_qx_{\bar{q}}\,S$.
Let us stress that the above expressions are over simplified for pedagogical reasons and do not
reflect the physics implemented in the \WINHAC{} event generation.
We have omitted the explicit dependence of the PDFs on the factorisation scheme, on the transverse 
momenta $\kT$ of annihilating partons present both in the ``$\kT$-non integrated'' PDFs, and in  
the partonic cross sections (via $\kT$ dependence of $\hat{s}$). 
All the above effects are present in \WINHAC{}.
In our analysis partons have both the primordial transverse momenta and the perturbative generated
ones as modeled by the initial state parton shower of the \Pythia{} generator.
\index{Pythia@\Pythia{} Monte Carlo event generator}
Their transverse momenta depend on the Bjorken $x$ of the annihilating (anti-)quark and, for 
heavy quarks (here $c$ and $b$), also on their masses (see Ref.~\cite{Sjostrand:2006za} for more 
details).

\index{W boson@$W$ boson!Mass charge asym@Mass charge asymmetry $\MWp-\MWm$|(}
As one can see in Eq.~(\ref{eq_sWp_sWm_ppar}), the charge asymmetry disappears for the $\ppbar$ 
collision mode if, as assumed in the presented studies $\MWp=\MWm$. 
Let us note that this equality holds no matter the level of the corrections embraced in the calculus.
Therefore, this collision scheme would be, on a theoretical level, the optimal one for measuring 
$\MWp-\MWm$. 
Any deviation from the equality of the masses would result in non zero asymmetries regardless
of the level of understanding of the flavour and momentum structure of the beam particles.
For $\pp$ collisions several effects, reflecting the present understanding of the partonic content 
of the beam particles --in particular, the understanding of the momentum distribution of valence 
quarks-- contribute to the charge asymmetry of the $\pTW$ distribution and may mimic $\MWp\neq\MWm$ 
effects.
\index{W boson@$W$ boson!Mass charge asym@Mass charge asymmetry $\MWp-\MWm$|)}

The asymmetries in $\pp$ collisions can be really reduced when going to isoscalar collisions. 
For example in $\dd$ collisions the asymmetries in the quark flavour can be deduced easily from 
Eq.~(\ref{eq_sWp_sWm_pp}) by considering here $u(x)=d(x)$. We obtain then
\begin{equation}
  \left(\sigma^{\plus} - \sigma^{\minus}\right)_{\dd}(S)
  \propto \iint dx_q\, dx_{\bar{q}}\;
  u^\val(x_q) \Big[ \Vckmsqr{u}{s}\,\bar s(x_{\bar{q}}) - \Vckmsqr{c}{d}\,\bar c(x_{\bar{q}})
    +\; \Vckmsqr{u}{b}\,\bar b(x_{\bar{q}}) \Big]\,
  \tilde\sigma_{q \bar{q}}(\hat{s}),
  \label{eq_sWp_sWm_dd}
\index{W boson@$W$ boson!Hadronic cross sections in Drell--Yan!LO expressions}
\end{equation}
where here the PDFs have to be understood as the parton distributions functions inside a deuteron
contrary to Eqs.~(\ref{eq_sWp_sWm_ppar}--\ref{eq_sWp_sWm_pp}) where they were the one related to the 
proton. This was omitted on purpose not to overload the expressions.
Here then, the asymmetry is driven by the Cabibbo suppressed difference of the distribution of the 
strange and charm quarks, weighted by the distributions of the valence quarks. 
The $\dd$ collision scheme is introduced at this point in order to analyse the relative importance 
of the valence quark and ``$s-c$'' effects.\index{Quarks!smc@$s-c$ asymmetry}

\begin{figure}[!ht] 
  \begin{center}
    \includegraphics[width=0.495\tw]{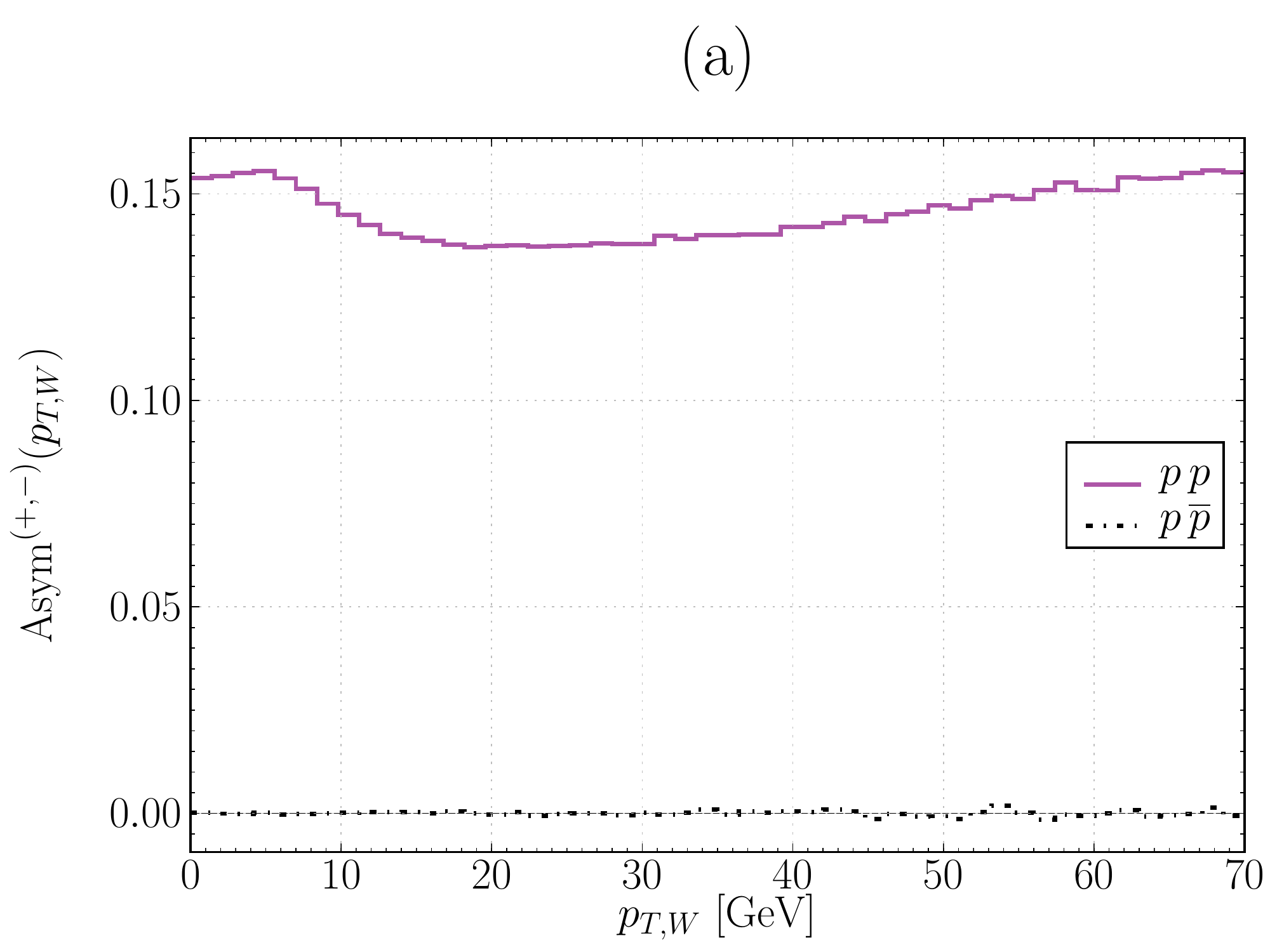}
    \hfill
    \includegraphics[width=0.495\tw]{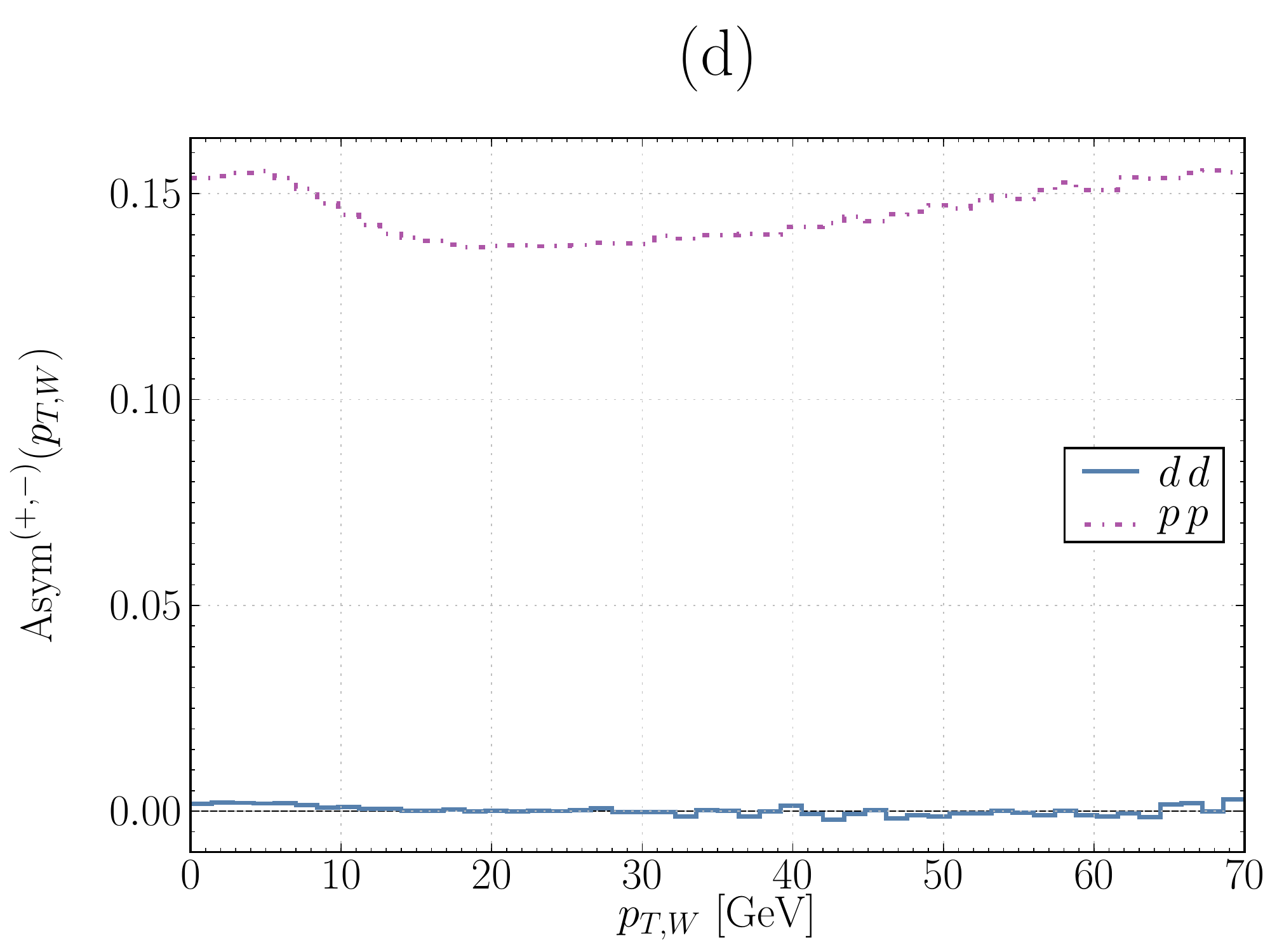}
    \vfill
    \includegraphics[width=0.495\tw]{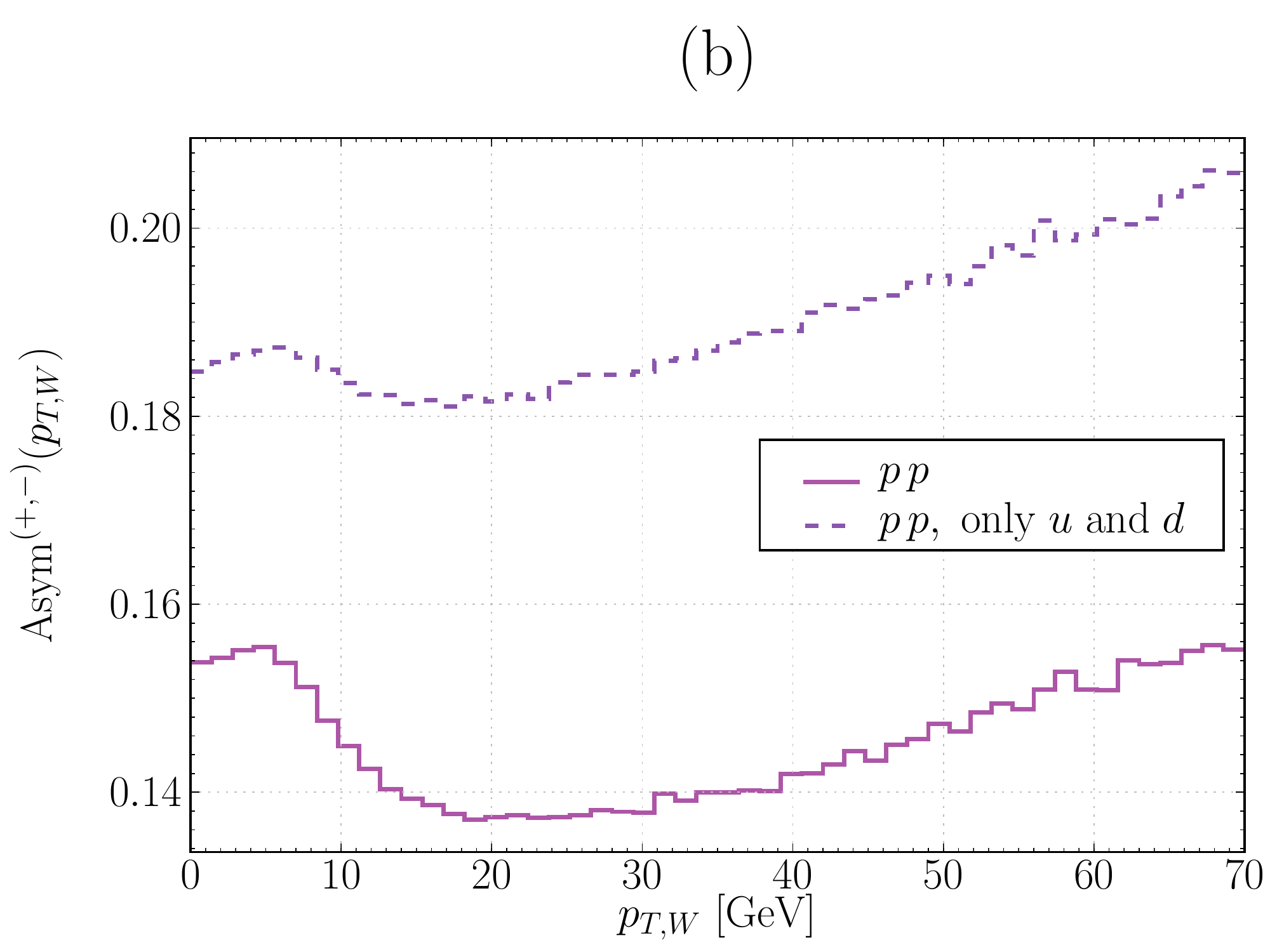}
    \hfill
    \includegraphics[width=0.495\tw]{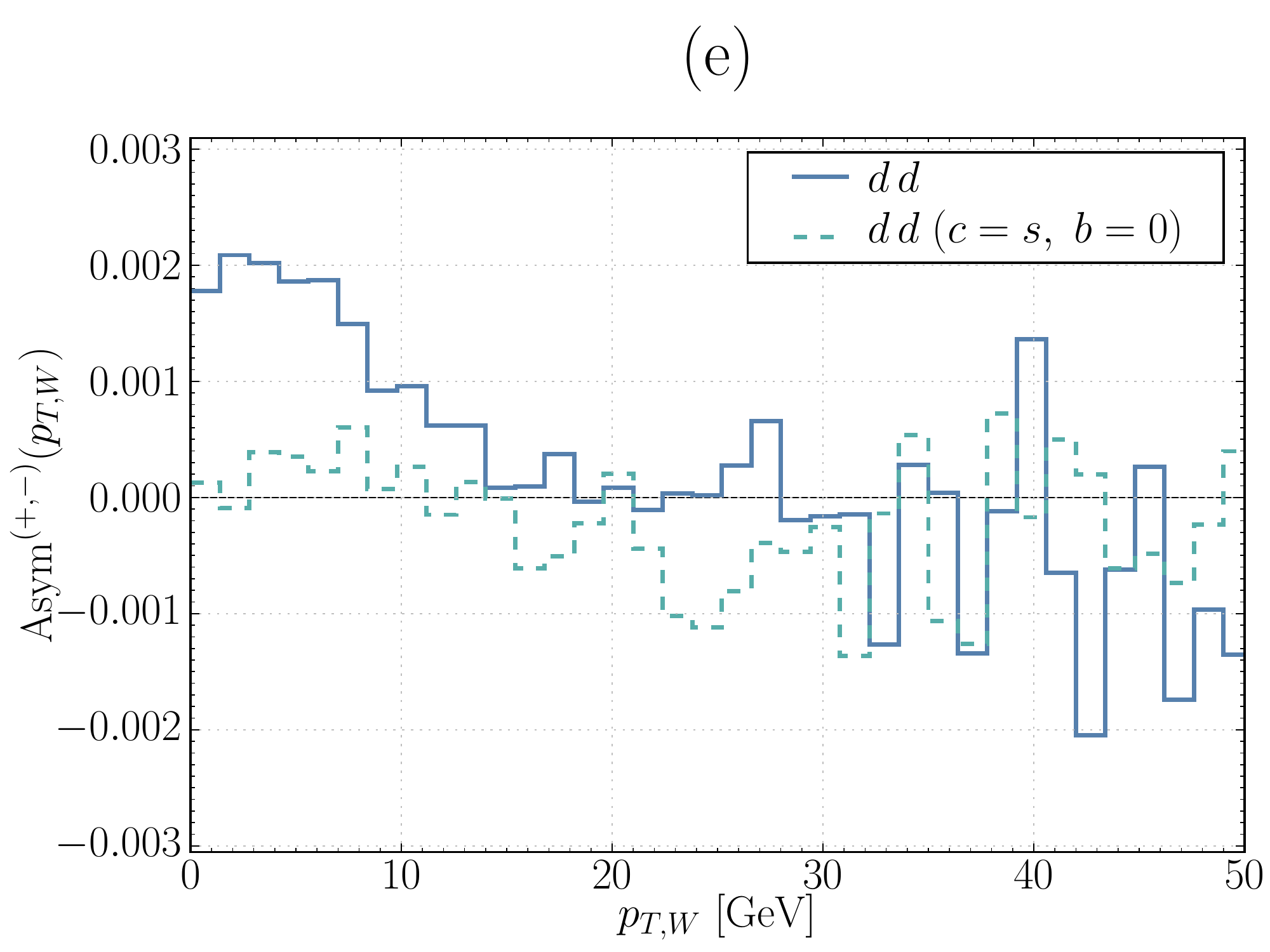}
    \vfill
    \includegraphics[width=0.495\tw]{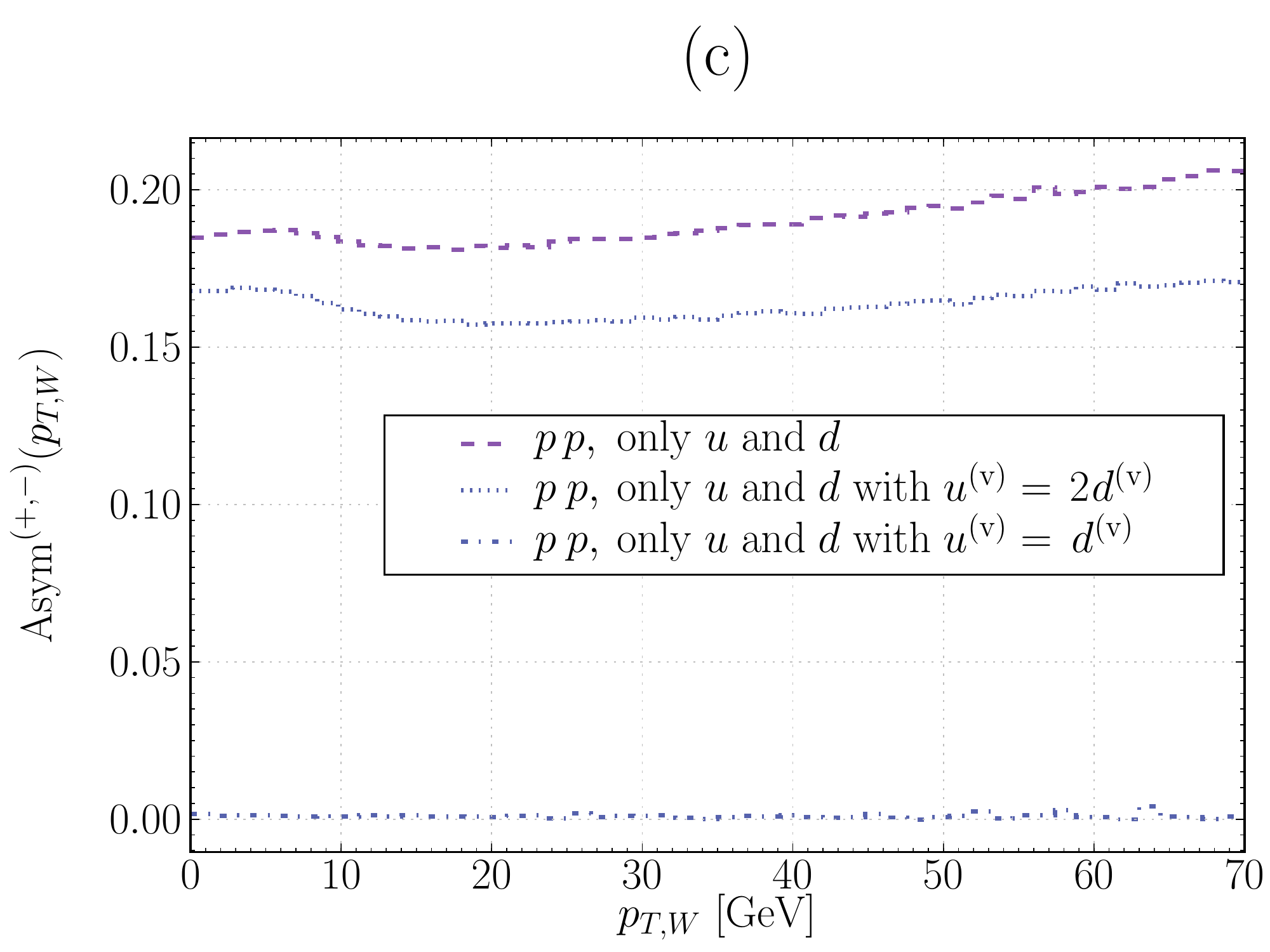}
    \hfill
    \includegraphics[width=0.495\tw]{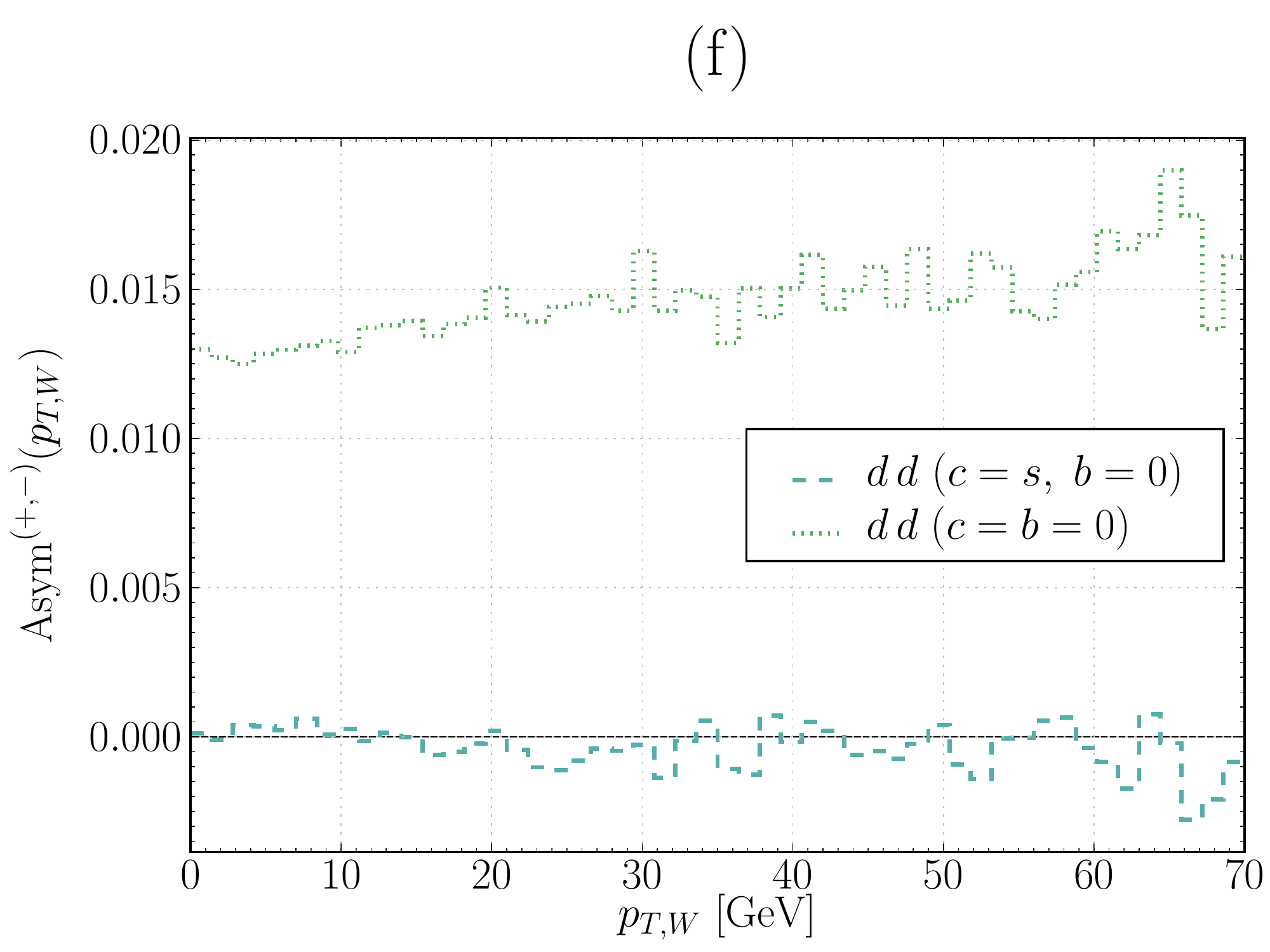}
    \caption[Charge asymmetries for $\pTW$ for several modification in the proton PDFs to 
      study the size of the flavour asymmetries for the $W$ production in $\pp$ and $\dd$ collisions]
            {\figtxt{The charge asymmetries for $\pTW$\,:\;
                $\pp$ vs. $\ppbar$ (a),
                $\pp$ vs. $\pp$ ($u$ and $d$ quarks only) (b),
                $\pp$ ($u$ and $d$ quarks only) vs. $\pp$ ($u$ and $d$ quarks only and 
                $u^\val=2\,d^\val$)(c), 
                $\dd$ vs. $\pp$ (d),
                $\dd$ vs. $\dd$ ($c=s$ and no $b$ contributions and with a focus on low $\pTW$) (e),
                and finally $\dd$ ($c=s$ and no $b$ contributions ) vs. $\dd$ with no $c$ nor $b$
                contributions) (f).}
            }
            \label{fig_W_prod_asym}
            \index{Quarks!uvmdv@$u^\val-d^\val$ asymmetry}
  \end{center} 
\end{figure}

In what follows the numerical importance of the various terms appearing in the above equations
is studied. For that purpose, to visualise the size coming from certain quarks contributions we
reject all the contributions which we are not interested in (basically weight these events by zero) 
and in some occasions we modify the predictions of the PDFs by hand using the framework described in
Fig.~\ref{fig_framework_generetion} of \S\,\ref{ss_gen_framework}.
The results are presented in Fig.~\ref{fig_W_prod_asym} by looking
each time at $\Asym{\pTW}$.

The first frame (a) compares $\pp$ collisions with the charge asymmetry free $\ppbar$ collisions.
The charge asymmetry for $\pp$ collisions is large and is varying with $\pTW$.

The next frame, (b), shows  the comparison of the $\pp$ asymmetry from the previous plot with the 
case where only the ``$u\to\leftarrow d$'' contributions ($u\,\dbar$ for $\Wp$ production and 
$d\,\ubar$ for $\Wm$ production) have been kept. 
The small gap between the two curves, compared to the size of the latter, reflects the small overall 
influence of $s$, $c$ and $b$ flavours in the asymmetry compared to the contribution of the $u$ and 
$d$ quarks.

\index{Quarks!uvmdv@$u^\val-d^\val$ asymmetry}
Frame (c) demonstrates this asymmetry is slightly reduced when in top of having only $u$ and 
$d$ quarks we have $u^\val=2\,d^\val$, \ie{} when the $u$ and $d$ valence quark PDFs are assumed to 
have the same shape and differ only by the normalisation factor corresponding to their number in 
the proton.
In addition, let us note that the asymmetry becomes flatter as a function of  $\pTW$ indicating 
the role of the relative $x$-shape of the $u$ and $d$ quarks PDFs. 
Finally, the asymmetry is reduced drastically when adding to the previous constraints 
$u^\val=d^\val$, as it would be the case when colliding isoscalar beams. \index{Quarks!Valence quarks}
This shows the consequences of $\ubar^\sea<\dbar^\sea$ for $10^{-4}<x$, which was found to be of the 
order of $\approx 0.05\percent$.\index{Quarks!Sea quarks}
In the rest of the plots the simplest isoscalar beam collision scheme, $\dd$, is considered.

The frame (d) shows that the charge asymmetry for $\dd$ is much smaller than the one for $\pp$ with
only small discrepancies mainly at low $\pTW$.
This results from the two following facts\,:\;(1) now all the terms contributing to the charge 
asymmetry include only off-diagonal CKM matrix elements and (2) contributions of the $s$, $c$ and $b$
quark PDFs are smaller than the ones coming from sea $u$ and $d$ quark PDFs. Hence this shows how in 
$\pp$ collisions the bulk of the charge asymmetries comes from the term $\propto\Vckmsqr{u}{d}$ in 
Eq.~(\ref{eq_sWp_sWm_pp}).

The remaining charge asymmetry is at the level of $0.002$ as can be seen in the zoomed
frame (e) and it can be reduced to a statistically negligible level when setting
$c=s$ and rejecting $b$ quarks contributions. The first constraint, because
$\Vckmsqr{u}{s}\approx\Vckmsqr{c}{d}$, assures to cancel the two first terms in the bracket of the
RHS of Eq.~(\ref{eq_sWp_sWm_dd}) while the second finalise the rise of any asymmetry.

Finally the frame (f) shows the effect of the deviation of the charge asymmetry due to the difference 
of the masses and of the momentum distributions of the strange and charm quarks in the extreme case 
where the $c$ contributions are rejected as well with the collisions involving $b$. 
In that case, the obtained asymmetry is significantly higher, at a level of $0.01-0.015$.
It is a factor of $\sim 10$ bigger than the asymmetry for the $\dd$ collisions assuming the present 
understanding of the relative asymmetry in the distribution of the strange and charm quarks.
\index{Electroweak!CKM matrix elements|)}

In the analysis presented so far the $\pTW$ distribution is inclusive, in other words it has been 
integrated over the full range of allowed $x_q$ and $x_{\bar{q}}$. 
In order to optimise the measurement strategy of the $W$ mass charge asymmetry, we now discuss the 
charge asymmetry $\pTW$ distributions restricted to selected kinematics regions.
The most obvious method to reduce the contributions of the valence quarks is to restrict the analysis 
to the $\yW\sim 0$ region, where $x_q\sim x_{\bar{q}} \approx 6\times 10^{-3}$, \ie{} where the 
valence quarks are largely outnumbered by the sea quarks.

\begin{figure}[!h] 
  \begin{center}
    \includegraphics[width=0.495\tw]{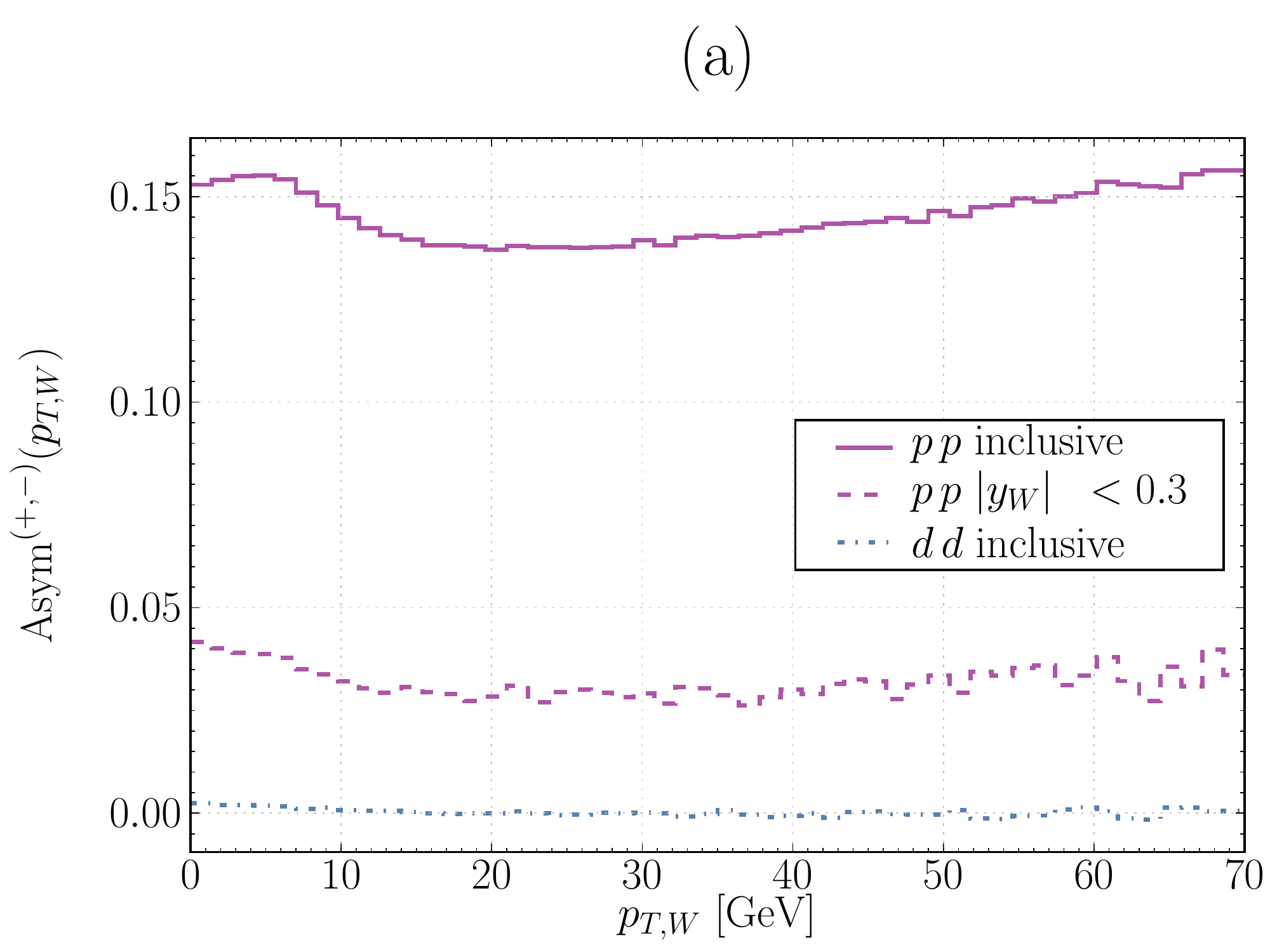}
    \hfill
    \includegraphics[width=0.495\tw]{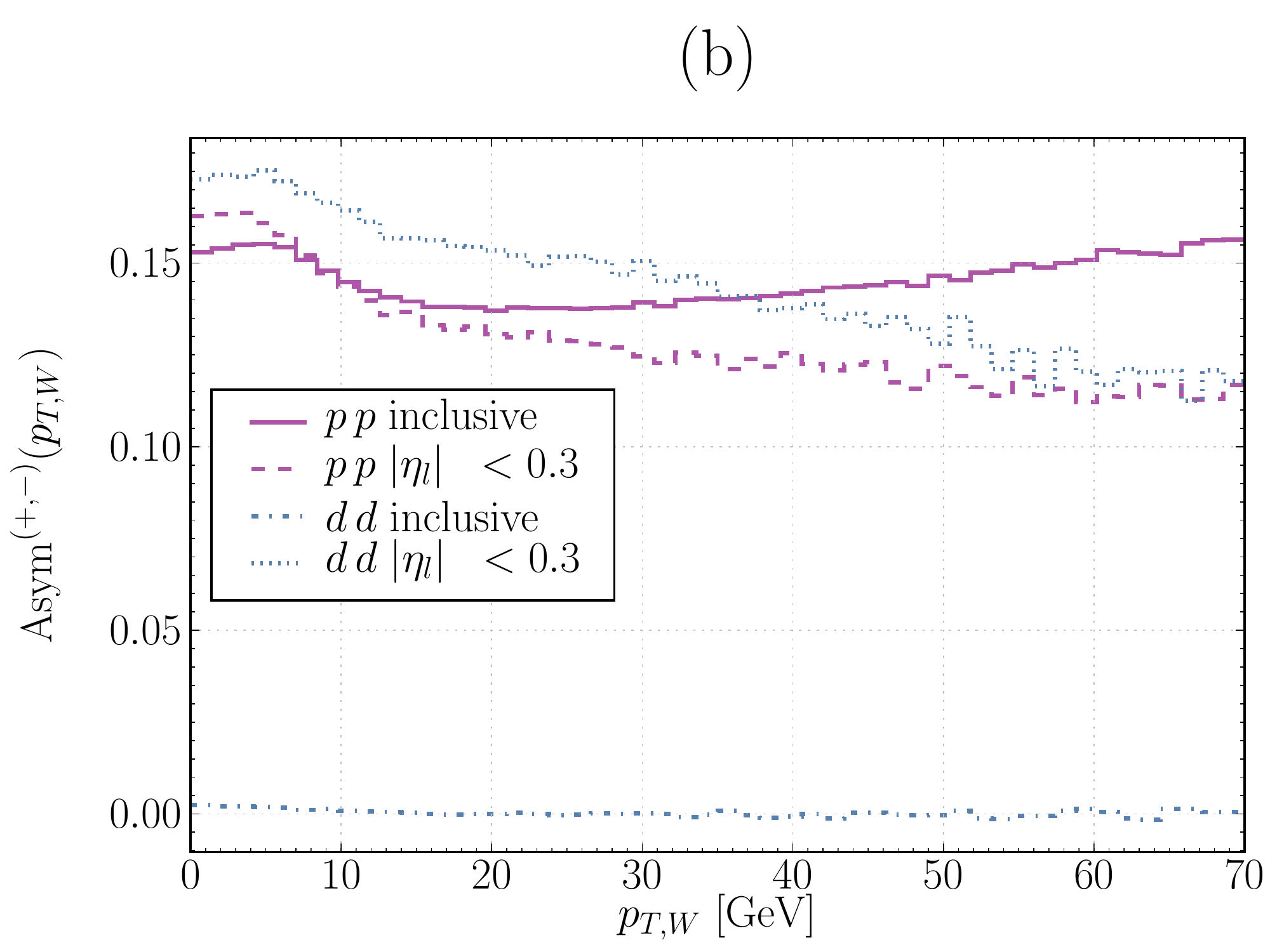}
    \caption[Charged asymmetries of $\pTW$ for inclusive $\pp$, $\dd$ collisions 
      and with $|\yW|<0.3$ for $pp$ and with $|\etal|<0.3$ for $\pp$ and $\dd$]
            {\figtxt{The charged asymmetries of $\pTW$ for inclusive $\pp,\,\dd$ collisions 
                     and with cuts\,:\; $|\yW|<0.3$ for $pp$ (a) and 
                     $|\etal|<0.3$ for $\pp$ and $\dd$ (b).}}
            \label{fig_W_prod_asym_centr_yW_etal_bin}
  \end{center}
\end{figure}
Figure~\ref{fig_W_prod_asym_centr_yW_etal_bin} presents the asymmetries
for $\pp$ and $\dd$ collisions for the narrow central bins in the $W$ rapidity (a), 
and in the lepton pseudo-rapidity (b). In the frame (a) the $\pp$ charge asymmetry is reduced and 
flattened by more than a factor of $3$ for the range of the $W$~boson rapidity $\yW<0.3$.
Now since $\yW$ cannot be measured directly it is natural to check if a comparable reduction can be 
obtained using the $\etal$ variable which is correlated with $\yW$.
Unfortunately, this is not the case, as shown in frame (b).
The $\etal$ variable has thus significantly lower discriminative power to reduce the valence quark 
contribution with respect to the $\yW$ variable. 
For the $\dd$ collisions the asymmetry restricted to the narrow $\etal$ bin increases considerably.
This observation draws our attention to the fact that the asymmetry in the decay mechanism of the 
$\Wp$ and $\Wm$ bosons will have an important impact on the asymmetry of leptonic observables. 
This will be discussed in detail in the next section. 

Let us conclude by noticing there can also be a contribution to the charge asymmetry coming from 
the QED radiation from quarks, as upper and lower components of quark doublets have different 
electric charges. 
However, it was found with the \Pythia{} parton shower model to be of the order of
$\Asym{\pTW}\approx 2.5\times 10^{-4}$, which is insignificant compared to the other previous 
contributions to the asymmetries for $\pp$ and $\dd$.\index{Pythia@\Pythia{} Monte Carlo event generator}
\index{W boson@$W$ boson!Production in pp@Production in $\pp$ collisions|)}
\index{W boson@$W$ boson!Transverse momentum|)}
\index{W boson@$W$ boson!Rapidity|)}

\section{Decays of $\BFWp$ and $\BFWm$ bosons}\label{ss_W_prod_decay}
\index{Charged lepton@Charged lepton from $W$ decay!Transverse momentum|(}
\index{Charged lepton@Charged lepton from $W$ decay!Pseudo-rapidity|(}
In this section the $W$~boson decay mechanism and its influence on the charge asymmetries are 
studied in both $\ppbar$ and $\pp$ collisions by looking at the charged lepton pseudo-rapidity 
and transverse momentum spectra.

First, in an overview, we see why in $\ppbar$ collisions we continue to observe perfectly symmetric 
features while in $\pp$ collisions we observe larger charge asymmetries than in the initial state
due to the decay properties of the leptons.
After that, to emphasise the arguments delivered in the previous overview a detailed description of the 
$\Wp$ and $\Wm$ decays in specific regions of $\yW$-space is presented.

\subsection{Overview in proton--anti-proton and proton--proton collisions}
\begin{figure}[!h] 
  \begin{center}
    \includegraphics[width=0.495\tw]{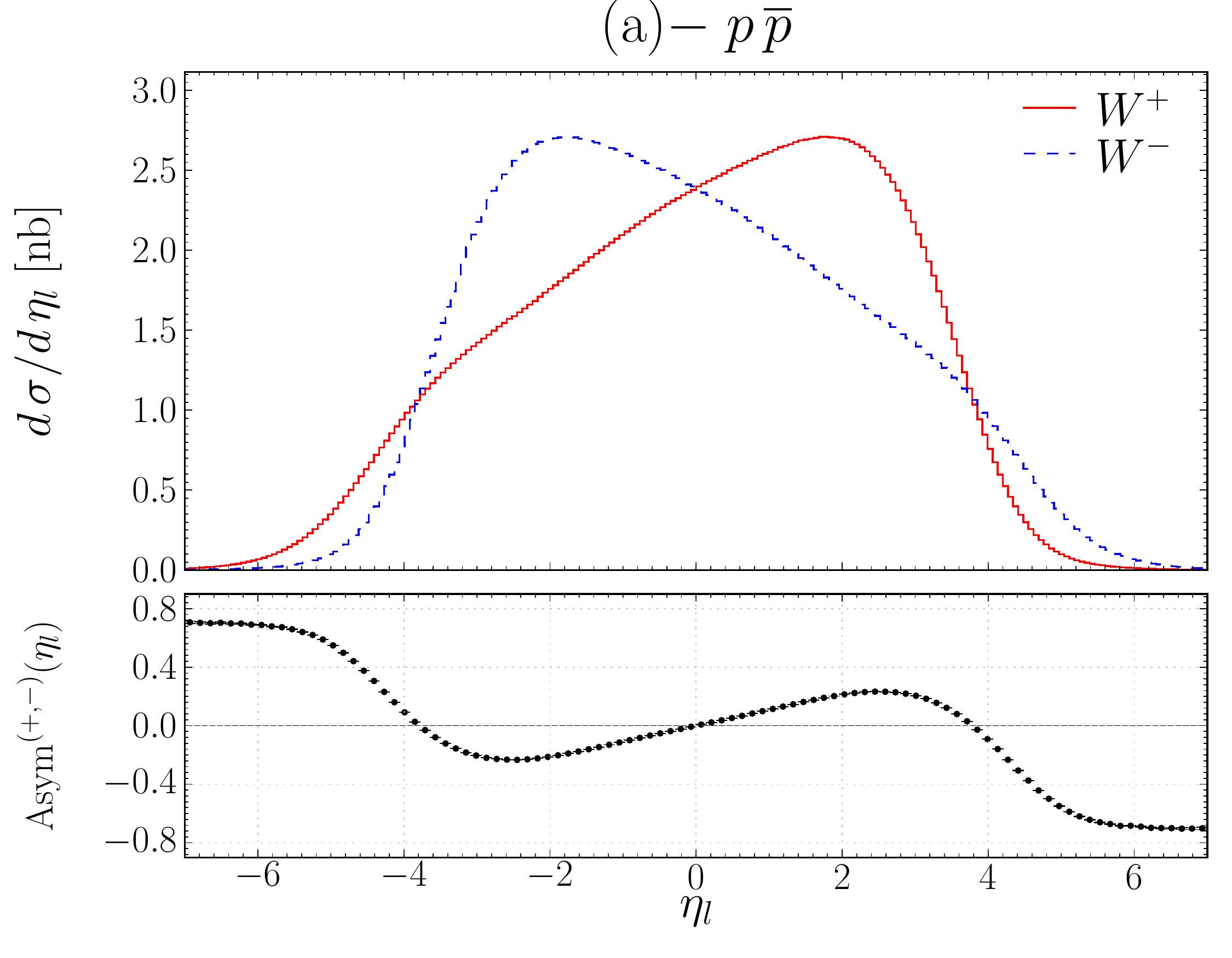}
    \hfill
    \includegraphics[width=0.495\tw]{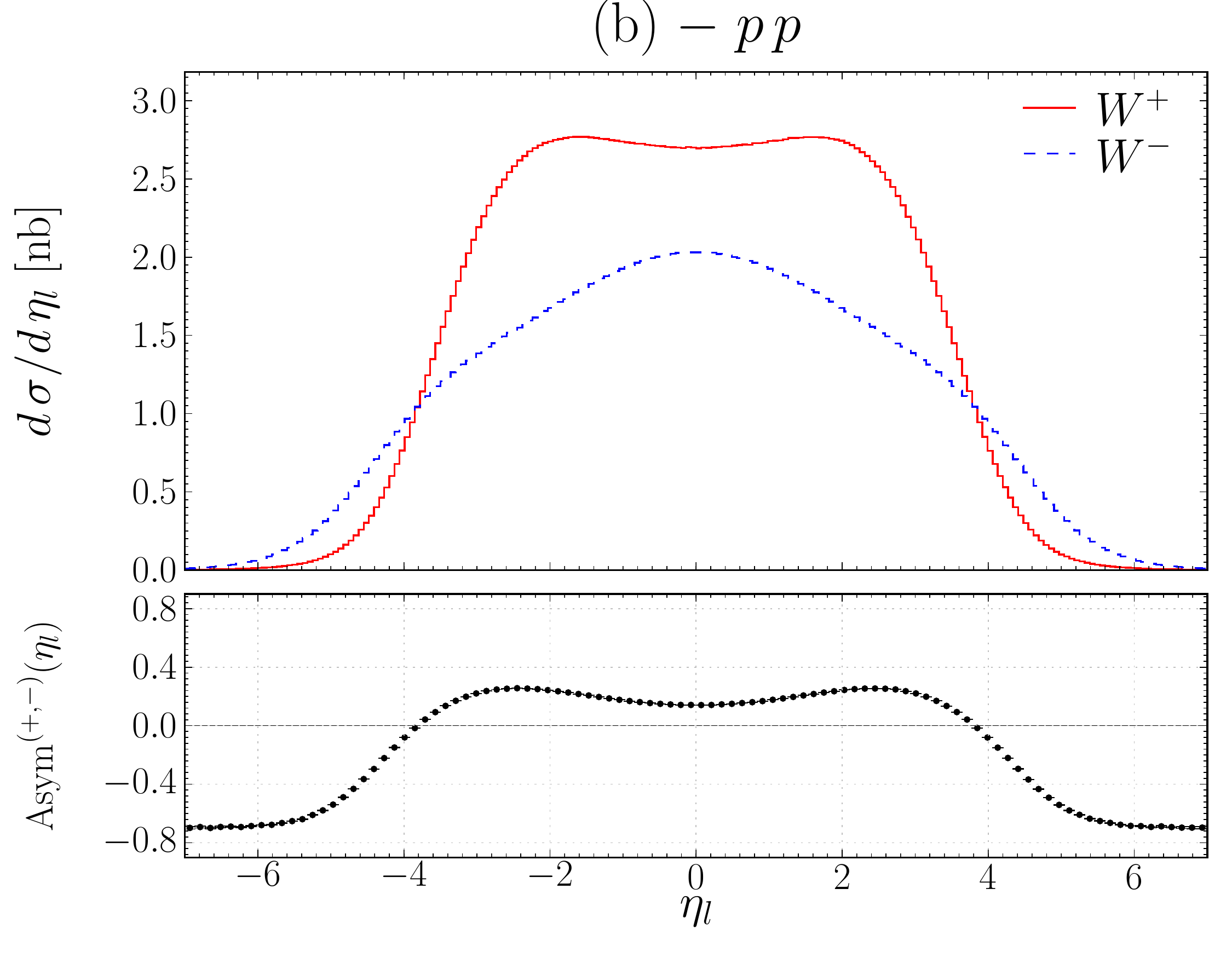}
    \vfill
    \includegraphics[width=0.495\tw]{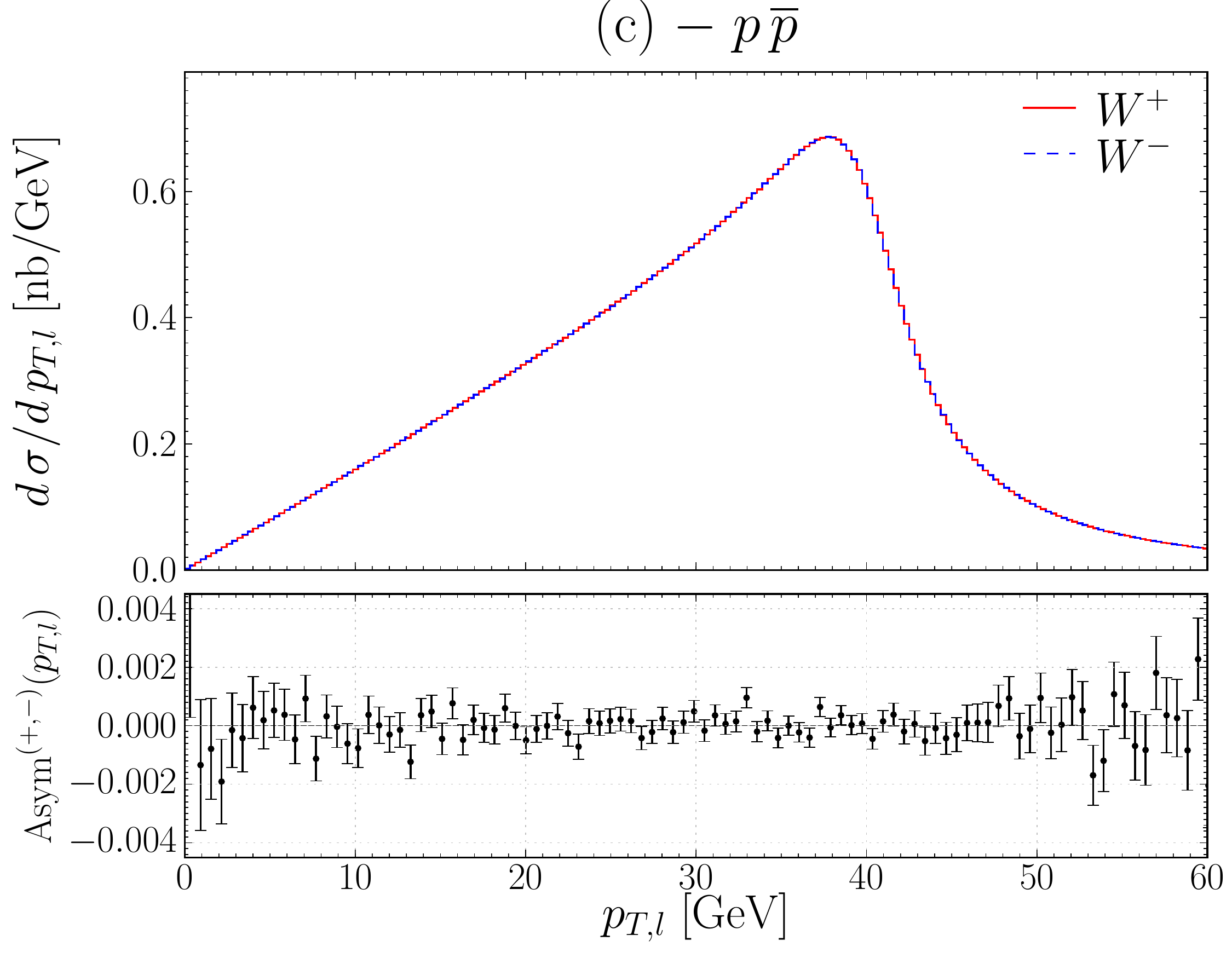}
    \hfill
    \includegraphics[width=0.495\tw]{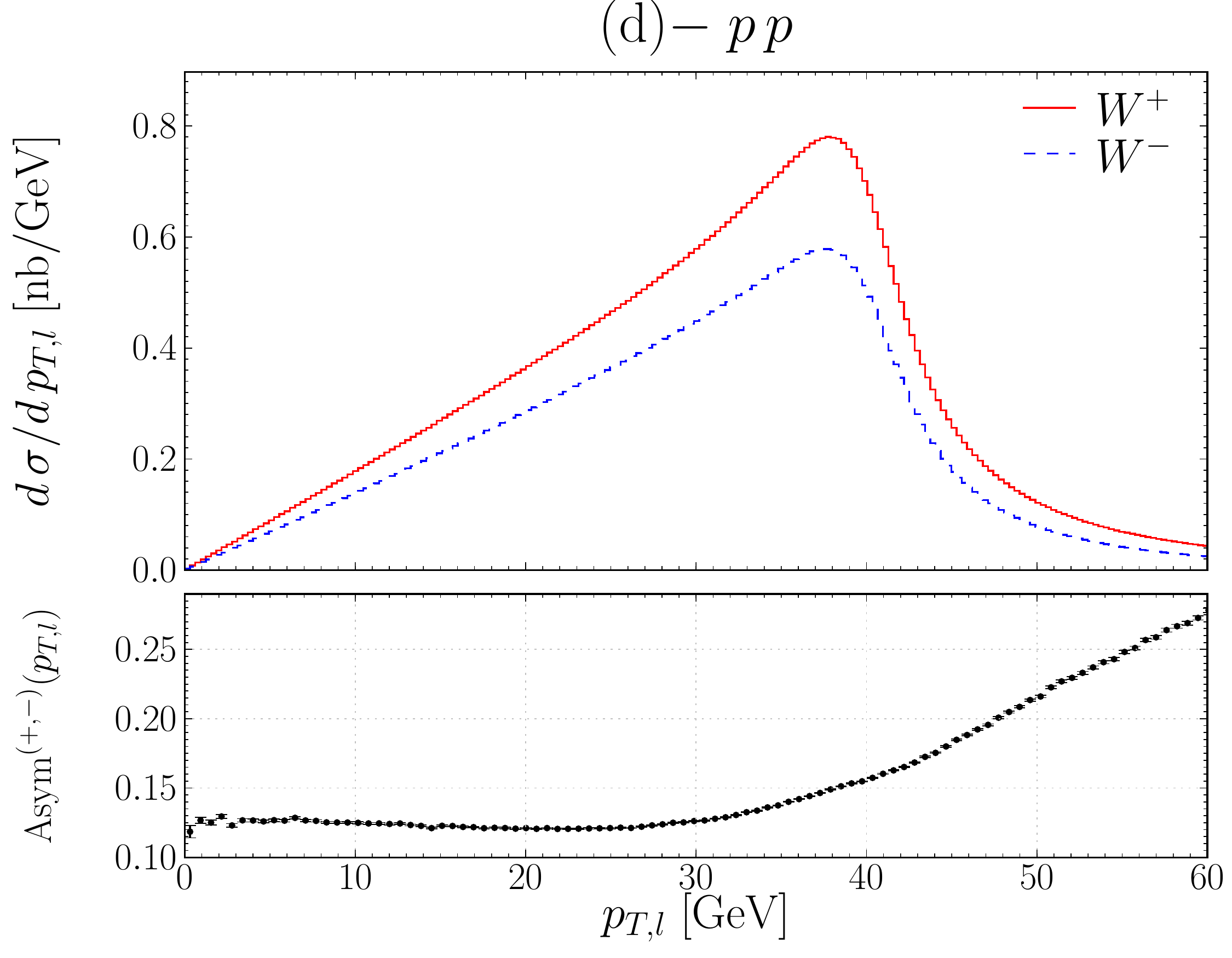}
    \caption[Pseudo-rapidity $\etal$ and transverse momentum $\pTl$ distributions for 
              charged leptons produced in $W$~boson decays for both $\ppbar$ and $\pp$ collisions]
            {\figtxt{The pseudo-rapidity $\etal$ and transverse momentum $\pTl$ distributions for 
              charged leptons produced in $W$~boson decays for the $\ppbar$~(a,b) 
              and $\pp$~(c,d) collisions.}
            }
            \label{fig_etal_pTl_ppb_pp}
  \end{center} 
\end{figure}

Figure~\ref{fig_etal_pTl_ppb_pp} presents the charged lepton $\etal$ and $\pTl$ distributions
along with their associated charge asymmetries.
Let us start by noticing that when studying the pattern of the $\etal$ spectra the influence of $\pTW$ 
can be neglected in the discussion since $\pTW\ll|\vec p_{z,W}|$.
More qualitatively, the improved LO affects the LO differential $\etal$ cross sections only to 
$\sim 4\percent$ maximum, thus --just in those cases-- we can consider $\vec p_{z,W}\sim\vec p_{W}$.

\paragraph{Proton--anti-proton collisions.}
\index{W boson@$W$ boson!Decay in ppbar@Decay in $\ppbar$ collisions|(}
Starting with $\ppbar$ collisions one can see the $\etal$ distributions (Fig.
\ref{fig_etal_pTl_ppb_pp}.(a)) are each other mirror reflection with respect to $\etal = 0$ this 
being a consequence of the symmetry of the problem as discussed in the previous section. 
Here it simply translates into that the whole $\lp$ event generation is exactly the same than 
the one of the $\lm$ up to the transformation $\theta_\lm=\pi-\theta_\lp$.

Once again this symmetry implies that projecting the charged leptons momenta on
the $r-\phi$ plane gives identical $\pTl$ distributions for positive and negative charges 
which is shown in Fig.~\ref{fig_etal_pTl_ppb_pp}.(c).
\index{W boson@$W$ boson!Decay in ppbar@Decay in $\ppbar$ collisions|)}

\paragraph{Proton--proton collisions.}
\index{W boson@$W$ boson!Decay in pp@Decay in $\pp$ collisions|(}
For $\pp$ collisions, on the other hand, at first sight the positively and negatively charged leptons
display highly non symmetric behaviour between each other. 
The pseudo-rapidity distributions indicates that $\lp$ leptons decay preferentially in the opposite 
direction of $\vec p_W$ while $\lm$ leptons display the inverse behaviour 
(Fig.~\ref{fig_etal_pTl_ppb_pp}.(b)).
The size of the charge asymmetry in $\pTl$ is much larger than the one in $\pTW$ and further 
investigations proved it is explained by the privileged decay of $\lp$ leptons in the direction of 
$\vec p_{T,W}$ while $\lm$ leptons decay in most cases in the opposite direction of $\vec p_{T,W}$
(Fig.~\ref{fig_etal_pTl_ppb_pp}.(d)).
\index{Helicity!Of the colliding quarks and the decaying leptons|(}
The reason for these two behaviours in $\etal$ and $\pTl$ finds its fundamental 
origin in the $V-A$ coupling of the leptons to the $W$~bosons. \index{Electroweak!VmA@$V-A$ coupling}
Indeed, as seen in Chapter~\ref{chap_theo}, at the level of the differential partonic cross section
the angular distribution in the \WRF{} of the final state charged lepton decaying from a $W$~boson 
of a given polarisation state can be expressed 
\begin{eqnarray}
  d\,\hat\sigma^{W_T^Q}/d\,\costhetaWlwrf 
  &\propto& \left(1 + \lambda\, Q\,\costhetaWlwrf\right)^2 \label{eq_WT_lep_decay},\\
  d\,\hat\sigma^{W_L^Q}/d\,\costhetaWlwrf 
  &\propto& \sin^2 \theta_{W,l}^{\ast},\label{eq_WL_lep_decay}
\end{eqnarray}
where, for reminders, $\thetaWlwrf$ is the charged lepton polar angle with respect to the direction
of the $W$ momentum $\vec p_W$ in the laboratory frame (Eq.~(\ref{eq_def_costhetaWlwrf})), 
$Q$ is the $W$~boson electric charge in units of $|e|$ and $\lambda=0,\pm 1$ is the $W$~boson helicity.
As can be seen, the angular distributions of $\lp$ and $\lm$ for longitudinally polarised $W$~bosons
($W_L$ for $\lambda=0$) are the same while for the transversely polarised $W$~bosons 
($W_T$ for $\lambda = \pm 1$), they depend upon the $W$~boson helicity.
\index{W boson@$W$ boson!Polarisation!Transverse states}
\index{W boson@$W$ boson!Polarisation!Longitudinal states}
To simplify the rest of the argumentation we will not consider the decay of longitudinally polarised
$W$. This leaves us with transversely polarised $W$ bosons which, due to the excess of 
matter from valence quarks\footnote{From another point of view the charge asymmetries can be seen as 
the consequence of a lack of anti-matter (valence anti-quarks) that cannot match this excess of matter 
from valence quarks.} 
in LHC collisions, are produced in larger proportions with negative helicities, \ie{} 
$W(\lambda = +1)<W(\lambda = -1)$.

\begin{figure}[!h] 
  \begin{center}
    \includegraphics[width=0.9\tw]{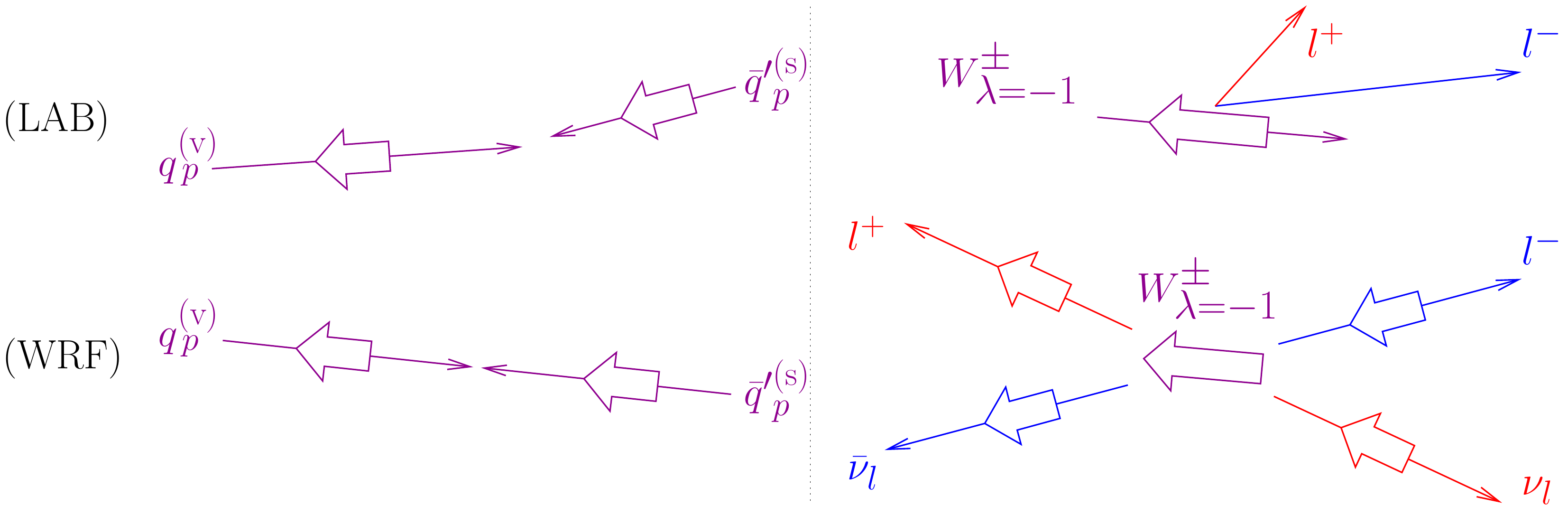}
    \caption[Valence quark and sea anti-quark annihilation producing a negative helicity $W$ boson]
            {\figtxt{Valence quark and sea anti-quark annihilation producing a negative helicity state 
                for the $W$.
                The collision is represented in both the laboratory frame (up) an in the \WRF{} 
                (down) and for both initial (left) and intermediate/final state (right).}}
            \label{fig_favored_qqbp_coll}
  \end{center}
\end{figure}

The consequences of this excess of $W(\lambda = -1)$ accounts for the behaviour of the 
pseudo-rapidity distributions illustrated in Fig.~\ref{fig_favored_qqbp_coll} 
in both \WRF{} and laboratory frame (LAB)\,: $\lm$ access larger pseudo-rapidity than the $\lp$.
Note this behaviour is already observed at leading order where $\pTW=0$ and is only slightly
affected by $\pTW$ in the improved LO.
The features observed in the longitudinal direction being understood we now turn our attention to the 
asymmetric pattern present in the transverse direction.

\index{Quarks!Transverse momenta in single W production@Transverse momenta in single $W$ production|(}
\begin{figure}[!h] 
  \begin{center}
    \includegraphics[width=0.6\tw]{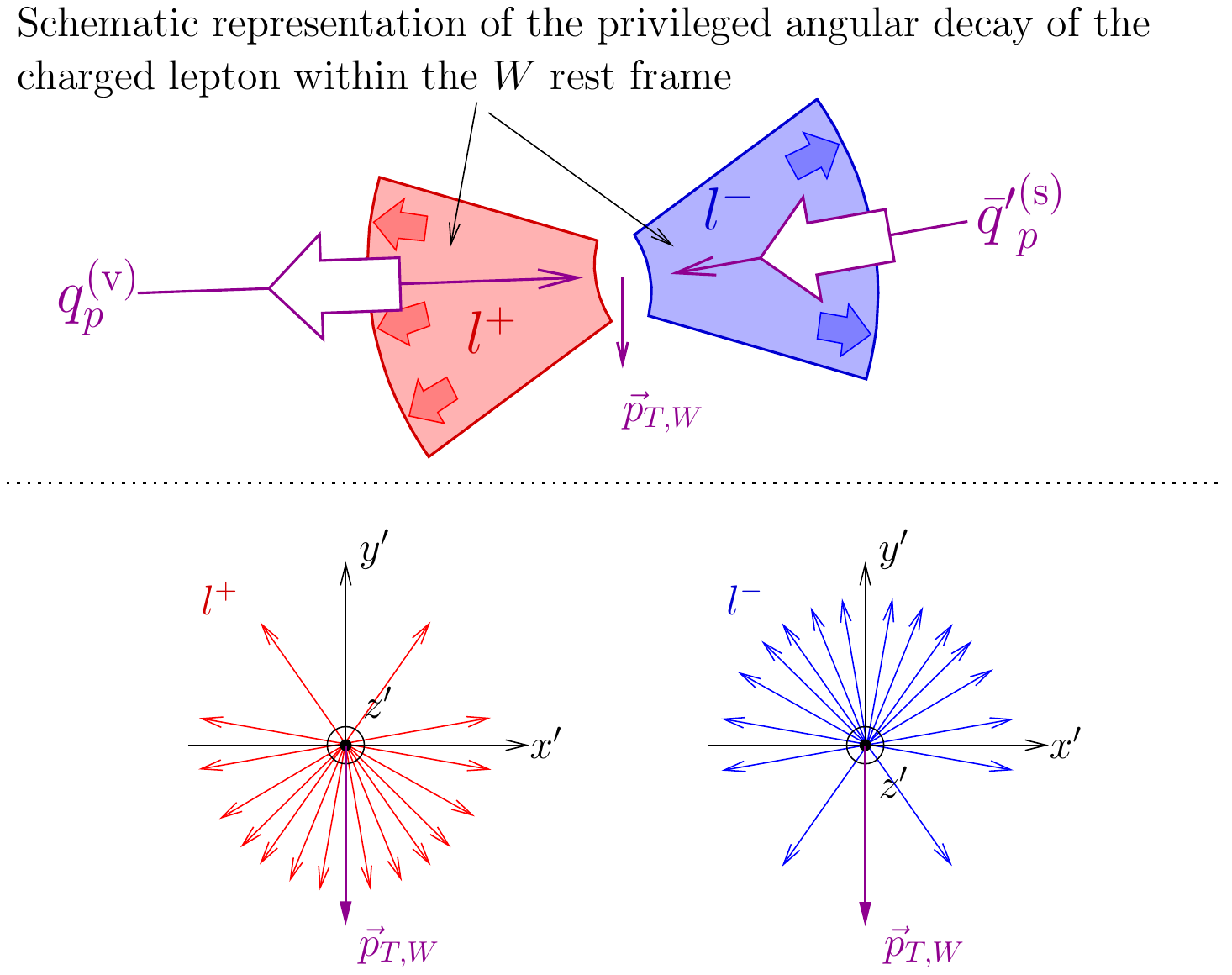}
    \caption[Schematic representation of a valence quark and sea anti-quark annihilation
      explaining the azimuthal anisotropy in the charged lepton decays with respect to the
      $\vec p_{T,W}$ direction]
            {\figtxt{Schematic representation of a valence quark and sea anti-quark annihilation
                seen in the laboratory frame but, for convenience, with the charged leptons decays 
                represented in the \WRF{} (up).
                Below, in the \WRF{}, the consequent azimuthal anisotropy with respect to the 
                $\vec p_{T,W}$ direction is represented for the decay of the charged leptons (down).
            }}
            \label{fig_app_phi_Wlrf_pp}
  \end{center} 
\end{figure}
As explained above a deeper investigation shows the $\lp$ leptons decay preferentially in the direction 
of $\vec p_{T,W}$ while the $\lm$ lepton display the inverse behaviour.\index{Electroweak!VmA@$V-A$ coupling|(}
This effect is the result of the $V-A$ coupling properties folded to the transverse motion of the $W$ 
or, more precisely, the relative size of the valence and sea (anti-)quarks in the collision.
Indeed, in a collision involving a valence quark the sea anti-quark carries in general a higher 
transverse motion (cf. \S\,\ref{app_ss_kT}) which, as can be understood 
using kinematic momentum conservation rules of thumb, constrains the charged lepton to privilege a 
certain azimuthal direction depending on the helicity it is holding.
Figure~\ref{fig_app_phi_Wlrf_pp} illustrates these effects, since the $\lp$ leptons have a positive
helicity, aligning it on the one of the sea quarks constrains it to decay in the same direction of
the ${\bar{q}{}'}^{\sea}$ anti-quark and by extension, since the latter carries most of the transverse motion,
with the one of $\vec p_{T,W}$. 
On the other hand negative helicity $\lm$ leptons, under the same 
kinematics, are constrained to prefer a decay in the opposite direction of the ${\bar{q}{}'}^{\sea}$ 
anti-quark/$\vec p_{T,W}$. Eventually it turns out the $\lp$ leptons intrinsic transverse momentum are amplified by 
$\pTW$ while in the case of the $\lm$ leptons they are reduced by $\pTW$.
Let us emphasise that even though this effect is ignited by the $V-A$ coupling attribute 
its amplitude depends of the relative unbalance in the transverse motion of the colliding valence 
quark and sea anti-quark, which is governed by the colliding energy $\sqrt S$ in the center of mass.
Also worth noticing is that even at the leading order there are already charge asymmetries
in the $\pTl$ spectra, still, due to their smaller size with respect to the one discussed above,
their study is treated in \S\,\ref{ss_app_pp}.\index{Electroweak!VmA@$V-A$ coupling|)}
\index{Quarks!Transverse momenta in single W production@Transverse momenta in single $W$ production|)}

In the rest of this section the previous distributions are reviewed for two narrow absolute rapidity 
regions, respectively in a central domain ($|\yW|<0.3$) and in a forward domain ($3.5<|\yW|<4.5$).
In the central region, the $W$~bosons are produced from quark--anti-quark pairs having 
$x_q\sim x_\qbar\approx 6\times 10^{-3}$ leading to small valence quarks contributions, and as a 
consequence, the assumed isospin symmetry in sea quarks flavours cannot produce charge asymmetries.
On the other hand, in the forward region the production of $W$~bosons involve most of the time a 
a valence (anti-)quark in $\ppbar$ collisions and a valence quark in $\pp$ collisions.
\index{W boson@$W$ boson!Decay in pp@Decay in $\pp$ collisions|)}

\subsection{Proton--anti-protons collisions}
\begin{figure}[!h] 
  \begin{center}
    \includegraphics[width=0.495\tw]{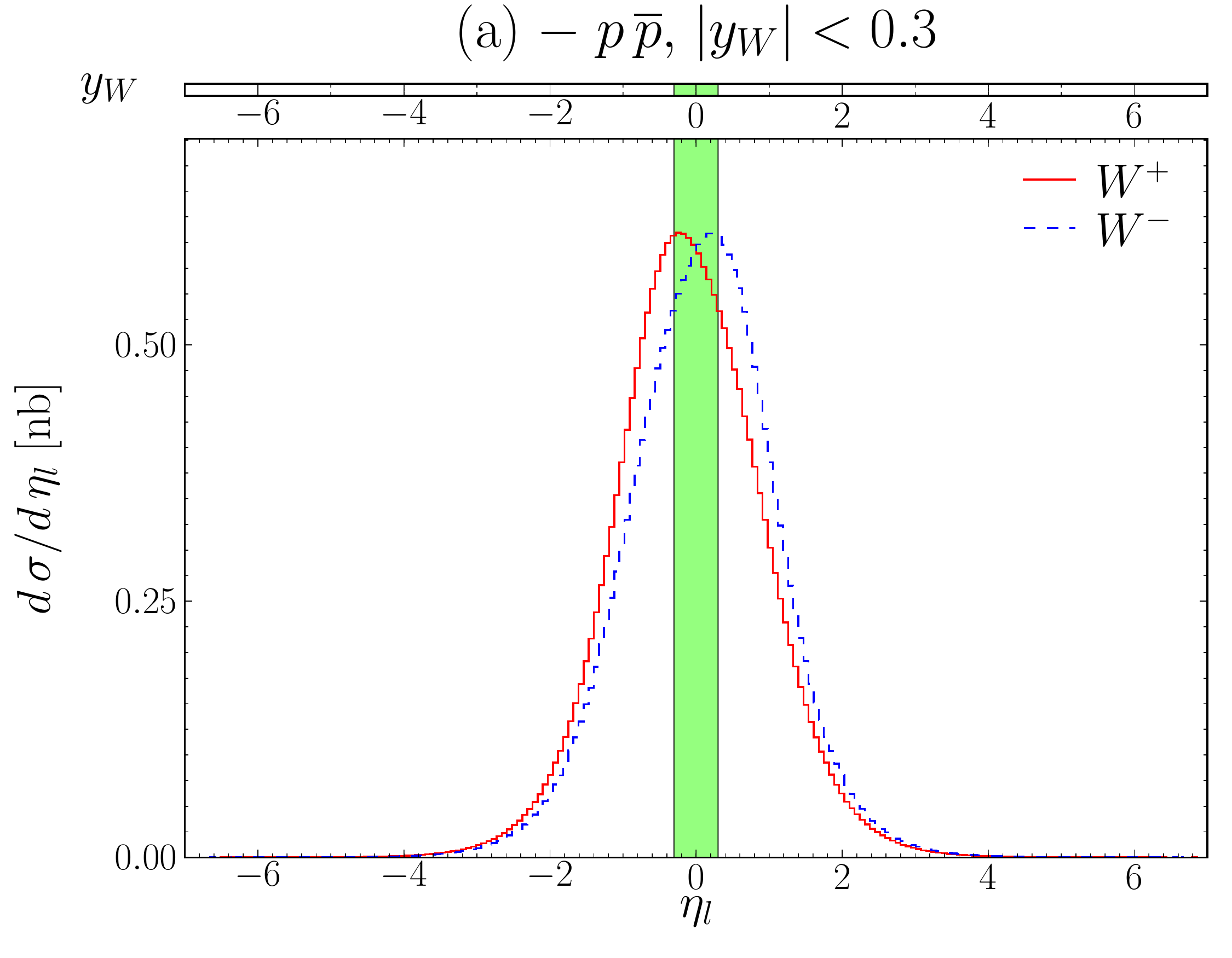}
    \hfill
    \includegraphics[width=0.495\tw]{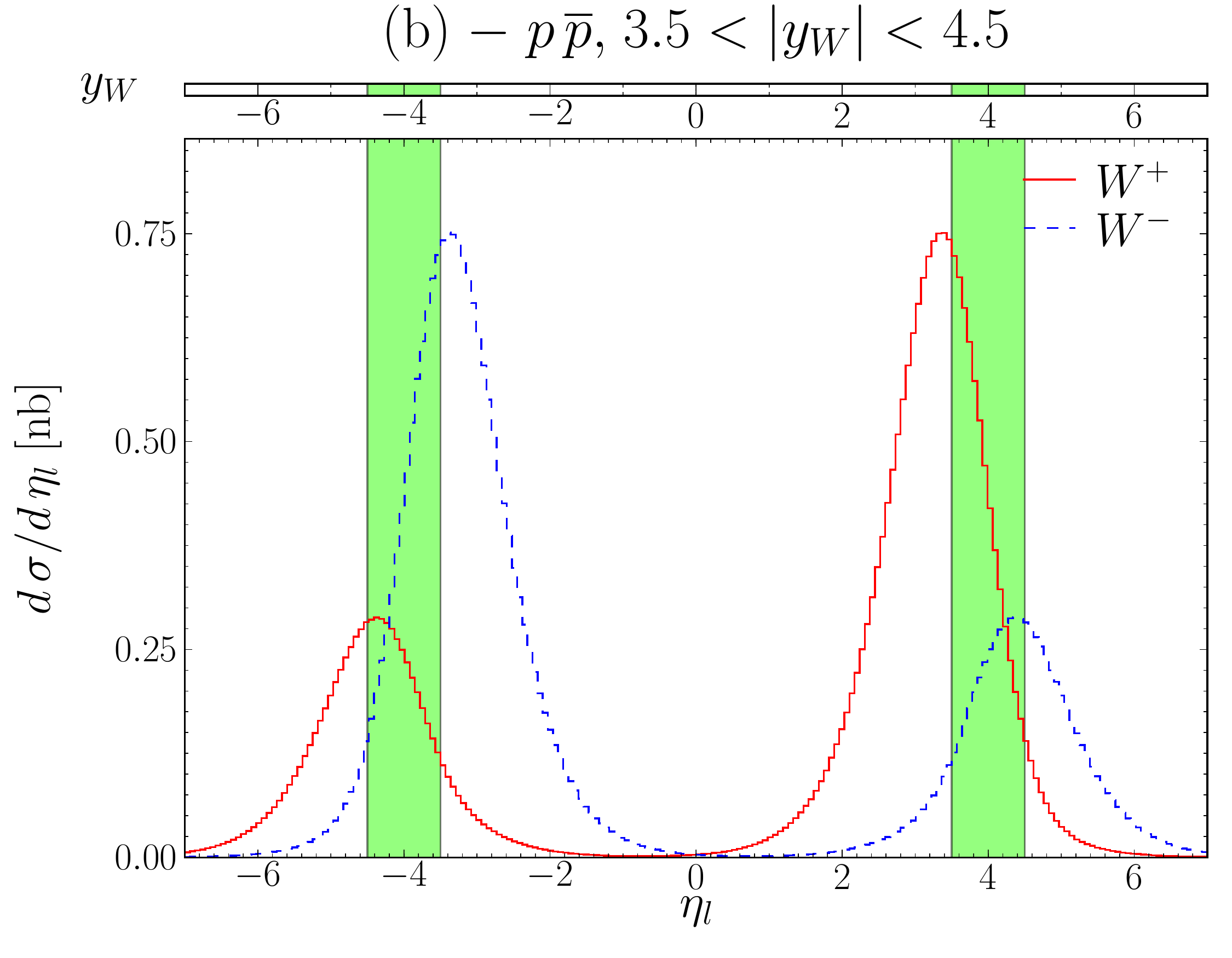}
    \vfill
    \includegraphics[width=0.495\tw]{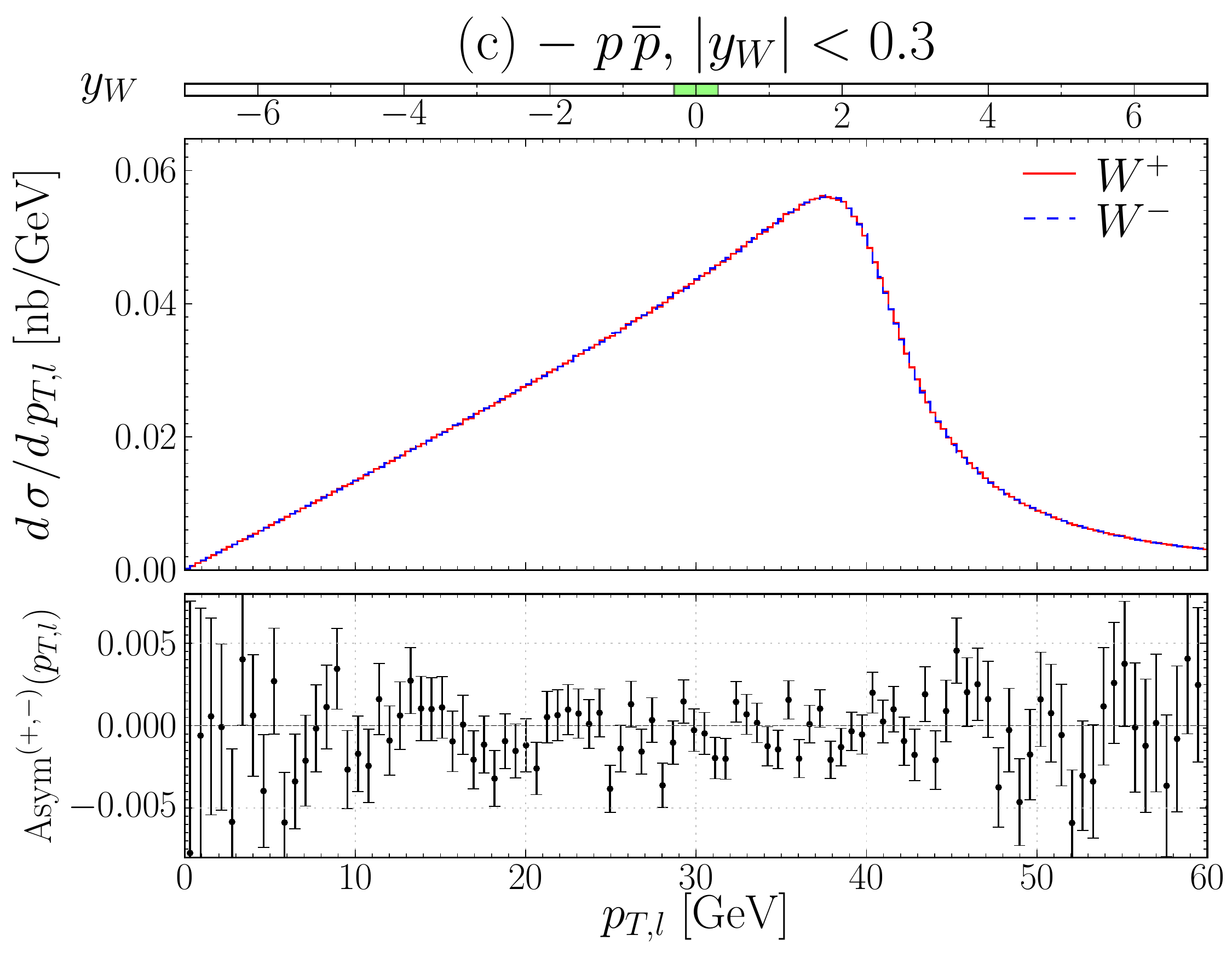}
    \hfill
    \includegraphics[width=0.495\tw]{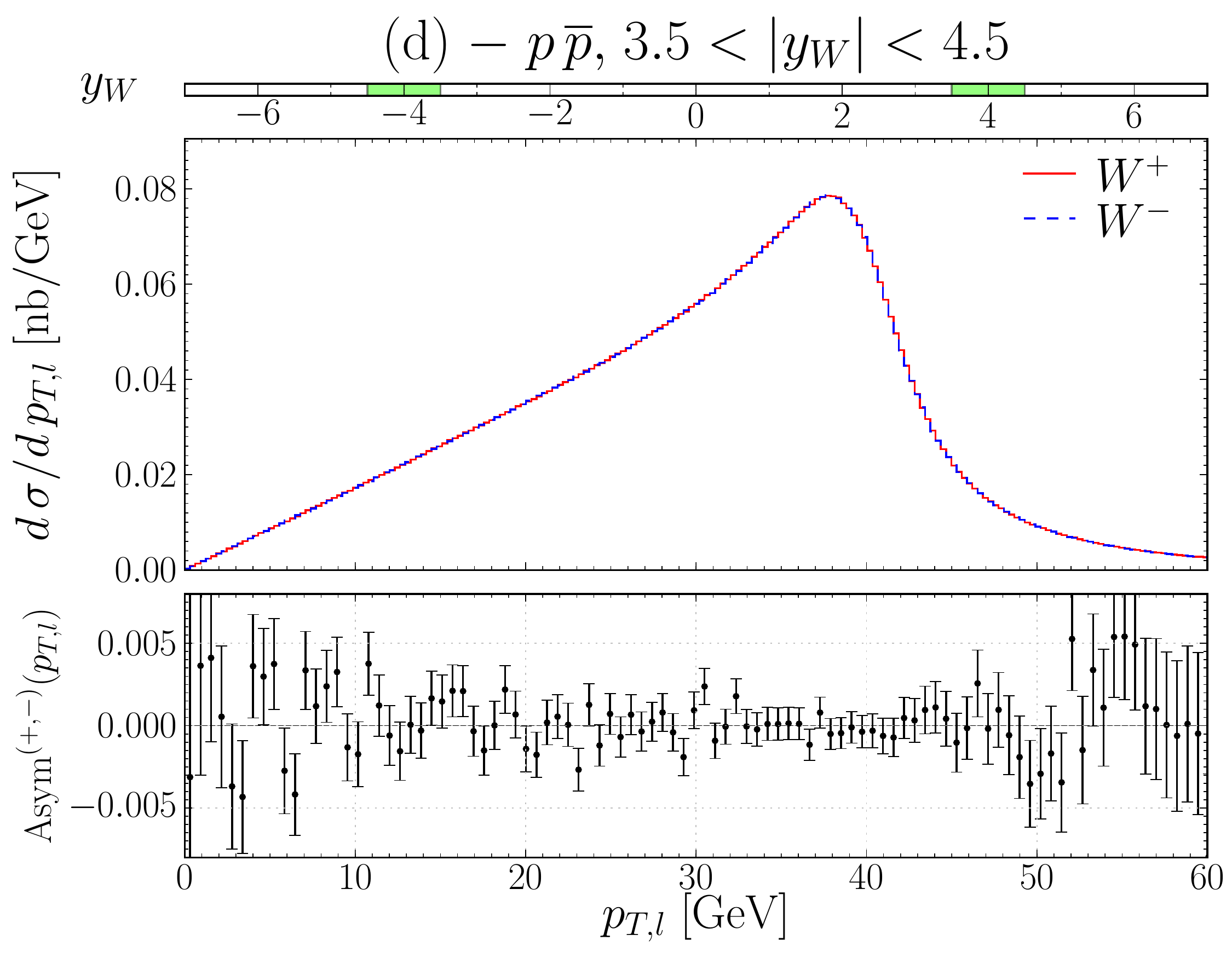}
    \caption[Positively and negatively charged lepton distributions of $\etal$ and $\pTl$ in 
      $\ppbar$ for two different ranges of the $W$ rapidity ($|\yW|<0.5$ and $3.5<|\yW|<4.5$]
            {\figtxt{Positively and negatively charged lepton pseudo-rapidity distributions for 
                $\ppbar$ collisions in two different ranges of the $W$ rapidity\,:\;%
                $|\yW|<0.3$ (a,c) and $3.5<|\yW|<4.5$ (b,d).}}
            \label{fig_ppbar_etal_pTl_in_yW_bin}
  \end{center} 
\end{figure}

\index{W boson@$W$ boson!Decay in ppbar@Decay in $\ppbar$ collisions|(}
The explanations in this subsection can be followed looking at 
Fig.~\ref{fig_ppbar_etal_pTl_in_yW_bin} showing the $\etal$, $\pTl$ and $\Asym{\pTl}$ 
distributions in $|\yW|<0.3$ and $3.5<|\yW|<4.5$ rapidity regions. 

\paragraph{$\BFWp$ and $\BFWm$ pseudo-rapidity distributions.}

In the narrow central rapidity region as it has been established valence (anti-)quark contributions are 
rather small, then $\lp$ and $\lm$ leptons display in a first approximation the same behaviour.
The discrepancies come from the small influence of valence (anti-)quarks.

In the forward rapidity region the valence (anti-)quarks influence creates large
local asymmetries in both $\etalp$ and $\etalm$ distributions.
Those behaviours can be understood looking at the pseudo-rapidity.
Using the LO approximation from Eq.~(\ref{eq_etal_yW_ylast}) gives
\begin{eqnarray}
\eta_\lp &\approx& \yW + y_\lp^\ast, \label{eq_etalp}\\
\eta_\lm &\approx& \yW + y_\lm^\ast,  \label{eq_etalm}
\end{eqnarray}
where $y_\lp^\ast$ and $y_\lm^\ast$ are respectively the intrinsic rapidity of the positively 
and negatively charged leptons.
The rapidity of the $W$ is much larger than the intrinsic $y_\lpm^\ast$ (cf. \S\,\ref{s_drell-yan}),
also the pattern of the distributions in Fig.~\ref{fig_ppbar_etal_pTl_in_yW_bin}.(b) can be seen as 
the consequence of\,:\;(1) the high longitudinal $p_{z,W}(\Leftrightarrow \yW)$ component of the boost 
that fixes roughly the center of the $\etal$ spectrum at $\etal\approx\yW$ and 
(2) the polar decay of the lepton --ruled by the $V-A$ coupling-- which smears the $\etal$ 
distribution from the central ``$\yW$-induced'' value.
Let us look at the case of the lepton $\lp$. In the region $3.5<\yW<4.5$, $\Wp$ bosons are produced 
mainly via $u_p^\val\,\dbar_\pbar^\sea$ pair collision, where for reminder $\val$ stands for valence and
$\sea$ for sea.\index{Quarks!Valence quarks}\index{Quarks!Sea quarks}\index{Electroweak!VmA@$V-A$ coupling}
In other words we have mostly $\Wp(\lambda=-1)$ which, combined to the $V-A$ coupling constrain the 
decaying $\lp$ to be preferentially emitted in the opposite direction to $\vec p_\Wp$.
This can be seen as most of the $\etalp$ distribution tends to shift to lower values of $\etal$.
On the other side, where $-4.5<\yW<-3.5$, bosons are produced mainly via $u_p^\sea\,\dbar_\pbar^\val$ 
annihilation which gives $\Wp(\lambda=+1)$. Here the decaying $\lp$ lepton prefers to be emitted 
following the $\vec p_\Wp$ direction. \index{W boson@$W$ boson!Polarisation!Transverse states}
Let us note the difference of height between two opposite $\etal$ bumps comes from the 
fact that $\dbar_\pbar^\val<u_p^\val$.
The same ideas can be applied to the case of the $\lm$ lepton decay, this time the opposite behaviour 
is observed with the exact same proportions

\paragraph{$\BFWp$ and $\BFWm$ transverse momenta distributions.}
The forward rapidity region is exclusively addressed as we know the issues related to the valence 
contributions are more important there.
In the \WRF{}, the spectrum of the relative angle between the $\lp$ and the $\Wp$ in $3.5<\yW<4.5$ 
is exactly the same than the one between the $\lm$ and the $\Wm$ in $-4.5<\yW<-3.5$ as shown
in Fig.~\ref{fig_ppbar_etal_pTl_in_yW_bin}.(b).
This is true independently of the relative unbalance in the colliding quark--anti-quark $\pT$,
thus the $W$ boost are identical and give eventually the same kinematics --up to a vertical flip-- for
the $\lp$ and $\lm$ from these opposite $\yW$-space domains.
The same idea can be used to account for the equality of the $\pTlp$ distribution in $-4.5<\yW<-3.5$ 
with the $\pTlm$ distribution in $3.5<\yW<4.5$. 
Then, as long the rapidity selection is achieved on the absolute value $|\yW|$
the positive and negative leptons transverse momenta show no asymmetries as displayed in 
Fig.~\ref{fig_ppbar_etal_pTl_in_yW_bin}.(d).
This property is obviously conserved when integrating over the whole phase space.
\index{W boson@$W$ boson!Decay in ppbar@Decay in $\ppbar$ collisions|)}

\subsection{Proton--proton collisions}
\index{W boson@$W$ boson!Decay in pp@Decay in $\pp$ collisions|(}
\begin{figure}[!h] 
  \begin{center}
    \includegraphics[width=0.495\tw]{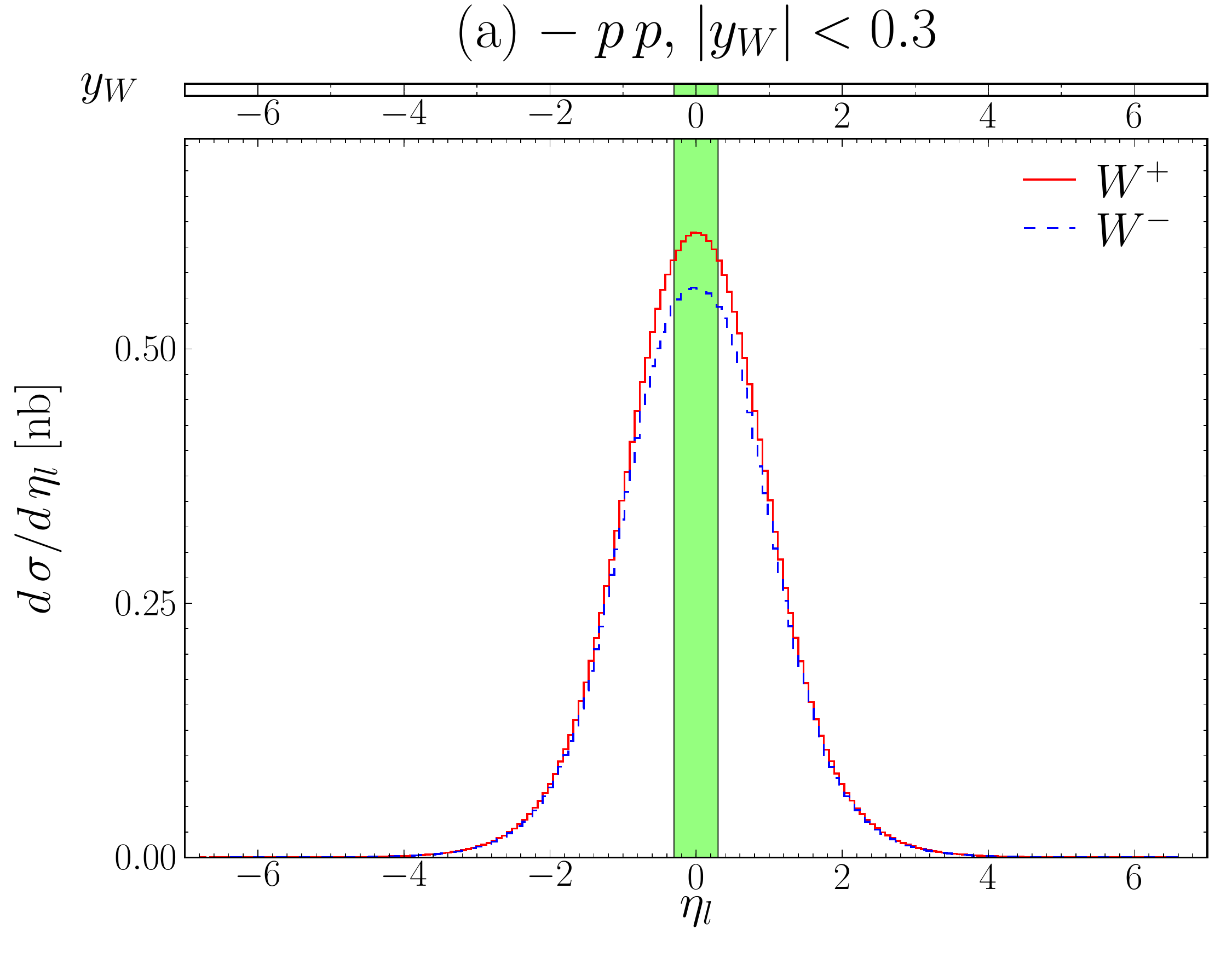}
    \hfill
    \includegraphics[width=0.495\tw]{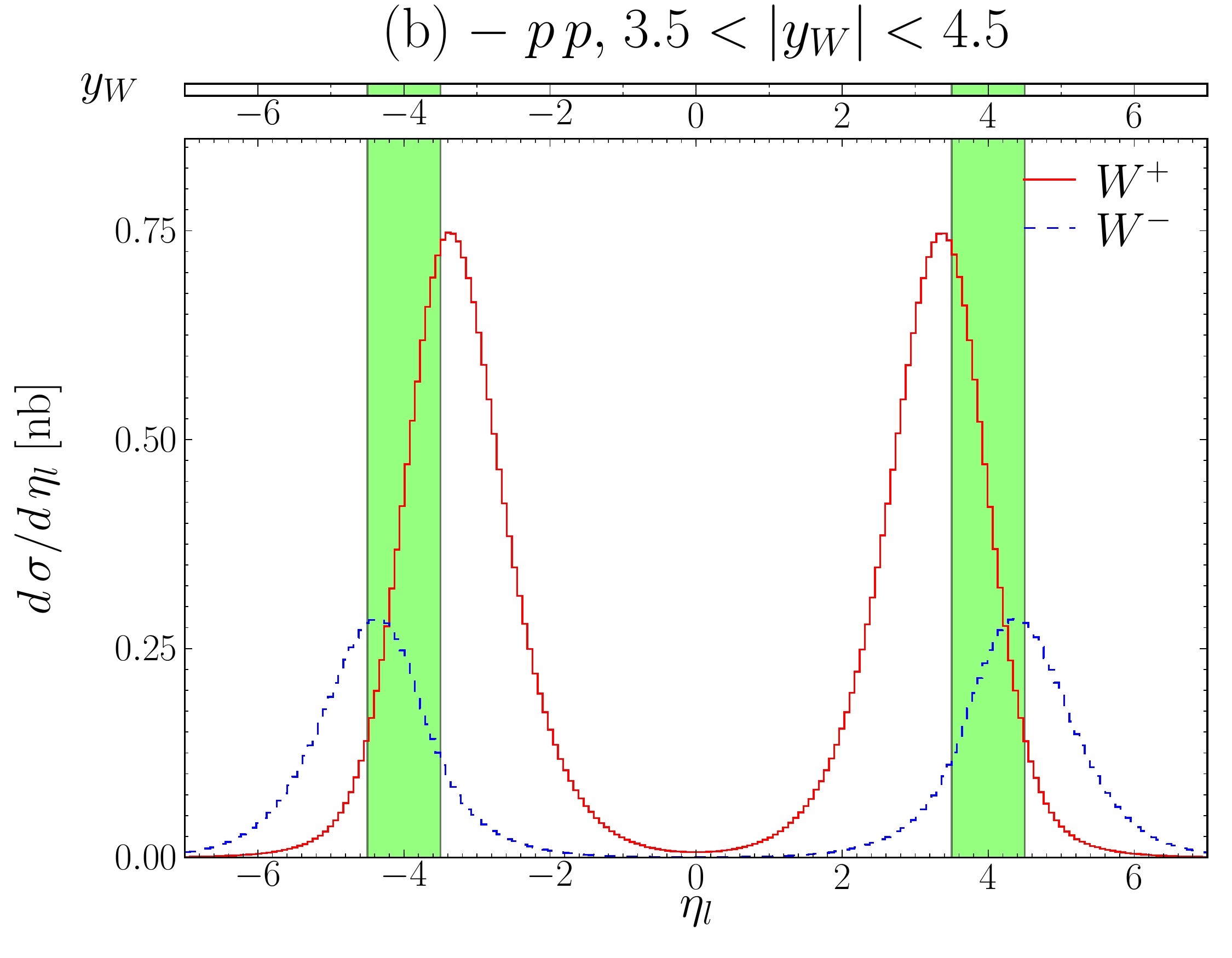}
    \vfill
    \includegraphics[width=0.495\tw]{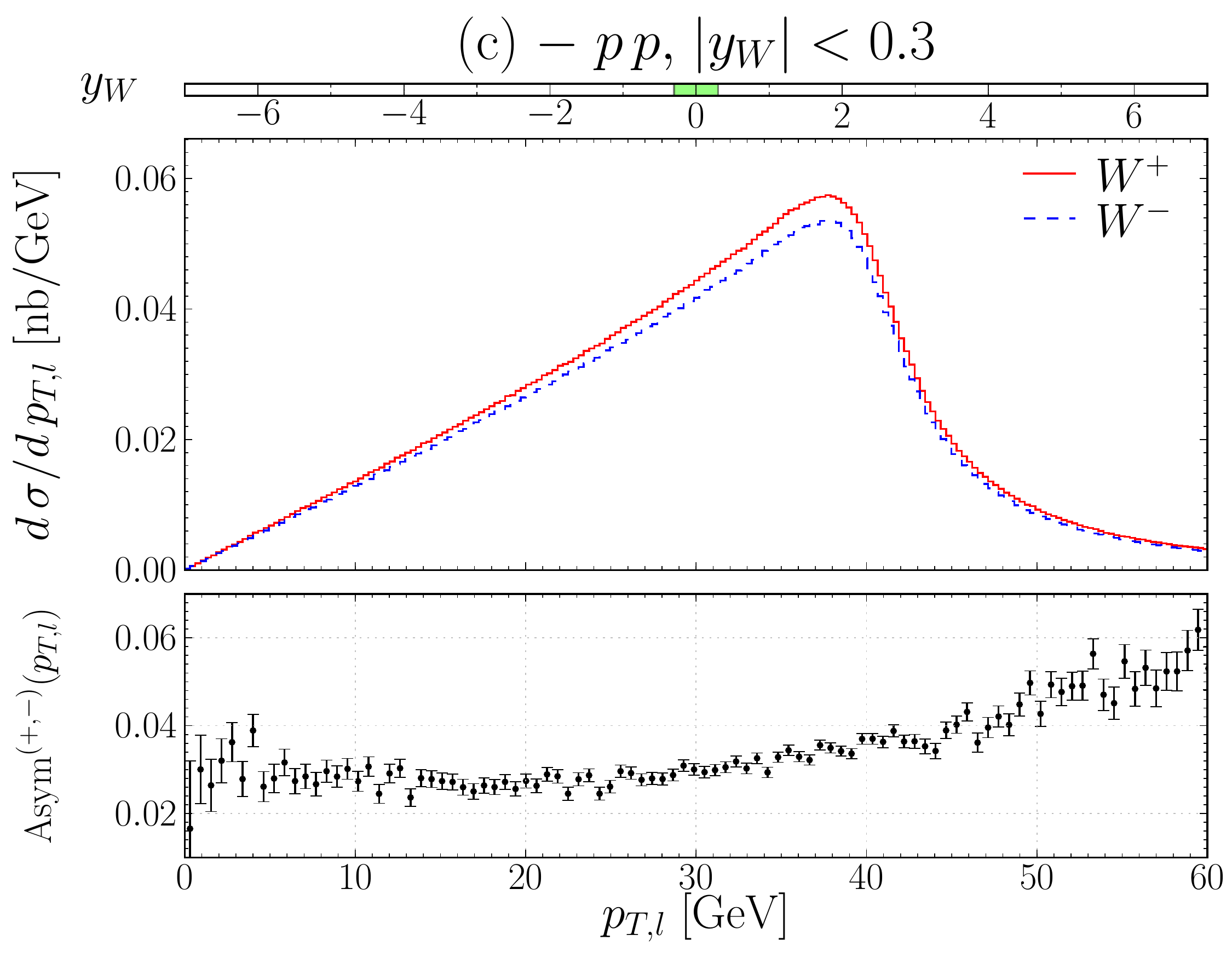}
    \hfill
    \includegraphics[width=0.495\tw]{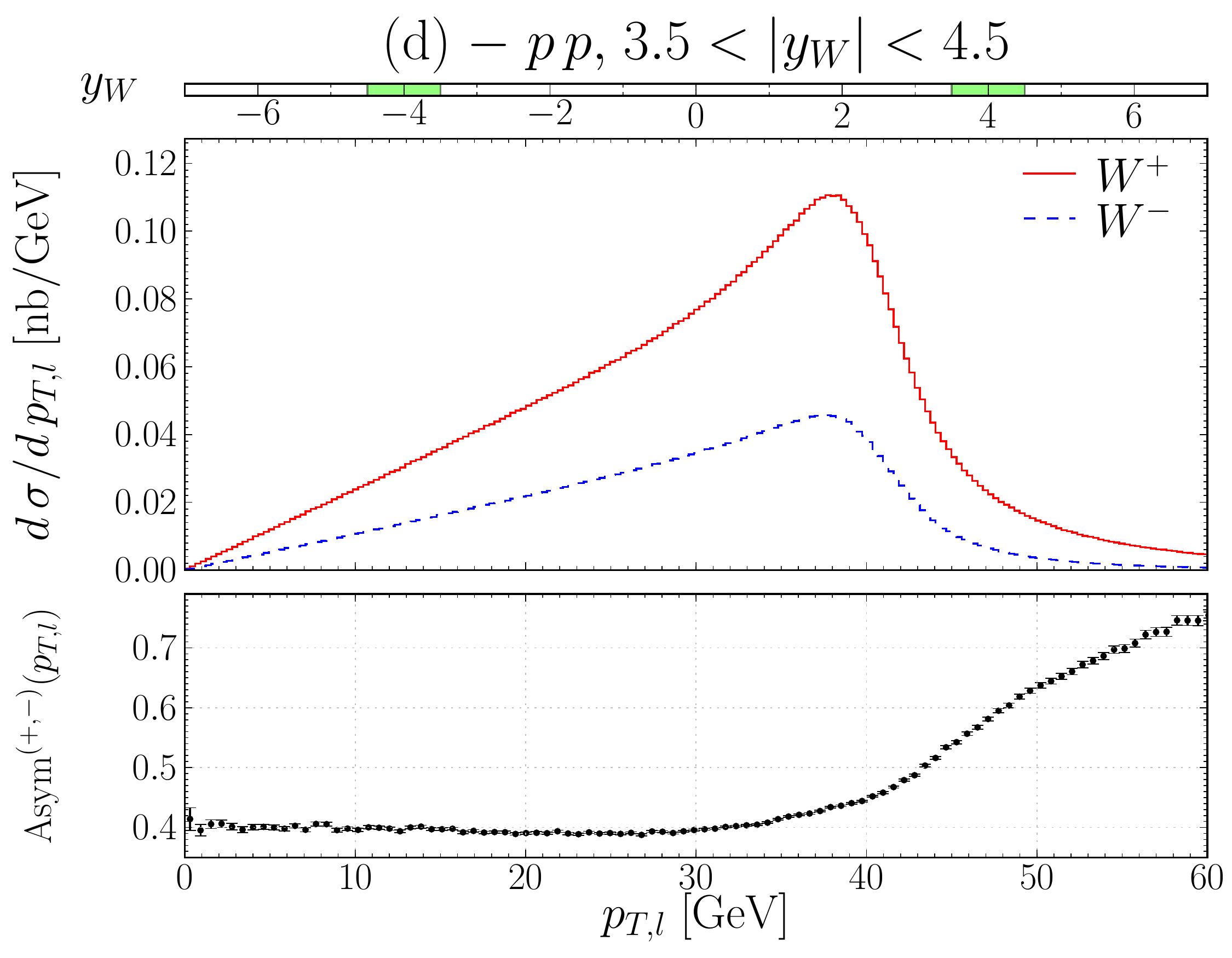}
    \caption[Positively and negatively charged lepton distributions of $\etal$ ($\ppbar$, $\pp$) 
      and $\pTl$ ($\pp$ only) for two different ranges of the $W$ rapidity ($|\yW|<0.5$ 
      and $3.5<|\yW|<4.5$]
            {\figtxt{Positively and negatively charged lepton pseudo-rapidity/transverse momenta 
                distributions for ranges of the $W$ rapidity\,:\;%
                $|\yW|<0.3$ (a,c) and $3.5<|\yW|<4.5$ (b,d).}}
            \label{fig_pp_etal_pTl_in_yW_bin}
  \end{center} 
\end{figure}
Considering now the case of $\pp$ collisions we repeat the same path of reasoning followed for 
the $\ppbar$ case. The $\etal$ and $\pTl$ distributions in $\pp$ collisions for the narrow central and
forward $W$ rapidity regions are shown in Fig.~\ref{fig_pp_etal_pTl_in_yW_bin}.

\paragraph{Pseudo-rapidity distributions.}
Once again going to $|\yW|<0.3$ allows to get rid of most of the valence quarks which, as it was
found out in the overview is the source of the asymmetries in $\pp$ collisions
(Fig.~\ref{fig_pp_etal_pTl_in_yW_bin}.(a)).
The small discrepancies are due to non negligible contributions from valence quarks.

Now moving to the $3.5<|\yW|<4.5$ region (Fig.~\ref{fig_pp_etal_pTl_in_yW_bin}.(b)) the general
pattern can be understood using Eqs.~(\ref{eq_etalp}--\ref{eq_etalm}).
Both $\etalp$ and $\etalm$ distributions are found in the vicinity of $|\etal|\approx |\yW|$ and 
the deviations from this central value are due to lepton decay in $\theta$ ruled by the $V-A$ coupling,
\ie{} only positive helicity $\lp$ and negative helicity $\lm$ can couple to $W$ bosons.
\index{Electroweak!VmA@$V-A$ coupling}

We start by analysing the positively charge lepton $\lp$.
Here, in both positive ($3.5<\yW<4.5$) and negative ($-4.5<\yW<-3.5$) rapidity domains the 
$\Wp$ are produced most likely from a $u_p^\val\,\overline d_p^\sea$ annihilation providing to the 
boson a negative helicity. 
Then the $\lp$ preferentially decay in the opposite direction of $\vec p_W$.
This explains why the $\etalp$ spectrum is slightly shifted to smaller absolute $\etal$ values. 
The same idea can be applied to the other charged lepton. One finds eventually that this time 
the $\lm$ decays in most cases in the same direction of $\vec p_{W}$ which translates here that the 
$\etalm$ spectrum is slightly shifted to higher absolute values of $\etal$.
This explains why we observed in Fig.~\ref{fig_etal_pTl_ppb_pp}.(b) a widening of the $\lm$ 
pseudo-rapidity distribution from $y_\Wm$ while the $\lp$ pseudo-rapidity tends to narrow from the 
inner $y_\Wp$ spectrum.
The relative size of the peaks is explained to the light that, as said previously, in a proton there
are more $u^\val$ than $d^\val$.

Let us stress that the differences between the $\yW$-bands width and the associated spread $\etal$
distributions demonstrates that $\etal$ has a significantly weaker resolution power for the 
momenta of annihilating quarks than $\yW$, and, as a consequence, weaker resolving power of the 
$W$~boson polarisation.
This gives already an idea of the loss of information one has to accept when trading the knowledge 
of the $W$ rapidity with the charged lepton pseudo-rapidity.

\paragraph{Transverse momenta distributions in forward rapidity region.}
The behaviour of the charged lepton transverse momenta are discussed exclusively in the forward 
rapidity region (Fig.~\ref{fig_pp_etal_pTl_in_yW_bin}.(d)).
For $3.5<\yW<4.5$, in the \WRF{} the relative angle between the $\lp$ and the $\Wp$
is different from the one of $\lm$ and the $\Wm$ due to the larger value of the sea quark transverse
momentum  with respect to the one of the valence quark. 
The $V-A$ coupling as shown in Fig.~\ref{fig_app_phi_Wlrf_pp}, by forcing the $\lp$ and $\lm$ to 
decay in opposite direction induce the $\lp$ to follow the transverse motion held mostly by the
sea quark while the $\lm$ decays in the opposite direction of the latter.
Taking into consideration the negative rapidity stripe concurs to roughly double the size of the 
asymmetry. The consequences of this effect on the pattern of the transverse momenta of the $\lp$ 
and $\lm$ has dramatic effects.

\begin{figure}[!h] 
  \begin{center}
    \includegraphics[width=0.6\tw]{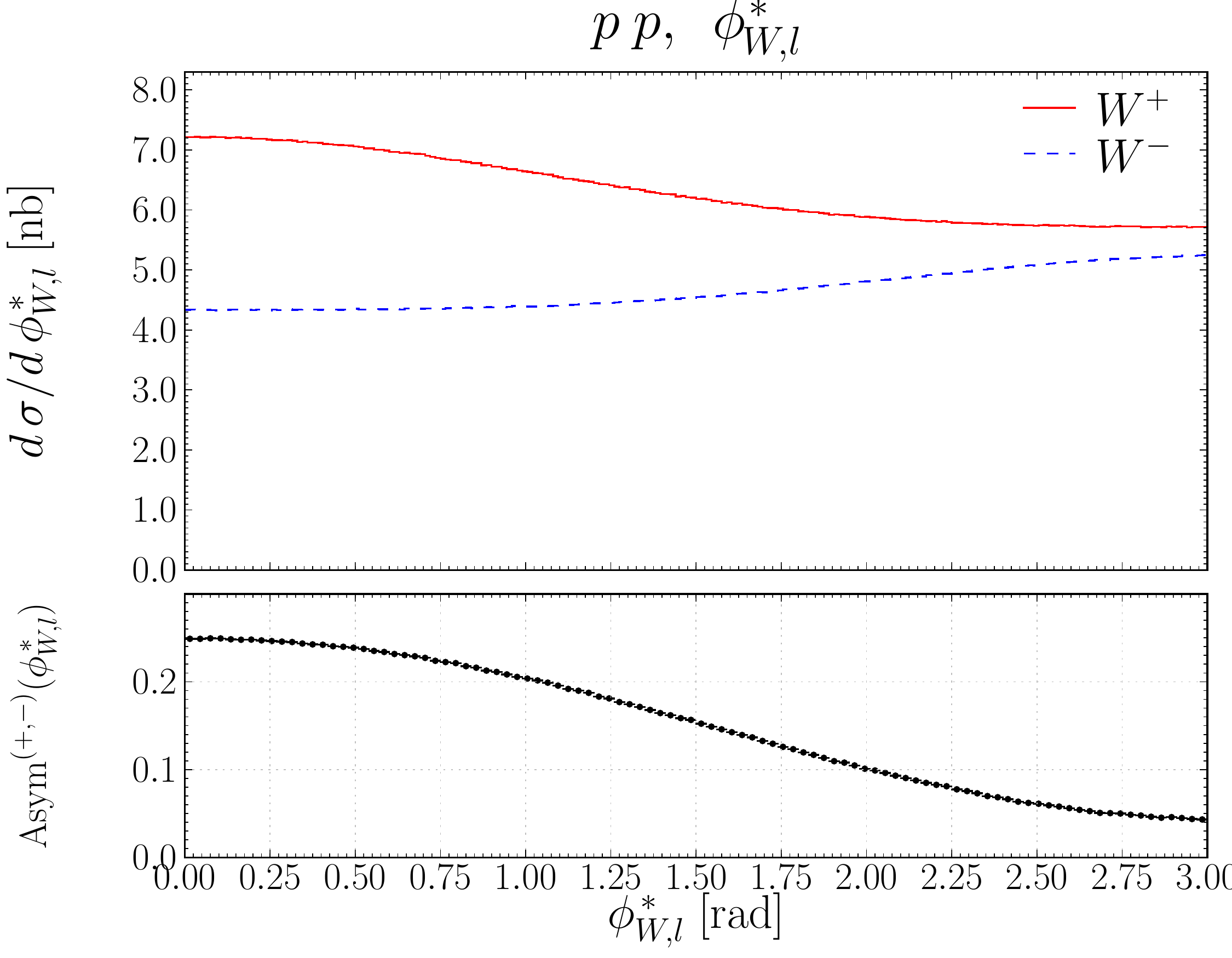}
    \caption[Distributions of $\phi_{W,l}^{\,\ast}$ in $\pp$ collisions]
            {\figtxt{Distributions of $\phi_{W,l}^{\,\ast}$ in $\pp$ collisions showing the azimuthal
                anisotropy in the charged leptons decays.}}
            \label{fig_pp_phi_histos}
  \end{center} 
\end{figure}

Figure~\ref{fig_pp_phi_histos} quantifies this important aspect at the inclusive level by displaying the
$\phi_{W,l}^{\,\ast}$ spectra which is defined as the azimuthal angle of the charged lepton in the \WRF{} with 
respect to the direction of the transverse momentum of the $W$ boson in the laboratory frame, that is
\begin{equation}
\phi^{\,\ast}_{W,l} \equiv \cos^{-1}\left(\frac{\vec p_{T,W} \cdot \vec p^{\,\ast}_{T,l}}
{|\vec p_{T,W}|\, |\vec p^{\,\ast}_{T,l}|}\right).
\end{equation}

In conclusion the absence of valence anti-quarks makes the production of $\Wp$ and $\Wm$ at the
LHC irretrievably different.
The increase of the transverse momenta of the $W$~bosons produced at high absolute rapidity 
amplifies the impact of the $W$~boson polarisation asymmetry on the charge asymmetry of its decay 
products. These effects will give rise to important measurement biases which were not present in the 
$\ppbar$ collision mode but will show up in the measurement of the $W$~boson properties in the LHC 
$\pp$ collision mode.

\begin{figure}[!h] 
  \begin{center}
    \includegraphics[width=0.495\tw]{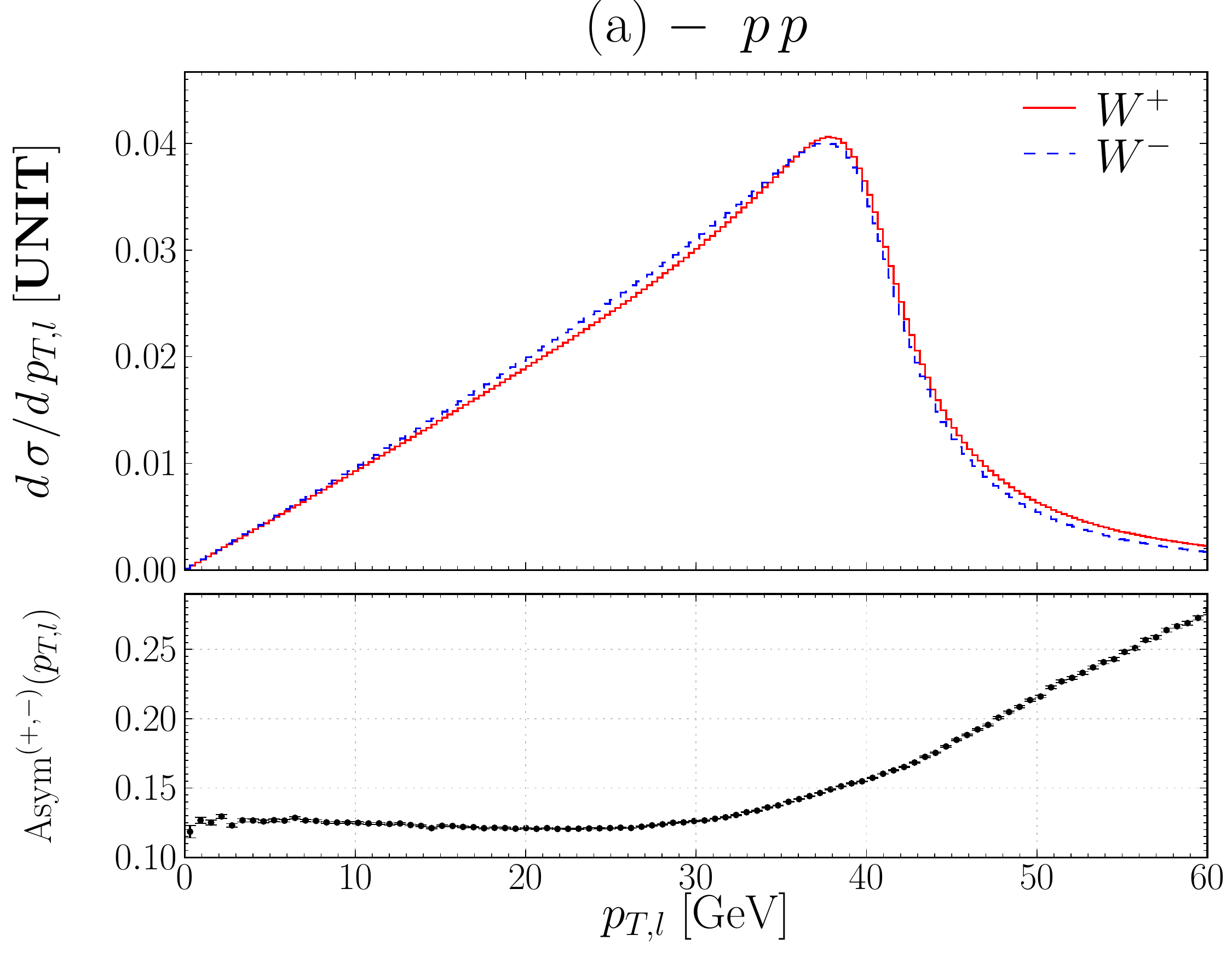}
    \hfill
    \includegraphics[width=0.495\tw]{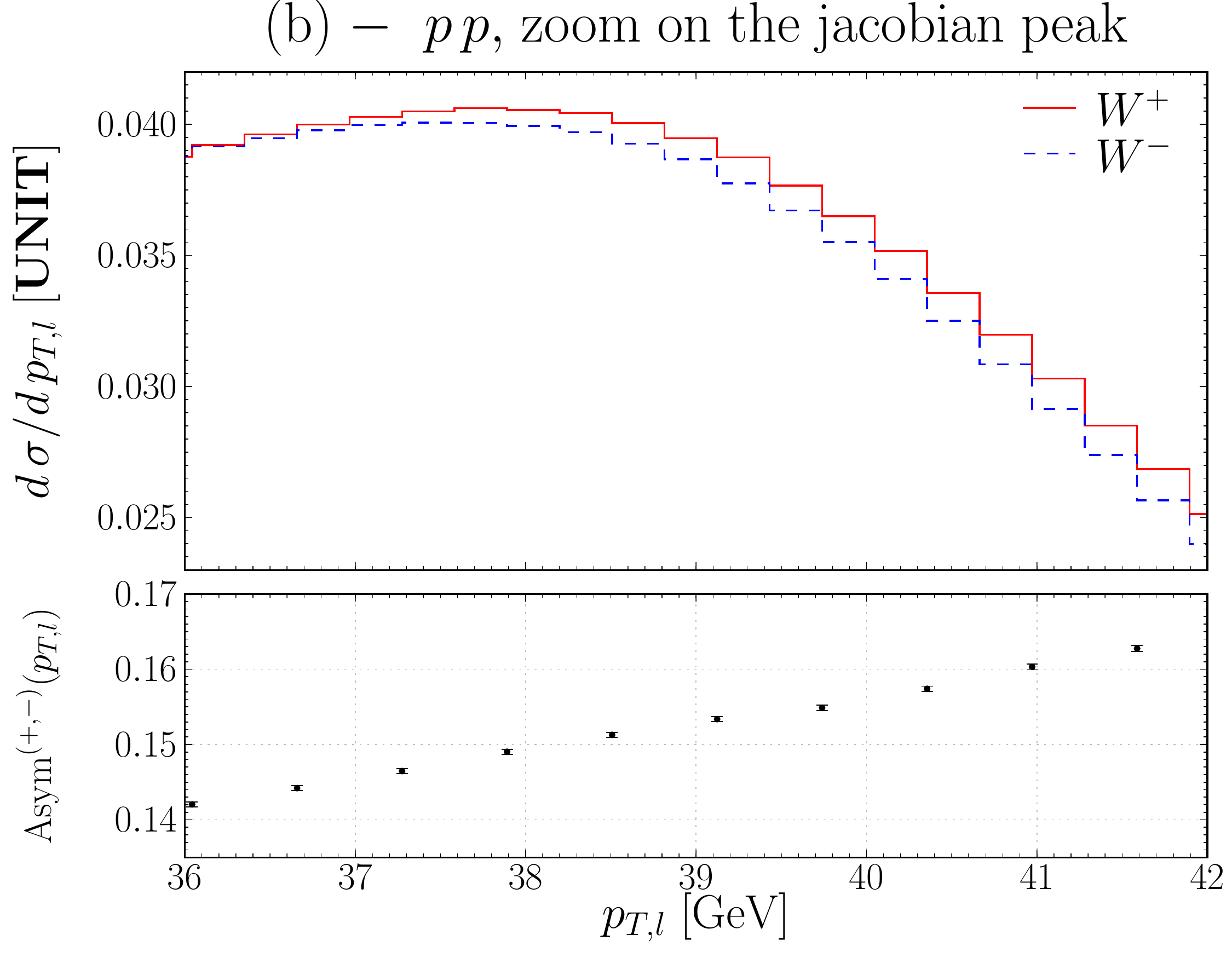}
    \caption[Positive and negatively charged leptons transverse momenta distributions normalised
    to unit]
            {\figtxt{Positive and negatively charged leptons transverse momenta distributions 
                normalised to unit, for the whole range in (a) and for a zoom on the jacobian peak 
                (b). Note the charge asymmetries are obtained from histograms normalised to nb.}}
            \label{fig_pTlp_pTlm_unit}
  \end{center} 
\end{figure}

In order to illustrate a little bit more the charge asymmetry in $\pTl$  Figure%
~\ref{fig_pTlp_pTlm_unit}.(a) shows the imperfect match between the $\flatDfDx{\sigma^+}{\pTl}$ and 
$\flatDfDx{\sigma^+}{\pTl}$ histograms when normalised to unit.
This demonstrates that the difference of scale being removed the discrepancies are important enough 
to be noticed by the eye. Especially frame (b) shows a zoom on the jacobian peak which shape is
of crucial importance for the extraction of the $W$ boson mass. In particular we can distinguish
a particular feature, that is at the LHC energies we are to expect\,:
\begin{equation}
\Mean{\pTlm}<\Mean{\pTlp},
\end{equation}
where $\Mean{\pTl}$ is the average $\pT$ of the charged lepton $l$.
To be more precise, using the \WINHAC{} Monte Carlo event generator for a statistic of $100,000$ inclusive
events we find
\begin{eqnarray}
\Mean{\pTlp} &\approx& 40.17\GeV, \label{eq_mean_pTlp_wh} \\ 
\Mean{\pTlm} &\approx& 39.98\GeV. \label{eq_mean_pTlm_wh}
\end{eqnarray}
\index{Charged lepton@Charged lepton from $W$ decay!Transverse momentum|)}
\index{Charged lepton@Charged lepton from $W$ decay!Pseudo-rapidity|)}

\subsection{More details on the leptons transverse momenta charge asymmetry in $\mbf{pp}$ collisions}
\begin{figure}[!h] 
  \begin{center}
    \includegraphics[width=0.495\tw]{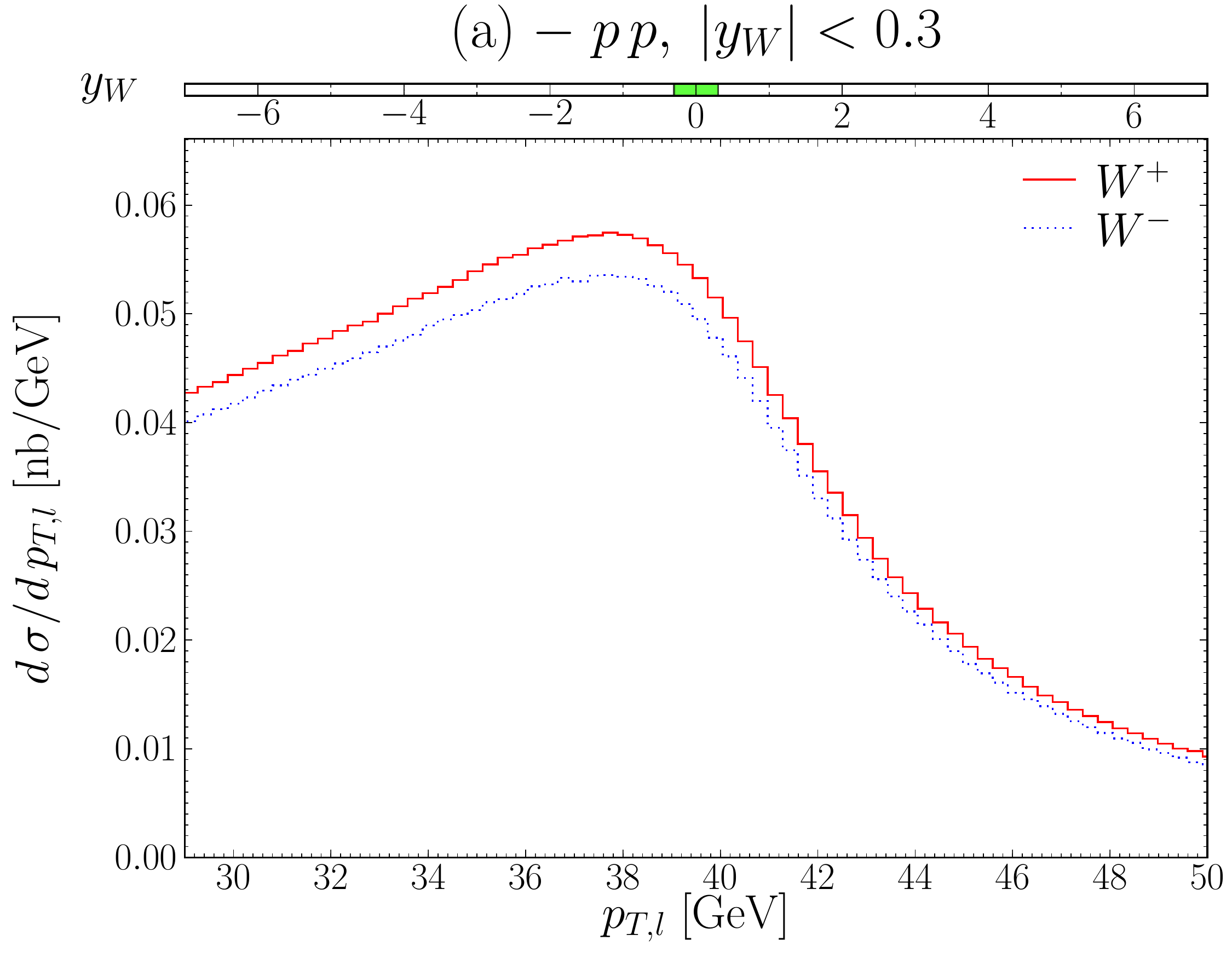}
    \hfill
    \includegraphics[width=0.495\tw]{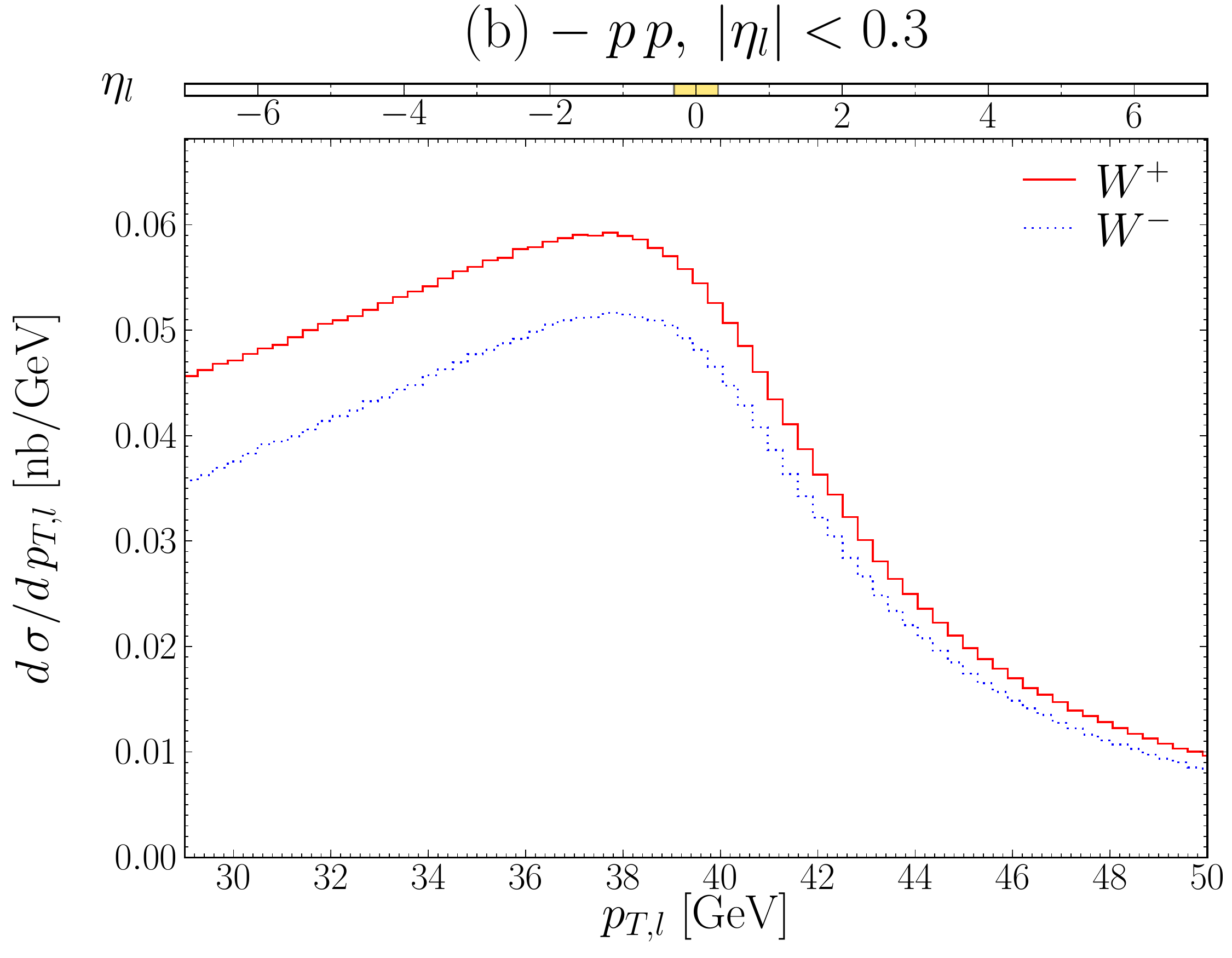}
    \vfill
    \includegraphics[width=0.495\tw]{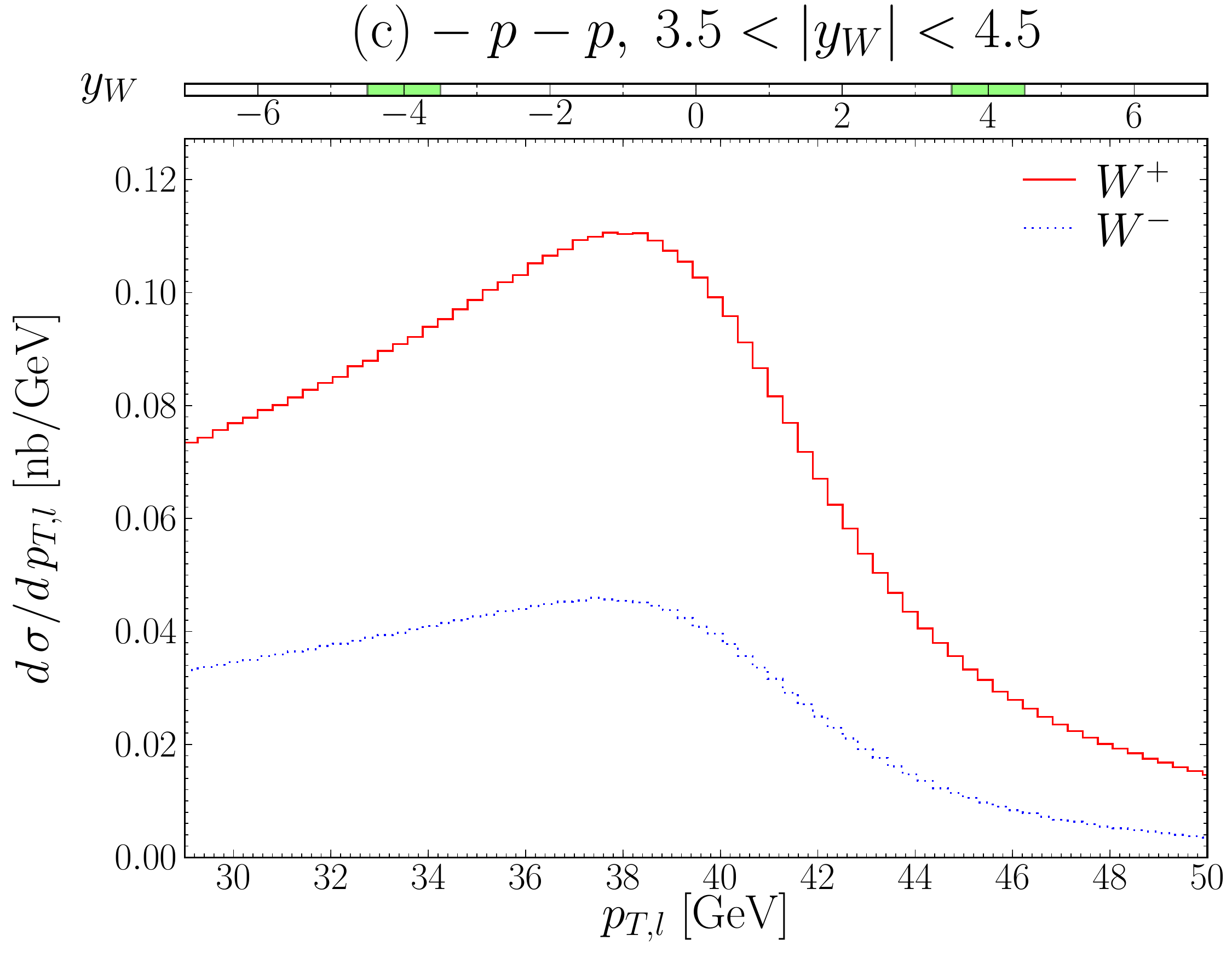}
    \hfill
    \includegraphics[width=0.495\tw]{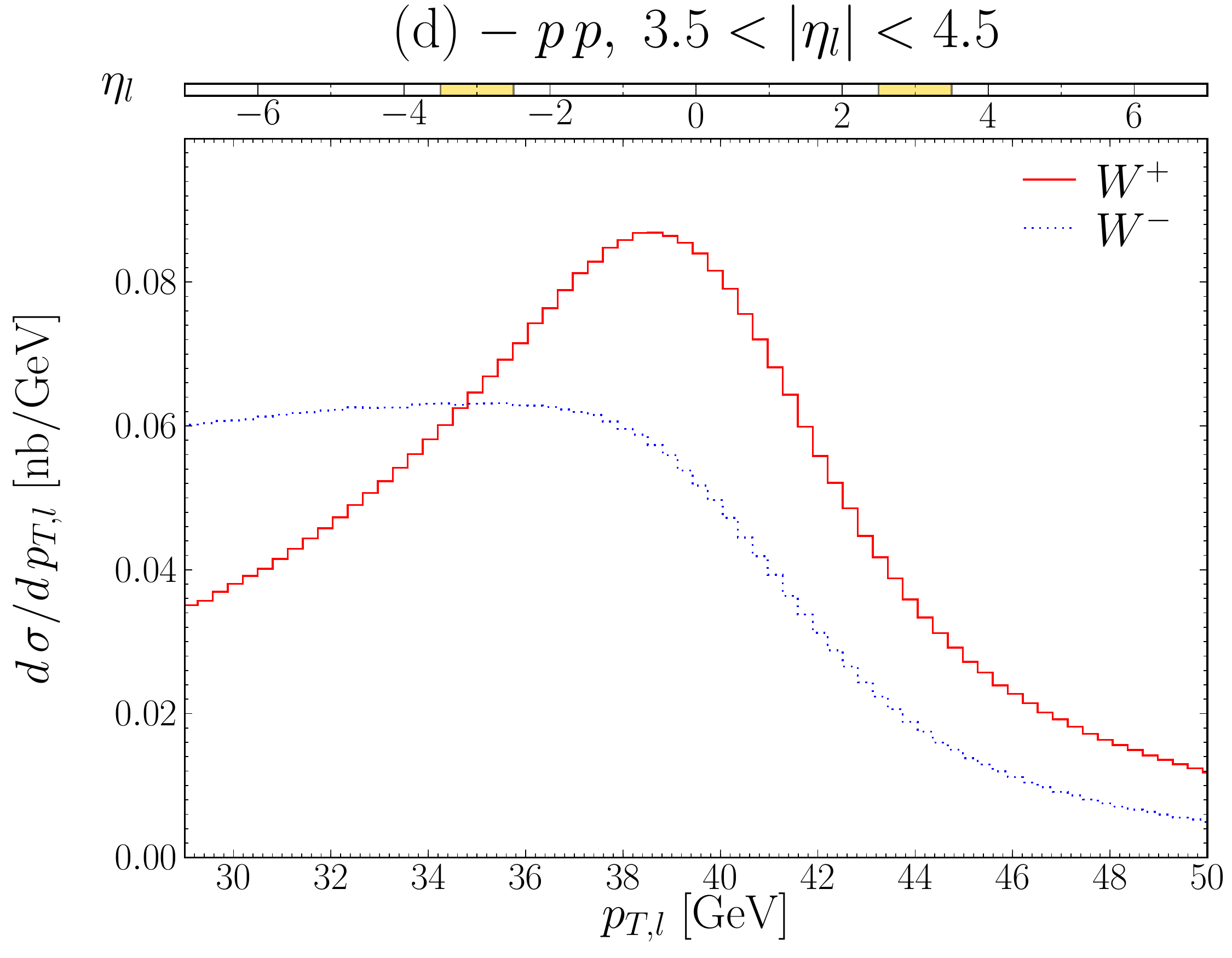}
    \caption[Charged lepton transverse momenta for central and forward region of $\yW$ 
      and a $\etal$ binning]
            {\figtxt{Charged lepton transverse momenta for central (a) and forward (b) 
                region of $\yW$ selection.}}
            \label{fig_pTl_JP_ZOOM}
  \end{center} 
\end{figure}

\index{Charged lepton@Charged lepton from $W$ decay!Transverse momentum|(}
To analyse a little bit more the relative position of the jacobian peaks between the two channels,
$\pTl$ distributions were studied for narrow central and forward $\yW$ and $\etal$,
respectively $|\yW|,|\etal|<0.3$ and $3.5<|\yW|,|\etal|<4.5$ selections as shown in 
Fig.~\ref{fig_pTl_JP_ZOOM}. The selection in $\etal$ being justified to show how the ambiguity
becomes large when making a selection with the latter. This gives complements on the
low $x$-selection resolution one have to deal when relying on $\etal$ rather than $\yW$.
To quantify the impact on the jacobian peak position we applied an empirical procedure consisting
into fitting $\pTlp$ and $\pTlm$ spectra by a polynomial of second order in the 
vicinity of the peak. This range was taken to be $37\GeV<\pTl<40\GeV$.
The $\pTl$ value corresponding to the two parabolas are noted $\varpi_+$ for $\pTlp$ and $\varpi_-$ 
for the one of $\pTlm$. The difference between these two position $\varpi_+-\varpi_-$ is considered
to quantify the variation of the peaks position. The results are gathered in the table below
\medskip
\begin{center}
\renewcommand\arraystretch{1.2}
\begin{tabular}{ccc}
  \hline
   phase space domain & $\varpi_+-\varpi_-$ [MeV]\\
  \hline\hline
  Inclusive           &   $170$         \\
  \arrayrulecolor{Greymin}
  \hline
  \arrayrulecolor{Black}
  $|\yW|  <0.3$       &  $\!\!\!\!-100$ \\
  $|\etal|<0.3$       &  $\!\!\!\!-240$ \\
  \arrayrulecolor{Greymin}
  \hline
  \arrayrulecolor{Black}
  $3.5<|\yW|  <4.5$   &   $300$         \\
  $3.5<|\etal|<4.5$   &  $\!\!\!2000$ \\
  \hline
\end{tabular}
\renewcommand\arraystretch{1.45}
\end{center}
\bigskip

As can be seen the difference are varying a lot depending on the cut compared to the ``reference''
from the inclusive case. Nonetheless let us remind those results correspond only to purely 
empirical attempt to quantify the $\pTl$ shapes relative variations observed in 
Fig.~\ref{fig_pTl_JP_ZOOM}.
\index{W boson@$W$ boson!Decay in pp@Decay in $\pp$ collisions|)}
\index{Charged lepton@Charged lepton from $W$ decay!Transverse momentum|)}

\subsection{Summary on the sources of charge asymmetries in proton--proton collisions}
To summarise, at the LHC, the difference between the positive and negative kinematics of the 
charged leptons result from the interplay of the three following effects\,:
\begin{itemize}
\item[(1)] the $V-A$ coupling \index{Electroweak!VmA@$V-A$ coupling} 
  of the $W$~boson to fermions in electroweak interactions, \ie{}
  only positive helicity $\lp$ and negative helicity $\lm$ couples to the $W$ bosons,
\item[(2)] the absence of valence anti-quarks to perfectly match via $CP$ the charge asymmetries 
  induced by the valence quarks,\index{Symmetry!CP@$CP$}
\item[(3)] the non-zero transverse momentum of the $W$~boson or, more precisely, the fact that
  in a $q^\val\,{\bar{q}{}'}^{\sea}$ collision the sea quarks carries most of the time the higher 
  transverse motion with respect to the valence quark.
\end{itemize}

These asymmetries will depend strongly on the choice of the kinematic region used in the analysis.
If expressed in terms of leptonic variables, the differences are amplified due to induced biases in 
the effective $x$-regions of partons producing positively and negatively charged $W$~bosons. 
As these differences could mimic the asymmetry in the masses of positively  and negatively charged 
$W$~boson, all these effects must be controlled to a high precision and/or, as advocated in our 
work, eliminated by using LHC dedicated measurement strategy.  

Finally let us remind the reader the above presentation is addressing only the essential points to
understand the original LHC features. The Appendix that follows takes back from the beginning this
presentation in a more pedestrian manner and provide much more details to understand thoroughly all
effects participating to this charge asymmetries.
\index{Helicity!Of the colliding quarks and the decaying leptons|)}

\cleardoublepage
\begin{subappendices}
\makeatletter\AddToShipoutPicture{%
\AtUpperLeftCorner{2cm}{2cm}{\ifodd\c@page\else\makebox[0pt]{\Huge$\bullet$}\fi}%
\AtUpperRightCorner{0cm}{2cm}{\ifodd\c@page\makebox[0pt]{\Huge$\bullet$}\else\fi}%
}\makeatother

\section{Detailed description of W in Drell--Yan for $\mathbf{p\bar p}$,
$\mathbf{p\,p}$ and $\mathbf{d\,d}$ collisions}\label{app_ppbar_pp_dd}
\setlength{\epigraphwidth}{0.7\tw}
\epigraph{
Mister Fantastic\,: 
``This device apparently caused sub-atomic particle dissociation, reducing us,
as we entered, to proto-matter, which it stored until it teleported us here, to pre-set
coordinates in space where it reassembled us inside a self generated life-support environment !''\\
The Hulk\,: ``That's obvious Richards !'' \\
Iron Man\,: ``Obvious ? What'd he say ?'' \\
The Torch\,: ``Just hang out Iron man !
Reed will get tired of talking in five dollar words in a minute, and then he'll explain in English !
Then he'll explain it again to the Thing in one-syllable words !''\\
The Thing\,: ``Hey Torch-- why don'tcha just shut up and look awestruck like the rest of us ?''
}%
{\textit{Marvel Superheroes Secret Wars \#1 - The War begins (May 1984)}}

This Appendix presents with more details the studies carried out to understand thoroughly 
the asymmetries between the $\Wp$ and $\Wm$ production in Drell--Yan for $\ppbar$, $\pp$ and $\dd$ 
collisions.
To reach this goal a few observables related to the $W$ boson and the charged leptons were considered,
each allowing to comprehend both dynamic and kinematic issues from different point of views.
Although some histograms below were not addressed in the core of the Chapter it has to be 
realised that the material presented here is the bulk of the work achieved which allowed to present
previously the gist of $W$ physics features in Drell--Yan at the LHC.

\subsection{Observables and context of the discussion}
The study the $W$ boson properties is made looking at its invariant mass $m_W$ defined like
\begin{equation}
m_W \equiv \sqrt{E_W^2-\vec{p}_W^{\;2}},
\end{equation}
its rapidity $\yW$ and its transverse momentum $\pTW$.

The study of the charged lepton properties are made by analysing its pseudo-rapidity $\etal$ and its 
transverse momentum $\pTl$. 
The study of the charged lepton angular decay in the \WRF{} relatively to the $W$ motion
is quantified with the angle $\theta^\ast_{W,l}$ defined as the opening angle between the 
charged lepton in the \WRF{} and the direction of the $W$ boson in the laboratory frame, which 
analytically reads
\begin{equation}
\cos\theta^\ast_{W,l} \equiv \frac{\vec p_W \cdot \vec p^{\,\ast}_l}{|\vec p_W|\, |\vec p^{\,\ast}_l|}.
\end{equation}
Due to the importance of the transverse direction for the extraction of the $W$ properties we consider
as well the angle $\phi^\ast_{W,l}$ corresponding to the opening angle of the charged lepton
transverse momentum in the \WRF{} with the direction of the $W$ transverse momentum in the laboratory
frame, which analytically translates to 
\begin{equation}
\phi^\ast_{W,l} \equiv \cos^{-1}\left(\frac{\vec p_{T,W} \cdot \vec p^{\,\ast}_{T,l}}
{|\vec p_{T,W}|\, |\vec p^{\,\ast}_{T,l}|}\right).
\end{equation}
Some of the previous variables are studied in bins of $\yW$ and $\etal$ to emphasise the 
variation of the charge asymmetries in function of the phase space region considered. 
Not to overload the discussion all those observables will not be discussed for each case but rather
the most important one to draw relevant conclusions.

These variables being reminded we state a few conventions adopted throughout the rest of the Appendix.
First, let us remind the notation $W$ indifferently stands for $\Wp$ or $\Wm$ and that 
respectively $l$ stands for $\lp$ or $\lm$.

Now, for convenience, the production of $W$ bosons is considered to be the addition of two type of 
contributions\,: (1) the contributions involving at least one valence (anti-)quark and 
(2) the contributions involving only sea quarks.
Due to the isospin symmetry $q^\sea=\qbar^\sea$ assumed in our model (CTEQ6.1M)
\index{Parton Distribution Functions (PDFs)!CTEQ}
the observed charged asymmetries are, as we will see, to be charged to the contributions of type (1).
In top of that, for argumentation, we will always consider regions of the phase space where these 
valence (anti-)quarks asymmetries contribute the most, \ie{} we will always implicitly assume in our 
examples the most probable cases where the valence (anti-)quark has a more larger longitudinal momentum
than the one of the sea (anti-)quark it enters in collisions with\,:  $x_{q,\qbar}^\val>x_{q,\qbar}^\sea$.
In the case of protons--anti-protons collisions valence quark and valence 
anti-quarks collisions happen too but they can be neglected in the discussion.
Indeed, at $\sqrt S =14\TeV$, if one of the valence (anti-)quark possesses a high fraction of the 
longitudinal momentum of the hadron $\smartpap$ it belongs to, say $10^{-1}\lesssim x_1$, then the other 
valence (anti-)quark in the collision must carry, due to the rough constrain $x_1\,x_2\,S\sim M_W^2$, 
a fraction of momentum verifying $x_2\lesssim 4\times 10^{-4}$. As Figure~\ref{fig_cteq61m} 
shows it, finding a valence (anti-)quark at such low $x$ is negligible.
Hence, in $\ppbar$ or LHC collisions, when addressing issues centered on valence (anti-)quark,
we will always discuss valence--sea collisions that hereafter will be mentioned as 
``valence'' collisions/contributions.
\index{Quarks!Valence quarks}\index{Quarks!Sea quarks}

Another important aspect in the upcoming discussion consists to address the decaying angle of the 
charged leptons with respect to the $W$ motion. 
For that matter the emphasis will be given to the transverse polarisation states of the $W$
which gives different decaying angle between the positively and negatively charged lepton
as seen in Eq.~(\ref{eq_WT_lep_decay}). 
For the longitudinal $W$ bosons (Eq.~(\ref{eq_WL_lep_decay})) the decays of the leptons
are charge independent, hence the charge discrepancies that can be observed are only due to
the difference between the kinematics of the $\Wp$ and the $\Wm$ specific to the collision mode
under consideration.
These charge asymmetries are of lesser importance compared to the one consequent to the decay of 
transversely polarised $W$ bosons and will maybe studied in future works.
\index{W boson@$W$ boson!Polarisation!Transverse states}

Also let us stress that some of the plots and explanations below are already present in the core of
the Chapter but, for convenience, they were reviewed to make this Appendix fully understandable and 
complete by itself.

In what follows the description of hadron--hadron collisions are considered for the cases of\,:
\begin{itemize}
\item[-] Hadron--hadron collisions with no valence quark inside the hadrons with $\sqrt S=14\TeV$.
\item[-] $\ppbar$ collisions with $\sqrt S=14\TeV$.
\item[-] $\pp$ collisions with $\sqrt S=14\TeV$.
\item[-] $\dd$ collisions with $\sqrt S=7\TeV$.
\end{itemize}
where each time $\sqrt S$ refers to the energy in the nucleon--nucleon center of mass inertia frame.
After those points a discussion on the relative transverse motion of the quarks and anti-quark is
provided.

\clearpage
\subsection{A study of the purely sea contributions}
\begin{figure}[!h] 
  \begin{center}
    \includegraphics[width=0.495\tw]{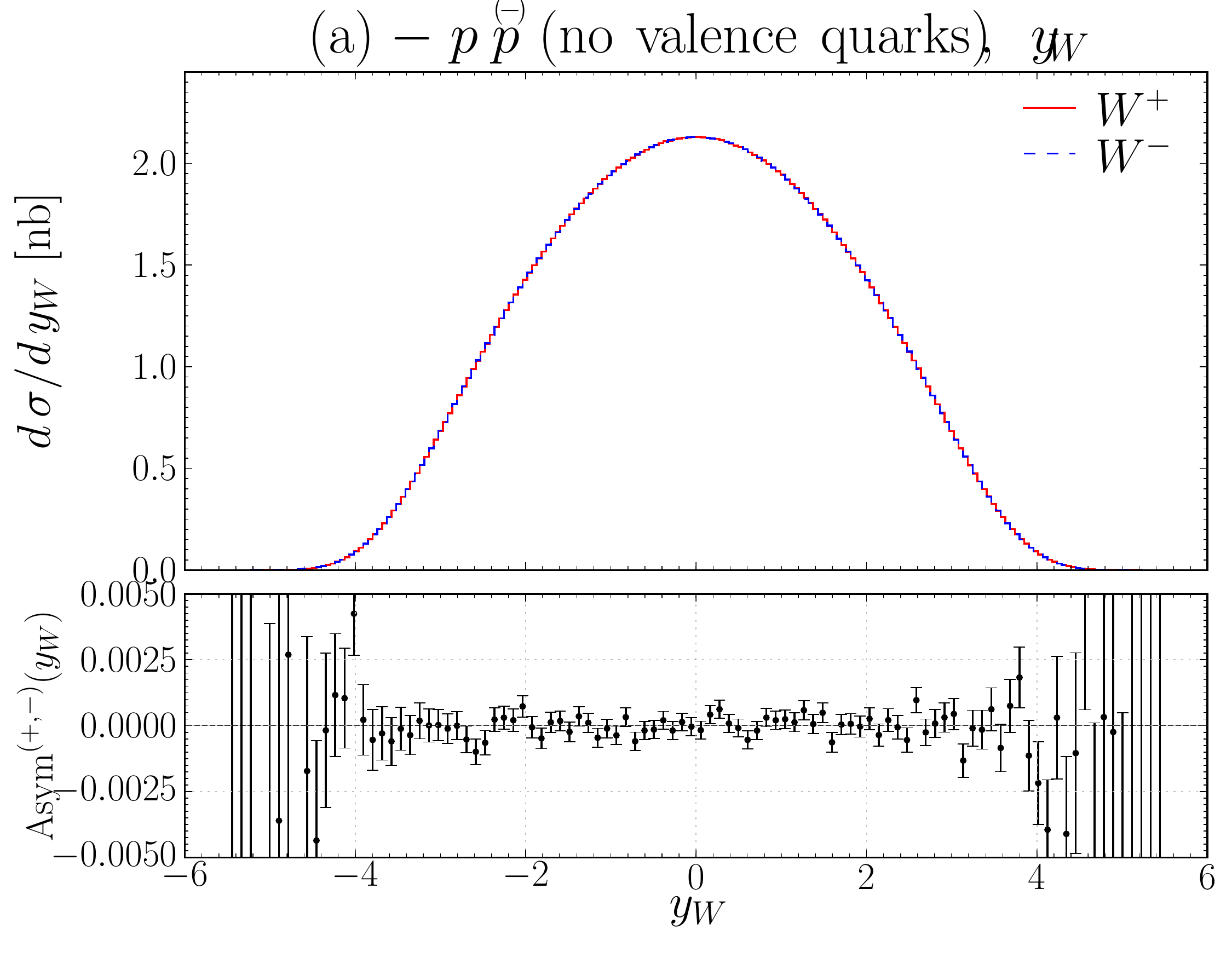}
    \hfill
    \includegraphics[width=0.495\tw]{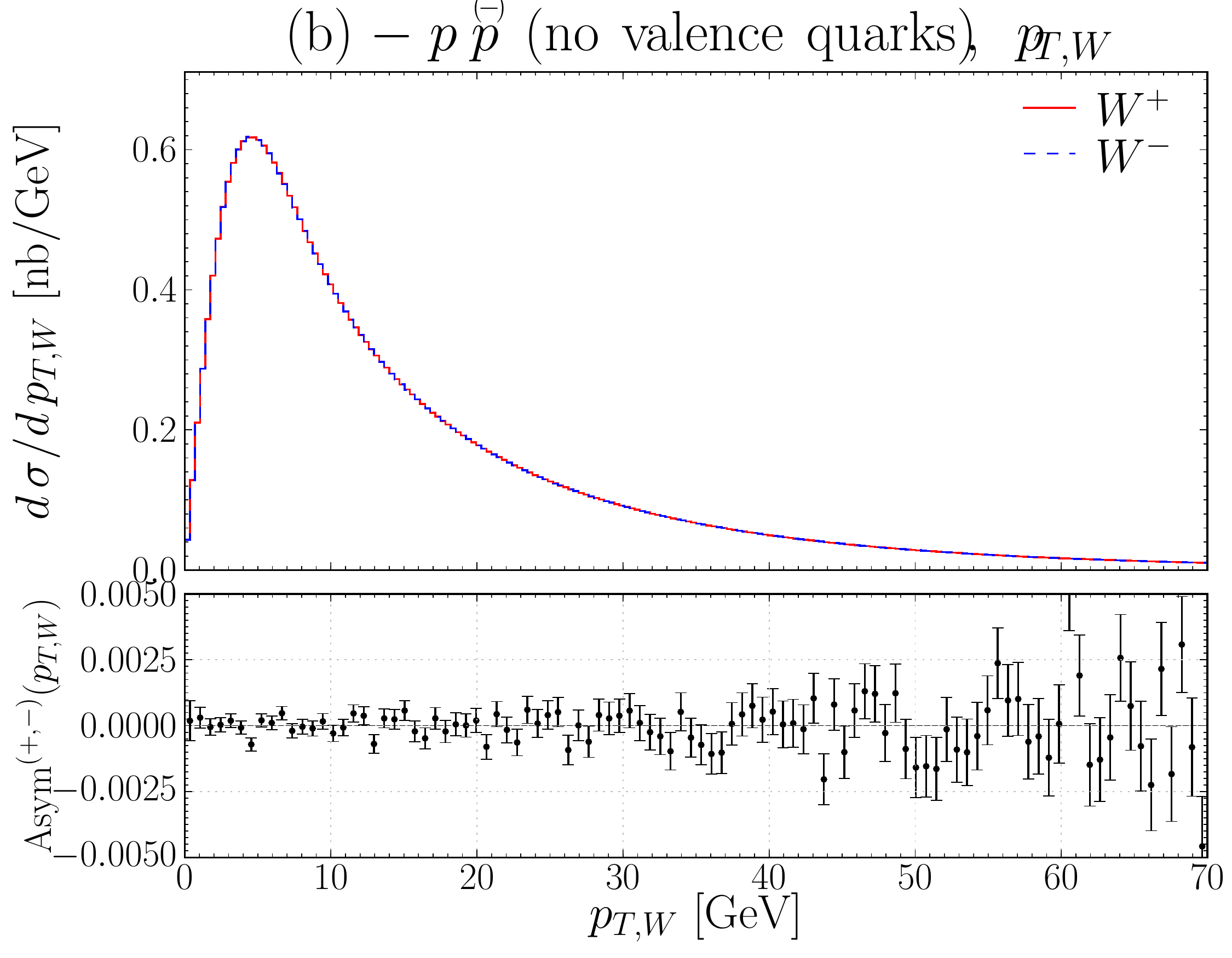}
    \vfill
    \includegraphics[width=0.495\tw]{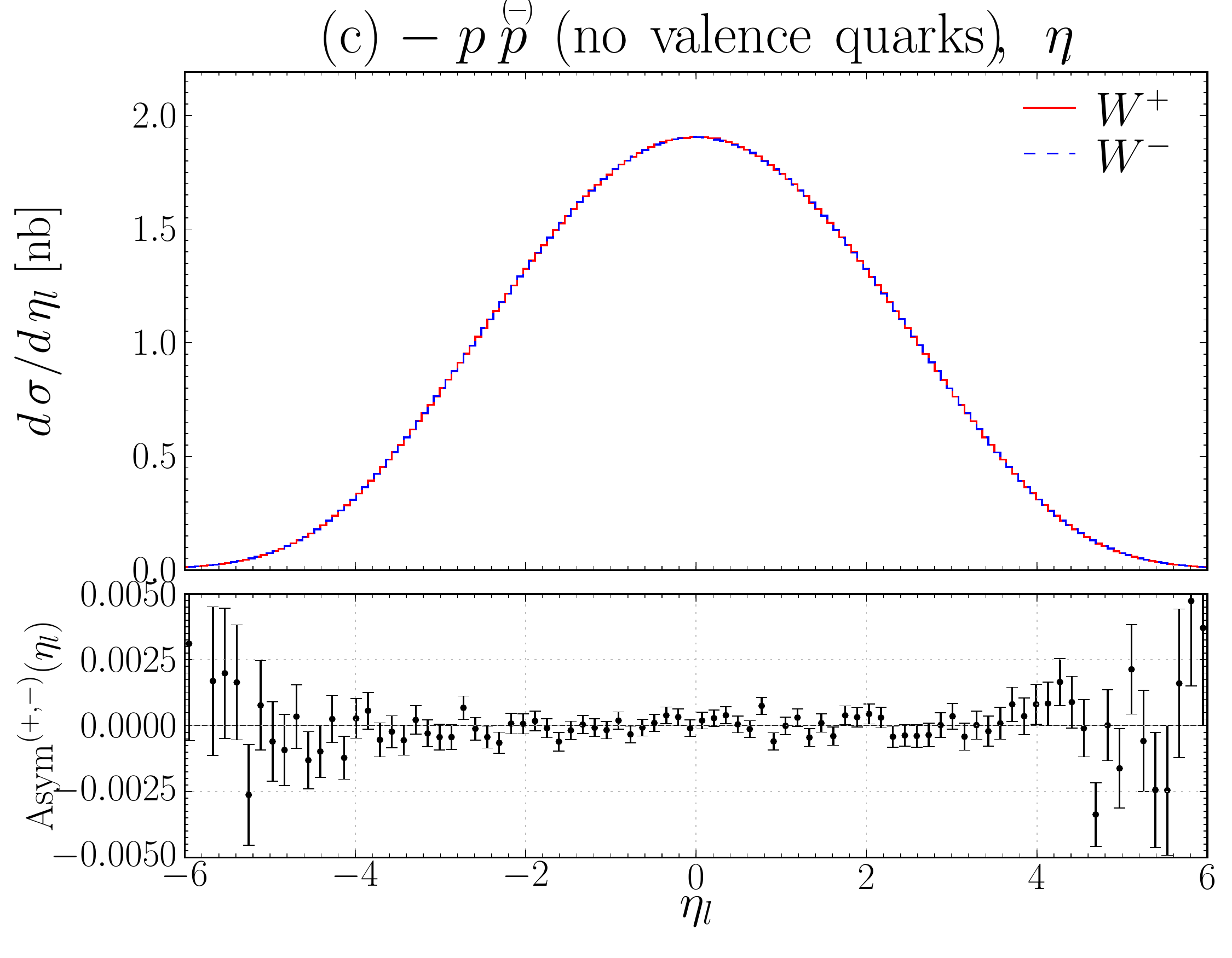}
    \hfill
    \includegraphics[width=0.495\tw]{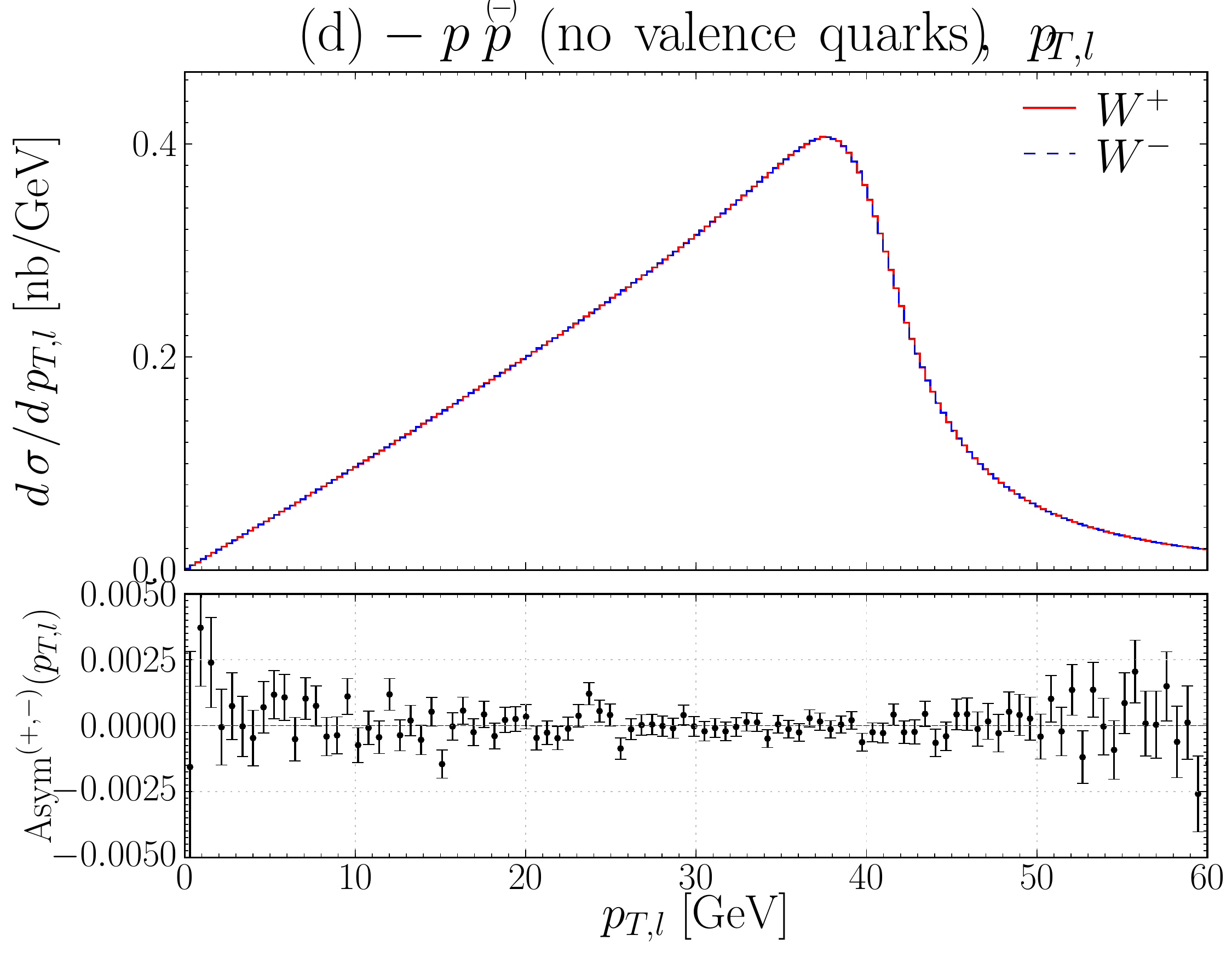}
    \caption[Distributions of the $W$ boson rapidity and transverse momentum along with the one of the charged lepton 
      pseudo-rapidity and transverse momentum in $\pp(\ppbar)$ collisions when all valence contributions are removed]
            {\figtxt{Distributions of the $W$ boson rapidity (a) and transverse momentum (b) 
                along with the one of the charged lepton pseudo-rapidity (c) and transverse momentum (d)
                in $\pp(\ppbar)$ collisions when all valence contributions are removed.}}
            \label{app_pp_NOVAL_yW_pTW_etal_pTl}
            \index{W boson@$W$ boson!Transverse momentum}
            \index{W boson@$W$ boson!Rapidity}
            \index{Charged lepton@Charged lepton from $W$ decay!Transverse momentum}
            \index{Charged lepton@Charged lepton from $W$ decay!Pseudo-rapidity}
  \end{center} 
\end{figure}

\index{W boson@$W$ boson!Production pp without valence quarks@
Production in $\pp$ or $\ppbar$ collisions without valence quarks contributions|(}
We start this discussion by looking at the fictitious case where there is no valence (anti-)quarks in 
the hadrons entering in collision.
This illustrates that in the assumed isospin symmetry there are absolutely no charge asymmetries.
For that matter we remove manually the valence terms from the CTEQ6.1M 
\index{Parton Distribution Functions (PDFs)!CTEQ}
previsions and run $\pp$ collisions which, in the present context, is equivalent to the $\ppbar$ mode, explaining the 
$p\smartpap$ notation afterward.
We obtain for the observables $\yW$, $\pTW$, $\etal$ and $\pTl$ the histograms shown in 
Fig.~\ref{app_pp_NOVAL_yW_pTW_etal_pTl}.

Here let us emphasise the charge symmetry is perfect, by that we mean that in a given point $d\,\Phi$ 
of the phase space the associated infinitesimal cross sections $d\,\sigma^+/d\,\Phi$ and 
$d\,\sigma^-/d\,\Phi$ are exactly equal. This allows us to get rid in the rest of the discussion.

\begin{figure}[!h] 
  \begin{center}
    \includegraphics[width=0.495\tw]{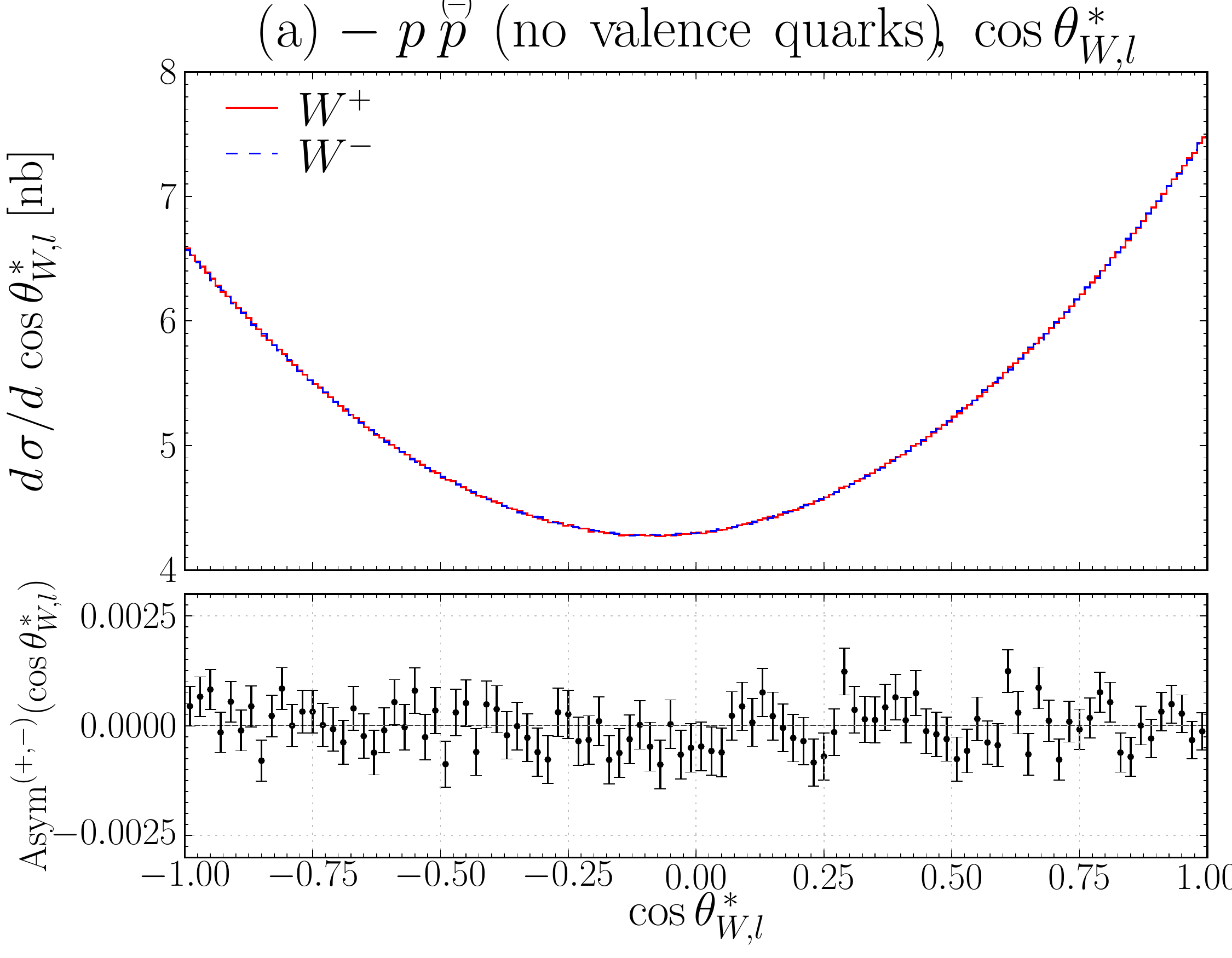}
    \hfill
    \includegraphics[width=0.495\tw]{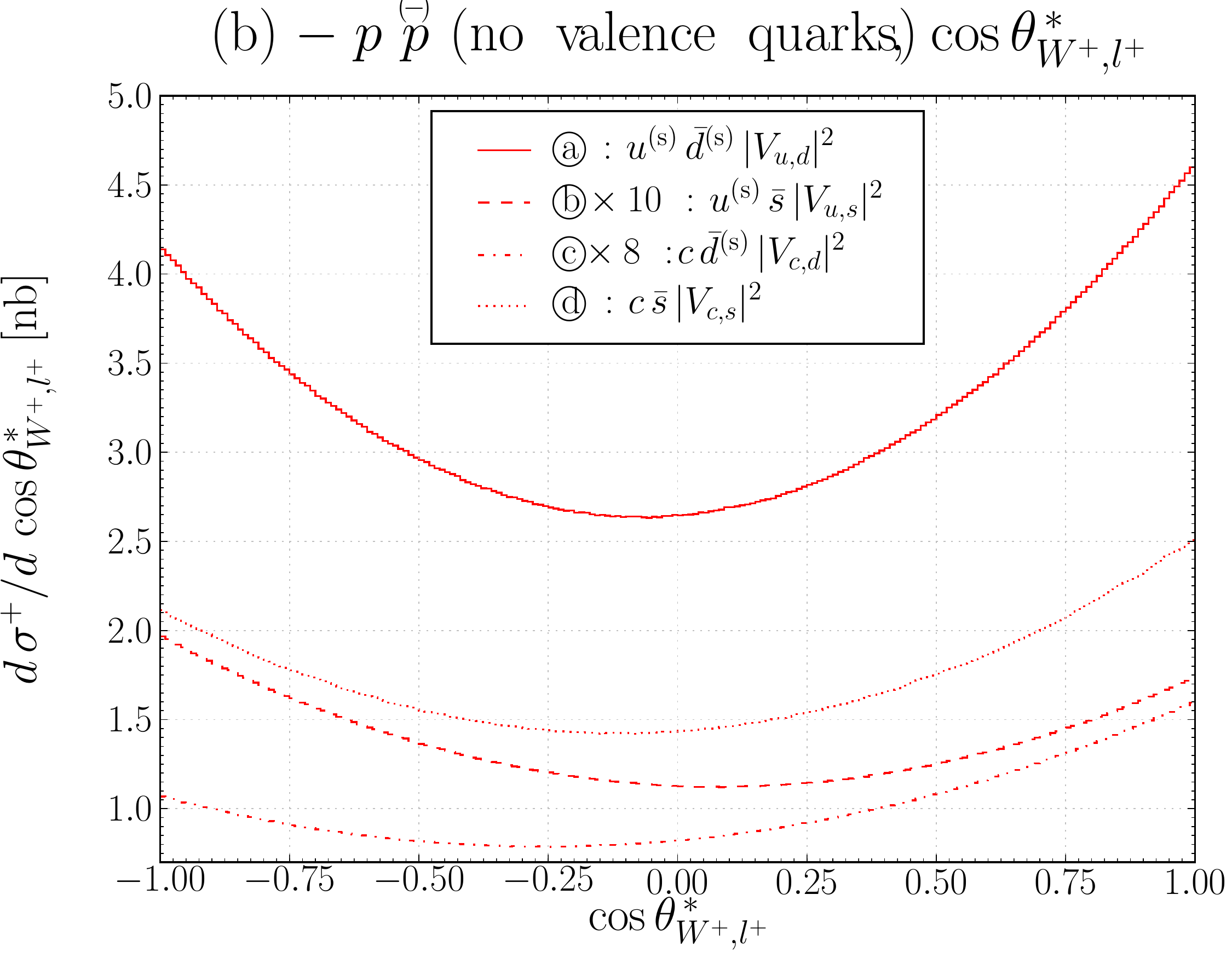}    
    \caption[Distributions of $\costhetaWlwrf$ in $\pp(\ppbar)$ collisions with no valence quarks
    in the hadrons, first with all the sea quarks contributions and then with all the contributions
    from the CKM point of view]
            {\figtxt{Distributions of $\costhetaWlwrf$ in $\pp(\ppbar)$ collisions with no valence 
                quarks in the hadrons, first with all the sea quarks contributions (a) and then with 
                the four main contributions from a CKM point of view (b).}}
            \label{app_pp_NOVAL_costhetaWlwrf}\index{Electroweak!CKM matrix elements}
  \end{center} 
\end{figure}

\index{W boson@$W$ boson!Polarisation!Transverse states|(}
Nonetheless, let us mention that the difference in the PDFs distributions induce an asymmetry between
the left and right polarisation states of the $W$. Indeed, as can be seen in 
Fig.~\ref{app_pp_NOVAL_costhetaWlwrf} the leptons preferentially decay in the direction of the 
momentum of the $W$ which means --using helicity conversation rule of thumb-- that 
$\Wp(\lambda=-1)<\Wp(\lambda=+1)$ and $\Wm(\lambda=-1)>\Wm(\lambda=+1)$.
This can be understood by considering the four main contributions to the $\Wp$ production. 
If we ignore the negligible contributions involving the $\bbar$ flavour we are left
with\,: $\textcircled{a}\,:\; u^\sea\,\dbar^\sea\,\Vckmsqr{u}{d}$,
$\textcircled{b}\,:\; u^\sea\,\sbar\,\Vckmsqr{u}{s}$,
$\textcircled{c}\,:\; c\,\dbar^\sea\,\Vckmsqr{c}{d}$,
$\textcircled{d}\,:\; c\,\sbar\,\Vckmsqr{c}{s}$.
In the case $\textcircled{b}$ the quark $u^\sea$ possesses in general a higher longitudinal momentum 
than the quark $\sbar$ it enters in collision with which means
$x_{u^\sea}>x_{\sbar} \Rightarrow \Wp(\lambda=-1)>\Wp(\lambda=+1)$. 
On the other hand in the cases $\textcircled{a}$, $\textcircled{c}$ and $\textcircled{d}$, 
it is the other way around\footnote{To a good approximation for $\textcircled{a}$ 
we have $x_{u^\sea}\sim x_{\dbar^\sea}$ except for $10^{-4}<x$ where $x_{u^\sea}\lesssim x_{\dbar^\sea}$.},
\ie{} $x_\qbar>x_q$ which means now $\Wp(\lambda=-1)<\Wp(\lambda=+1)$.
This hypothesis is proved when looking at the individual behaviour of these four contributions  
in Fig.~\ref{app_pp_NOVAL_costhetaWlwrf}.(b), 
where, to fit the present range in coordinate, the $\textcircled{b}$
and $\textcircled{c}$ contributions are respectively multiplied by factors $10$ and $8$.
Actually this shows the predominance of right $\Wp$ is essentially the consequence of the dominating
$\textcircled{a}$-terms that overrule the other ones.
\index{W boson@$W$ boson!Polarisation!Transverse states|)}

\begin{figure}[!h] 
  \begin{center}
    \includegraphics[width=0.5\tw]{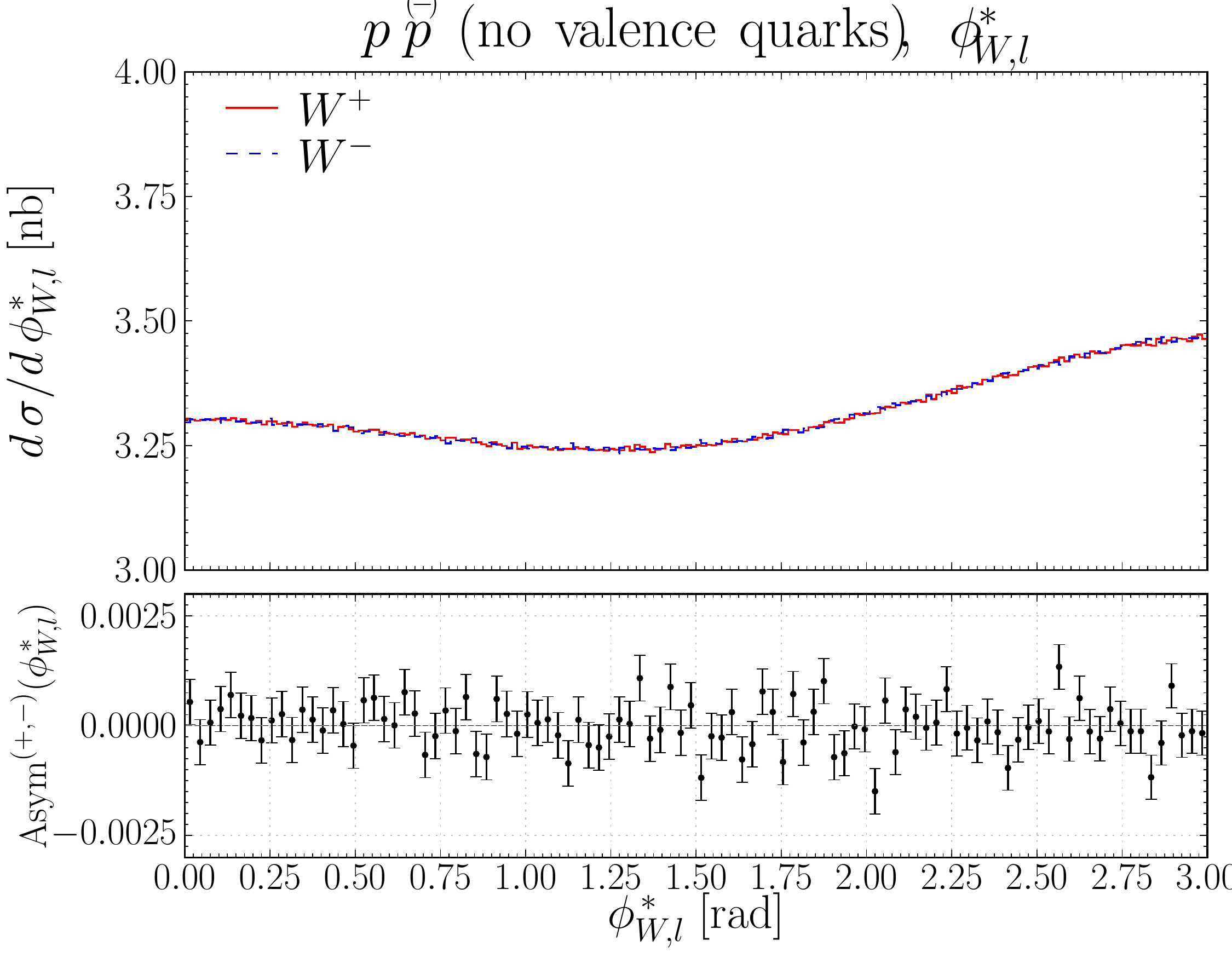}
    \caption[Distributions of $\phi_{W,l}^{\,\ast}$ in $\pp(\ppbar)$ collisions with no valence quarks
      in the hadrons]
            {\figtxt{Distributions of $\phi_{W,l}^{\,\ast}$ in $\pp(\ppbar)$ collisions with no valence 
                quarks in the hadrons}}
            \label{app_pp_NOVAL_phiWlwrf}
  \end{center} 
\end{figure}

Also worth mentioning is the angle $\phi_{W,l}^{\,\ast}$ which is not, as displayed by Fig.~%
\ref{app_pp_NOVAL_phiWlwrf}, flat along the whole $[0,\pi]$ range. The reasons for that comes from
the $V-A$ coupling \index{Electroweak!VmA@$V-A$ coupling} of the leptons to the $W$ and to the presence of non-zero 
and asymmetric $\pT$ of the colliding quarks. Still the explanations to amend this pattern is reported 
to the study of the $\ppbar$ and $\pp$ collisions since in those cases the effects are much higher and 
then easier to explain (note the important zoom in coordinates show the distribution is actually quite flat).
The important point here is that just like for $\costhetaWlwrf$ there are no charge asymmetries 
which means no impact for a study of $\MWp-\MWm$.
\index{W boson@$W$ boson!Mass charge asym@Mass charge asymmetry $\MWp-\MWm$}
\index{W boson@$W$ boson!Production pp without valence quarks@
Production in $\pp$ or $\ppbar$ collisions without valence quarks contributions|)}

\clearpage
\subsection{Proton--anti-proton collisions}
\index{W boson@$W$ boson!Production in ppbar@Production in $\ppbar$ collisions!Detailed|(}
Now we review the $\ppbar$ collisions mode reminding the features that makes it perfectly symmetric 
between the production of $\Wp$ and $\Wm$. Let us remind that in our conventions the proton impinges in
the $+z$ direction while the anti-proton does from the $-z$ direction.

\begin{figure}[!h] 
  \begin{center}
    \includegraphics[width=0.5\tw]{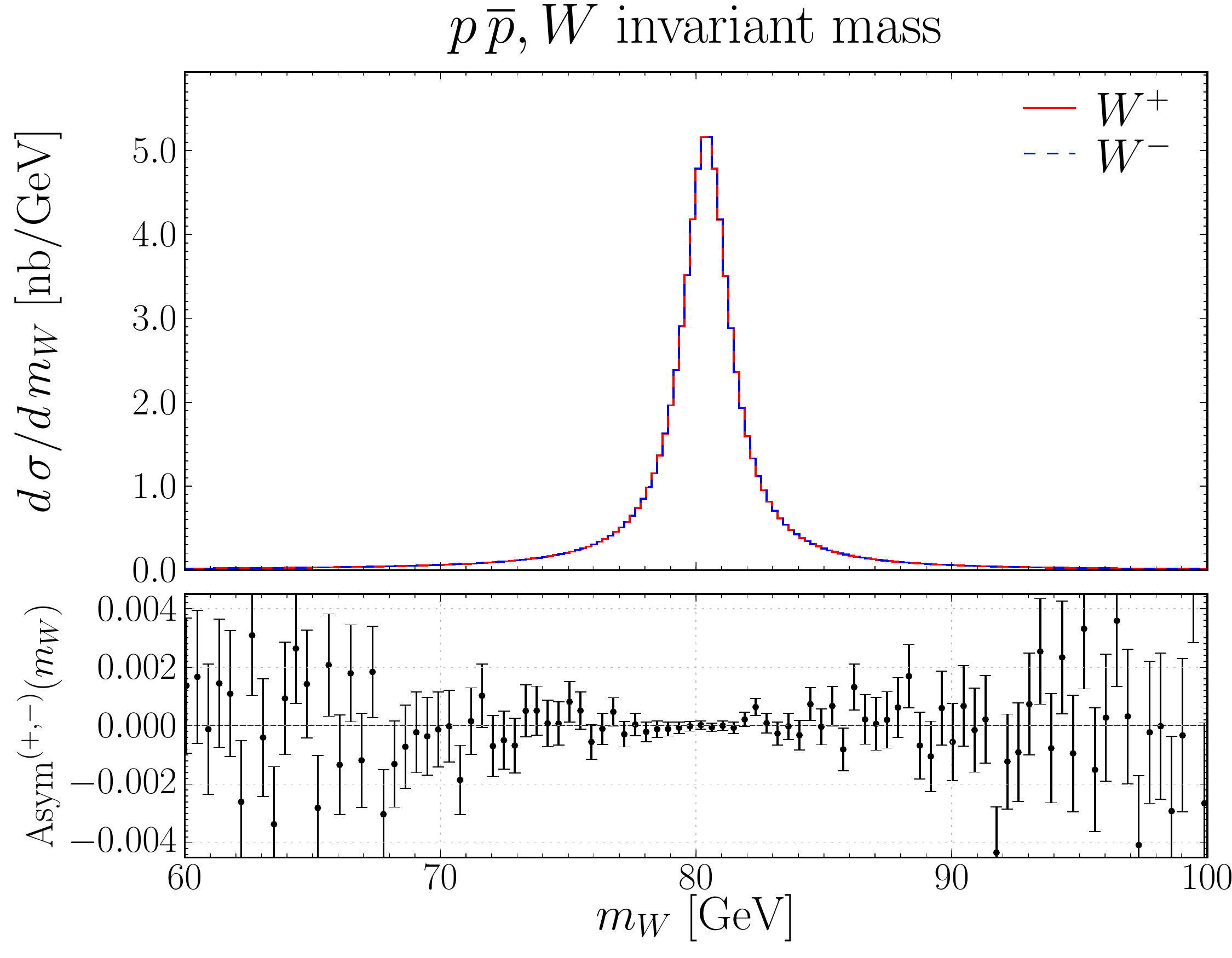}
    \caption[Invariant mass of the $W$ boson in $\ppbar$ collisions]
            {\figtxt{Invariant mass of the $W$ boson in $\ppbar$ collisions.}}
            \label{app_ppb_mW}
  \end{center} 
\end{figure}
First we turn our attention to the invariant mass $m_W$ of the positively and negatively charged $W$ 
as shown in Fig.~\ref{app_ppb_mW}. 
Note these invariant masses are to be understood as the result of the production of $W$ bosons as 
Breit--Wigner resonances weighted by the PDFs, which adds up to the former behaviour the probability of 
occurrence of a given mass $m_W$.
Hence, even if here $m_\Wp=m_\Wm$ such an equality is non trivial and has to be 
seen as the consequence that $\Wp$ and $\Wm$ are produced with the same dynamics in the very 
particular case of $\ppbar$ collisions.

\begin{figure}[!h] 
  \begin{center}
    \includegraphics[width=0.495\tw]{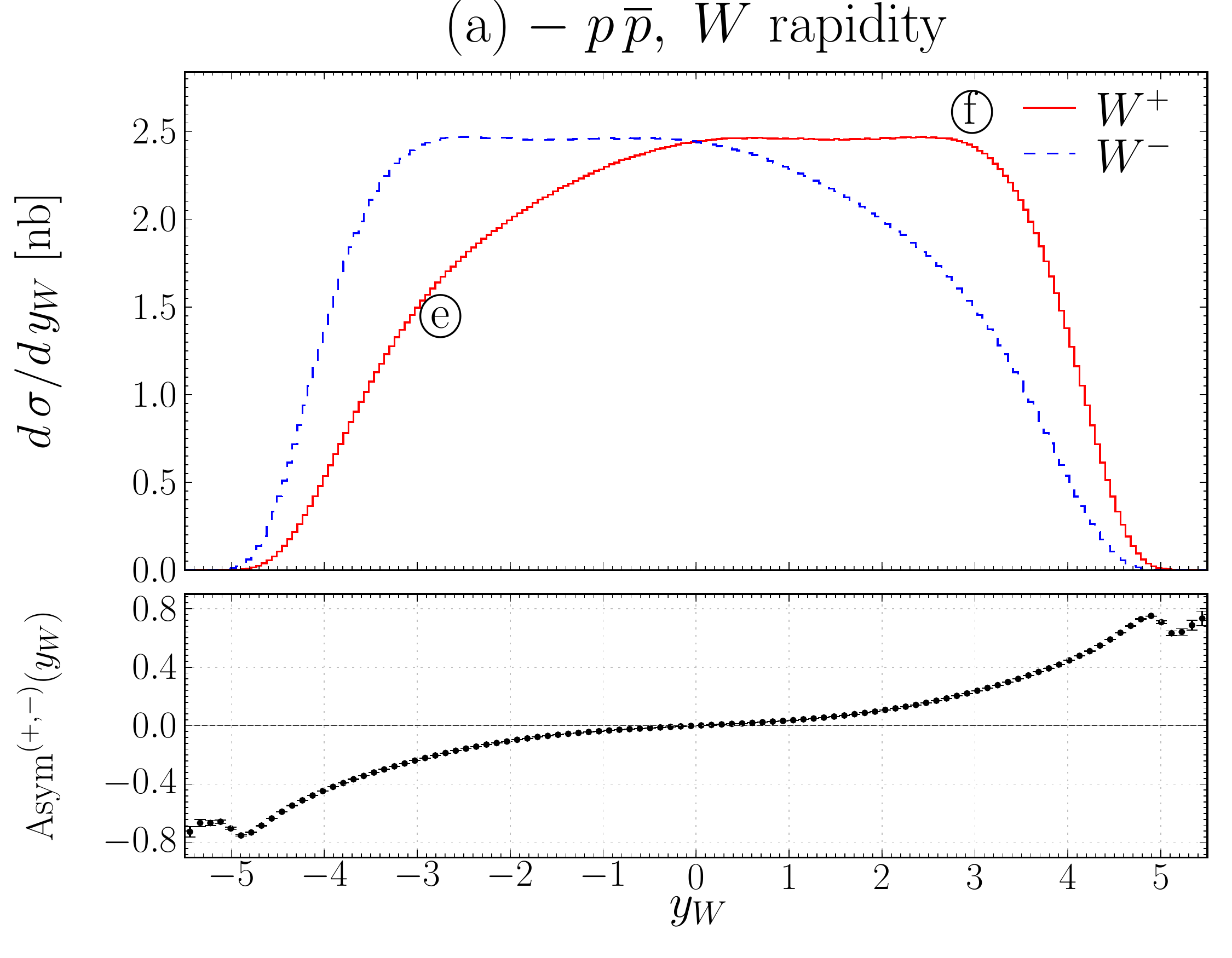}
    \hfill
    \includegraphics[width=0.495\tw]{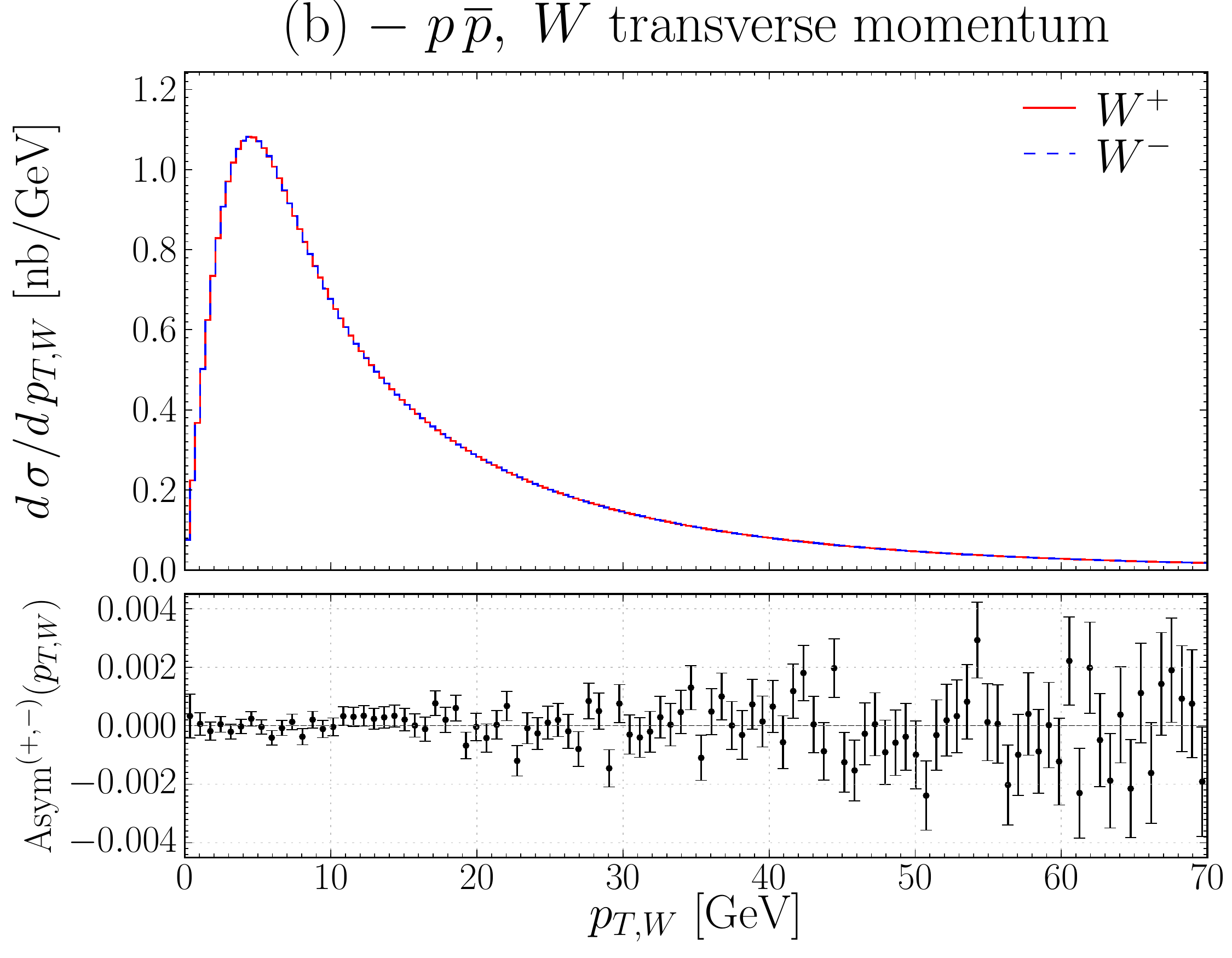}
    \vfill
    \includegraphics[width=0.495\tw]{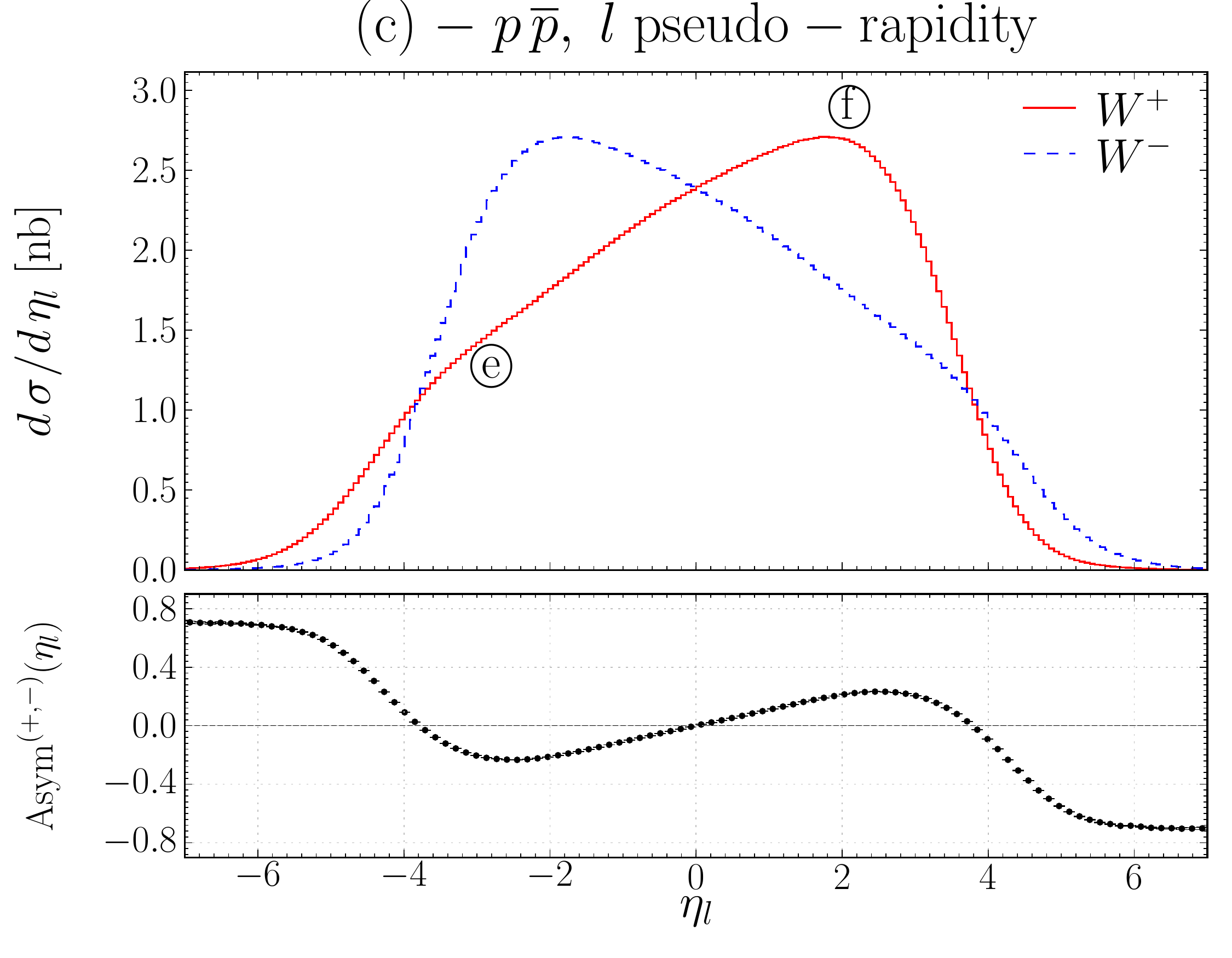}
    \hfill
    \includegraphics[width=0.495\tw]{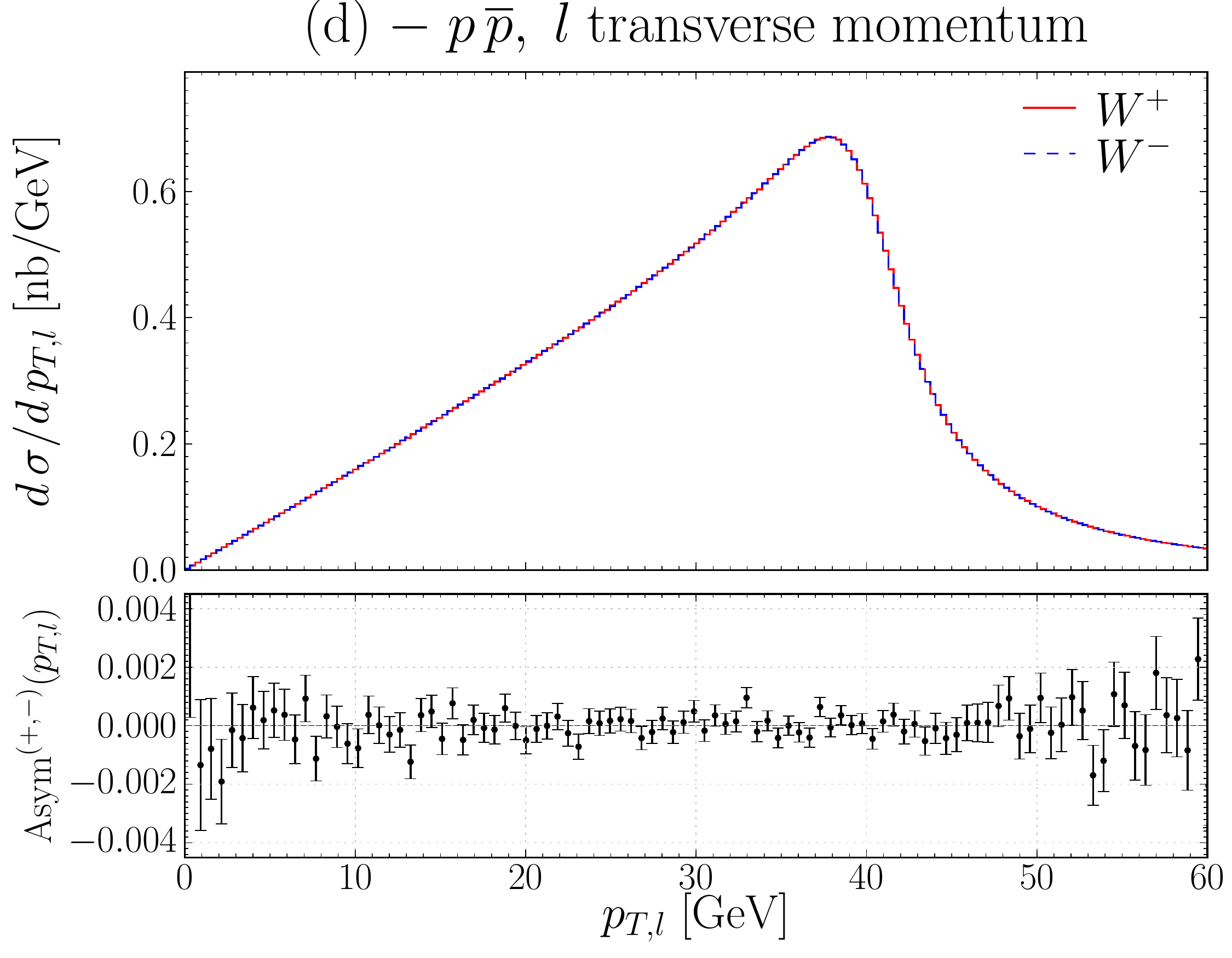}
    \caption[Distributions of the $W$ boson rapidity and transverse momentum along with the one of the
      charged lepton pseudo-rapidity and transverse momentum in $\ppbar$ collisions]
            {\figtxt{Distributions of the $W$ boson rapidity (a) and transverse momentum (b) 
                along with the one of the charged lepton pseudo-rapidity (c) and transverse momentum (d)
                in $\ppbar$ collisions.}}
            \label{app_ppb_yW_pTW_etal_pTl}
            \index{W boson@$W$ boson!Transverse momentum}
            \index{W boson@$W$ boson!Rapidity}
            \index{Charged lepton@Charged lepton from $W$ decay!Transverse momentum}
            \index{Charged lepton@Charged lepton from $W$ decay!Pseudo-rapidity}
  \end{center} 
\end{figure}

The observables $\yW$, $\pTW$, $\etal$ and $\pTl$ are shown in Fig.~\ref{app_ppb_yW_pTW_etal_pTl}.
In $\ppbar$ collisions the production of the $\Wp$ and of the $\Wm$ are produced
with the same dynamics since to every element of the phase-space of, say the $\Wp$, exists
the perfect $CP$ \index{Symmetry!CP@$CP$} match in an other phase-space point of $\Wm$.
This symmetry translates in coordinate space into that the $\Wp(\lp)$ and $\Wm(\lm)$ kinematics are the 
same up to a vertical flip with respect to the interaction point. 
This behaviour can be seen in the $\yW$ and $\etal$ distributions but is no longer decipherable when 
projecting the $W$ or the charged lepton transverse momenta on the transverse plane.
Note in the $\yW$ and $\etal$ distributions the labels $\textcircled{e}$ and $\textcircled{f}$ 
attached to the $\Wp$ data. 
Focusing on the main ``valence'' term $u\,\dbar\,\Vckmsqr{u}{d}$,
in each of these regions the $\Wp$ are produced respectively via 
$\textcircled{e}\,:\;u_p^\sea\to\longleftarrow\dbar_\pbar^\val$ 
and $\textcircled{f}\,:\;u_p^\val\longrightarrow\leftarrow\dbar_\pbar^\sea$
for which the inequality $\textcircled{e}<\textcircled{f}$ materialises the fact that since
$\dbar_\pbar^\val\equiv d_p^\val$ we have
\begin{equation}
d_p^\val < u_p^\val.\label{eq_uval_sup_dbarval}
\index{Quarks!uvmdv@$u^\val-d^\val$ asymmetry}
\end{equation}
This last inequality explains the presence of the exact same asymmetry for the case of the $\Wm$ 
production.

\begin{figure}[!h] 
  \begin{center}
    \includegraphics[width=0.495\tw]{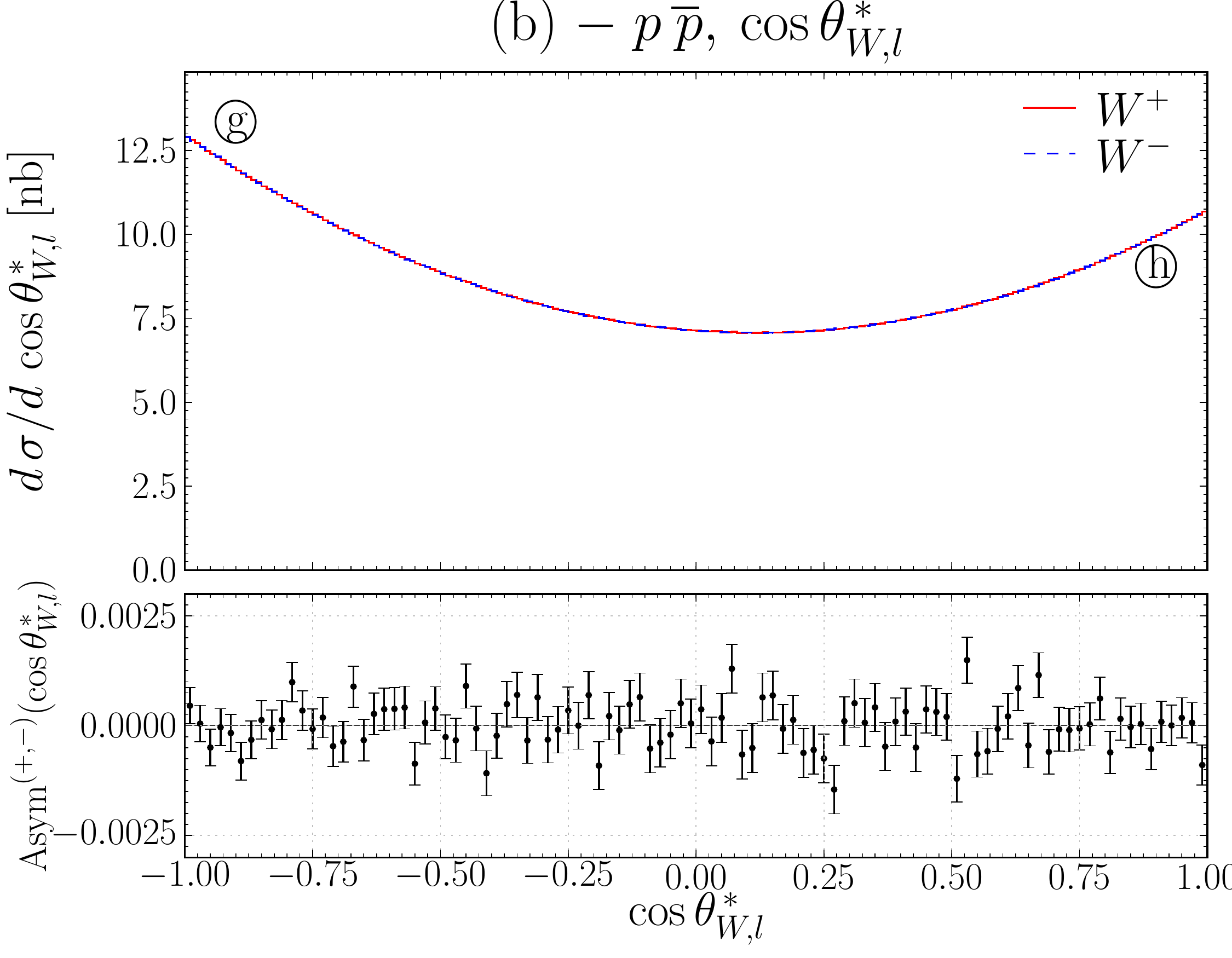}
    \hfill
    \includegraphics[width=0.495\tw]{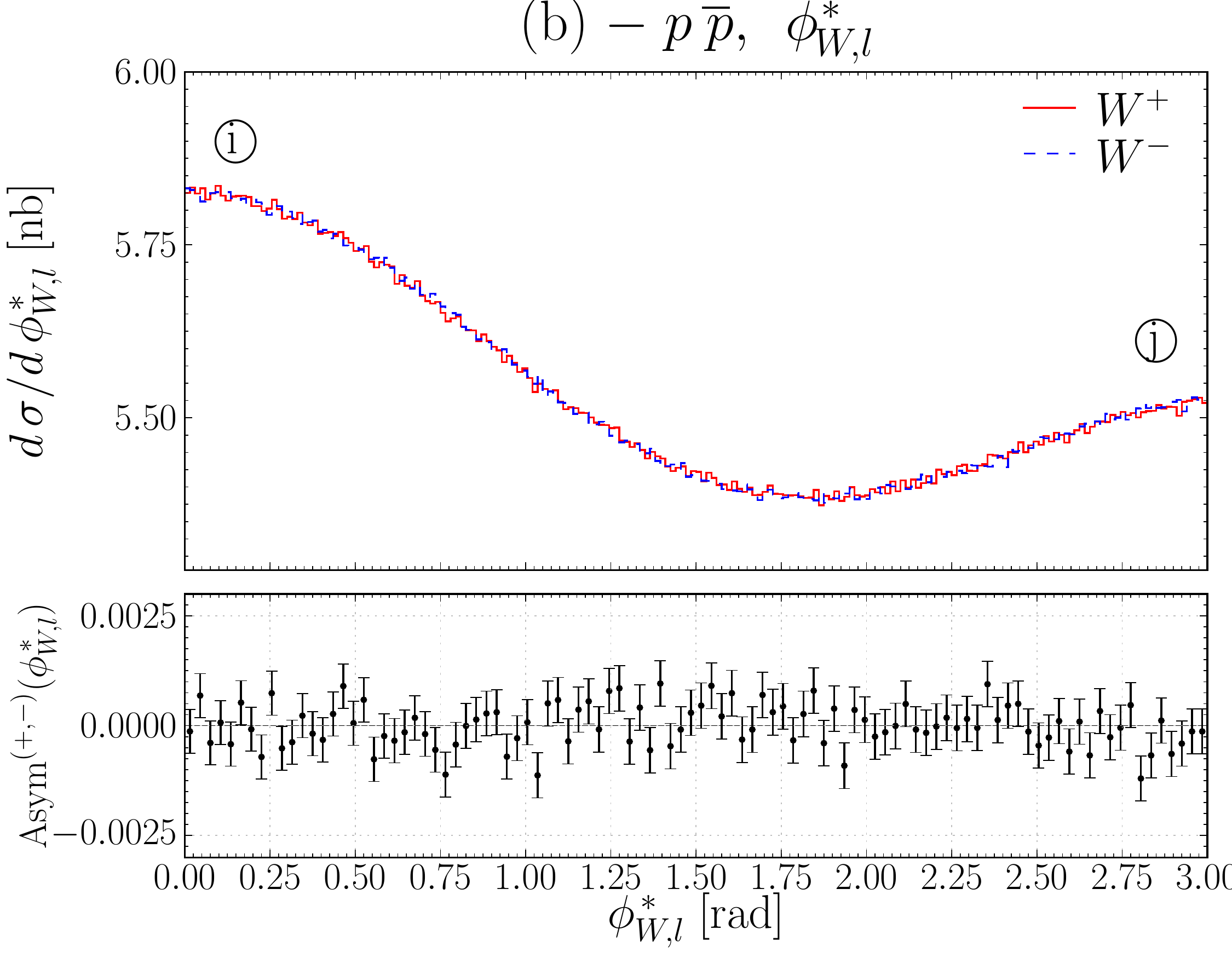}
    \caption[Distributions of $\costhetaWlwrf$ and $\phi_{W,l}^{\,\ast}$ in $\ppbar$ collisions]
            {\figtxt{Distributions of $\costhetaWlwrf$ (a) and $\phi_{W,l}^{\,\ast}$ (b) in $\ppbar$ 
                collisions.}}
            \label{app_ppb_cos_histos}
  \end{center} 
\end{figure}

The angular distributions of $\costhetaWlwrf$ and $\phi_{W,l}^{\,\ast}$ are shown in 
Fig.~\ref{app_ppb_cos_histos}.
We start by looking at $\costhetaWlwrf$.
Both $\Wp$ and $\Wm$ distributions are superimposed and display a small asymmetry with respect to the
origin. This pattern can be explained by focusing on the $\Wp$ and the two labels 
$\textcircled{g}$ and $\textcircled{h}$ attached to it in Fig.~\ref{app_ppb_cos_histos}.(b).
The configurations where the $\Wp$ are produced via 
$\textcircled{g}\,:\;u_p^\val\,\dbar_\pbar^\sea\to\Wp(\lambda=-1)$ 
are such that the $\lp$ will decay preferentially in the opposite direction of the $\Wp$ momentum. 
In the cases where the boson is produced like
$\textcircled{h}\,:\;u_p^\sea\,\dbar_\pbar^\val\to\Wp(\lambda=+1)$ the $\lp$ will decay preferentially 
in the same direction of the $\Wp$ momentum.
Then, the $\textcircled{g}>\textcircled{h}$ pattern shares the same origin than the asymmetries seen 
in the $\yW$ and $\etal$ distributions and written in Eq.~(\ref{eq_uval_sup_dbarval}).
Previously, while studying the pure sea contributions, the inverse unbalance was observed, but now the
main ``valence'' term  $u\,\dbar\,\Vckmsqr{u}{d}$ overrules the latter.

\begin{figure}[!h] 
  \begin{center}
    \includegraphics[width=0.9\tw]{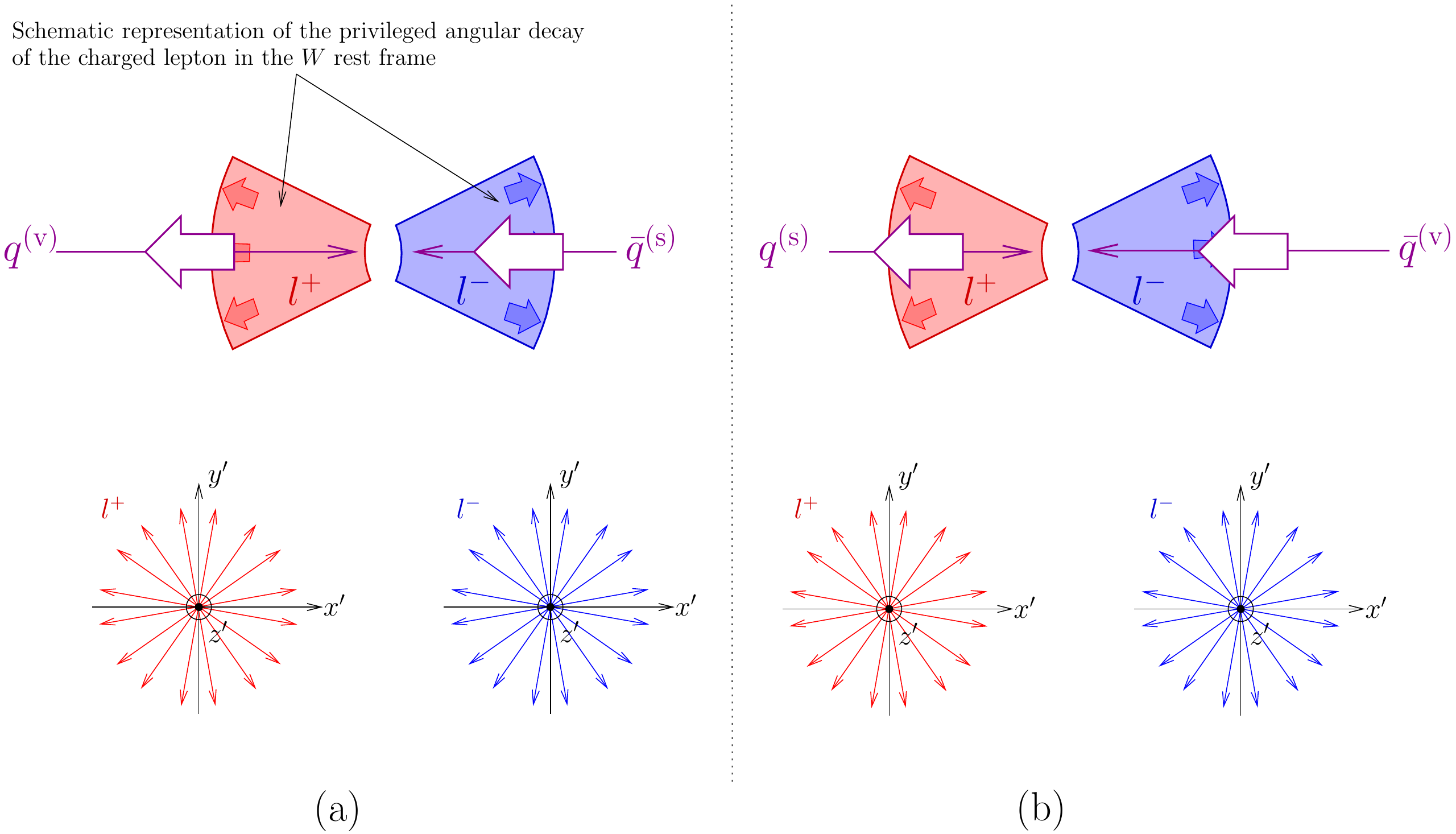}
    \caption[Schematic representations of the decay of the charged leptons at the leading order]
            {\figtxt{Schematic representations of the decay of the charged leptons at the leading order.
                For convenience, even though the quarks are represented in the laboratory frame the
                decay of the charged leptons are symbolised in the \WRF{}.
                Frame (a) represents the case of ``valence'' contribution involving a quark of valence
                while frame (b) displays the case of ``valence'' contribution involving an anti-quark 
                of valence.}}
            \label{app_phi_Wl_wrf_LO}
  \end{center} 
\end{figure}

\index{W boson@$W$ boson!Polarisation!Transverse states|(}
\index{Helicity!Of the colliding quarks and the decaying leptons|(}
Now even more interesting is the fact that Fig.~\ref{app_ppb_cos_histos} shows $\lp$ and $\lm$
preferentially decays in the same direction of $\pTW$ which, as we know, must have dramatic 
consequences on the smearing of the $\pTl$ distributions.
Before explaining what happens we take a step back to the LO order and understand how the lack of 
transverse motion in this approximation induce no asymmetry in the azimuthal decay of the leptons.
Figure~\ref{app_phi_Wl_wrf_LO} shows the collinear quark--anti-quark collisions of $q^\val\,\qbar^\sea$
(left) and $q^\sea\,\qbar^\val$ (right). 
In both cases the valence quarks hold most of the longitudinal 
momentum and the privileged direction of decays of the positively and negatively charged leptons 
(symbolised in the \WRF{}) which, even if opposite with each other, do not induce any anisotropic 
behaviour in $\phi$.
\begin{figure}[!h] 
  \begin{center}
    \includegraphics[width=0.9\tw]{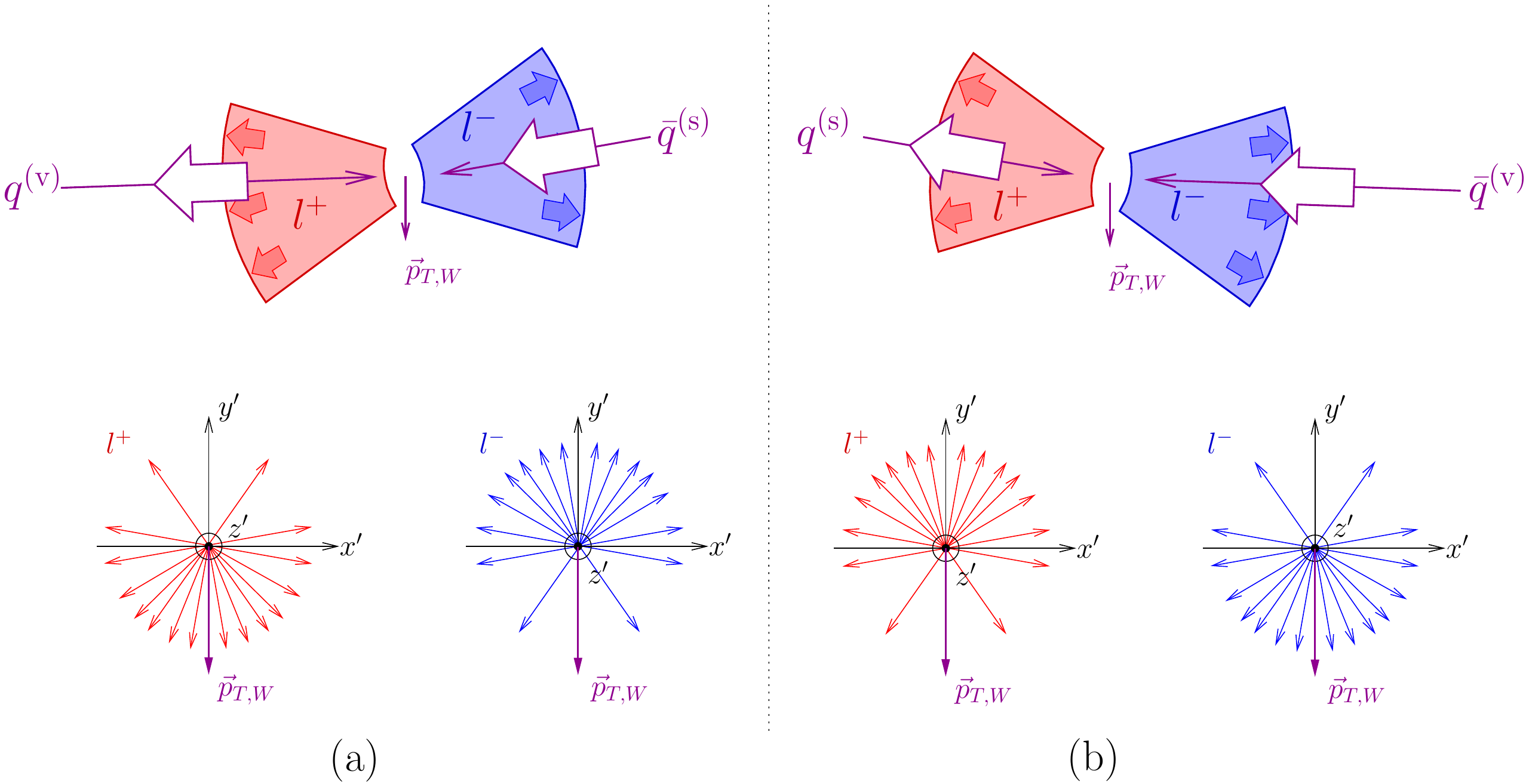}
    \caption[Schematic representations of the decay of the charged leptons at the improved leading order]
            {\figtxt{Schematic representations of the decay of the charged leptons at the improved leading 
                order.
                For convenience, even though the quarks are represented in the laboratory frame the
                decay of the charged leptons are symbolised in the \WRF{}.
                Frame (a) represents the case of ``valence'' contribution involving a quark of valence
                while frame (b) displays the case of ``valence'' contribution involving an anti-quark 
                of valence.}}
            \label{app_phi_Wl_wrf_iLO}
  \end{center} 
\end{figure}
Moving to the improved leading order picture, 
now quarks and anti-quarks possess a transverse initial motion and radiate gluons and photons on their
way to the collision. We end up this time with ``valence'' cases with the valence (anti-)quark
that bears most of the longitudinal momentum and on the other side the sea (anti-)quark which bears 
most of the transverse motion.
In this context the higher transverse momentum of the sea (anti-)quark constrain the leptons to 
modify their privileged decaying angle to satisfy the kinematic momentum conservation as shown in
Fig.~\ref{app_phi_Wl_wrf_LO}
Hence for $q^\val\,\qbar^\sea$ events the $\lp$ decays preferentially along the direction of $\pTW$
while the $\lm$ decays preferentially in the opposite direction of $\pTW$.
In the case of $q^\sea\,\qbar^\val$ it is the exact opposite.
Now, in the context of the $\Wp$ production, focusing again on the main ``valence'' term 
$u\,\dbar\,\Vckmsqr{u}{d}$ Figure~\ref{app_phi_Wl_wrf_iLO}.(a) translates to 
$u_p^\val\longrightarrow\leftarrow\dbar_\pbar^\sea$ while (b) translates to 
$u_p^\sea\to\longleftarrow\dbar_\pbar^\val$. Because of the $d<u$ inequality 
(Eq.~(\ref{eq_uval_sup_dbarval})) (a)-cases overrule (b)-cases and the $\lp$ prefers, in general, 
to decay in the direction of $\pTW$. 
The same argument can explain the asymmetry observed in the negative channel.
So far, while the fact the valence (anti-)quark hold most of the longitudinal momentum is well known,
we did not justified the fact that the sea (anti-)quark hold most of the transverse motion.
The very last section of this Appendix address this issue. For now we can justify it heuristically by 
stating that the sea (anti-)quark having a small fraction $x$ is the result of the loss of energy
on the way to the collision by gluon/photon emission which confers to the (anti-)quark more transverse
motion than the valence (anti-)quark.
\index{W boson@$W$ boson!Polarisation!Transverse states|)}
\index{Helicity!Of the colliding quarks and the decaying leptons|)}

\begin{figure}[!h] 
  \begin{center}
    \includegraphics[width=0.495\tw]{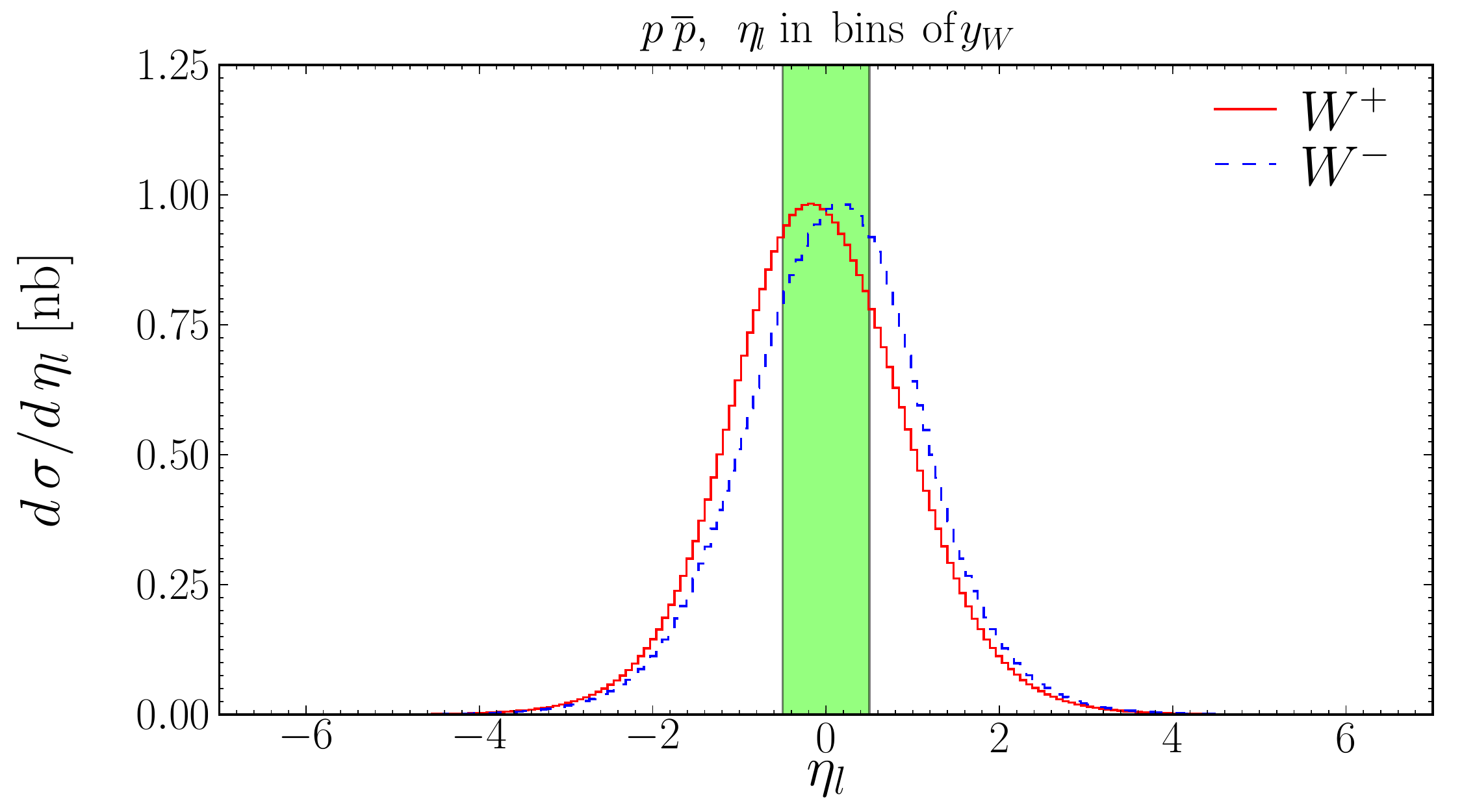}
    \hfill
    \includegraphics[width=0.495\tw]{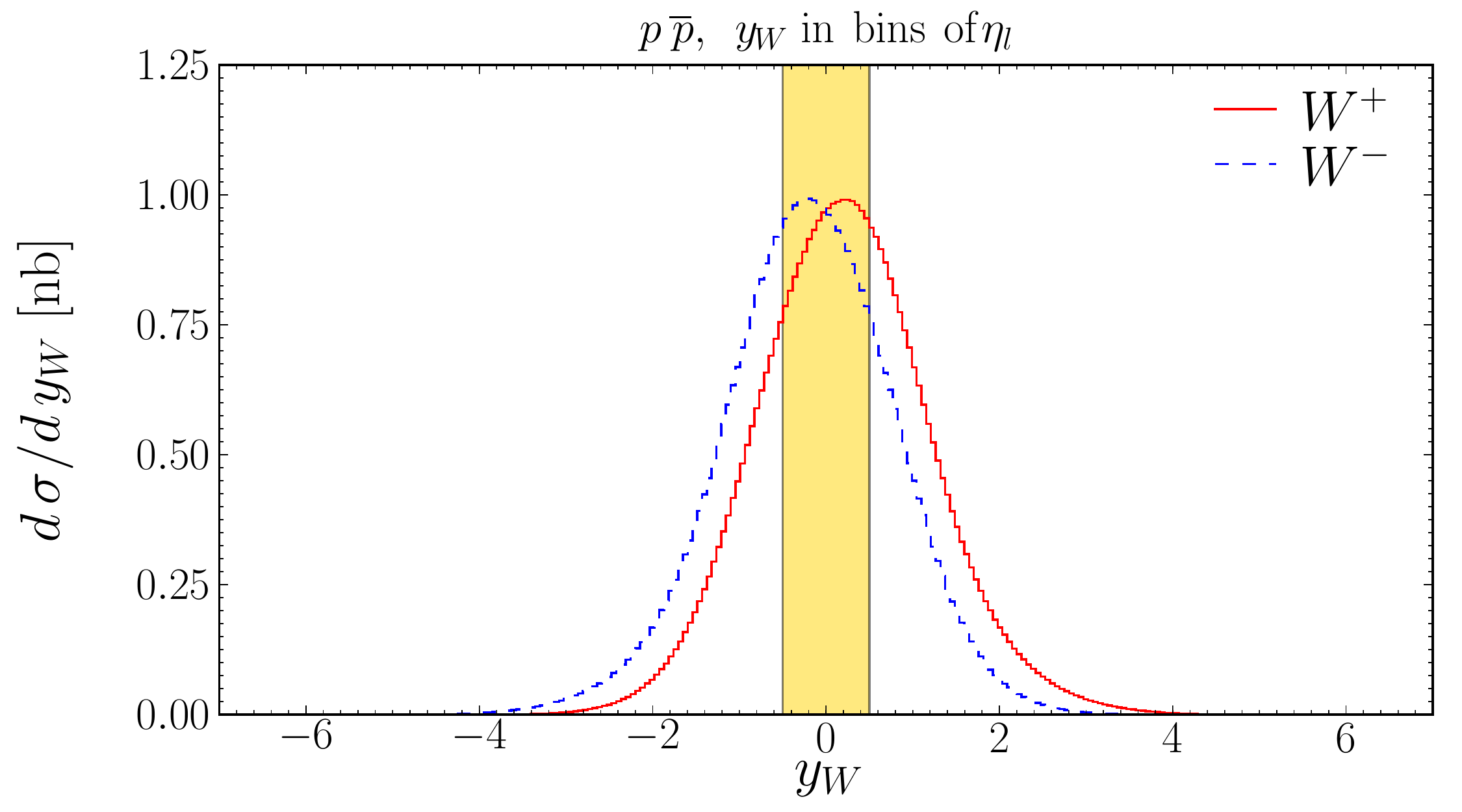}
    \vfill   
    \includegraphics[width=0.495\tw]{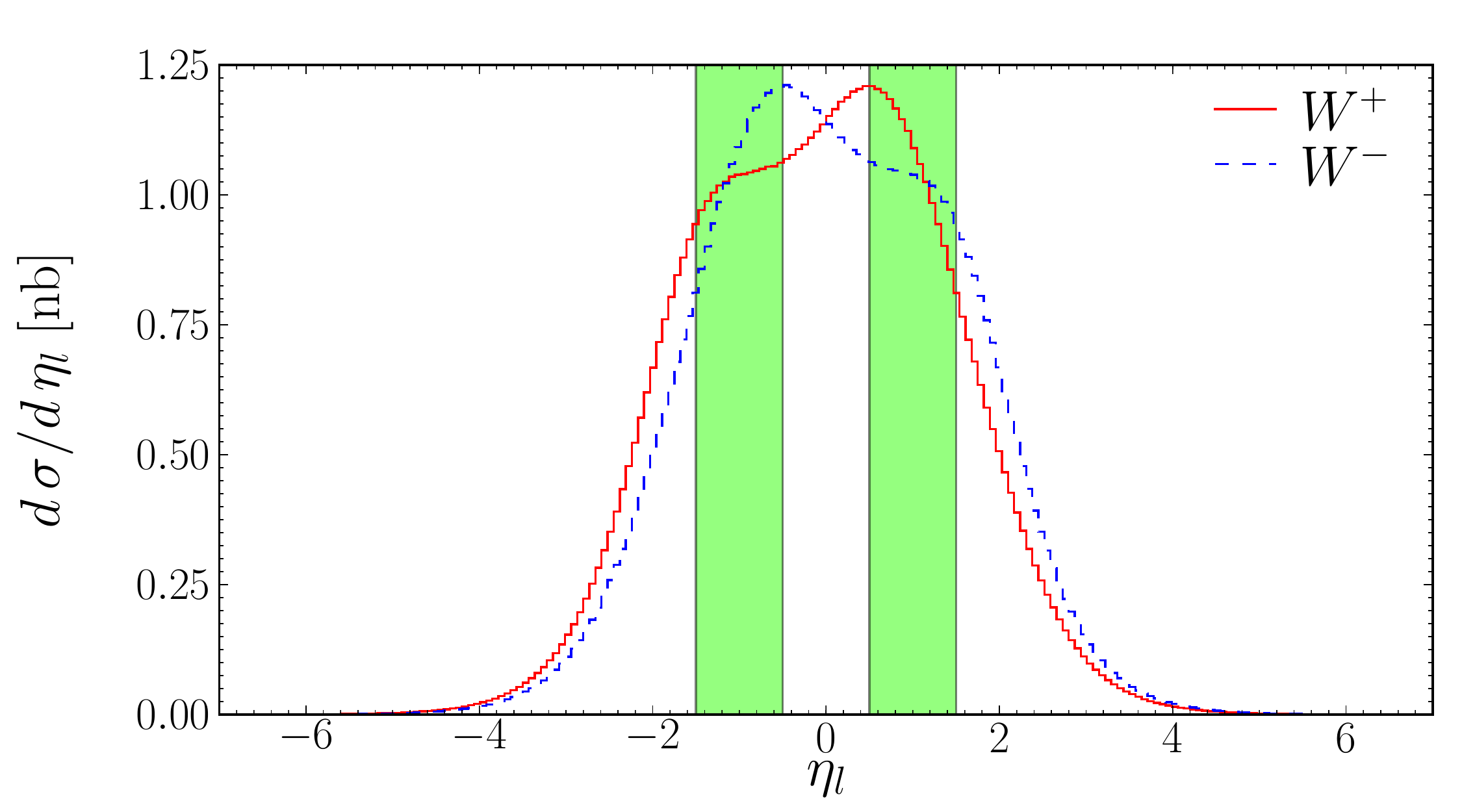}
    \hfill
    \includegraphics[width=0.495\tw]{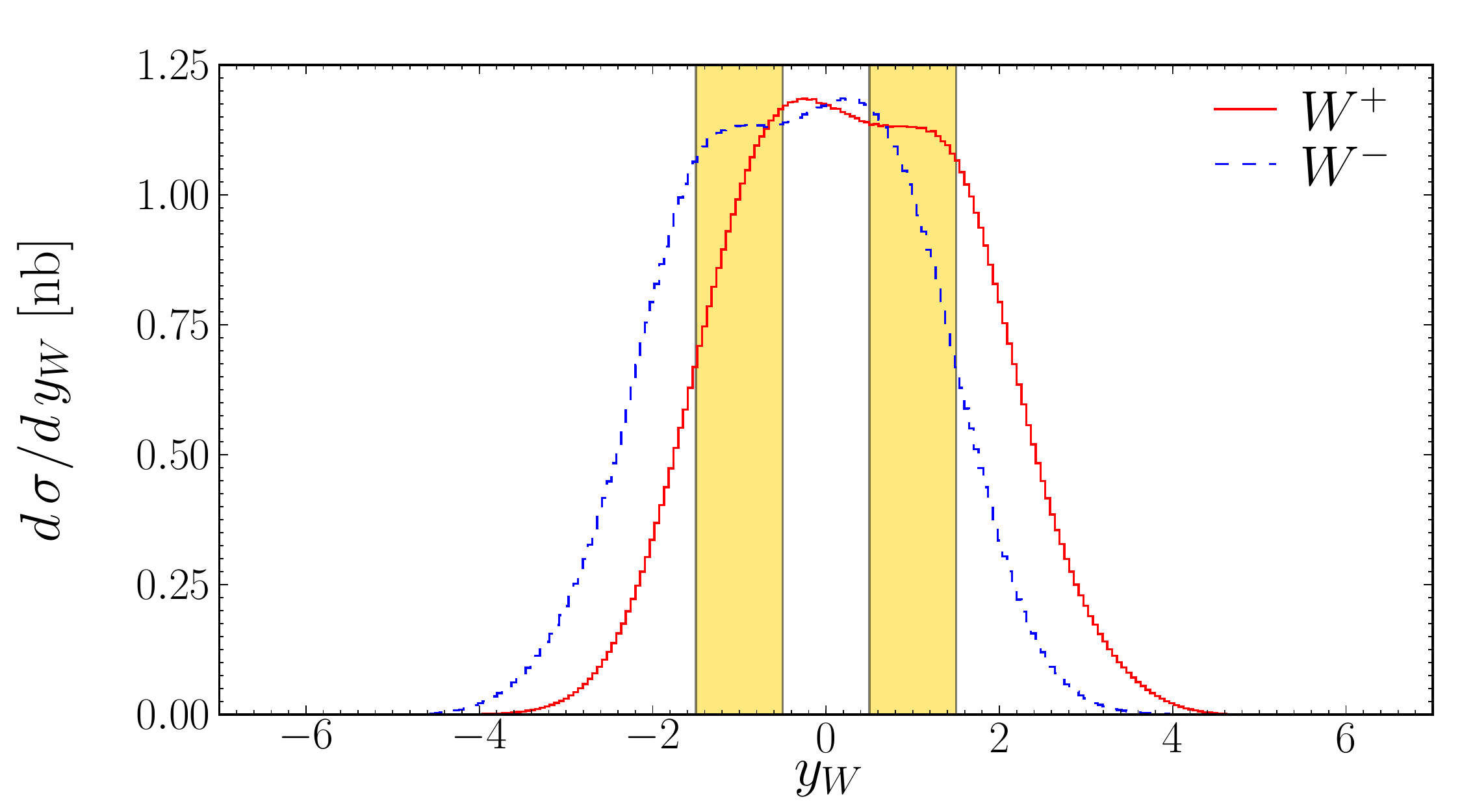}
    \vfill
    \includegraphics[width=0.495\tw]{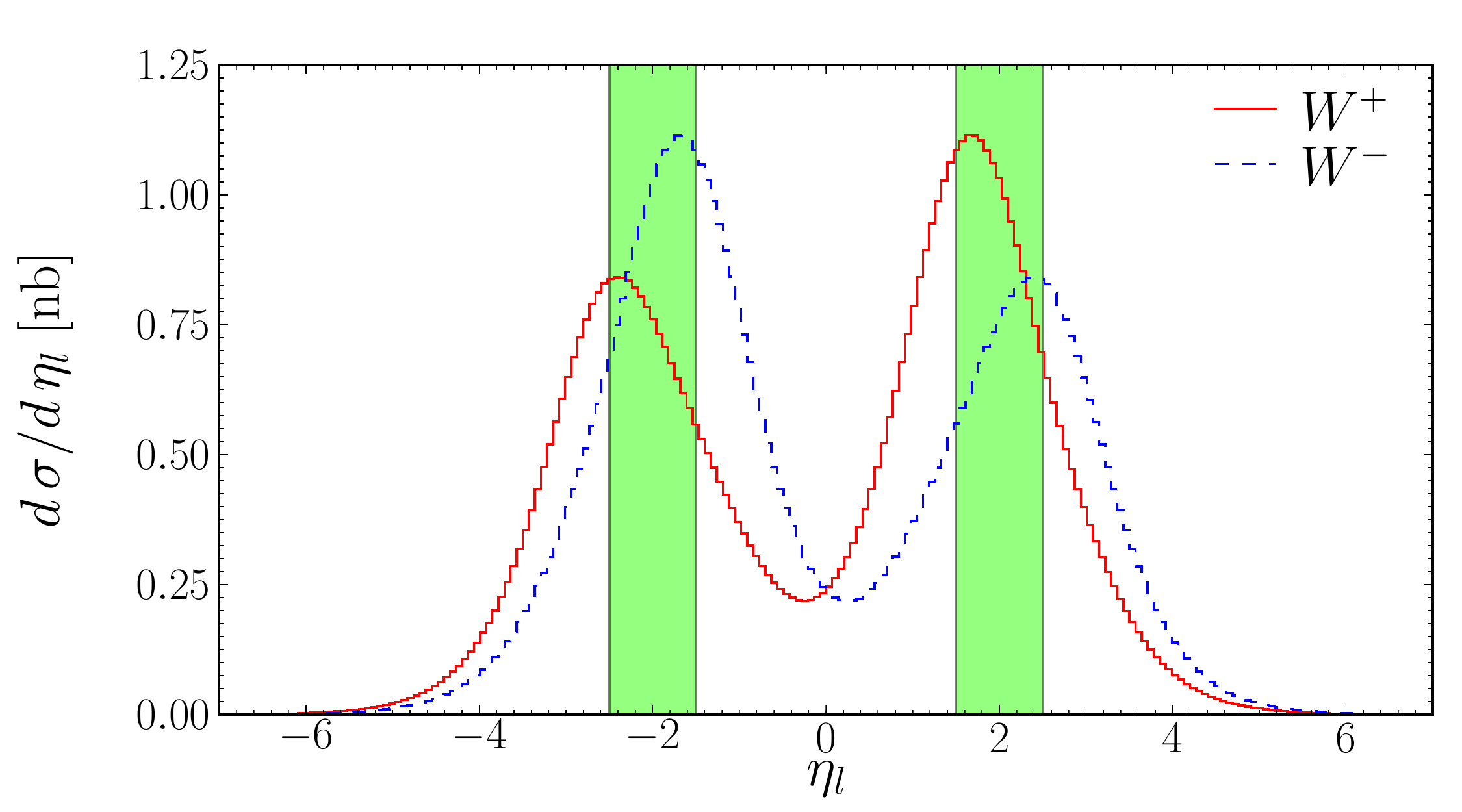}
    \hfill
    \includegraphics[width=0.495\tw]{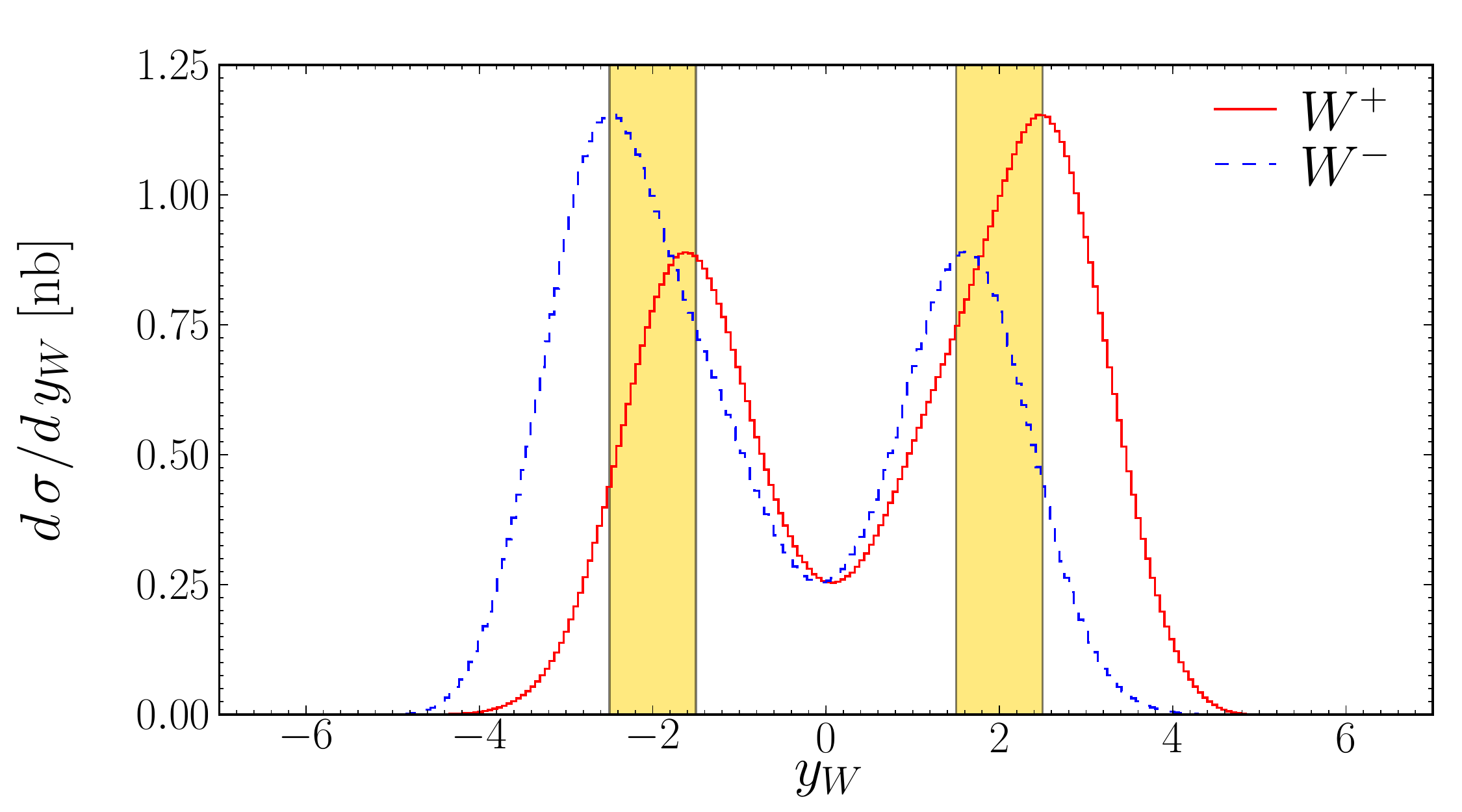}
    \vfill
    \includegraphics[width=0.495\tw]{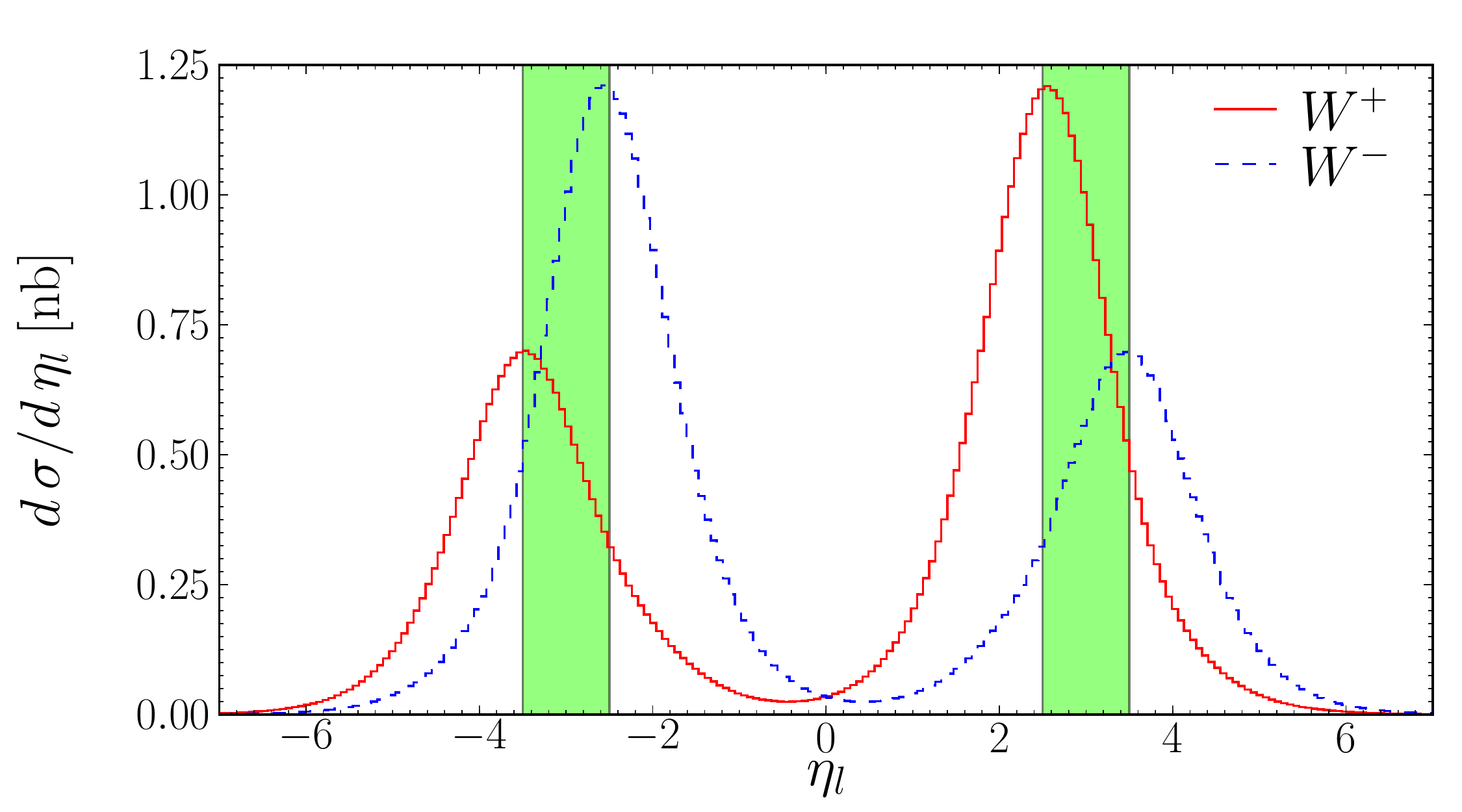}
    \hfill
    \includegraphics[width=0.495\tw]{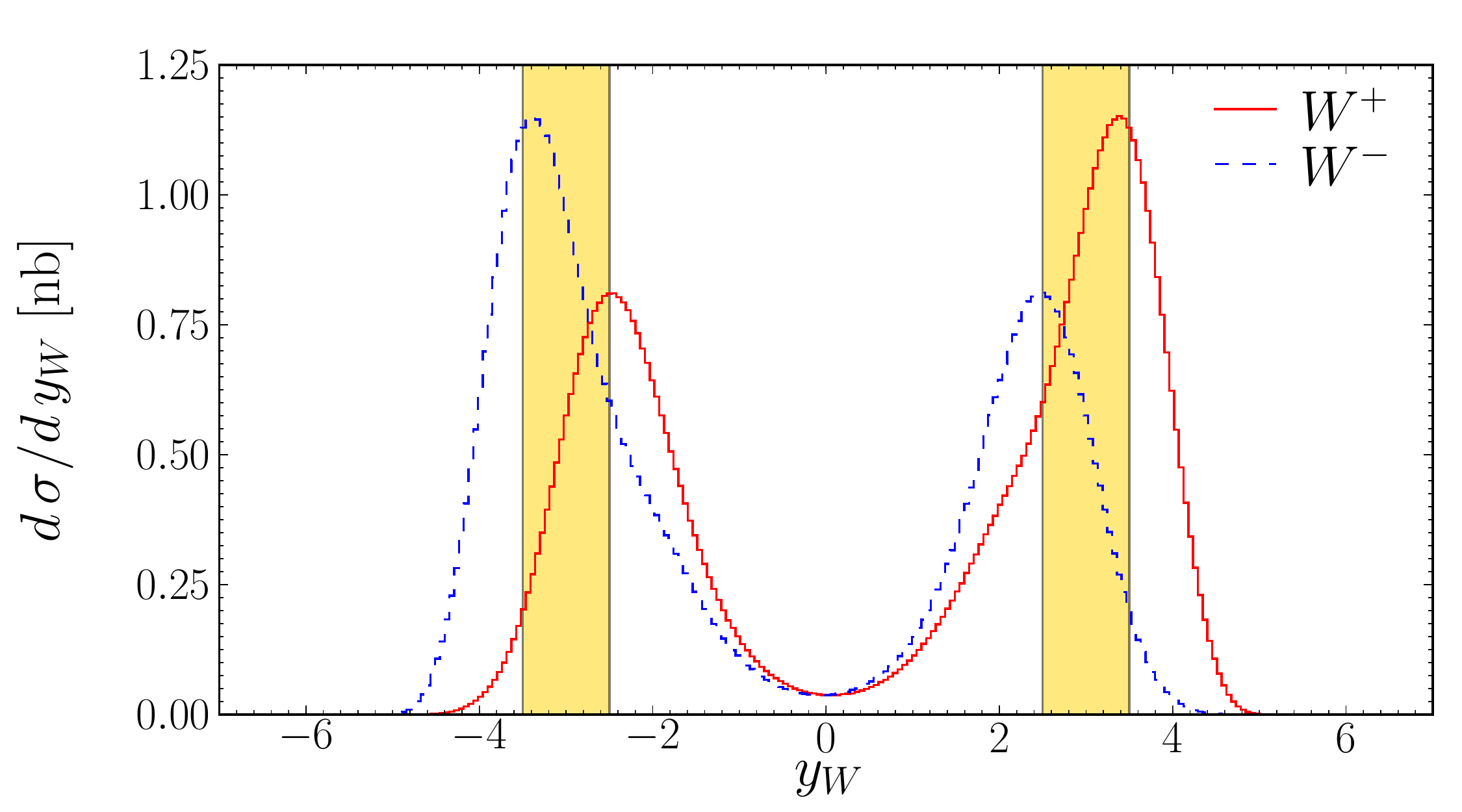}
    \vfill
    \includegraphics[width=0.495\tw]{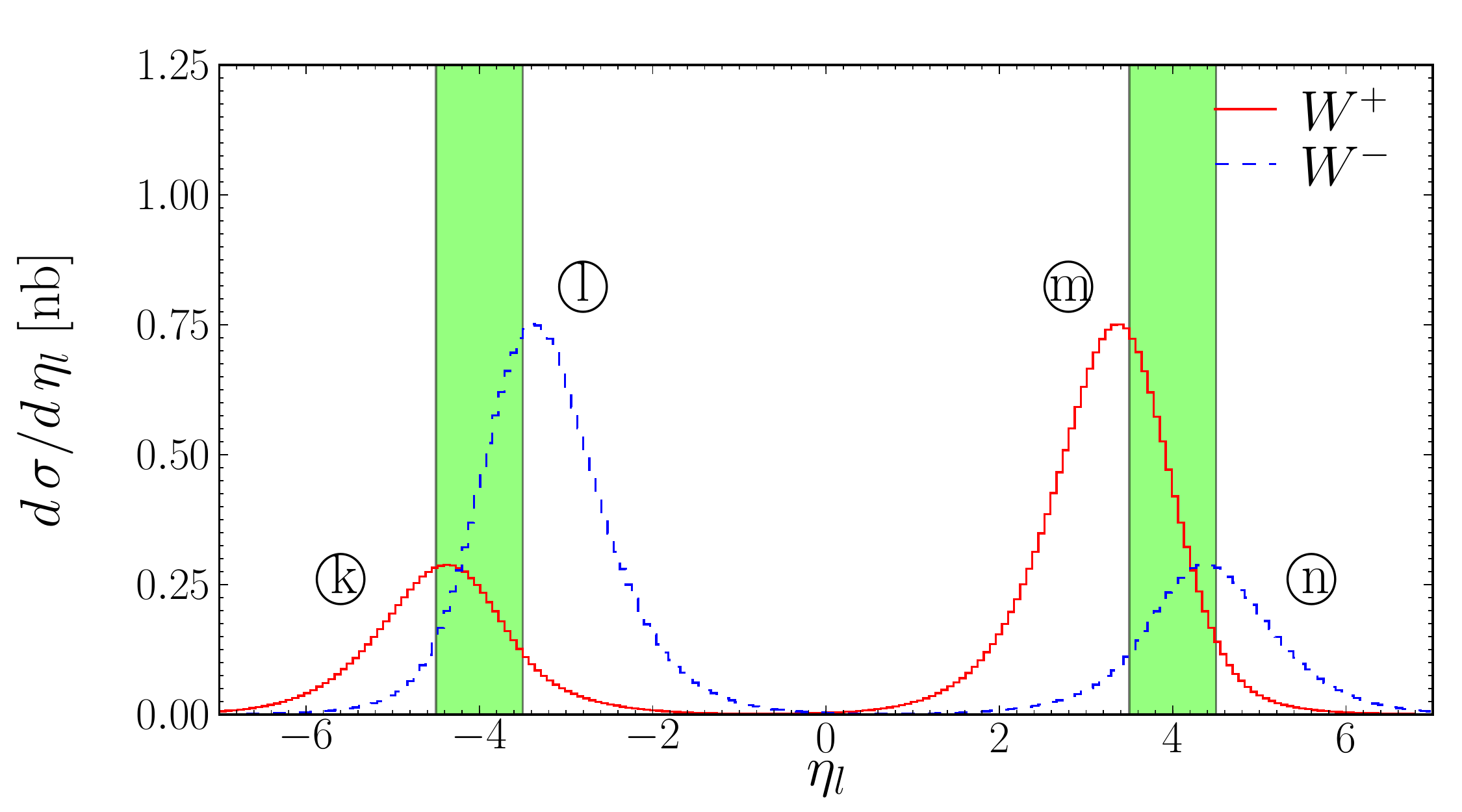}
    \hfill
    \includegraphics[width=0.495\tw]{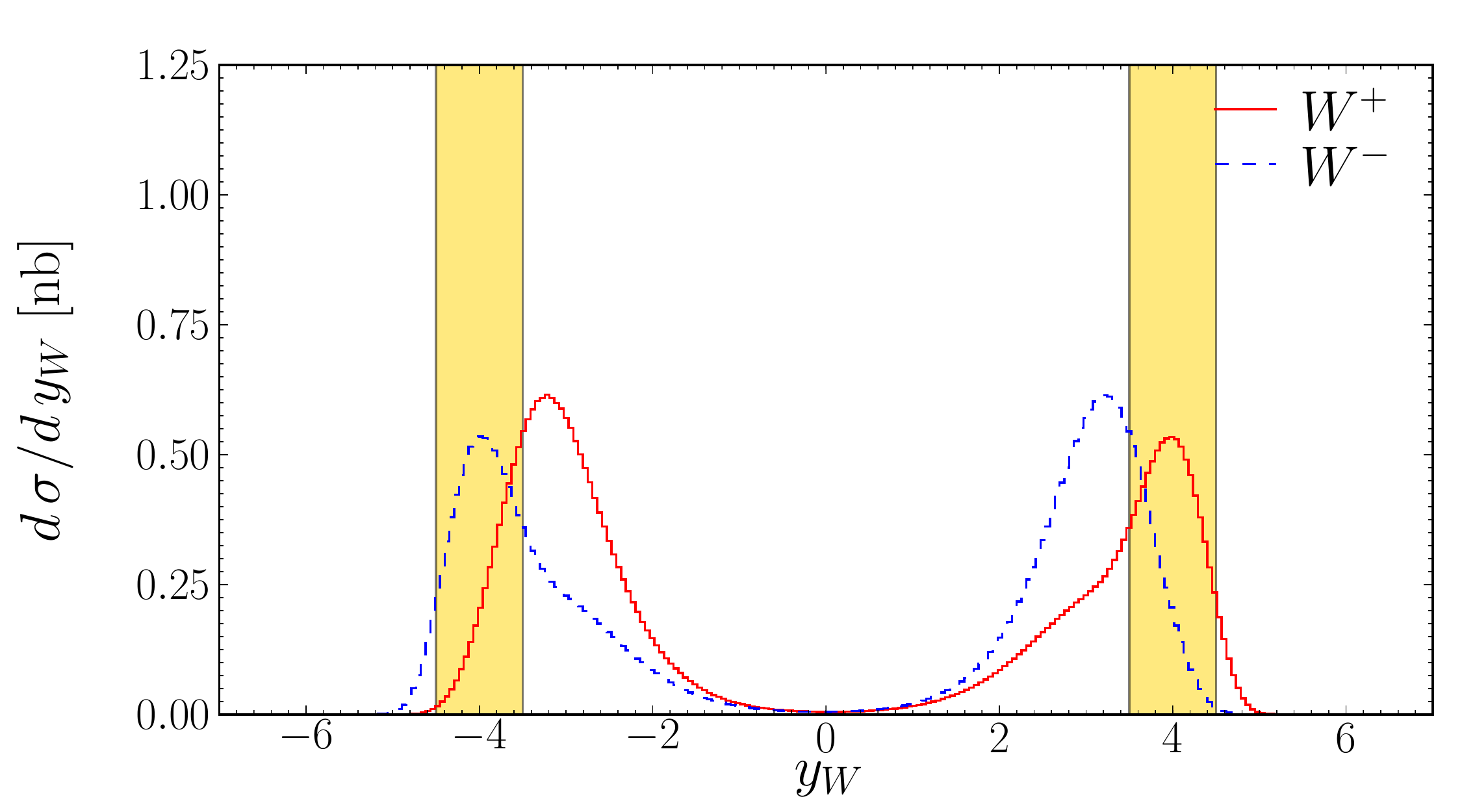}
    \caption[$W$ boson rapidity distributions in bins of the charged lepton pseudo-rapidity and 
      \textit{vice versa} in the case of $\ppbar$ collisions]
            {\figtxt{$W$ boson rapidity distributions in bins of the charged lepton pseudo-rapidity 
                (left) and \textit{vice versa} (right) in the case of $\ppbar$ collisions. 
                In each plot the corresponding $\yW$ or $\etal$ selection is materialised by the 
                colored stripe(s).}}
            \label{app_ppb_etal_yW_in_yW_etal_bins}
  \end{center} 
\end{figure}
Figure~\ref{app_ppb_etal_yW_in_yW_etal_bins} by showing the $\etal$ distribution in bins of $\yW$ 
and \textit{vice versa} illustrates the importance of the ``valence'' contributions in function of the 
fraction of momentum $x_{q,\qbar}$.
Starting with $\etal$ in bins of $\yW$ (left) we see the pseudo-rapidity distribution is the addition 
of two quantities\,: (1) the rapidity $\yW$ inherited from the $W$ boost and that confers to the 
charged lepton most of its longitudinal motion and (2) the decay of the charged lepton governed by the 
$V-A$ coupling \index{Electroweak!VmA@$V-A$ coupling} of leptons to $W$ which eventually gives this bump representing
the angular coverage of the charged leptons $\theta_l\in[0,\pi]$.
Let us remind $\pTW$ participates in the shape of (2) but can be neglected in this discussion since 
$\pTW\ll|\vec p_{z,W}|$.
In the narrow central bin region where all events verify $|\yW|<0.5$ the small charge asymmetry 
is due to non-negligible ``valence'' quark contributions.
As the selection in rapidity increases, the ``valence'' contributions rise displaying in consequence
large charge asymmetries in the lepton decay for a local value of $\etal$.
Turning to $3.5<|\yW|<4.5$, for the negative values of $\etal$ the ``valence'' contributions
$\textcircled{k}\,:\;u_p^\sea\to\longleftarrow\dbar_\pbar^\val$ give $\Wp(\lambda=+1)$ bosons while
$\textcircled{l}\,:\;d_p^\sea\to\longleftarrow\ubar_\pbar^\val$ give $\Wm(\lambda=+1)$ bosons. 
The ``valence'' terms contribute then in both cases to produce right $W$ bosons, which imply
the $\lp(\lm)$ decays preferentially in the same(opposite) direction of the $W$ momentum.
Going to the positive value of $\etal$ is achieved by taking the $CP$ transformation \index{Symmetry!CP@$CP$}
of the previous example, this time we have $\textcircled{m}\,:\;u_p^\val\longrightarrow\leftarrow\dbar_\pbar^\sea$
giving $\Wp(\lambda=-1)$ bosons
and $\textcircled{n}\,:\;d_p^\val\longrightarrow\leftarrow\ubar_\pbar^\sea$ $\Wm(\lambda=-1)$ bosons. 
Note that considering at the same time those two opposite $\yW$-phase space regions completely 
reestablish the equivalence in the dynamic production of the $\Wp$ and the $\Wm$ bosons  which,
using the label notation, would write 
$\textcircled{k}+\textcircled{m}\equiv\textcircled{l}+\textcircled{n}$.

On the right of Fig.~\ref{app_ppb_etal_yW_in_yW_etal_bins} the $\yW$ distributions in bins of $\etal$
hold more or less the same information than the previous plots. Here though, looking at the $\yW$
variable allows to look at the smearing of the distributions from the $W$ boost point of view while
earlier the smearing was essentially the one arising from the $V-A$ coupling \index{Electroweak!VmA@$V-A$ coupling}
of the charged leptons to the $W$. 
Also now we can see that a given slice of $\etal$ receives contributions from an important 
range of different $\yW$ rapidity. Although it is not important here as no $\ppbar$ collisions can 
be made
at the LHC it will give insight for LHC collisions at which level the narrow $\etal$ selection in our
analysis are contaminated by high rapidity region and hence from ``valence'' contributions.

\begin{figure}[!h] 
  \begin{center}
    \includegraphics[width=0.495\tw]{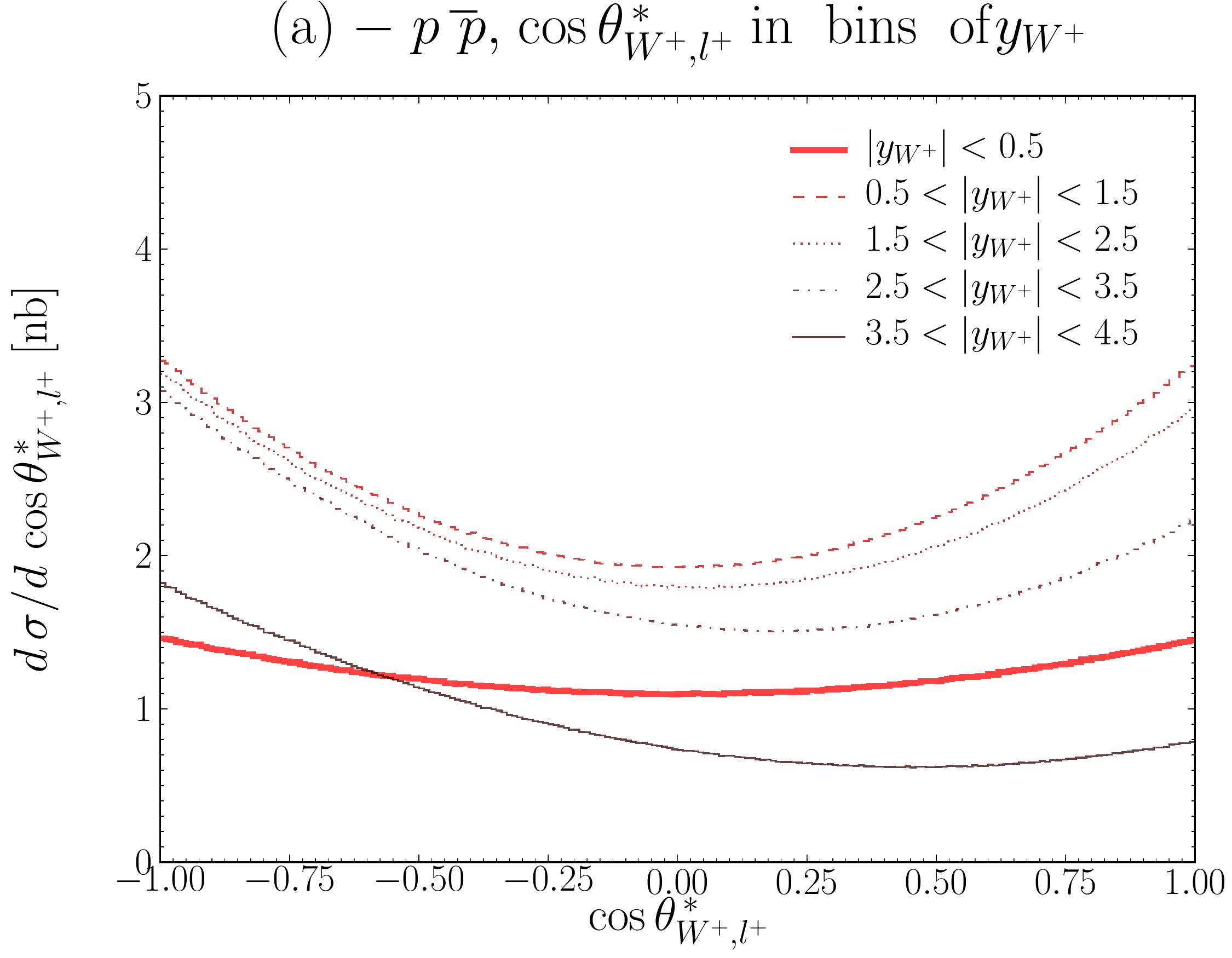}
    \hfill
    \includegraphics[width=0.495\tw]{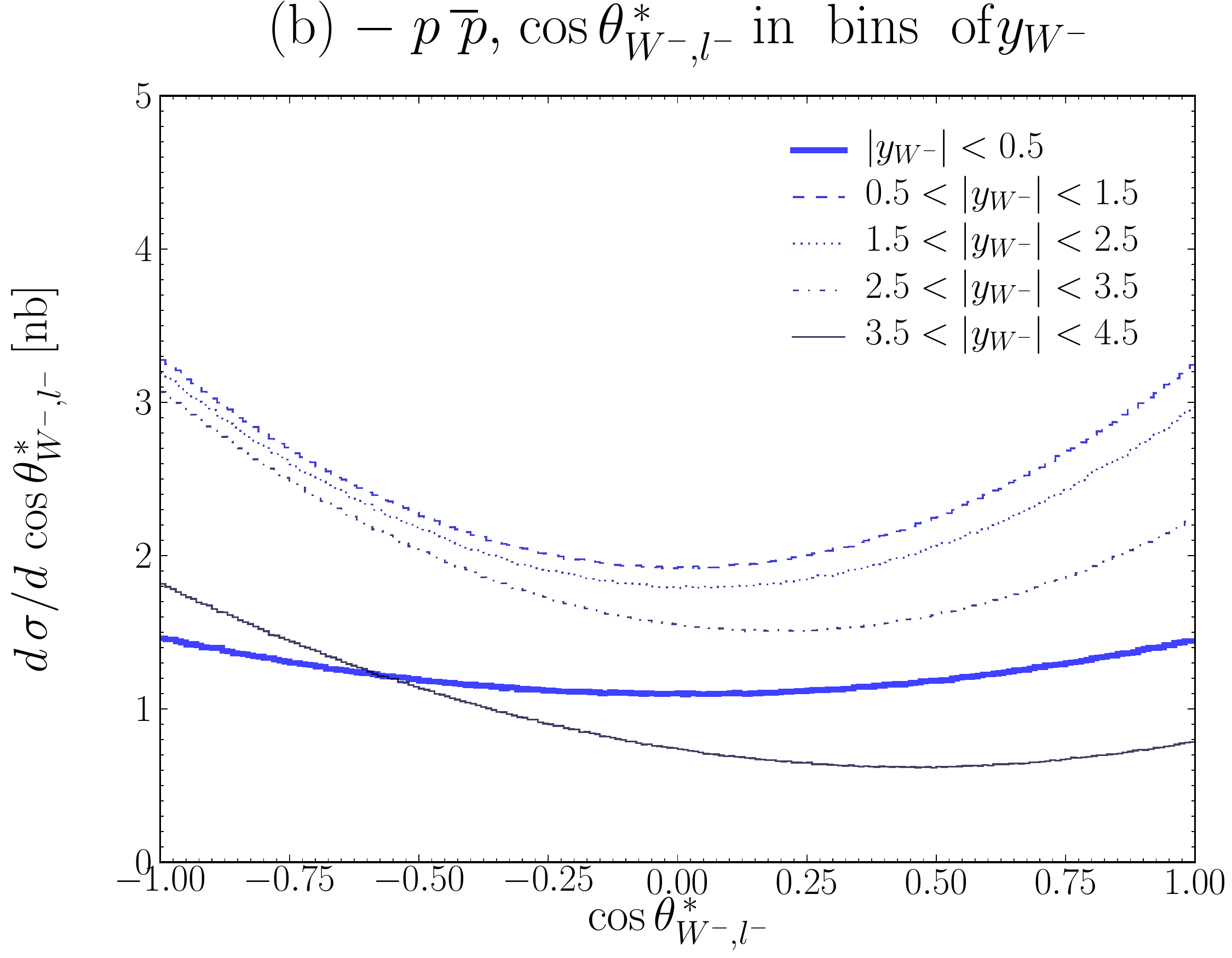}
    \vfill
    \includegraphics[width=0.495\tw]{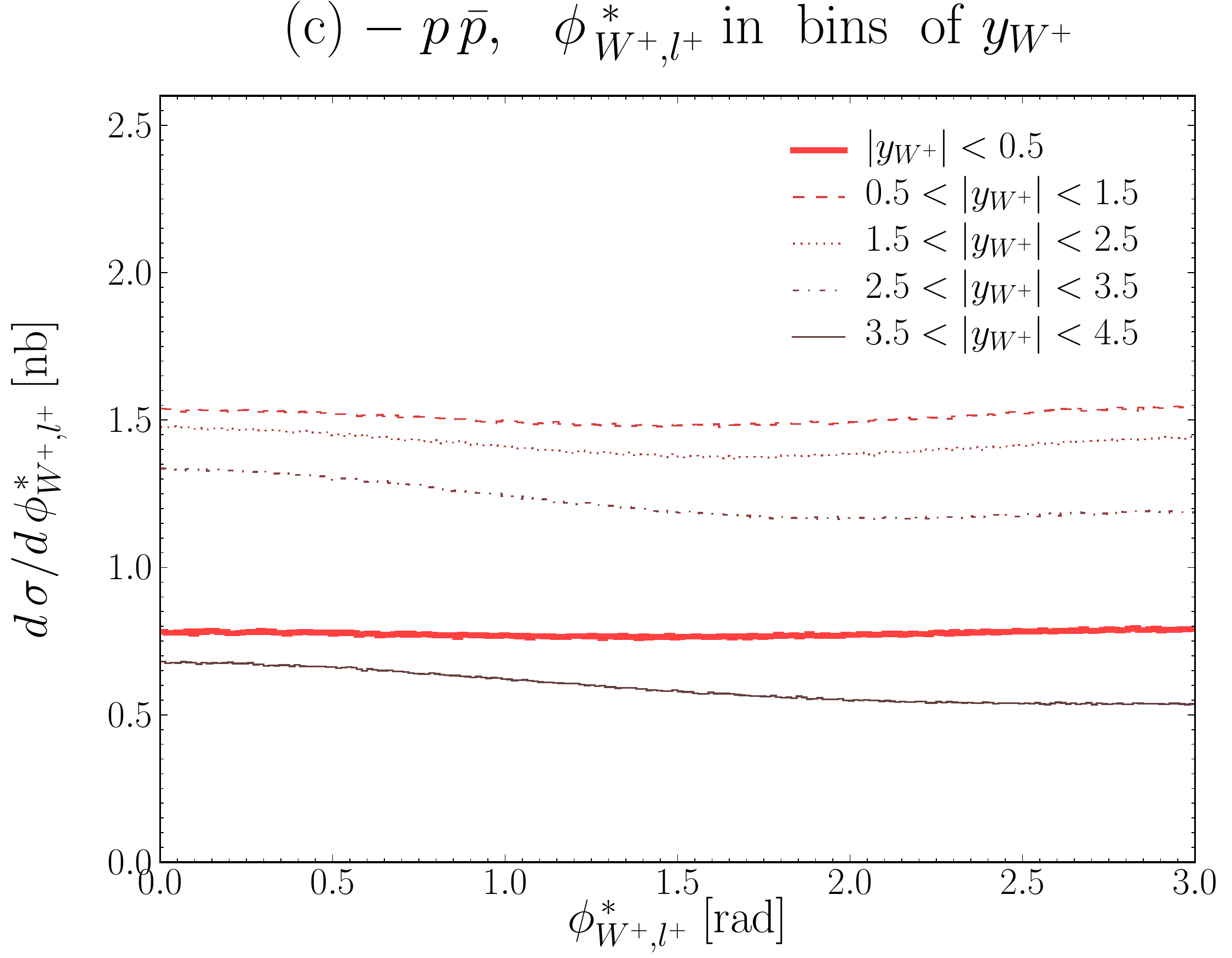}
    \hfill
    \includegraphics[width=0.495\tw]{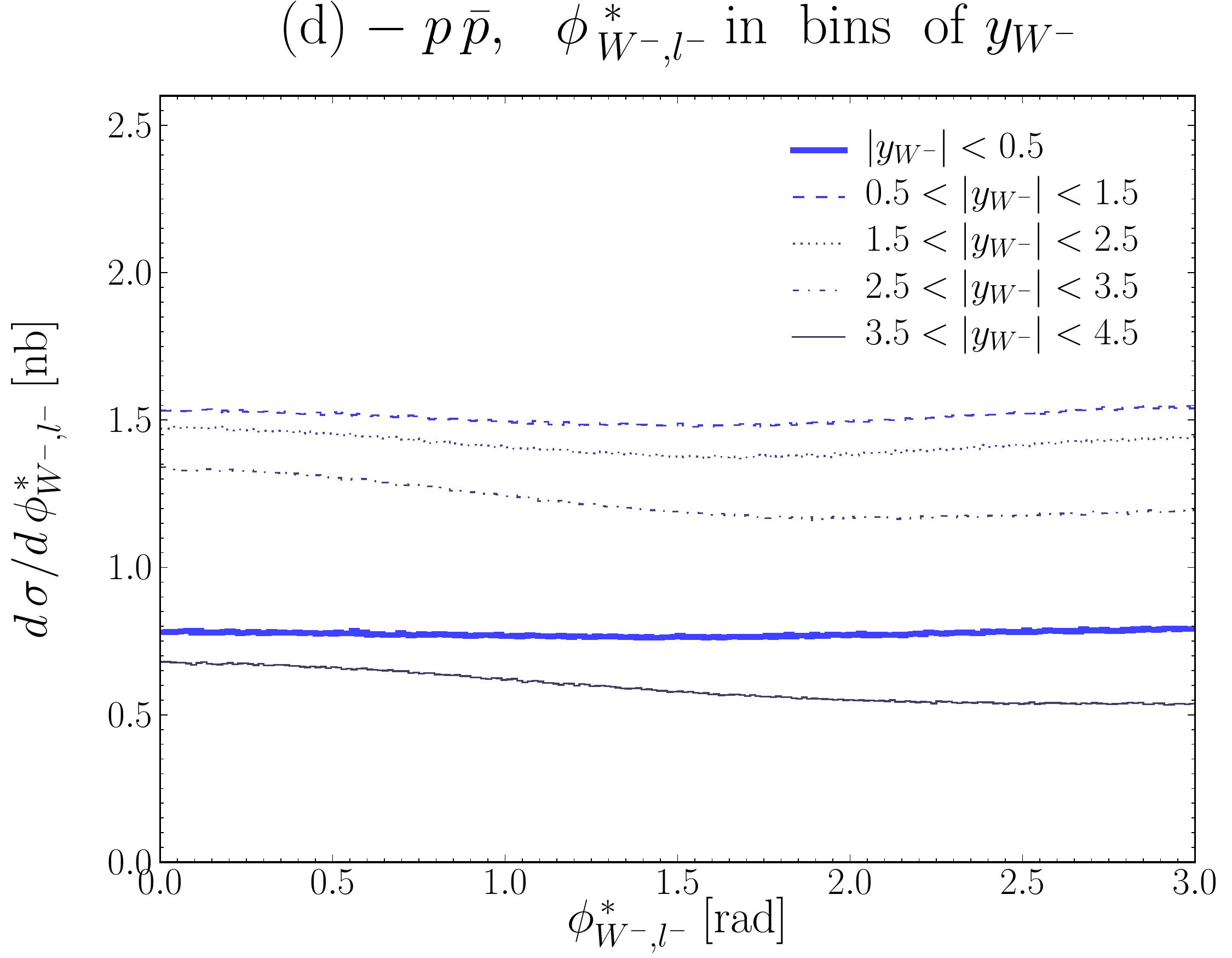}
    \caption[Distributions of $\costhetaWlwrf$ and $\phi_{W,l}^{\,\ast}$ in bins of $\yW$ for the positively and negatively charged 
      channel in $\ppbar$ collisions]
            {\figtxt{Distributions of $\costhetaWlwrf$ and $\phi_{W,l}^{\,\ast}$ in bins of $\yW$ for the positively (a,c) and 
                negatively (b,d) charged channel in $\ppbar$ collisions.}}
            \label{app_ppb_costhetaWlwrf_YWx}
  \end{center} 
\end{figure}

\clearpage
Finally we show the behaviour of the ``valence'' contribution in bins of $\yW$ from the point of view
of  $\costhetaWlwrf$ and $\phi_{W,l}^{\,\ast}$ in Fig.~\ref{app_ppb_costhetaWlwrf_YWx} where both charged 
channels show the same
behaviour with almost no asymmetries in $|\yW|<0.5$ because of the negligible ``valence'' contributions.
As the ``valence'' contributions increases with the selection in rapidity we see that the 
supremacy of left(right) $\Wp(\Wm)$ which was observed inclusively with 
$\textcircled{g}>\textcircled{h}$ is back in Fig.~\ref{app_ppb_cos_histos}.
\index{W boson@$W$ boson!Production in ppbar@Production in $\ppbar$ collisions!Detailed|)}

\clearpage
\subsection{Proton--proton collisions}\label{ss_app_pp}
\index{W boson@$W$ boson!Production in pp@Production in $\pp$ collisions!Detailed|(}
As we have seen previously even in the case of the $\ppbar$ collisions there are different asymmetries
appearing in the decay of the lepton but eventually no charge asymmetries are observed.
Here in $\pp$ collisions, the absence of anti-quarks of valence will open the door to the previous 
asymmetries to fully express themselves and create important asymmetries between the
positive and negative kinematics, especially in the final state.

First of all, before looking at the $W$ boson and charged leptons kinematics in specific domain 
of the phase-space we quantify the global charge asymmetry at the inclusive level due to the fact that 
in overall we produce more $\Wp$ than $\Wm$. Taking the numerical values of the inclusive cross 
sections for $\pp$ collision (Table~\ref{table_xtot}) and writing $A_\pp$ the global charge asymmetry 
we evaluate it like 
\begin{eqnarray}
A_\pp &\equiv& \frac{\sigma_\mm{incl.}^+-\sigma_\mm{incl.}^-}
                    {\sigma_\mm{incl.}^+-\sigma_\mm{incl.}^-},\\
     &\approx& 0.1478.
\end{eqnarray}
Then, when looking further one at the discrepancies between the two charged channels in the observable $a$ 
via $\Asym{a}$, all deviations from the horizontal axis $A_\pp$ have to be understood as the
result --in the considered point of phase space-- of dynamic issues creating noticeable differences 
between the positive and negative kinematics in top of the global larger production of $\Wp$ with
respect to $\Wm$.

\begin{figure}[!h] 
  \begin{center}
    \includegraphics[width=0.5\tw]{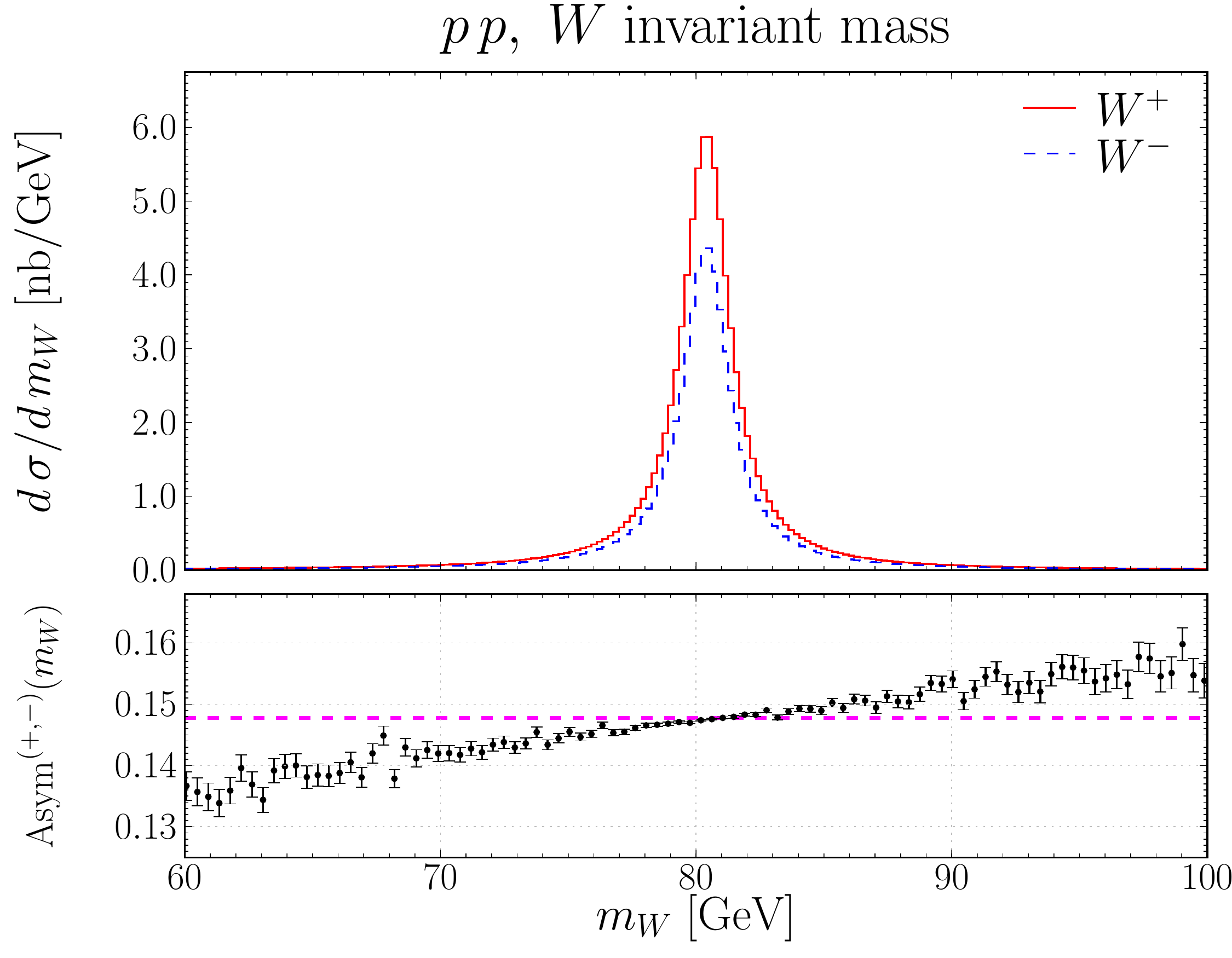}
    \caption[Invariant mass of the $W$ boson in $\pp$ collisions]
            {\figtxt{Invariant mass of the $W$ boson in $\pp$ collisions.}}
            \label{app_mw_pp}
  \end{center} 
\end{figure}

This remark can be applied to the very first example in Fig.~\ref{app_mw_pp} where we can see the
peak of the invariant mass of the $\Wp$ is slightly centered to a higher mass than the one of the 
$\Wm$. We observe $m_\Wp-m_\Wm\sim \mathcal O(10\MeV)$. 
Let us emphasise this behaviour is already present at LO.
We explain this effect heuristically with the following idea. 
In the quark--anti-quark collision the more energy is brought, the more the mass of the $W$ can 
take its share for its mass $m_W$.
The cases which fall under this category are the one involving a valence $u$ quark, for a $\Wp$,
and a valence $d$ quark for a $\Wm$.
Since in general $u$ quarks carry more longitudinal momentum than the $d$ quarks it turns out they
have more opportunities to participate to the production of a high mass $W$.
Hence, eventually, at the hadronic level $\Wp$ are produced with higher masses with respect to 
the $\Wm$.

\begin{figure}[!h] 
  \begin{center}
    \includegraphics[width=0.495\tw]{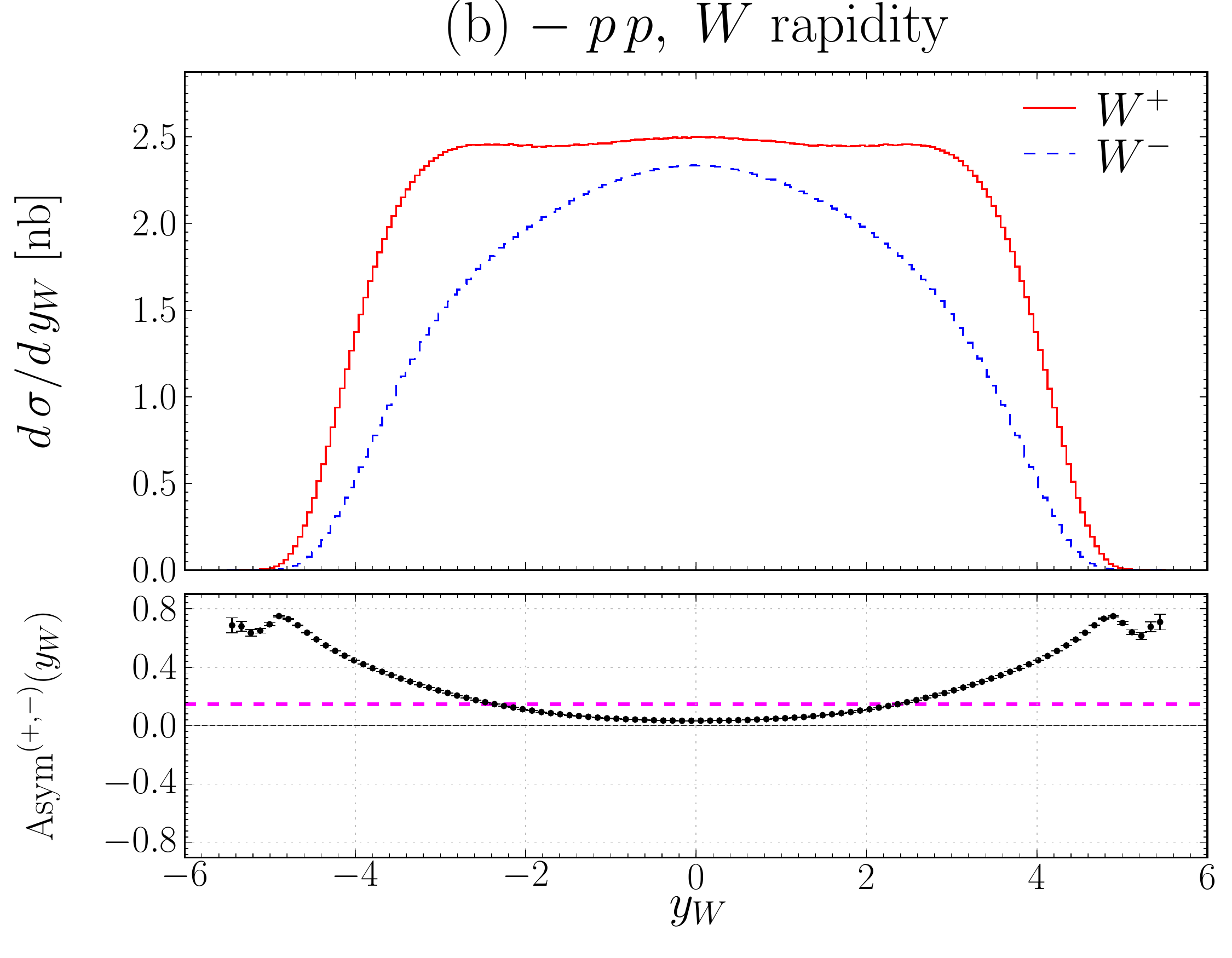}
    \hfill
    \includegraphics[width=0.495\tw]{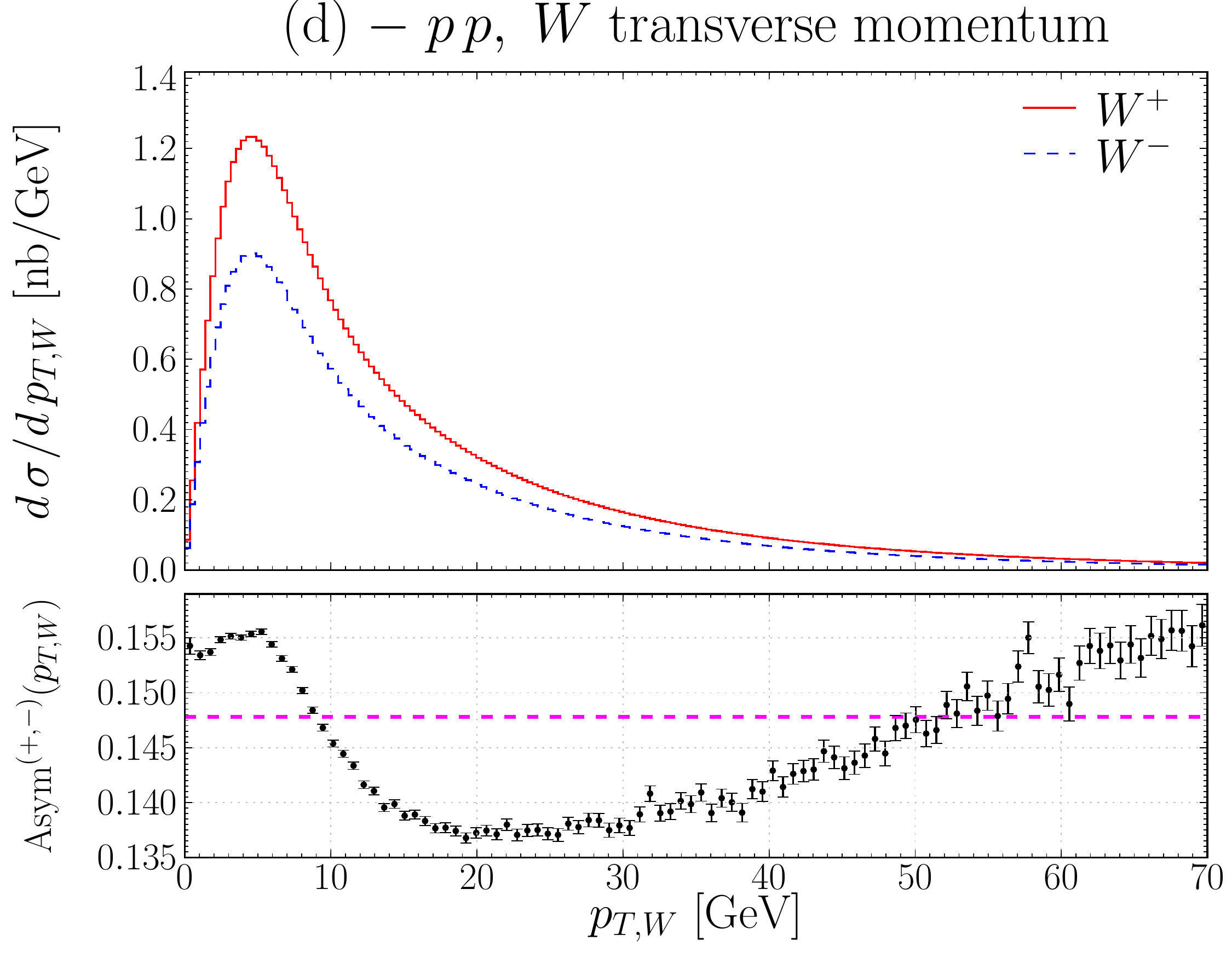}
    \vfill
    \includegraphics[width=0.495\tw]{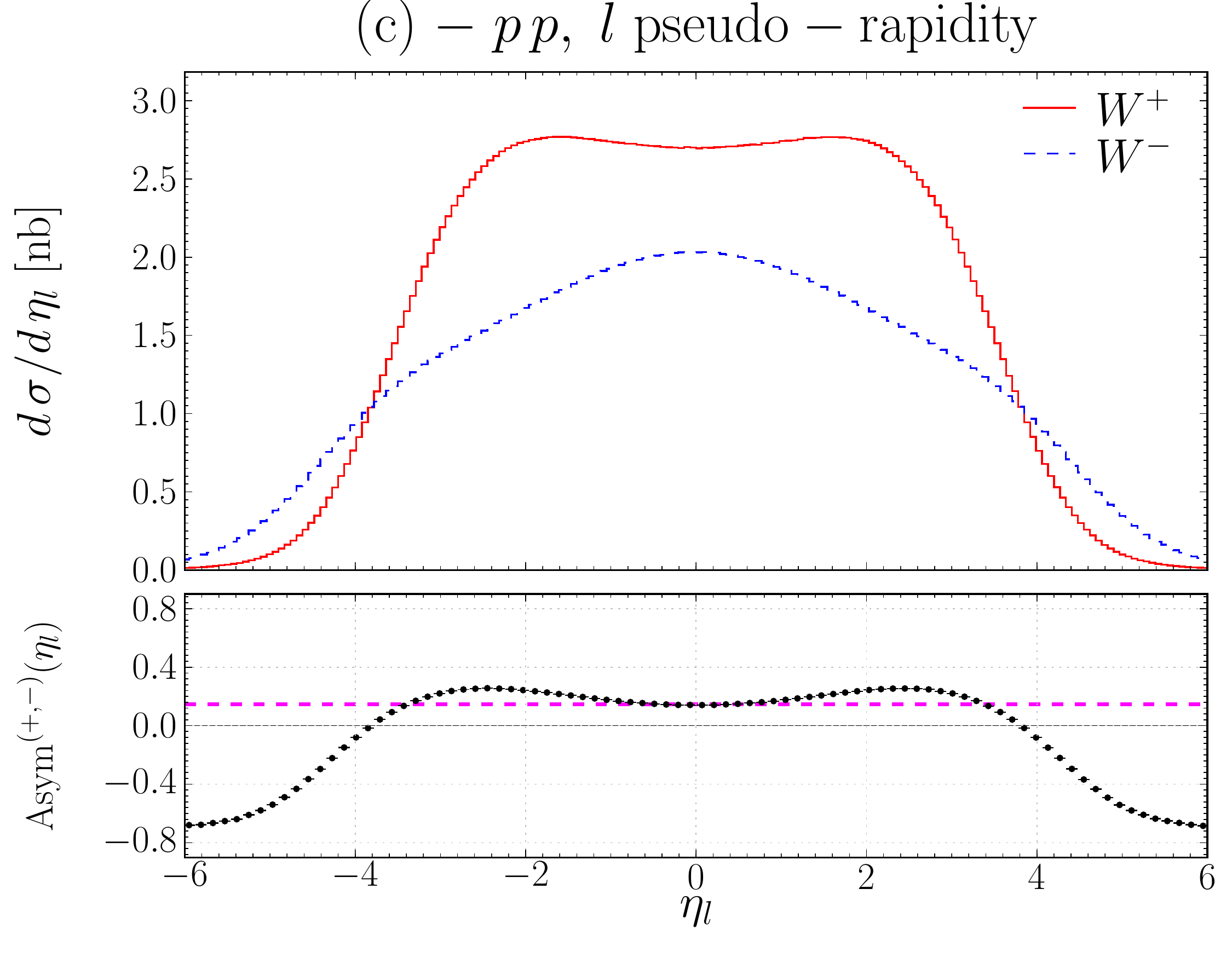}
    \hfill
    \includegraphics[width=0.495\tw]{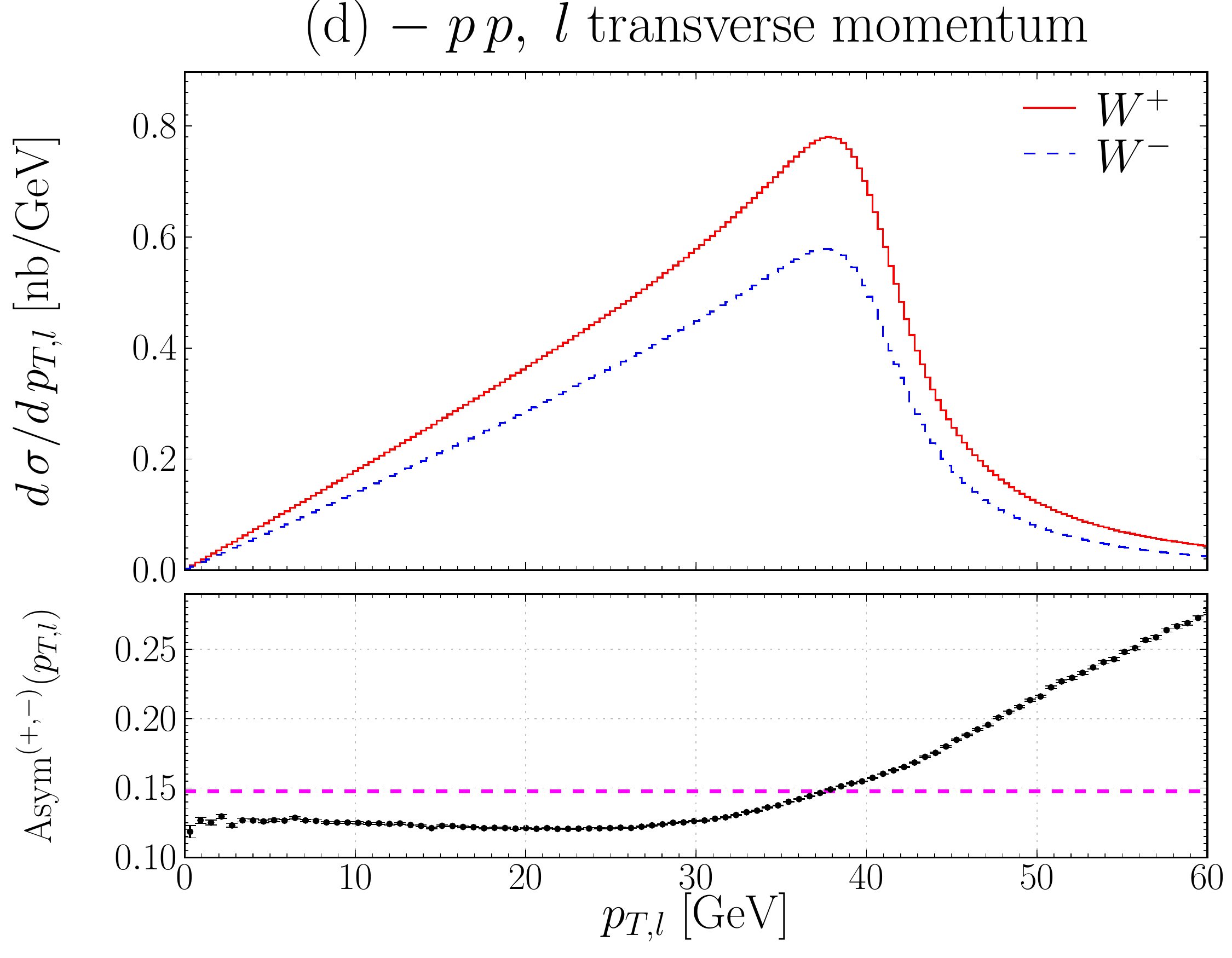}
    \caption[Distributions of the $W$ boson rapidity and transverse momentum along with the one of the 
      charged lepton pseudo-rapidity and transverse momentum in $\pp$ collisions]
            {\figtxt{Distributions of the $W$ boson rapidity (a) and transverse momentum (b) 
                along with the one of the charged lepton pseudo-rapidity (c) and transverse momentum (d)
                in $\pp$ collisions.}}
            \label{app_pp_yW_pTW_etal_pTl}
            \index{W boson@$W$ boson!Transverse momentum}
            \index{W boson@$W$ boson!Rapidity}
            \index{Charged lepton@Charged lepton from $W$ decay!Transverse momentum}
            \index{Charged lepton@Charged lepton from $W$ decay!Pseudo-rapidity}
  \end{center} 
\end{figure}
\begin{figure}[!h] 
  \begin{center}
    \includegraphics[width=0.495\tw]{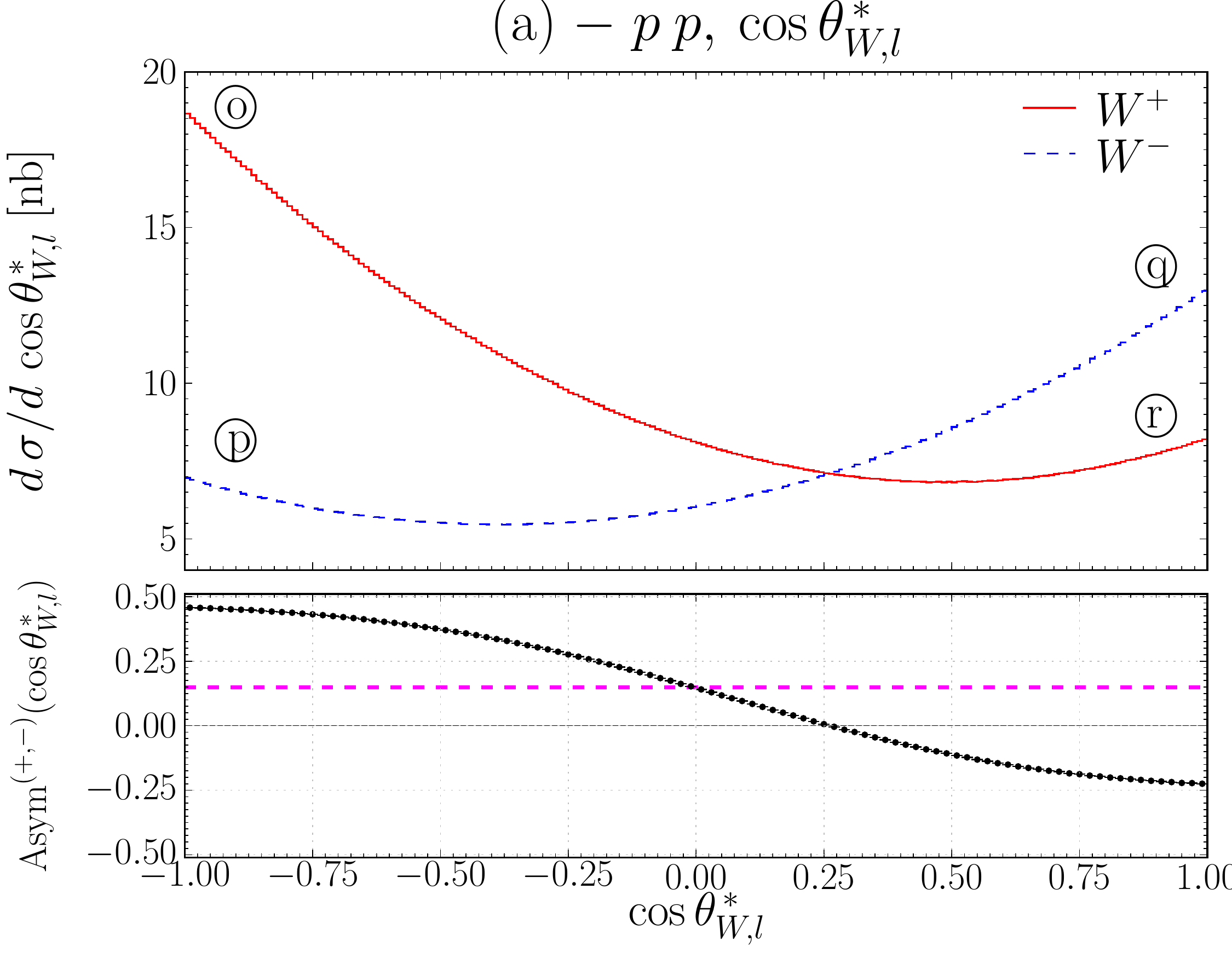}
    \hfill
    \includegraphics[width=0.495\tw]{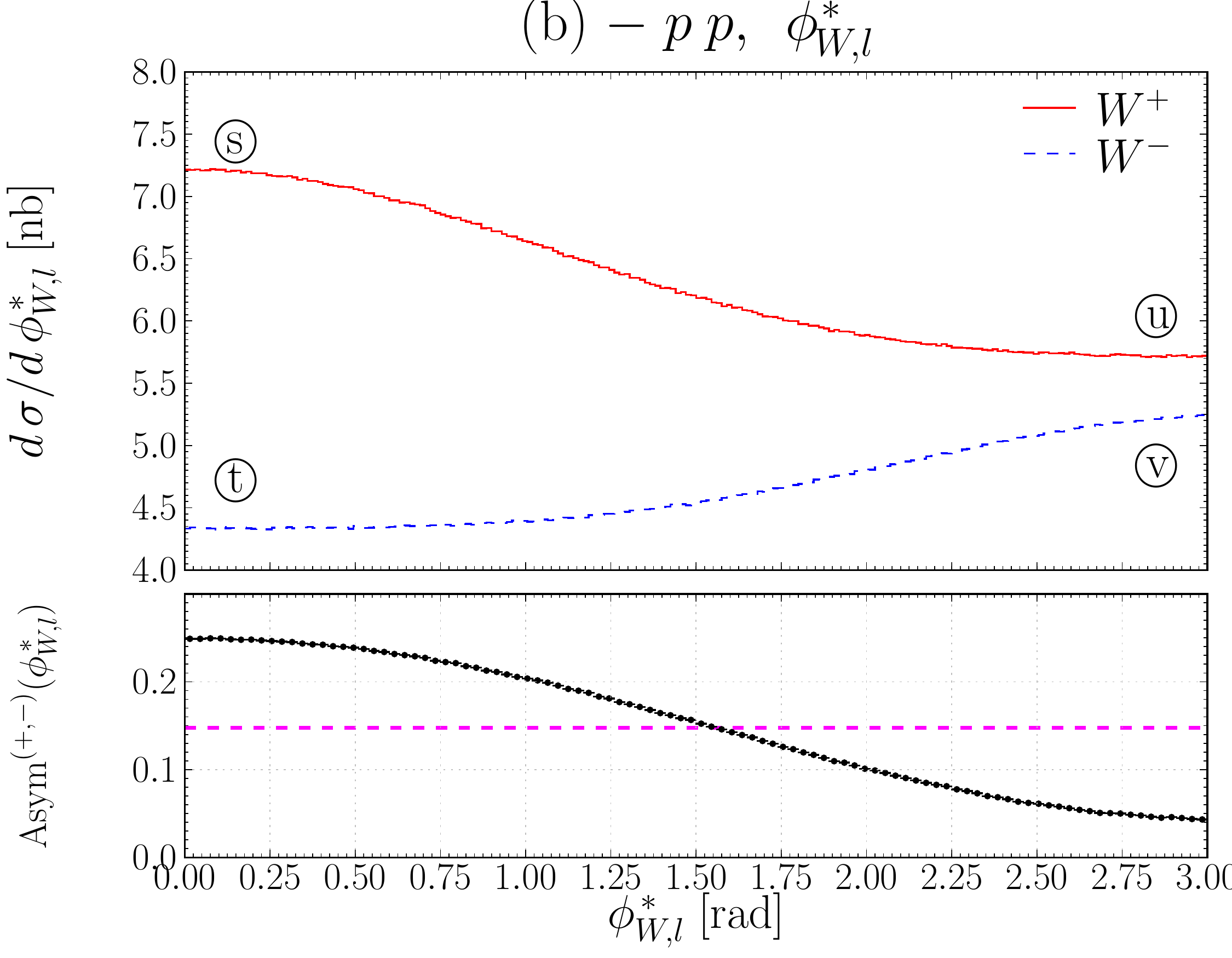}
    \caption[Distributions of $\costhetaWlwrf$ and $\phi_{W,l}^{\,\ast}$ in $\pp$ collisions]
            {\figtxt{Distributions of $\costhetaWlwrf$ (a) and $\phi_{W,l}^{\,\ast}$ (b) in $\pp$ 
                collisions.}}
            \label{app_pp_cos_histos}
  \end{center} 
\end{figure}

Figure~\ref{app_pp_yW_pTW_etal_pTl} shows now the kinematics of the $W$ bosons and of the charged leptons.
The rapidity displays two striking features, the first and most trivial, is a difference in the scale 
which simply relates that we have roughly twice many $u$ than $d$ in $\pp$ collisions. 
The second feature shows the $\Wp$ rapidity extends to a wider range than the $\Wm$ which means
the $\Wp$ can gain a more important longitudinal motion than the $\Wm$ which is the consequence than
the $u$ quarks carry most of the proton momentum.
The charge asymmetries in $\pTW$ comes from the difference in the production of the $\Wp$ and $\Wm$.
Because of the important implications in terms of systematic errors in the study of $\MWp-\MWm$ this
specific topic was discussed in the core of the Chapter (cf. \S\,\ref{ss_W_prod}).
Turning now to the charged lepton kinematics we observe that the pseudo-rapidity distribution follows
roughly the pattern of the rapidity with the particular feature that the $\lm$ tends to decay in the 
direction of the $W$ momentum direction while it is the inverse behaviour for the $\lp$.
This pattern will be detailed further while looking at the $\etal$ distributions in $\yW$ bins, 
at this stage we can already understand that the excess of matter in $\pp$ collisions implies
\begin{equation}
\qbar< q \qquad\Rightarrow\qquad W(\lambda=+1) < W(\lambda=-1).
\end{equation}
To the light of the known behaviour of the charged lepton in their decay under the constrain of the 
$V-A$ coupling in EW interactions, this explains respectively the narrowing and flattening of the
$\eta_\lp$ and $\eta_\lm$ distributions. This can be seen more directly in 
Fig.~\ref{app_pp_cos_histos}.(a) looking at the $\costhetaWlwrf$ distributions.
To the purely sea pattern seen in Fig.~\ref{app_pp_NOVAL_costhetaWlwrf}.(a), 
we add up the ``valence'' terms but consider for simplicity in
the discussion only the main term $u_p^\val\,\dbar_p^\sea\,\Vckmsqr{u}{d}$ for the $\Wp$ and 
$d_p^\val\,\ubar_p^\sea\,\Vckmsqr{u}{d}$ for the $\Wm$.
Events such as $u^\val_p\longrightarrow\leftarrow \dbar_p^\sea$ give $\Wp(\lambda=-1)$ bosons from which the 
$\lp$ leptons decay preferentially in the opposite direction of $\vec p_\Wp$ materialising eventually 
as $\textcircled{o}$ in the Figure.
For $\Wm$ bosons, this time $d_p^\val\longrightarrow\leftarrow u^\sea_p$ create 
$\Wm(\lambda=-1)$ bosons from which the $\lm$ leptons decay most of the time in the direction of $\vec p_\Wm$.
Eventually the inequality in the gaps ``$\textcircled{o}-\textcircled{r}$'' $>$
``$\textcircled{q}-\textcircled{p}$'' is explained once again by $d<u$, 
\ie{} from the common pattern of the purely sea contribution the predominance of $u$ quarks in the 
collision adds up a larger asymmetry in $\Wp$ relatively to the one for $\Wm$.

\begin{figure}[!h] 
  \begin{center}
    \includegraphics[width=0.5\tw]{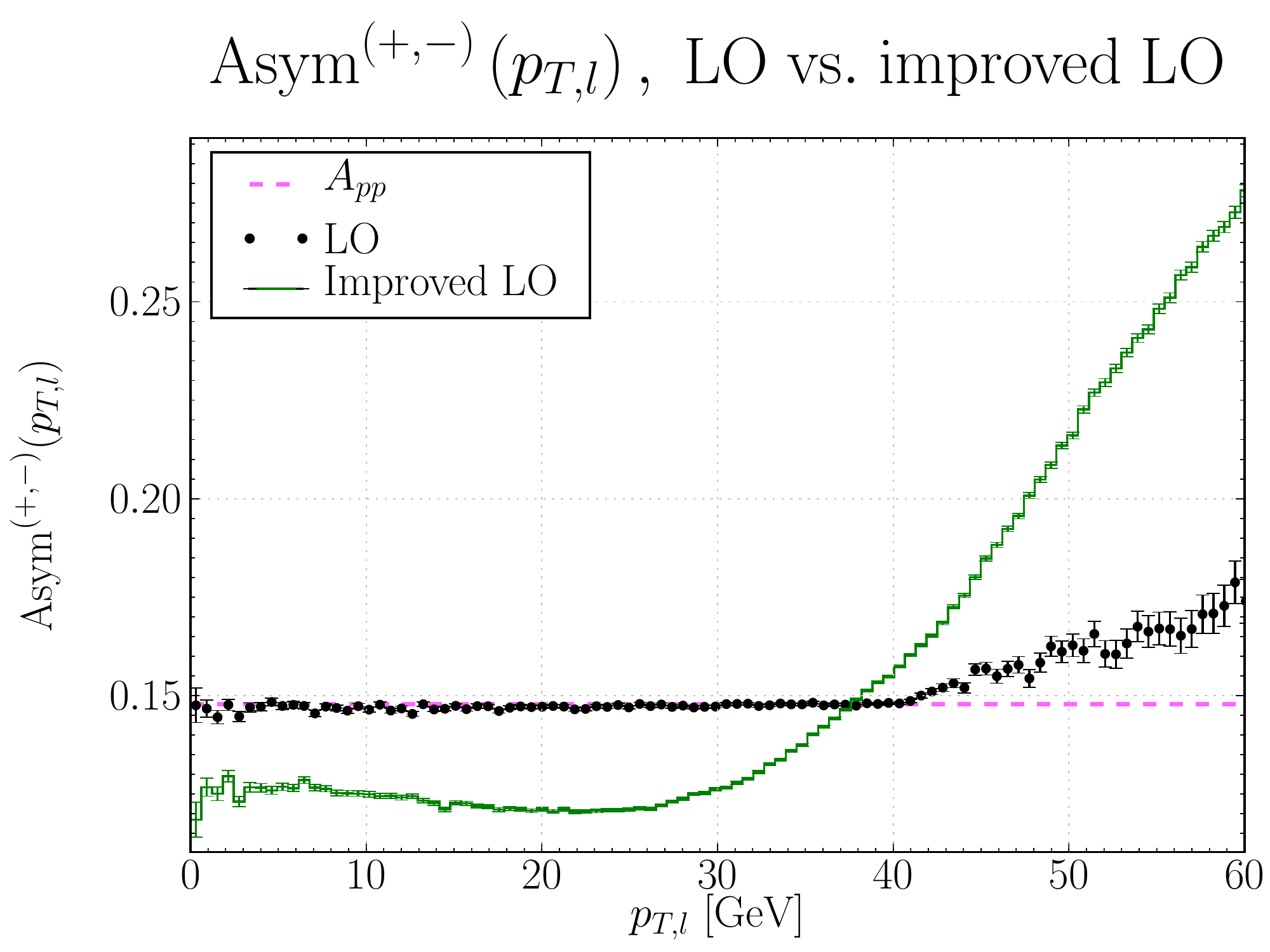}
    \caption[Final state charge asymmetry at the leading order and improved leading order]
            {\figtxt{Final state charge asymmetry at the leading order and improved leading order.}}
            \label{app_asym_pTl_LO_iLO}
  \end{center} 
\end{figure}

Coming back to the charged lepton kinematics the transverse momentum of the charged lepton shows 
very important charge asymmetries. The sources for those are\,: 
(1) $m_\Wp>m_\Wm$, (2) $p_{T,\Wp}\neq p_{T,\Wm}$ and (3) the non isotropic azimuthal decay of the charged
leptons with respect to the direction of $\pTW$ feature which has been already discussed in the case of 
$\ppbar$ collisions.
The first source cannot be the one responsible for the size of these effects.
Figure~\ref{app_asym_pTl_LO_iLO} proves it by showing $\Asym{\pTl}$ at LO
where $m_\Wp>m_\Wm$ is already present while (2) and (3) are not, 
and at the improved LO where all three effects are present.

The reason for these large asymmetries in the $\pTl$ distributions is explained by the 
predisposition of the $\lp$ leptons to decay preferentially in the direction of $\pTW$ while it is the 
over way around for the $\lm$ leptons. This can be seen directly in Fig.~\ref{app_pp_cos_histos}.(b)
showing the $\phi_{W,l}^{\,\ast}$ distribution.
The explanations of this effect can be followed looking at Fig.~\ref{app_phi_Wl_wrf_iLO}, where this
time the total absence of the (b) contribution involving valence anti-quarks open the door to effects 
(a) to fully express their charge dependent discrepancies.
We start by addressing the case of the $\Wp$ through the main ``valence'' term 
$u^\val_p\longrightarrow\leftarrow \dbar_p^\sea$ where, the higher transverse motion held by the 
$\dbar$ forces the $\lp$ to decay most of the time so that $\phi_{\Wp,\lp}^{\,\ast}\approx 0$.
The case of the $\Wm$ is identical but this time the opposite helicity of the $\lm$ makes it so 
it decays preferentially in the opposite direction of $\vec p_{T,W}$, \ie{} 
$\phi_{\Wm,\lm}^{\,\ast}\approx \pi$.
Once again the relative size of the ``$\textcircled{s}-\textcircled{u}$'' and 
``$\textcircled{t}-\textcircled{v}$'' gaps is justified by $d<u$.

\begin{figure}[!h] 
  \begin{center}
    \includegraphics[width=0.495\tw]{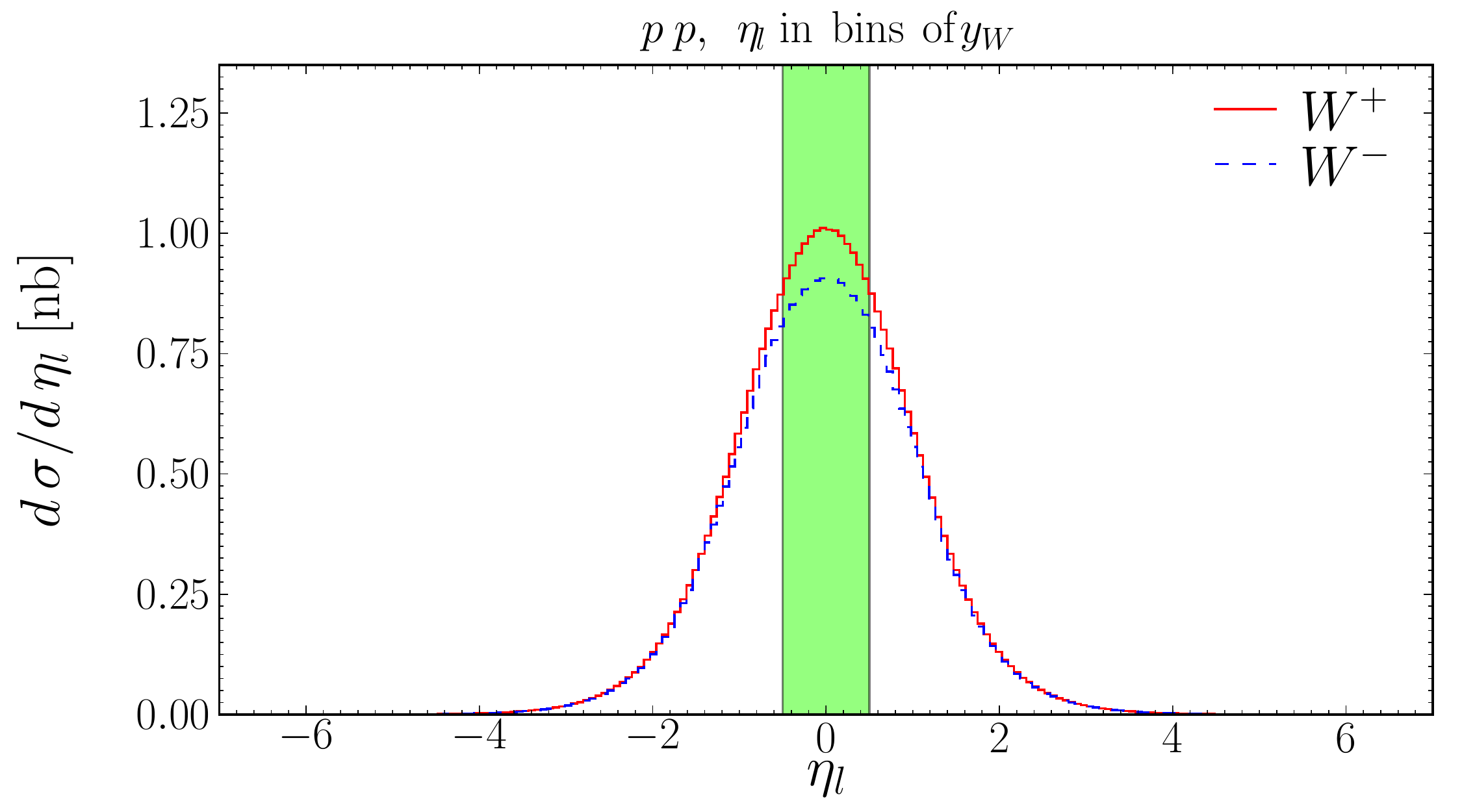}
    \hfill
    \includegraphics[width=0.495\tw]{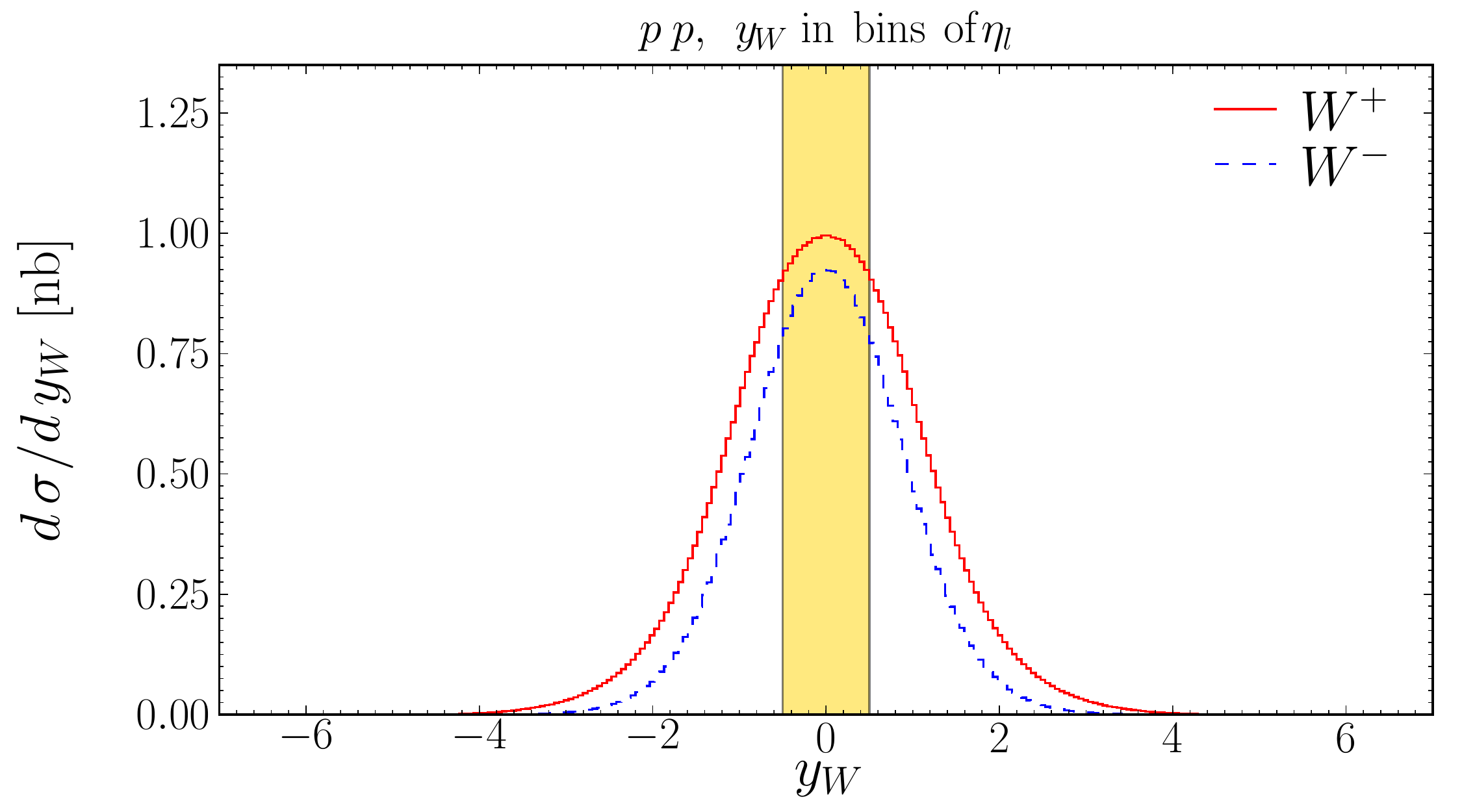}
    \vfill   
    \includegraphics[width=0.495\tw]{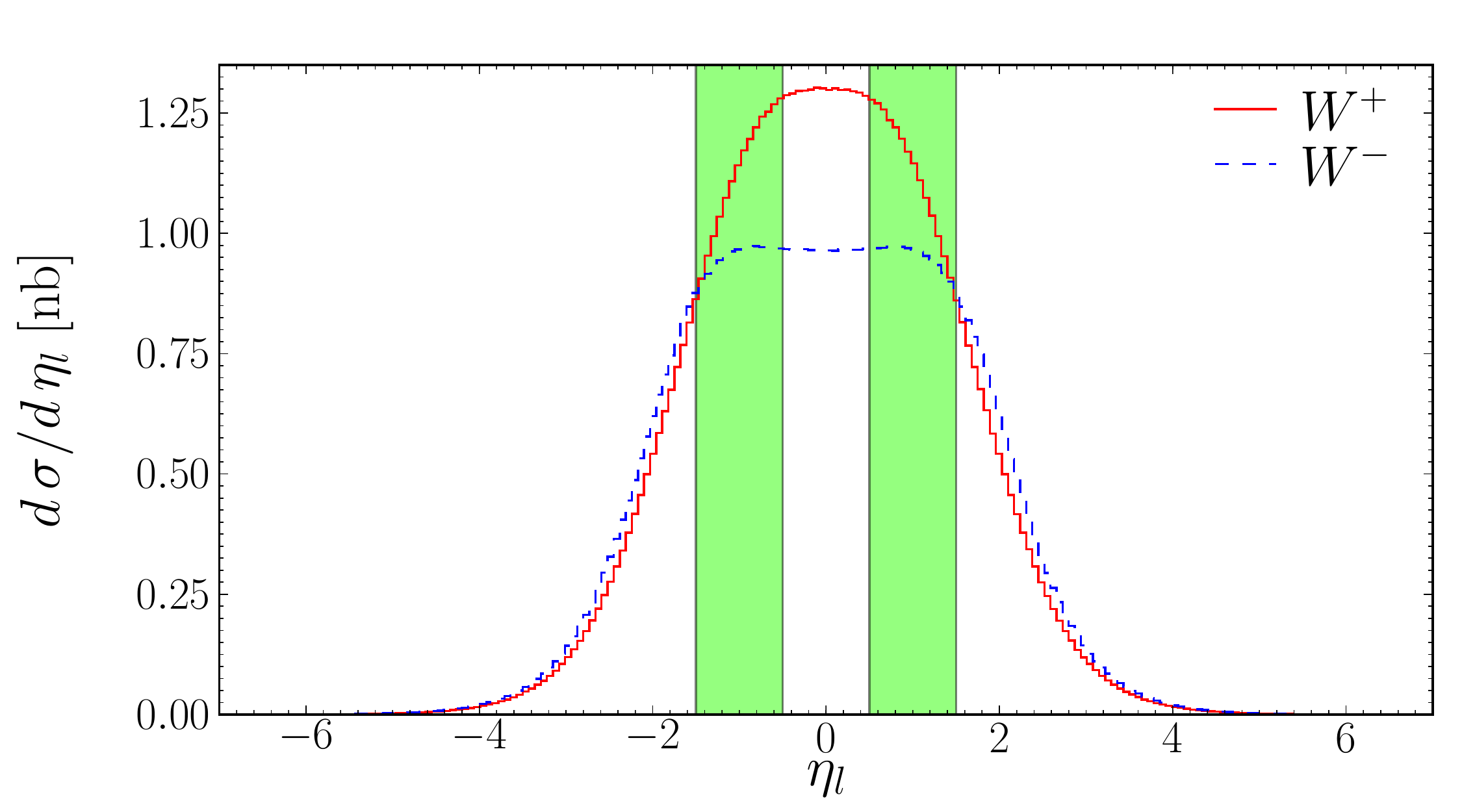}
    \hfill
    \includegraphics[width=0.495\tw]{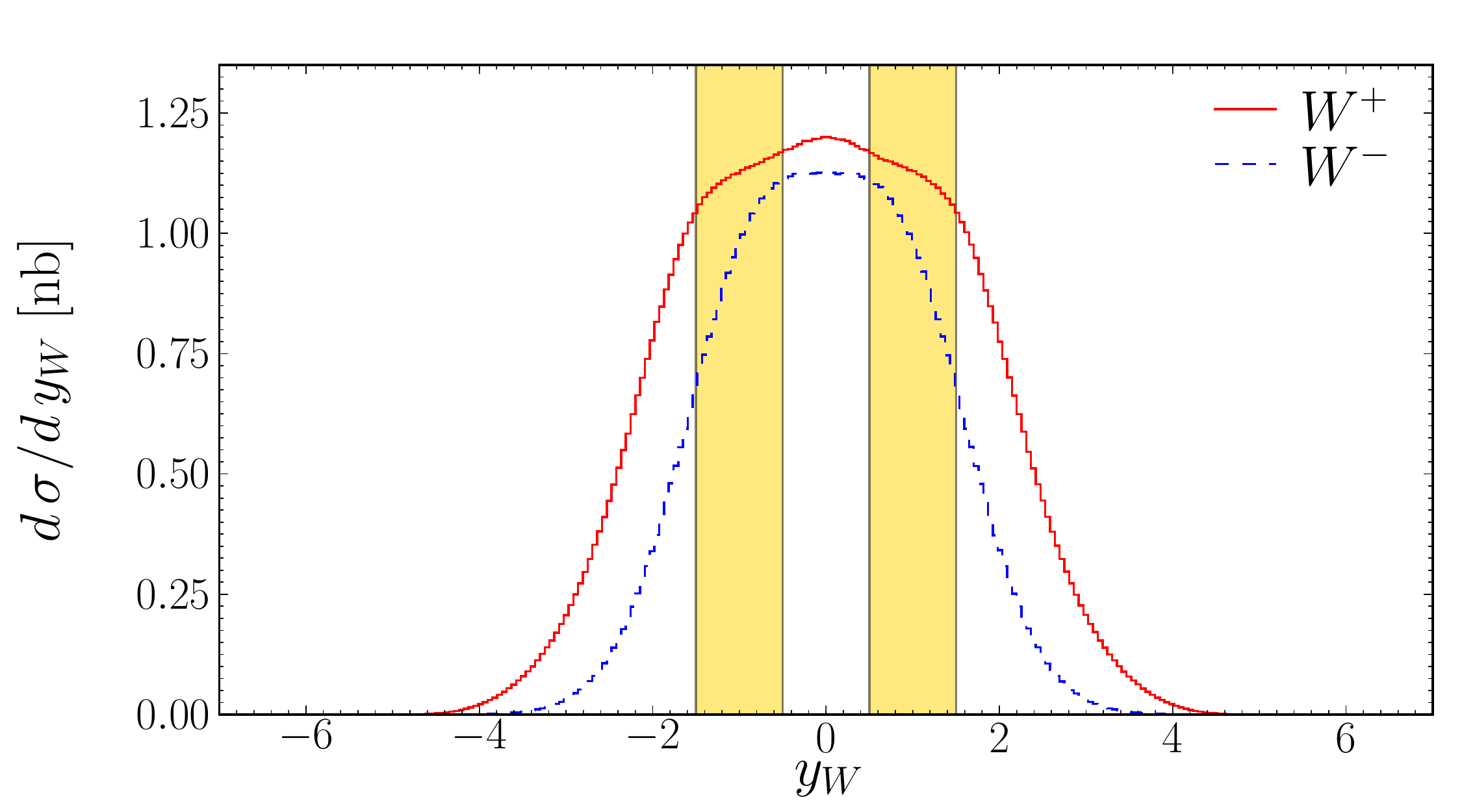}
    \vfill
    \includegraphics[width=0.495\tw]{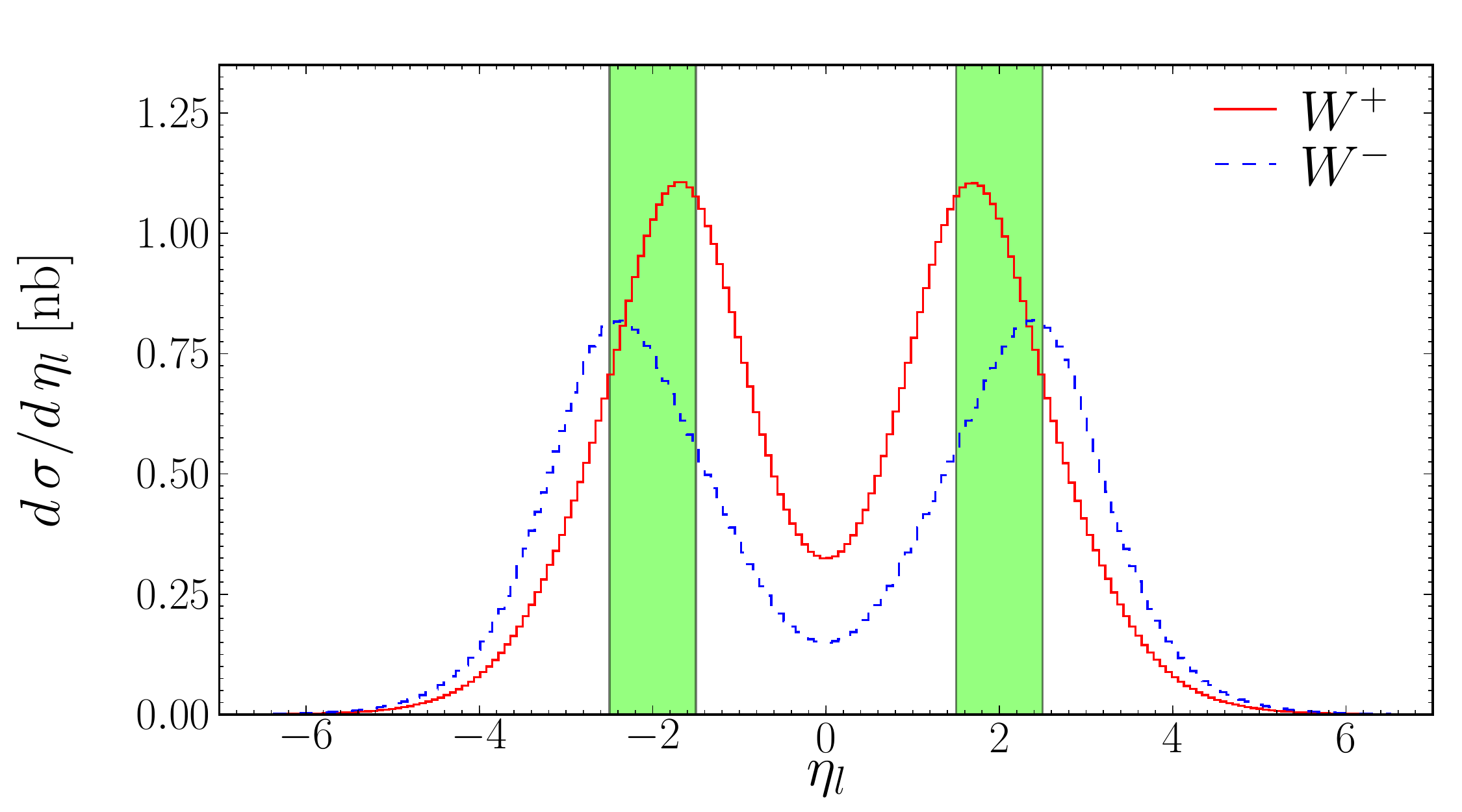}
    \hfill
    \includegraphics[width=0.495\tw]{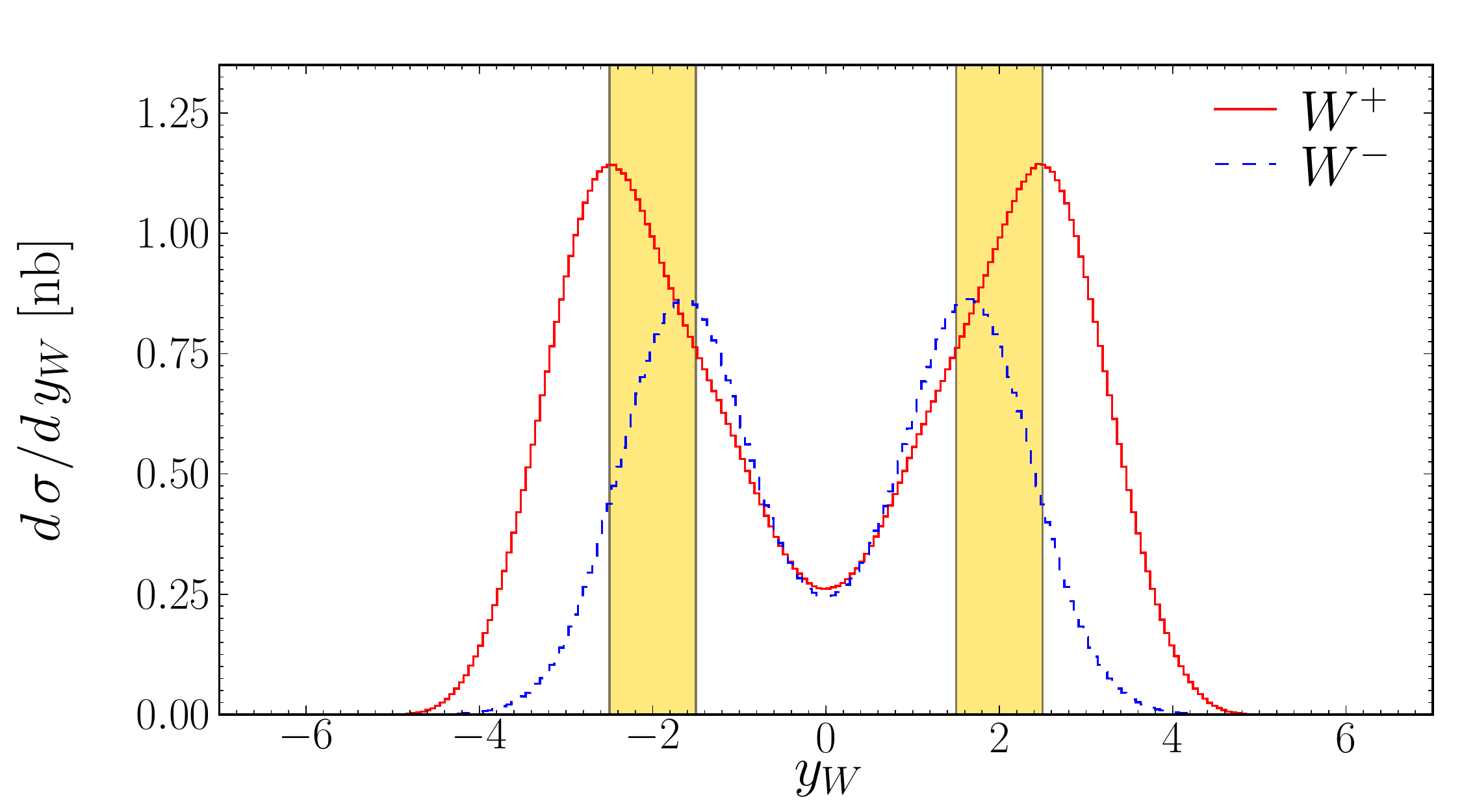}
    \vfill
    \includegraphics[width=0.495\tw]{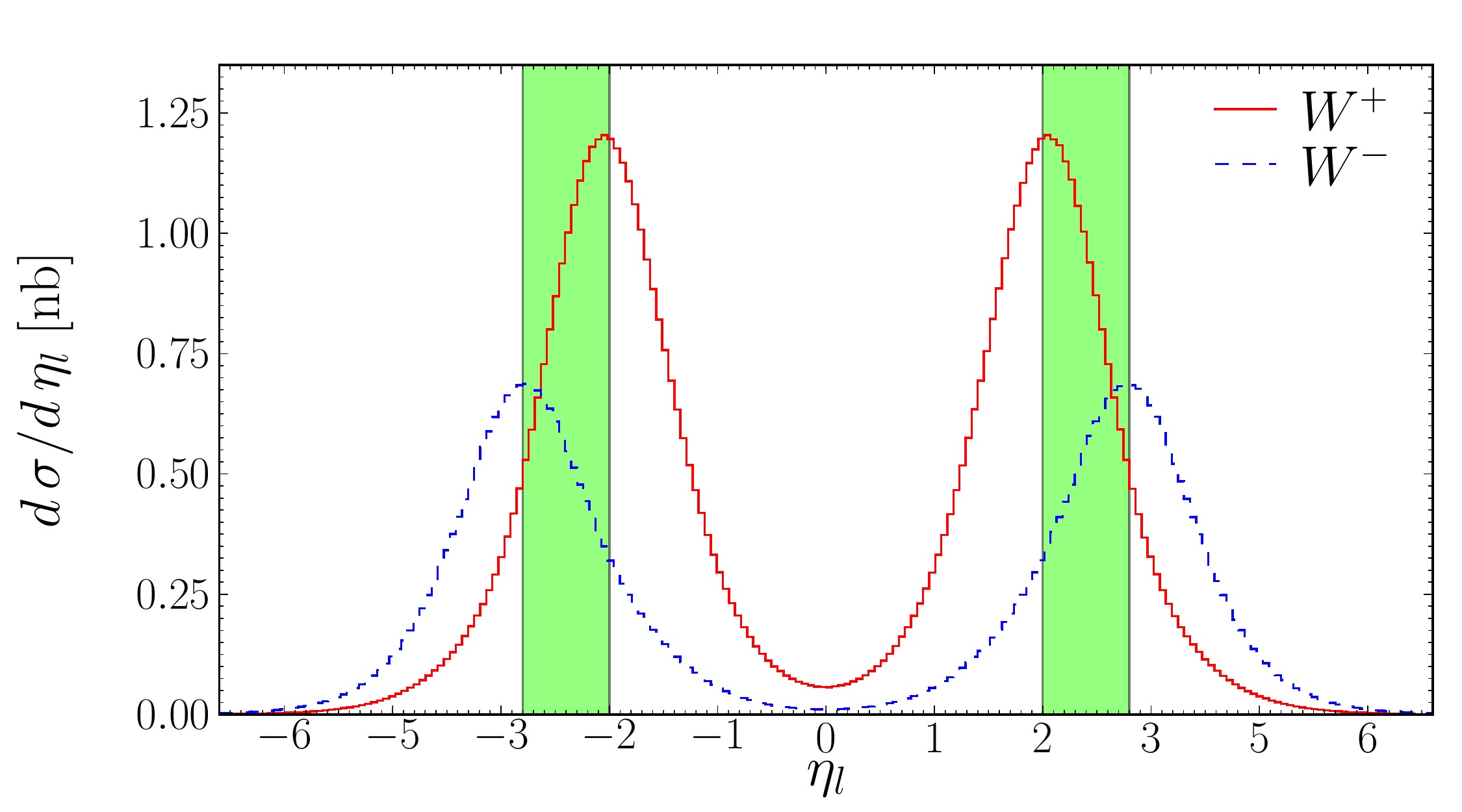}
    \hfill
    \includegraphics[width=0.495\tw]{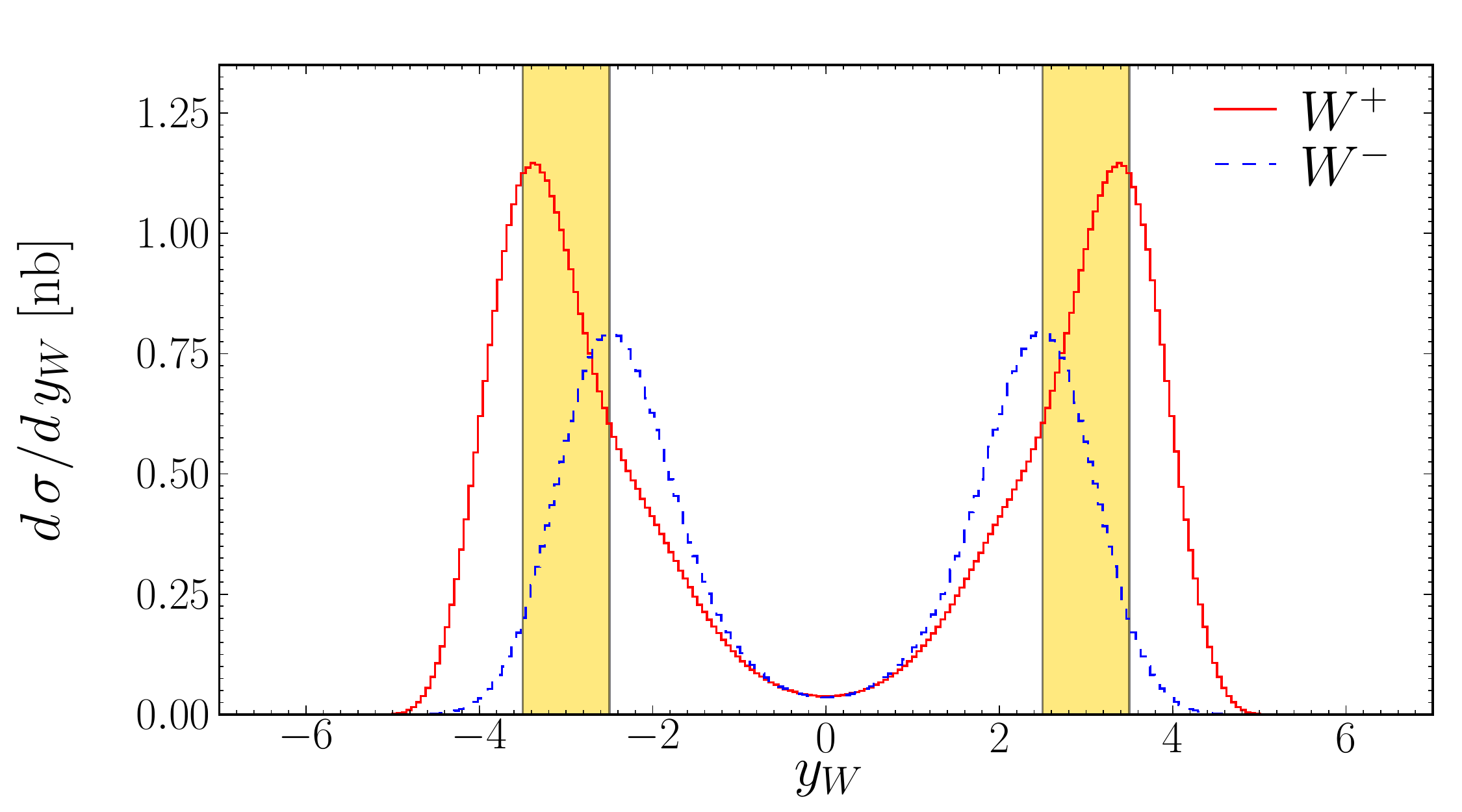}
    \vfill
    \includegraphics[width=0.495\tw]{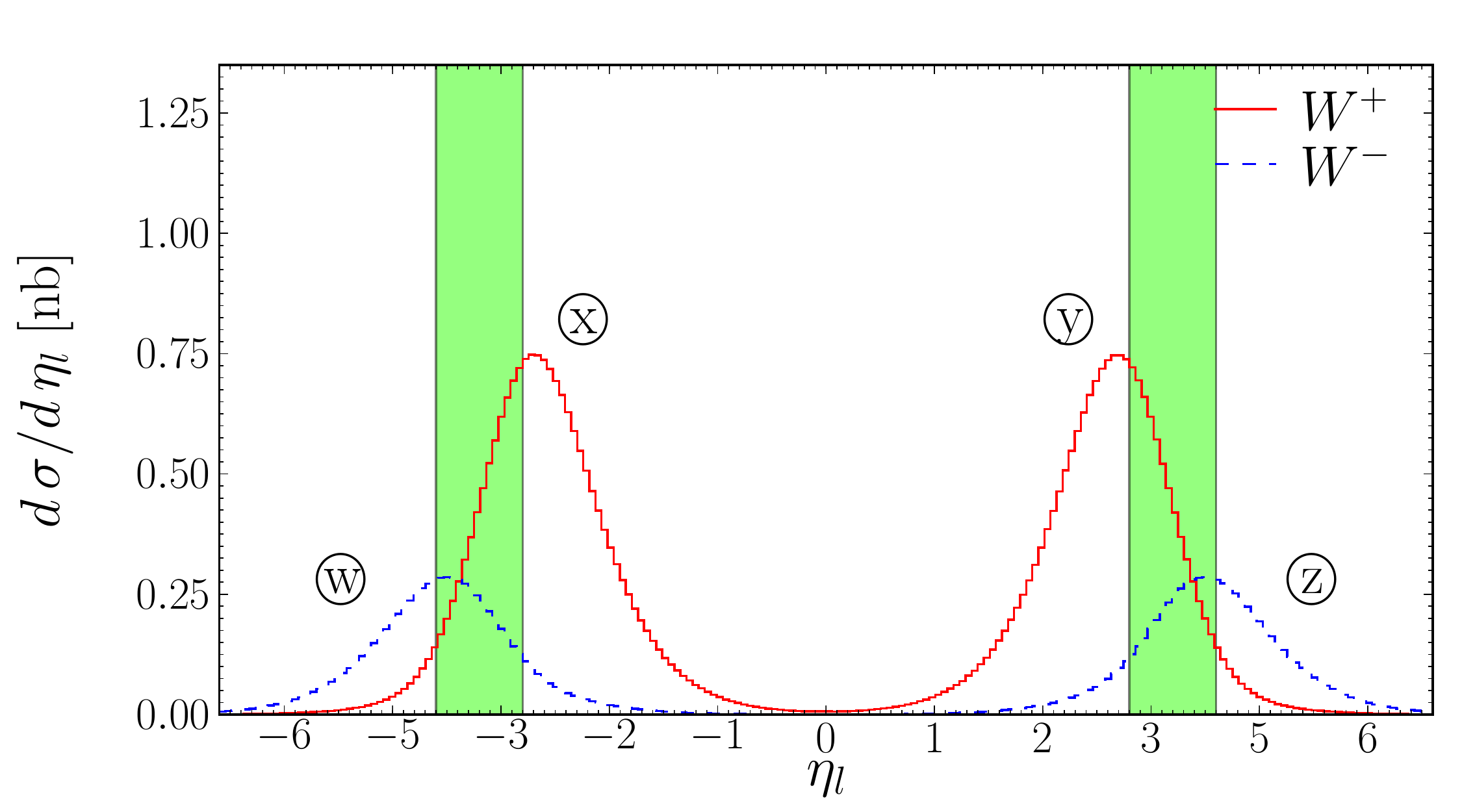}
    \hfill
    \includegraphics[width=0.495\tw]{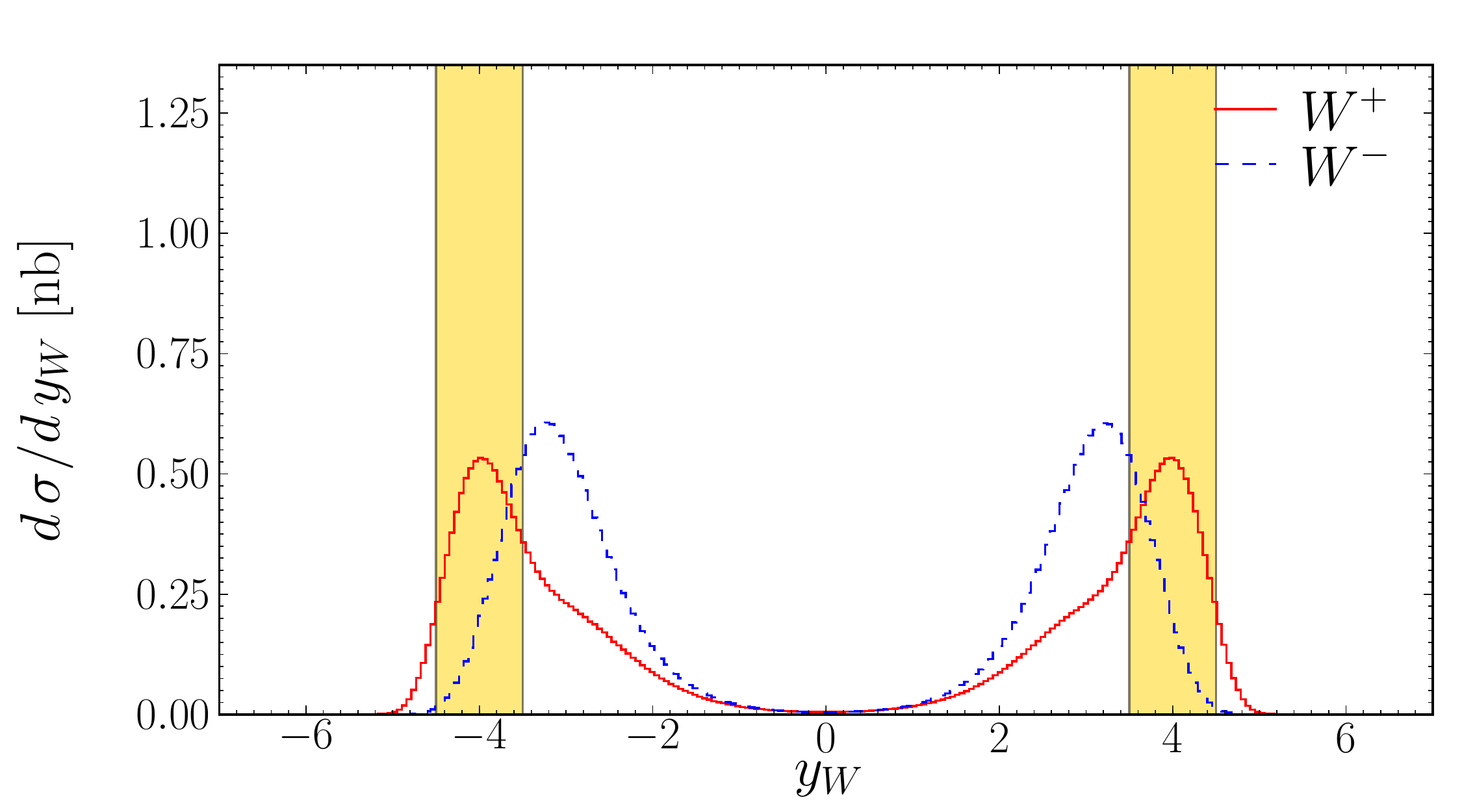}
    \caption[$W$ boson rapidity distributions in bins of the charged lepton pseudo-rapidity and \textit{vice versa}
      in the case of $\pp$ collisions]
            {\figtxt{$W$ boson rapidity distributions in bins of the charged lepton pseudo-rapidity (left) and 
                \textit{vice versa} (right) in the case of $\pp$ collisions. 
                In each plot the corresponding $\yW$ or $\etal$ selection is materialised by the colored stripe(s).}}
            \label{app_pp_etal_yW_in_yW_etal_bins}
  \end{center} 
\end{figure}

Figure~\ref{app_pp_etal_yW_in_yW_etal_bins}.(left) allows now to see the difference in the behaviour of
the decaying leptons in bins of $\yW$ through their pseudo-rapidity. 
Just like in $\ppbar$ collisions the asymmetry due to the
``valence'' contributions is almost negligible in the central narrow $\yW$ selection and then rise 
with the rapidity except that this time the asymmetries are not matched between $\Wp$ and $\Wm$ when 
summing the two opposite $\yW$ slices contributions but rather magnified.
In the case of $3.5<|\yW|<4.5$, the excess of $\Wp(\lambda=-1)$ is such that in average most of the 
$\lp$ leptons decay in the opposite direction of $\vec p_W$ as shown by the bumps in $\textcircled{x}$ 
and $\textcircled{y}$. On the other hand the $\lm$, due to the excess of $\Wm(\lambda=-1)$, follow the 
$W$ bosons momentum direction as displayed by the $\textcircled{w}$ and $\textcircled{z}$ bumps. 

This large influence of the valence quarks in the high rapidity region suggests that for a measurement
of $\MWp-\MWm$, selecting charged leptons emitted from low $W$ rapidity would at least allow to reduce 
the systematics uncertainties related to the valence or sea quarks.
\index{W boson@$W$ boson!Mass charge asym@Mass charge asymmetry $\MWp-\MWm$}
By going to the narrow region $|\etal<|0.5$ one can see the size of the ``valence'' contamination in 
this region is non negligible.
This is the reason why in the core of the Chapter the narrow selection was made rather to $|\etal|<0.3$
since it was found to be the best compromise for a realistic measurement prospect in Chapter~\ref{chap_W_mass_asym}. 

\begin{figure}[!h] 
  \begin{center}
    \includegraphics[width=0.495\tw]{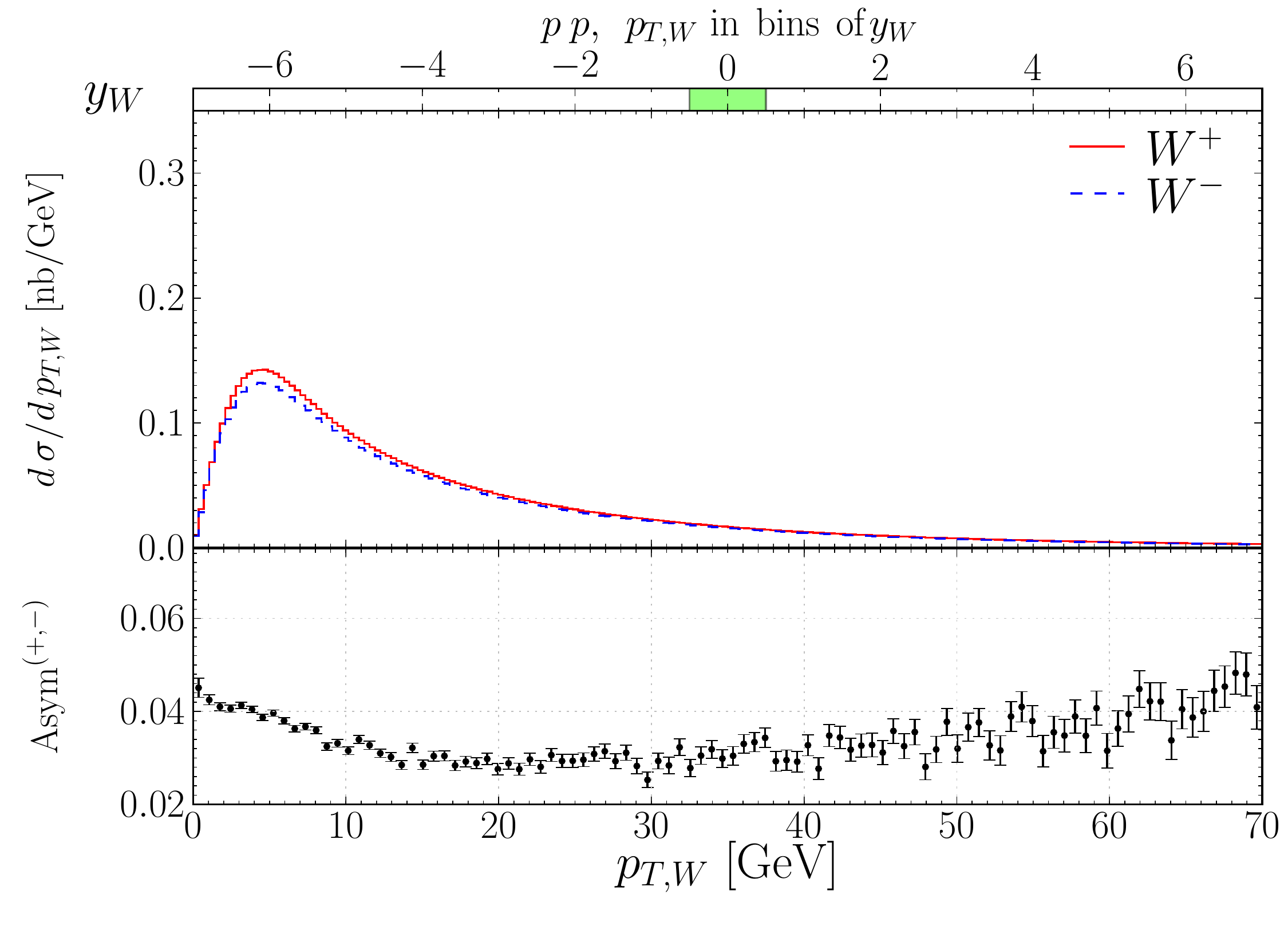}
    \hfill
    \includegraphics[width=0.495\tw]{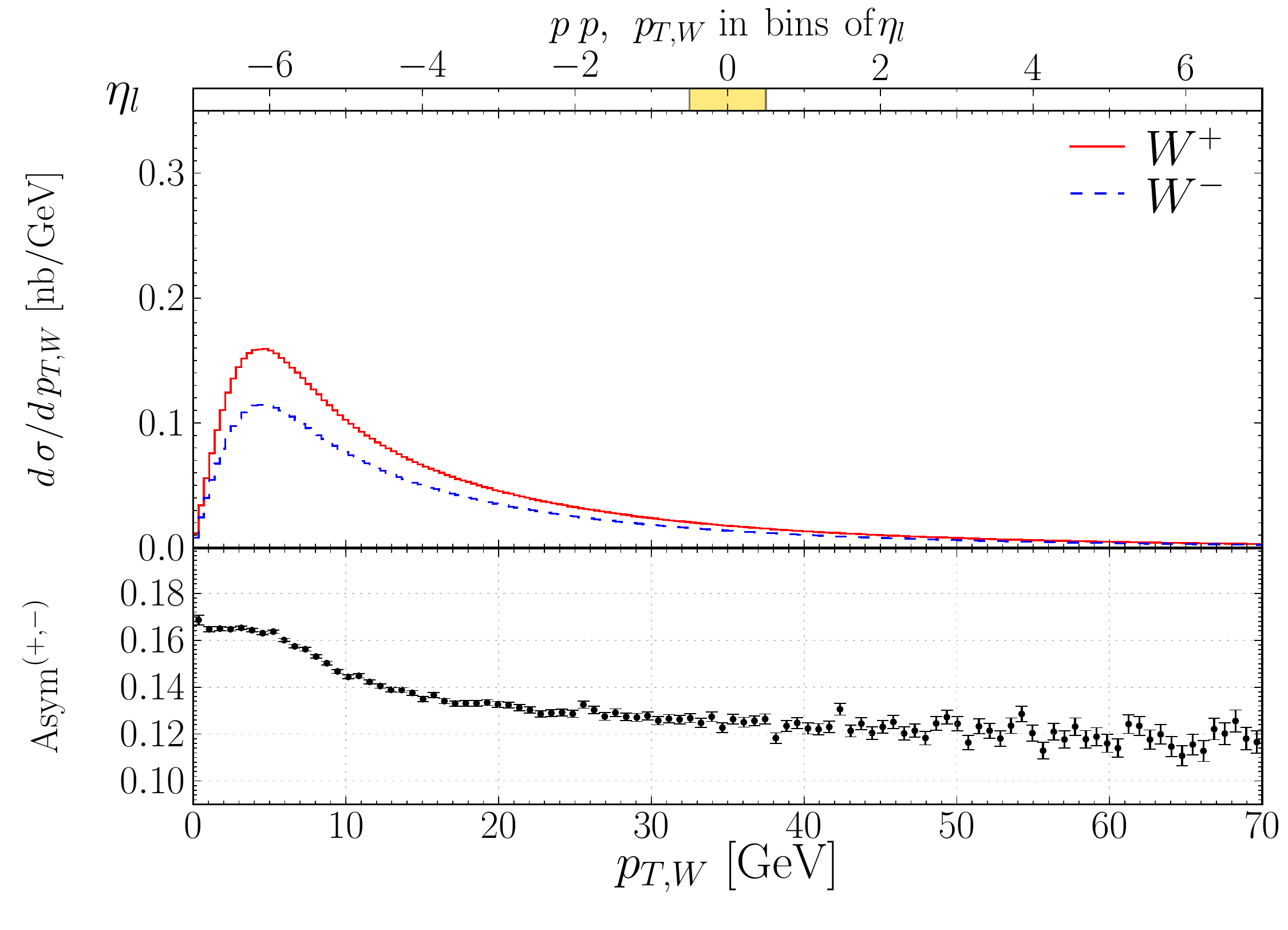}
    \vfill   
    \includegraphics[width=0.495\tw]{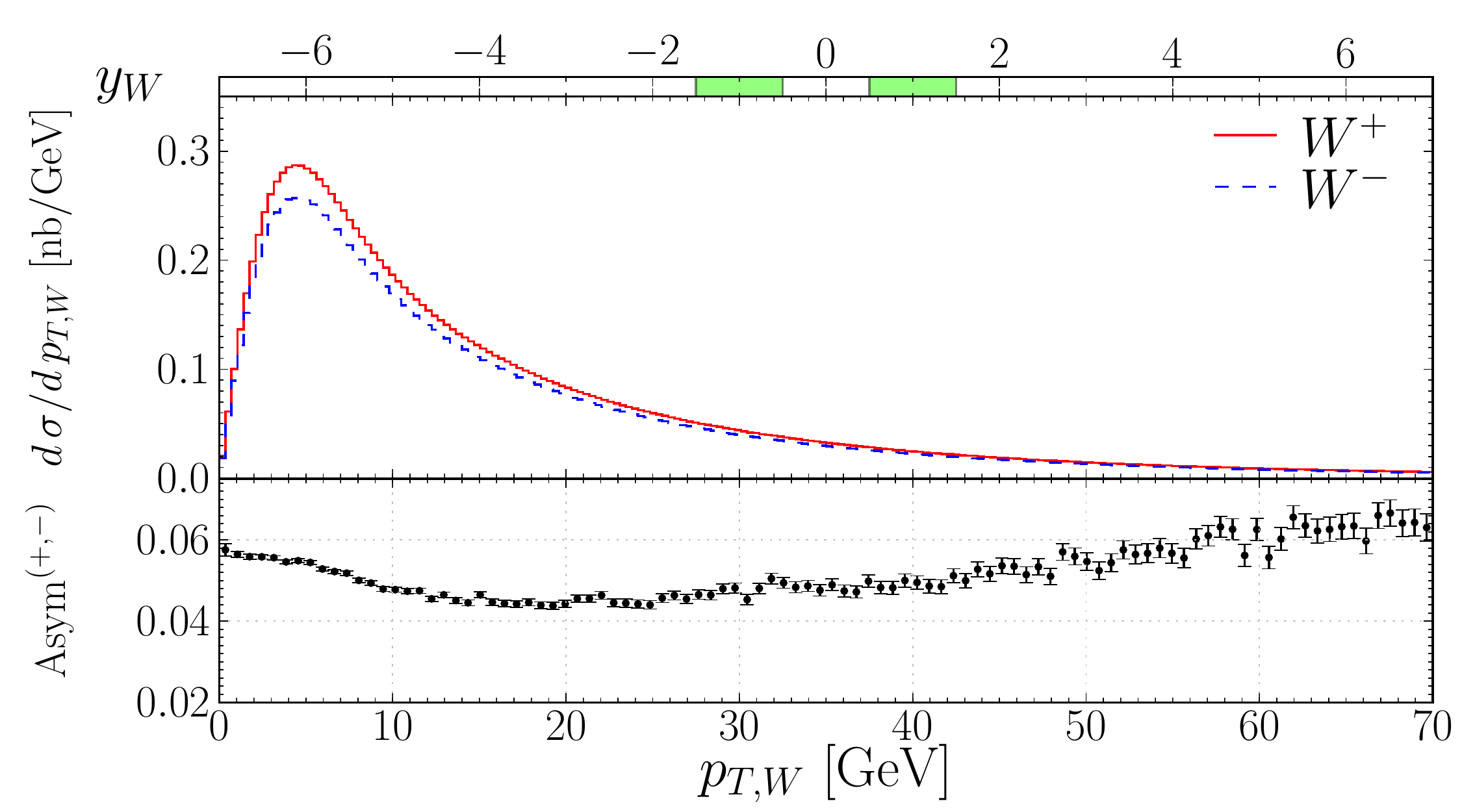}
    \hfill
    \includegraphics[width=0.495\tw]{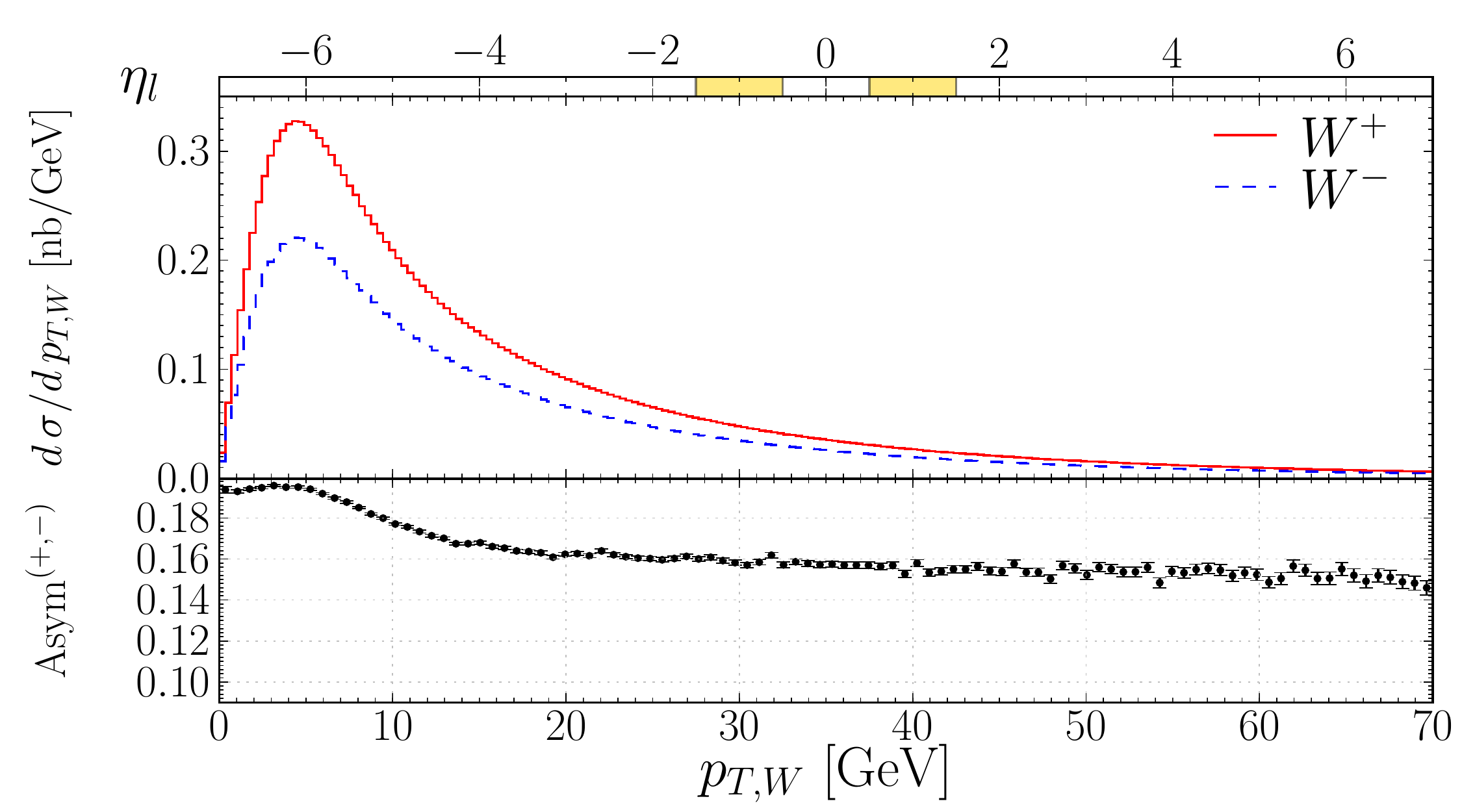}
    \vfill
    \includegraphics[width=0.495\tw]{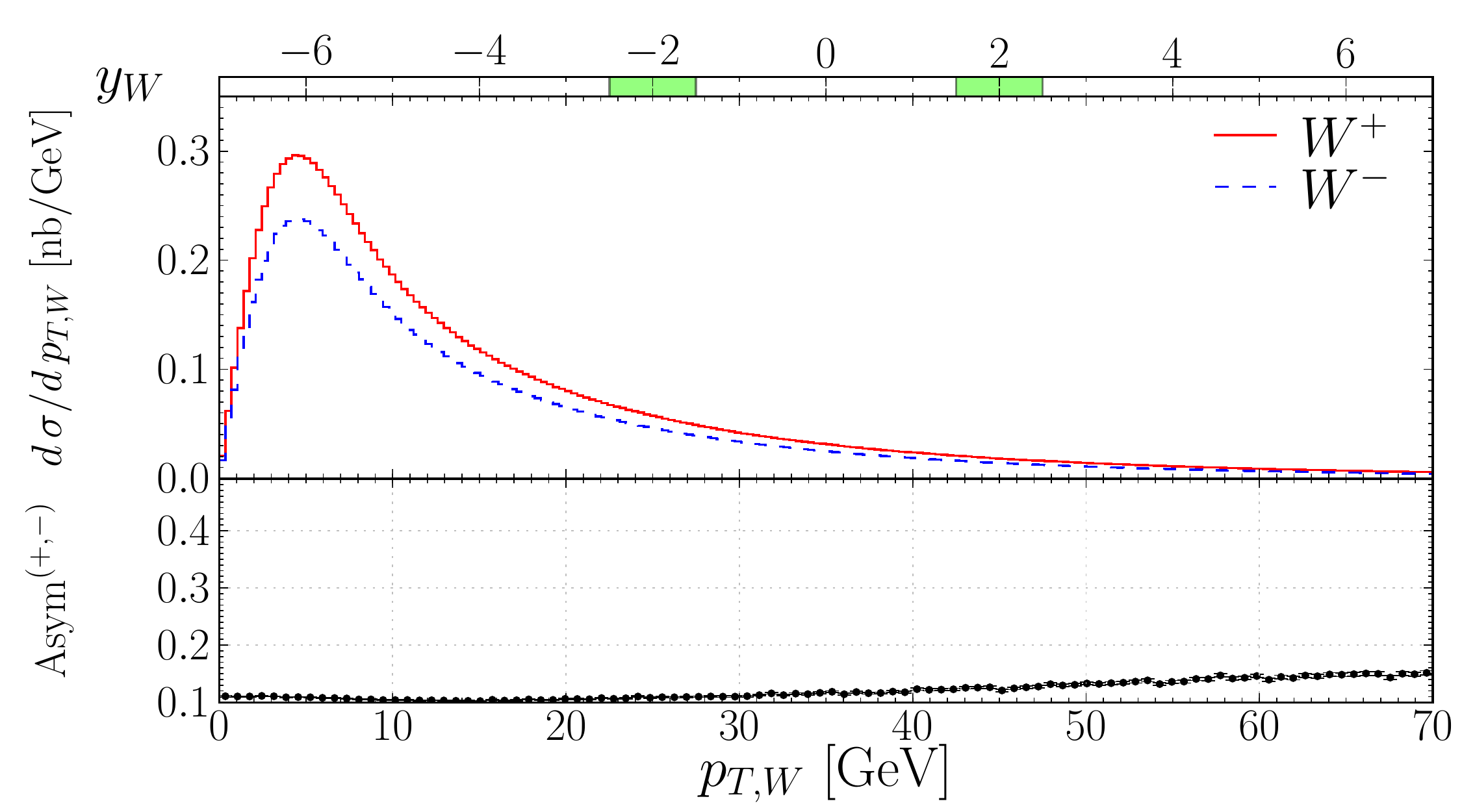}
    \hfill
    \includegraphics[width=0.495\tw]{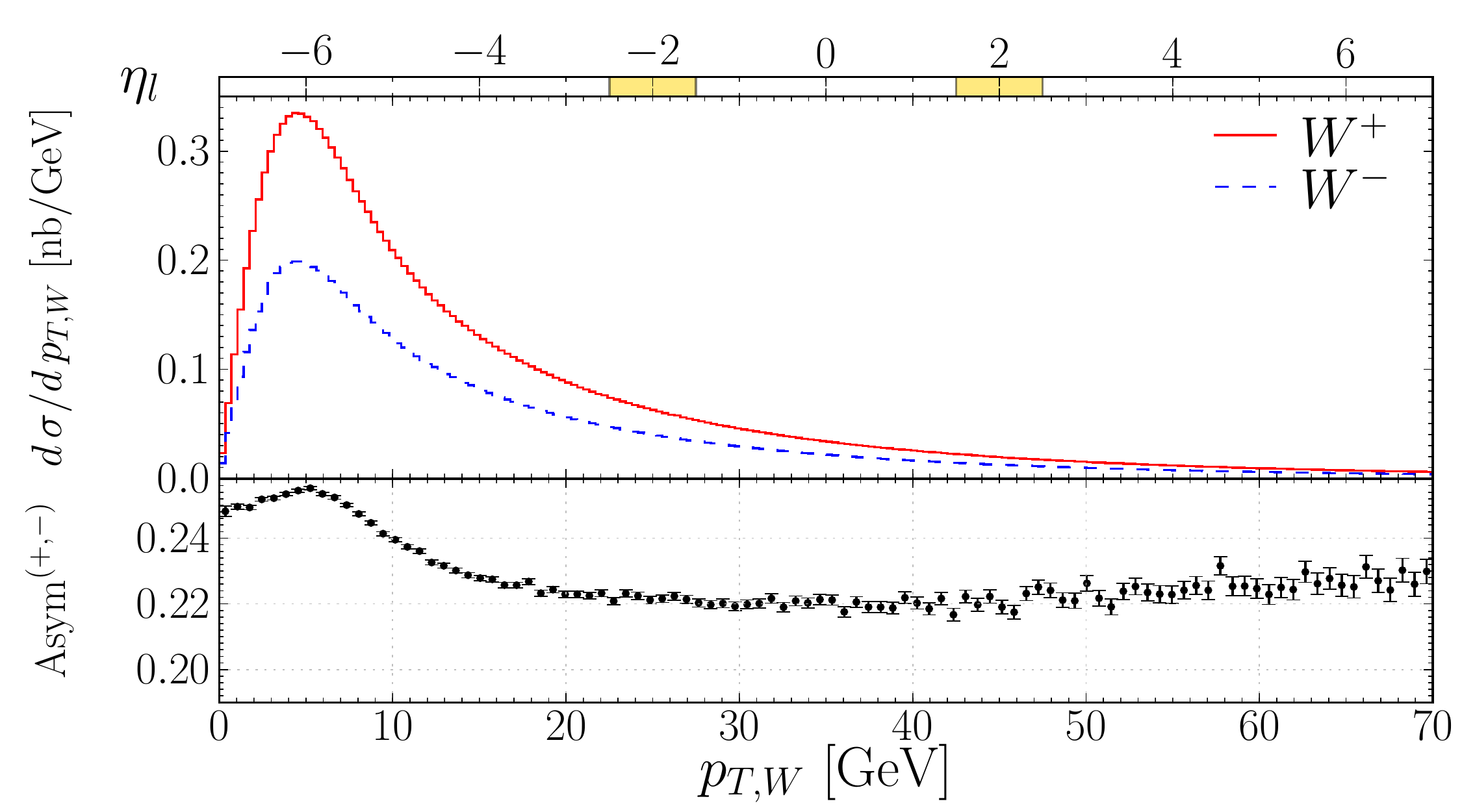}
    \vfill
    \includegraphics[width=0.495\tw]{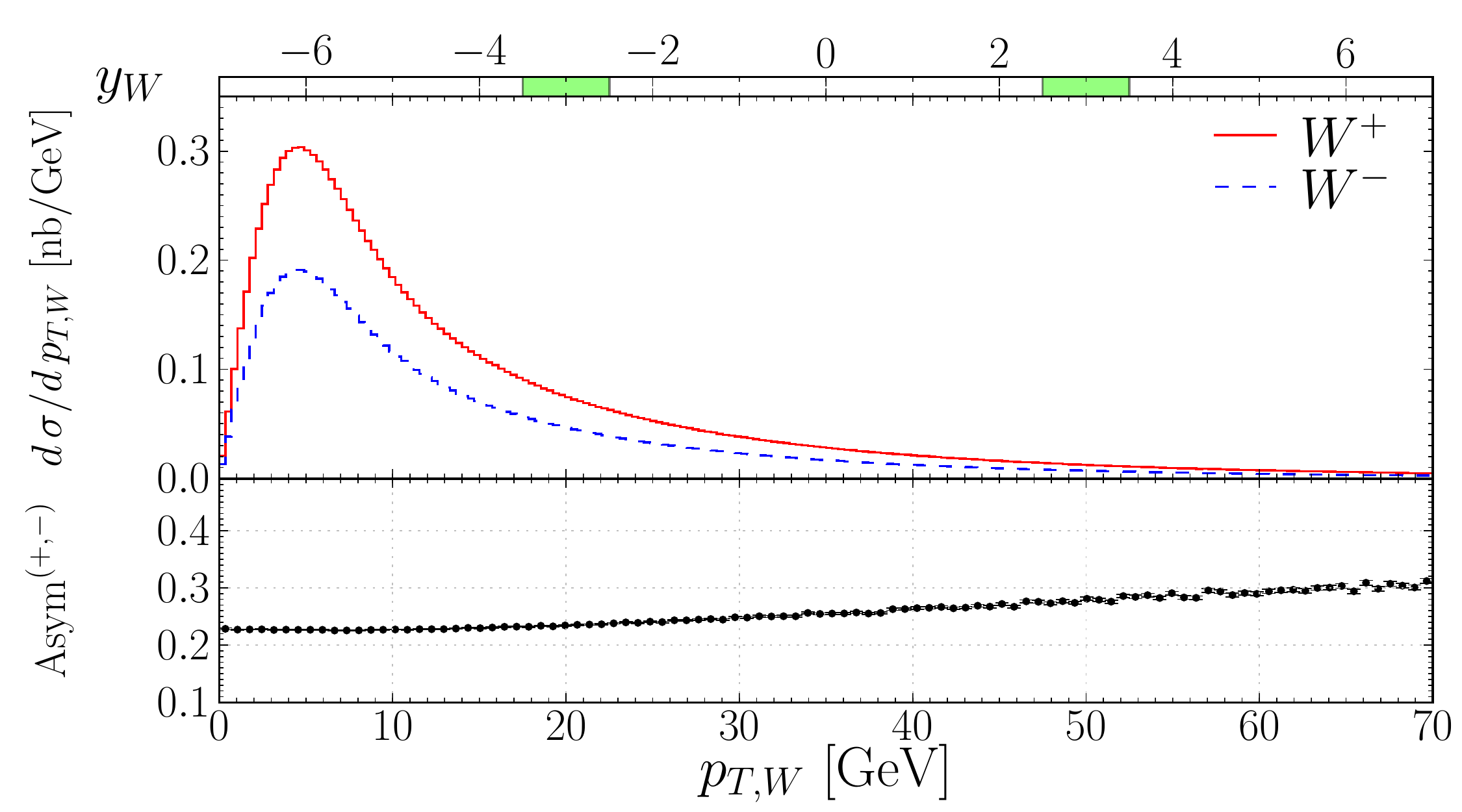}
    \hfill
    \includegraphics[width=0.495\tw]{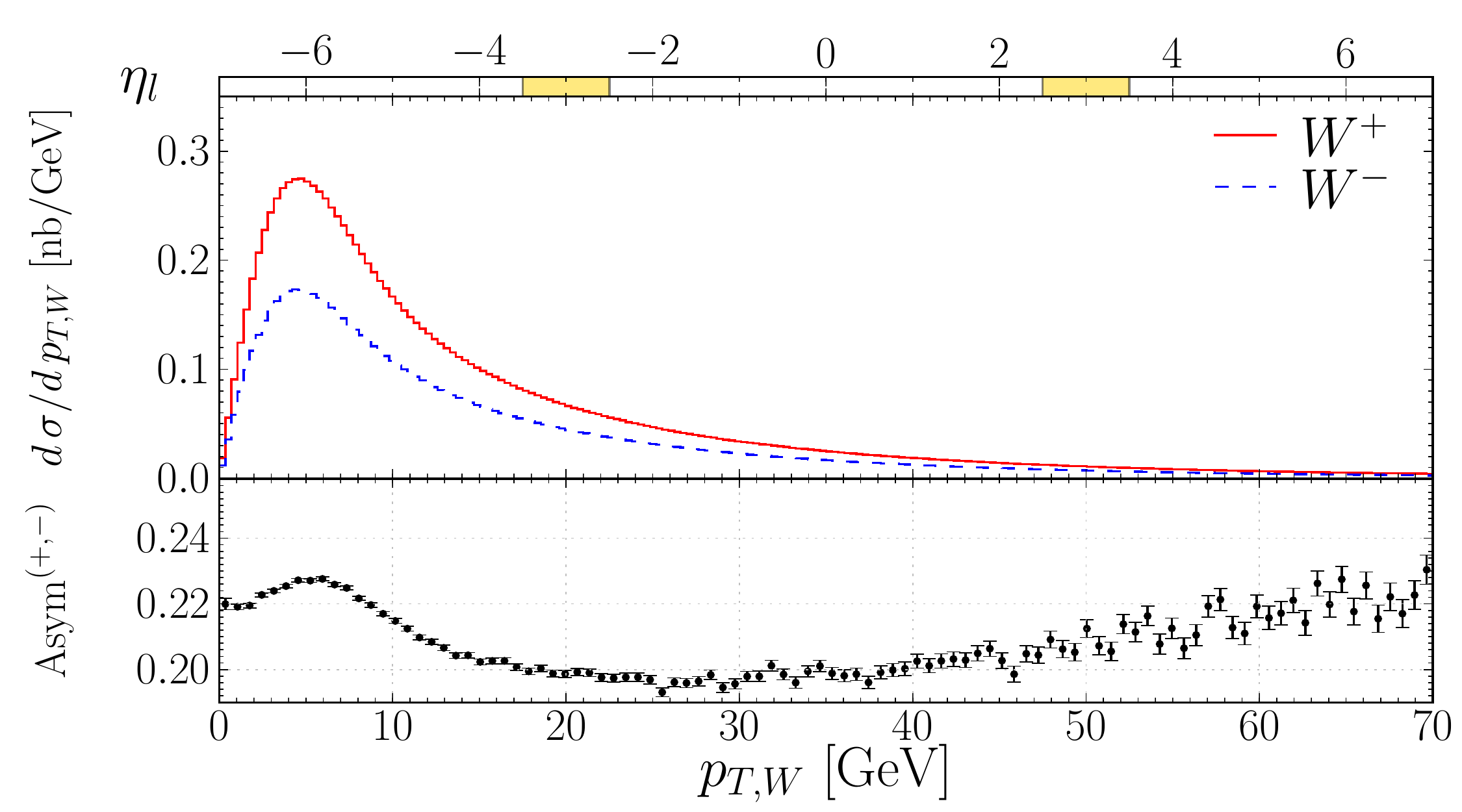}
    \vfill
    \includegraphics[width=0.495\tw]{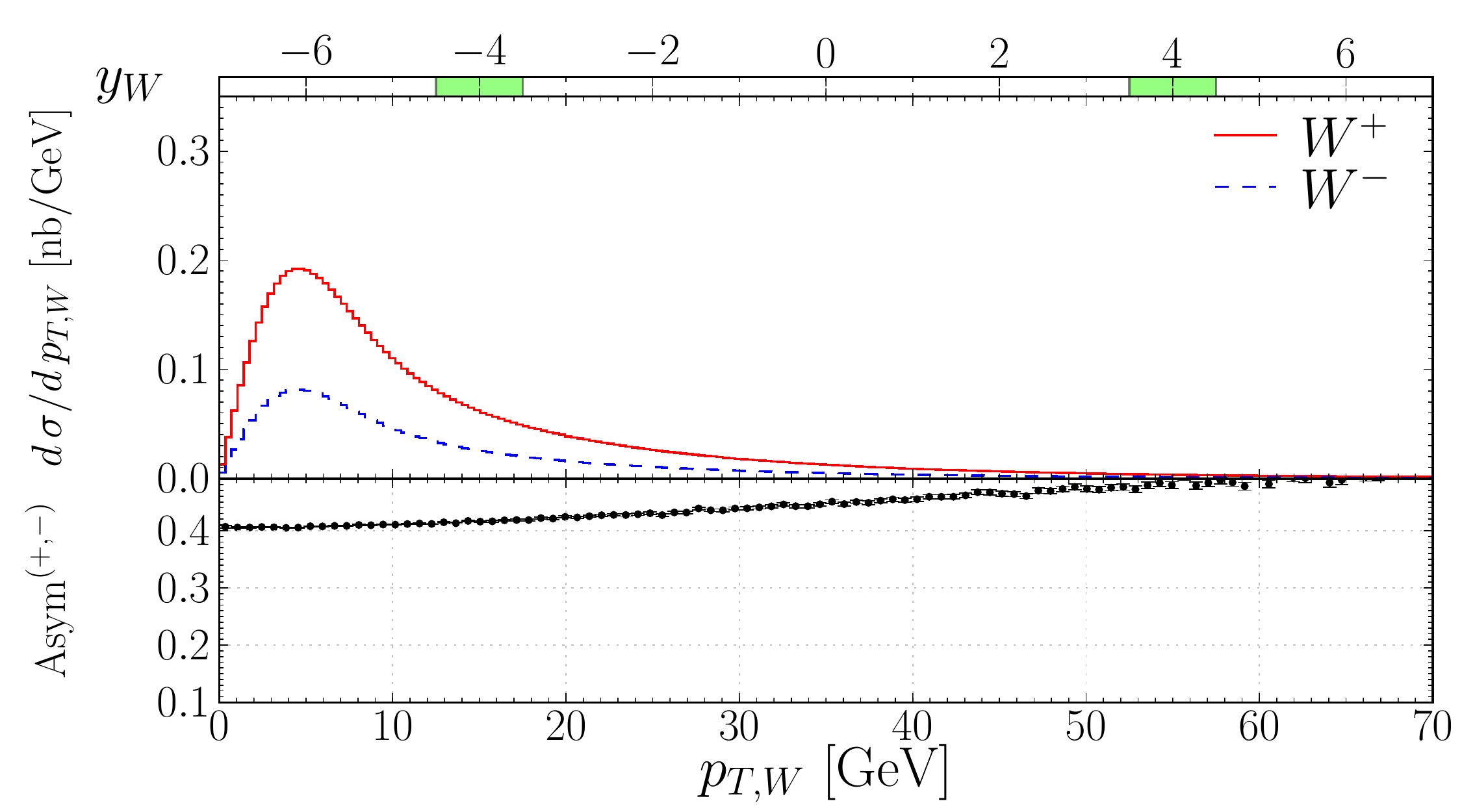}
    \hfill
    \includegraphics[width=0.495\tw]{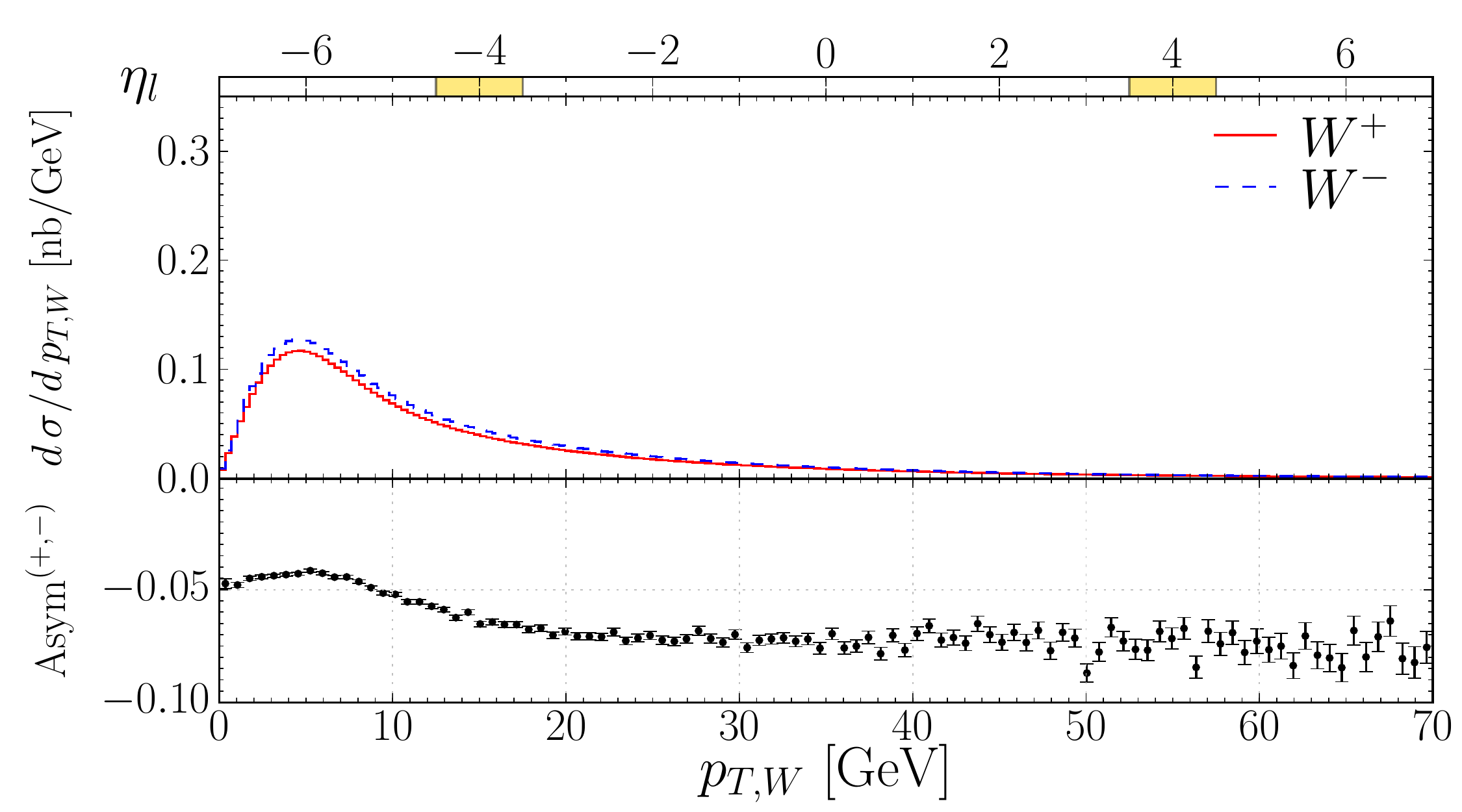}
    \caption[$W$ boson transverse momentum in bins of $\yW$ and in bins of $\etal$ in $\pp$ collisions]
            {\figtxt{$W$ boson transverse momentum in bins of $\yW$ (left) and in bins of $\etal$ (right) 
                in $\pp$ collisions.}}
            \label{app_pp_pTW_in_yW_etal_bins}
  \end{center} 
\end{figure}
Figure~\ref{app_pp_pTW_in_yW_etal_bins} shows the size of the initial state charge asymmetries by 
looking at the transverse momentum of the $W$ boson.
\begin{figure}[!h] 
  \begin{center}
    \includegraphics[width=0.495\tw]{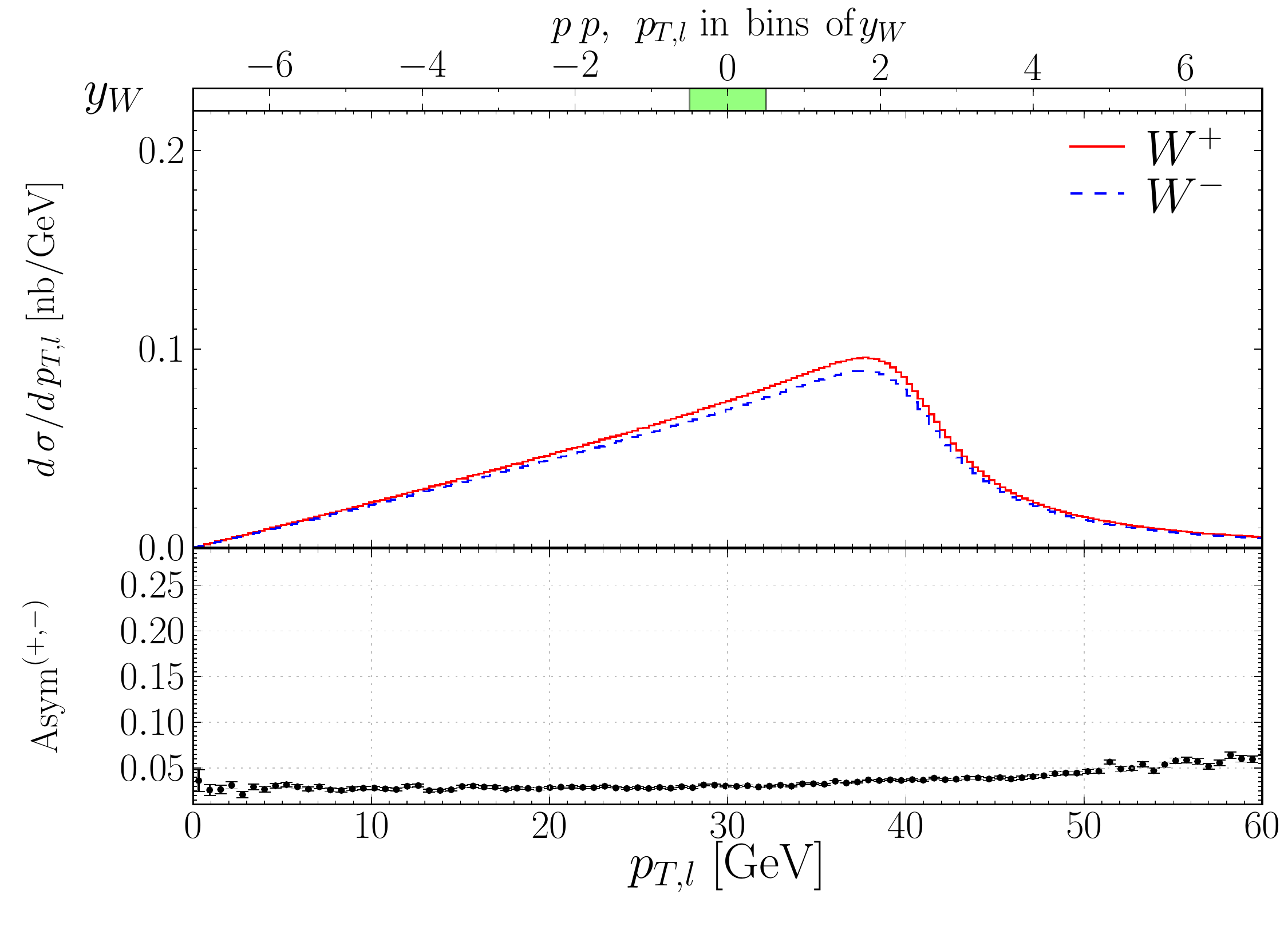}
    \hfill
    \includegraphics[width=0.495\tw]{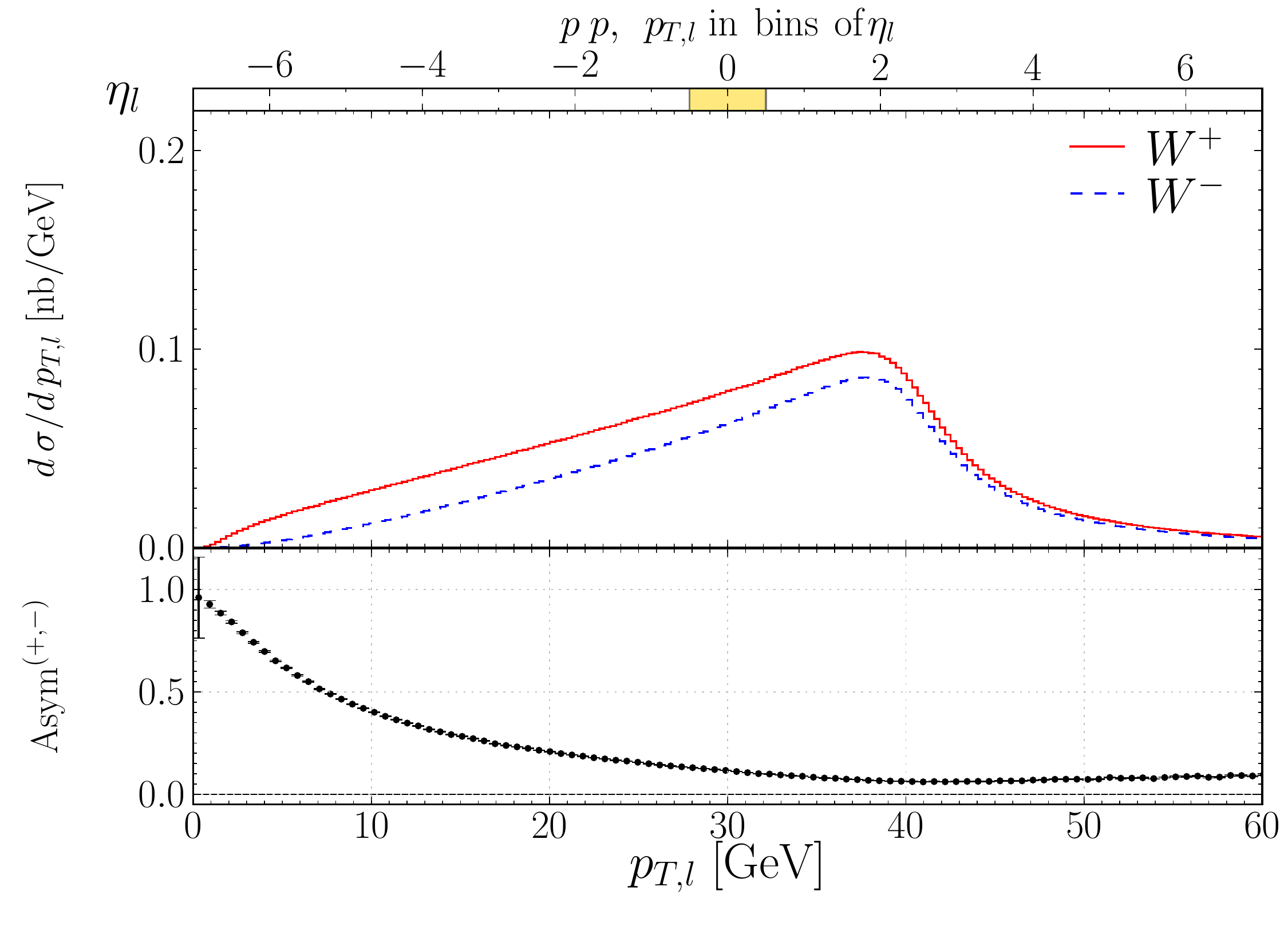}
    \vfill   
    \includegraphics[width=0.495\tw]{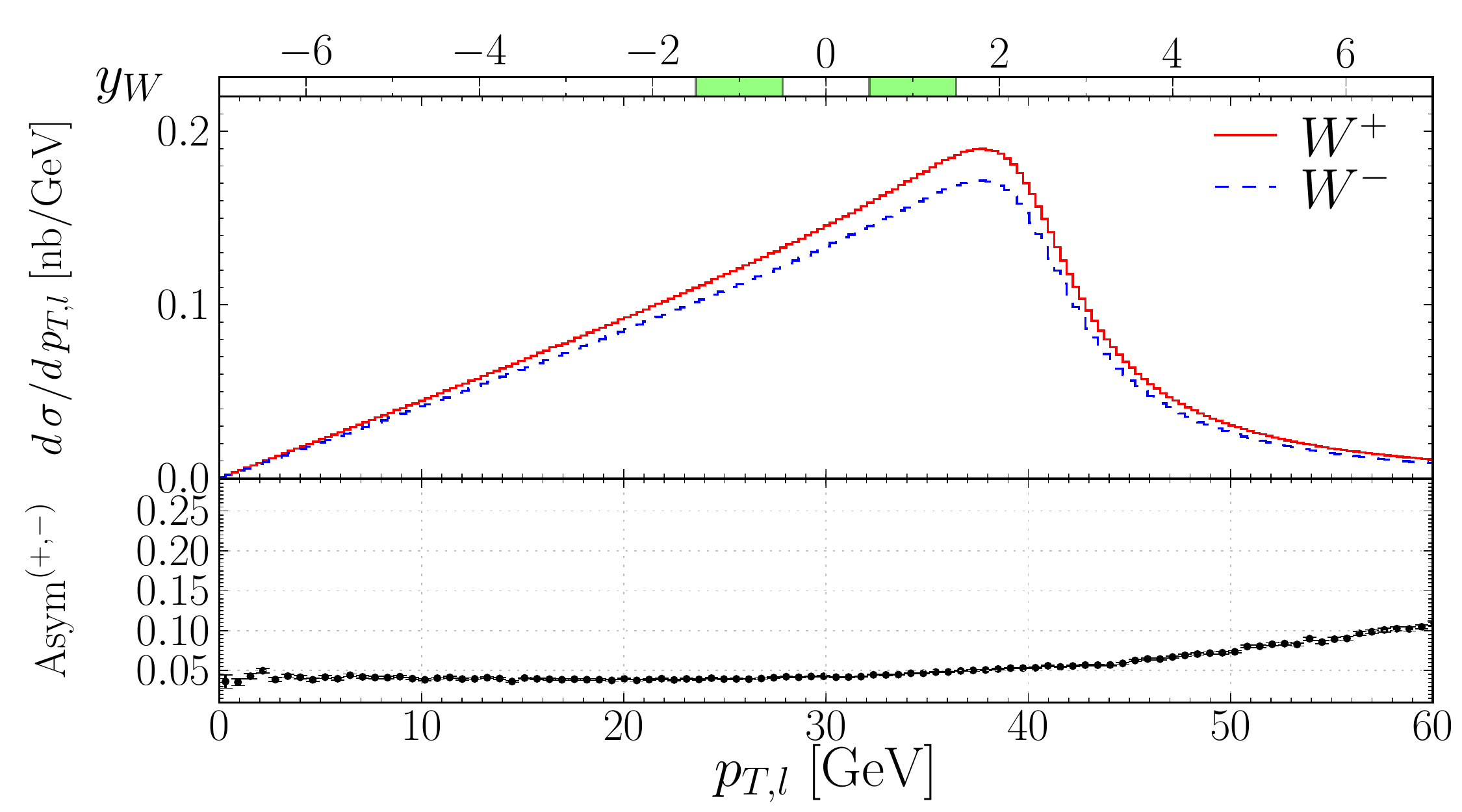}
    \hfill
    \includegraphics[width=0.495\tw]{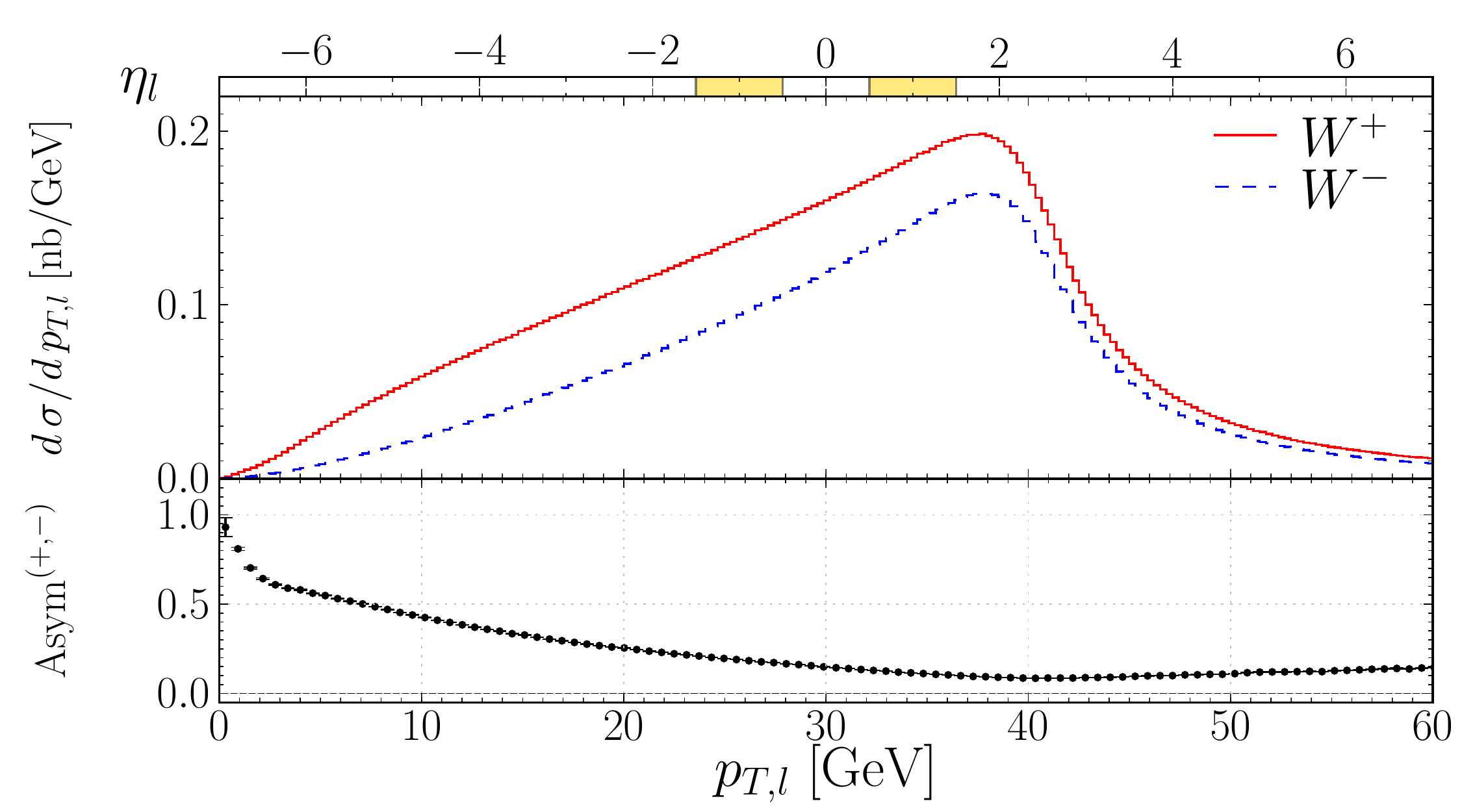}
    \vfill
    \includegraphics[width=0.495\tw]{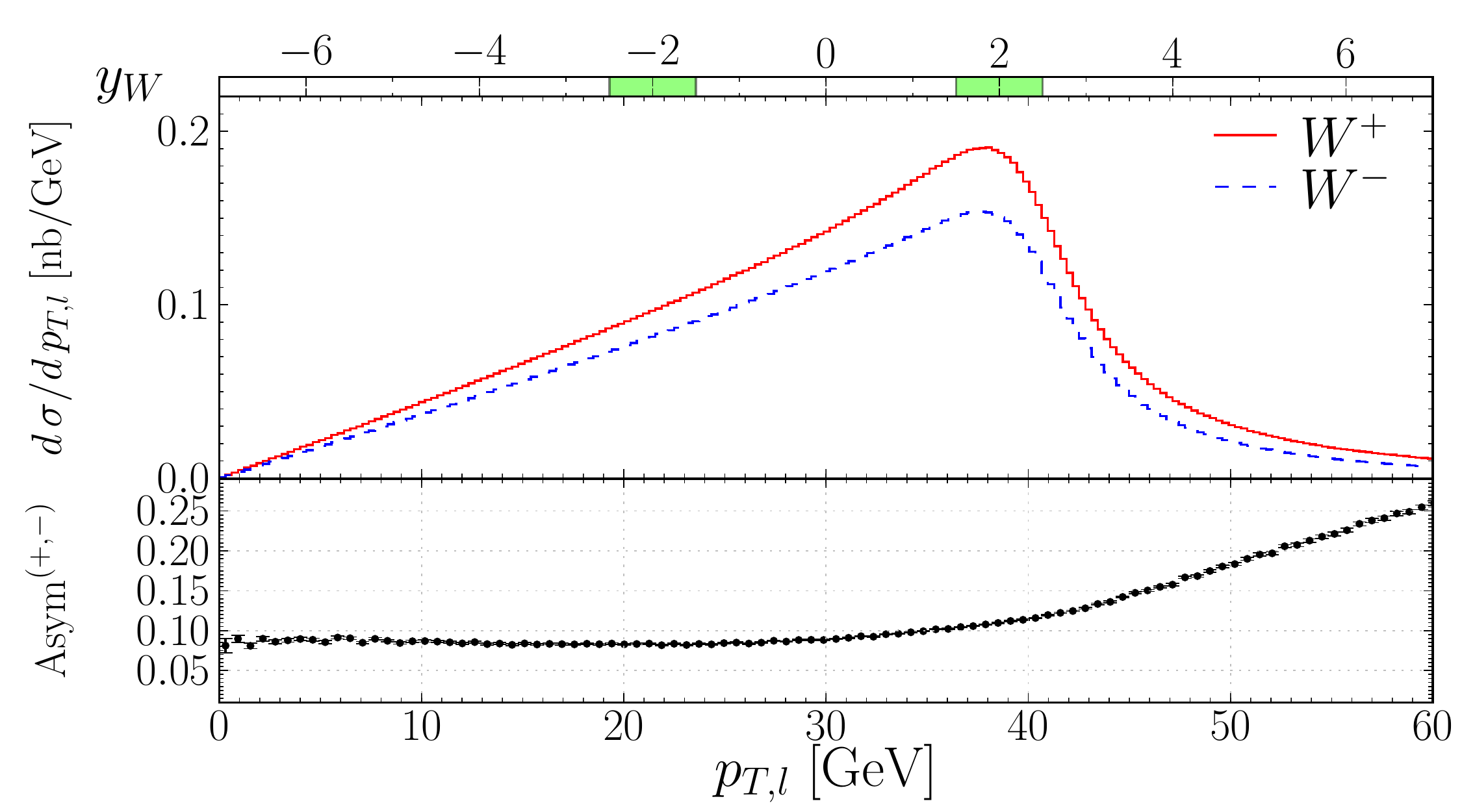}
    \hfill
    \includegraphics[width=0.495\tw]{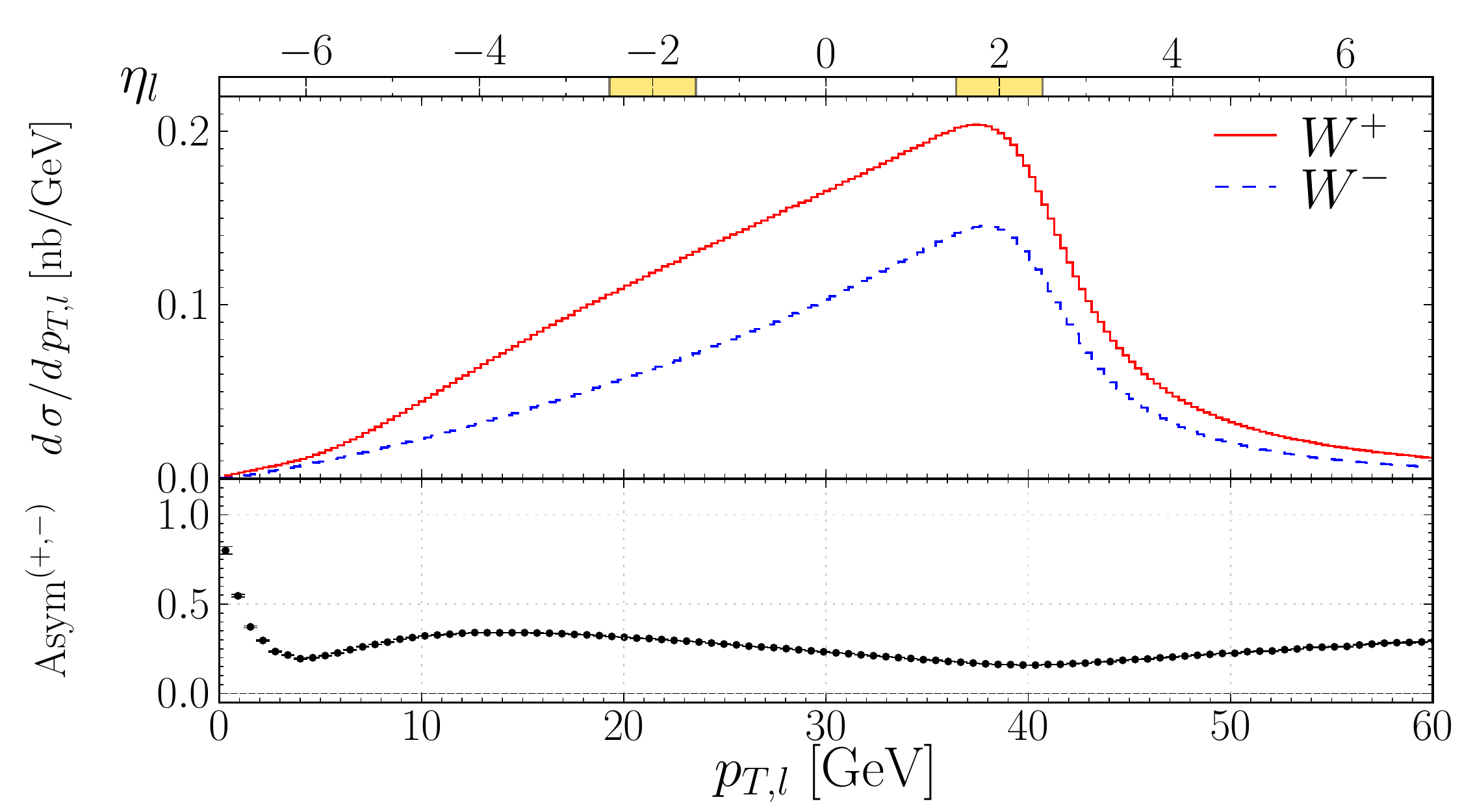}
    \vfill
    \includegraphics[width=0.495\tw]{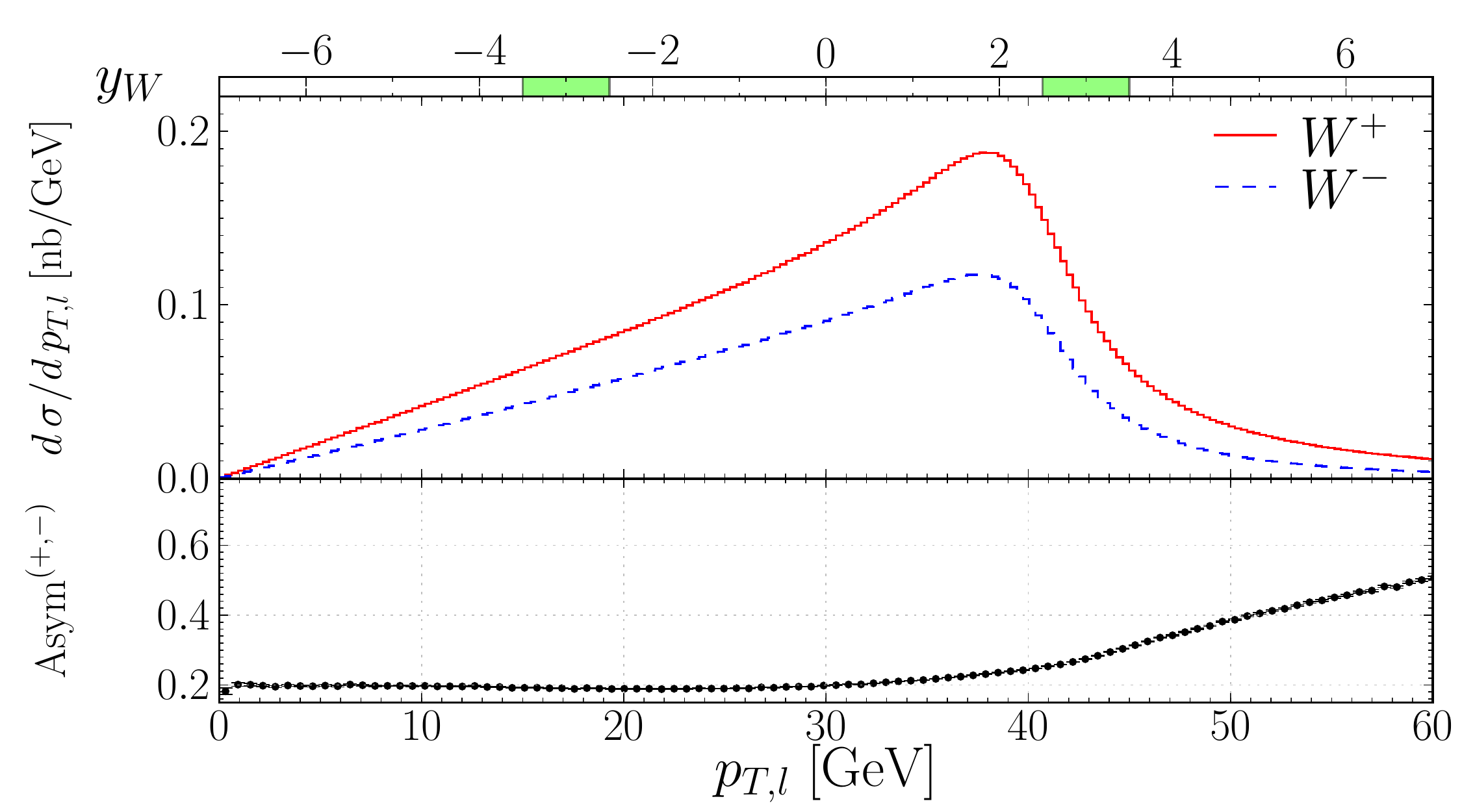}
    \hfill
    \includegraphics[width=0.495\tw]{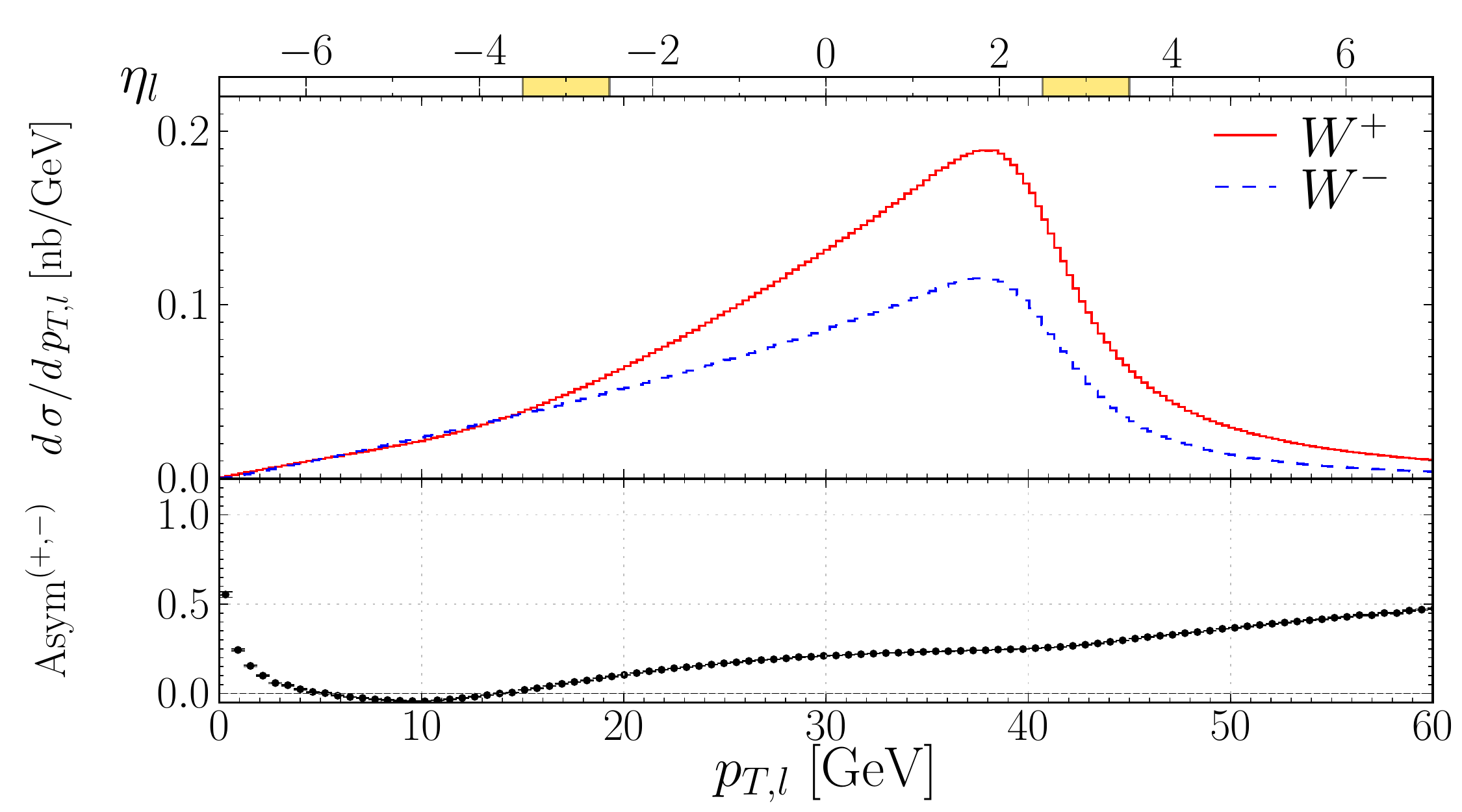}
    \vfill
    \includegraphics[width=0.495\tw]{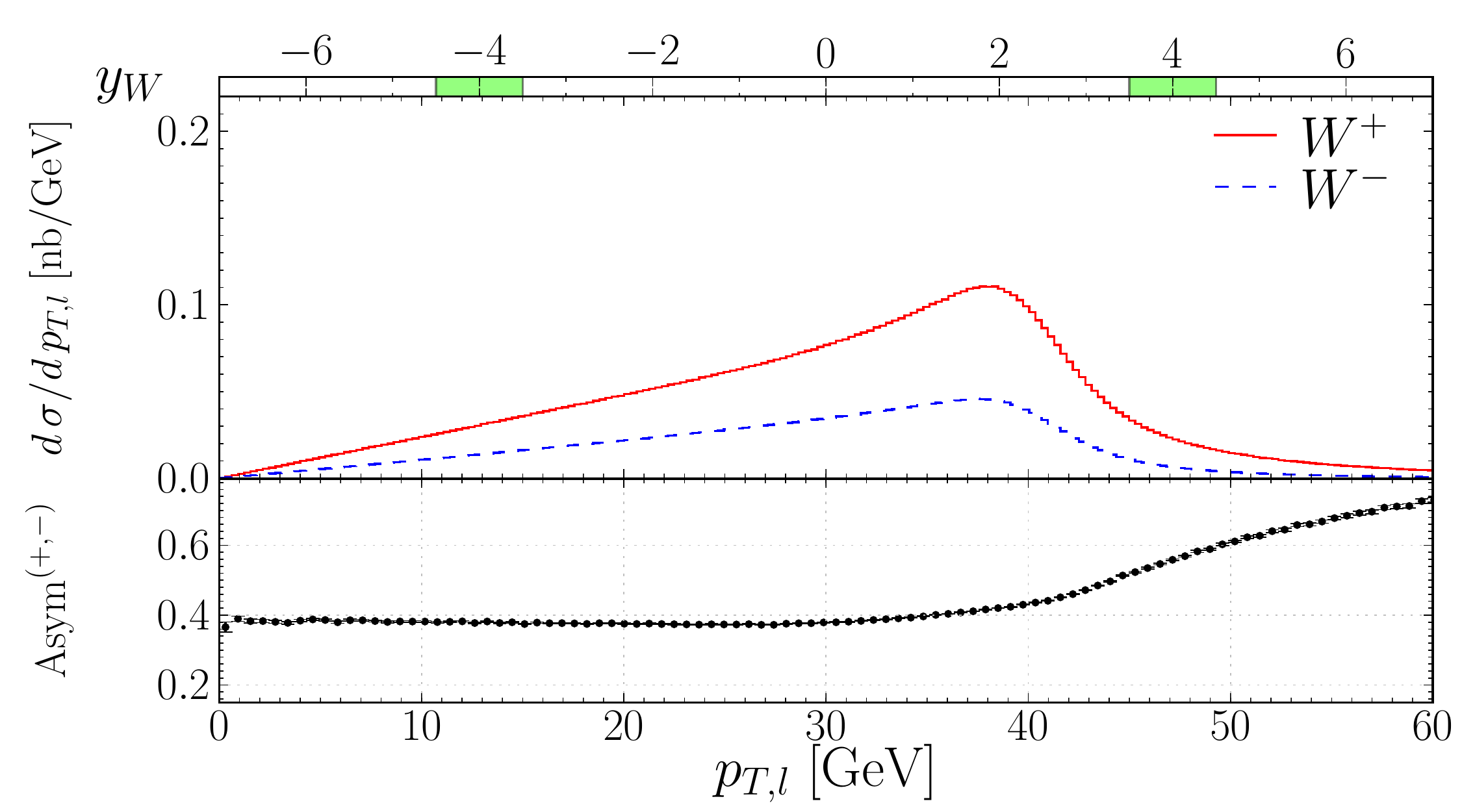}
    \hfill
    \includegraphics[width=0.495\tw]{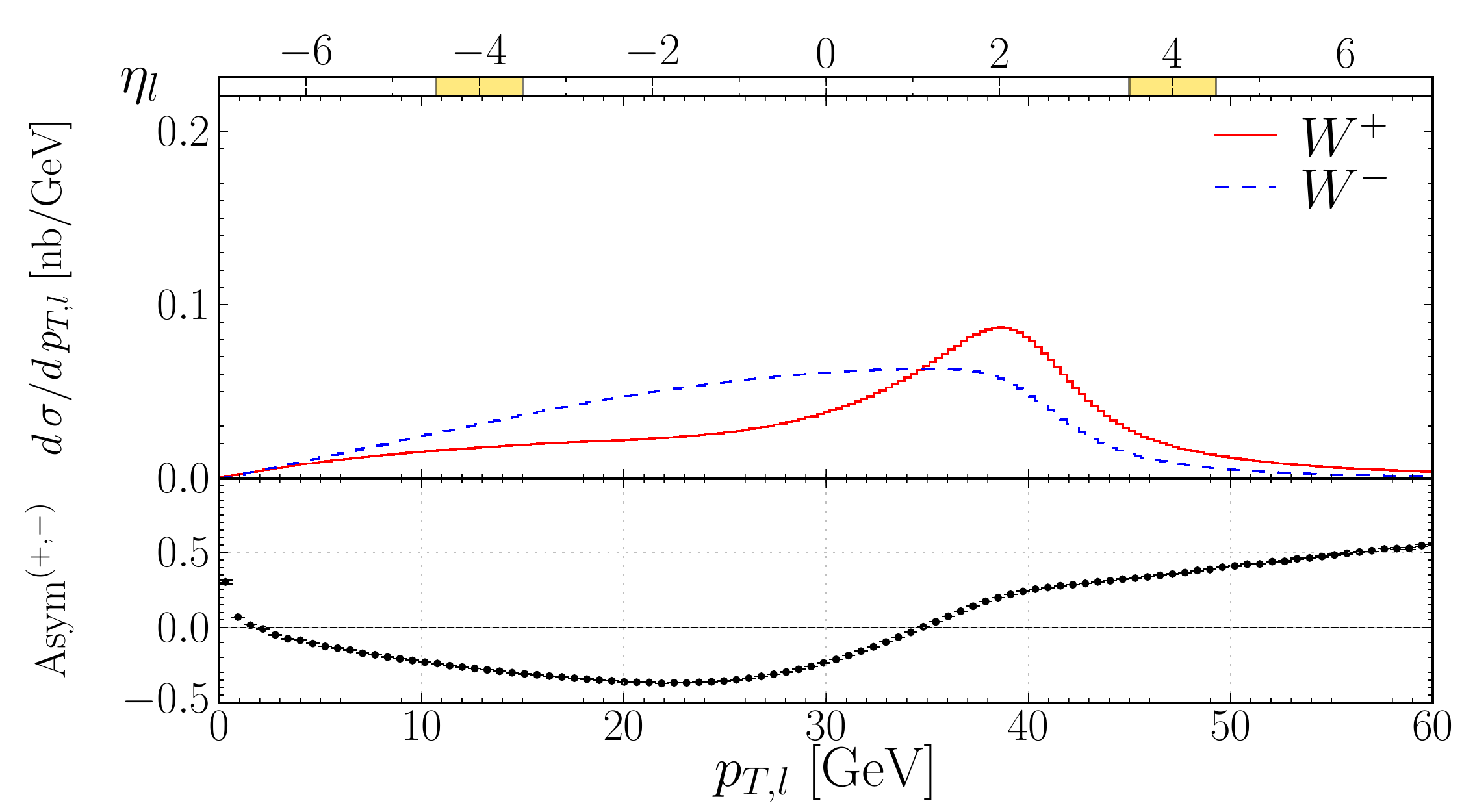}
    \caption[Charged lepton transverse momentum in bins of $\yW$ and in bins of $\etal$ in $\pp$ collisions]
            {\figtxt{Charged lepton transverse momentum in bins of $\yW$ (left) and in bins of $\etal$ (right) 
                in $\pp$ collisions.}}
            \label{app_pp_pTl_in_yW_etal_bins}
  \end{center} 
\end{figure}
Figure~\ref{app_pp_pTl_in_yW_etal_bins} shows the increase of the final state charge asymmetry in bins 
of $\yW$ and $\etal$.
\begin{figure}[!h] 
  \begin{center}
    \includegraphics[width=0.495\tw]{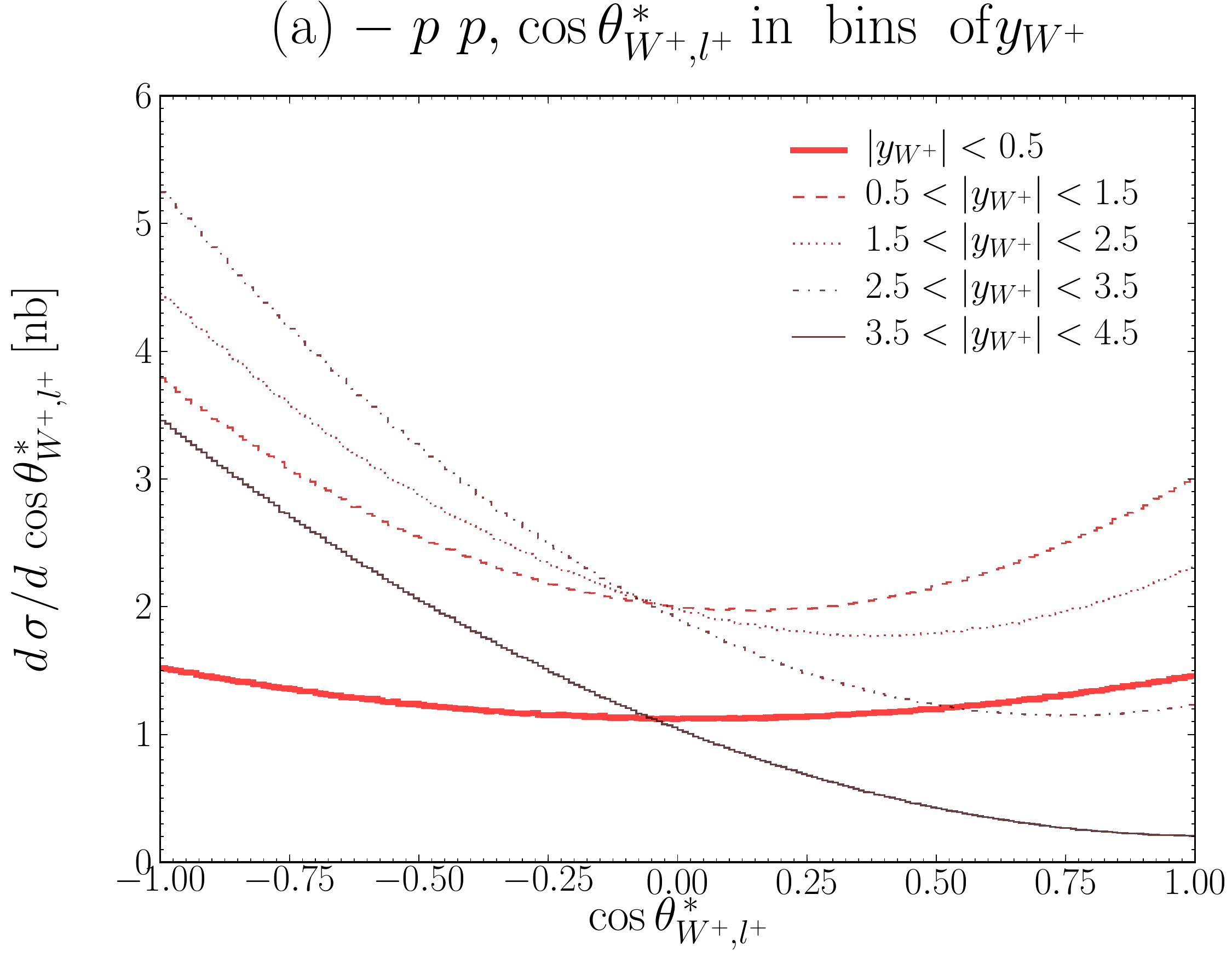}
    \hfill
    \includegraphics[width=0.495\tw]{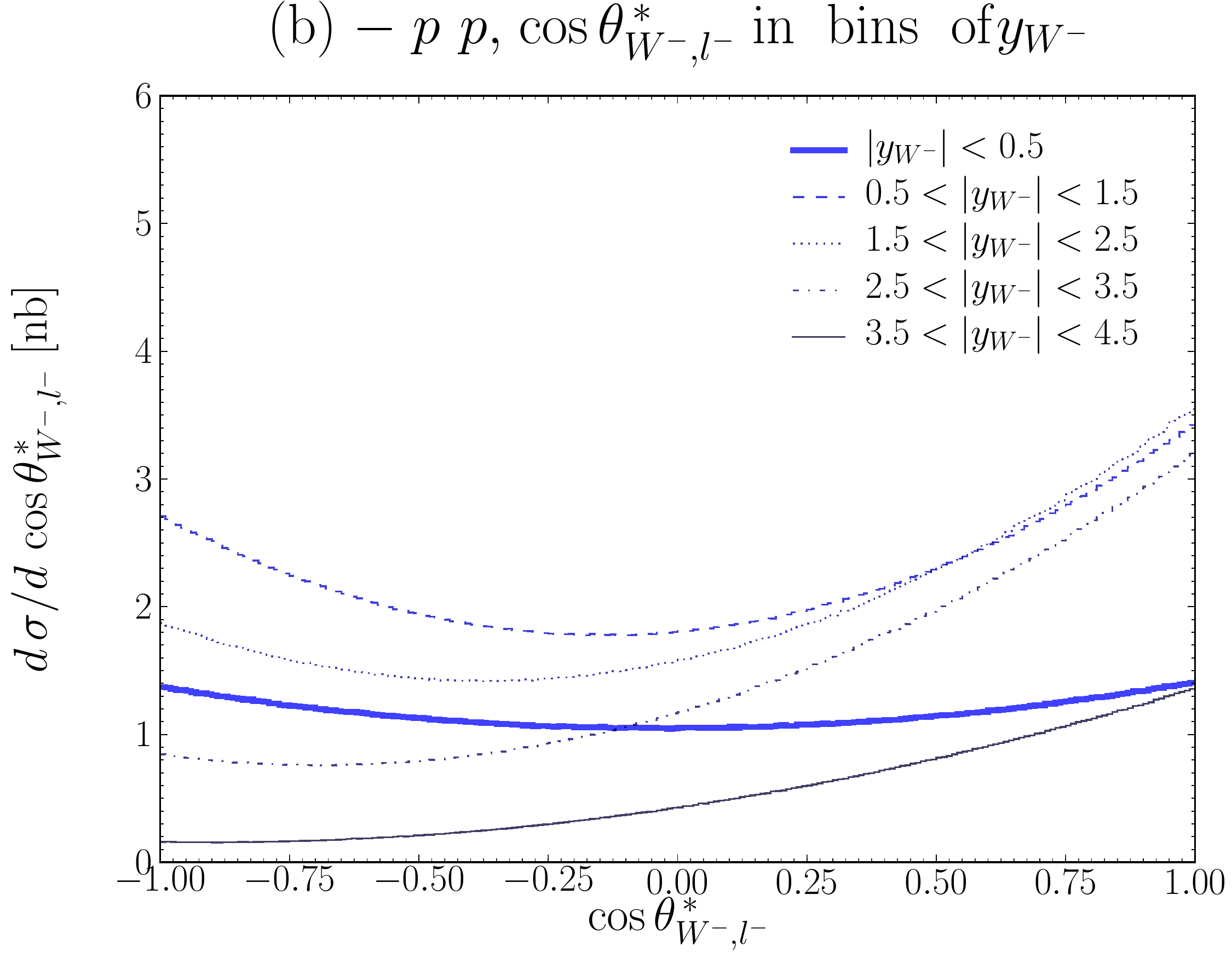}
    \vfill
    \includegraphics[width=0.495\tw]{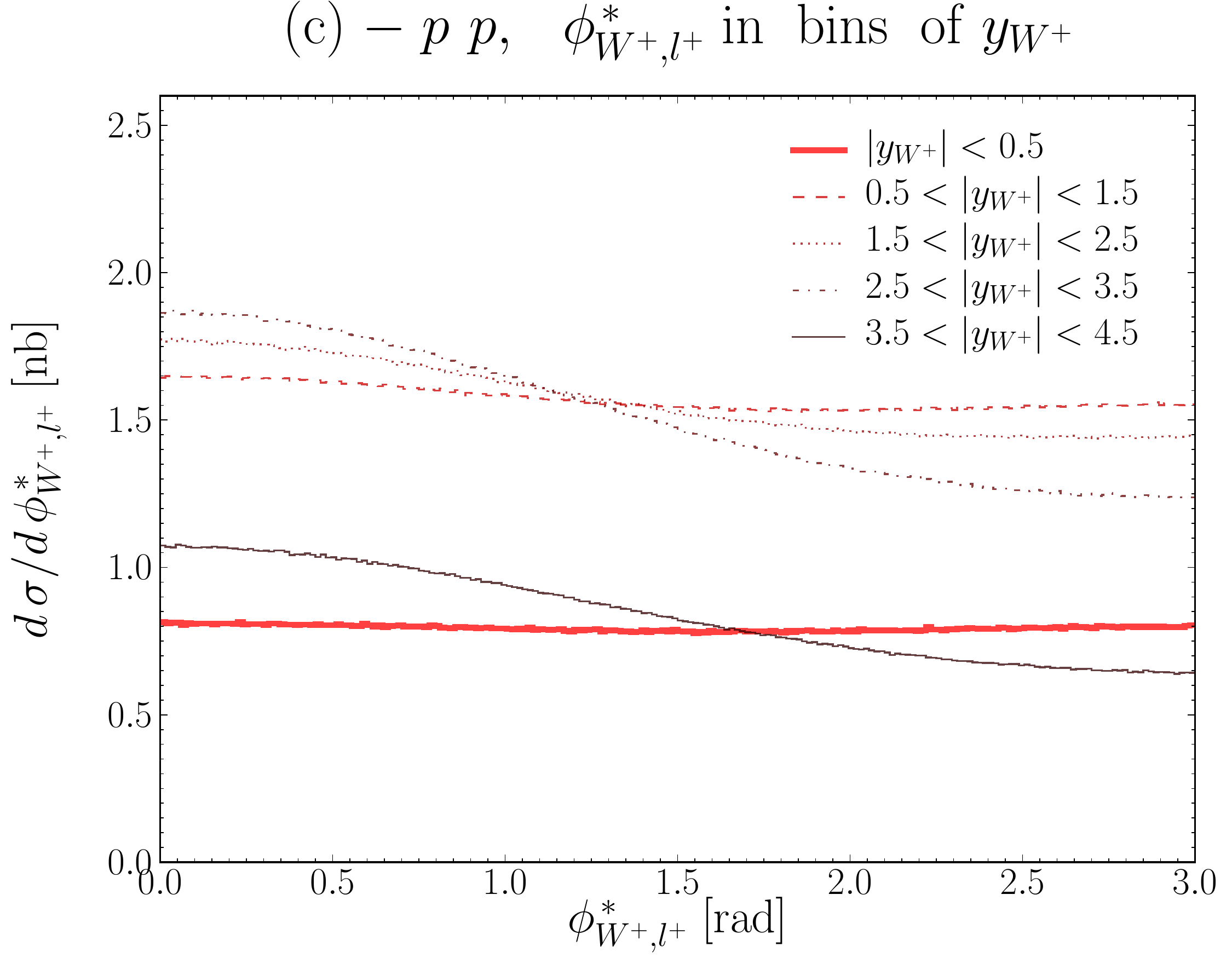}
    \hfill
    \includegraphics[width=0.495\tw]{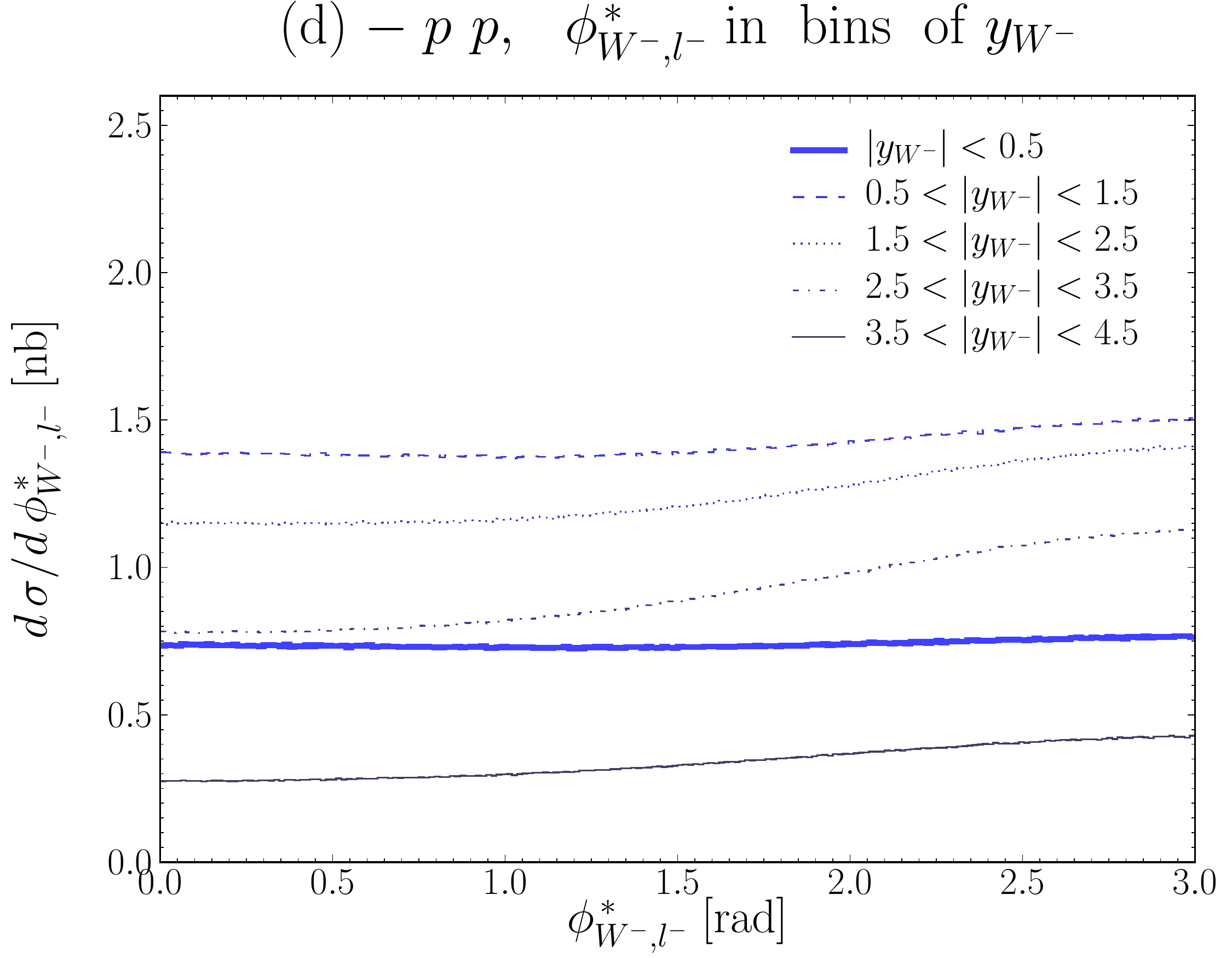}
    \caption[Distributions of $\costhetaWlwrf$ and $\phi_{W,l}^{\,\ast}$ in bins of $\yW$ 
      for the positively and negatively charged channel in $\pp$ collisions]
            {\figtxt{Distributions of $\costhetaWlwrf$ and $\phi_{W,l}^{\,\ast}$ 
                in bins of $\yW$ for the positively (a,c) and negatively (b,d)
                charged channel in $\pp$ collisions.}}
            \label{app_pp_costhetaWlwrf_YWx}
  \end{center} 
\end{figure}
Another vision of the growth of the ``valence'' contributions in function of $\yW$ can be made looking
at the $\costhetaWlwrf$ and $\phi_{W,l}^{\,\ast}$ distributions in bins of $\yW$ 
(Fig.~\ref{app_pp_costhetaWlwrf_YWx}). 
In the central region the asymmetry between left and right $W$ is hardly decipherable 
(contrary to the same phenomenon seen in 
$\etal$-space).
\index{W boson@$W$ boson!Production in pp@Production in $\pp$ collisions!Detailed|)}

\clearpage
\subsection{Deuteron--deuteron collisions}
\index{W boson@$W$ boson!Production in dd@Production in $\dd$ collisions!Detailed|(}
Here, the overall charge asymmetry based on the inclusive cross sections from 
Table~\ref{table_xtot} gives
\begin{equation}
A_\dd \approx 0.0015.
\end{equation}

The predictions for $\dd$ collisions can be understood from the previous study of $\pp$ 
collisions.
Indeed, here the novelty is that reaching the equality $u^\val=d^\val$ allows to cancel out 
the most important differences between the $\Wp$ and $\Wm$ at the level of the production.

\begin{figure}[!h] 
  \begin{center}
    \includegraphics[width=0.5\tw]{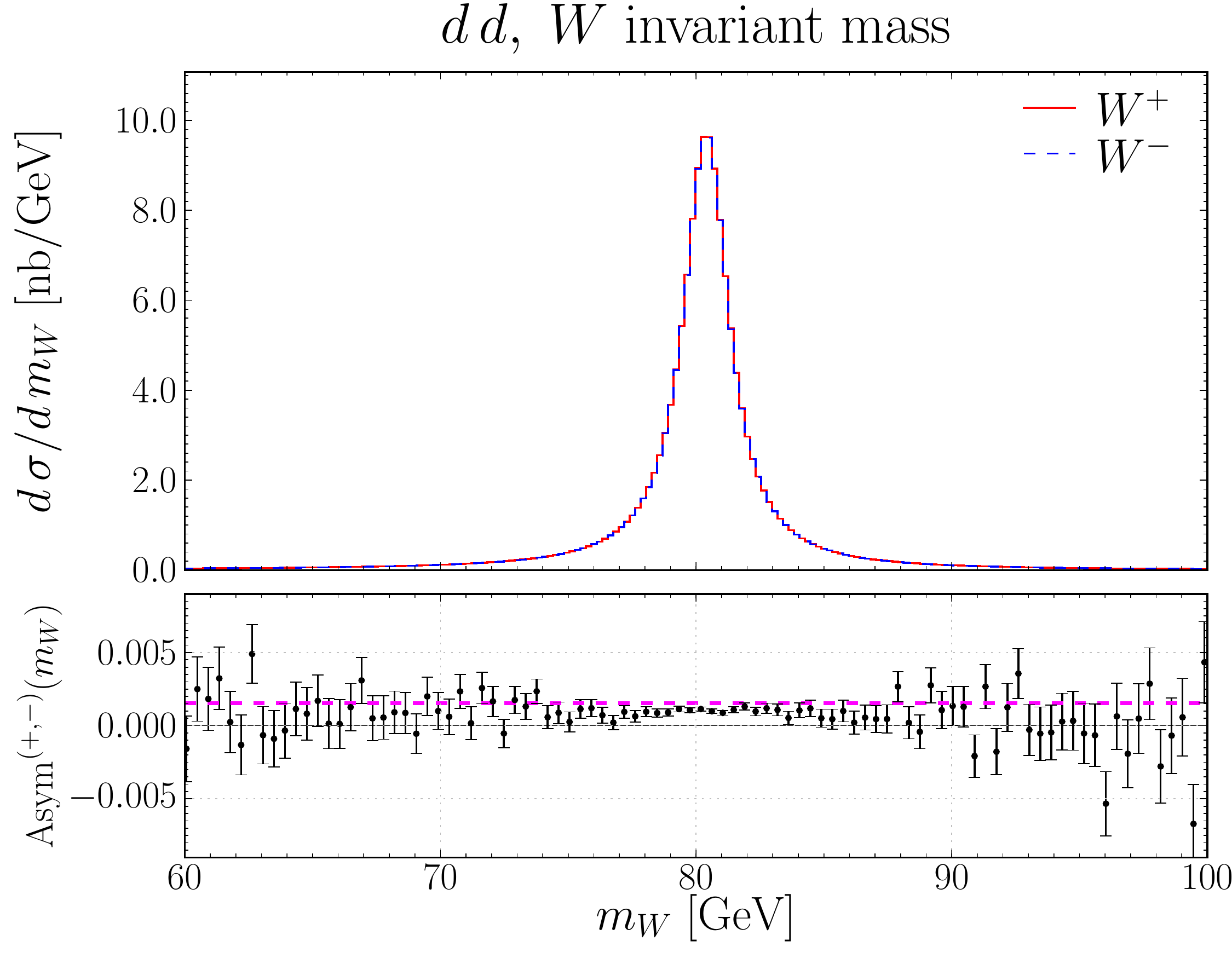}
    \caption[Invariant mass of the $W$ boson in $\dd$ collisions with $\sqrt{S_{n_1\,n_2}}=7\TeV$]
            {\figtxt{Invariant mass of the $W$ boson in $\dd$ collisions with 
                $\sqrt{S_{n_1\,n_2}}=7\TeV$.}}
            \label{app_dd_mW}
  \end{center} 
\end{figure}

The invariant masses of the $\Wp$ and $\Wm$ in Fig.~\ref{app_dd_mW} are now superimposed up to the
very small $A_\dd$ offset. 
Hence here the $\Wp$ and $\Wm$ masses occur with almost the same probabilities, 
we find $m_\Wp-m_\Wm\sim \mathcal{O}(0.1\MeV)$.

\begin{figure}[!h] 
  \begin{center}
    \includegraphics[width=0.495\tw]{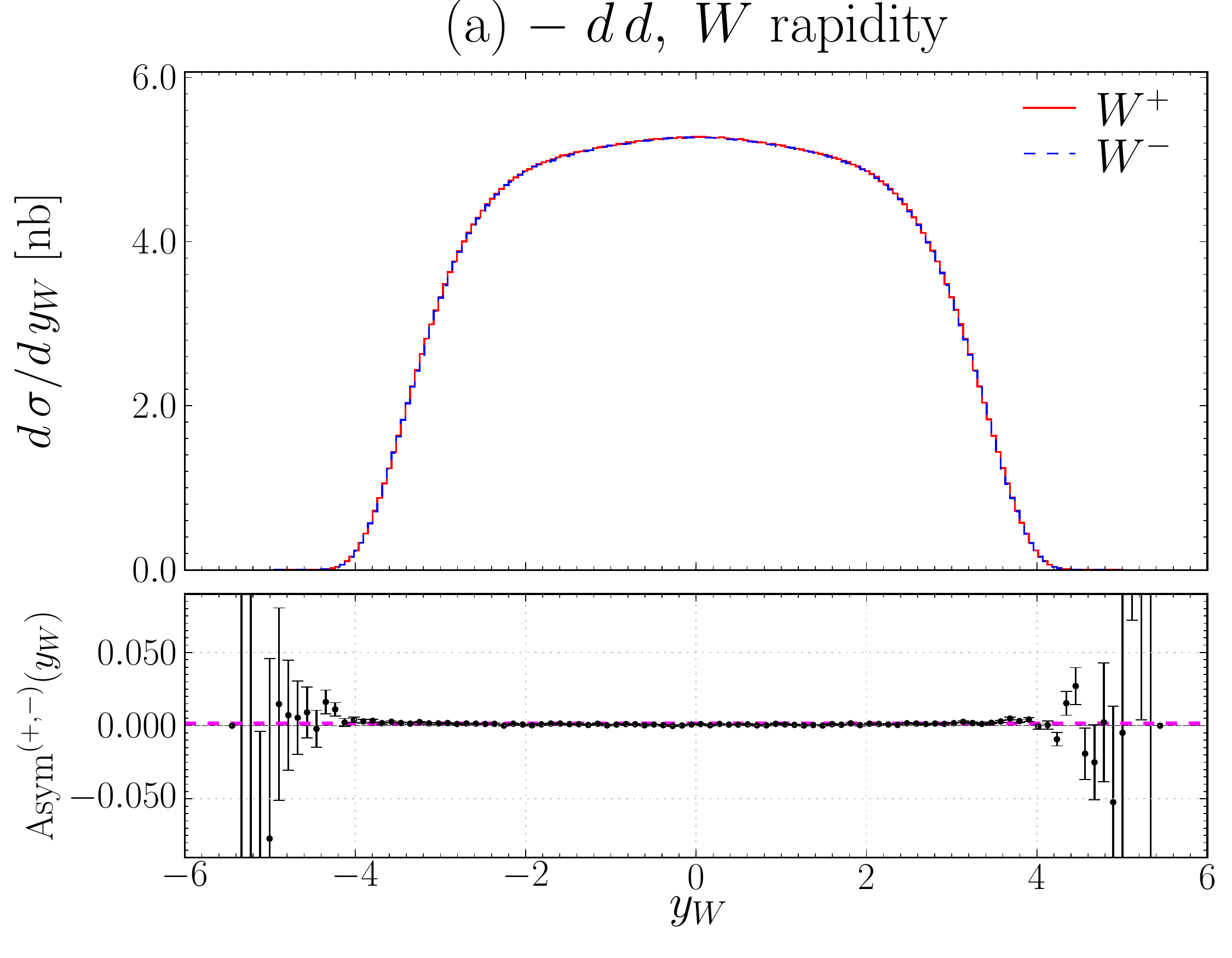}
    \hfill
    \includegraphics[width=0.495\tw]{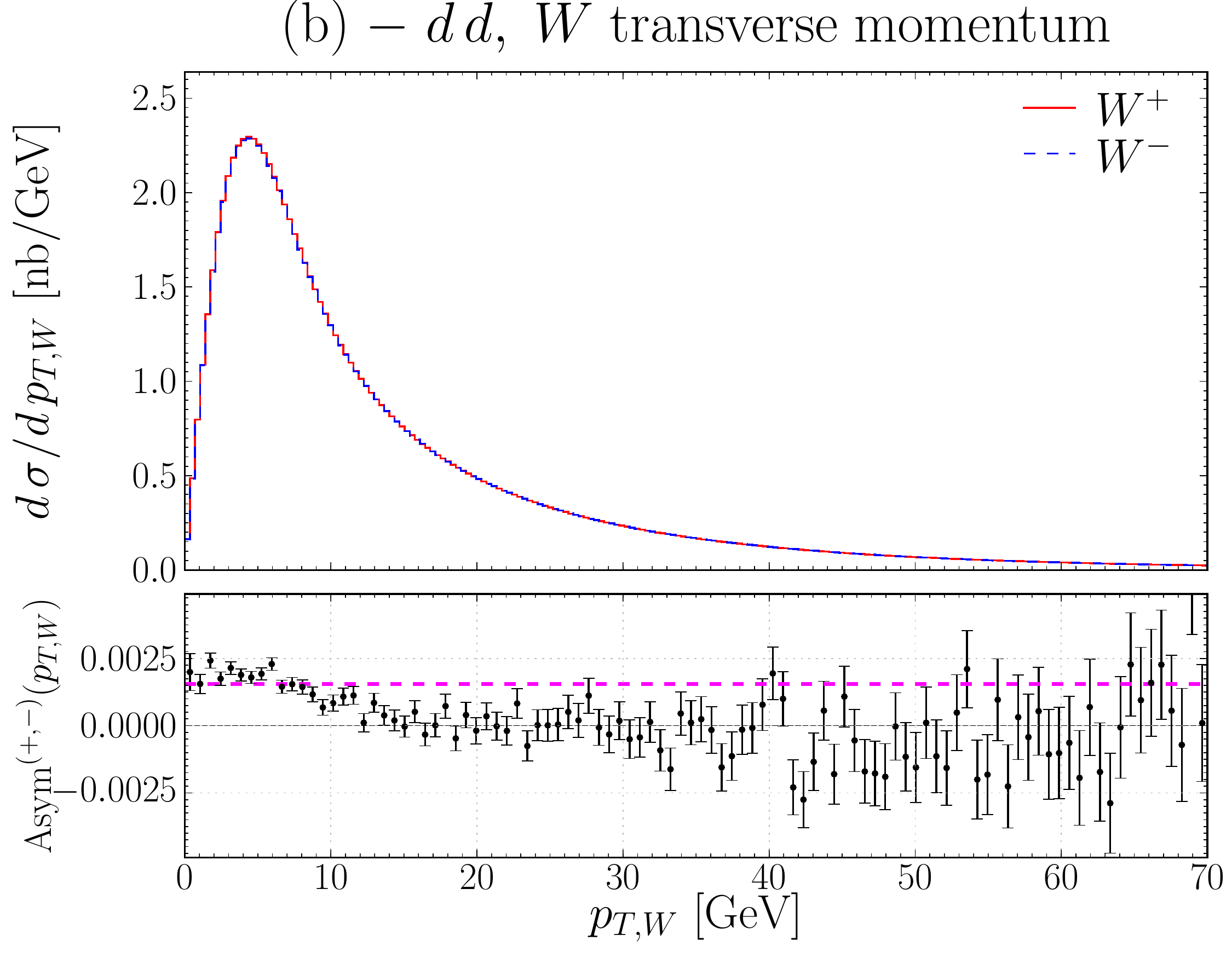}
    \vfill
    \includegraphics[width=0.495\tw]{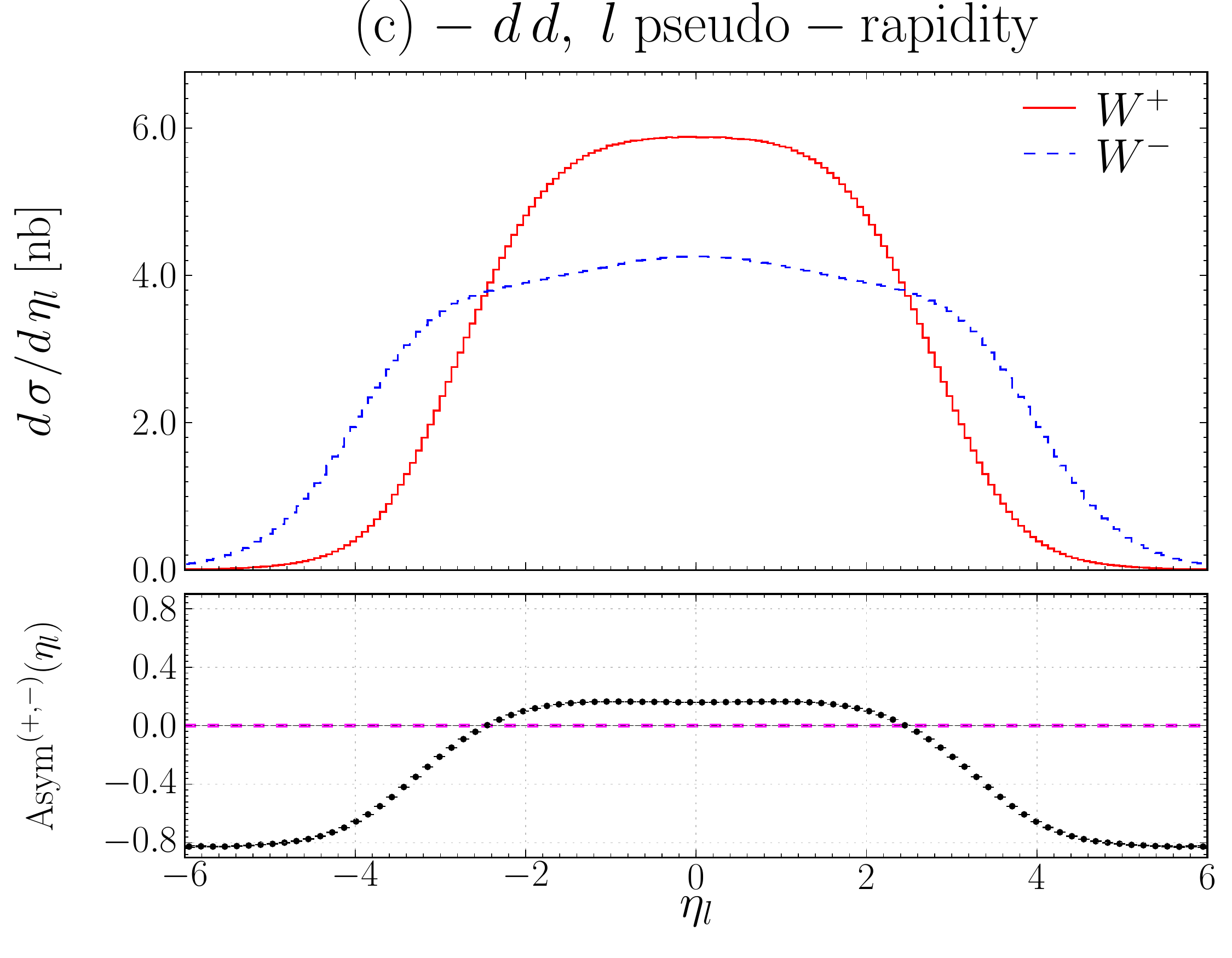}
    \hfill
    \includegraphics[width=0.495\tw]{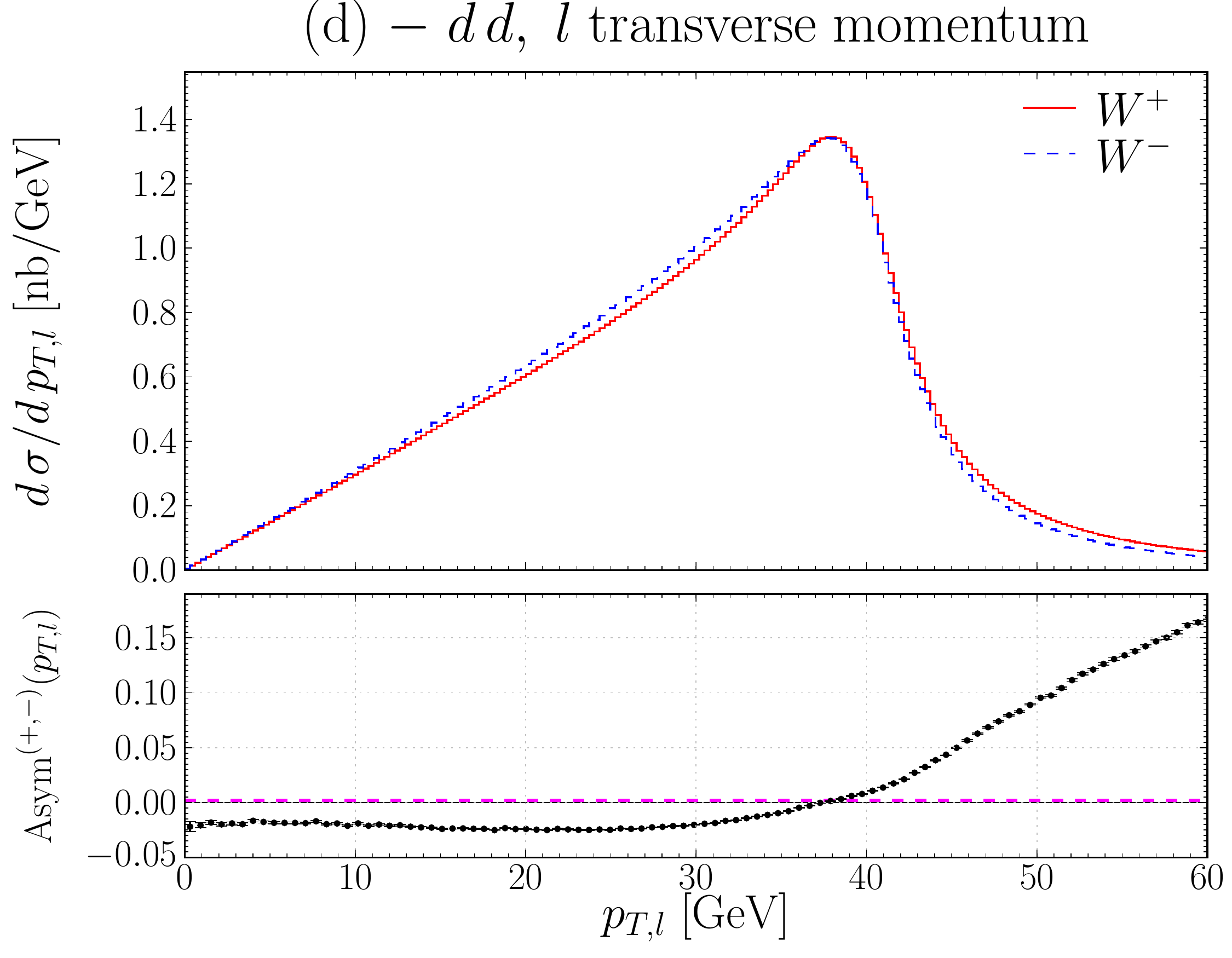}
    \caption[Distributions of the $W$ boson rapidity and transverse momentum along with the one of the
      charged lepton pseudo-rapidity and transverse momentum in $\dd$ collisions with $\sqrt{S_{n_1\,n_2}}=7\TeV$]
        {\figtxt{Distributions of the $W$ boson rapidity (a) and transverse momentum (b) 
            along with the one of the charged lepton pseudo-rapidity (c) and transverse momentum (d)
            in $\dd$ collisions with $\sqrt{S_{n_1\,n_2}}=7\TeV$.}}
            \label{app_dd_yW_pTW_etal_pTl}
            \index{W boson@$W$ boson!Transverse momentum}
            \index{W boson@$W$ boson!Rapidity}
            \index{Charged lepton@Charged lepton from $W$ decay!Transverse momentum}
            \index{Charged lepton@Charged lepton from $W$ decay!Pseudo-rapidity}
  \end{center} 
\end{figure}

Concerning the main kinematics $\yW$, $\pTW$, $\etal$ and $\pTl$ shown in Fig.~%
\ref{app_dd_yW_pTW_etal_pTl} the most striking feature is the gain of charge symmetry in the initial 
state with respect to $\pp$ collisions. 
Because now the production of the $\Wp$ and $\Wm$ is almost the same in proportions the rapidity 
and transverse momentum distributions almost perfectly match each other.
On the other hand in the final state the $V-A$ \index{Electroweak!VmA@$V-A$ coupling} effect folded to the issues
related to the transverse motion of the $W$ gives unchanged features such as\,: 
(1) the flattening of the $\eta_\lm$ and narrowing of the $\eta_\lp$
distributions with respect to $\yW$ and 
(2) the privileged decay of the $\lp$($\lm$) in the same(opposite) direction of $\vec p_{T,W}$.
These two effects can be seen respectively more directly from the angular point of view with 
$\costhetaWlwrf$ and $\phi_{W,l}^{\,\ast}$ in Fig.~\ref{app_pp_cos_histos}.
Let us note as well the size of $\Asym{\pTl}$ which is different from $\pp$ collisions. 
This is due to the energy in the collision which, being lower here, gives more importance to the 
valence quarks while at $\sqrt S=14\TeV$ the latter are drowned in the large purely sea contributions.
This variation of $\Asym{\pTl}$ as a function of $\sqrt S$ is treated in the next sub-section.

\begin{figure}[!h] 
  \begin{center}
    \includegraphics[width=0.495\tw]{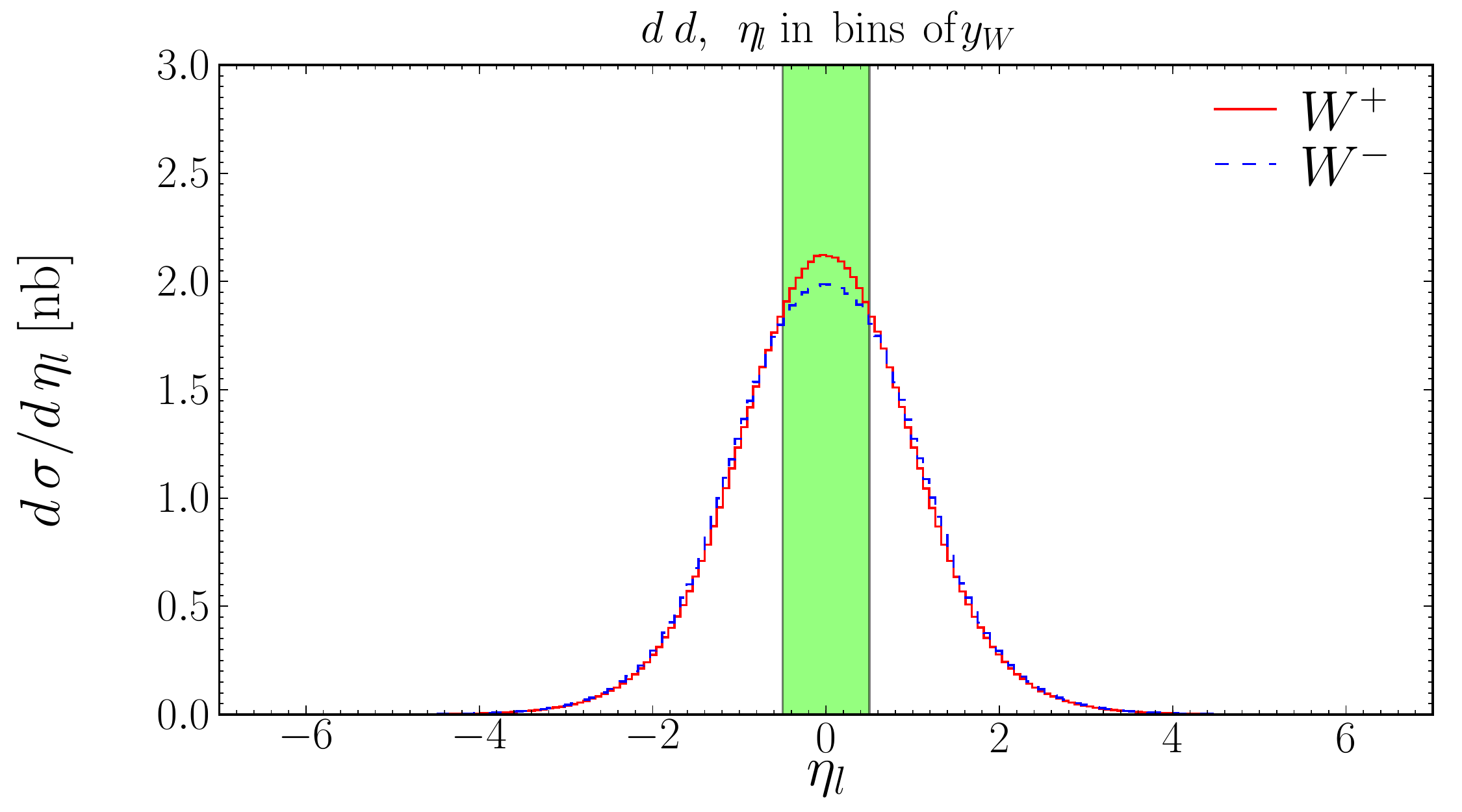}
    \hfill
    \includegraphics[width=0.495\tw]{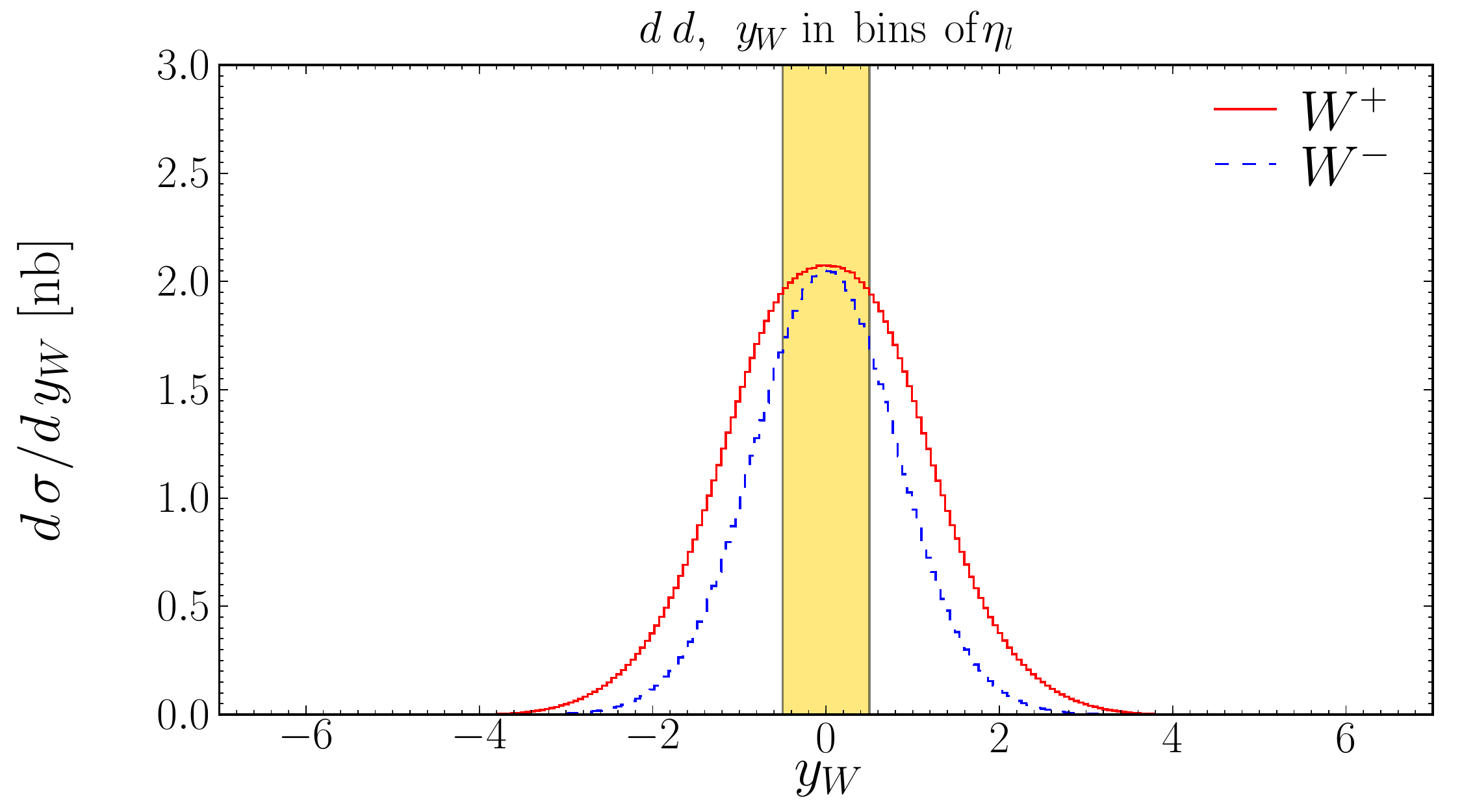}
    \vfill   
    \includegraphics[width=0.495\tw]{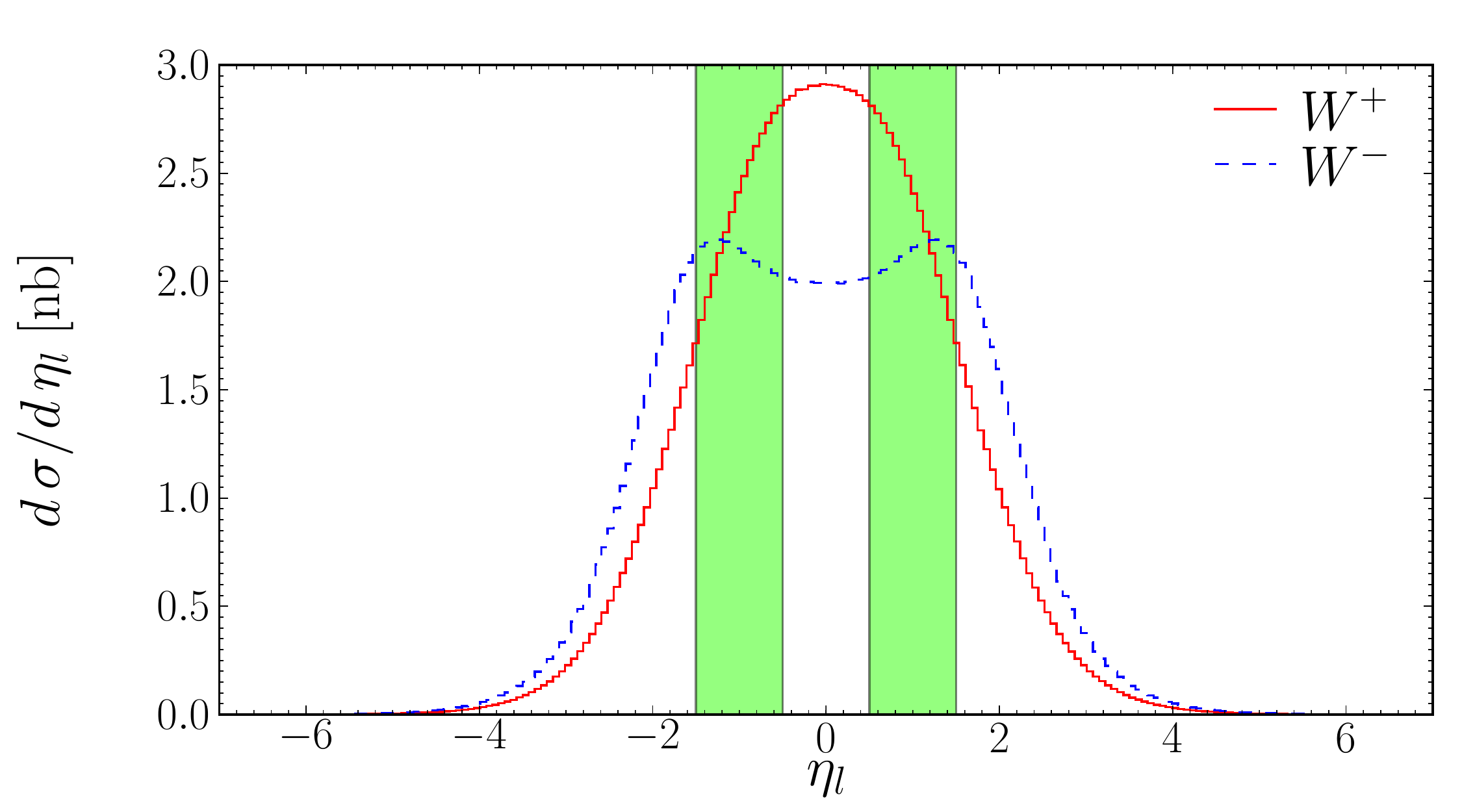}
    \hfill
    \includegraphics[width=0.495\tw]{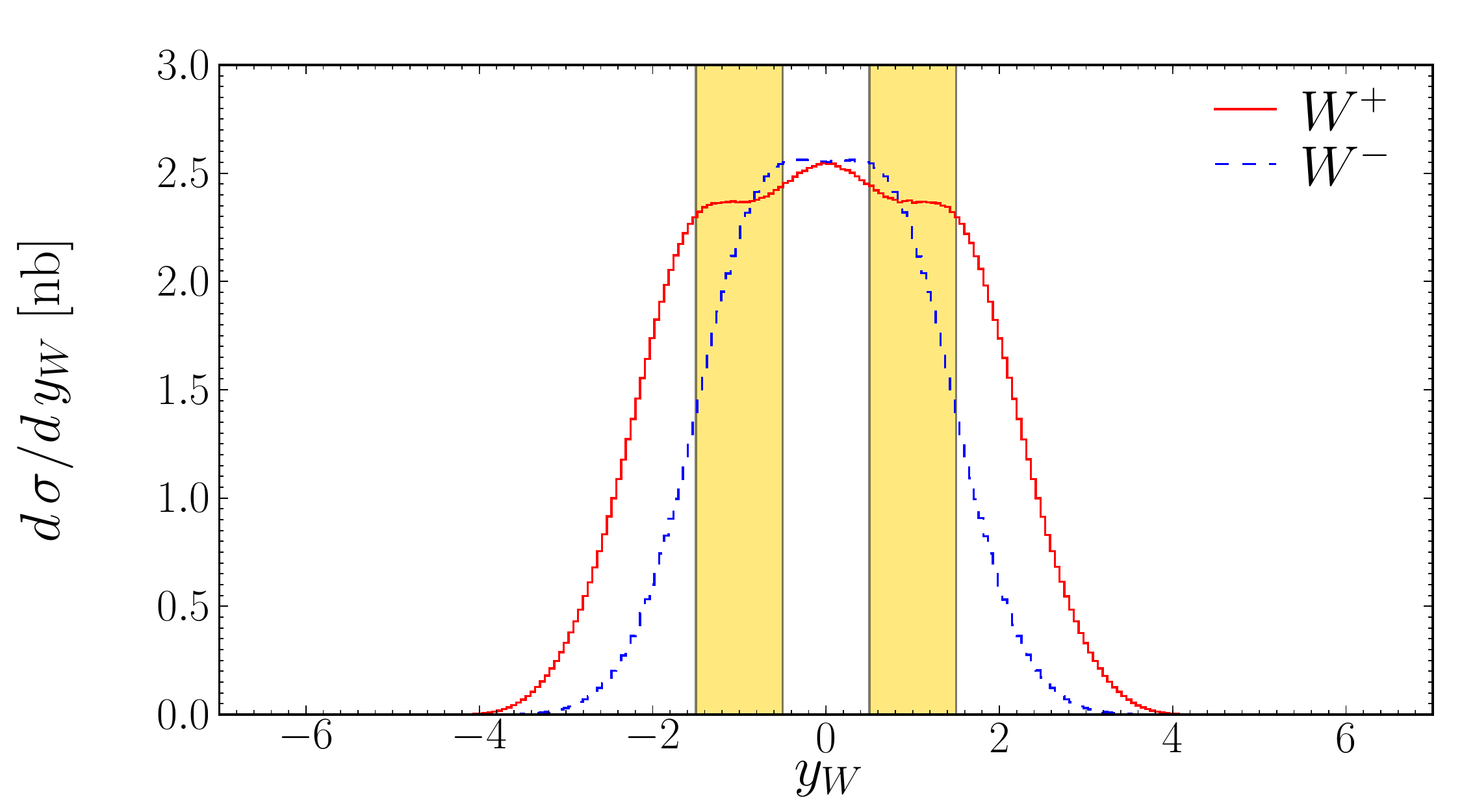}
    \vfill
    \includegraphics[width=0.495\tw]{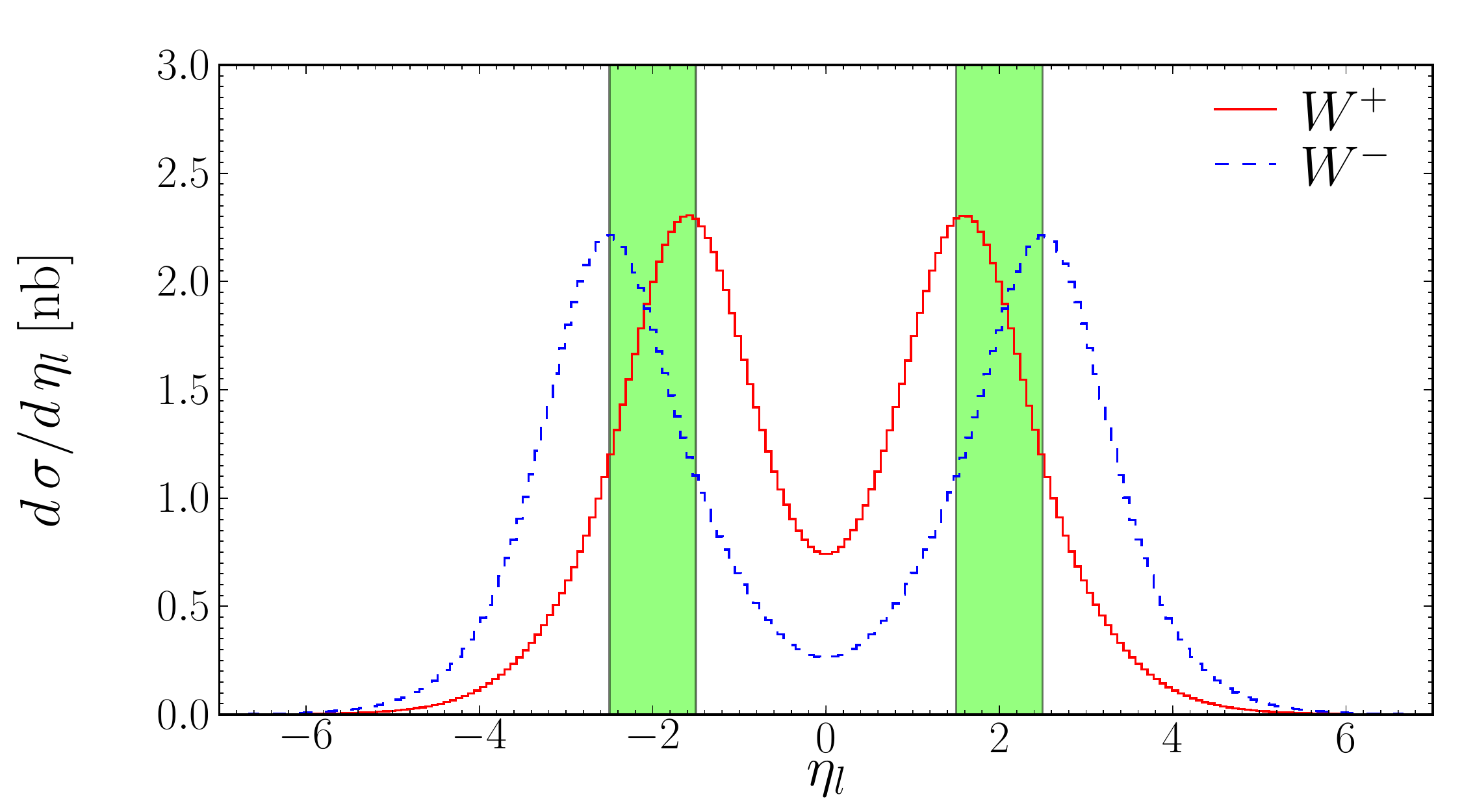}
    \hfill
    \includegraphics[width=0.495\tw]{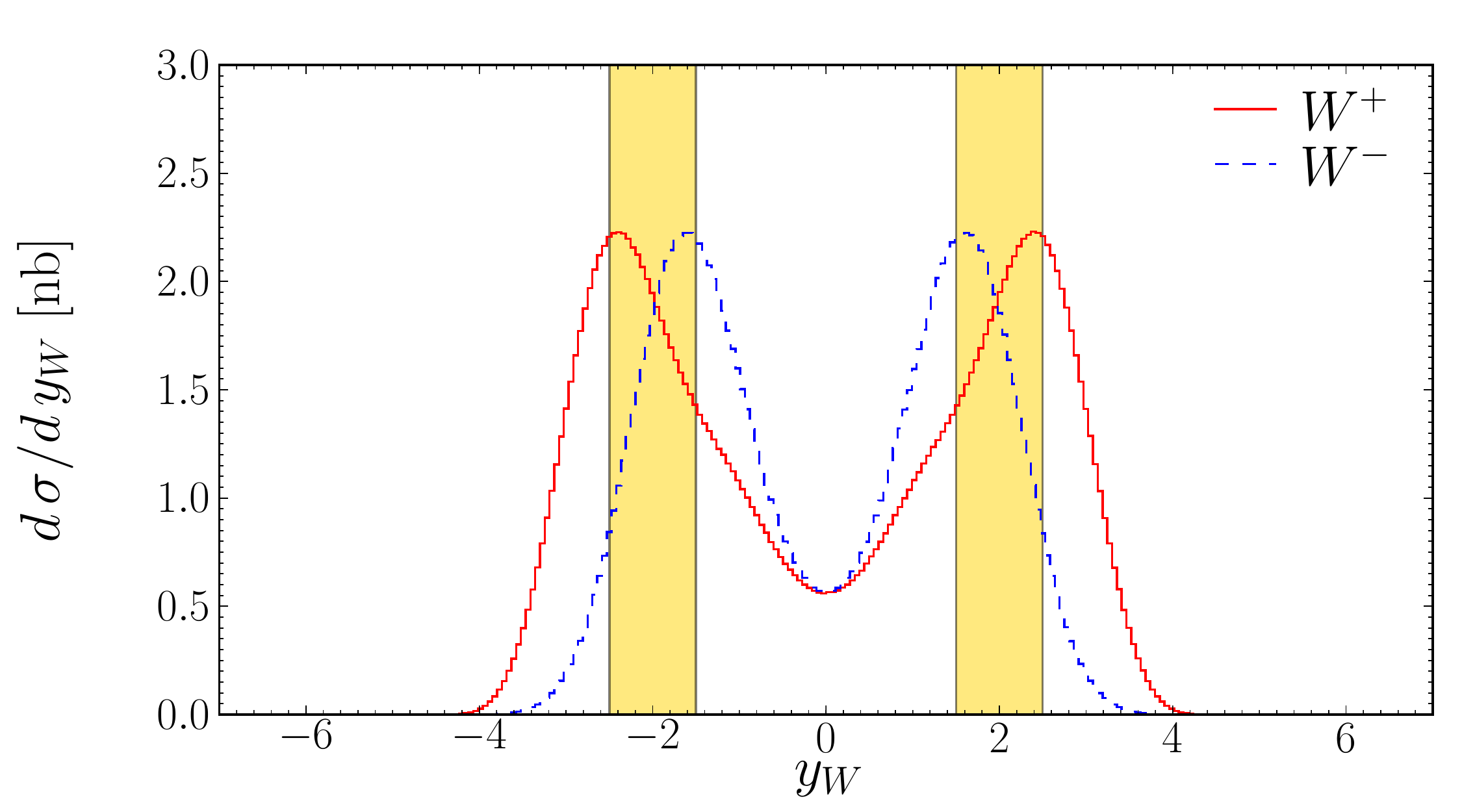}
    \vfill
    \includegraphics[width=0.495\tw]{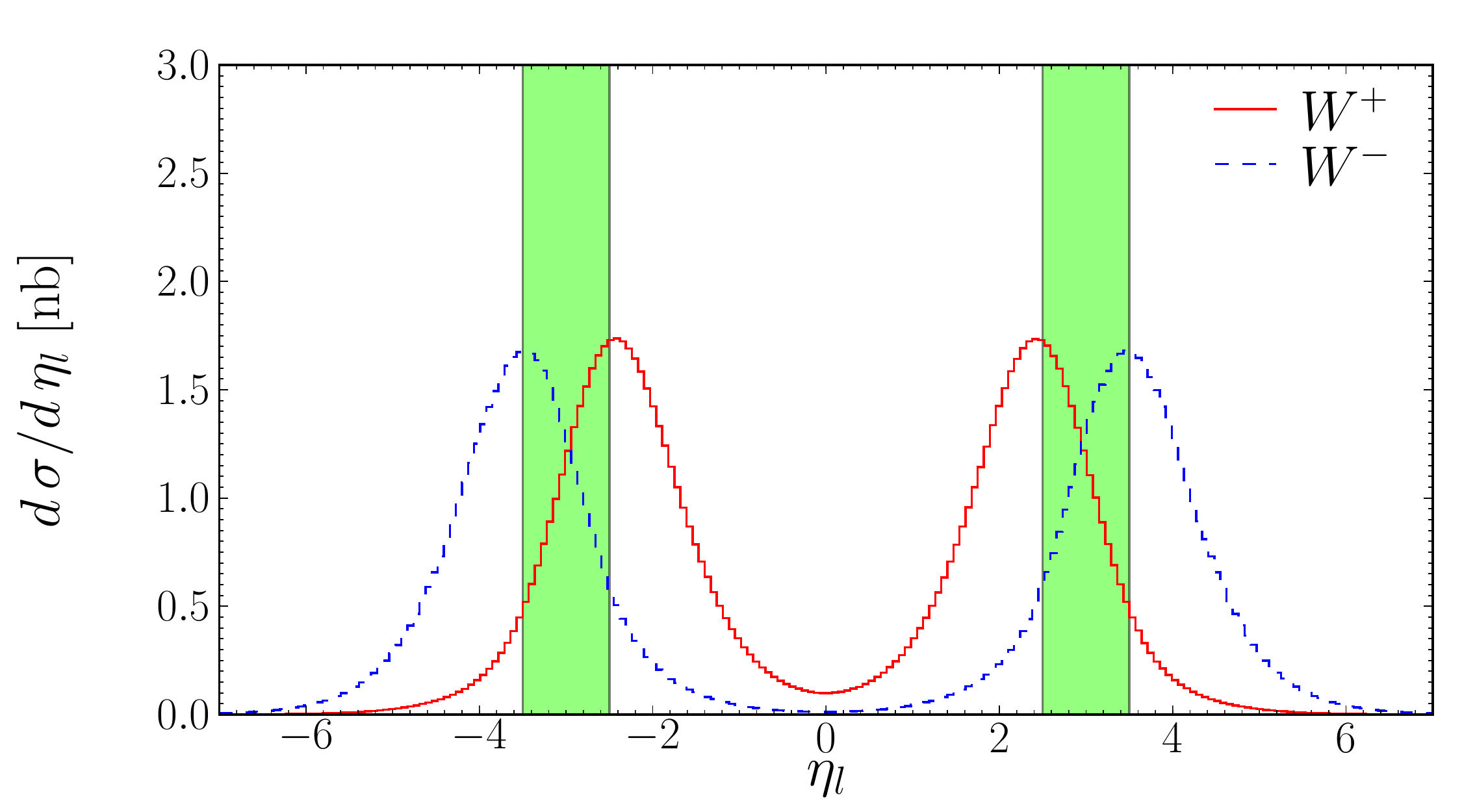}
    \hfill
    \includegraphics[width=0.495\tw]{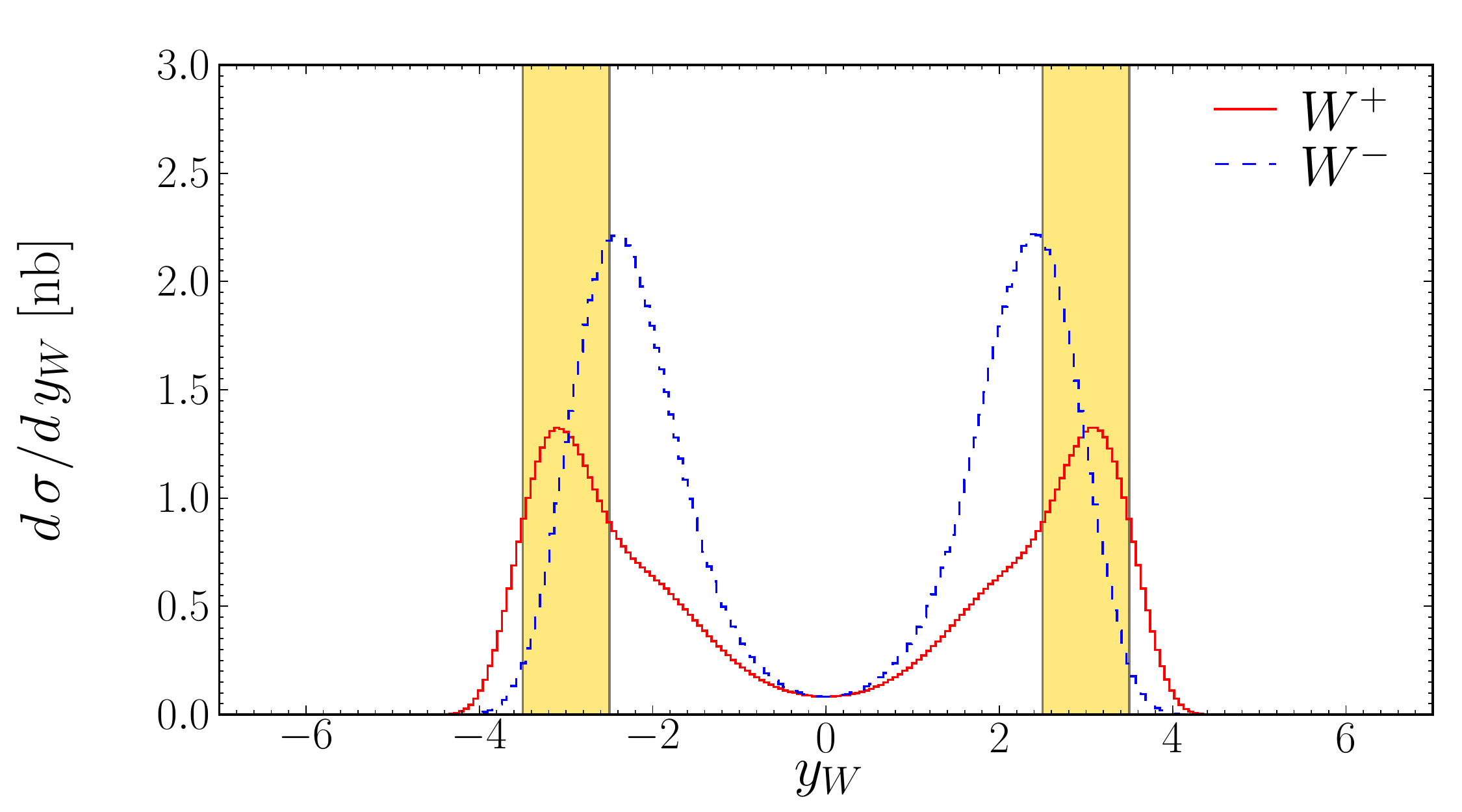}
    \vfill
    \includegraphics[width=0.495\tw]{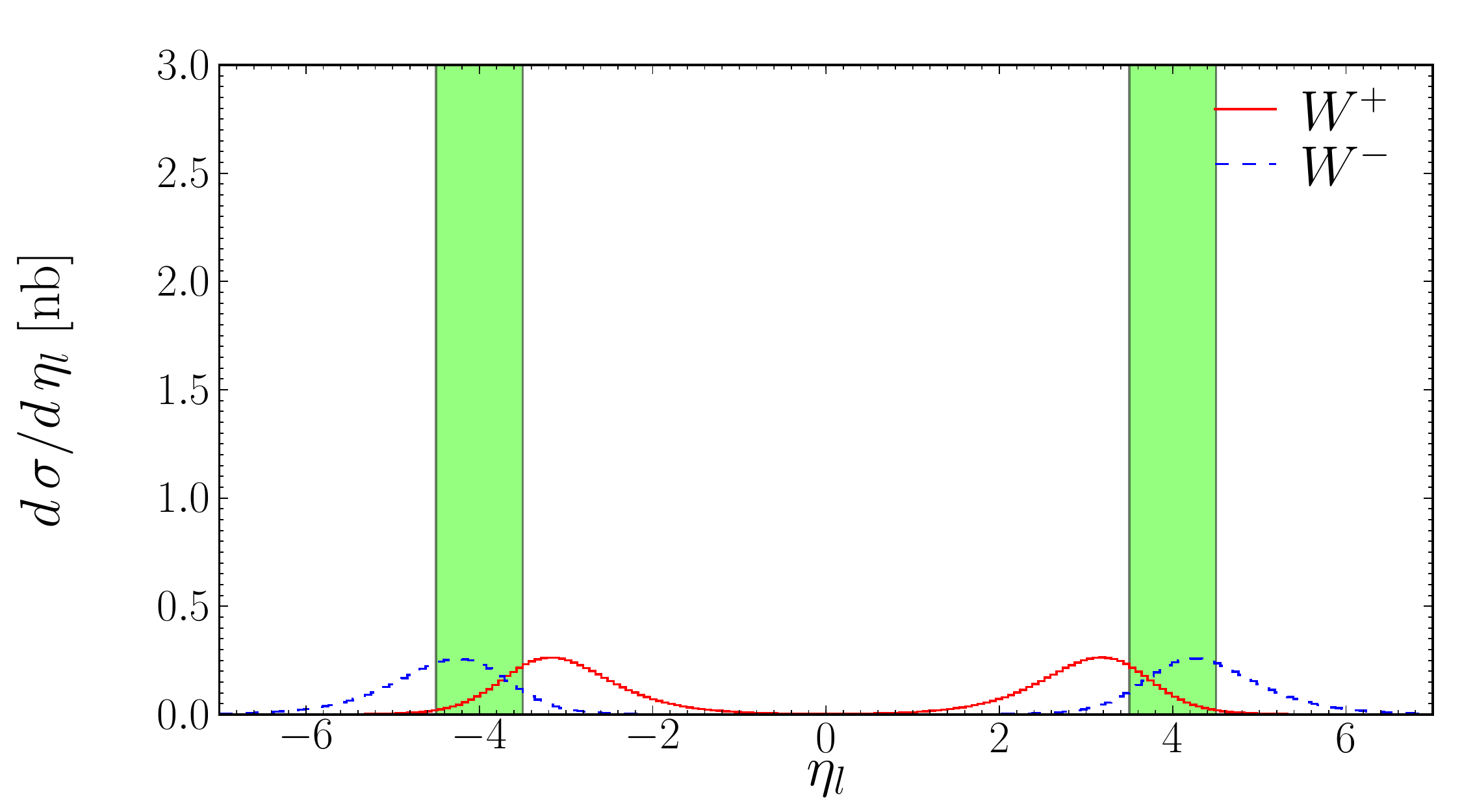}
    \hfill
    \includegraphics[width=0.495\tw]{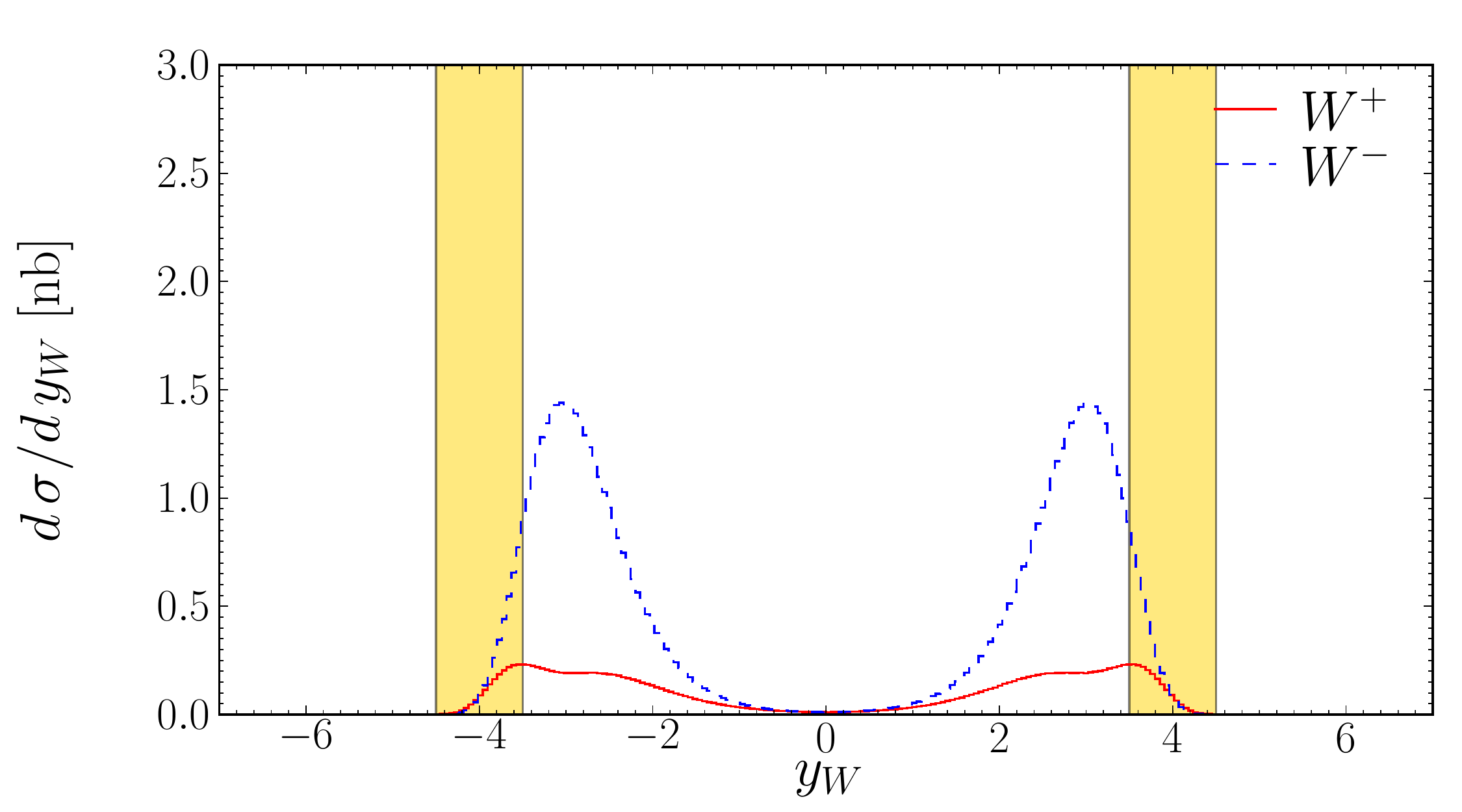}
    \caption[$W$ boson rapidity distributions in bins of the charged lepton pseudo-rapidity and 
      \textit{vice versa} in the case of $\dd$ collisions with $\sqrt{S_{n_1\,n_2}}=7\TeV$]
            {\figtxt{$W$ boson rapidity distributions in bins of the charged lepton pseudo-rapidity 
                (left) and \textit{vice versa} (right) in the case of $\dd$ collisions with
                $\sqrt{S_{n_1\,n_2}}=7\TeV$.
                In each plot the corresponding $\yW$ or $\etal$ selection is materialised by the 
                colored stripe(s).}}
            \label{app_dd_etal_yW_in_yW_etal_bins}
  \end{center} 
\end{figure}

\begin{figure}[!h] 
  \begin{center}
    \includegraphics[width=0.495\tw]{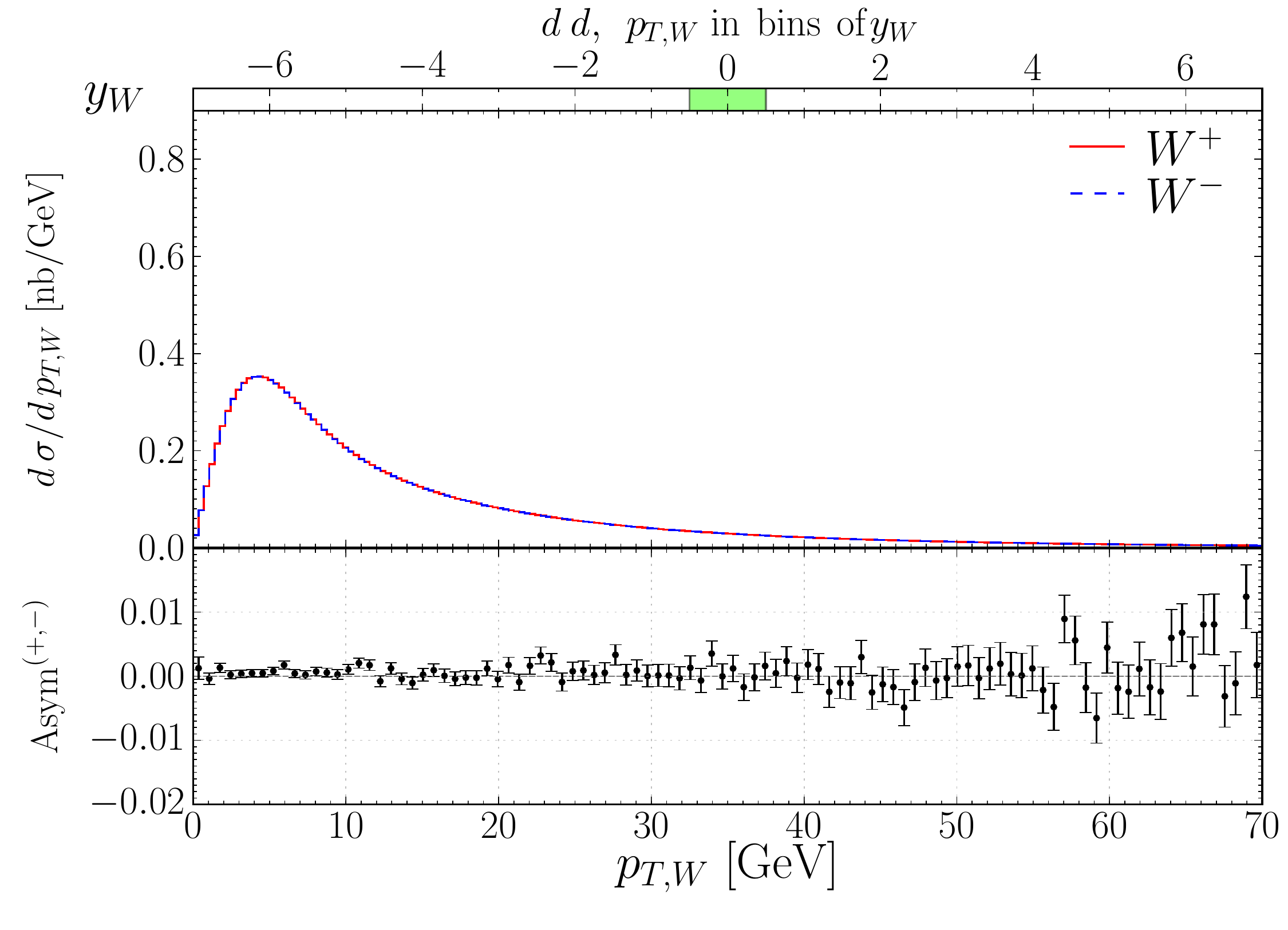}
    \hfill
    \includegraphics[width=0.495\tw]{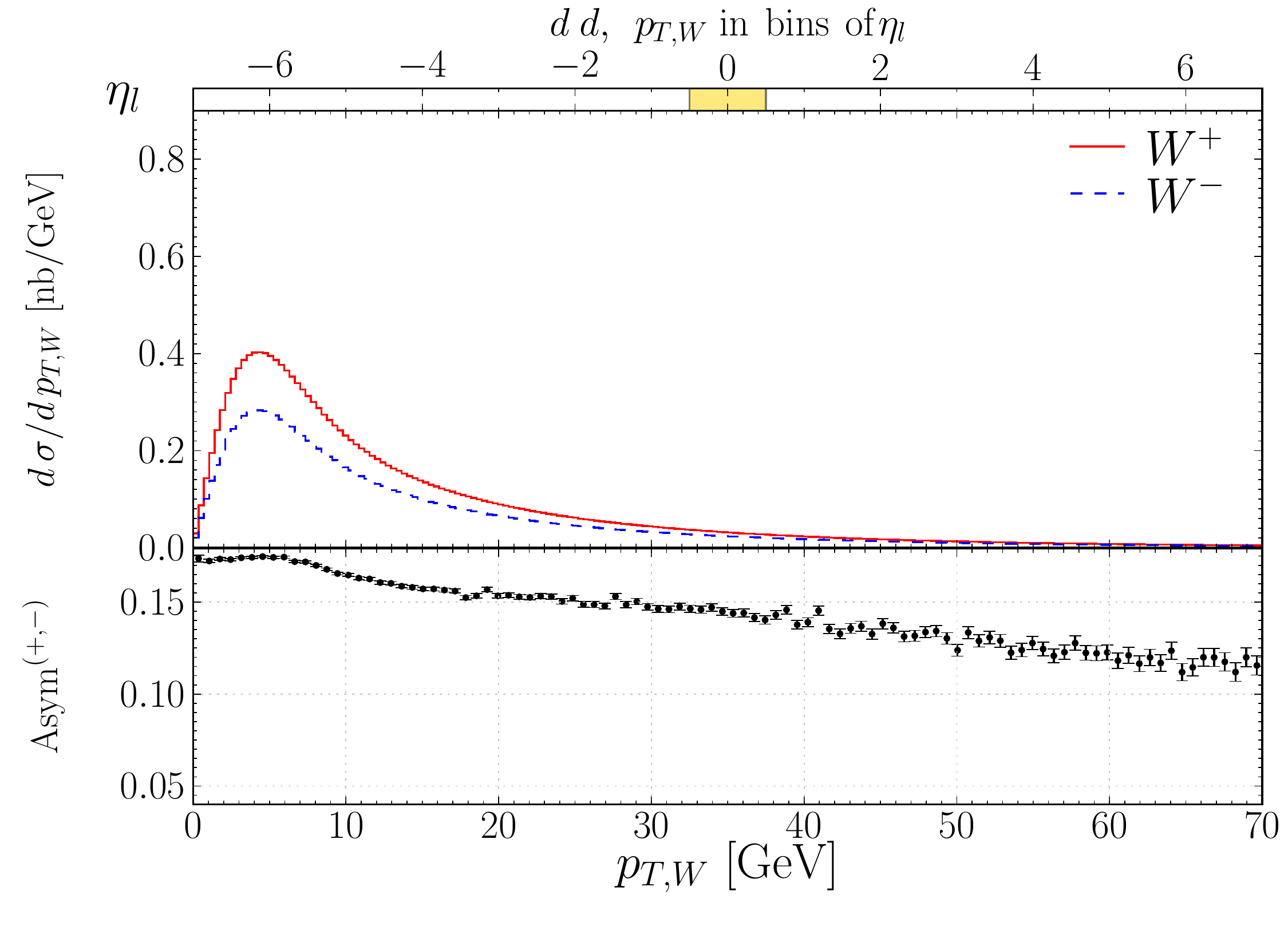}
    \vfill   
    \includegraphics[width=0.495\tw]{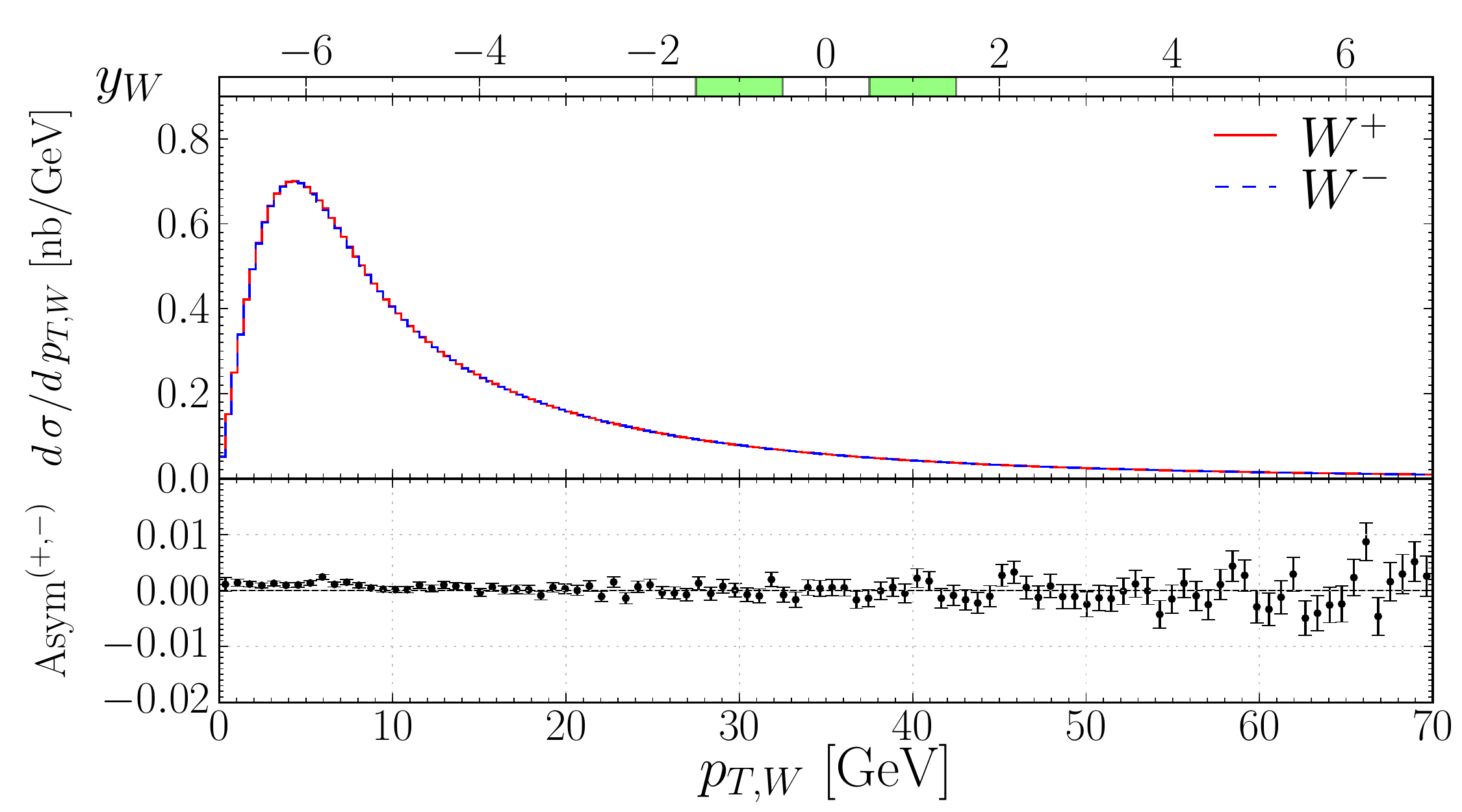}
    \hfill
    \includegraphics[width=0.495\tw]{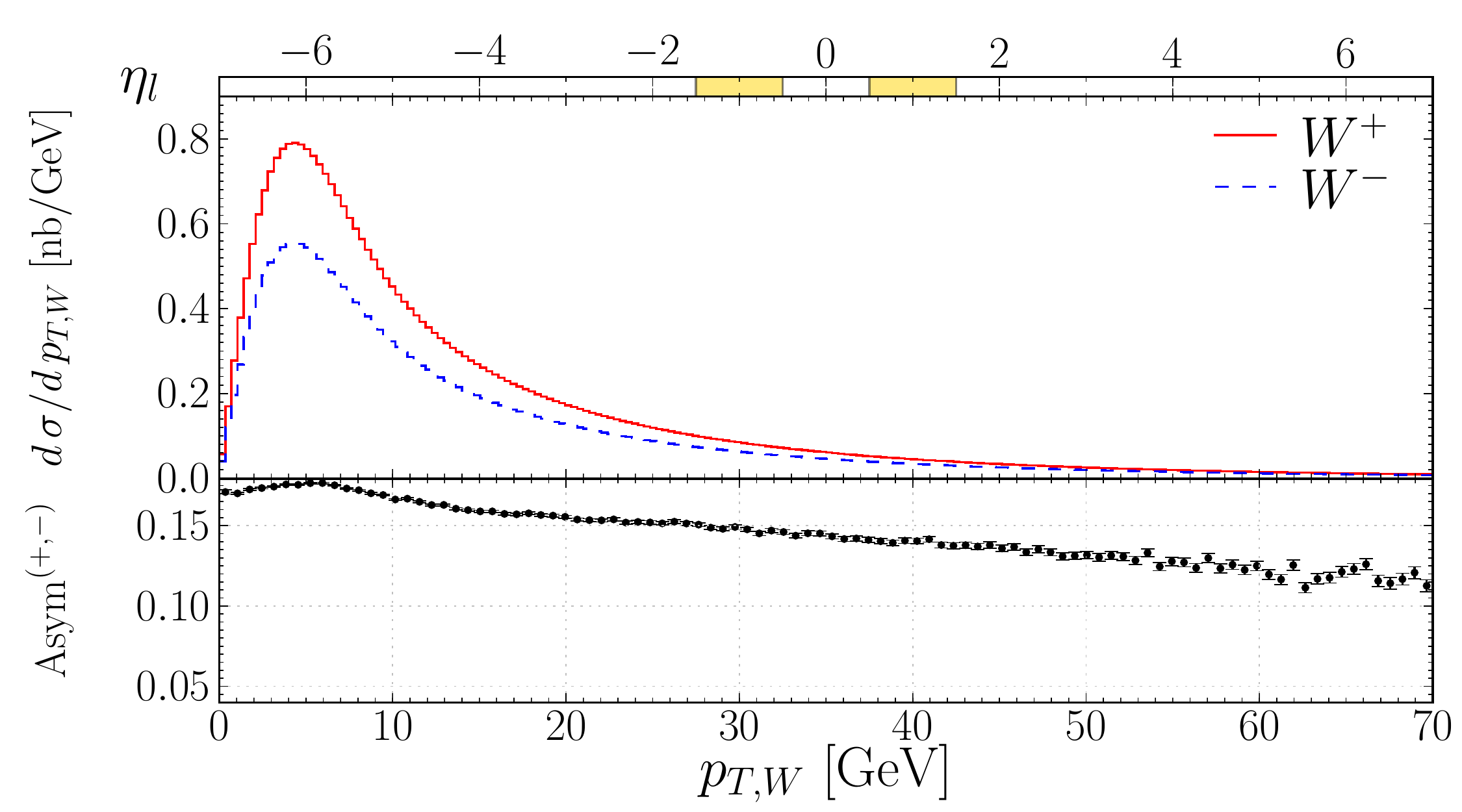}
    \vfill
    \includegraphics[width=0.495\tw]{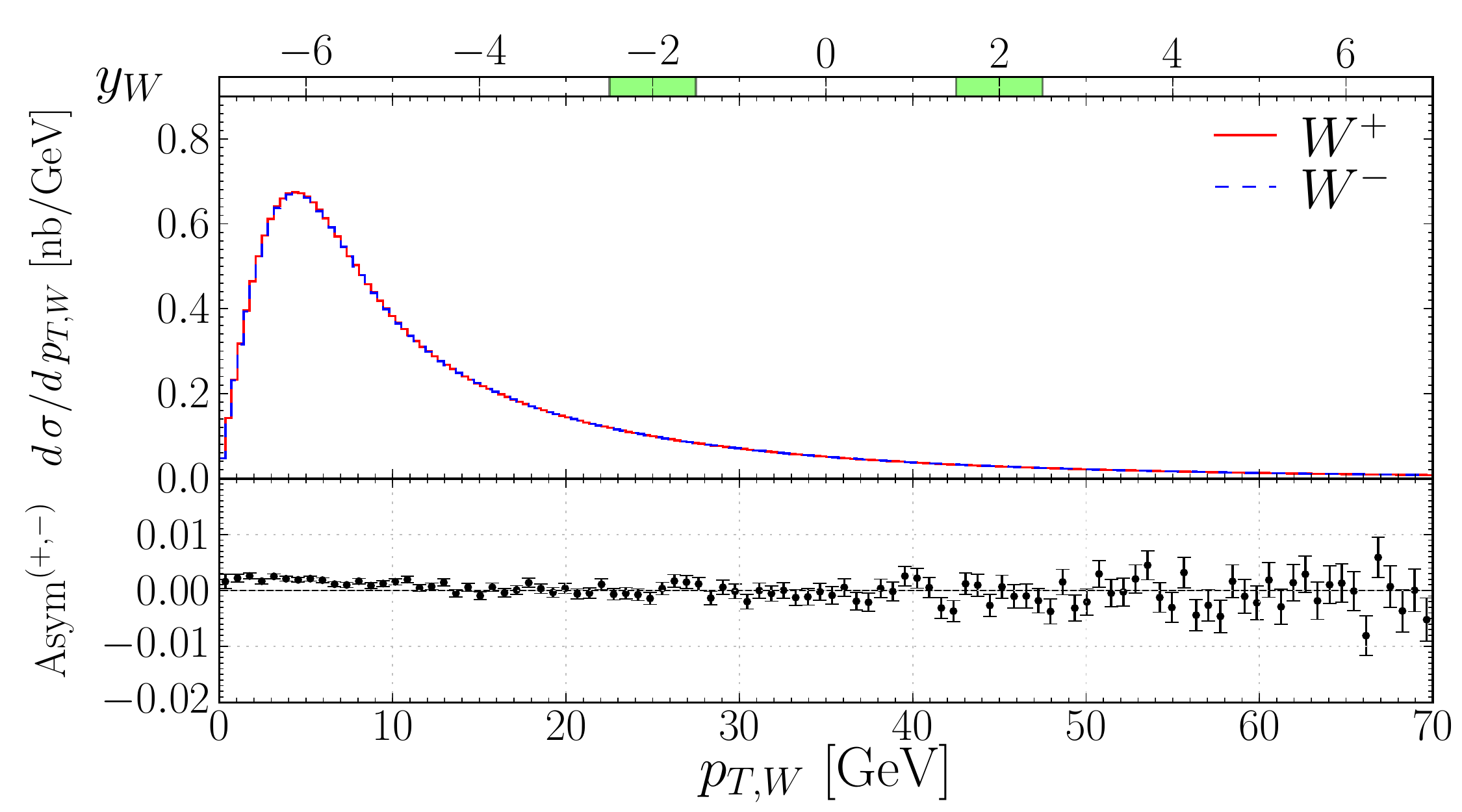}
    \hfill
    \includegraphics[width=0.495\tw]{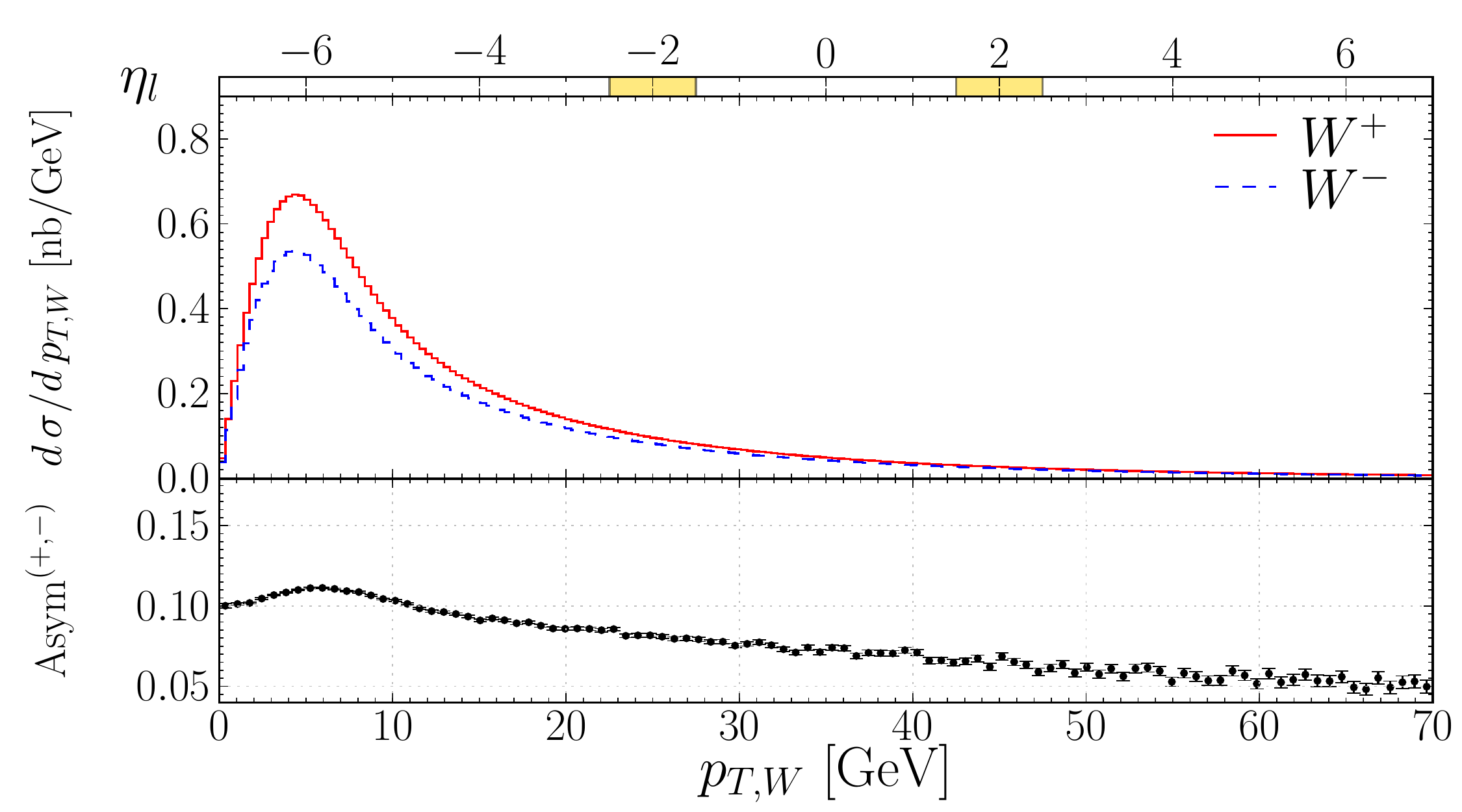}
    \vfill
    \includegraphics[width=0.495\tw]{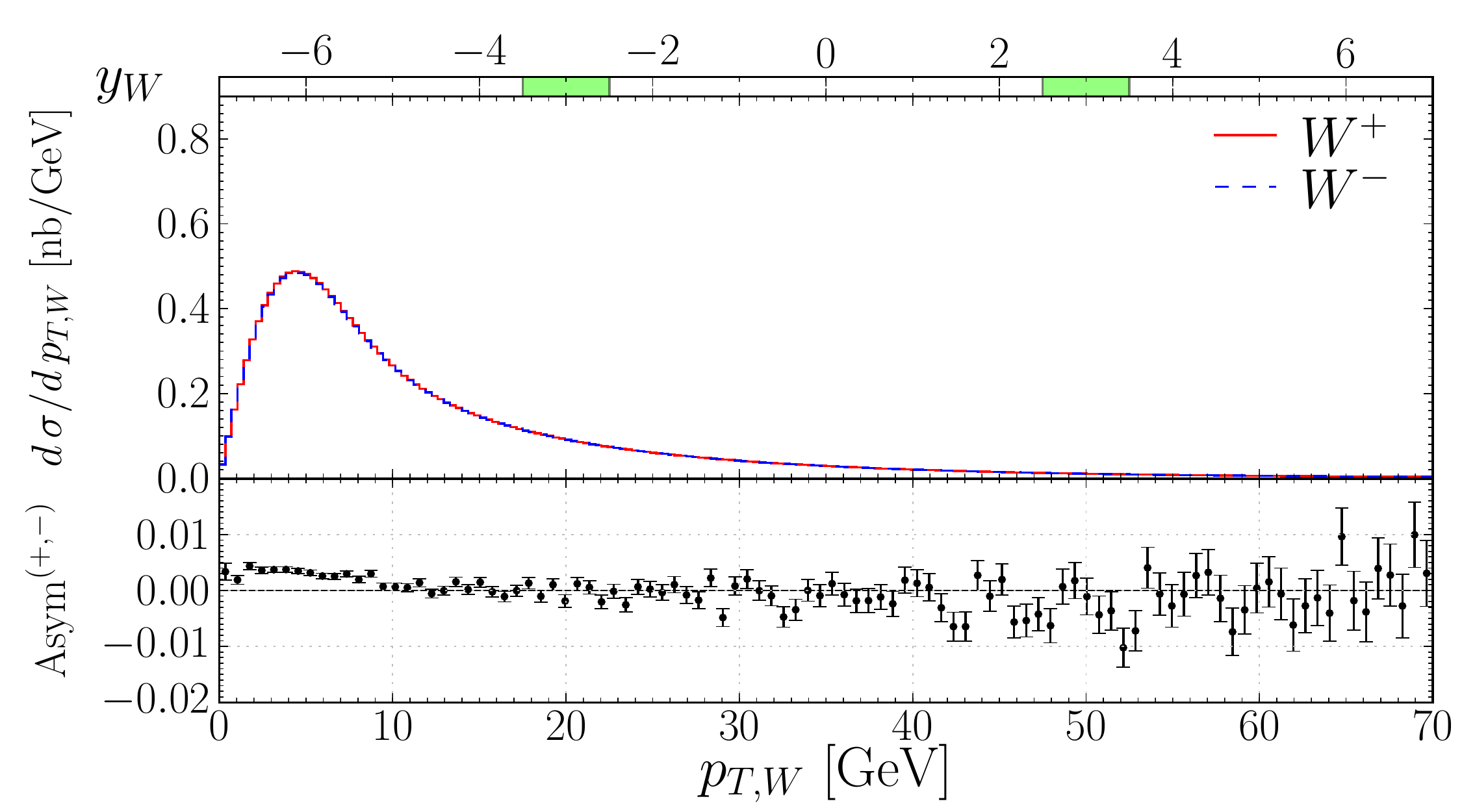}
    \hfill
    \includegraphics[width=0.495\tw]{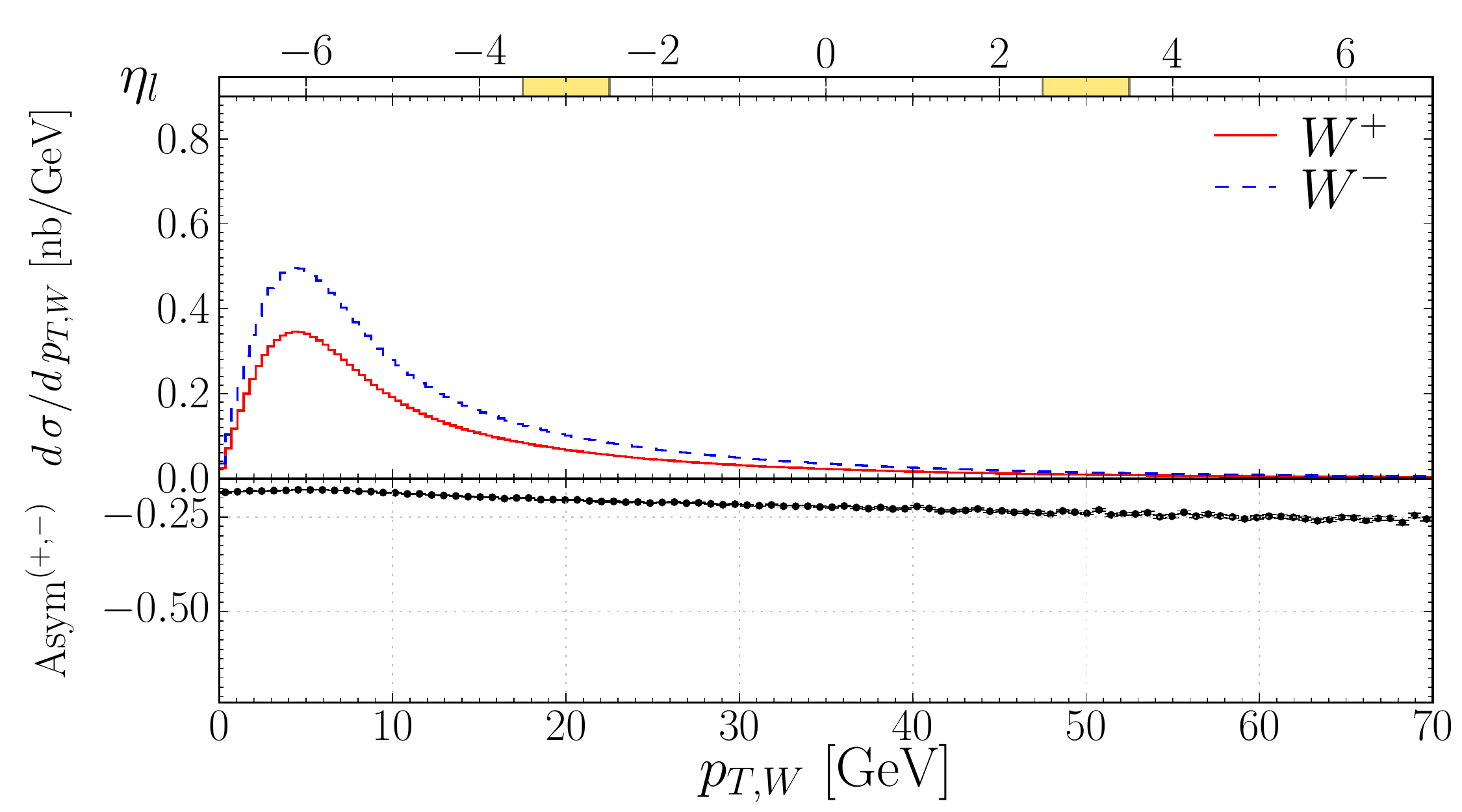}
    \vfill
    \includegraphics[width=0.495\tw]{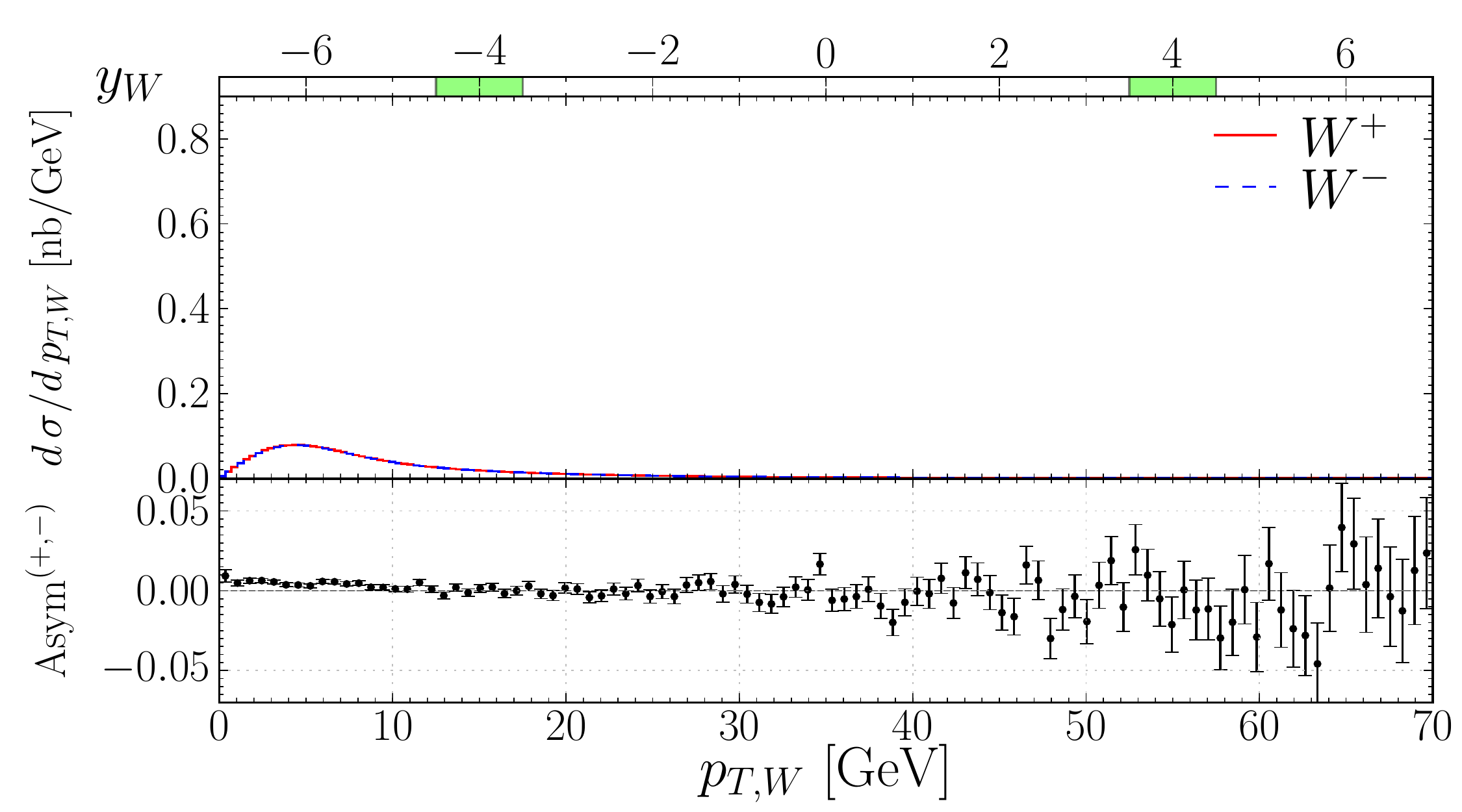}
    \hfill
    \includegraphics[width=0.495\tw]{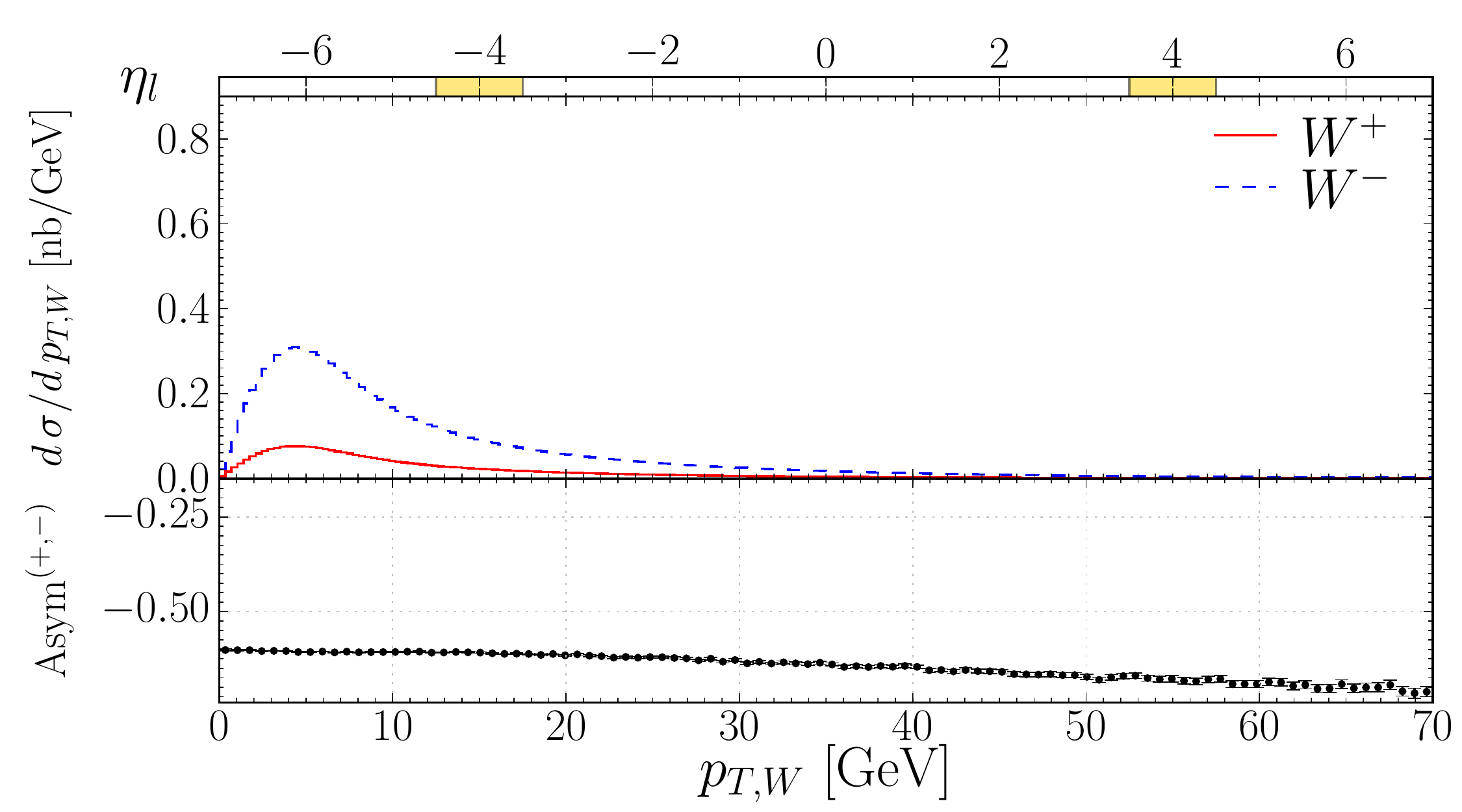}
    \caption[$W$ boson transverse momentum in bins of $\yW$ and in bins of $\etal$ in $\dd$ collisions
    with $\sqrt{S_{n_1\,n_2}}=7\TeV$]
            {\figtxt{$W$ boson transverse momentum in bins of $\yW$ (left) 
                and in bins of $\etal$ (right) in $\dd$ collisions with $\sqrt{S_{n_1\,n_2}}=7\TeV$.}}
            \label{app_dd_pTW_in_yW_etal_bins}
  \end{center} 
\end{figure}

\begin{figure}[!h] 
  \begin{center}
    \includegraphics[width=0.495\tw]{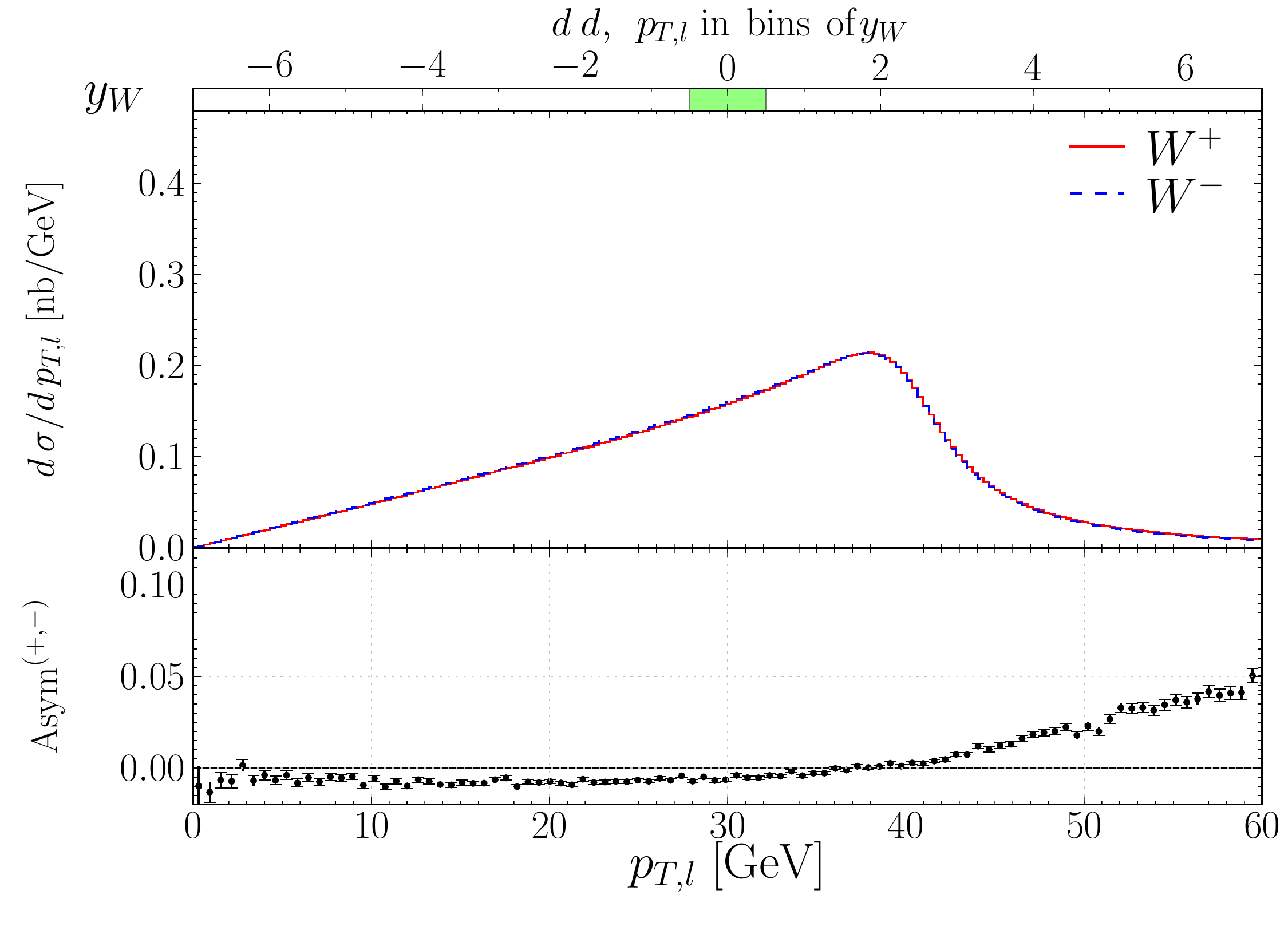}
    \hfill
    \includegraphics[width=0.495\tw]{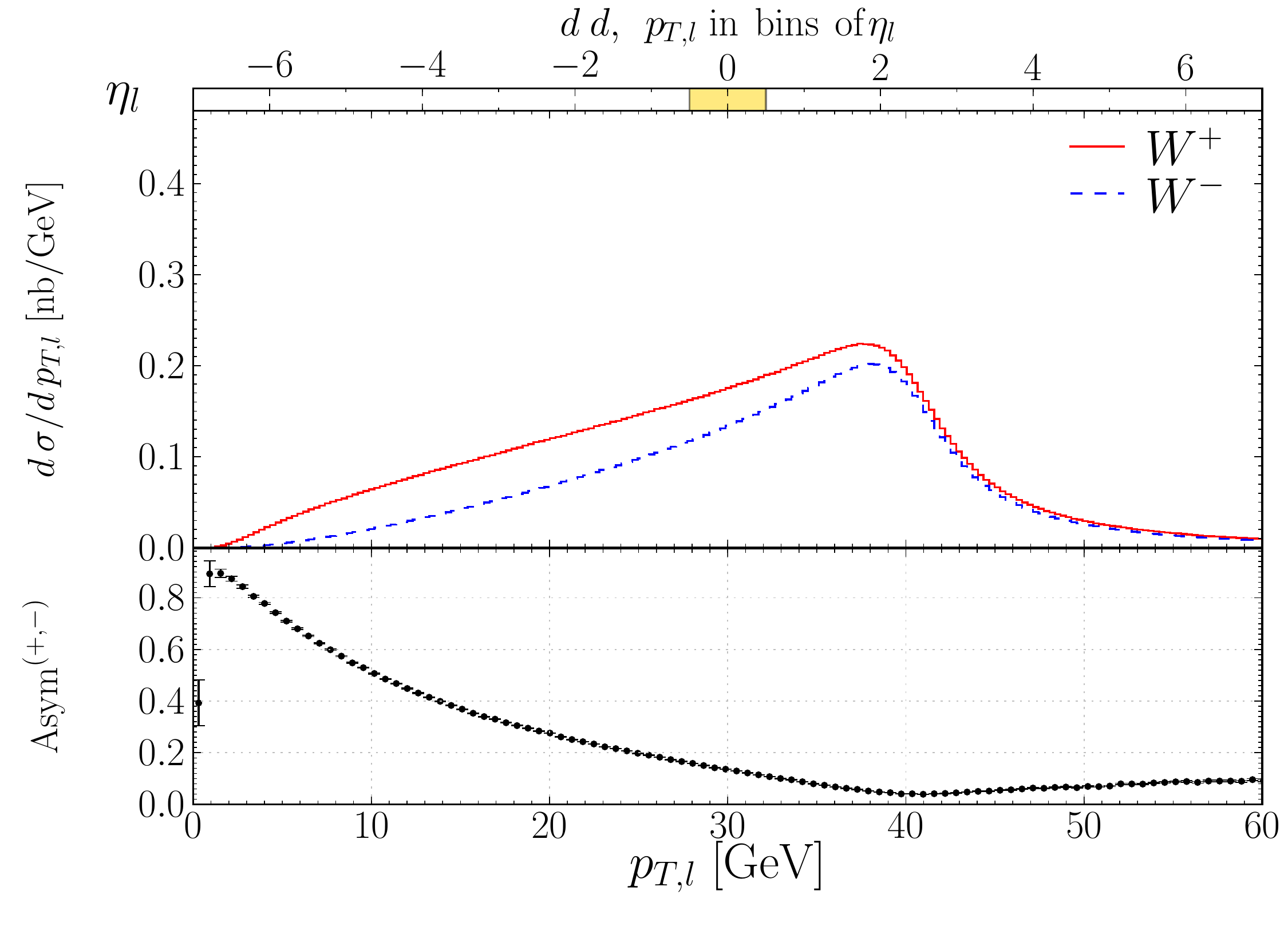}
    \vfill   
    \includegraphics[width=0.495\tw]{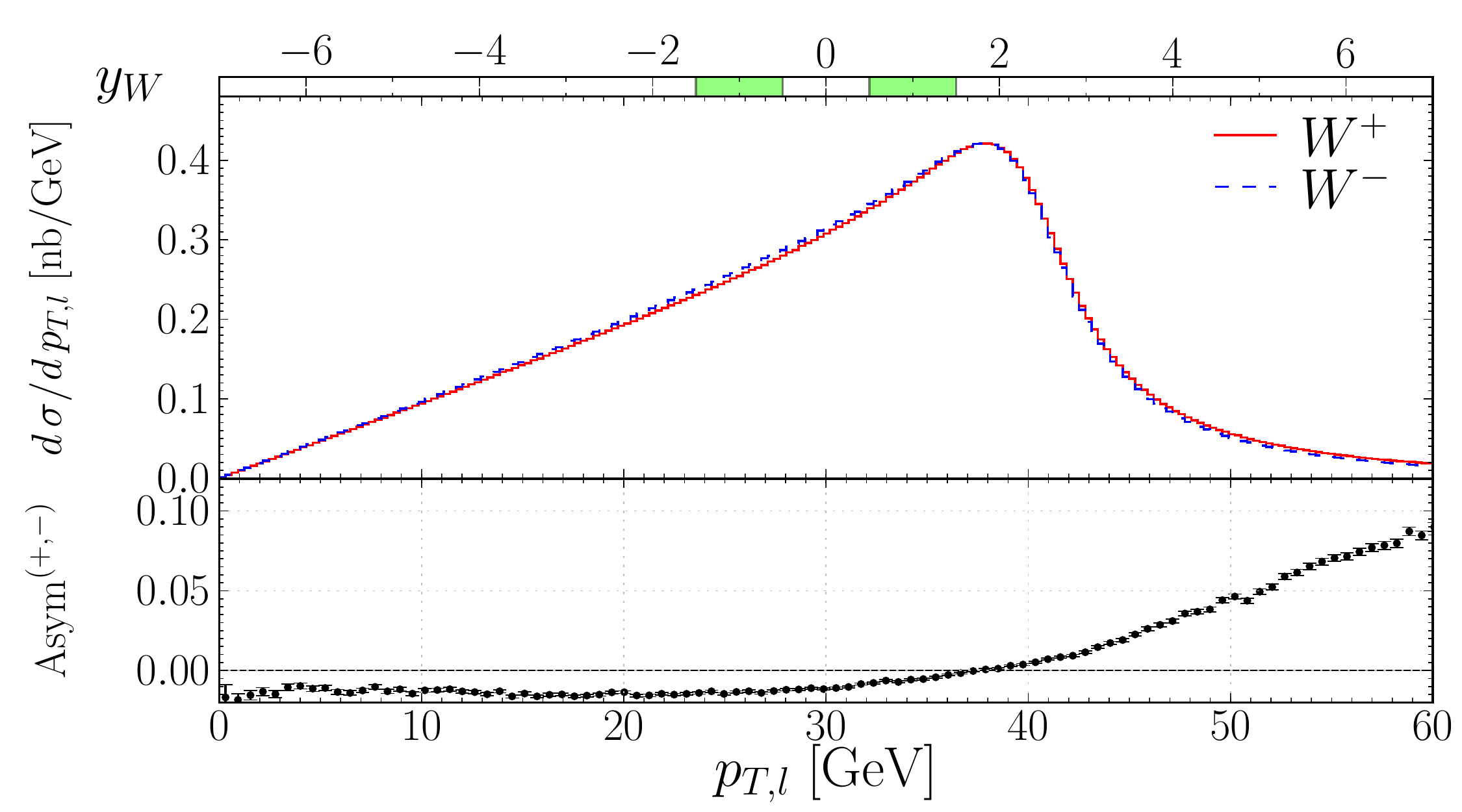}
    \hfill
    \includegraphics[width=0.495\tw]{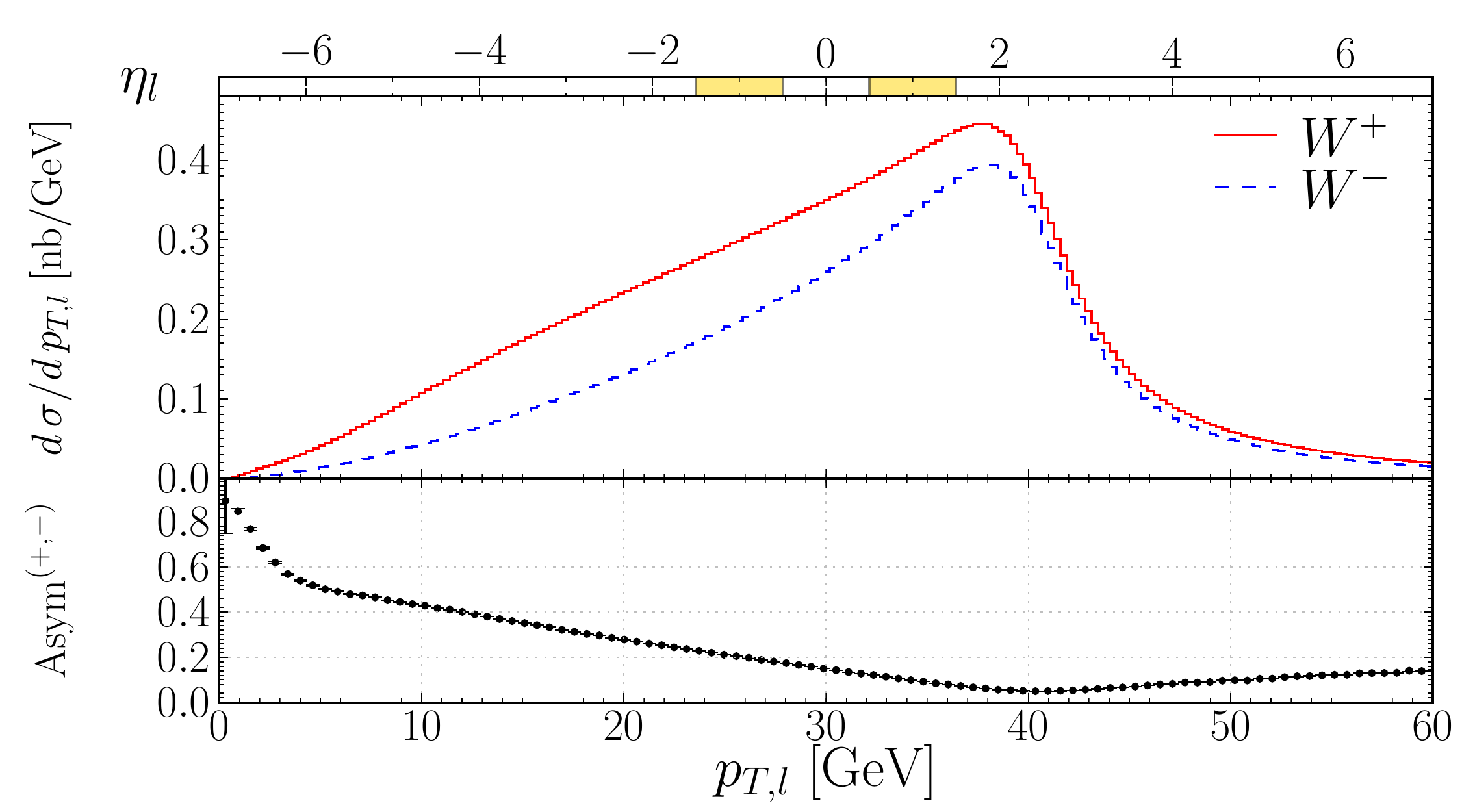}
    \vfill
    \includegraphics[width=0.495\tw]{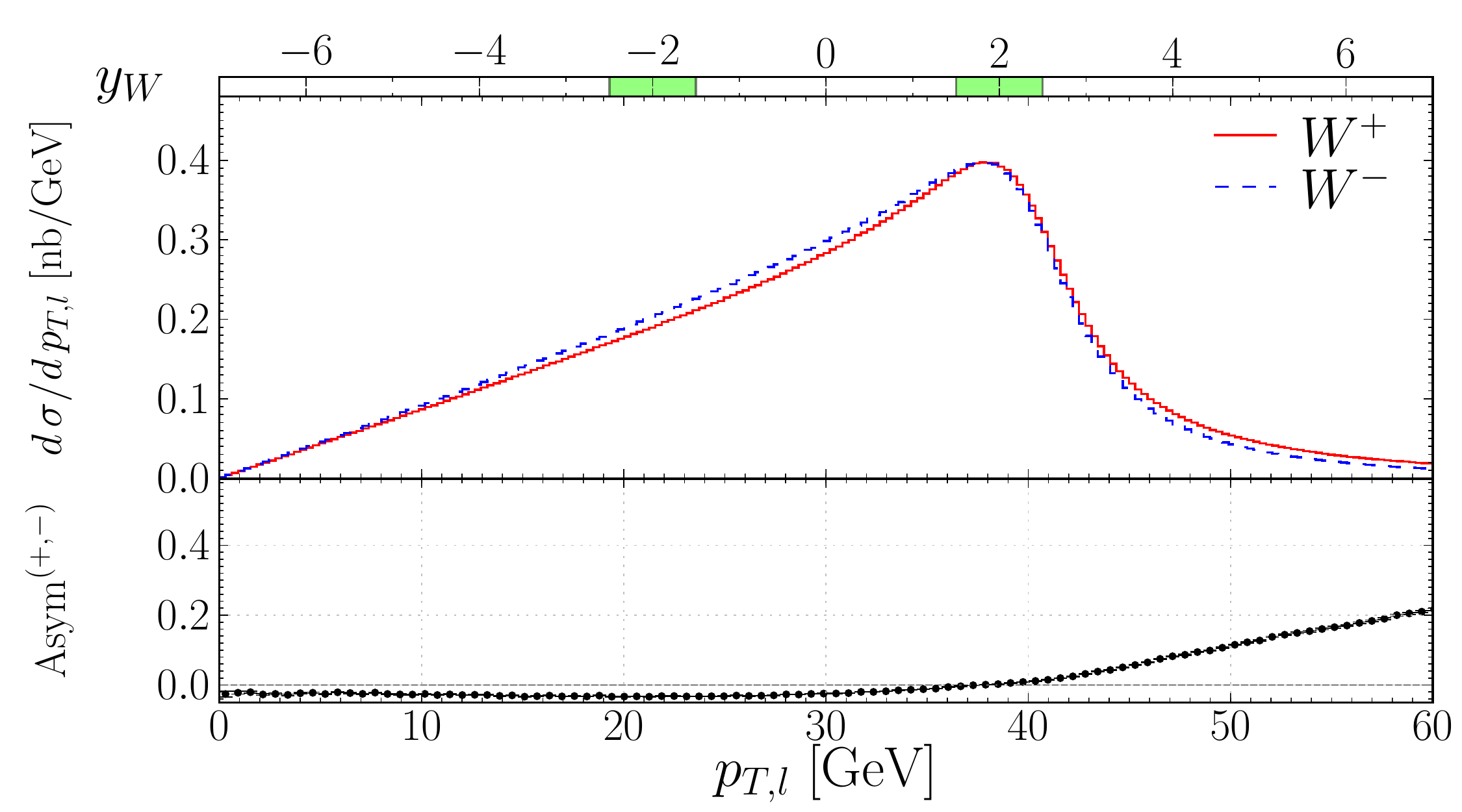}
    \hfill
    \includegraphics[width=0.495\tw]{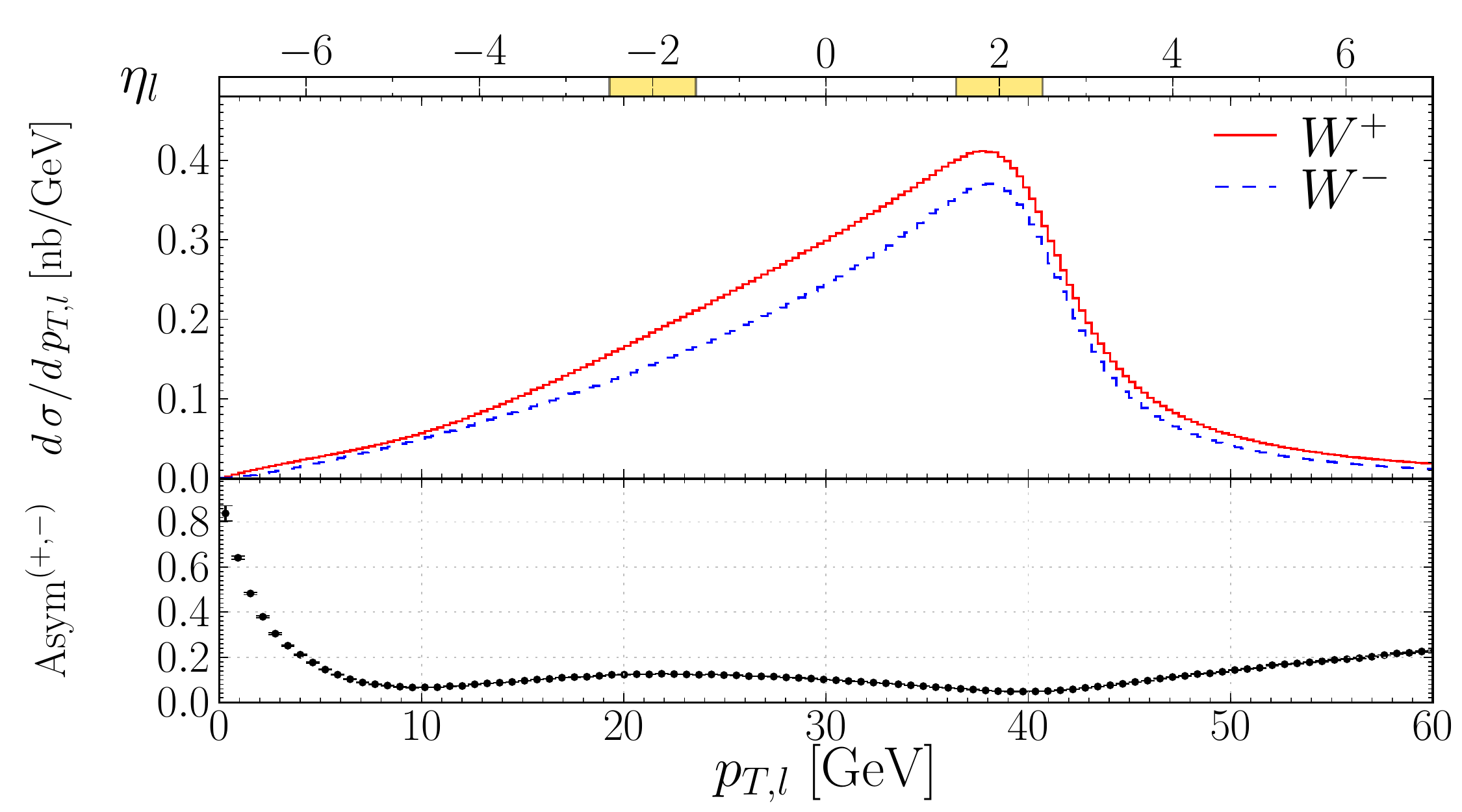}
    \vfill
    \includegraphics[width=0.495\tw]{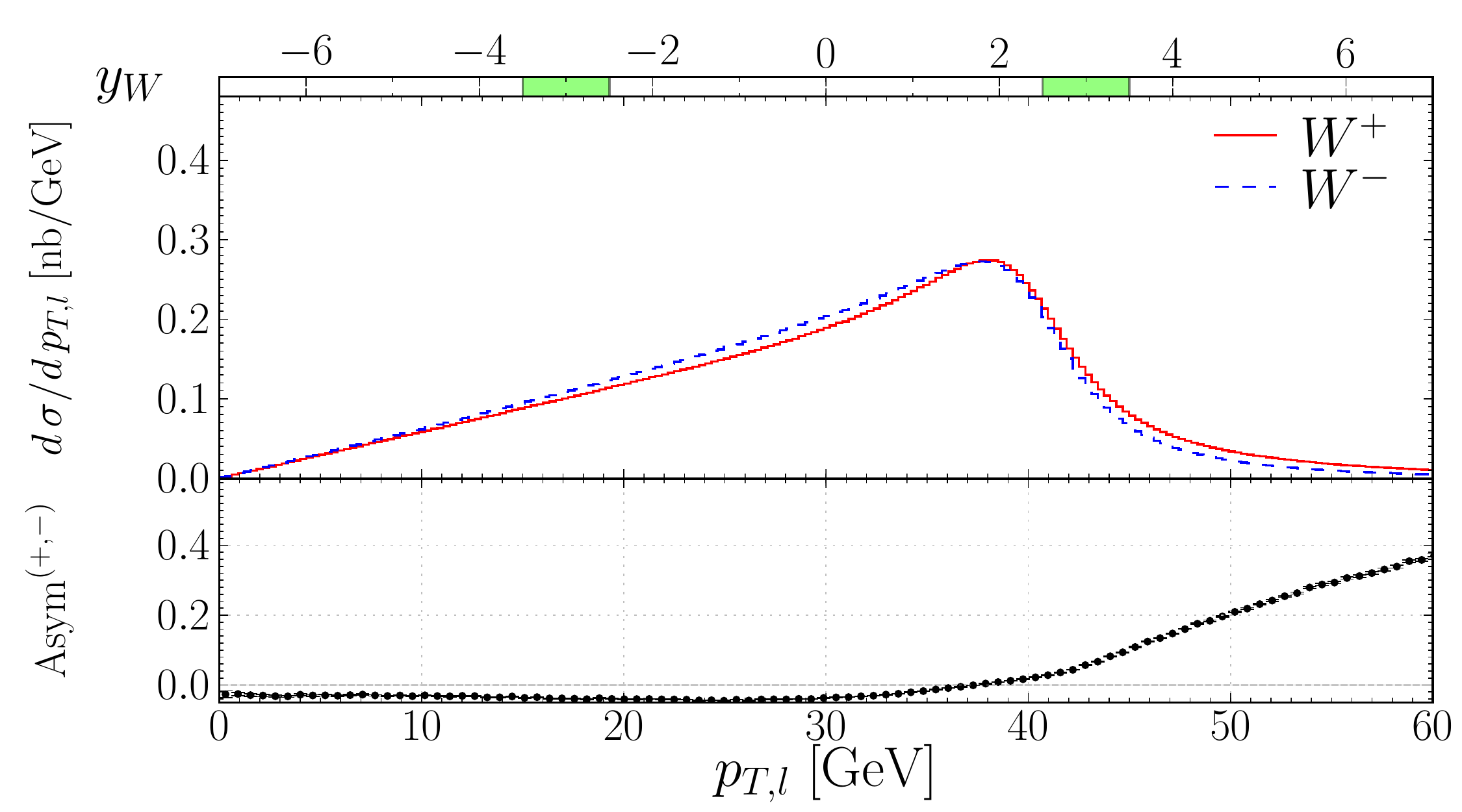}
    \hfill
    \includegraphics[width=0.495\tw]{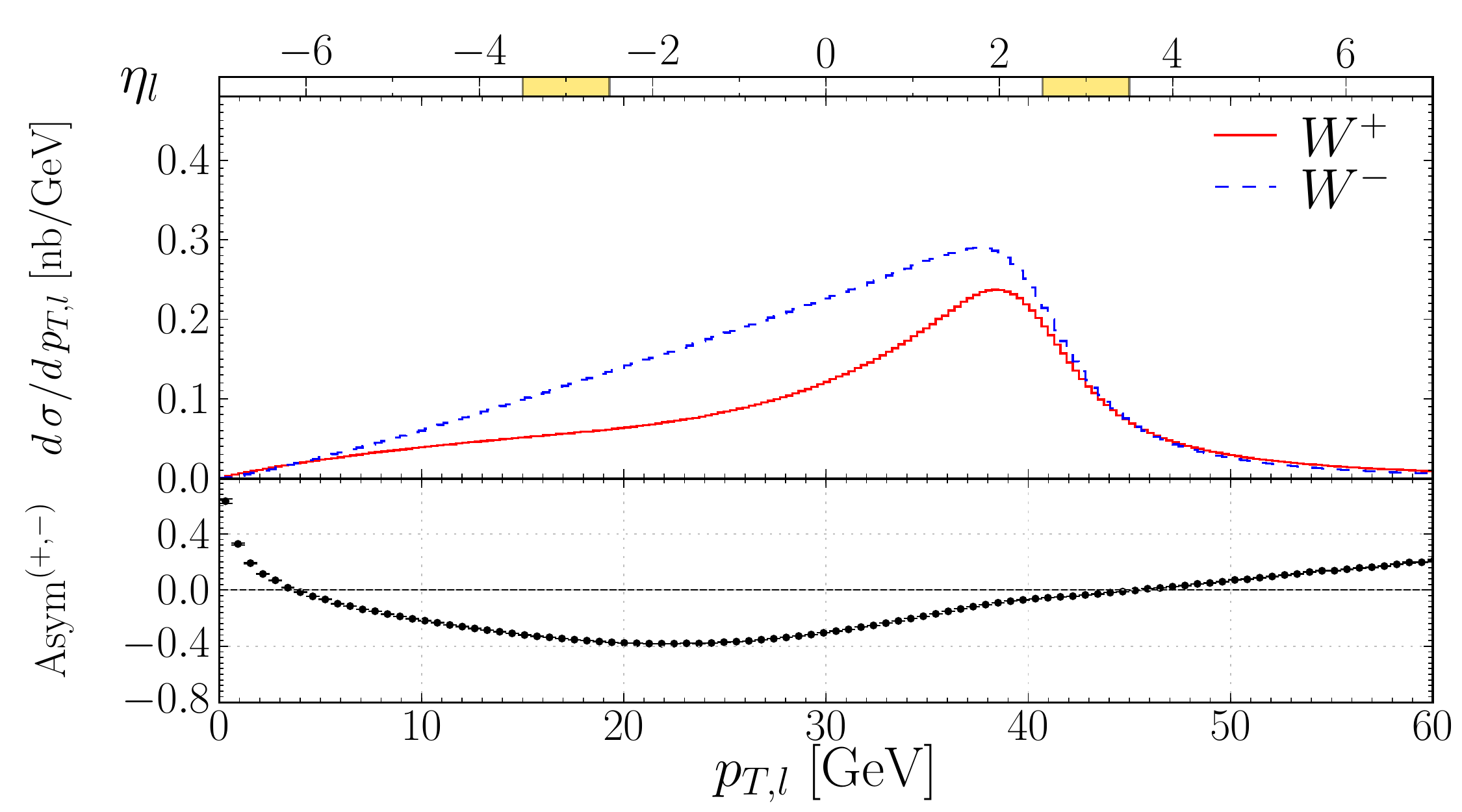}
    \vfill
    \includegraphics[width=0.495\tw]{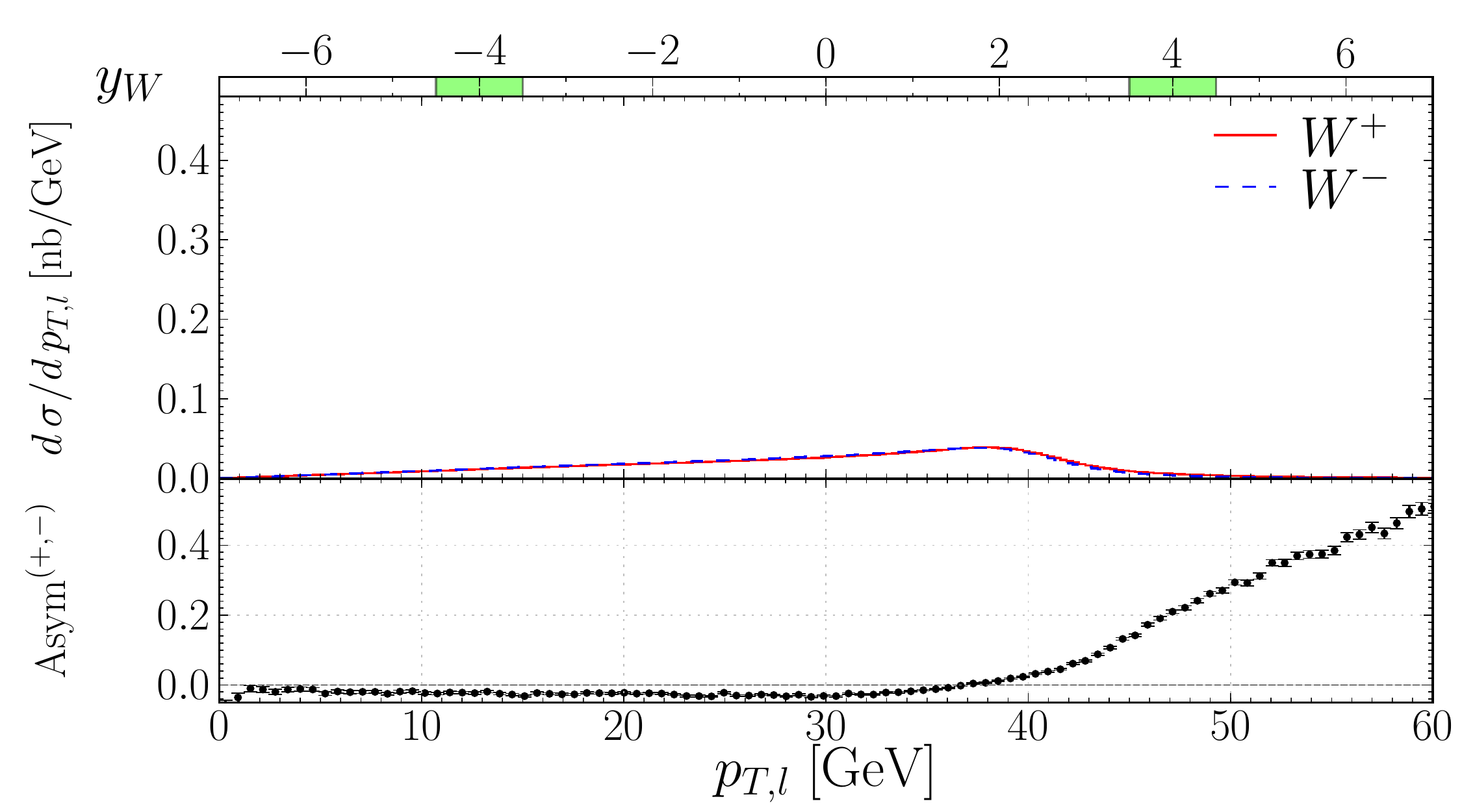}
    \hfill
    \includegraphics[width=0.495\tw]{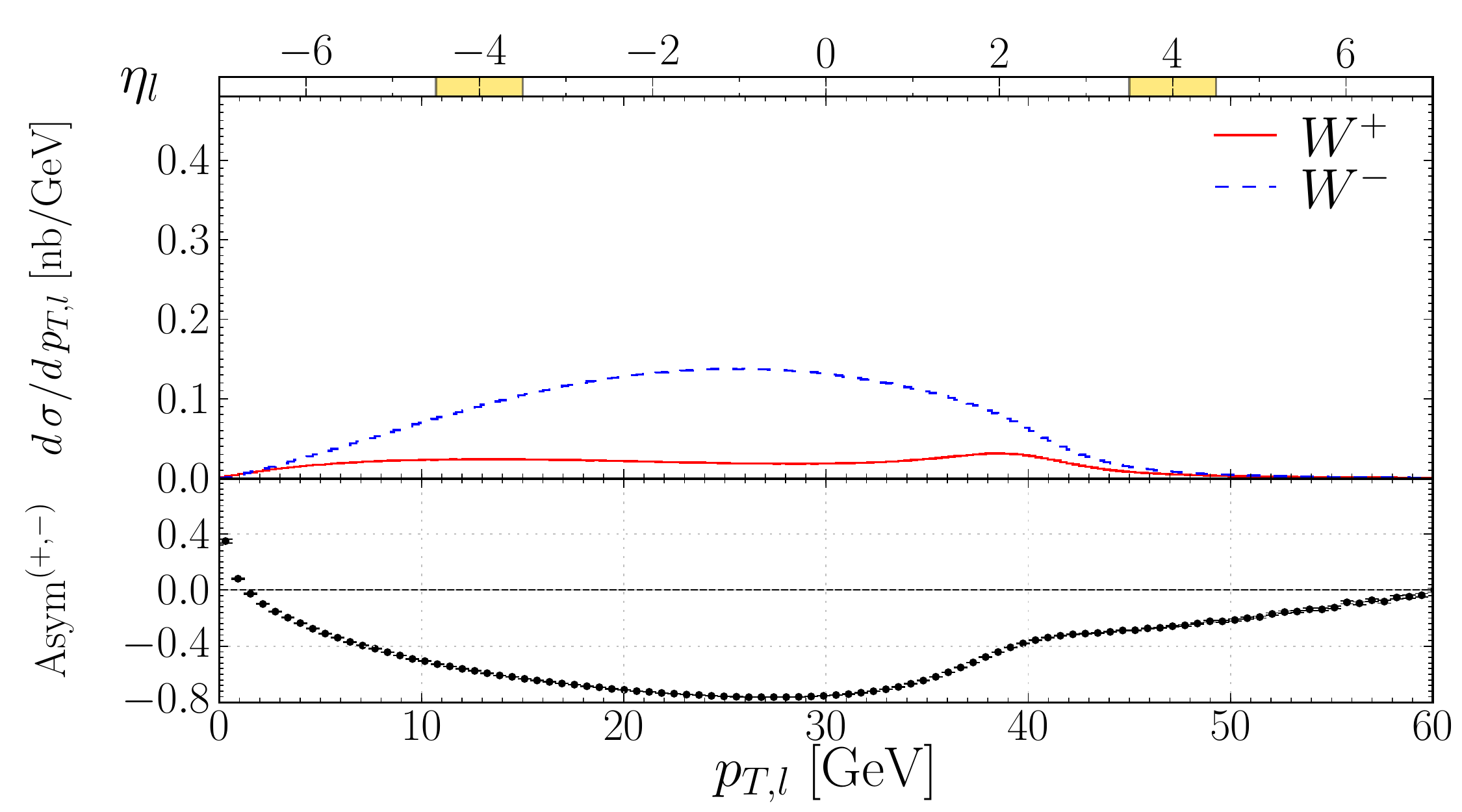}
    \caption[Charged lepton transverse momentum in bins of $\yW$ and in bins of $\etal$ in 
      $\dd$ collisions with $\sqrt{S_{n_1\,n_2}}=7\TeV$]
            {\figtxt{Charged lepton transverse momentum in bins of $\yW$ (left) 
                and in bins of $\etal$ (right) in $\dd$ collisions with $\sqrt{S_{n_1\,n_2}}=7\TeV$.}}
            \label{app_dd_pTl_in_yW_etal_bins}
  \end{center} 
\end{figure}

\begin{figure}[!h] 
  \begin{center}
    \includegraphics[width=0.495\tw]{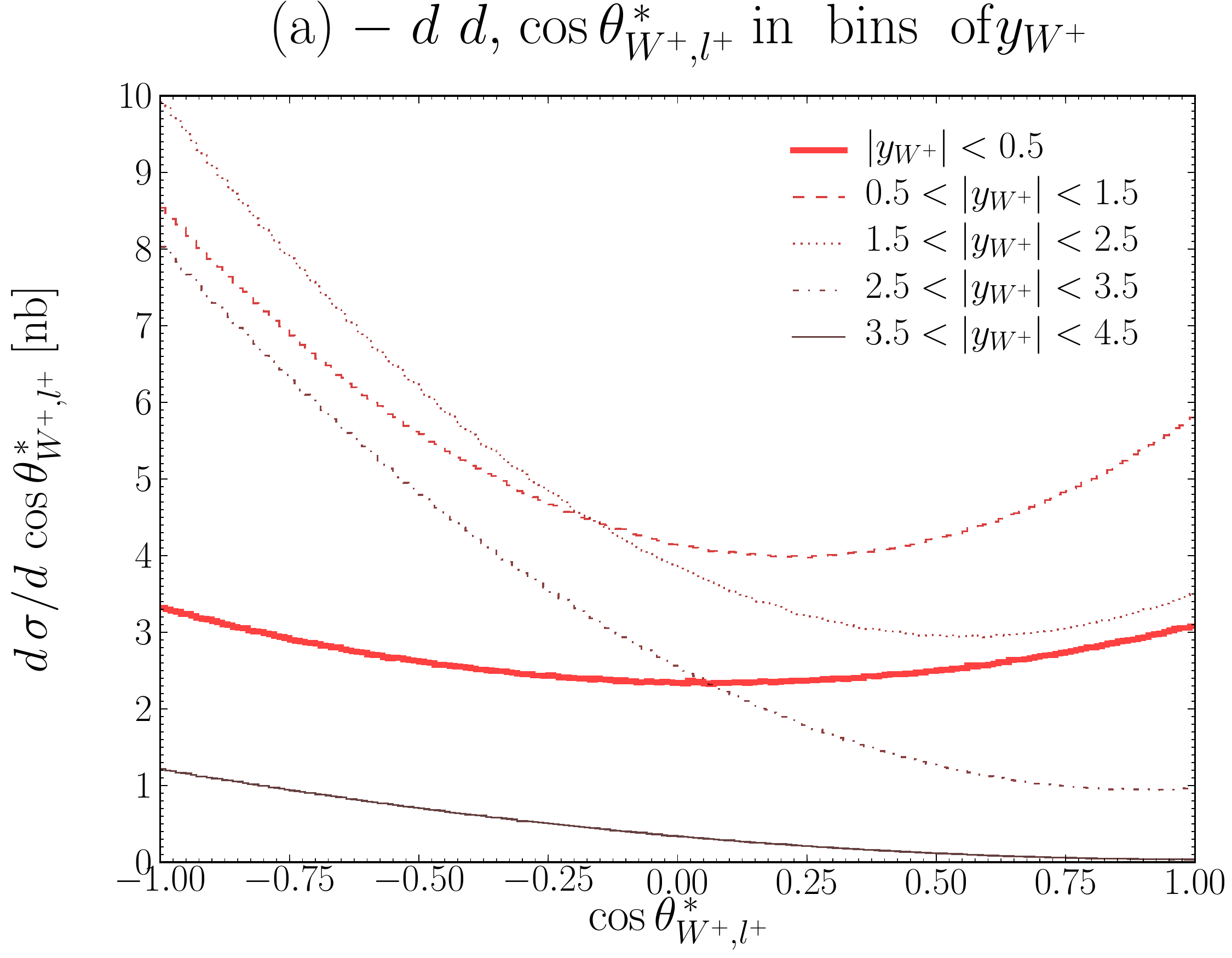}
    \hfill
    \includegraphics[width=0.495\tw]{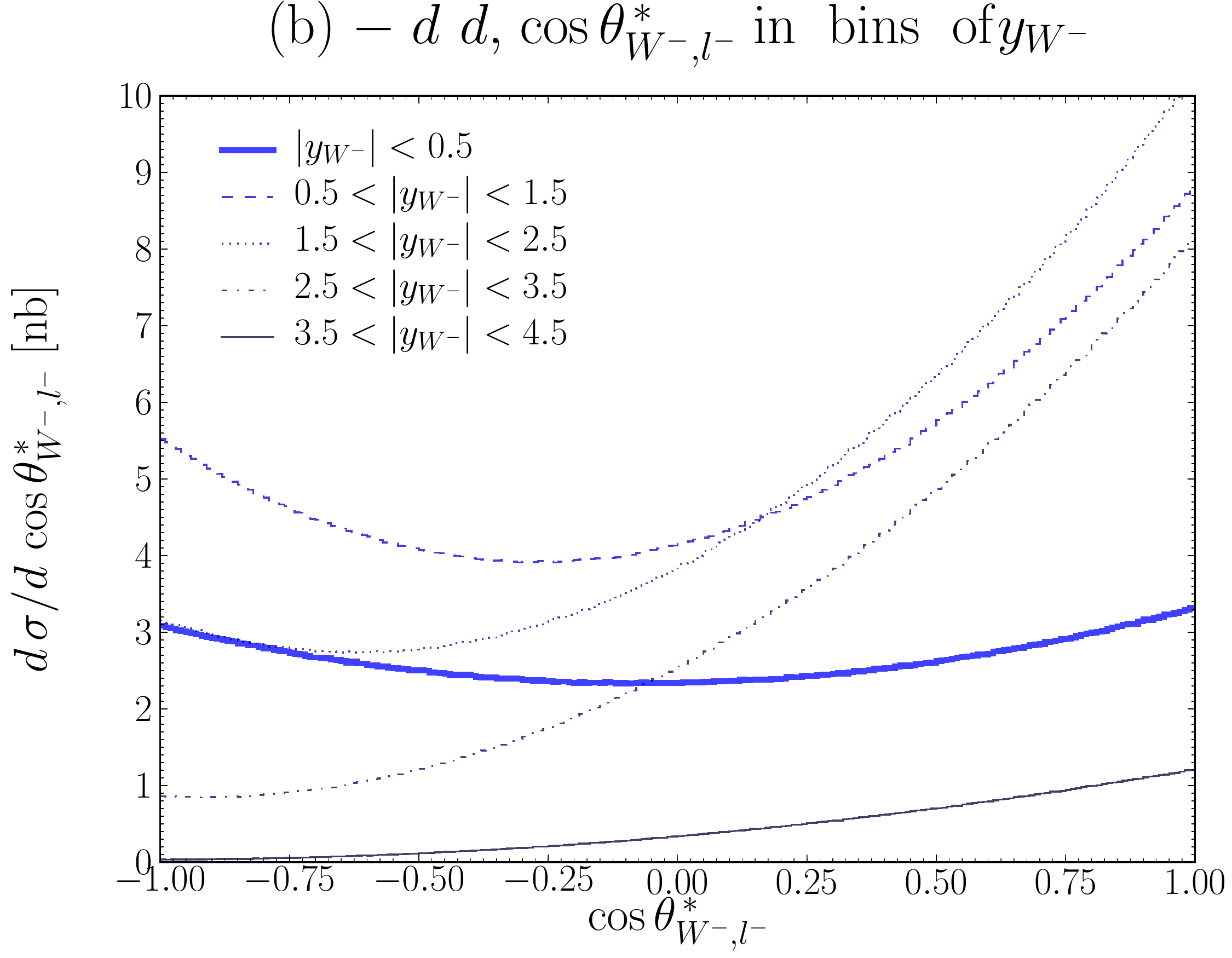}
    \vfill
    \includegraphics[width=0.495\tw]{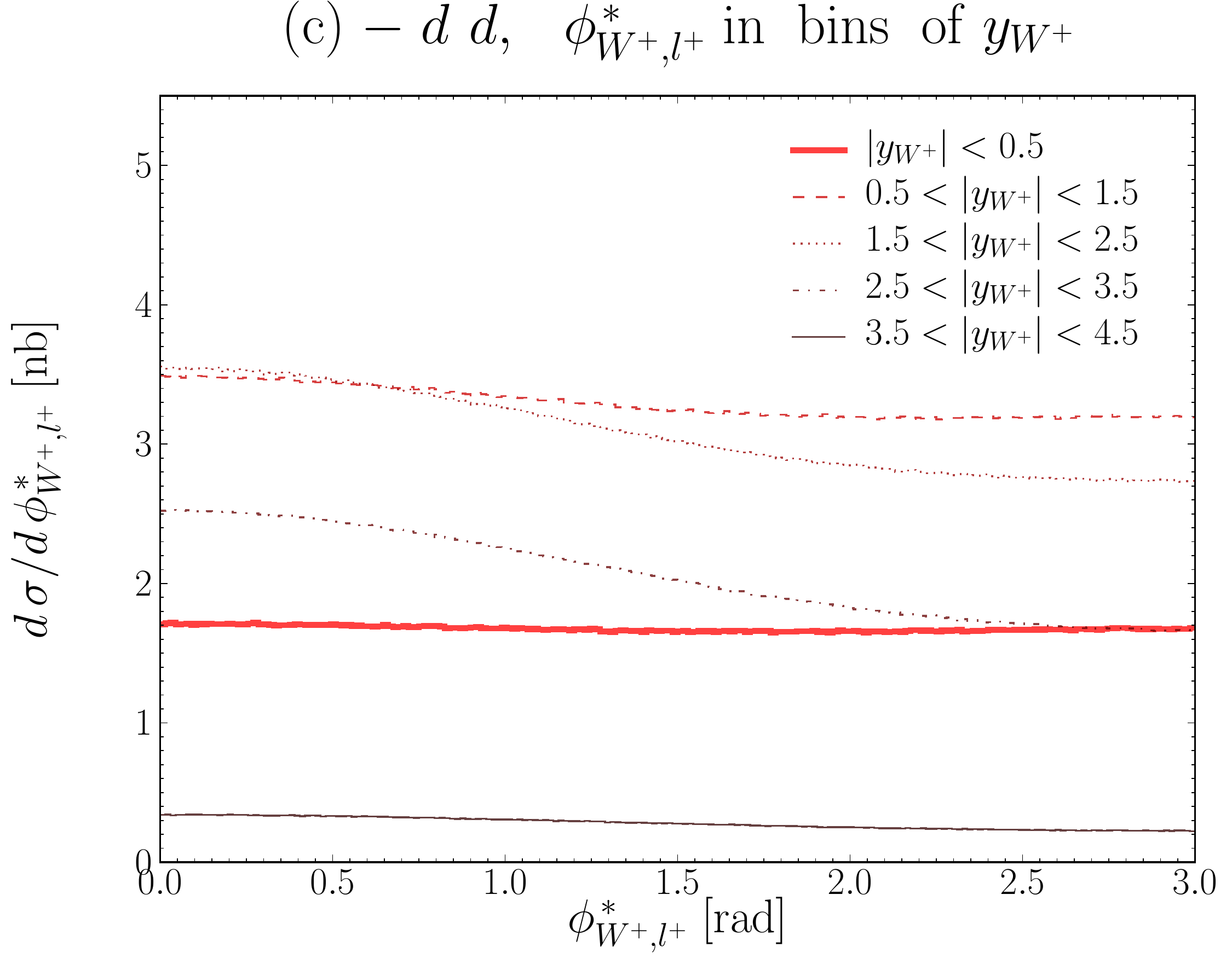}
    \hfill
    \includegraphics[width=0.495\tw]{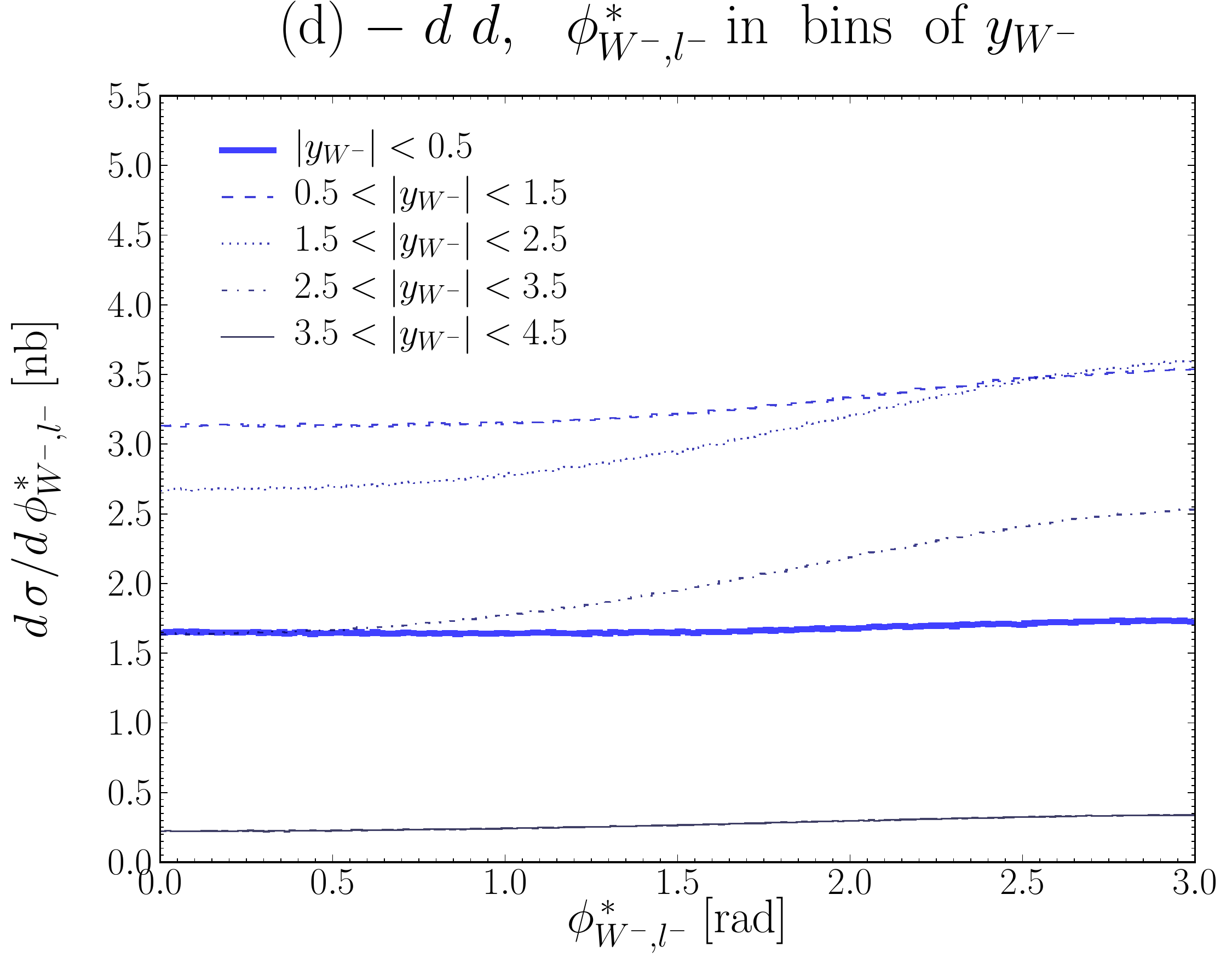}
    \caption[Distributions of $\costhetaWlwrf$ and $\phi_{W,l}^{\,\ast}$ in bins of $\yW$ for the positively and negatively 
      charged channel in $\dd$ collisions with $\sqrt{S_{n_1\,n_2}}=7\TeV$]
            {\figtxt{Distributions of $\costhetaWlwrf$ and $\phi_{W,l}^{\,\ast}$ in bins of $\yW$ for the positively (a,c) 
                and negatively (b,d) charged channel in $\dd$ collisions with 
                $\sqrt{S_{n_1\,n_2}}=7\TeV$.}}
            \label{app_dd_costhetaWlwrf_YWx}
  \end{center} 
\end{figure}

Figure~\ref{app_dd_etal_yW_in_yW_etal_bins} shows the $\etal$ distributions in bins of $\yW$ and
\textit{vice versa}. Note the charged leptons behaviour are the same than in $\pp$ collisions but
now the proportions of $\Wp$ and $\Wm$ are the same, which is particularly noticeable in the 
last two lines where the positive and negative $\etal$-pieces of the plots do not interpenetrate each other.
Figure~\ref{app_dd_pTW_in_yW_etal_bins} shows the transverse momentum of the $W$ in bins of 
$\yW$ and $\etal$. As can be seen, even if here the discrepancies are induced by the $s$, $c$ and $b$
flavours since they are weighted by the valence quarks (Eq.~(\ref{eq_sWp_sWm_dd})) it is in the 
forward rapidity region they are the most important.
Figure~\ref{app_dd_pTl_in_yW_etal_bins} displays the charge asymmetry of the charged leptons 
transverse momenta in bins of $\yW$ (left) and $\etal$ (right). 
Again we observe the growth of the charge asymmetry as $\yW$ increases and in general using a crude 
$\etal$ selection we witness a rather poor correlation with respect to the reference $\yW$ selection
results.
The $\costhetaWlwrf$ and $\phi_{W,l}^{\,\ast}$ distributions in bins of $\yW$ are presented
in Fig.~\ref{app_dd_costhetaWlwrf_YWx}.

\begin{figure}[!h] 
  \begin{center}
    \includegraphics[width=0.495\tw]{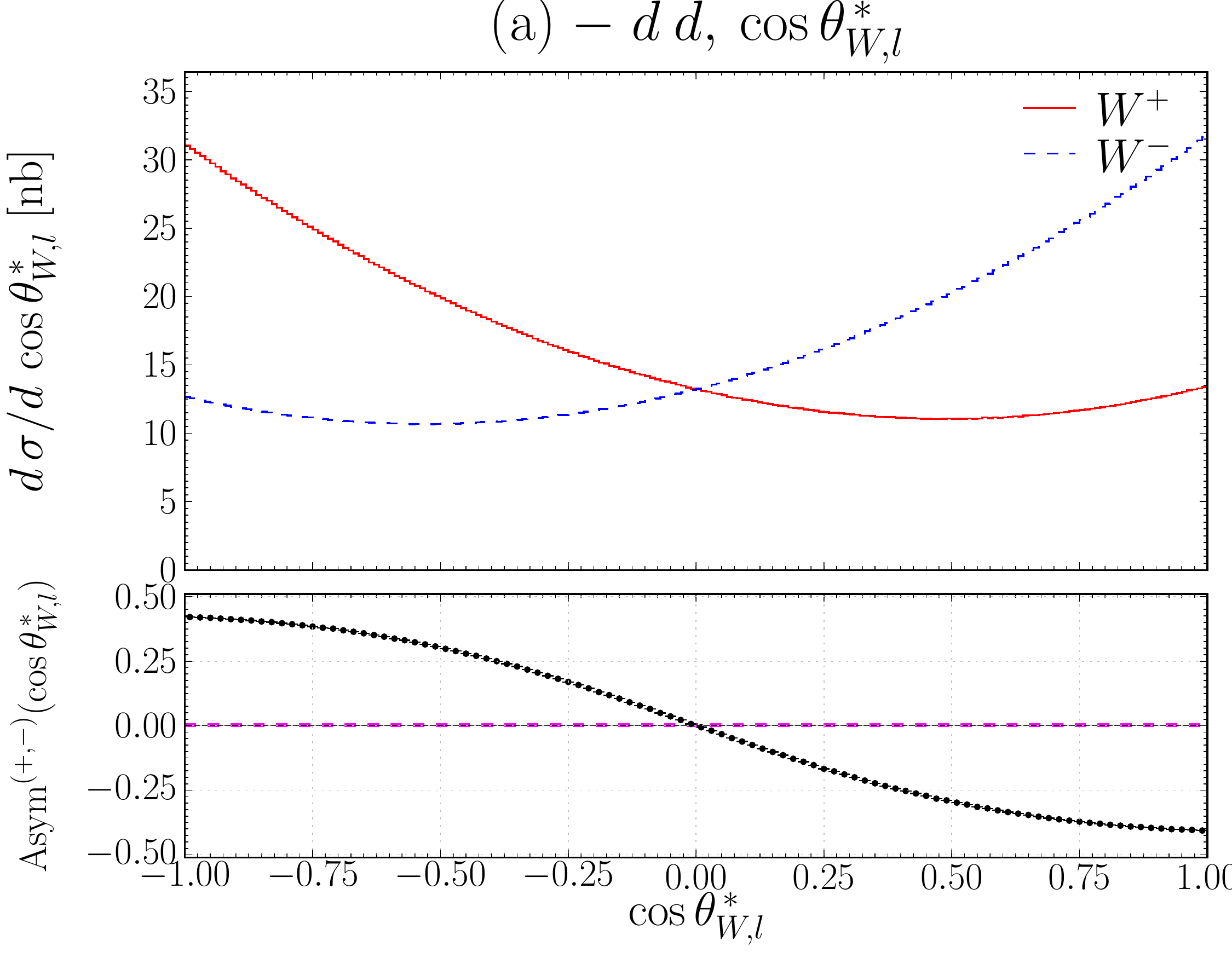}
    \hfill
    \includegraphics[width=0.495\tw]{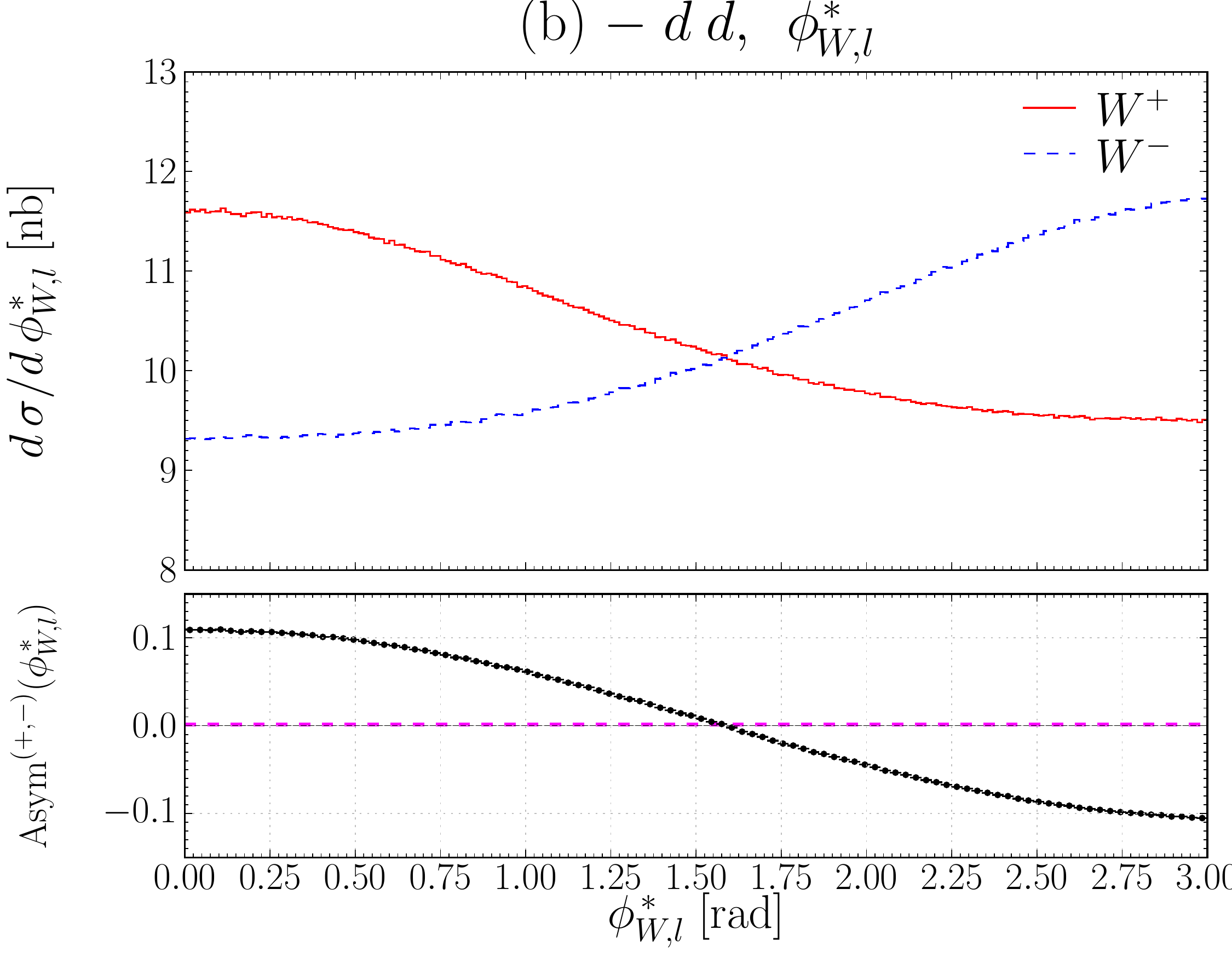}
    \caption[Distributions of $\costhetaWlwrf$ and $\phi_{W,l}^{\,\ast}$ in $\dd$ collisions 
      with $\sqrt{S_{n_1\,n_2}}=7\TeV$]
            {\figtxt{Distributions of $\costhetaWlwrf$ (a) and $\phi_{W,l}^{\,\ast}$ (b) in $\dd$ 
                collisions with $\sqrt{S_{n_1\,n_2}}=7\TeV$.}}
            \label{app_dd_cos_histos}
            
  \end{center} 
\end{figure}

\index{W boson@$W$ boson!Production in dd@Production in $\dd$ collisions!Detailed|)}

\clearpage
\subsection{Transverse momentum of the (anti-)quarks for $\BFW$ in Drell--Yan}\label{app_ss_kT}
\index{Quarks!Transverse momenta in single W production@Transverse momenta in single $W$ production|(}
This sub-section presents more explanations on the transverse momenta unbalance between the
valence quark and the sea anti-quark.

In most cases the valence quarks brings most of the energy in the collision.
These $q^\val\,{\bar{q}{}'}^{\sea}$ configurations with high $x_q$ for the valence quark and low 
$x_\qbp$ for the sea anti-quark are already well know.
Now the thing is that a parton with a small $x$ will have in general a higher transverse motion 
gained on its way to the collision by radiating gluons and photons (note also the low $x$ 
(anti-)quark can be as well the product of the radiation of a gluon via $g\to q'\qbp$).

Finally because of the CKM \index{Electroweak!CKM matrix elements} missing angles the other flavours such as 
$\sbar$, $\cbar$ and $\bbar$ have their transverse motion influenced as well.
To emphasise this aspect let us note that in the case of the production of $Z$ bosons since the 
previous flavours create $Z$ bosons only via $s\,\sbar,\;c\,\cbar,\;b\,\bbar$ annihilation we observe
$p_{T,q}=p_{T,\qbar}$ in those specific cases (we used for that the generator \WINHAC{} which as said
earlier can produce $Z$ bosons, cf. \S\,\ref{ss_tools_WINHAC}).

\begin{figure}[!h] 
  \begin{center}
    \includegraphics[width=0.495\tw]{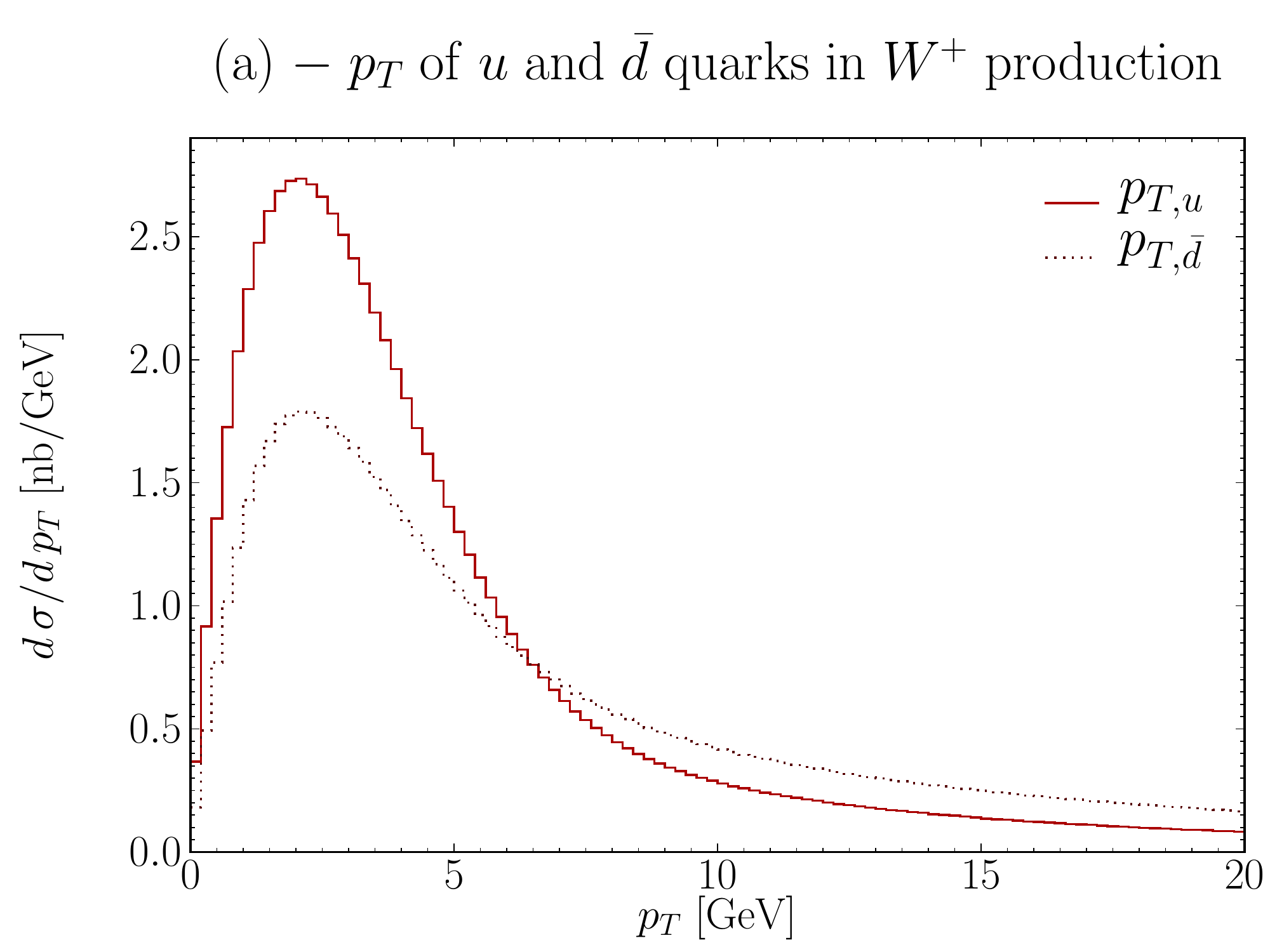}
    \hfill
    \includegraphics[width=0.495\tw]{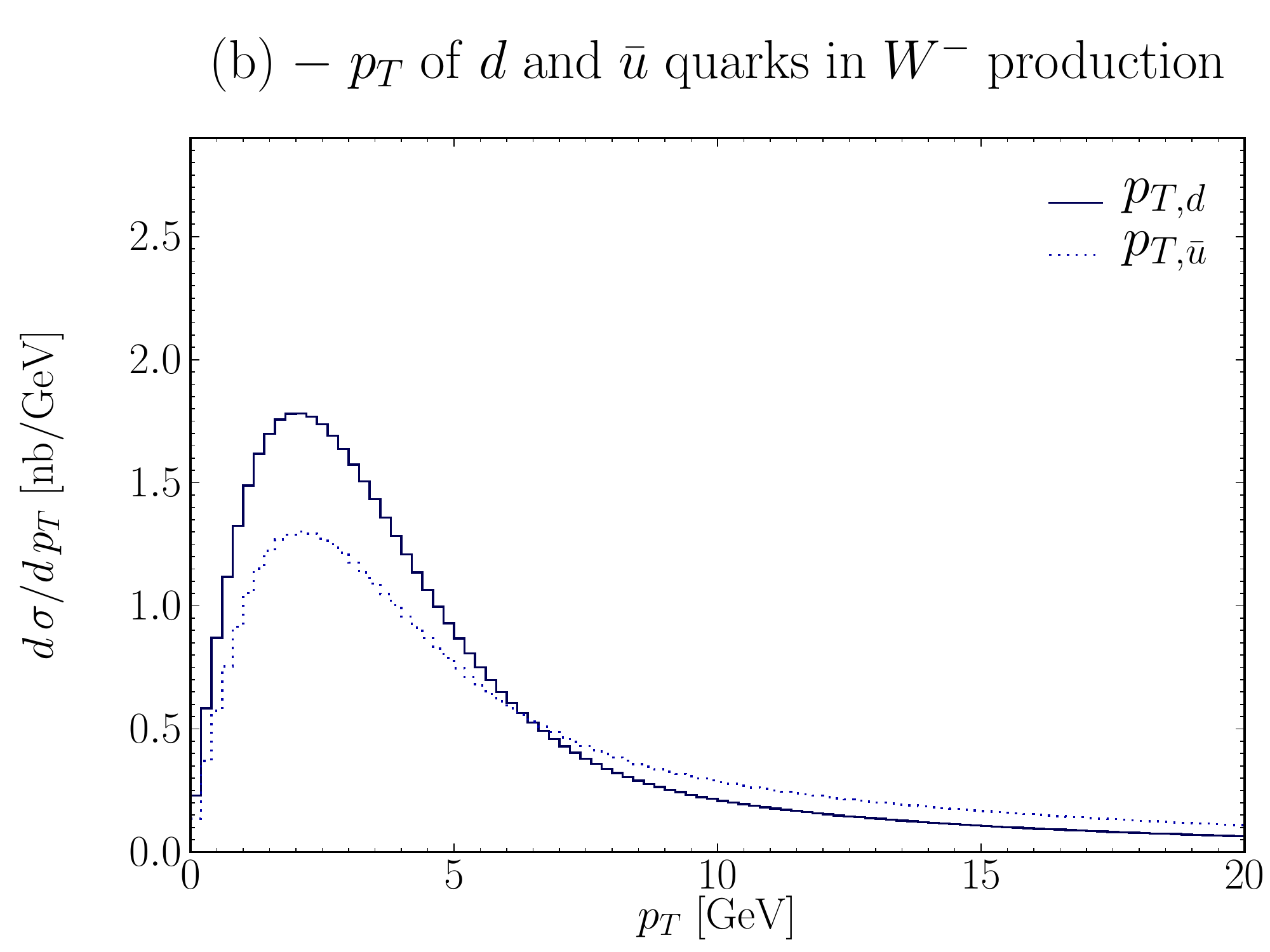}
    \caption[Transverse momenta of the $u$ and $\dbar$ quarks in $\Wp$ production along with the
      one of the $d$ and $\ubar$ quarks in $\Wm$ production]
            {\figtxt{Transverse momenta of the $u$ and $\dbar$ quarks in $\Wp$ production (a) 
                along with the one of the $d$ and $\ubar$ quarks in $\Wm$ production (b).}}
            \label{app_kT_udbar_dubar}
  \end{center} 
\end{figure}

The previous ideas are illustrated in Fig.~\ref{app_kT_udbar_dubar} with the histograms of the 
inclusive transverse momenta of the $u$ and $\dbar$ flavours in the $\Wp$ production (a) and the one 
of the $d$ and $\ubar$ flavours in the $\Wm$ production (b). In both cases the sea quark show larger 
probability to hold a high $\pT$ while the valence quark hold a low $\pT$.
These histograms were obtained, like the others, for a statistic of $200$ millions of weighted events.

The corresponding estimation of the average $\pT$ of the $u,\,\ubar$ and $d,\,\dbar$ quarks are\:
\begin{eqnarray}
\Mean{p_{T,u}} &\approx& 7.05\GeV,\qquad  \Mean{p_{T,\dbar}} \;\approx\; 12.58 \GeV 
\qquad\mbox{in $\Wp$ production}\\
\Mean{p_{T,d}} &\approx& 7.65\GeV,\qquad  \Mean{p_{T,\ubar}} \;\;\approx\; 11.77\GeV 
\qquad\mbox{in $\Wm$ production}
\end{eqnarray}

Anyhow in the previous values $u^\sea$ and $d^\sea$ values intervene, to factorise in a better
way the asymmetries induced by the cases where there is a valence quark involved in the collisions
we repeat the previous study putting to zero the sea contributions $u^\sea$ and $d^\sea$, in that 
context, the average $\pT$ of the quarks in $q^\val\,{\bar{q}{}'}^{\sea}$ ``valence'' contributions
for $\Wp$ production of
\begin{eqnarray}
\Mean{p_{T,u^\val}} \approx 4.99\GeV \longleftrightarrow
\left\{
\begin{array}{c@{\;\approx\;}l}
\Mean{p_{T,\dbar}} & 14.91\GeV \\
\Mean{p_{T,\sbar}} & 15.77\GeV \\
\Mean{p_{T,\bbar}} & 21.97\GeV
\end{array}\right.,
\end{eqnarray}
and for $d^\val\,{\bar{q}{}'}^{\sea}$ ``valence'' contributions in $\Wm$ production of
\begin{eqnarray}
\Mean{p_{T,d^\val}} \approx\; 5.03\GeV\longleftrightarrow
\left\{
\begin{array}{c@{\;\approx\;}l}
\Mean{p_{T,\ubar}} & 14.64\GeV \\
\Mean{p_{T,\cbar}} & 16.88\GeV
\end{array}\right.
\end{eqnarray}
In both cases we observe that the valence quark holds a very low transverse motion with respect to
the anti-quark. Note also that the transverse motion of the sea anti-quark increases with the mass 
of the quark.

Now these average $\pT$ of the quarks for purely sea contribution are\,:
\begin{eqnarray}
\Mean{p_{T,u^\sea}} \approx 9.87\GeV \longleftrightarrow
\left\{
\begin{array}{c@{\;\approx\;}l}
\Mean{p_{T,\dbar}} & 9.04\GeV \\
\Mean{p_{T,\sbar}} & 10.71\GeV  \\
\Mean{p_{T,\bbar}} & 16.19\GeV
\end{array}\right.
\end{eqnarray}
for the $\Wp$ production. Because $u^\sea<d^\sea$ for $10^{-4}<x$ we observe
$\Mean{p_{T,u^\sea}}>\Mean{p_{T,\dbar}}$ while it was the other way around for ``valence''
contributions. Also let us remark again that the higher the mass of a quark is the higher its
transverse motion is.
In the case of the $\Wm$ production we have
\begin{eqnarray}
\Mean{p_{T,d^\sea}} \approx 9.29\GeV \longleftrightarrow
\left\{
\begin{array}{c@{\;\approx\;}l}
\Mean{p_{T,\ubar}} & 9.97\GeV \\
\Mean{p_{T,\cbar}} & 12.69\GeV
\end{array}\right.
\end{eqnarray}
Let us remark that averaging the ``valence'' and purely sea contributions does not give 
exactly the values estimated when taking all contributions. 
This can be explained by the fact the proton's integrity cannot be exactly factorised into
``valence'' and purely sea contributions.
\index{Quarks!Transverse momenta in single W production@Transverse momenta in single $W$ production|)}

\subsection{Amplitude of the final state charge asymmetries in function of the energy in the 
collision}\label{ss_asmy_pTl_sqrtS}
\begin{figure}[!h] 
  \begin{center}
    \includegraphics[width=0.6\tw]{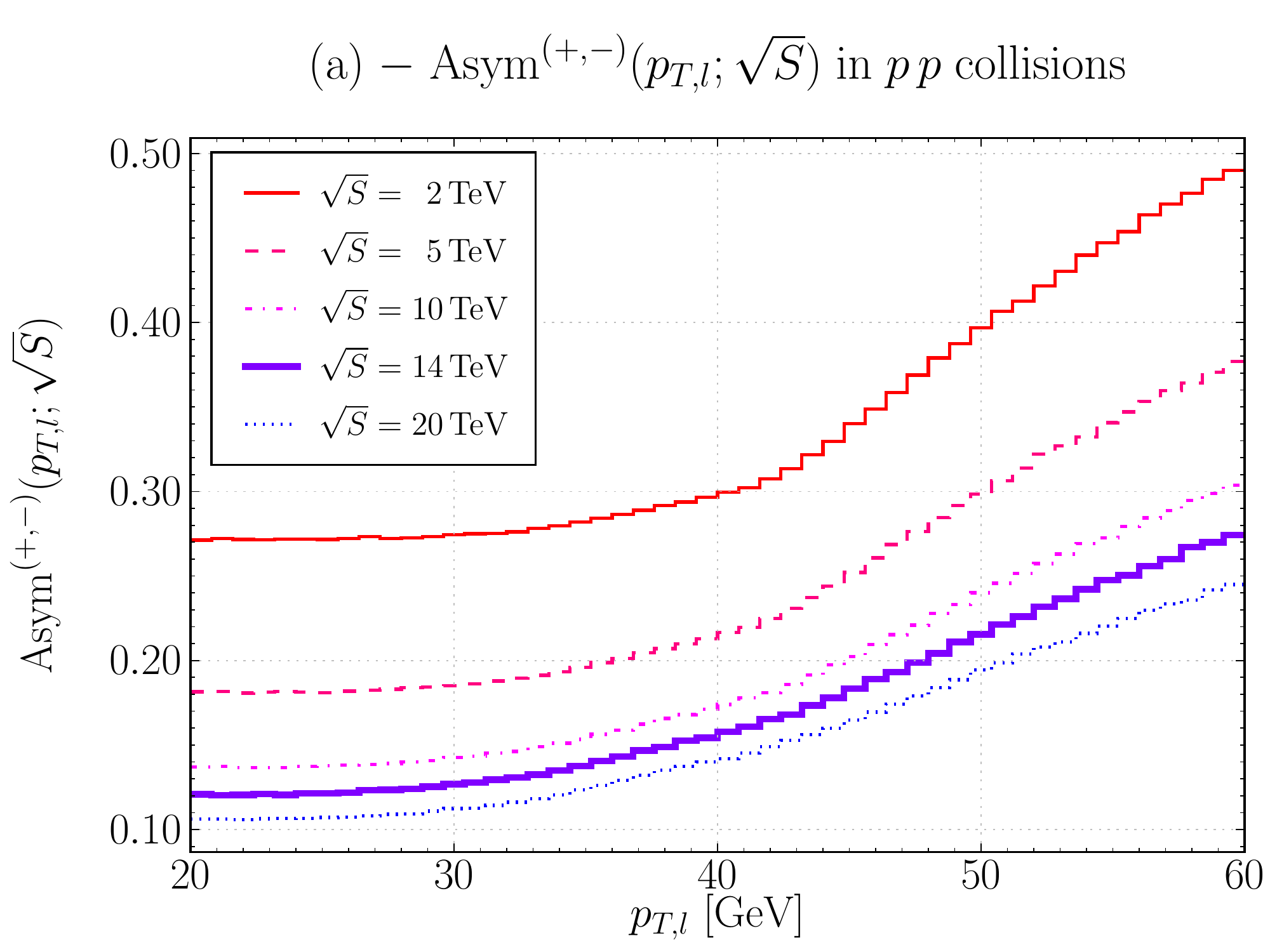}
    \vfill
    \includegraphics[width=0.6\tw]{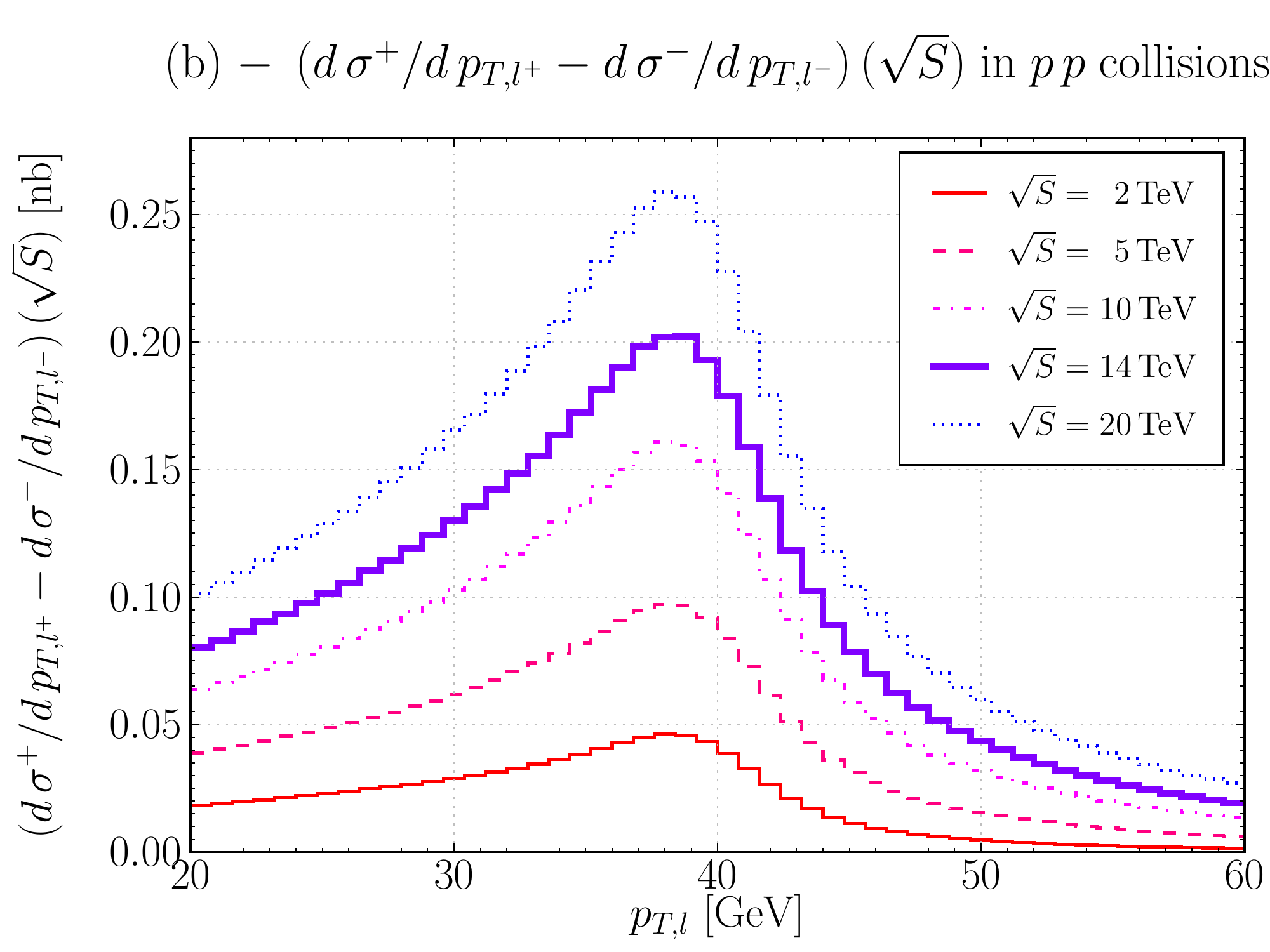}
    \caption[Charge asymmetry and bare difference of the transverse momenta 
                distributions of the charged leptons for several values of $\sqrt S$]
            {\figtxt{Charge asymmetry (a) and bare difference (b) of the transverse momenta 
                distributions of the charged leptons for several values of $\sqrt S$.}}
            \label{fig_asym_pTl_S_var}
  \end{center} 
\end{figure}

As observed in the study of deuteron--deuteron collisions the final state charge asymmetries
in the observable $\pTl$ is larger than for the case of proton--proton collisions.
The energy in the collision is the parameter responsible for those effects.
To see how this works we study $\Asym{\pTl}$ as a function of the energy $\sqrt S$ in the center of 
mass frame (Fig.~\ref{fig_asym_pTl_S_var}.(a)).
As $\sqrt S$ increases we witness the decrease of the charge asymmetry. Knowing the charge asymmetry
comes from ``valence'' contributions as the energy rise the charge symmetric purely sea contributions
increases and we end up with a lower and lower proportion of asymmetry due to ``valence'' terms.
Nonetheless, the asymmetry in the quarks $\pT$ for ``valence'' contributions 
$q^\val\,{\bar{q}{}'}^{\sea}$ increases with $\sqrt S$.
We find for example in the two extreme cases and the nominal $\sqrt S=14\TeV$ collision mode\,:
\begin{eqnarray}
\mbox{for } \sqrt S=2\TeV \quad&:&\quad \Mean{p_{T,u^\val}}\approx 4.73\GeV 
\;\;\longleftrightarrow\;\; \Mean{p_{T,\dbar}}\approx 8.43\GeV \\
\mbox{for } \sqrt S=14\TeV \quad&:&\quad \Mean{p_{T,u^\val}}\approx 4.99\GeV 
\;\;\longleftrightarrow\;\; \Mean{p_{T,\dbar}}\approx 15.77\GeV \\
\mbox{for } \sqrt S=20\TeV \quad&:&\quad \Mean{p_{T,u^\val}}\approx 4.99\GeV 
\;\;\longleftrightarrow\;\; \Mean{p_{T,\dbar}}\approx 15.90\GeV
\end{eqnarray}
To emphasise the increase of the asymmetry in absolute as $\sqrt S$ grows we look at the
bare difference of the positively and negatively charged histograms. This is shown in 
Fig.~\ref{fig_asym_pTl_S_var}.(b), the purely sea contributions giving exactly the same contributions
we are left with the difference between the ``valence'' terms only.

Anyhow, the important idea here is to understand that the energy in the collision which is lower in the
hypothetical case of $\dd$ collisions runs at the LHC forces us to deal with larger charge 
asymmetries with respect to the seen in $\pp$ collisions runs. This will explain the 
different behaviour of the systematic analysis prospect between $\dd$ and $\pp$ collisions in the next 
and final Chapter.

\end{subappendices}
\cleardoublepage
\ClearShipoutPicture

\chapter{Strategy for a reduction of the systematic errors on $\mbf{\MWp-\MWm}$} 
\label{chap_W_mass_asym}
\setlength{\epigraphwidth}{0.7\tw}
\epigraph{
``But then \dots'' I venture to remark, ``you are still far from the solution. \dots''\\
``I am very close to one,'' William said, ``but I don't know which.''\\
``Therefore you don't have a single answer to your questions ?''\\
``Adso, if I did I would teach theology in Paris.''\\
``In Paris do they always have the true answer ?''\\
``Never,'' William said, ``but they are very sure of their errors.''
}%
{\textit{The Name of the Rose} \\\textsc{Umberto Ecco}}

Here we finally present the dedicated strategies that were devised --based on the knowledge built 
throughout Chapter~\ref{chap_w_pheno_in_drell-yan}-- to prepare a measurement of $\MWp-\MWm$ at the 
LHC. As already mentioned
in the first Chapter while at the Tevatron the statistical and systematic errors where of the same 
order at the LHC the first one will drop to $\approx 5\MeV$ for just one year of harvesting data
at low luminosity. On the other hand the systematic error will be 
still important and we show how using the LHC and ATLAS detector capabilities we could eventually 
decrease those systematic errors and thus enhance by a factor $20$ the actual precision on 
$\MWp-\MWm$. The most important results below were presented in Ref.~\cite{Fayette:2008wt}.

The Chapter is divided as follow.
The first Section presents the strategies and observables developed in our collaboration which, 
according to us, should be more robust to address this particular measurement at the LHC. 
Following that a presentation of the general context for studying $W$ in Drell--Yan is
briefly reminded.
The second Section describes the principle of the analysis to extract the mass parameter for both 
the classic and the new suggested methods.
The third Section addresses the sources of theoretical and experimental systematic errors that should
be taken into account in the estimation of the error on this measurement.
Finally in the last Section our strategies robustness with respect to both theoretical and apparatus
unknowns addressed in our analysis are improved on the example of the simulation of one year of 
data taking at low luminosity.

\section{Measurement strategies}\label{s_measurement_method} 

\subsection{Event selection}\label{ss_event_selection}
As mentioned earlier the extraction of the $W$ and $Z$ bosons properties 
\index{Z boson@$Z$ boson!Properties extraction} is achieved by studying
their electronic and muonic decays, that is respectively $W\to l\,\nul$ and $Z\to \lp\lm$.
The choice for leptonic decays is motivated by the fact the di-jet background is of several orders 
of magnitude more important than the $W\to q\,\qbp$ and $Z\to q\,\qbar$ signals we would be 
interested in while the leptonic decays provide clean processes (\ie{} absence of QCD in the 
final state) with large cross sections.
The decays to the tau channel is not considered though because the mass of the charged $\tau$ is
such that it decays within $\approx 300\times 10^{-15}\,\mm{s}$, which leads to at least two neutrinos
to deal with for the detection of the final state.

The particular study of the $\Wp$ and $\Wm$ bosons is made by splitting the data collected for
the $W$ bosons into positively and negatively charged leptons in the final state to get the 
corresponding $\Wp$ and $\Wm$ data.
The main criteria used in hadronic collider to study $W$ bosons are to have a high energy 
charged lepton in the high precision detector volume of the detector associated with high 
transverse missing energy $\ETmiss$ that betrays the presence of the neutrino.
Also, to reduce the background, rejection are made using constraints on the recoil $\vec u$
defined as the sum of all momenta recoiling against the $W$ system
\begin{eqnarray}
\vec u &\equiv& - \vec p_W, \\
       &=&      - \vec p_l - {\slashiv{\vec p}}_\nul,
\end{eqnarray}
but again, due to the non possibility to measure the longitudinal component of the particles escaping
the beam-pipe, its transverse component $u_T$ is considered instead.

From the total number of events produced inclusively (Eq.~(\ref{eq_N_evt_sigma_lumi})) now must be 
taken into account the effects of the acceptance selection, the reconstruction efficiency and the
background processes whose kinematics can lure the trigger to recognise them as lepton pair 
decaying from single $W$.
Taking that into account, eventually the number of $W$ candidates $N_W^\mm{(candidates)}$ from 
Drell--Yan are
\begin{equation}
N_W^\mm{(candidates)} \equiv N_W^\mm{(acc.)}+N_W^\mm{(bckg.)},\label{eq_Ncand_evts}
\end{equation}
where $N_W^\mm{(acc.)}$ is the number of accepted $W$ events by the cuts 
and $N_W^\mm{(bckg.)}$ is the number of background events.
The expression of $N_W^\mm{(acc.)}$ reads
\begin{equation}
N_W^\mm{(acc.)} = L\,\sigma^\mm{(cut)}_W\,\es,\label{eq_Nacc_evts}
\end{equation}
where $L$ the integrated luminosity, $\sigma^\mm{(cut)}_W$ the cross section that the $W\to l\nul$ 
events fulfill the selection requirements and $\es$ is the efficiency reconstruction for the signal 
in the given fiducial region.
The number of background events can be expressed like
\begin{equation}
N_W^\mm{(bckg.)} = L\,\sigma^\mm{(cut)}_{W\,\mm{bckg.}}\,\es + B,
\end{equation}
where $\sigma^\mm{(cut)}_{W\,\mm{bckg.}}$ is the cut cross section of all the processes that displays
kinematics in their final state that mimic the features of the final state leptons from Drell--Yan $W$
and $B$ is the contribution term due to the electronic noise, the cavern background, \etc{}.
Below, more details on these criteria are given for the case of CDF, ATLAS and finally for our 
stand-alone analysis prospect.

\paragraph{Event selection in CDF.}
\index{CDF detector!Event selection@$W$ in Drell--Yan event selection|(}
In CDF II~\cite{Aaltonen:2007ps} a Drell--Yan $W$ is recognised by having at least one lepton 
candidate in the central region ($|\etal|\lesssim 1$) and by applying the following narrowing
selection on the kinematics\,:
\begin{itemize}
\item[-] The charged lepton $\pT$ must verify\,:\;$30\GeV<\pTl<55\GeV$.
\item[-] The neutrino $\pT$, deduced from the measurement of $\ETmiss$, must 
  verify\,:\;$30\GeV<\slashiv{p}_{T,\nul}<55\GeV$.
\item[-] The leptons transverse mass \index{Transverse mass of the lepton pair!In CDF II @In CDF II $\MW$ analysis} 
must verify\,:\;$60\GeV<\mTlnu<100\GeV$
\item[-] The transverse recoil must verify\,:\;$u_T<15\GeV$.
\end{itemize}
Additional cuts are made to reject background from $Z$ leptonic decays.
Eventually the measurement of the $W$ properties in Ref.~\cite{Aaltonen:2007ps} based on these
selection criteria provided $\approx 50,000$ candidates events for $W\to\mu\,\numu$ 
($L\approx 191\,\mm{pb}^{-1}$) and $\approx 64,000$ candidates events for $W\to e\,\nue$ 
($L\approx 220\,\mm{pb}^{-1}$).
\index{CDF detector!Event selection@$W$ in Drell--Yan event selection|)}

\paragraph{Expected event selection in ATLAS.}
In ATLAS, although no data is there yet to confirm which criteria would fit the best the LHC context
the adopted requirements taken from Ref.~\cite{ATL-COM-PHYS-2008-243} are\,:
\begin{itemize}
\item[-] Have a charged muon or isolated electron with $\pTl>20\GeV$ in the precision physics 
  acceptance volume ($|\etal|<2.5$).
\item[-] Have a missing transverse energy of $\ETmiss>20\GeV$.
\item[-] No jets in the event with $\pT>30\GeV$ to reject backgrounds from jets and $t\tbar$ events.
\item[-] A recoil $u_T<50\GeV$ to avoid to much the influence of the $\pTW$ smearing on the leptons.
observables
\end{itemize}
In ATLAS, these constraints should be such that from the $\approx 350$ millions of $W$~bosons produced
for an integrated luminosity of $L=10\,\mm{fb}^{-1}$ only $60$ millions should be accepted 
(no matter the charge or the decay channel).\index{Luminosity!At the LHC (expected)!Integrated} 

\paragraph{Event selection for the present study.}
The $W$ boson production events used in our studies are selected by requiring 
that the simulated decaying charged electron or muon satisfy the following requirements\,: 
\begin{equation}
\pTl>20\GeV  \quad {\rm and} \quad |\etal|<2.5.
\label{pTl-etal-cuts}
\end{equation} 
We assume that by using a suitable $\ETmiss$ cut the impact of the uncertainty in the background 
contribution on the measurement of the charge asymmetry of the $W$ boson mass at the LHC 
can be made negligible. This allows us to skip the generation and simulation of the background 
event samples for our studies. 
It has to be stressed that the above assumption  is weaker for the measurement of 
the charge asymmetry than for the measurement of the average $W$ boson mass
\cite{Buge:2006dv, Besson:2008zs} because, to a good approximation, only the difference of the background
for the positive and negative lepton samples will bias the measurement.
The presentation of the results of our studies is largely simplified by noticing that they are 
insensitive to the presence of the $\ETmiss$ cut in the signal samples.
Therefore, the results are based on selection of events purely on the basis of the reconstructed 
charged lepton kinematics variables as shown in Eq.~(\ref{pTl-etal-cuts}).
Studies have shown that these results will remain valid whatever $\ETmiss$
cut will be used at the LHC to diminish the impact of the background contamination at the required 
level of precision. To be more precise, in our simple model this $\ETmiss$ cut was 
emulated by a direct $\slashiv{p}_{T,\nul}$ cut.
Nonetheless, in the specific purpose of a cut, this approximation is believed to be fully
justified.

The detector response is based on the ATLAS tracker which, for reminder, was emulated using
Gaussian resolutions for $\pT$ and $\cotan\theta$ of the charged electron and muons according
to Eqs.~(\ref{eq_rho_smearing}--\ref{eq_theta_smearing}).

Concerning the statistic, in absence of background and imperfect reconstruction
efficiencies, we simply have
\begin{equation}
N_W^\mm{(acc.)} = L\,\sigma^\mm{(cut)}_W.\label{eq_Nacc_evts_our_model}
\end{equation}
The cut cross section corresponding to the requirements of Eq.~(\ref{pTl-etal-cuts}) 
and for the expected LHC energies for $\pp$ (Eq.~(\ref{eq_pp_Ecm_lumi})) 
and $\dd$ (Eq.~(\ref{eq_dd_Ecm_lumi})) collisions are compiled in Table~\ref{table_xtot_2}.
Table~\ref{table_xtot_3} gives the correspondence in terms of number of events obtained by
using the respective integrated luminosities for $\pp$ and $\dd$ collisions (%
Eqs.~(\ref{eq_pp_Ecm_lumi}--\ref{eq_dd_Ecm_lumi})) in Eq.~(\ref{eq_Nacc_evts_our_model}).
\index{Luminosity!At the LHC (expected)!Integrated}

\begin{table}[]
\begin{center}
\renewcommand\arraystretch{1.5}
\begin{tabular}{c|ccc|ccc}
  \hline
  Collisions  & $\sigma^\mm{(incl.)}_\Wp$ [nb] & $\sigma^\mm{(incl.)}_\Wm$ [nb] 
                                            & $\sigma^\mm{(incl.)}_W$ [nb] 
            & $\sigma^{(\cut)}_\Wp$ [nb] & $\sigma^{(\cut)}_\Wm$ [nb] & $\sigma^{(\cut)}_W$ [nb]  \\
  \hline\hline
  $\pp$     & $19.81$ & $14.72$ & $34.53$ & $10.95$ &  $7.92$   & $18.87$   \\
  $\dd$     & $32.52$ & $32.44$ & $64.96$ & $22.14$ & $17.75$ & $39.89$  \\
  \hline
\end{tabular}
\renewcommand\arraystretch{1.45}
\caption[Total and cut hadronic cross section $\sigma_\Wp$ and $\sigma_\Wm$ for $\ppbar$ and $\pp$ 
  with $\sqrt S=14\TeV$ and for $\dd$ with $\sqrt S=7\TeV$ in the nucleon--nucleon center of mass 
  energy]
        {\figtxt{Inclusive ($\sigma^\mm{(incl.)}$) and 
            cut ($\sigma^\mm{(cut)}$) cross sections $\sigma_\Wp$, $\sigma_\Wm$ and $\sigma_W$
            for $\pp$ ($\sqrt S=14\TeV$) and for $\dd$ ($\sqrt S_{n_1n_2}=7\TeV$ 
            in the $n_1\,n_2$ nucleon--nucleon center of mass energy).}}
\label{table_xtot_2}
\end{center}
\end{table}

\begin{table}[]
\begin{center}
\renewcommand\arraystretch{1.5}
\begin{tabular}{c|ccc|ccc}
  \hline
  Collisions  & $N_\Wp^{\mm{(incl.)}}$ & $N_\Wm^{\mm{(incl.)}}$ & $N_W^{\mm{(incl.)}}$
            & $N_\Wp^{\mm{(acc.)}} $ & $N_\Wm^{\mm{(acc.)}} $ & $N_W^{\mm{(acc.)}}$ \\
  \hline\hline
  $\pp$     & $198.1$ & $147.2$ & $345.3$ & $109.5 $ & $79.2$ & $188.7$ \\
  $\dd$     &  $81.3$ & $ 81.1$ & $162.4$ &  $55.3 $ & $44.4$ &  $99.7$ \\
  \hline
\end{tabular}
\renewcommand\arraystretch{1.45}
\caption[Produced inclusive and accepted millions of events (for $\pTl>20\GeV$ and $|\etal|,2.5$) for 
  $\Wp$, $\Wm$ and $W$ for an integrated luminosity of$L=10\,\mm{fb}^{-1}$ ]
        {\figtxt{Produced inclusive and accepted millions ($\times 10^6$) of events 
            (for $\pTl>20\GeV$ and $|\etal|,2.5$) for $\Wp$, $\Wm$ and $W$ for one year of 
            data collection at low luminosity 
            ($L=10\,\mm{fb}^{-1}$ for $\pp$ and $L=2.5\,\mm{fb}^{-1}$ for $\dd$).}}
\label{table_xtot_3}\index{Luminosity!At the LHC (expected)!Integrated}
\end{center}
\end{table}

\subsection{Observables}\label{ss_observables}
The values of the $\Wp$ and $\Wm$ boson masses can be unfolded
from the measured lepton charge distributions $\pTl$, $\slashiv{p}_{T,\nul}$ and
from the transverse mass $\mTlnu$\index{Transverse mass of the lepton pair}. 
We discuss only the methods based on the measurement of $\pTl$. 
These methods are almost insensitive to the detector and modeling 
biases in the reconstructed values of the neutrino transverse momentum $\pTnu$.
We are aware that, for the measurement of the average mass of the $W$ boson, this merit is 
outbalanced by the drawback of their  large sensitivity to the precise understanding 
of the distribution of the $W$ boson transverse momentum $\pTW$ (Fig.~\ref{fig_pTl_hadr_look}).
However, for the measurement of the charge asymmetry of the masses this 
is no longer the case because QCD radiation --which drives  
the shape of the $\pTW$ distribution-- is independent of the charge of the produced $W$ boson.
In our view the $\pTl$ based methods will be superior with respect to 
the $\mTlnu$ based ones, in particular for the first measurements of the $W$ mass charge 
asymmetries at the LHC. 

\subsubsection{Commonly used observables}
\index{Charged lepton@Charged lepton from $W$ decay!Transverse momentum}
In the context of a $\pTl$ based method for the extraction of $\MWp-\MWm$, the most natural method 
is to analyse separately the $\lp$ and $\lm$ event samples and determine independently the masses 
of the $\Wp$ and $\Wm$ bosons.
This method, that will be investigated, is based on independent measurements of the 
$\FlatDsigmaDobs{\pTlp}$ and $\FlatDsigmaDobs{\pTlm}$ distributions. 
It will be called hereafter the classic method and actually corresponds to what physicists did 
at with the CDF detector to provide measurements of $\MWp-\MWm$.

\subsubsection{The charge asymmetry}
\index{Charge asymmetry!Used for extraction of MW@used for the extraction of $\MW$|(}
A new method proposed and evaluated in our collaboration is based on the 
measurement of the charge asymmetry (Eq.~(\ref{eq_def_charge_asym})) of the $\pTl$ distribution.
This method will be called hereafter the charge asymmetry method. 
The distribution of $\FlatAsym\pTl$ is, by definition,
robust with respect to those of systematic measurement effects and those of model dependent 
effects that are independent of the lepton charge. The acceptance, and the lepton selection 
efficiency corrections for this observable will, in the leading order approximation, 
reflect only their lepton charge dependent asymmetries.
In addition, the $\FlatAsym\pTl$ observable  is expected to be robust with respect 
to the modeling uncertainty of the QCD and QED radiation processes.

If extrapolated from the experience gained at the Tevatron (cf. Appendix~\ref{cdf_tracker}), 
the precision of {the charge asymmetry method}
will be limited by the understanding of relative biases in the reconstructed
transverse momenta for positively and negatively charged particles (cf. \S\,\ref{s_weak_modes}).
These biases, contrary to the lepton charge averaged  biases, cannot be controlled 
using the $J/\psi$, $\Upsilon$ and $Z$ ``standard candles'' \index{Z boson@$Z$ boson!To calibrate the
leptons energy scale} that helps only to correct for the absolute
energy scale. This problem leads us to the third and final observable for our study\,:\:%
the double charge asymmetry.
\index{Charge asymmetry!Used for extraction of MW@used for the extraction of $\MW$|)}

\subsubsection{The double charge asymmetry}
\index{Double charge asymmetry!Definition|(}
The double charge asymmetry is defined like
\begin{equation}
  \DAsym{\rhoTl} \;\equiv\; \frac{1}{2} \left[ 
    \mathrm{Asym}^{(+,-)}_{\vec B =  B\,\vec e_z}\left(\rhoTl\right) + 
    \mathrm{Asym}^{(+,-)}_{\vec B = -B\,\vec e_z}\left(\rhoTl\right) 
    \right],
\label{eq_def_dble_charge_asym}
\end{equation}
where the variable $\rhoTl$ defined like $\rhoTl\equiv1/\pTl$ (cf.~\S\,\ref{notations_conventions}),
represents the radius of the track curvature at the $W$ boson production vertex in the plane 
$r-\phi$ perpendicular to the beam collision axis and $B$ is the strength of the 
solenoidal magnetic field bathing the inner tracker.
The choice on $\rhoTl$ was cast to follow the notation adopted in our former work \cite{Krasny:2007cy}.
The measurement method using the $\DAsym{\rhoTl}$ distribution will be called hereafter 
the double charge asymmetry method.
\index{Double charge asymmetry!Definition|)}

The $\DAsym{\rhoTl}$ distribution is expected to be robust with respect to the charge dependent 
track measurement biases if the following two conditions are fulfilled\,:\;(1) the inversion of 
the $z$-component of the magnetic field in the tracker volume can be controlled to a requisite 
precision and (2) the $\vec E \times \vec B$ Lorentz drift relative corrections to the 
reconstructed hit positions for the two magnetic field configurations, in the silicon tracker could 
be determined to a requisite precision.
Discussing these two conditions in details is beyond the scope of this work. Nonetheless
the second point, in view of the ATLAS detector design was considered qualitatively in our work%
~\cite{DiscussionPawelBruckmanDeRestrom} with regard to the specificity of the pixel, SCT and TRT 
barrel modules.

\begin{figure}[!ht] 
  \begin{center}
    \includegraphics[width=0.9\tw]{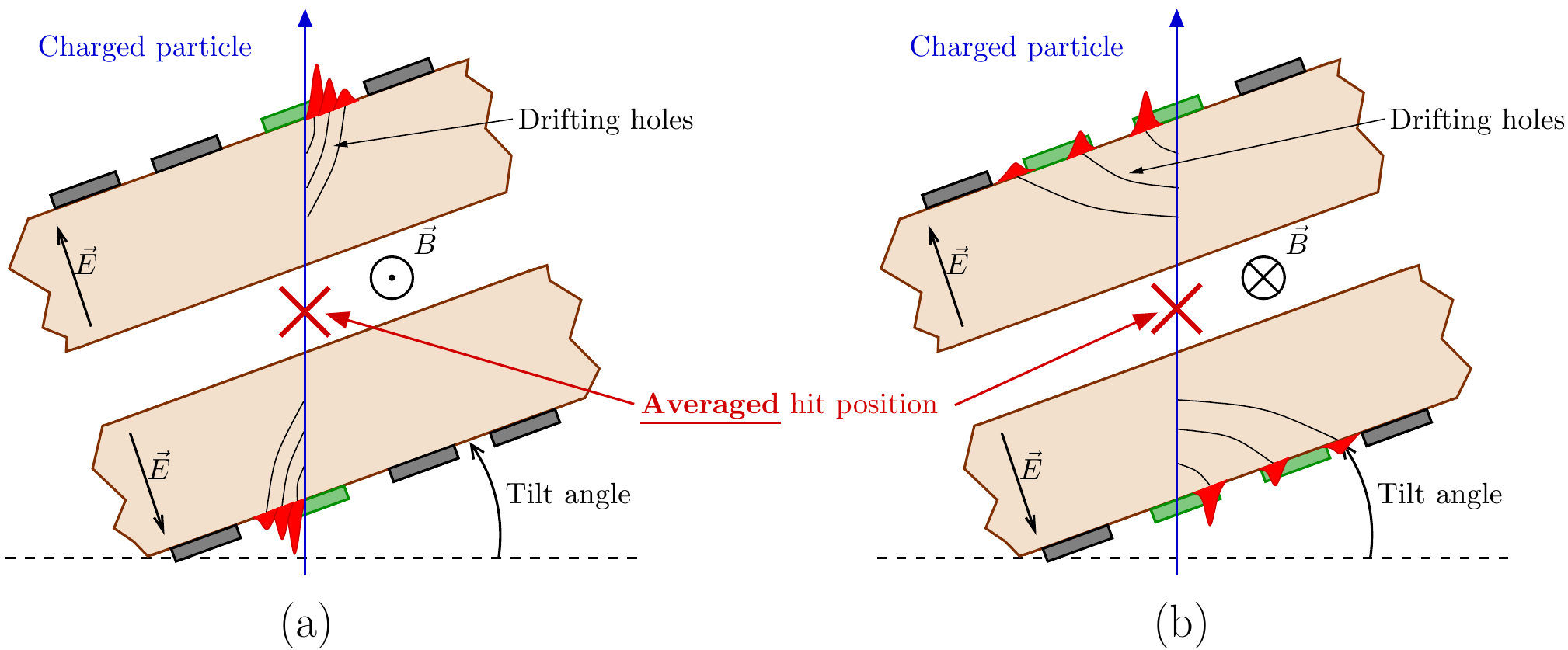}
    \caption[Schematic representation of the reconstructed hit in a barrel module of the SCT 
      detector with two opposite magnetic field configurations]
            {\figtxt{Schematic representation of the reconstructed hit in a barrel module of the SCT 
                detector for the nominal (a) and inverted (b) configurations of the 
                solenoidal magnetic field.}}
            \label{fig_dasym_lorentz_angle}
  \end{center} 
\end{figure}

First, the inversion of the magnetic field should \textit{a priori} have no influence on the TRT 
intrinsic capabilities since the measurement of the hit is based solely on the time it takes for 
electrons to drift to the center wire as shown in Fig.~\ref{fig_atlas_trt}.(b).
Then, reverting the magnetic field should reverse the drift paths of the electron but not affect the
drift time.
Concerning the SCT barrel modules the tilt angle for the invert magnetic field configuration is now
worsening the nominal hit resolution as depicted in Fig.~\ref{fig_dasym_lorentz_angle}.
In the best case scenario the Lorentz drift angle made by the drifting electrons is well under 
control and the resolution of the SCT for the inverted magnetic field should be well accounted.
If that would not be the case then the resolution of the SCT modules will change. Nonetheless
the double layer feature of silicon wafers, by providing an average offline hit position, should 
help to tackle such problems.
This is no longer the case for the pixel detector whose modules posses only one layer. 
As seen previously the pixel hits are important as they provide data on the particles before they 
start to loose energy to the apparatus and also because of the important lever arm they have for the
track fitting reconstruction procedure. 
Here, to each event, looking at the origins of the other particles tracks should help for a possible
correction of the production vertex of the $W$ leptonic decay.

\begin{figure}[!ht] 
  \begin{center}
    \includegraphics[width=0.495\tw]{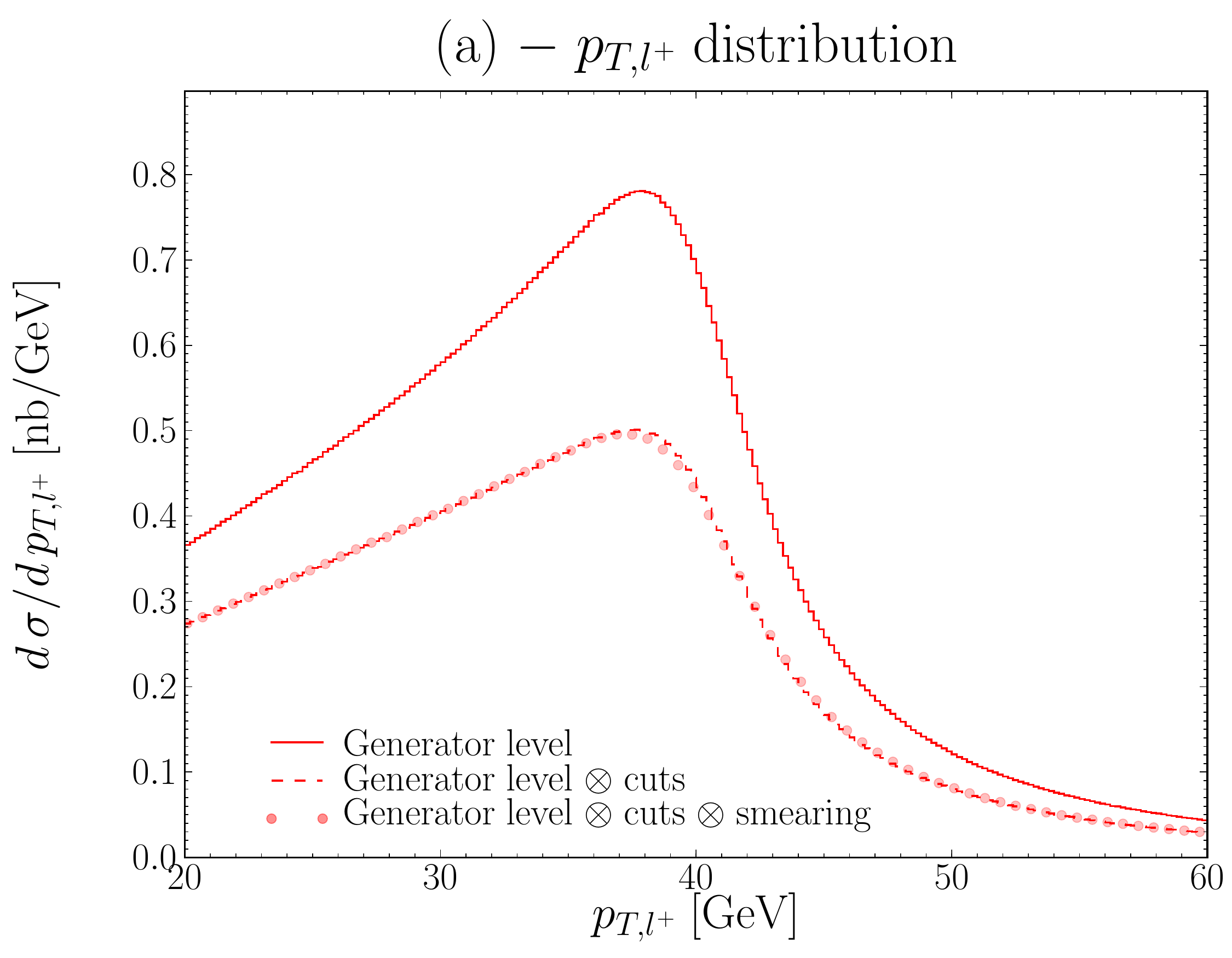}
    \hfill
    \includegraphics[width=0.495\tw]{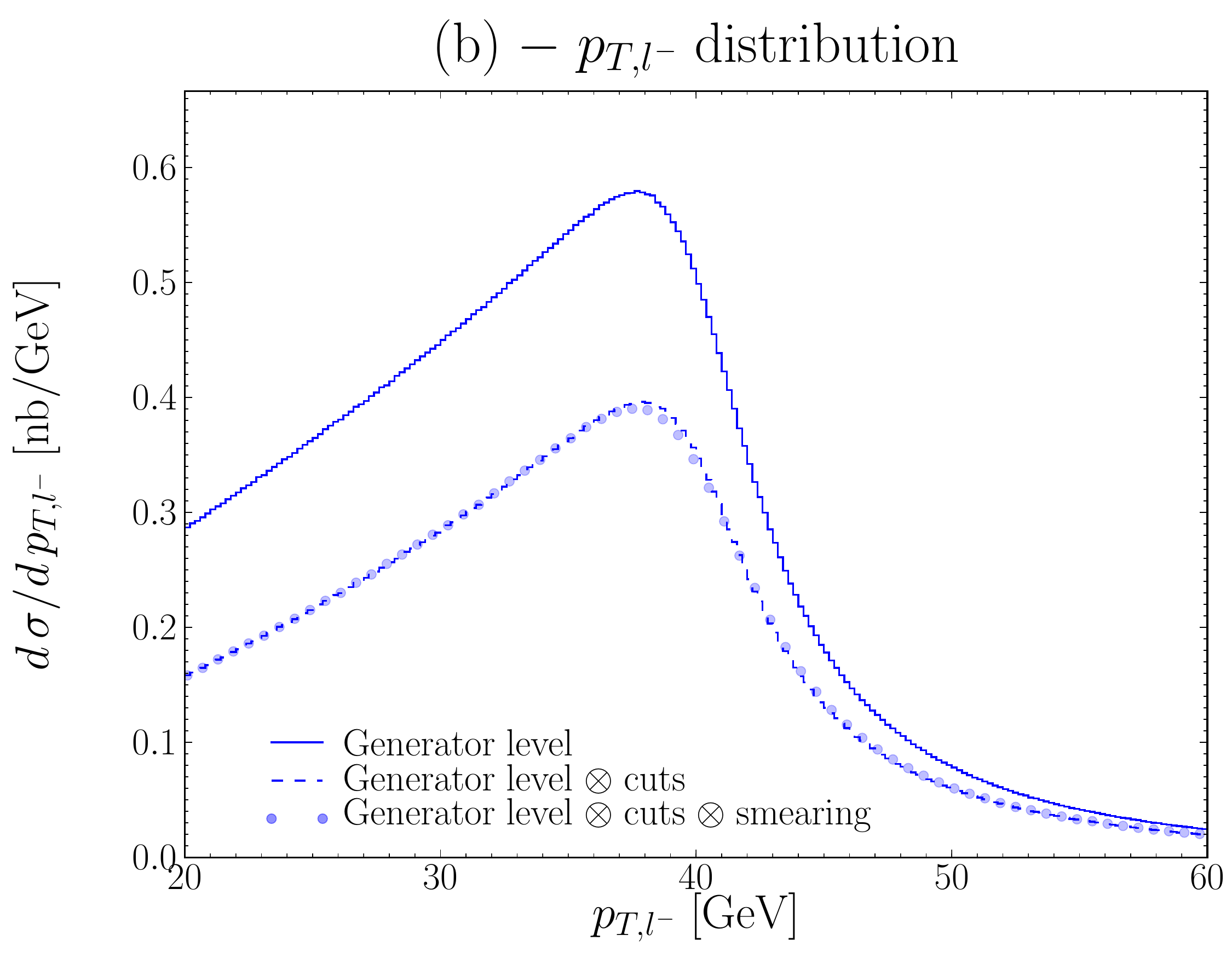}
    \vfill
    \includegraphics[width=1.\tw]{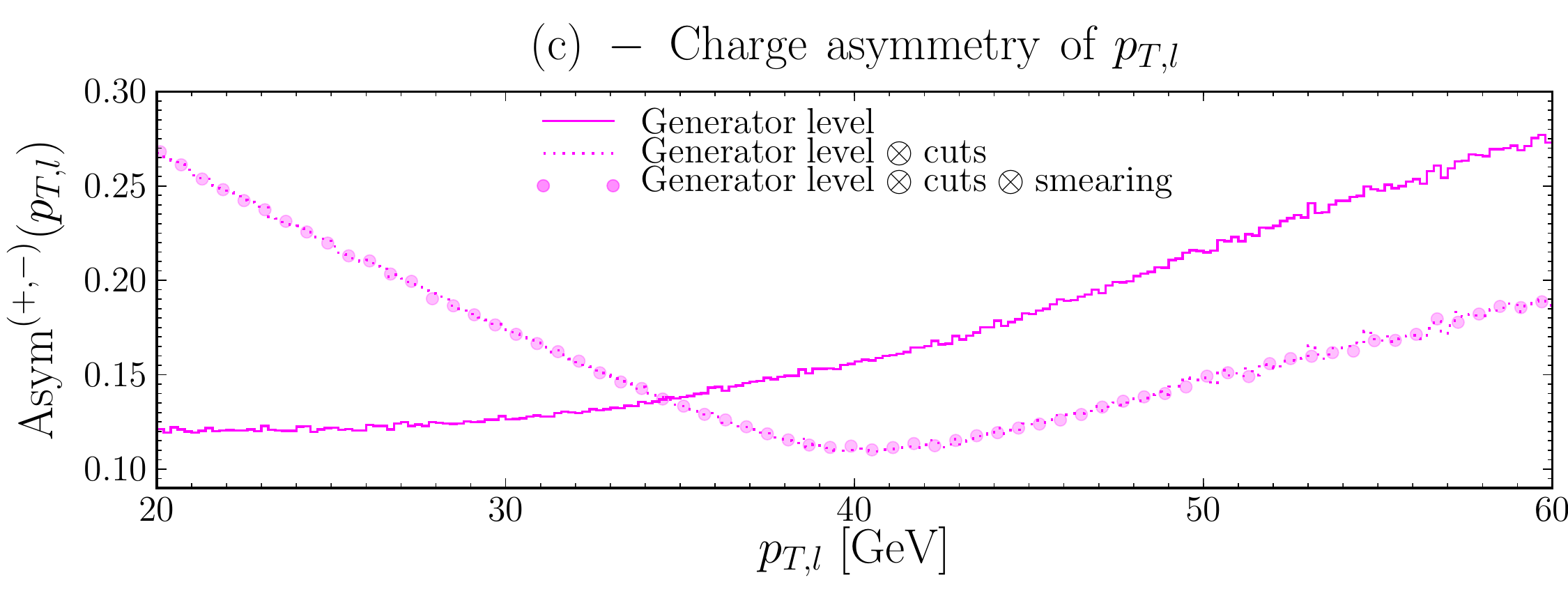}
    \caption[The $\pTlp$, $\pTlm$ and $\Asym{\pTl}$ distributions at
                the generator level, after the cuts ($\pTl>20\GeV$ and $|\etal|<2.5$) and
                finally, by adding the inner detector smearing]
            {\figtxt{The $\pTl$ distributions for the positive (a) and negative (b) leptons 
                and $\Asym{\pTl}$ at the generator level, after the cuts ($\pTl>20\GeV$ and 
                $|\etal|<2.5$), and finally, by adding the inner detector smearing.}
            }
            \label{fig_apparatus}
            \index{Charged lepton@Charged lepton from $W$ decay!Transverse momentum}
  \end{center} 
\end{figure}

To conclude, in  Fig.~\ref{fig_apparatus} we show the  distributions of $\FlatDsigmaDobs{\pTl}$ and 
$\Asym{\pTl}$ for\,: \index{Charged lepton@Charged lepton from $W$ decay!Transverse momentum}
(1) the generated and unselected sample of events, 
(2) the generated and selected sample of events and 
(3) the unbiased simulated detector response and selected sample of events.
The corresponding histograms in $\rhoTl$-space can be in seen in Figs.~\ref{fig_app_rhol}.(a), 
(b) and (c) of Appendix~\ref{app_validation}.
The analysis of the systematic biases affecting these distributions allows to evaluate the 
precision of the measurement methods discussed in this section.    

\subsection{The machine and the detector settings }\label{ss_configuration}
The primary goal of the ATLAS and other LHC experiments is to search for new phenomena at the 
highest possible collision energy and machine luminosity.
It is obvious that, initially, the machine and the detector operation modes will be optimised  
for the above research program.
The main target presented here is to evaluate the precision of the 
measurement of the $W$ mass charge asymmetry which is achievable in such a phase of the detector 
and machine operation.  

A natural extension of this work is to go further and investigate if, and to which extent, the 
precision of measurement of the Standard Model parameters could be improved in dedicated machine
and detector setting runs. 
In our first work~\cite{Krasny:2007cy} we discussed the role of\,:\; 
(1) dedicated runs with reduced beam collision energies, 
(2) dedicated runs with isoscalar beams and 
(3) runs with dedicated detector magnetic field settings\,;\;in optimising the use of the $Z$ boson
production processes as a ``standard candle'' for the $W$ boson processes. 
Here some of these dedicate settings were found to be of interest but the most important ones were
devised in the particular context of this measurement of $\MWp-\MWm$.
More precisely, we discuss the possible improvement in the measurement precision of the charge 
asymmetry of the $W$ boson mass which can be achieved (1) by replacing the proton beams with light 
isoscalar--ion beams and (2) by running the detector for a fraction of time with the inverse 
direction of the solenoidal magnetic field. 
These and other dedicated operation modes could be tried  in the 
advanced ``dedicated measurement phase'' of the LHC experimental program. 
Such a phase, if ever happens, could start following the running period  when the collected 
luminosity will become a linear function of the running time and 
the gains/cost ratio  of its further increase will be  counterbalanced by the gains/cost  
ratio of running dedicated machine and detector operation modes.

\section{The analysis method}\label{s_analysis_strategy}
In  this section we present the technical aspects of the analysis method used in the evaluation of the 
achievable precision of the measurement of the charge asymmetry of the $W$ boson masses, 
$\MWp-\MWm$, denoted sometimes for a matter of convenience $\DeltaPM$.    

The shapes of the experimental distribution (generator level $\otimes$ cuts $\otimes$ smearing) 
shown in Fig.~\ref{fig_apparatus} are sensitive to\,:\;%
(1) the assumed values of $\MWp$ and $\MWp$, 
(2) the values of the other parameters of the Standard Model, 
(3) the modeling parameters of the $W$ boson production processes and 
(4) the systematic measurement biases. 
Our task is to evaluate the impact of the uncertainties of (2), (3) and (4) on the extracted 
values of $\MWp$ and $\MWm$ for each of the proposed measurement methods. 
It is done using a likelihood analysis
of the distributions for the pseudo-data ($\cal{PD}$) event samples and those for the mass template
($\cal{MT}$) event samples.
Each $\cal{PD}$ sample represents a specific measurement or modeling bias, implemented 
respectively in the event simulation or event generation process. Each $\cal{MT}$ sample
was generated by assuming a specific $\MWp$ ($\MWm$) value or a value of their charge asymmetry
$\DeltaPM$.
The $\cal{MT}$ samples are simulated using the unbiased detector response and fixed values of all 
the parameters used in the modeling of the $W$ boson production and decays except for the mass 
parameters.   
The likelihood analysis, explained below in more detail, allows us\,:\;(1) to find out which of the 
systematic measurement and modeling errors, could be falsely absorbed into the measured value of the
$W$ boson masses and (2) to evaluate quantitatively the corresponding measurement biases.

\subsection{Likelihood analysis}\label{ssec_chiTwo}
\index{Chi2 Likelihood analysis@Chi-2 ($\chiD$) likelihood analysis!Principle@Principle (based on the extraction of $\MW$)|(}
Let us consider, as an example, the impact of a systematic effect $\xi$ on the bias in the 
measured value of the $\Wp$ mass determined from the likelihood analysis of the 
$\FlatDsigmaDobs{\pTlp}$ distributions. Since some of the upcoming results were carried based on 
$\rhoTl$ observable as well we mention already the correspondence in terms of resolution and range.

The simulation of the pseudo-data event sample, $\cal{PD}$, representing a given systematic bias 
$\xi$, is carried out  for a fixed value of the mass $M_{W}^{\mathrm{(ref.)}}$. Subsequently, 
a set of the $2\,\mathcal{N}+1$ unbiased (\ie{} $\xi=0$) template data samples $\cal{MT}$ is simulated. 
Each sample $n$ of the $\cal{MT}$ set corresponds to a given value of 
\begin{equation}
M_\Wp^{(n)} = M_{W}^{\mathrm{(ref.)}} + \delta M_\Wp^{(n)},
\end{equation}
with $n=\{-\mathcal{N},\ldots,\mathcal{N}\}$ and where for the rest of the document 
$M_{W}^{\mathrm{(ref.)}}=80.403\GeV$.
The likelihood between the binned $\FlatDsigmaDobs{\pTlp}$ distributions for the 
$n^\mathrm{th.}$ $\cal{MT}$ sample and the $\xi$-dependent $\cal{PD}$ sample is quantified
in terms of the $\chiD$ value\,: 
\begin{equation}
  \chiD(\pTlp;\,\xi, n ) \, = \, \sum^{N^\mm{tot.}_{\mm{bins}}}_i 
  \frac{\left(d\,\sigma_{i;\,\xi} - d\,\sigma_{i;\,\xi=0, n}\right)^2}
       { \left(\Delta d\,\sigma_{i;\,\xi}\right)^2 + \left(\Delta d\,\sigma_{i;\,\xi=0,n}\right)^2},
\label{eq:chi2}
\end{equation}
where $d\,\sigma_i\equiv d\,\sigma_i/d\,\pTlp$ is the content of the $i^\mathrm{th.}$ bin of the 
histogram entering in the analysis and $\Delta d\,\sigma_i$ is the corresponding statistical error.
In total, $N^\mm{tot.}_\mm{bins}$ bins enters in the analysis, it represents the degrees of freedom 
(dof), $\dof\equiv N^\mm{tot.}_\mm{bins}$.
The bulk of the results presented has been obtained using a bin size corresponding to 
$\delta\,\pTl =200\MeV$ (respectively $\delta\,\rhoTl =0.2\,\mm{MeV}^{-1}$) which is approximately 
the anticipated measurement resolution of the track curvature%
~\cite{AtlasInnerDetector:1997fs} and the summation range satisfying the following 
condition for $\pTl$ ($\rhoTl$)
\begin{equation}
30\GeV <\pTl<50\GeV \qquad (0.02\,\mm{GeV}^{-1} <\rhoTl<0.03\,\mm{GeV}^{-1}),
\end{equation}
which corresponds to degrees of freedom of $\dof_\pTl=100$ ($\dof_\rhoTl=49$).
This choice is justified as the CDF II measurements were realised for the $\pTl$ observable in the range 
$32\GeV <\pTl<48\GeV$ to get rid as much of the background by focusing on the jacobian peak region.
\index{CDF detector!W analysis@$W$ analysis}

The $\chiD(\pTlp;\,\xi, n)$ dependence upon $\delta M_\Wp^{(n)}$ is fitted by a polynomial of 
second order. The position of the minimum, $\MWp(\xi)_{\min}$, of the fitted function determines the 
systematic mass shift $\Delta\MWp(\xi) =\MWp(\xi)_{\Min} - M_{W^+}^{\mathrm{(ref.)}}$ due to the 
systematic 
effect $\xi$. If the systematic effect under study can be fully absorbed by a shifted value of 
$\Wp$, then the expectation value of $\chiDmin/\dof$ is $\approx 1$
and the error on the estimated value of the mass shift, $\delta\,\left(\Delta\MWp(\xi)\right)$, 
can be determined from the condition 
$\chiD(\MWp(\xi)_{\Min}+\delta\,\left(\Delta\MWp(\xi)\right)) \equiv\chiDmin+1$
(see Ref.~\cite{Amsler:2008zz} for further details on this topic).

Of course, not all the systematic and modeling effects can be absorbed into the variation 
of a single parameter, even if the likelihood is estimated in a narrow bin-range,
purposely chosen to have the highest sensitivity to the mass parameters. In such cases 
the value of $\chiDmin/\dof$ can be substantially larger than $1$, and 
$\delta\,\left(\Delta\MWp(\xi)\right)$ looses its statistical meaning. 
This can partially be recovered by introducing additional degrees of freedom 
(the renormalisation of the $\cal{PD}$ samples, discussed later on in \S\,\ref{ss_scaling_trick}, 
is an example of such a procedure). However, even in such a case the estimated value of 
$\delta\,\left(\Delta\MWp(\xi)\right)$ will remain slightly dependent upon the number of the  
$2\,\mathcal{N}+1$ $\cal{MT}$ samples, more precisely their $\MWp$ spacing in the vicinity of the 
minimum and the functional form of the fit.
Varying these parameters in our analysis procedure in a $\xi$-dependent 
manner would explode the PC farm CPU time and was abandoned. 
Instead, we calibrated the propagation of statistical bin-by-bin errors into the 
$\delta\,\left(\Delta\MWp(\xi)\right)$ error, and checked the biases of all the aspects of the 
above method using the statistically independent ``$\cal{PD}$-calibration samples'' in which, 
instead of varying the $\xi$ effects, we varied the values of $\MWp$.    
\index{Chi2 Likelihood analysis@Chi-2 ($\chiD$) likelihood analysis!Principle@Principle (based on the extraction of $\MW$)|)}

\subsection{The $\mbf{\cal{MT}}$ and $\mbf{\cal{PD}}$ event samples}
\index{Chi2 Likelihood analysis@Chi-2 ($\chiD$) likelihood analysis!Used for the extraction of MWpmMWm@
Used for the extraction of $\MWp-\MWm$|(}
\subsubsection{Classic method}
\label{sss:classic-method}
In the classic method the bias of $\DeltaPM(\xi)$ resulting from the systematic effects 
$\xi$ is determined in the following three steps\,:
\begin{enumerate}
\baselineskip 1pt
\item Determine  $\Delta\MWp(\xi)\,\pm\,\delta\,\left[\Delta\MWp(\xi)\right]$
  using $\chiD(\flatDfDx{\sigma}{\pTlp};\,\xi)$. 
\item Determine  $\Delta\MWm(\xi)\,\pm\,\delta\,\left[\Delta\MWm(\xi)\right]$ 
  using $\chiD(\flatDfDx{\sigma}{\pTlm};\,\xi)$.
\item Combine these results and derive respectively the central error for $\DeltaPM(\xi)$ 
and the associated error $\delta\,\left[\DeltaPM(\xi)\right]$ 
by adding up quadratically the errors on each charged channel
\begin{eqnarray}
  \DeltaPM(\xi)  &\equiv& \Delta\MWp(\xi) - \Delta\MWm(\xi),\\
  \delta\,\left[\DeltaPM(\xi)\right]  &\equiv&\sqrt{ 
    \left(\delta\,\left[\Delta\MWp(\xi)\right]\right)^2 + 
    \left(\delta\,\left[\Delta\MWm(\xi)\right]\right)^2}
  \label{eq:Deltapm} 
\end{eqnarray}
\end{enumerate}

In the generation of the $\cal{PD}$ samples we assumed $M_{W^+}^\PD=M_{W^-}^\PD=M_{W}^{\mathrm{(ref.)}}$.
The $\cal{MT}$ samples have been generated for
$\delta  M_\Wp^{(n)}  = \delta M_\Wm^{(n)}= n \times 5\MeV$ with $n=\pm\, 1,\ldots,\pm\,6$
and for   
$\delta M_\Wp^{(n)}=\delta M_\Wm^{(n)}= \pm\, 40, \pm\, 50, \pm\, 75, \pm\, 100, \pm\, 200\MeV$
with $ n=\pm\, 7,\ldots,\pm\, 11$. In total $46$ $\cal{MT}$ samples, corresponding to $\mathcal{N}=11$,
have been generated. Table~\ref{table_asym_mt} recaptures in the left column the shifted values from
the $M_W^{\mathrm{(ref.)}}$ used to generate both positive and negatively $W$ bosons $\MT$.

\begin{figure}[!ht] 
  \begin{center}
    \includegraphics[width=0.495\tw]{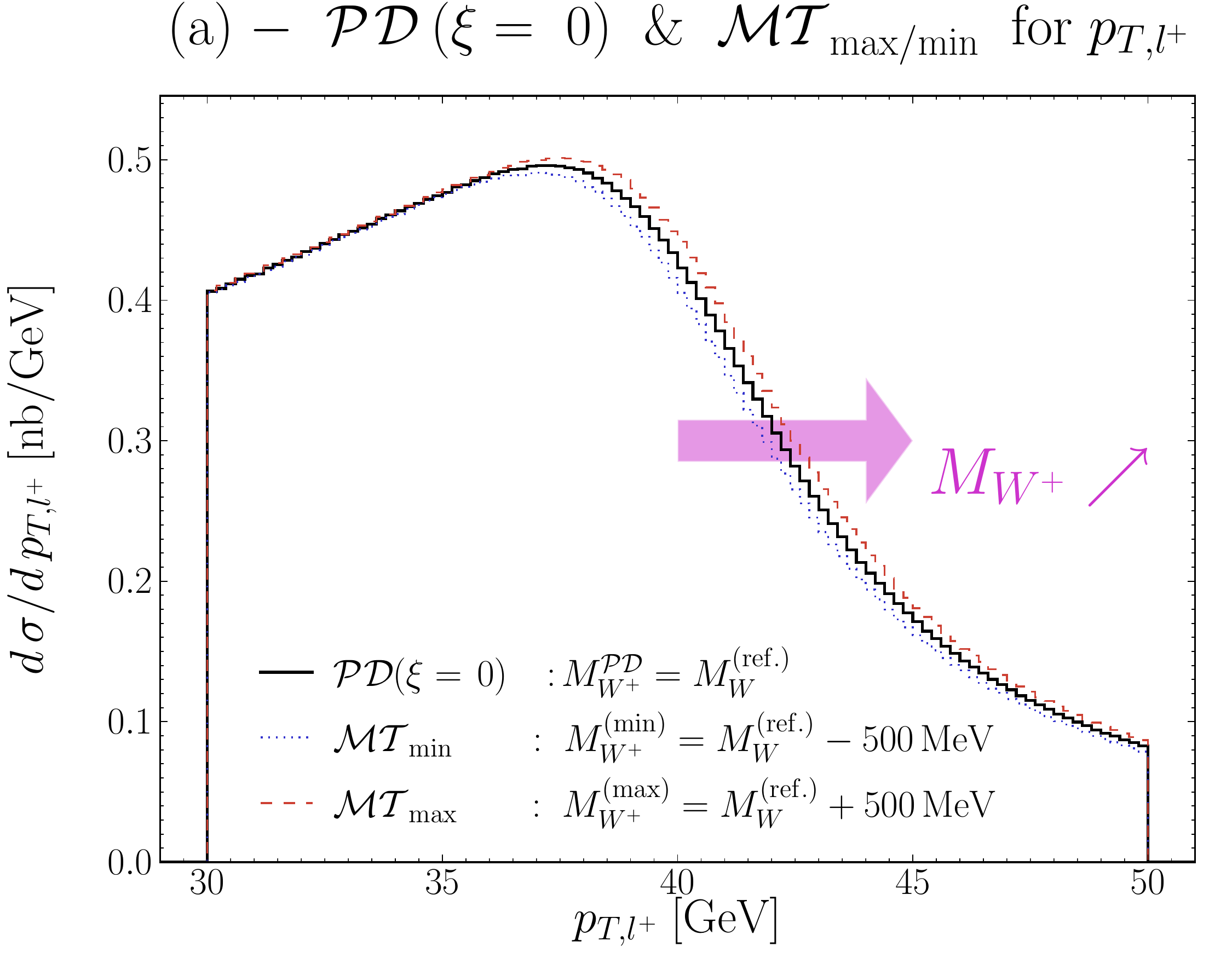}
    \hfill
    \includegraphics[width=0.495\tw]{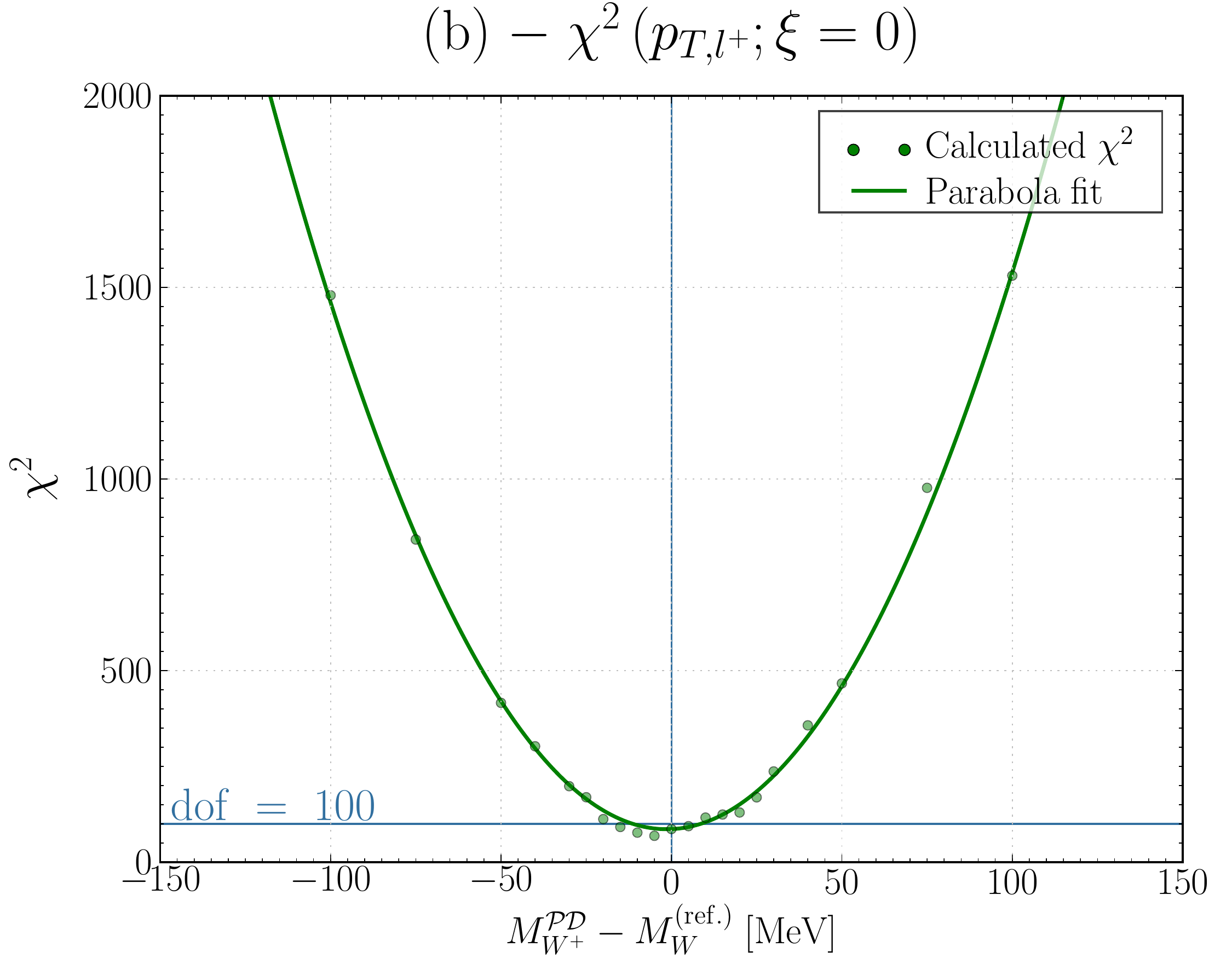}
    \caption[Distribution of the transverse momentum of the positively charged lepton for the three values of 
      $W$ mass and the $\chiD$ dependence in $\DeltaPM$]
            {\figtxt{Distribution of the transverse momentum of the positively charged lepton
                for the three values of $\delta\,M=-500,\,0,\,+500\MeV$ with respect to $80.403\GeV$
                (a) and the $\chiD$ dependence for each $\{\PD,\,\MT\}$ couple (points) and their 
                associated polynomial fit (line) in function of $\DeltaPM$ (b).}
            }
            \label{fig_chi2_exp_cent_pTlp}
            \index{Charged lepton@Charged lepton from $W$ decay!Transverse momentum}
  \end{center} 
\end{figure}

This procedure is illustrated in Fig.~\ref{fig_chi2_exp_cent_pTlp} where, for convenience,
and just for this specific figure, the extreme mass templates where generated with $\pm\, 500\MeV$ 
to make it possible to see the difference between the three distributions. As expected since there is
no bias the parabola fit is centered on $M_\Wp^\mathcal{PD}-M_W^\mm{(ref.)}=0$, more precisely
$M_\Wp^\mathcal{PD}-M_W^\mm{(ref.)}=(-1.4~\pm~2.7)\MeV$ with $\chiDmin/\dof=0.86$.
The same procedure applied for the negatively charged channel gives the result
$M_\Wm^\mathcal{PD}-M_\Wm^\mm{(ref.)}=(-2.2~\pm~3.1)\MeV$ with $\chiDmin/\dof=0.98$.
These two results, once combined, give
\begin{equation}
M_\Wp^\mathcal{PD}-M_W^\mm{(ref.)}=(0.8~\pm~5.8)\MeV
\end{equation}
It shows that no systematic biases are introduced by the proposed analysis method and give
an idea on the size of the statistical errors.

\subsubsection{Charge asymmetry method}
\label{sss:charge-asym}
\index{Charge asymmetry!Used for extraction of MWpmMWm@used for the extraction of $\MWp-\MWm$|(}
In this method the biases $\DeltaPM(\xi)$ resulting from the systematic effects 
are  determined by a direct analysis of the $\Asym{\pTl}$ distributions for the $\cal{MT}$ and 
$\cal{PD}$ event samples.
For the $\MT$ samples, combining the event sample $-\mathcal{N}$ and $\mathcal{N}$ produced shifts for 
the $\DeltaPM$ parameter as shown in the second column of Table~\ref{table_asym_mt}. 
Even though the basics $\MT$ allowed us to spread the range of the analysis to $\pm 400\MeV$, 
most of the values for the systematic errors are such that it proved to be reasonable to stay in the 
$\pm 200\MeV$ range.
\begin{table}[]
\begin{center}
\renewcommand\arraystretch{1}
\begin{tabular}{rcr}
  \hline
  $\delta M_W^{(n)}$ [MeV] & &
  $\Delta_{(+,-)}^{(n)}$ [MeV] \\
  \hline\hline
  $0$        &$\qquad\quad\to$&    $ 0$ \\
  $\pm\, 5 $   &$\qquad\quad\to$&    $\pm\, 10$ \\
  $\pm\, 10$   &$\qquad\quad\to$&    $\pm\, 20$ \\
  $\pm\, 15$   &$\qquad\quad\to$&    $\pm\, 30$ \\
  $\pm\, 20$   &$\qquad\quad\to$&    $\pm\, 40$ \\
  $\pm\, 25$   &$\qquad\quad\to$&    $\pm\, 50$ \\
  $\pm\, 30$   &$\qquad\quad\to$&    $\pm\, 60$ \\
  $\pm\, 40$   &$\qquad\quad\to$&    $\pm\, 80$ \\
  $\pm\, 50$   &$\qquad\quad\to$&   $\pm\, 100$ \\
  $\pm\, 75$   &$\qquad\quad\to$&   $\pm\, 150$ \\
  $\pm\, 100$  &$\qquad\quad\to$&   $\pm\, 200$ \\
  $\pm\, 200$  &$\qquad\quad\to$&   $\pm\, 400$ \\
  \hline
\end{tabular}
\renewcommand\arraystretch{1.45}
\caption[Mass templates generated for $\Wp$ and $\Wm$ and the corresponding 
available $\MWp-\MWm$ mass templates]
{\figtxt{Basic mass templates $\MT$ generated for $\Wp$ and $\Wm$ used
with the classic method (left column) 
and the corresponding $\Delta_{(+,-)}^{(n)}$ $\MT$ samples (right column) 
constructed for the study of the charge asymmetry method. Note though 
$\Delta_{(+,-)}=\pm\,400\MeV$ was not used so in both classic and charge asymmetry
methods the $\MT$ span the range $\pm\,200\MeV$.}}
\label{table_asym_mt}
\end{center}
\end{table}

In preliminary studies the charge asymmetry method was first verified using two charge symmetric procedures. 
In the first one the variation of $\DeltaPM$ was made by fixing $\MWp=M_{W}^{\mm{(ref.)}}$ 
and by changing $\MWm$.
In the second one we inverted the role of $\MWp$ and $\MWm$.
The results obtained with these two charge symmetric methods were found to agree within the 
statistical errors.
 
The first measurement of the charge asymmetry of the $W$ boson masses at the LHC 
will have to use, as the first iteration step, the best existing constraints on the $W$ boson masses.
The best available constraint is the average mass of the $\Wp$ and $\Wm$ bosons\,:\;%
$M_W=M_W^{\mm{(ref.)}}$.
To mimic the way how the measurement will be done at the LHC, we thus fixed
$M_\Wp^\PD+M_\Wm^\PD=M_W^{\mathrm{(ref.)}}$ value and varied, in a correlated way, 
both the $\MWp$ and $\MWm$ values when constructing the $\DeltaPM$ dependent $\cal{MT}$ samples.
For that matter the templates generated previously for the classic method prospect where used in
the creation of these new templates for the charge asymmetry. 
In top of saving CPU time it allows
as well to share the same data and then improve the safety in our analysis. For that reason the
pseudo-data were also shared among all methods.
 
\begin{figure}[!ht] 
  \begin{center}
    \includegraphics[width=0.495\tw]{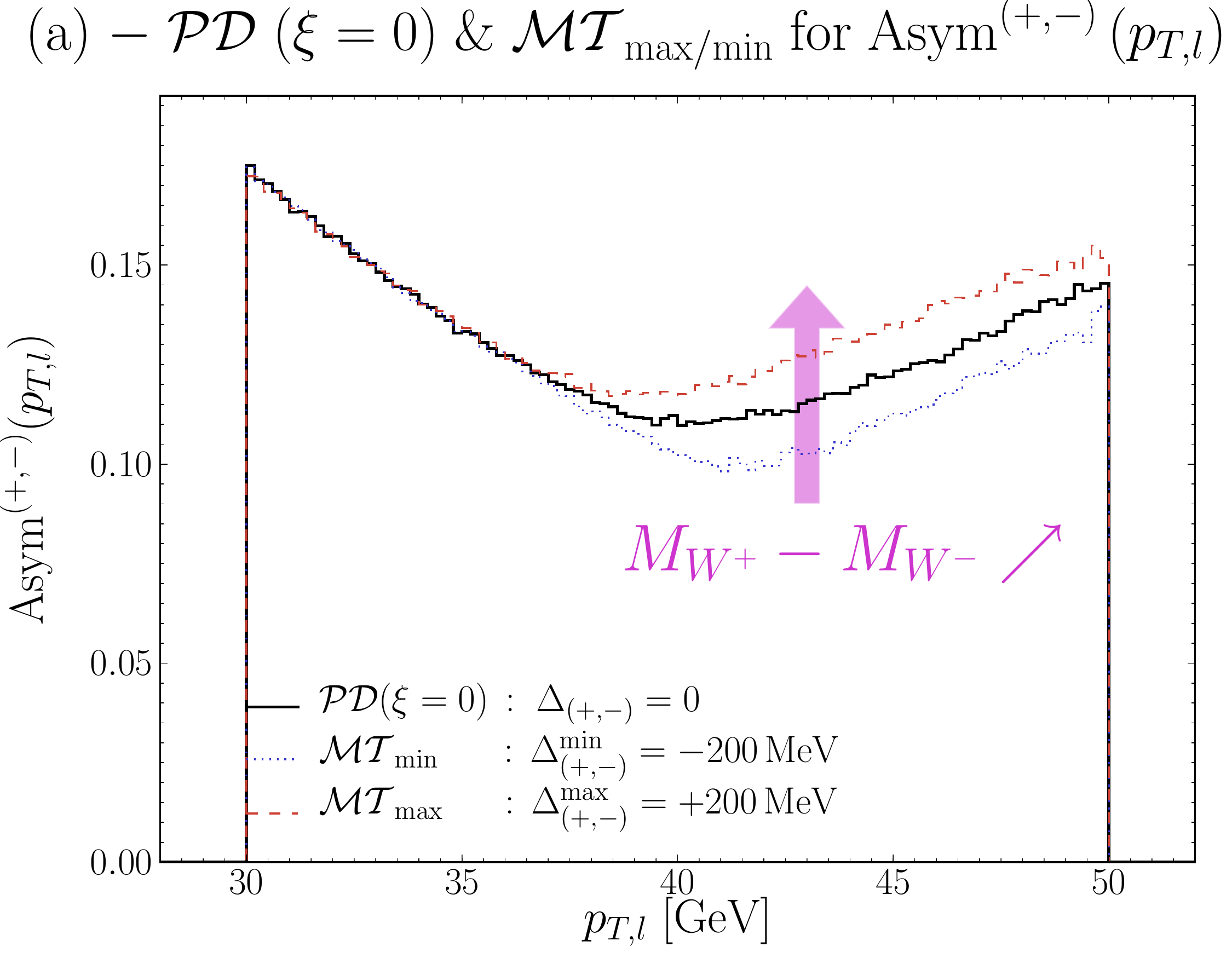}
    \hfill
    \includegraphics[width=0.495\tw]{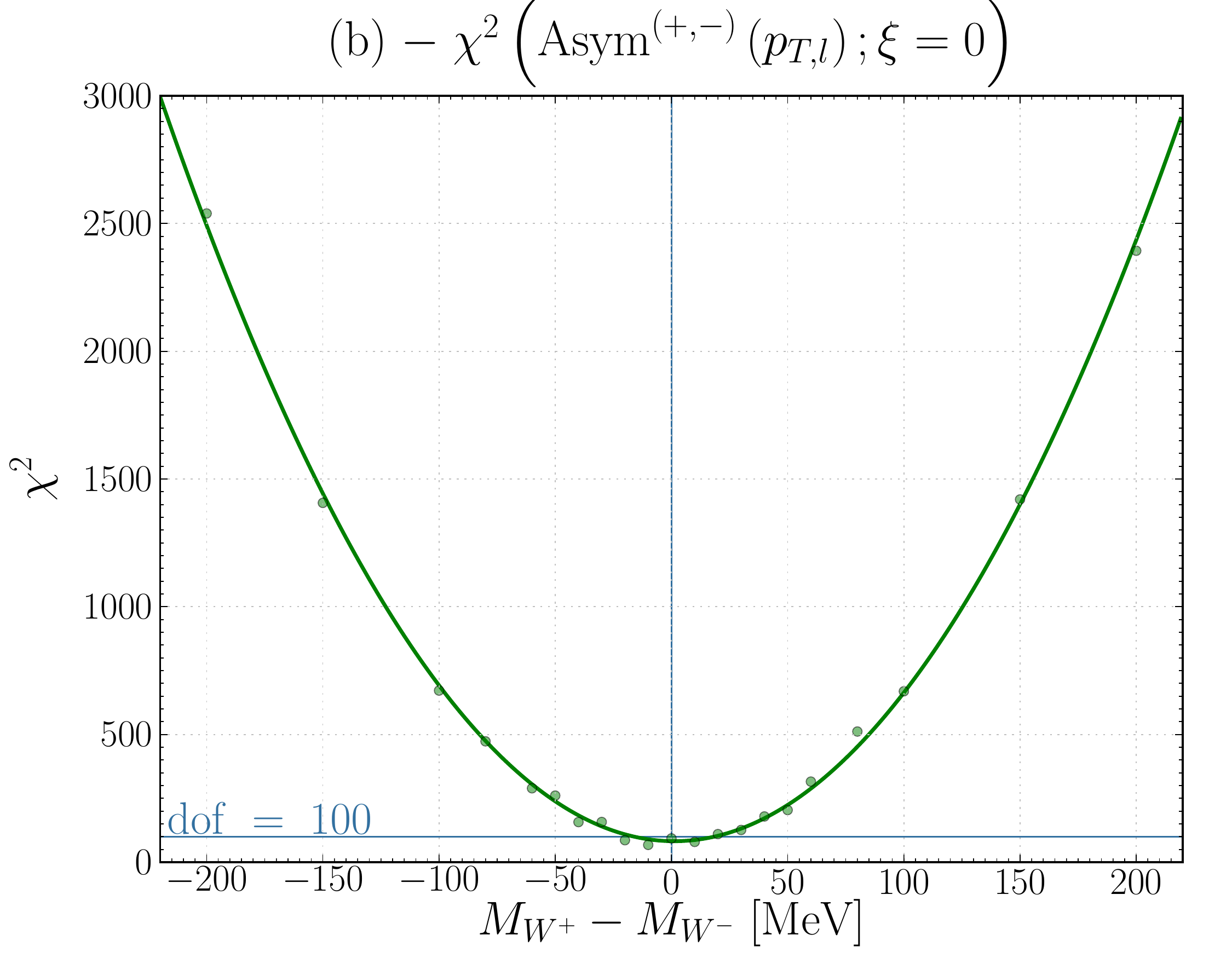}
    \caption[Charge asymmetry of $\pTl$ for the three values of $\DeltaPM$ and 
              the $\chiD$ dependence in $\DeltaPM$]
            {\figtxt{The charge asymmetry of $\pTl$ for the three values of $\DeltaPM$\,:\;%
                $-200,~0,~+200\MeV$ (a)
                and the $\chiD$ dependence for each $\{\PD,\,\MT\}$ couple (points) and their 
                associated polynomial fit (line) in function of $\DeltaPM$ (b).}
            }
            \label{fig_chi2_exp_cent_asym}
  \end{center} 
\end{figure}

This procedure is illustrated in Fig.~\ref{fig_chi2_exp_cent_asym}
for the case of the charge asymmetry method.
The $\Asym{\pTl}$ distribution is plotted in Fig.~\ref{fig_chi2_exp_cent_asym}.(a) as a function of 
$\pTl$ for three values $\DeltaPM$. This plot illustrates the sensitivity of 
the $\Asym{\pTl}$ distribution to the $\DeltaPM$ value.
In Fig.~\ref{fig_chi2_exp_cent_asym}.(b) the $\chiD$ variable is plotted 
for the $\cal{PD}$-calibration sample corresponding to $\Delta_{(+,-)}^{\mathrm{(ref.)}} = 0$ and to an 
unbiased detector response, as a function of $\DeltaPM$. 
The position of the minimum is 
\begin{equation}
\DeltaPM(\xi=0)= (1.2~\pm~4.1)\MeV,\label{eq_DeltaPM_asym_std}
\end{equation}
with a convergence of $\chiDmin/\dof =0.82$. This plot illustrates the calibration procedure.
It calibrates the statistical precision of the measurement for the integrated luminosity of 
\index{Luminosity!At the LHC (expected)!Integrated!Corresponding statistical error on MWpmMWm@Corresponding statistical error on $\MWp-\MWm$}
$10\,\mathrm{fb}^{-1}$, and based on the error obtained for the classic method in the rest of the
Chapter we round all results to the MeV.
Our goal will be to reduce the systematic biases in the measurement of $\DeltaPM$ with a 
comparable precision than the statistical error.    
\index{Charge asymmetry!Used for extraction of MWpmMWm@used for the extraction of $\MWp-\MWm$|)}

\subsubsection{Double charge asymmetry method}
\label{sss:double-asym}\index{Double charge asymmetry!Used for the extraction of MWpmMWm@
Used for the extraction of $\MWp-\MWm$}
The $\cal{MT}$ event samples for this methods are exactly the same as for the charge asymmetry 
method. The $\cal{PD}$ event samples have been simulated  in two steps corresponding to the two 
half-a-year periods of data taking corresponding to the two magnetic field configurations.  
\index{Chi2 Likelihood analysis@Chi-2 ($\chiD$) likelihood analysis!Used for the extraction of MWpmMWm@
Used for the extraction of $\MWp-\MWm$|)}

\subsection{Scaling distributions for quarks flavors systematics}
\label{ss_scaling_trick}
Most of systematic measurement and modeling biases discussed in this work lead to a distortion of 
the distributions and do not change their overall normalisation. The notable 
exception,  discussed in more detail in the next section, are the biases driven by the PDFs 
uncertainties. These biases cannot be ``absorbed'' by  the corresponding $\DeltaPM(\xi)$ shifts 
and require an adjustment of the event/nb normalisation of the corresponding $\MT$ samples
to obtain acceptable $\chiD$ values. 

The most natural method would be to extend the one-dimensional analysis presented in this 
section into two-dimensional analysis of both the mass and the normalisation parameters. Such 
an analysis would have, however, ``squared'' the necessary computing time of the 
$\cal{MT}$ samples and, therefore, was not feasible in our time-scale. 
Instead,  we have tried  to un-correlate the adjustment of the normalisation 
parameter and the mass parameters. As can be seen in 
Fig.~\ref{fig_chi2_exp_cent_asym}.(a) the $\Asym{\pTl}$ distribution is, in the region of small 
$\pTl$ ($\pTl\lesssim 35\GeV$), independent of the variations of $\DeltaPM$ over the range discussed here. 
We use this observation and modify correspondingly the likelihood  analysis method. 
Before calculating $\chiD$, the $\cal{PD}$ and $\cal{MT}$ distributions
are integrated in the range\,:
\begin{equation}
\int_{20\GeV }^{35\GeV } d\,\pTl\; \left(\DfDx{\mathrm{Asym}^{(+,-)}}{\pTl}\right),
\label{eq:interal-pTl}
\end{equation}
giving respectively two scalars\,:\;$\alpha$ and $\beta^{(n)}$.
Then we re-normalise the $n^\mm{th.}$ $\cal{MT}$ distribution by the factor $\alpha/\beta^{(n)}$ and 
calculate the $\chiD$ values for the rescaled distributions. 
We have checked that the above procedure improves significantly the resulting $\chiD$ values for 
each of the three analysis methods. By changing the integration region we have verified that the 
above procedure does not introduce significant biases in the estimated $\DeltaPM(\xi)$ values. 

\begin{figure}[!ht] 
  \begin{center}
    \includegraphics[width=0.495\tw]{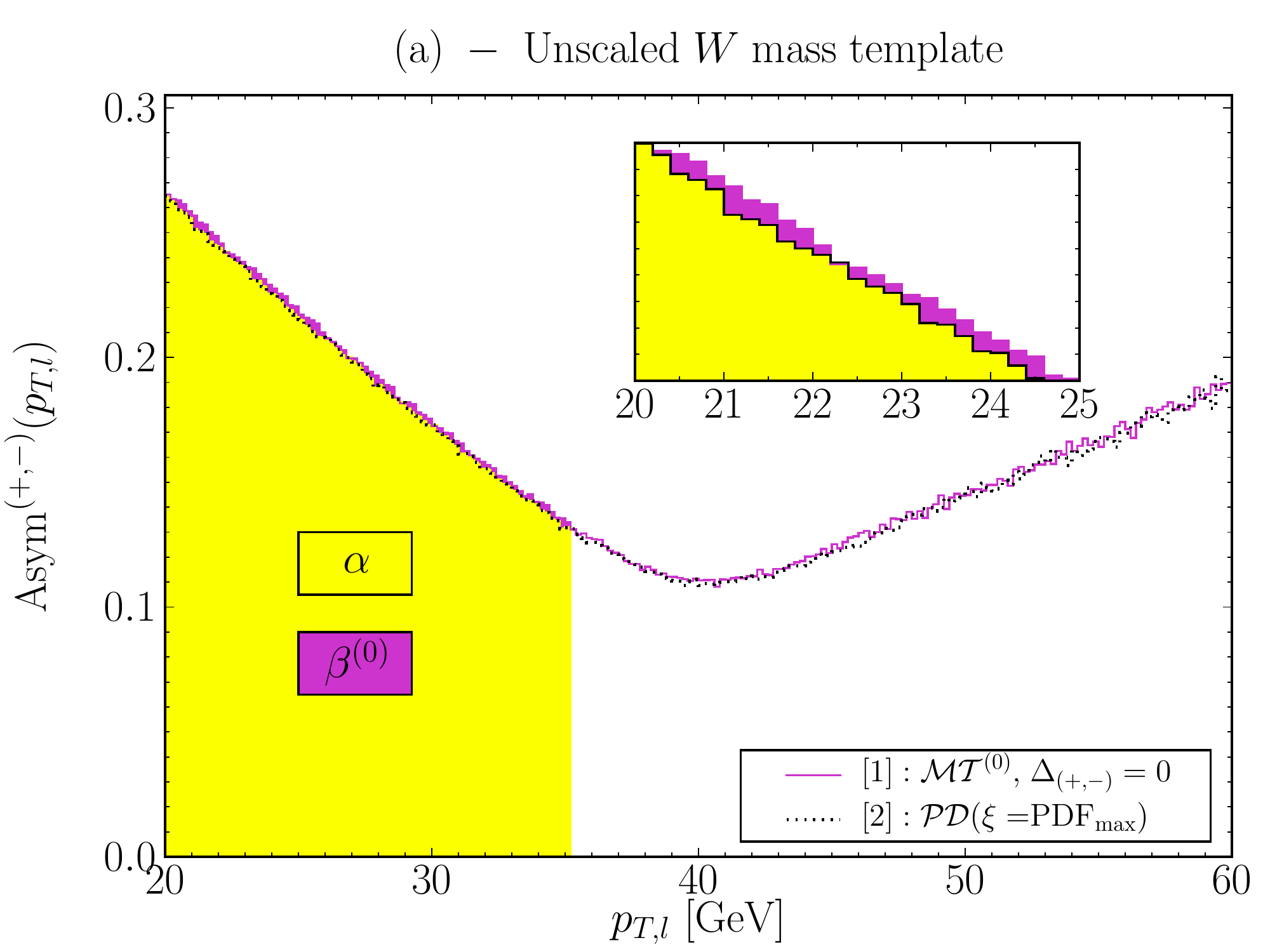}
    \hfill
    \includegraphics[width=0.495\tw]{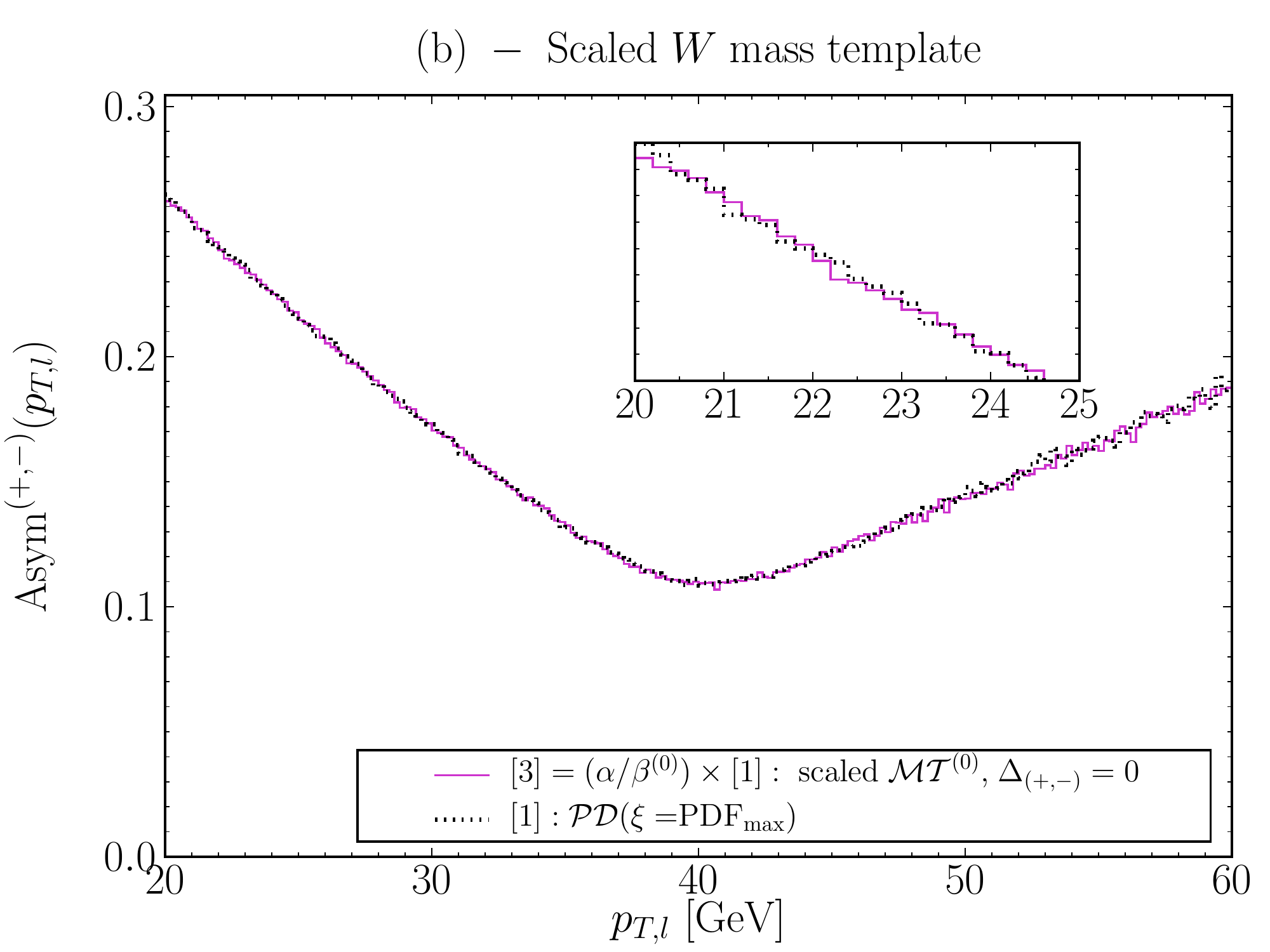}
    \vfill
    \includegraphics[width=0.495\tw]{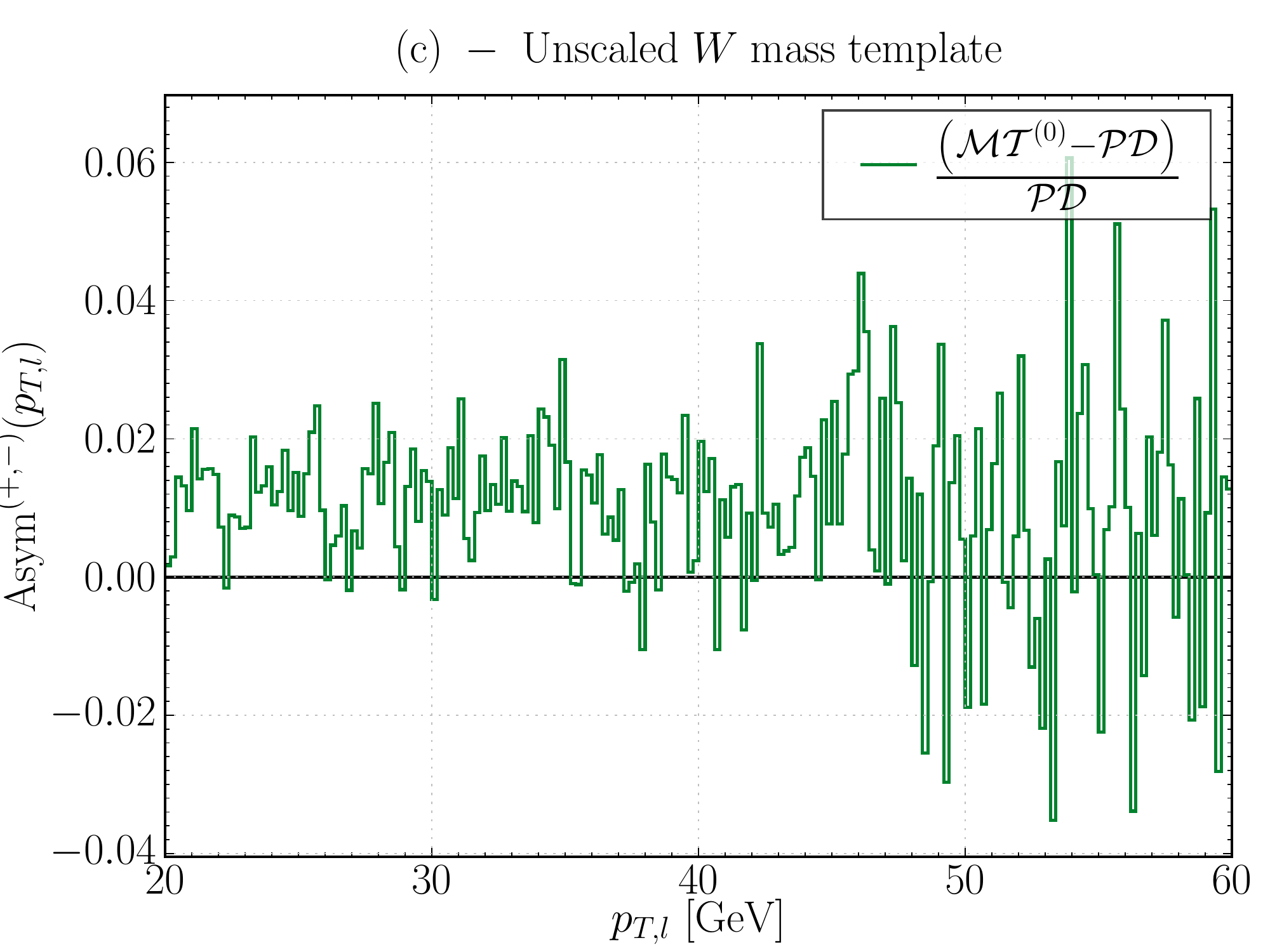}
    \hfill
    \includegraphics[width=0.495\tw]{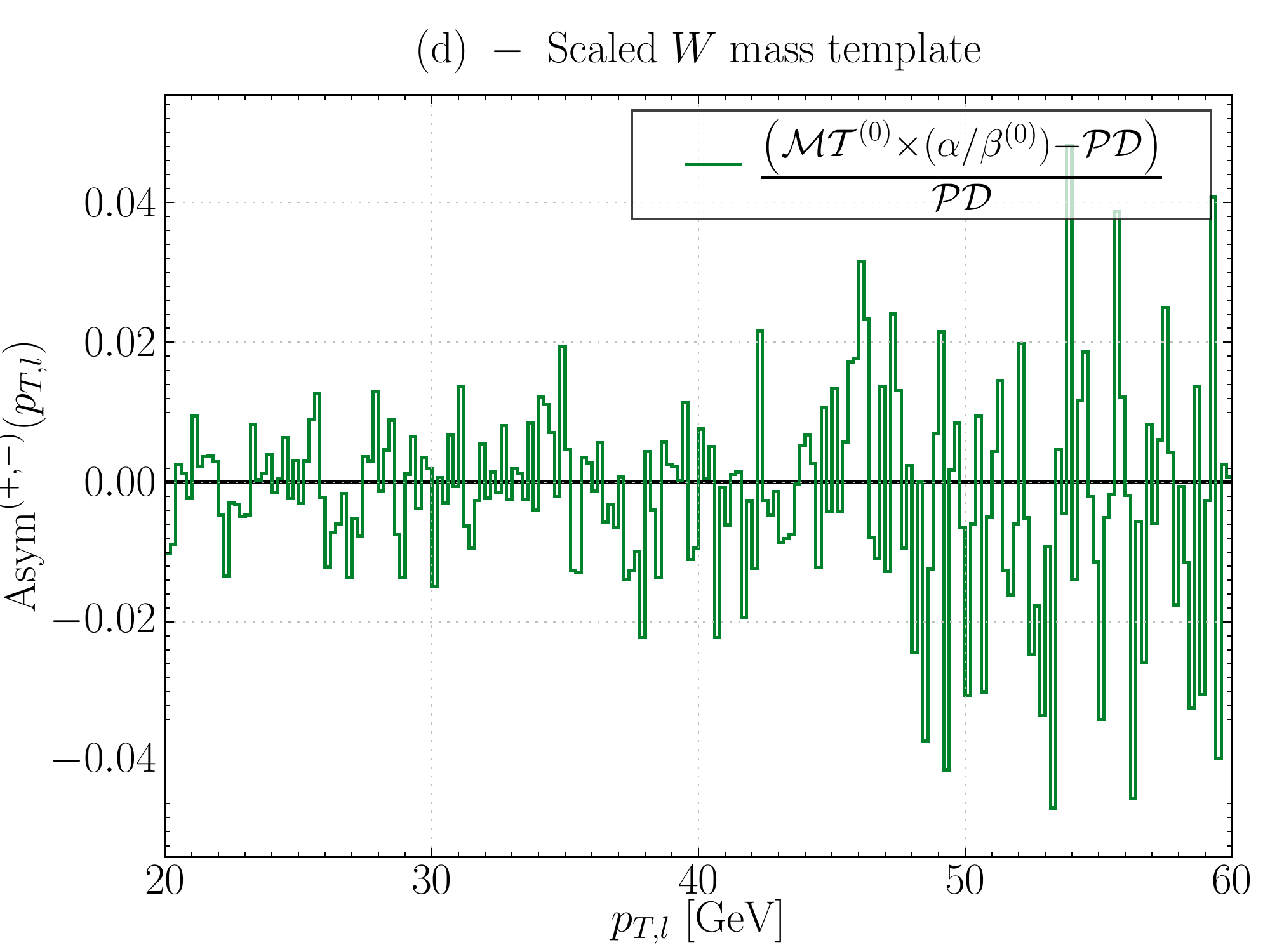}
    \caption[Scaling of a mass template distribution with a biased pseudo-data histogram]
            {\figtxt{Central mass template $\mathcal{MT}^{(0)}$ ($\DeltaPM=0$)distribution with a 
                biased pseudo-data histogram $\mathcal{PD}_\Max$ 
                (overestimation of the PDFs weights) before scaling (a) and after scaling (b)
                and refined visualisation of the latter in (c) and (d) by looking at the normalised difference 
                $\left(\MT-\PD\right)/PD$.}}
            \label{fig_chi2_exp_cent_asym2}
  \end{center} 
\end{figure}

The effect of this trick can be visualised on a concrete example where we consider the incriminated
bias to reflect an overestimation of the PDF. The consequences of the different cross section 
between the $\PD$ with respect to the one of the central $\MT^{(0)}$ ($\DeltaPM=0$) histogram can be 
seen directly in Fig.~\ref{fig_chi2_exp_cent_asym2}.(a) and more finely in frame (c) where the 
normalised difference between the two histograms have been drawn. In (c) then, we observe an overall 
scaling factor --constant below the jacobian peak-- between the two plots that will ultimately wreck 
any chance for the $\chiD$ test to 
converge and then give a relevant result. Computing the scalars $\alpha$ and $\beta^{(0)}$ allows
to scale $\MT^{(0)}$ by multiplying it with $\alpha/\beta^{(0)}$ as shown in 
Figs.~\ref{fig_chi2_exp_cent_asym2}.(b) and (d).

\section{Systematic error sources}\label{s_model_impact_sys} 

In this section we identify and model the systematic error sources that will limit the precision of
the $\MWp-\MWm$ measurement at the LHC. Each of these error source will be modeled and reflected 
in the corresponding $\cal{PD}$ sample of events.
   
These error sources are of two kinds\,:\;(1) those reflecting  
uncertainties in modeling of the $W$ boson production and decay 
processes, (2) those reflecting the event selection and 
event reconstruction biases. 
A large fraction of the error sources have been 
identified \cite{Aaltonen:2007ps}  and reevaluated in the context of the measurement 
of the average mass of the $W$ boson at the LHC \cite{Buge:2006dv,Besson:2008zs}. 
We focus our discussion on the dominant errors
for the measurement of the $W$ mass charge asymmetry, in particular
on those that are specific to the LHC environment and have not been identified
in the earlier studies. We shall not discuss here\,:\;(1) the measurement errors 
reflecting the uncertainties in the background estimation and in the efficiency 
of the events selection, (2) other measurement uncertainties which can be studied to the 
required level of precision only once the real data are collected. As 
demonstrated  in the analysis of the Tevatron data \cite{Aaltonen:2007ps}, 
they are of secondary importance.     

\subsection{Phenomenological modeling uncertainties}\label{s_model_impact_sys_pheno} 
The uncertainties in the modeling of the production and decay of the $W$ bosons include\,:\;%
(1) the uncertainties in modeling of non-perturbative effects, 
(2) the approximations present in theoretical 
modeling of the perturbative EW and QCD effects,
(3) the uncertainties in the parameters of the Standard Model and (4) a possible  
presence of the Beyond Standard Model (BSM) effects, 
affecting both the production and decay mechanisms of the $W$ bosons. 
The two first items of them limit  our present understanding of the Wide-Band-partonic-Beam (WBpB) 
at the LHC.

\subsubsection{WBpB at LHC}
\label{sss:WBpB}

The measurement precision of the $W$ mass charge asymmetry  
will depend upon the level of understanding 
of the flavour structure, the momentum spectrum and the emittance 
of the WBpB at the LHC collision energy. 
The hard scale dependent emittance of the WBpB
is defined here, in analogy to the emittance of the parent hadron beam,
in terms of its transverse momentum distribution
and in terms of its transverse and longitudinal beam-spot sizes.      
The above dynamic properties of the WBpB are highly correlated.
Only their scale dependence can be controlled by the 
Standard Model perturbative methods. In addition,  
several aspects of such a control, in particular the 
precise modeling  of the 
correlations between the flavour, the longitudinal and the transverse
momentum degrees of freedom of the WBpB  have not so far been 
implemented in the  Monte Carlo generators available 
for the initial  phase of the LHC experimental program.\index{Monte Carlo}

The \textit{statu quo} of understanding of the WBpB at the LHC
is driven by the presently available Monte Carlo generator tools. 
Within this \textit{statu quo}, the flavour dependent longitudinal momentum distribution 
of the WBpB, specified by `collinear' PDFs, is fed to one of the available  
parton shower MC generators. The transverse momentum distribution of the WBpB  is 
then derived from the assumed longitudinal one. This procedure 
depends upon a particular evolution scheme dependent form of the parton shower
and upon the order of the perturbative expansion.
It depends as well upon the modeling method of the quark flavour (quark mass) 
effects in the parton shower generation.   
The effects of the flavour dependence 
of the beam size in the transverse plane are partially controlled 
using  auxiliary, impact parameter
dependent re-summation procedures. Finally, the scale dependent evolution of 
the longitudinal beam spot size is presently assumed to be driven by the 
DGLAP evolution equations~\cite{Gribov:1972ri,Altarelli:1977zs,Dokshitzer:1977sg}.

It is obvious that the precision of the present understanding of the WBpB at the LHC 
is difficult to asses within the 
above modeling environment. Since its impact on the precision measurements
of electroweak processes will be significantly higher for the LHC WBpB
with respect to the Tevatron one, some novel measurement and/or modeling  schemes
must be developed. They must assure either better theoretical control  of the WBpB 
parameters or, as proposed, reduce  
their impact on the measured observables
to such an extent that their detailed modeling becomes irrelevant.   
For the latter strategy it is sufficient to rely on crude modeling methods of 
the WBpB at the LHC which are available within the \WINHAC{} generator.

\subsubsection{Uncertainty of PDFs}
\label{sss:PDFuncertainty}

The uncertainties in PDFs are, most often, 
propagated to the measurement errors of 
the physics observables by varying  
the PDF sets chosen  in
the event generation process. Alternatively, the 
uncertainties of the QCD fit parameters of a given PDF set are propagated 
by re-weighting the generated events
with ``min'' and ``max'' weights,  
$\mathrm{PDF}_{\Max/\Min}=\mathrm{PDF}_\Cen \pm \delta\mathrm{PDF}$,
where $\mathrm{PDF}_\Cen$ are the central value distributions of 
a given PDF set and
$\delta\mathrm{PDF}$ is computed according to the 
method described in Ref.~\cite{Pumplin:2002vw}.
We followed the latter method, mostly because  
of computing time constraints. 
We have used the CTEQ6.1M PDF set~\cite{CTEQ6.1:2003}\index{Parton Distribution Functions (PDFs)!CTEQ}
in modeling of the standard PDFs uncertainties. 
The above methods, in our view, largely underestimate the influence of the 
PDFs uncertainty on the measurement precision of the $W$ boson mass.
 
As discussed in the last Chapter, the charge 
asymmetry of the $W$ boson production and decay processes is 
sensitive\,:\;(1) the presence of valence quarks in the \index{Quarks!Valence quarks}
WBpB, (2) the flavour asymmetry of their 
longitudinal momentum distribution (called hereafter  
the $u^\val-d^\val$ asymmetry) and (3) \index{Quarks!uvmdv@$u^\val-d^\val$ asymmetry}
the asymmetry in the relative momentum distribution of the strange and 
charm quarks (the $s-c$ asymmetry). The corresponding uncertainties 
must be modeled directly using the existing (future)  experimental constraints rather 
than be derived from the uncertainty of the PDF set parameters. 
This is because they are driven almost entirely 
by the non perturbative effects, and because the QCD evolution
effects are, except for the quark mass dependency, flavour independent.

\subsubsection{Uncertainty of $\mbf{u^\val-d^\val}$  asymmetry.}
\label{sss:ud-asym}\index{Quarks!Valence quarks}
\index{Quarks!uvmdv@$u^\val-d^\val$ asymmetry|(}
We assume the following two ways of modeling 
the uncertainty in the $u^\val-d^\val$ asymmetry\,:
\begin{equation}
u^\val_{\Max/\Min} = u^\val \pm 0.05\,u^\val,\qquad
d^\val_{\Min/\Max} = d^\val \mp 0.05\,u^\val,
\label{eq:ud1}
\end{equation}
and 
\begin{equation}
u^\val_{\Max/\Min} = u^\val \pm 0.02\,u^\val,\qquad
d^\val_{\Min/\Max} = d^\val \mp 0.08\,d^\val.
\label{eq:ud2}
\end{equation}
The first one preserves the sum of the distribution of the $u$ and $d$ quarks and is 
constrained, to a good precision, by the measured singlet structure function  
in neutrino and anti-neutrino Deep Inelastic Scattering (DIS) of  isoscalar 
nuclei. At the LHC the sum of the distributions will be constrained
by the rapidity distribution of the $Z$ bosons 
($d$ quarks and $u$ quarks contribute with similar weights).
The second one assures  the correct propagation 
of the measurement errors of the sum of the charge square weighted distributions of 
the $u$ and $d$ quarks, constrained by the high precision charged lepton beam 
DIS data, to the uncertainty of the individual distributions. 

While the sums of the distributions are well controlled by the existing and future data,
their mutually compensating shifts are not. The only experimental constraints on  
such shifts come from\,: (1) the measurements of   
the ratio of the cross sections for deep inelastic scattering of charged leptons 
on proton and deuteron targets and (2) the measurements 
of the ratio of the neutrino--proton  to anti-neutrino--proton  
DIS cross sections. They determine the present uncertainty range of the  $u^\val-d^\val$
asymmetry. Improving this uncertainty range in the standard $\pp$ LHC colliding mode will 
be difficult and ambiguous. It will require simultaneous unfolding of the   
momentum distribution and the charge asymmetry of the sea quarks.      
\index{Quarks!uvmdv@$u^\val-d^\val$ asymmetry|)}

\subsubsection{Uncertainty of $\mbf{s-c}$ asymmetry}
\label{sss:cs-asym}
\index{Quarks!smc@$s-c$ asymmetry|(}
The $cs$ annihilation represent $\approx 7\percent$ of the total contribution to the $W$ boson 
production cross section at the Tevatron collision energy. 
At the LHC collision energy it rises to $\approx 25\percent$ and becomes charge asymmetric\,: 
$\approx 21\percent$ for the $\Wp$ boson and $\approx 28\percent$ for the $\Wm$ boson. 
The uncertainty in the relative distribution of the strange and charm quarks becomes an 
important source of systematic measurement errors of both the average $W$ boson mass and 
its charge asymmetry. 

We assume the two following ways of modeling 
the uncertainty in the $s-c$ asymmetry\,:
\begin{equation}
s_{\Max/\Min} = s \pm \gamma\,c,\qquad
c_{\Min/\Max} = c \mp \gamma\,c
\end{equation}
with $\gamma = \{0.2,\,0.1\}$ representing respectively  the present and future  uncertainty 
range for the relative shifts in the $s$ and $c$ quark distributions. 
As in the case of the $u^\val-d^\val$ asymmetry, we have assumed that the sum of the 
distribution of the $s$ and $c$ quarks will  be controlled to a very good precision 
by the $Z$ boson rapidity distribution. Therefore we have introduced only  unconstrained, mutually 
compensating modifications of the $s$ and $c$ quark distributions%
\footnote{In reality, the $s$ and $c$ quarks 
couple to the $Z$ boson with slightly different strength but the resulting effect will play no 
important role in the presented analysis.}.  
As seen previously in Eqs.~(\ref{eq_sWp_sWm_pp}) and  (\ref{eq_sWp_sWm_dd}), 
the valence quarks 
excess magnifies the contribution of the $s-c$ uncertainty to the 
measurement precision of $W$ boson mass.  
\index{Quarks!smc@$s-c$ asymmetry|)}

\subsubsection{WBpB emittance}
\label{sss:WBpBemittance}
The $u^\val-d^\val$ and  $s-c$ longitudinal momentum asymmetries 
would have no effect on the measured
$W$ boson mass asymmetry in the case of the collinear partonic beams.
The angular divergence (transverse momentum smearing) of the WBpB at the LHC
is driven by the gluon radiation. Its parton shower Monte Carlo model  
determines the relationship between  the longitudinal and the transverse 
degrees of freedom of the WBpB. It gives rise to the parton shower model dependent 
asymmetries of the $\Wp$ and $\Wm$ boson transverse momenta. 

\index{Quarks!Intrinsic transverse momenta|(}
Instead of trying to estimate 
the uncertainties related to the precision of the parton shower
modeling of the the quark flavour dependent effects, we allow 
for exceedingly large uncertainty in the size of the flavour independent 
primordial transverse momentum Gaussian smearing of the WBpB\,:\;%
$\Mean{\kT}= 4^{+3}_{-2}\GeV $ (the $\cal{PD}$ samples have been simulated for the following
values of the sigma parameter of the Gaussian smearing\,: 
$\Mean{\kT}=2,3,4,5,6,7\GeV $).
Such a large uncertainty range, easily controllable using 
the $Z$ boson transverse momentum distribution, represents the effect of   
amplifying (small values of $\Mean{\kT}$) or smearing out 
(large values of $\Mean{\kT}$) the flavour dependent asymmetries
of the WBpB transverse momentum. 
The range has been chosen to be large enough to cover 
the uncertainties due to\,:\;(1) non perturbative effects, \eg{} those discussed 
in~\cite{Gieseke:2007ad}, (2) the quark mass effects and (3) re-summation effects.  
\index{Quarks!Intrinsic transverse momenta|)}

\subsubsection{EW radiative corrections}
\label{sss:EWcor}
\index{Electroweak!Radiative corrections in W in Drell--Yan@Radiative corrections in $W$ in Drell--Yan}
Out of the full set of the EW radiative corrections implemented in the \WINHAC{}
generator, those representing the emission of real photons could contribute
to the measured $W$ mass charge asymmetry. Two effects need to be 
evaluated\,:\;the charge asymmetric interference terms between the photon emission in the 
initial and final states, and the radiation of the photons in the $W$ boson decays
in the presence of the $V-A$ couplings.\index{Electroweak!VmA@$V-A$ coupling}
The above corrections are described to a high precision by \WINHAC{}, as has been shown
in Refs.~\cite{CarloniCalame:2004qw,Bardin:2008fn}. Therefore, their influence on the
$W$ mass charge asymmetry measurement can be modeled very accurately.  
We did not considered these effects, leaving a detailed study for our future
works.

\subsection{Experimental uncertainties}
\subsubsection{Energy scale (ES) of the charged lepton}
\label{sss:ES}
The uncertainty in the lepton energy scale is the most important source of 
the $\MW$ measurement error for the Tevatron experiments.
At the LHC, the production of unequal numbers of $\Wp$ and $\Wm$ bosons,
its impact on the overall measurement precision will be amplified. 
For the measurement methods discussed the lepton 
energy scale error will be determined\,:\;(1) by the curvature radius
measurement errors, (2) by the uncertainties in the magnetic field maps 
within the tracker volume and (3) by the modeling precision 
of the physics processes which drive the link between the measurements
of the particle hits in the tracker and the reconstructed particle 
momentum. While the first two sources of the 
measurement error are independent of the lepton flavour, 
the third one affects the electron and muon samples differently. 
In the following we shall assume, on the basis of the Tevatron 
experience, that modeling of physics processes of particle 
tracking will be understood at the LHC to the required level of precision, 
on the basis of dedicated auxiliary measurements\footnote{For example,  
the energy loss of the electrons in the dead material within the tracker 
volume will be understood using a conjugate process of the photon conversion.}.
This simplification allows us to discuss  the muon and electron track measurement 
simultaneously. We assume as well that the solenoid magnetic field strength in the 
volume of the tracker will be understood to better than $0.1\percent$ of its 
nominal value. We base this assumption on the precision of $0.01\percent$ achieved 
\eg{} by the H1 experiment at HERA \cite{H1:1990} and by the ALEPH experiment at LEP
\cite{ALEPH:1995}. If this condition is fulfilled, 
the energy scale error $\es_{l}$ is driven by the curvature radius measurement error
\begin{equation}
\rho_{T,l}^\rec = \rho_{T,l}^\smear \,(1+\es_{l}),
\end{equation}
where $\rho_{T,l}^\rec$ and $\rho_{T,l}^\smear$ are, respectively, the reconstructed   
and the true curvature smeared by the unbiased detector response function.

Based on the initial geometrical surveys, the initial scale 
of $\rho_{T,l}$ will be known to the precision of $0.5\percent$.
This precision will have to be improved at least by a factor of 10 
to match the precision of the Tevatron experiments,
if the same measurement strategy is applied. 
To achieve such a precision,
the local alignment of the tracker elements and/or average biases of 
the reconstruction of the trackers space-points must be known to the $\approx 3\, \mu\mm{m}$
precision. In addition, the global deformation of the tracker elements 
assembly must be controlled to a precision  which is beyond the reach  
of the survey methods. 

Several modes of the global deformations can be considered in a first approximation
as discussed in Ref.~\cite{Brown:2006zz}. 
Below we make a sum up of the conclusions drew from \S\,\ref{s_weak_modes} and 
Appendix~\ref{cdf_tracker}.
The main difference between the measurements of the $W$ boson properties at the Tevatron 
and the LHC boils down to their sensitivity to the different types of the global deformation 
modes.
Both for the Tevatron and LHC measurements the $\Delta z$ translations are of no consequences since 
they  do not  affect the shape of the 
transverse projection of the particle helix. The $\Delta r$ deformations
(the radial expansion $r\,\Delta r$, the elliptical flattening $\phi\,\Delta r$ and the bowing 
$z\,\Delta r$)
give rise to common biases for positive and negative particle tracks. 
On the other hand, 
the $\Delta\phi$ curl and twist deformations\index{Weak modes!In ATLAS tracker|(}
give rise to biases which are opposite for negative and positive particles.
In the case of the Tevatron $\ppbar$ collisions, producing equal numbers 
of  the $\Wp$ and $\Wm$ bosons,  the 
dominant effect of $\pm z$-coherent curling of the outer tracker layers with 
respect to the inner tracker layers has  residual influence on the uncertainty of the
average $W$ boson mass, leaving the residual effect of relative twist of the $+z$ and $-z$  
sides of the tracker volume as the principal source of the measurement error. 
For the LHC $\pp$ collisions, producing unequal numbers of the $\Wp$ and $\Wm$ bosons, 
both deformation modes influence the measurement biases of the average $W$ boson mass.
In the case of the LHC there is no escape from the necessity of precise understanding 
of the lepton charge dependent biases on top of the lepton charge independent biases.  

In the presence of the above two sources of biases the energy scale bias $\es_{l}$
can be expressed in the limit of small deformations 
(cf. Eqs.~(\ref{eq_rhoTlprec}--\ref{eq_rhoTlmrec})) as follow
\begin{eqnarray}
\es_\lp &=& \:\:\; \es_{\Delta\phi} + \es_{\Delta r},\\
\label{eq:epsilonp}
\es_\lm &=& - \es_{\Delta\phi} + \es_{\Delta r},
\label{eq:epsilonm}
\end{eqnarray}
where $\es_{\Delta\phi}$ represents the particle charge dependent  $\Delta\phi$-type bias
and $\es_{\Delta r}$  represents the charge independent $\Delta r$-type bias.

While the $\es_{\Delta r}$-type biases can be controlled with the help of the $Z$ boson, 
$\Upsilon$ and
$J/\Psi$ ``standard candles'', \eg{} using the CDF procedures, \index{CDF detector}
the global charge dependent 
and symmetric $\es_\mathrm{curl}$ biases cannot. At the Tevatron these biases were investigated 
using the electron samples by  studying the charge dependent $E/p$ distribution, where $E$ is 
the energy of the electron (positron) measured in the calorimeter and $p$ is its  reconstructed
momentum. The relative scale error of positive and negative electrons was re-calibrated using 
the mean values of the  $E/p$ distributions. The achieved precision was the principal 
limiting factor of the measurement of $\DeltaPM$.  Even if the statistical precision 
of such a procedure can be improved significantly at the LHC, this method is no longer unbiased.
This is related to the initial asymmetry of the transverse momentum distribution for 
positive and negative leptons in the selected $W$ boson decay samples. 
As a consequence, both the positive and negative lepton events, 
chosen for the calibration on the basis of the energy deposited in the calorimeter, will 
no longer represent charge unbiased samples of tracks. The biases will be driven both by the 
influence of distribution shape and by the migration in and out of the chosen energy range. 
A partial remedy consists of using a statistically less precise sample of positive and negative 
lepton tracks in a selected sample of $Z$ boson decays. However, due to the different 
weights of the $V-A$ and $V+A$ couplings of the $Z$ boson to leptons, even these track samples
are biased. In both cases these biases can be corrected for, but the correction factor will 
be sensitive to the uncertainty in the momentum spectra of the valence quarks.

Given the above sources of the uncertainties, we assume the following two values for the 
size of the biases, both for the charge independent and charge dependent scale shifts
\begin{eqnarray}
\es_\lp=+\es_\lm&=&\pm 0.5\percent,\,\pm 0.05\percent,\\
\label{eq:epspval}
\es_\lp=-\es_\lm &=&\pm 0.5\percent,\,\pm 0.05\percent. 
\label{eq:epsmval}
\end{eqnarray}
The first value corresponds to the precision which can be achieved on the basis of the initial 
geometrical survey and 
the initial measurement of the field maps. The second one corresponds to what, in our view,
can be achieved using the above data based on the calibration methods --given all 
the LHC specific effects-- which make this procedure more difficult at the LHC 
than at the Tevatron.  
\index{Weak modes!In ATLAS tracker|)}

\subsubsection{Resolution (RF) of the charged lepton track parameters }
\label{sss:resolution}
The finite resolution of measuring the lepton track 
parameters  may lead to biases in the measured value of 
$\MWp-\MWm$. We model the possible biases introduced by the 
ambiguity in the assumed size of the $\sigma_{1/\pT}$ (Eq.~(\ref{eq_rho_smearing}))
and $\sigma_{\cotan\theta}$ (Eq.~(\ref{eq_theta_smearing})) smearing by decreasing or increasing the 
widths of their Gaussian distributions by the factor $\mathrm{RF}=0.7,\,1.3$.

\section{In search for the optimal measurement strategy}\label{sec_results}

\subsection{Validation of the framework and behaviour of the likelihood analysis}
Before presenting the results, now that we have a global idea of the challenge we present a few more
details on the validation of the analysis framework from both technical and physical point of views
and look also at the influence on the results of some input parameters to the analysis process. 
The following tests were performed for trivial ($\xi=0$) as well as non trivial results 
($\xi\neq 0$) when necessary, \ie{} when $\DeltaPM(\xi)_\Min\gg 0$ and/or $\chiDmin/\dof\gg 1$.
Since most of these tests are quite redundant they were not included in the core of the Chapter in 
the aim not to break the flow of the expose. Rather than that, the gist of it was compiled at the
end of the present Chapter.
Hence, in what follows the corresponding sections in the Appendix are indicated so that the reader 
wishing to have quantitative results on a particular topic knows where to find them.

The first test consisted to cross check the framework using for all three methods the observable 
$\rhoTl$ (\S\,\ref{app_1}.\ref{app_val1}). All results ($\xi=0$ and $\xi\neq 0$) confirmed the 
calculus using $\pTl$-based methods.

Since trivial validation tests ($\xi=0$) displays $\MWp-\MWm=0$ the relevancy of the sign of the
result cannot be checked, also we generated $\PD$ with a shift of $+\,100\MeV$ with respect to the 
mass of reference of the $W$ bosons and saw that indeed the signs were coherent
(\S\,\ref{app_1}.\ref{app_val2}). 

Using these shifted masses in the $\PD$ allow on a technical level to see at which level the 
convergence of a result depends of the number and localisation of $\MT$ from the minimum of the 
parabola. Thus, by playing with the number of the $\MT$ samples participating to the analysis and
choosing them such that all $\MT$ masses are far and on one side only from the minimum we can show
that the convergence, and in consequence $\DeltaPM$, suffers in such extreme configurations 
(\S\,\ref{app_val_infl}.\ref{app_val3}). Nonetheless the order of the size of the error is respected.
Also, despite our reduced number of $\MT$ samples, in what follows results displaying large values of 
$\DeltaPM(\xi)_\Min$ can be trusted.

The influence of the detector resolution on the results was investigated by increasing the size of 
the bin widths of the $\PD$ and $\MT$ histograms before performing the analysis. It proved that the
$\chiD$ test is quite robust so that values are very close to the one obtained with the nominal
resolution, excepted for the convergence accuracy which suffers much more from lower resolution when 
$\DeltaPM(\xi)_\Min\gg 0$ (\S\,\ref{app_val_infl}.\ref{app_val4}).

The charge asymmetry method being new we considered a spread in the analysis range to perform the 
$\chiD$ analysis which is not possible in the classic method without a very good knowledge of the 
background at low $\pT$ and of the influence of $\GamW$ at high $\pT$.
Tests based on $\Asym{\pTl}$ performed in the range $20\GeV<\pTl<50\GeV$ and then
in $20\GeV<\pTl<60\GeV$ displayed no real enhancements on the results apart from slightly increasing
the precision of the convergence $\chiDmin/\dof$ (\S\,\ref{app_val_infl}.\ref{app_val5}). 
This can be understood as the mass of the $W$ has a real impact on the region where the jacobian 
peak arise.
Nonetheless with real data this could provide a good test to see a dependence of the results from
the background and $W$ width influence on the extraction of the mass with this method.

Last but not least the test of the influence of a $\ETmiss$ cut on the result was performed by
doing at the stage of the generation directly a cut on $\pTnu$. The consequences on the results
are absolutely not visible and it is believed that a realistic cut based on $\ETmiss$ would display
the same feature (\S\,\ref{app_val_infl}.\ref{app_val6}). 
This test justify then the non treatment of realistic $\ETmiss$ cuts and recoil modeling in our 
analysis.

To conclude, from all these tests the one having the most striking effect on the calculus was found
to be the lack of $\MT$ samples thinly separated and covering a large range of masses.
As stated previously the CPU was prohibitive to refine the analysis and would not have brought much 
enhancement as most of the systematic error $\DeltaPM(\xi)$ absolute values, as it will be shown, 
are such that $|\DeltaPM(\xi)|<200\MeV$.
Also, to preserve a better clarity of the discussion --as independent as possible from the choice of 
the input parameters-- we have chosen that all the $\chiD$ presented in the core of the Chapter were
made for mass templates $\MT$ covering the range $\pm 200\MeV$. 

Finally, just out of curiosity and to insist on the difference between the kinematics of the charged 
lepton decaying from a $\Wp$ or a $\Wm$ at the LHC we considered the case where --not aware of the
$V-A$ coupling of fermions in electroweak interactions-- 
the $\chiD$ test is made between pseudo-data related to the $\Wp$ information and the mass templates 
related to the $\Wm$ information.
Performing such a naive test assuming similar kinematics for the positive and negative charged 
leptons gives absolutely non relevant results, such that $\MWp-\MWm(\xi=0)\sim 1\GeV$ and 
$\chiDmin/\dof\sim 10,000$ (cf.~Appendix~\ref{app_how_not_to}).

\subsection{Reducing impact of systematic measurement errors}
\index{Charge asymmetry!Used for extraction of MWpmMWm@used for the extraction of $\MWp-\MWm$|(}
In Section~\ref{s_measurement_method} three measurement methods 
of $\DeltaPM$ have been presented\,:\;the classic, charge asymmetry and double charge asymmetry 
methods. The basic merits of the two latter 
methods is that they use the dedicated observables which are meant to be  largely insensitive 
to the precise understanding of the event selection and reconstruction efficiency, the background 
contamination level, understanding to the absolute calibration and the biases
of the reconstruction of the neutrino transverse momentum, the internal and 
external (dead-material) radiation. It will remain to be proved, using the data 
collected at the LHC, that all these error sources have negligible impact on the 
precision of the $\DeltaPM$ measurement. At present, such a statement must rely on  
the extrapolation of the Tevatron experience.
In what follows, links to detailed comments and/or graphics for each addressed systematic error are 
made to the Appendix~\ref{app_detailed_chi2_res} since for clarity we kept only the essential 
information in the core of the Chapter.

The impact of the remaining measurement errors specified in the previous Section 
and quantified using the analysis methods discussed in Section~\ref{s_analysis_strategy}  
is presented in Table~\ref{table_exp_sys_classic_vs_casym_vs_dcasym}.

The precision of estimating the systematic shifts of $\DeltaPM(\xi)$ for each of 
the systematic effect $\xi$ and each measurement method is assessed using the validation procedures 
described in Section~\ref{s_analysis_strategy}. The resulting $\delta\left[\DeltaPM(\xi)\right]$ 
of $\approx 5\MeV$ corresponds the collected luminosity of $10\,\mathrm{fb}^{-1}$. 
The first observation is that the precise understanding of the measurement 
smearing RF of the track parameters does not introduce any bias in the 
measured values of $\DeltaPM$. The impact of the energy scale errors 
on the $\DeltaPM$ biases differs for each of the discussed methods.
\index{Double charge asymmetry!Used for the extraction of MWpmMWm@
Used for the extraction of $\MWp-\MWm$|(}
\begin{table}[]
\begin{center}
\renewcommand\arraystretch{1.45}
\begin{tabular}{|c|c||r@{\kern\tabcolsep}>{\kern-\tabcolsep}l|r@{\kern\tabcolsep}>{\kern-\tabcolsep}l|r@{}l|}
  \cline{3-8}
  \multicolumn{2}{c|}{} & \multicolumn{6}{c|}   {$\MWp-\MWm\quad[\mm{MeV}]$} \\ 
  \cline{2-8}
  \multicolumn{1}{c}{}  & \multicolumn{1}{|c||} {Systematic $\xi$}
                        & \multicolumn{2}{c|}   {``Classic'' Method}
                        & \multicolumn{2}{c|}   {$\Asym{\pTl}$} 
                        & \multicolumn{2}{c|}   {$\DAsym{\rhoTl}$}  \\

  \hline
  \multicolumn{1}{|c}{MC truth}   & \multicolumn{1}{|c||}{$\xi=0$} 
  & \multicolumn{2}{c|}{$\!\!\!\!\!-2 \,\pm\, 3$ }
  & \multicolumn{2}{c|}{$\!\!\!\!\!-1 \,\pm\, 3$ }
  & \multicolumn{2}{c|}{$ 0 \,\pm\, 3 $} \\
  \multicolumn{1}{|c}{Cent. Exp.} & \multicolumn{1}{|c||}{$\xi=0$} 
  & \multicolumn{2}{c|}{$1 \,\pm\, 4$ } 
  & \multicolumn{2}{c|}{$1 \,\pm\, 4$ }
  & \multicolumn{2}{c|}{$0 \,\pm\, 4$ } \\
  \hline\hline
  \multirow{8}{*}{ES [\%]}
  & $\es_\lp=+\es_\lm=+0.05\,\%$ 
  &   \hspace*{1.25cm} $3$& 
  &   \hspace*{1.15cm} $2$&
  &   \multicolumn{2}{c|}{} \\
  & $\es_\lp=+\es_\lm=-0.05\,\%$ 
  &   $-2$& 
  &    $0$&
  &   \multicolumn{2}{c|}{ \multirow{2}{*}{$\huge\times$}} \\
  \arrayrulecolor{Greymin}
  \cline{2-6}
  \arrayrulecolor{Black}
  & $\es_\lp=+\es_\lm=+0.50\,\%$ 
  &   $16$&
  &    $8$& 
  &   \multicolumn{2}{c|}{}  \\
  & $\es_\lp=+\es_\lm=-0.50\,\%$ 
  &  $-36$&
  &   $-6$& 
  &   \multicolumn{2}{c|}{}  \\
  \arrayrulecolor{Greymax}
  \cline{2-8}
  \arrayrulecolor{Black}
  & $\es_\lp=-\es_\lm=+0.05\,\%$ 
  &  $-56$& 
  &  $-57$& 
  &  \multicolumn{2}{c|}{\multirow{2}{*}{1}} \\
  & $\es_\lp=-\es_\lm=-0.05\,\%$ 
  &   $57$& 
  &   $57$& 
  &  \multicolumn{2}{c|}{} \\
  \arrayrulecolor{Greymin}
  \cline{2-8}
  \arrayrulecolor{Black}
  & $\es_\lp=-\es_\lm=+0.50\,\%$
  & $-567$& 
  & $-611$& 
  & \multicolumn{2}{c|}{\multirow{2}{*}{-1}} \\ 
  & $\es_\lp=-\es_\lm=-0.50\,\%$
  &  $547$& 
  &  $515$&
  &  \multicolumn{2}{c|}{} \\ 
  \hline\hline
  \multirow{2}{*}{RF}
  & $0.7$ 
  &  $1$& 
  & $-2$& 
  & \multicolumn{2}{c|}{ \multirow{2}{*}{$\huge\times$}} \\
  & $1.3$ 
  & $-3$& 
  &  $3$& 
  & \multicolumn{2}{c|}{} \\
  \hline
\end{tabular}
\renewcommand\arraystretch{1.45}

  \caption[Experimental systematic errors for the classic method, the charge asymmetry and the 
    double charge asymmetry]
          {\figtxt{Experimental systematics errors for the classic method, the charge asymmetry 
              and the double charge asymmetry. 
              In Appendix~\ref{app_detailed_tables}, 
              Table~\ref{table_app_exp_sys_classic_vs_casym_vs_dcasym} 
              reproduces the present results with more details.
            }}
          \label{table_exp_sys_classic_vs_casym_vs_dcasym} 
\index{Chi2 Likelihood analysis@Chi-2 ($\chiD$) likelihood analysis!Results}
\end{center}
\end{table}

For the lepton charge independent shift even the ``initial'' ($0.5\percent$)
scale error has no statistically significant impact on the measurement precision 
for the charge asymmetry method. For the classic method the scale error 
has to be reduced to the ``ultimate'' value of ($0.05\percent$) to achieve 
a comparable  measurement precision of $\DeltaPM$.

\begin{figure}[!h] 
  \begin{center}
    \includegraphics[width=0.495\tw]{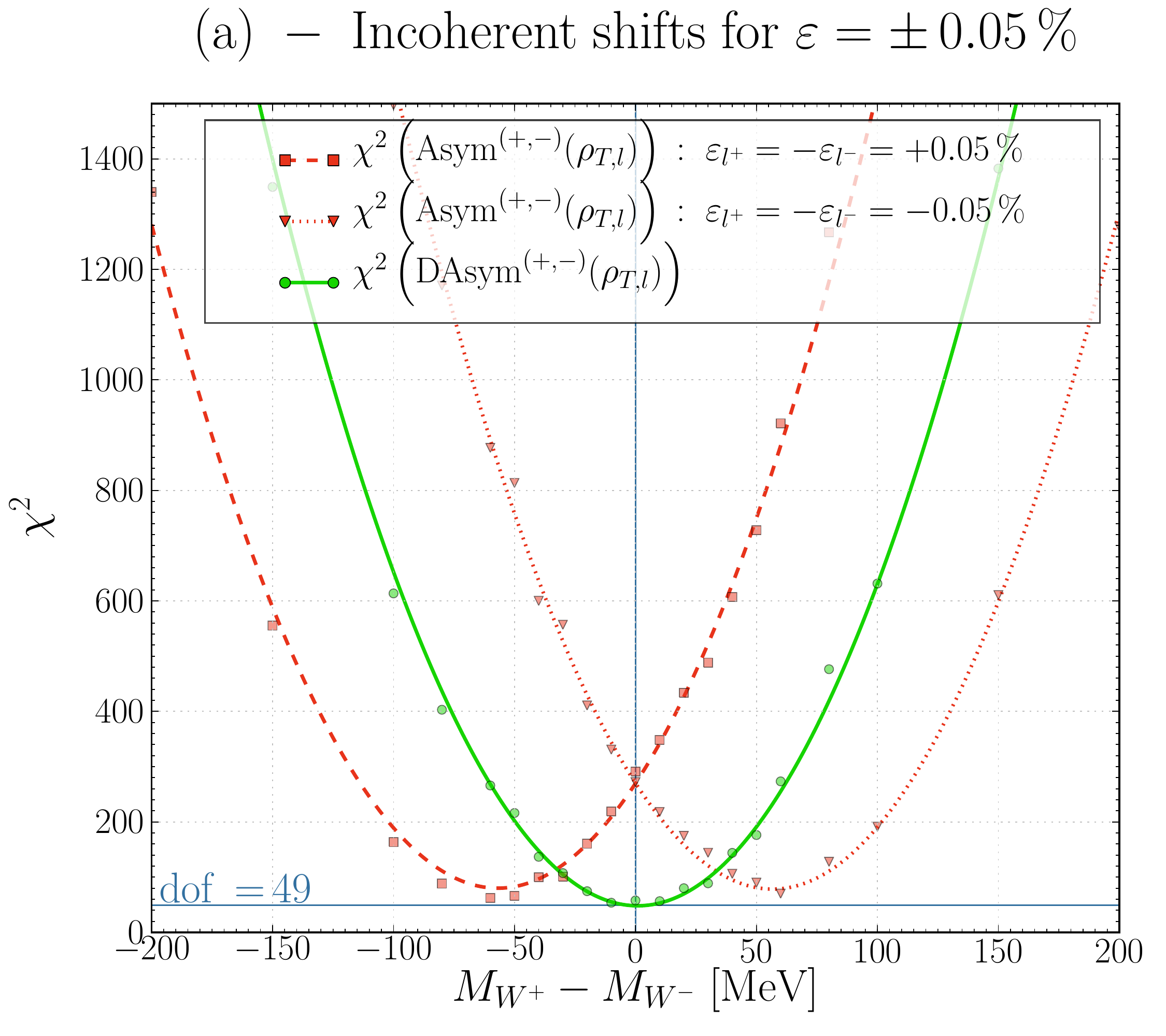}
    \hfill
    \includegraphics[width=0.495\tw]{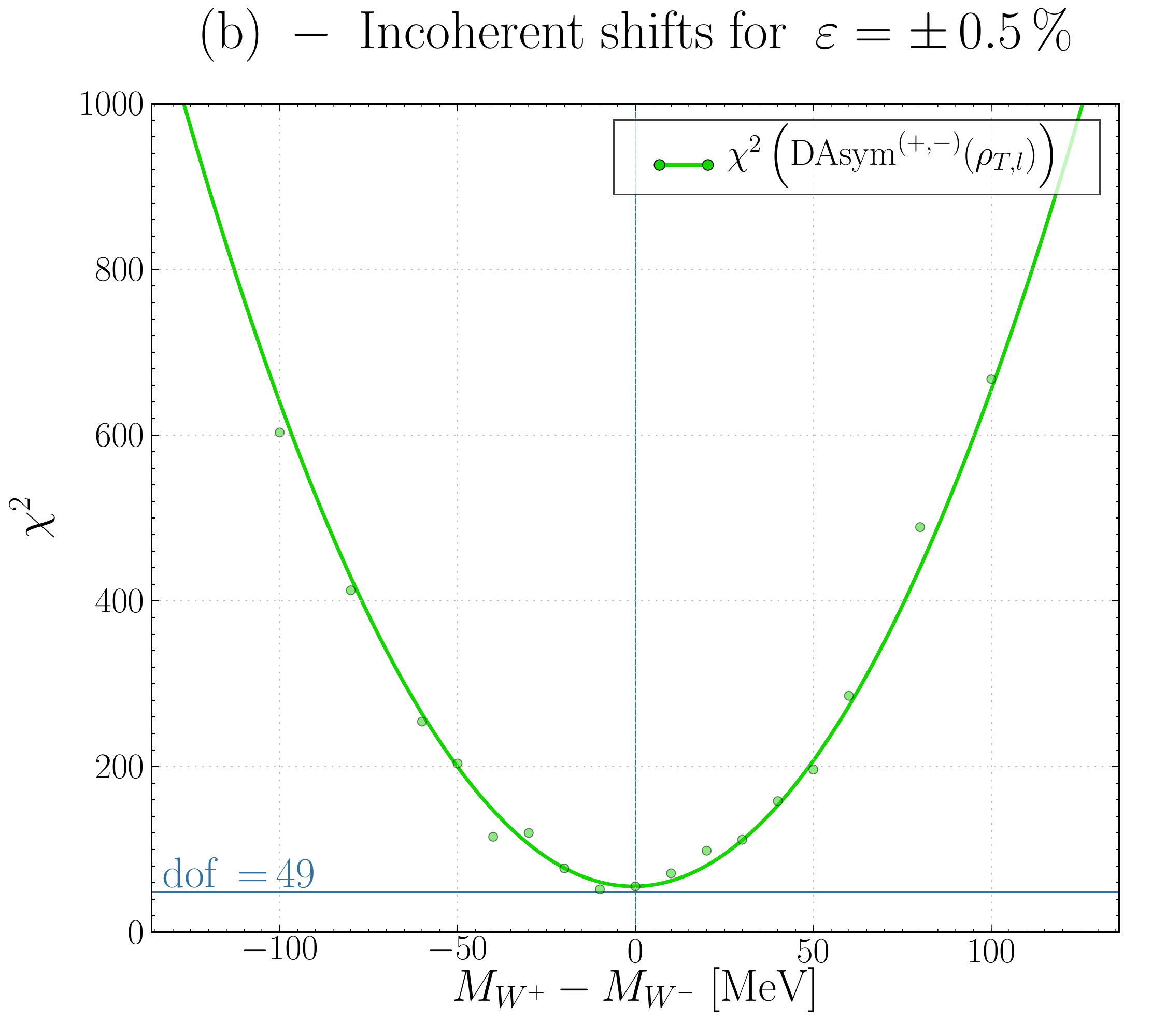}
    \caption[$\chiD$ results for the incoherent shifts of the energy scale
                for both the asymmetry and the double charge asymmetry $\varepsilon=\pm 0.05\%$
                and double charge asymmetry only for $\varepsilon=\pm 0.5\,\%$]
            {\figtxt{The $\chiD$ results for the incoherent shifts of the energy scale
                for both the asymmetry and the double charge asymmetry using 
                $\varepsilon=\pm 0.05\%$ in (a), and only for the double charge asymmetry 
                with $\varepsilon=\pm 0.5\,\%$ ($\es_\lp=-\es_\lm>0$ for the first half year and
                $\es_\lp=-\es_\lm<0$ during the over half) in (b).
              }
            }
            \label{fig_chi2_ES_asym_dasym}
  \end{center} 
\end{figure}

For the lepton charge dependent shifts the classic and asymmetry methods
provide similar measurement precision. The measurement error remains to be 
of the order of $\approx 60\MeV$ even if the ultimate precision of controlling the 
energy scale biases to $0.05\percent$ is reached. The double charge asymmetry method reduces 
the measurement error to the extend that the resulting  bias is statistically 
insignificant, even for the initial scale uncertainty. This is illustrated 
in Figures~\ref{fig_chi2_ES_asym_dasym}.(a) and \ref{fig_chi2_ES_asym_dasym}.(b). 
These plots show the comparison of the $\chiD$ fits for the charge asymmetry method 
and the double charge asymmetry method for the lepton charge dependent scale error of 
$\es_l=\pm 0.05\percent$ (a), and the $\chiD$ fit corresponding to $\es_l=\pm 0.5\percent$ for 
the double charge asymmetry method (b). 
The results for the double charge asymmetry correspond to $\es_\lp=-\es_\lm>0$ for the first 
running period with the standard magnetic field configuration and $\es_\lp=-\es_\lm<0$ for the 
running period with the inverted direction of the $z$-component of the magnetic field.

The above reduction of the measurement sensitivity to the energy scale error can be achieved for 
the initial survey precision of the tracker alignment. Such a survey will have to be made at the 
beginning of each of the two running periods. A special 
care will have to be taken to understand the relative curl and twist deformations
induced by reversing the current in the solenoid. 
It has to be stressed that the precision of the double charge asymmetry method
is insensitive to the relative $\vec E\times \vec B$ biases of the reconstructed 
hit positions for the two data taking periods, provided that they are 
not larger than $10$ times the average hit reconstruction precision achieved
in the standard field configuration running period. 
Worsening of the hit position resolution for the inverted field configuration, 
driven by the geometrical layout of the silicon tracker modules,  
have no significant effect on the measurement precision. 
Similarly, the required level of precision of understanding the hysteresis effects, leading to 
inequality of the absolute field strength in the two running periods, corresponding 
to reverse solenoid current directions can be achieved with the standard field 
mapping methods. Note that the precision required for the asymmetry measurement may be up to 
$10$ times worse with respect to the one needed for the measurement of the average $W$ boson mass.   
The reduced sensitivity to all the above effects 
is due to the fact that the impact of each of these effects is, to a large 
extent, canceled in each of the running periods. This is done in the same way as canceling the  
time-dependent effects of the detector response, calibration and alignment procedures.   
Note, that the residual impact of all the above effects can be reduced further
(if necessary)  using the 
$\vec B$-field configuration dependent analyses of straight track residua and/or 
the position of the reconstructed $Z$ boson mass peak.
\index{Double charge asymmetry!Used for the extraction of MWpmMWm@
Used for the extraction of $\MWp-\MWm$|)}

Detailed comments and graphics on the impact of 
the energy scale ES of the charged lepton and of
the resolution smearing RF of its track parameters
are presented respectively in \S\,\ref{app_detailed_chi2_res}.\ref{ss_a1} and 
\S\,\ref{app_detailed_chi2_res}.\ref{ss_a2} of Appendix~\ref{app_detailed_chi2_res}.

\subsection{Reducing impact of systematic modeling errors}
As discussed in the previous section, by using the charge asymmetry 
(double charge asymmetry) methods the systematic measurement precision of 
$\DeltaPM$ could  be reduced to the level of ${\cal O}(10)\,\MeV$. 
In this section we discuss the impact of the 
modeling uncertainties described in Subsection~\ref{s_model_impact_sys_pheno} 
on the measurement 
precision of $\DeltaPM$ for the charge asymmetry method%
\footnote{From the point of view of the modeling uncertainties,
the charge asymmetry and the double charge asymmetry methods are equivalent
and the discussed results are the same for both methods.}.
The detailed discussion and graphics related to the systematic errors due to the uncertainties
on the intrinsic $\kT$, global PDF, the valence $u-d$ and sea $s-c$ asymmetry are present
from \S\,\ref{ss_a3} through \S\,\ref{ss_a6} of
Appendix~\ref{app_detailed_chi2_res}.

In Table~\ref{table_charge_asym_results} we show, in the first column, the expected measurement 
biases of $\DeltaPM$ due to the dominant modeling uncertainties, discussed in the 
previous section, for $\pp$ collisions at the LHC energy. 
We do not see a significant impact of the coherent shifts 
of the partonic distributions, defined in the previous section and denoted as 
the PDFs uncertainty. It would be, however,  misleading to conclude 
prematurely that the  $\DeltaPM$ biases are  insensitive to the uncertainties in the partonic 
distributions.  

\begin{table}[]
\begin{center}
\begin{TableSize}
\renewcommand\arraystretch{1.5}
\begin{tabular}
{|c|l||r@{\kern\tabcolsep}>{\kern-\tabcolsep}l|r@{\kern\tabcolsep}>{\kern-\tabcolsep}l|r@{\kern\tabcolsep}>{\kern-\tabcolsep}l|r@{\kern\tabcolsep}>{\kern-\tabcolsep}l|}
  \cline{3-10}
  \multicolumn{2}{c|}{} & \multicolumn{8}{c|}  {$\MWp-\MWm\;[\mm{MeV}]$ results using the Charge Asymmetry of $\pTl$} \\ \cline{2-10}
  \multicolumn{1}{c|}{} & \multicolumn{1}{c||} {Systematic $\xi$}
                        & \multicolumn{2}{c|}   {$\pp$ - $|\etal|<2.5$}
                        & \multicolumn{2}{c|}   {$\pp$ - $|\etal|<0.3$  } 
                        & \multicolumn{2}{c|}   {$\pp$ - $|\yW|<0.3$}
                        & \multicolumn{2}{c|}   {$\dd$ - $|\etal|<2.5$}  \\

  \cline{2-10}\hline
  \multicolumn{1}{|c|}{MC truth}   
  & \multicolumn{1}{c||}{$\xi=0$} 
  & \multicolumn{2}{c|}{$\!\!\!\!-1 \,\pm\, 3$ }
  & \multicolumn{2}{c|}{$0  \,\pm\, 1$ }
  & \multicolumn{2}{c|}{$0  \,\pm\, 1$ }
  & \multicolumn{2}{c|}{$0 \,\pm\, 5$ } \\
  \multicolumn{1}{|c|}{Cent. Exp.} 
  & \multicolumn{1}{c||}{$\xi=0$} 
  & \multicolumn{2}{c|}{$1 \,\pm\, 4$ }
  & \multicolumn{2}{c|}{$0 \,\pm\, 4$ }
  & \multicolumn{2}{c|}{$1 \,\pm\, 4$ }
  & \multicolumn{2}{c|}{$4 \,\pm\, 5$ } \\
  \hline\hline
  \multirow{5}{*}{$\Mean{\kT}$ [GeV]}
  & \multicolumn{1}{c||}{$2$} 
  & \hspace*{.8cm} $8$&
  & \hspace*{.8cm} \cellcolor{hl}{$0$}&\cellcolor{hl}{} 
  & \hspace*{.9cm} \cellcolor{hl}{$2$}&\cellcolor{hl}{} 
  & \hspace*{.65cm} $28$& \\
  & \multicolumn{1}{c||}{$3$}
  &  $7$&
  &  \cellcolor{hl}{$3$}&\cellcolor{hl}{}
  & \cellcolor{hl}{$-2$}&\cellcolor{hl}{} 
  & $20$& \\
  & \multicolumn{1}{c||}{$5$}
  & $-4$&
  & \cellcolor{hl}{$-3$}&\cellcolor{hl}{}
  & \cellcolor{hl}{$-6$}&\cellcolor{hl}{} 
  &$-15$& \\
  & \multicolumn{1}{c||}{$6$}
  & $-8$&
  &  \cellcolor{hl}{$2$}&\cellcolor{hl}{}
  & \cellcolor{hl}{$-5$}&\cellcolor{hl}{} 
  &$-35$&  \\
  & \multicolumn{1}{c||}{$7$}
  & $-16$& 
  &   \cellcolor{hl}{$2$}&\cellcolor{hl}{}
  &  \cellcolor{hl}{$-8$}&\cellcolor{hl}{} 
  & $-49$& \\
  \hline\hline
  \multirow{2}{*}{PDF$^{(\ast)}$}
  & \multicolumn{1}{c||}{Min.} 
  & $-4$&
  &  $6$&
  &  $0$& 
  & $-3$& \\
  & \multicolumn{1}{c||}{Max.}
  &  $4$&
  & $-8$&
  &  $5$&
  &  $8$& \\
  \hline\hline
  \multirow{8}{*}{$u^\val,\,d^\val{}^{(\ast)}$}
  & \twolinebox{${u^\val_\Max=1.05\,u^\val}$}{${d^\val_\Min\,=d^\val-0.05\,u^\val}$} 
  & $115$&
  &  $69$&
  & $-38$&
  &  \cellcolor{hl}{$3$}&\cellcolor{hl}{} \\
  \arrayrulecolor{Greymin}
  \cline{2-10}
  \arrayrulecolor{Black}
  & \twolinebox{${u^\val_\Min\,=0.95\,u^\val}$}{${d^\val_\Max=d^\val+0.05\,u^\val}$} 
  & $-139$&
  &  $-87$&
  &   $60$&
  &  \cellcolor{hl}{$5$}&\cellcolor{hl}{} \\
  \arrayrulecolor{Greymax}
  \cline{2-10}
  \arrayrulecolor{Black}
  & \twolinebox{${u^\val_\Max=1.02\,u^\val}$}{${d^\val_\Min\,=0.92\,d^\val}$} 
  &  $84$& 
  &  $53$& 
  & $-31$& 
  &   \cellcolor{hl}{$1$}&\cellcolor{hl}{} \\
  \arrayrulecolor{Greymin}
  \cline{2-10}
  \arrayrulecolor{Black}
  & \twolinebox{${u^\val_\Min\,=0.98\,u^\val}$}{${d^\val_\Max=1.08\,d^\val}$}
  & $-89$&
  & $-57$& 
  &  $44$& 
  &  \cellcolor{hl}{$6$}&\cellcolor{hl}{} \\
  \hline\hline
  \multirow{6}{*}{$s,\,c^{(\ast)}$}
  & \twolinebox{${c_\Min\,=0.9\,c},$}{${s_\Max=s+0.1\,c}$}
  & $17$&
  & $10$&
  & \cellcolor{hl}{$7$}&\cellcolor{hl}{}
  & $20$& \\
  \arrayrulecolor{Greymin}
  \cline{2-10}
  \arrayrulecolor{Black}
  & \twolinebox{${c_\Max=1.1\,c},$}{${s_\Min\,=s-0.1\,c}$}
  & $-11$& 
  & $-10$& 
  &  \cellcolor{hl}{$0$}&\cellcolor{hl}{} 
  & $-16$& \\
  \arrayrulecolor{Greymax}
  \cline{2-10}
  \arrayrulecolor{Black}
  & \twolinebox{${c_\Min\,=0.8\,c},$}{${s_\Max=s+0.2\,c}$}
  & $39$& 
  & $25$& 
  &  \cellcolor{hl}{$6$}&\cellcolor{hl}{} 
  & $38$& \\
  \arrayrulecolor{Greymin}
  \cline{2-10}
  \arrayrulecolor{Black}
  & \twolinebox{${c_\Max=1.2\,c},$}{${s_\Min\,=s-0.2\,c}$}
  & $-29$& 
  & $-24$& 
  & \cellcolor{hl}{$1$}&\cellcolor{hl}{}
  & $-34$& \\
\hline
\end{tabular}
\renewcommand\arraystretch{1.45}

\end{TableSize}
  \caption[The shifts of the $W$-mass charge asymmetry
    corresponding to various modeling effects using the charge asymmetry of $\pTl$ for the
    analysis]
          {\figtxt{The shifts of the $W$-mass charge asymmetry
                   corresponding to various modeling effects.
                   The systematic labeled $\ast$ are obtained using the scaling trick mentioned in 
                   \S\,\ref{ss_scaling_trick}.
                   In Appendix~\ref{app_detailed_chi2_res} \S\,\ref{app_detailed_tables} 
                   Table~\ref{table_app_charge_asym_results} presents a detailed reproduction of
                   these results.
          }}
          \label{table_charge_asym_results}
          \index{Quarks!smc@$s-c$ asymmetry}
          \index{Quarks!uvmdv@$u^\val-d^\val$ asymmetry}
          \index{Quarks!Intrinsic transverse momenta}
          \index{Chi2 Likelihood analysis@Chi-2 ($\chiD$) likelihood analysis!Results}
\end{center}
\end{table}

\index{Quarks!uvmdv@$u^\val-d^\val$ asymmetry|(}
\index{Quarks!Valence quarks|(}
Indeed, the  present uncertainty of the relative distribution of the 
the $u$ and $d$ valence quarks (the  $u^\val-d^\val$ asymmetry) 
leads to large shifts in the $\DeltaPM$ values. These shifts 
are specific to the LHC $\pp$ collider and are largely irrelevant for the Tevatron $\ppbar$ 
collisions. This might explain why they were neglected in the previous 
studies~\cite{Buge:2006dv,Besson:2008zs}, in spite that they concern the average $W$ boson mass 
measurement. 
There are three origins for these shifts. The effects due to each of them add up 
and result in the amplification of the biases. The discussion and the numbers provided in 
\S\,\ref{app_ss_kT} of Chapter~\ref{chap_w_pheno_in_drell-yan} might be helpful to understand the
following ideas.
Firstly, increasing the $u^\val$ content 
of the proton shifts downwards the average momentum of the $\bar d$ anti-quarks.
This leads to an increase of the average transverse momentum $\Mean{p_{T,\dbar}}$ of the 
$\bar d$ anti-quarks producing $\Wp$, mimicking the increase of the $\Wp$ boson mass. 
Simultaneous decreasing of the $d^\val$ acts in the opposite direction for $W^-$,
\ie{} this time the average $\pT$ of the colliding sea quark decreases which  
lead to large and positive values of $\DeltaPM$. Secondly, 
at the LHC, contrary to the Tevatron, the presence of the $d^\val$ quarks 
leads to an asymmetry in the production rate of the $W$ boson from the $c$
quarks and $\bar c$ anti-quarks. Since the average transverse 
momentum of the charm quarks is higher with respect to the 
light quarks, this asymmetry shows up in the relative shifts 
in the $\pTl$ distributions for positive and negative leptons.
Increasing the density of the $d^\val$ quarks mimics 
thus the effect of increasing the mass of the $\Wm$ with respect to the 
$\Wp$ boson. 
The above two effects are amplified by the bias in 
the degree of the transverse polarisation 
of $\Wm$ with respect to $\Wp$,   
induced by the event selection procedure based on the 
lepton kinematics. The relative movements of the $d^\val$ 
and $u^\val$ amplify (attenuate) the initial event selection 
procedure bias.  
What must be stressed is that if the $d^\val$ shifts are  
compensated by the corresponding shifts of the $u^\val$ distributions,
they cannot be constrained to a better precision by the present data,
and they will not affect the rapidity distributions of the $Z$ boson. 
Thus, it will be difficult to pin them down using the standard 
measurement procedures. 
\index{Quarks!uvmdv@$u^\val-d^\val$ asymmetry|)}
\index{Quarks!Valence quarks|)}

The uncertainties of the relative density of the strange and charm quarks, \index{Quarks!smc@$s-c$ asymmetry}
the $s-c$ asymmetry, gives rise to smaller but significant biases  in the $\DeltaPM$ values,
as shown in the first column of  Table~\ref{table_charge_asym_results}. 
Since the transverse momentum of the $c$ quarks is significantly higher than the corresponding 
momentum for the $s$ quarks, this effect, even if  Cabbibo suppressed,  cannot be neglected. 
What must be stressed again is that if the $c$ shifts are \index{Electroweak!CKM matrix elements}
compensated by the corresponding shifts of the $s$ distributions,
they will not affect the rapidity distributions of the $Z$ bosons. 
Thus, it will be difficult to pin them down using the standard 
measurement procedures. This asymmetry can be constrained unambiguously only  
by using dedicated measurements, \eg{} by measuring the associated production of the $W$ bosons 
and charmed hadrons.  

Compared to the above, the biases corresponding to the 
uncertainties in the flavour independent smearing of the intrinsic transverse 
momentum distribution of partons are smaller in magnitude and can be neglected, 
if the intrinsic transverse momentum of partons is controlled to the precision of 
$2\GeV$.\index{Quarks!Intrinsic transverse momenta}

It is obvious from the above discussion that using the standard measurement 
procedures, the modeling uncertainties will be the dominant  source of the 
measurement errors of the $W$ boson mass asymmetry, already for the collected luminosity 
$100$ times smaller than the one considered here.
In order to diminish the impact of the modeling errors on the measurement of $\DeltaPM$
to a level comparable to statistical and experimental measurement errors, some dedicated 
measurement methods must be applied. Two such procedures are proposed and evaluated below\,:\;%
(1) the narrow bin method and (2) the isoscalar beams method.

\subsubsection{Narrow bin method}
\label{sss:narrowbin}
As discussed previously, the dominant source of  
large uncertainties in  $\DeltaPM$ comes from the presence of the 
valence quarks in the WBpB and from the uncertainties in their 
flavour dependent momentum distributions. In order to reduce this effect 
we propose to profit from the large centre of mass energy of the LHC and 
measure  $\DeltaPM$ using a selected fraction of the $W$ bosons which are 
produced predominantly by the sea rather than by the valence quarks. These 
$W$ bosons are produced with small longitudinal momentum in the laboratory frame.
 
Two methods of selecting such a sample are discussed below. The first
is based on restricting the measurement region to  $|\etal|<0.3$. 
The merit of the $\etal$-cut based selection is that it uses a directly 
measurable kinematic variable. Its drawback is that  rather broad 
spectrum of the longitudinal momenta of annihilating partons is accepted due to 
the large mass of the $W$ boson. The second is based on
restricting the measurement region to $|\yW|<0.3$. Here, only a narrow 
bin of the longitudinal momenta of annihilating partons is accepted in the region 
where the sea quarks outnumber the valence quarks. However, $\yW$ cannot 
be measured directly. It has to be unfolded from the measured transverse momentum 
of the charged lepton and the reconstructed transverse momentum of the neutrino.
The unfolding procedure \cite{Bodek:2007cz} neglects the width of the 
$W$ boson and depends upon the initial assumption of the relative momentum spectra
of the valence and sea quarks. However, in  the selected kinematic region 
the above approximation are expected to lead to a negligible measurement bias.
It has to be stressed that the narrow bin measurements will require a $10$ times
higher luminosity to keep the statistical error of $\DeltaPM$ at the level of 
$5\MeV$. Therefore, the results presented below for the narrow bin method
correspond to an integrated luminosity of $100\, \mathrm{fb}^{-1}$ and  
are based on the dedicated set of the simulated mass template and pseudo-data 
samples. Each sample contains $N_\Wp=1.74\times 10^9$ and $N_\Wm=1.14\times 10^9$
simulated (weighted) events, respectively.
\index{Luminosity!At the LHC (expected)!Integrated!High}

The systematic biases of $\DeltaPM$ due to modeling uncertainties discussed 
in the previous sections are presented in columns 2 and 3 of  
Table~\ref{table_charge_asym_results} respectively for the $|\etal|<0.3$
and $|\yW|<0.3$ selections. The $\etal$-cut based method reduces slightly  
the biases related to the uncertainties in the $u^\val-d^\val$ and $s-c$ asymmetries.
The gain in the measurement precision is clearly seen for the $\yW$-cut based method
which reduces to a negligible level the $s-c$ biases. \index{Quarks!smc@$s-c$ asymmetry}
It is interesting to note that the  $u^\val-d^\val$ shifts in $\DeltaPM$ change their signs 
for the above two methods, reflecting the importance of the $W$ boson polarisation effects 
discussed earlier.
\index{Quarks!uvmdv@$u^\val-d^\val$ asymmetry}

The narrow bin method allows, thus to reduce the impact of the $W$ boson modeling 
uncertainties on the $\DeltaPM$ biases to the level comparable to the statistical 
precision for all the effects, except for the $u^\val-d^\val$ asymmetry effect.
Here another remedy has to be found.\index{Quarks!Valence quarks}

\subsubsection{Isoscalar beams}
\label{sss:isobeams}

Isoscalar targets have been successfully used in most of  the previous fix-target
deep inelastic scattering experiments at SLAC, FNAL and CERN,
but this aspect has been rarely discussed in the context of the electroweak physics at the LHC.
The merits of the ion beams for the generic searches 
of the electroweak symmetry breaking mechanism at the LHC have been 
discussed in~\cite{Krasny:2005cb}. Their use as carriers of the parasitic electron 
beam, to measure the emittance of the WBpB at the LHC,  
has been proposed in~\cite{Krasny:2004ue}.
In our work we strongly advocate 
the merits of the isoscalar beams in improving the measurement 
precision of the parameters of the Standard Model. In this 
section we discuss their role in increasing the precision of
the measurement of $\DeltaPM$. We shall 
consider light ions\,: deuterium or helium.
As far as the studies of the $W$ boson asymmetries are concerned,
they are equivalent because  
shadowing corrections are quark-flavour independent.
The energies of the LHC ion beams satisfy the equal magnetic 
rigidity condition. For the isoscalar beams the nucleon energy 
is thus two times lower that the energy of the proton beam.
In the presented studies we assume that the ion--ion luminosity 
is reduced by the factor $A^2$ with respect to the $\pp$ luminosity
(cf. \S\,\ref{ss_lhc}).

In column 4 of Table~\ref{table_charge_asym_results}
we present the impact of the modeling uncertainties on the 
$\DeltaPM$ biases. The isoscalar beams allow to 
reduce the measurement biases due to the $u^\val-d^\val$ asymmetry effect
to a negligible level. This colliding beam configuration          
allows to profit from the isospin symmetry of the strong interactions 
which cancels the relative biases in the momentum distribution 
of the $u$ and $d$ quarks. It is interesting to note that, 
as expected, the $s-c$ biases are similar for the proton and 
for the light isoscalar beams. On the contrary, the biases 
related to the flavour independent intrinsic momentum of the 
quarks and anti-quarks are amplified due to the reduced centre of mass
collision energy%
\footnote{In order to amplify this effect, we have kept 
the same central value of the intrinsic transverse momentum 
smearing in the reduced collision energy as for the nominal collision
energy.}.   
Indeed, as seen already in \S\,\ref{ss_asmy_pTl_sqrtS} we came up to the
conclusion that at lower energies in the center of mass the charge asymmetry 
in the final state is much more visible since the charge symmetric sea 
contributions are not large enough.

\subsection{Two complementary strategies}

Two complementary strategies to achieve the ultimate measurement precision
of $\DeltaPM$ will certainly be tried. The first one will be based on 
an attempt to reduce the size of the systematic measurement and 
modeling uncertainties, discussed in the previous section. 
In our view, such a strategy will quickly reach the precision brick wall
-- mostly due to the a lack of data-driven constraints on modeling the 
flavour dependent $W$ boson production at the LHC energy. 
The second one, instead of reducing the size of the uncertainties, 
attempts to reduce their impact on the systematic error of the measured quantity by 
applying the dedicated methods. Such a strategy requires
running the dedicated machine and detector configurations. It is thus time 
and luminosity consuming. However, in our view, only such a strategy 
allows to measure the $W$ mass charge asymmetries at the precision comparable 
to the one achieved in the muon decay experiments.  
  
Let us recollect the main elements of the proposed dedicated measurement strategy
that allow to reduce the systematic errors to the level shown in 
the shaded areas of Table~\ref{table_charge_asym_results}\,:
\begin{itemize}
\item[-]
The charge asymmetry method allows 
to reduce the impact of most of the systematic measurement errors, 
except for the relative momentum scale errors for the positive and negative leptons.
If they cannot be experimentally controlled to the level of ${\cal O} (10^{-4})$,
their impact can be drastically reduced  in the dedicated LHC running periods  
using the double charge asymmetry method. 
\item[-]
The impact of the uncertainty in modeling of the 
intrinsic transverse momentum of the WBpB can  be reduced to a negligible level  using the 
narrow bin measurement method. 
\item[-]
The impact of the $s-c$ uncertainty can be attenuated  
using the $\yW$ selection based method in narrow bin. \index{Quarks!smc@$s-c$ asymmetry}
\item[-]
Finally, the impact of the $u^\val-d^\val$
uncertainty can be annihilated in the dedicated LHC runs with light isoscalar beams. 
\end{itemize}
\index{Charge asymmetry!Used for extraction of MWpmMWm@used for the extraction of $\MWp-\MWm$|)}

\cleardoublepage
\begin{subappendices}
\makeatletter\AddToShipoutPicture{%
\AtUpperLeftCorner{2cm}{2cm}{\ifodd\c@page\else\makebox[0pt]{\Huge$\bullet$}\fi}%
\AtUpperRightCorner{0cm}{2cm}{\ifodd\c@page\makebox[0pt]{\Huge$\bullet$}\else\fi}%
}\makeatother

\section{Validation tests and miscellanea technical variations of the analysis}
\label{app_validation}
\setlength{\epigraphwidth}{0.7\tw}
\epigraph{
Buster\,:\;``Actually, I'm studying cartography now, the mapping of uncharted territories.''\\
Michael\,:\;``Sure. Hasn't everything already sort of been discovered, though, by, like Magellan and 
Cort\'es? NASA,\dots you know ?'' \\
Buster\,:\;``Oh, yeah, yeah\dots Those guys did a pretty great job.''\\
Lucile\,:\;(appearing between her sons) ``Never hurts to double-check.''
}%
{\textit{Arrested Development - Extended Pilot}}

This Appendix presents in details the validations tests that were achieved to improve
the analysis framework. The points addressed are the following. 

The first part presents trivial results carried using the observable $\rhoTl$ to crosscheck the
one made with $\pTl$. Other trivial tests were made as well with $\pTl$-based distributions 
using shifted $W$ masses in the pseudo-data.
The second part treats of the steadiness of the results with respect to certain input parameters for
the analysis, such as the influence of the number of templates and their localisation with respect 
to the $\chiD$ minimum, the number of bins of the histograms, the window range for
the analysis and the emulation of $\ETmiss$ cuts.

In the rest of this Appendix unless stated otherwise the tests are made preferentially using 
$\pp$ collisions selected with the requirements shown in Eq.~(\ref{pTl-etal-cuts}) and in the
conventional range of $30\GeV<\pTl<50\GeV$.

\subsection{Details on the validation of the analysis framework}\label{app_1}
\subsubsection{Cross check of the analysis using the observable $\mbf{\rhoTl}$}\label{app_val1}
\begin{figure}[!ht] 
  \begin{center}
    \includegraphics[width=0.495\tw]{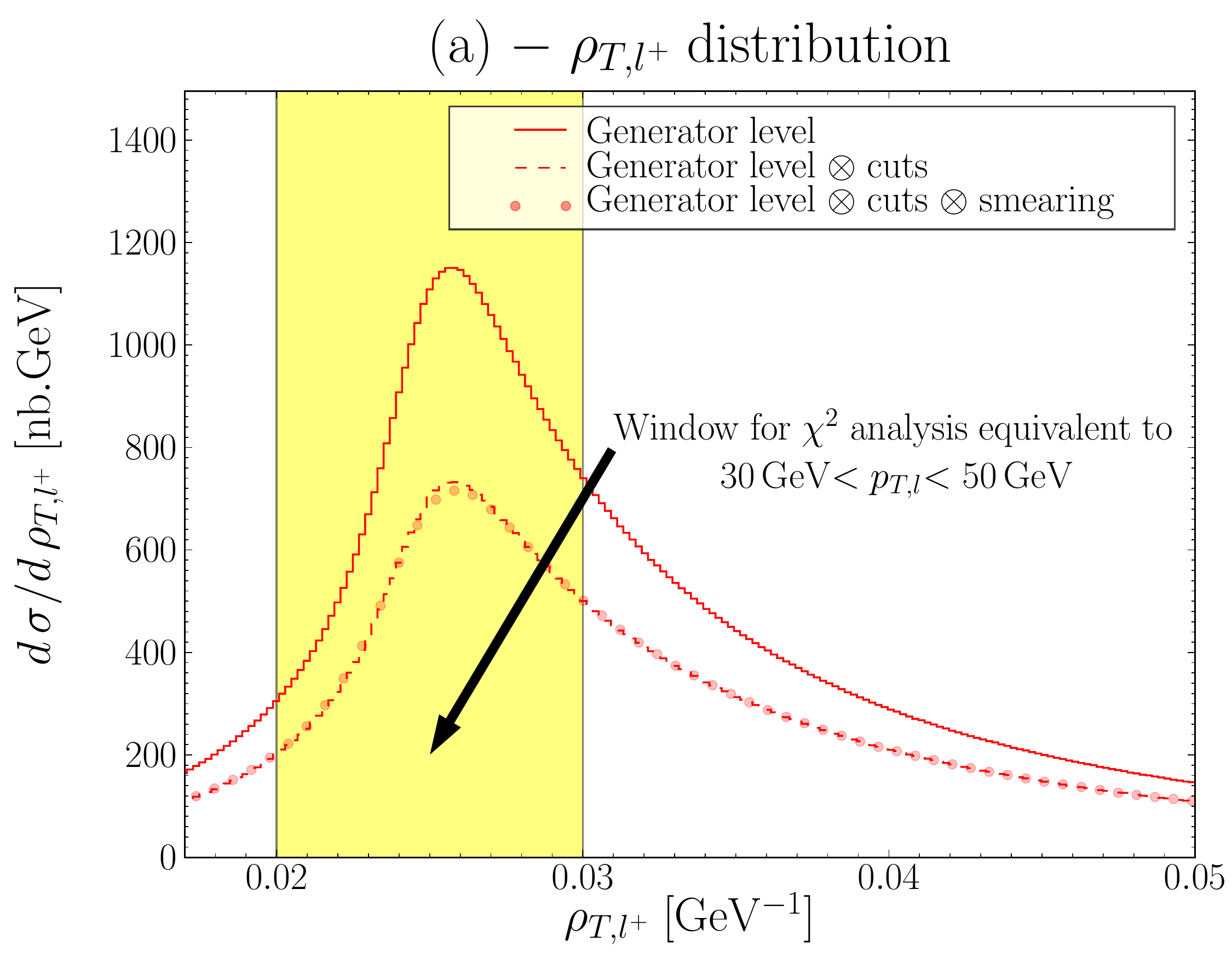}
    \hfill
    \includegraphics[width=0.495\tw]{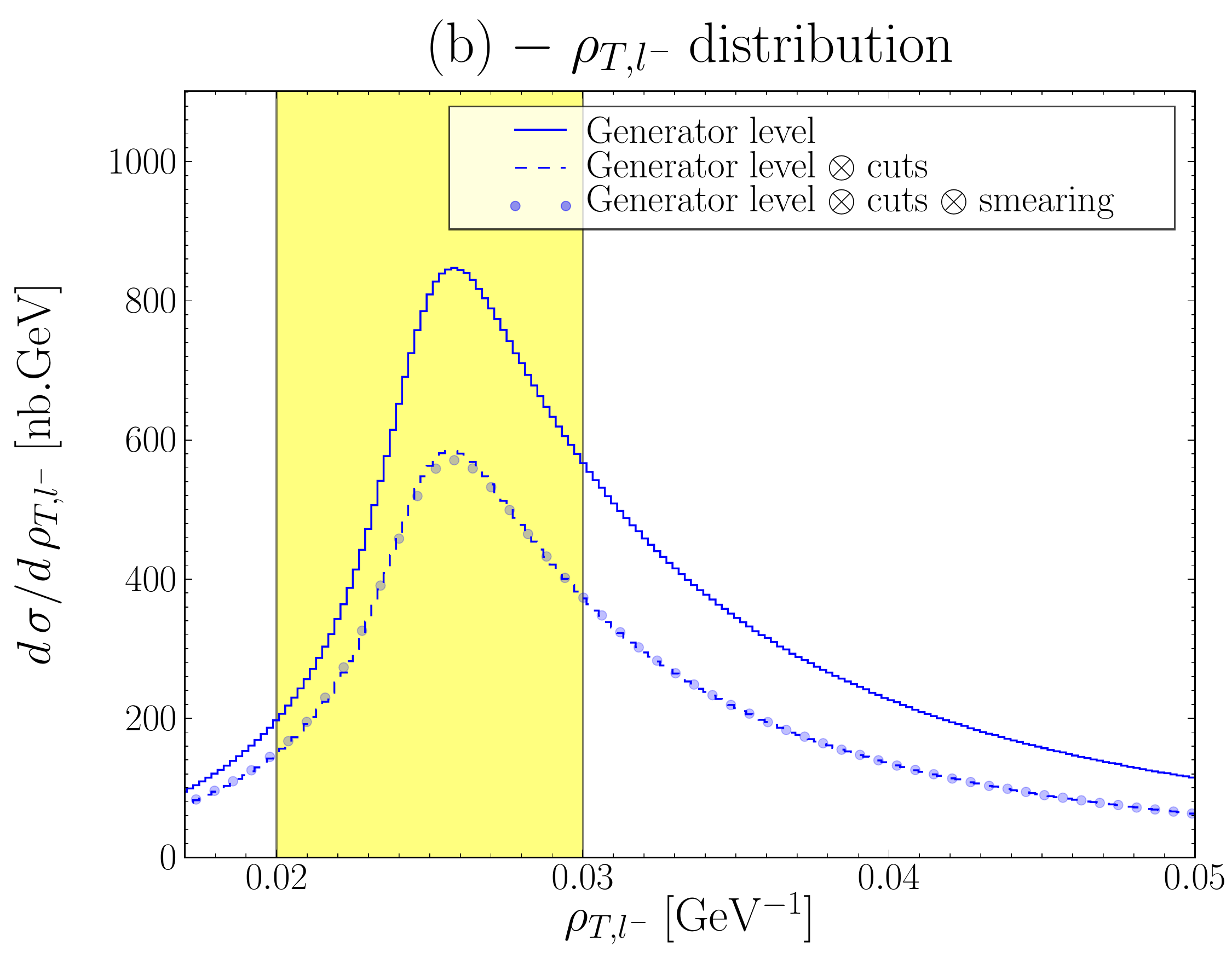}
    \vfill
    \includegraphics[width=0.495\tw]{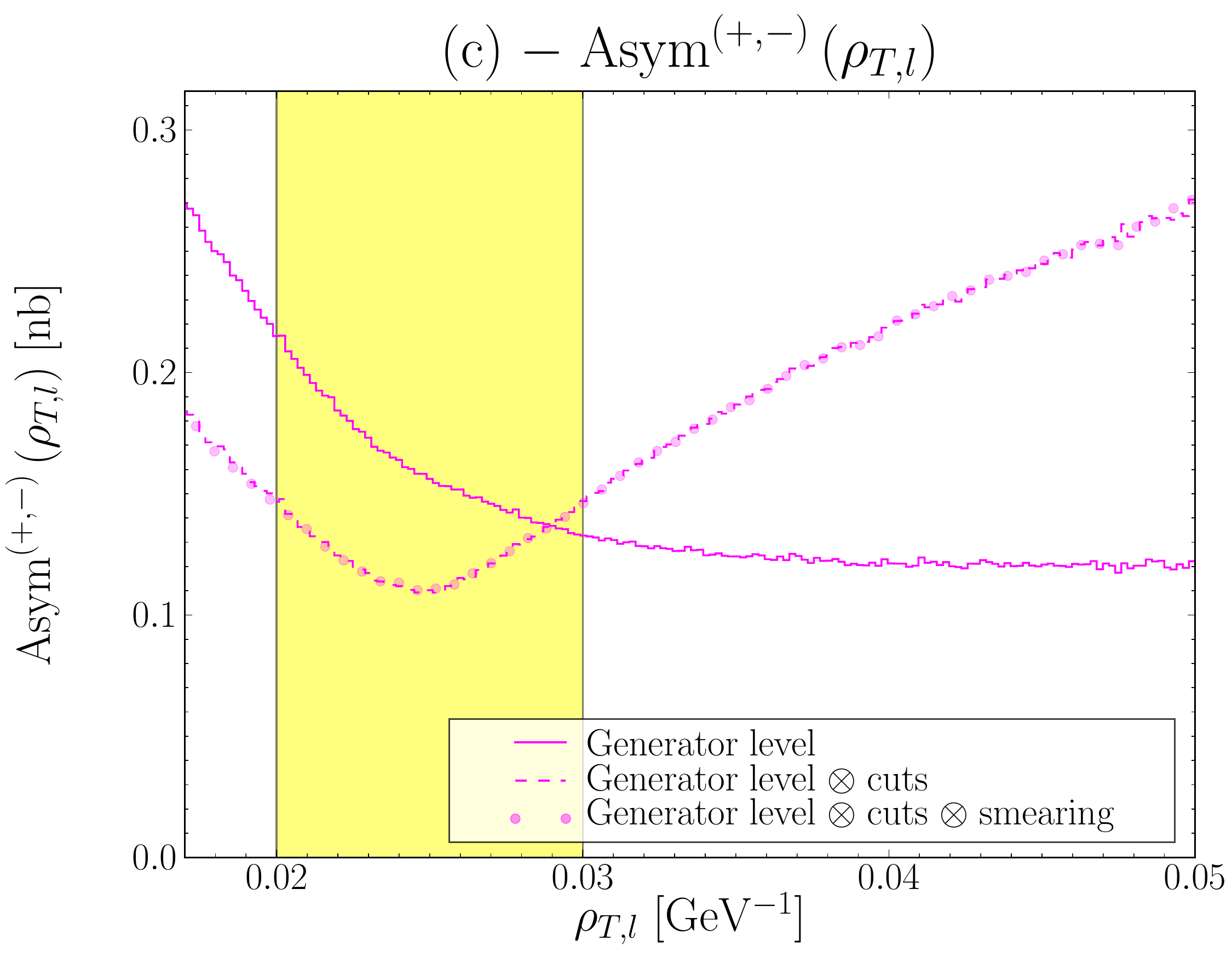}
    \hfill
    \includegraphics[width=0.495\tw]{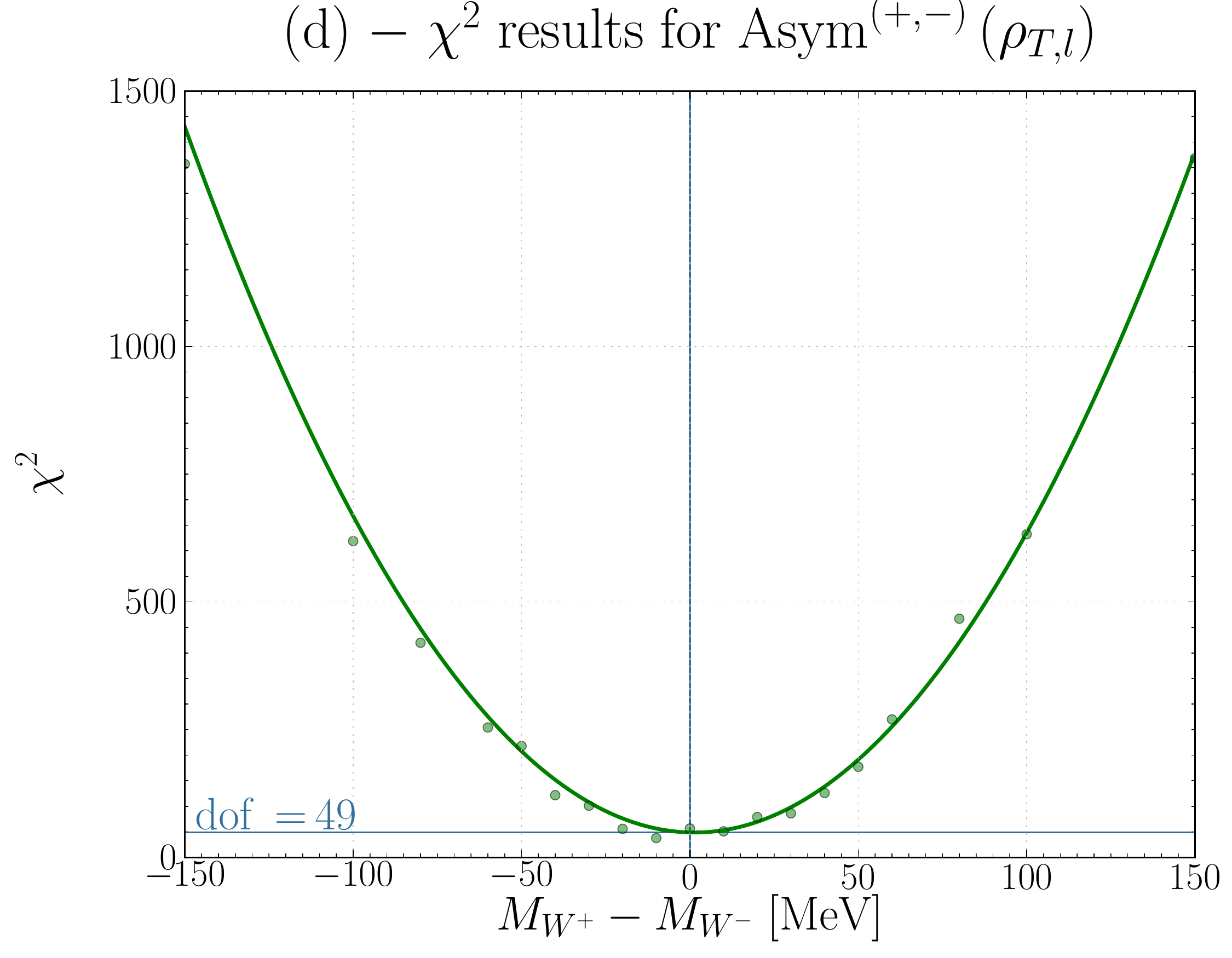}
    \caption[Visualisation of the inverse of the positive and negative charged lepton transverse 
      momenta and associated charge asymmetry at the generator level, adding up cuts and detector 
      smearing \& validation $\chiD$ analysis for the trivial unbiased pseudo-data]
            {\figtxt{Visualisation of the inverse of the positive (a) and negative (b) charged 
                lepton transverse momenta and associated charge asymmetry (c) at the generator level,
                adding up cuts and detector smearing. 
                Frame (d) shows the trivial $\chiD$ results for unbiased pseudo-data ($\xi=0$).}}
            \label{fig_app_rhol}
  \end{center} 
\end{figure}

Since the observable $\rhoTl$ is booked at the time of the generation it is pertinent to
cross check the framework with $\rhoTl$-based methods for the case of trivial ($\xi=0$) as well
as non trivial ($\xi\neq 0$) tests.
Besides, this simple studies allows as well to get accustomed to the behaviour of the variable
$\rhoTl$ and $\Asym{\rhoTl}$.
Starting with that last remark, Figure.~\ref{fig_app_rhol} presents in frames (a), (b) the behaviour
of the observable $\rhoTl$ for\,:\;%
(1) the generated and unselected sample of events, 
(2) the generated and selected sample of events and 
(3) the unbiased simulated detector response and selected sample of events.
In the each frame the range for the analysis $0.02\GeV<\rhoTl<50\GeV$ 
--equivalent to $30\GeV<\pTl<50\GeV$-- is 
highlighted.
Just like for the $\pTl$ distribution we can see that the cuts are responsible of a drastic change
in the behaviour of the $\rhoTl$ spectrum in particular in the region of the jacobian slope.
In Fig.~\ref{fig_app_rhol}.(d) a trivial $\chiD$ test was performed for an unbiased pseudo-data using
$\Asym{\rhoTl}$. As expected the value $\DeltaPM=0$ is found.
The results for $\xi\neq 0$ were found to be in a complete agreement 
--up to non avoidable numerical discrepancies-- with the one provided by $\pTl$-methods.

\subsubsection{Validation using shifted mass for the W boson mass in the pseudo-data}\label{app_val2}
Even though the previous trivial tests were found to be relevant, as long as the $\DeltaPM=0$ value 
doesn't have a sign it is --so far-- impossible to fully check the relevancy of the values. 
Also, the next logical step was to simulate unbiased pseudo-data but with shifted masses with respect
to $M_W^{\mm{(ref.)}}$ for the boson and see if in the analysis we rediscover those shifts. 
Three tests were carried for that purpose using both classic and charge asymmetry methods. 
In each of these tests, the central values for the masses of the $\Wp$ and $\Wm$ bosons were 
fixed for the generation of the pseudo-data to\,:
\begin{itemize}
\item[1.] $M_\Wp^\PD=M_W^\mm{(ref.)}+100\MeV \quad \& \quad M_\Wm^\PD=M_W^\mm{(ref.)}$.
\item[2.] $M_\Wp^\PD=M_W^\mm{(ref.)}         \quad \& \quad M_\Wm^\PD=M_W^\mm{(ref.)}+100\MeV$.
\item[3.] $M_\Wp^\PD=M_W^\mm{(ref.)}+100\MeV \quad \& \quad M_\Wm^\PD=M_W^\mm{(ref.)}+100\MeV$.
\end{itemize}
while the $\MT$ remained the same \ie{} generated with no biases and with 
$\MWp=\MWm=M_W^\mm{(ref.)}$ and where for reminder $M_W^\mm{(ref.)}=80.403\GeV$.

\begin{figure}[!ht] 
  \begin{center}
    \includegraphics[width=0.495\tw]{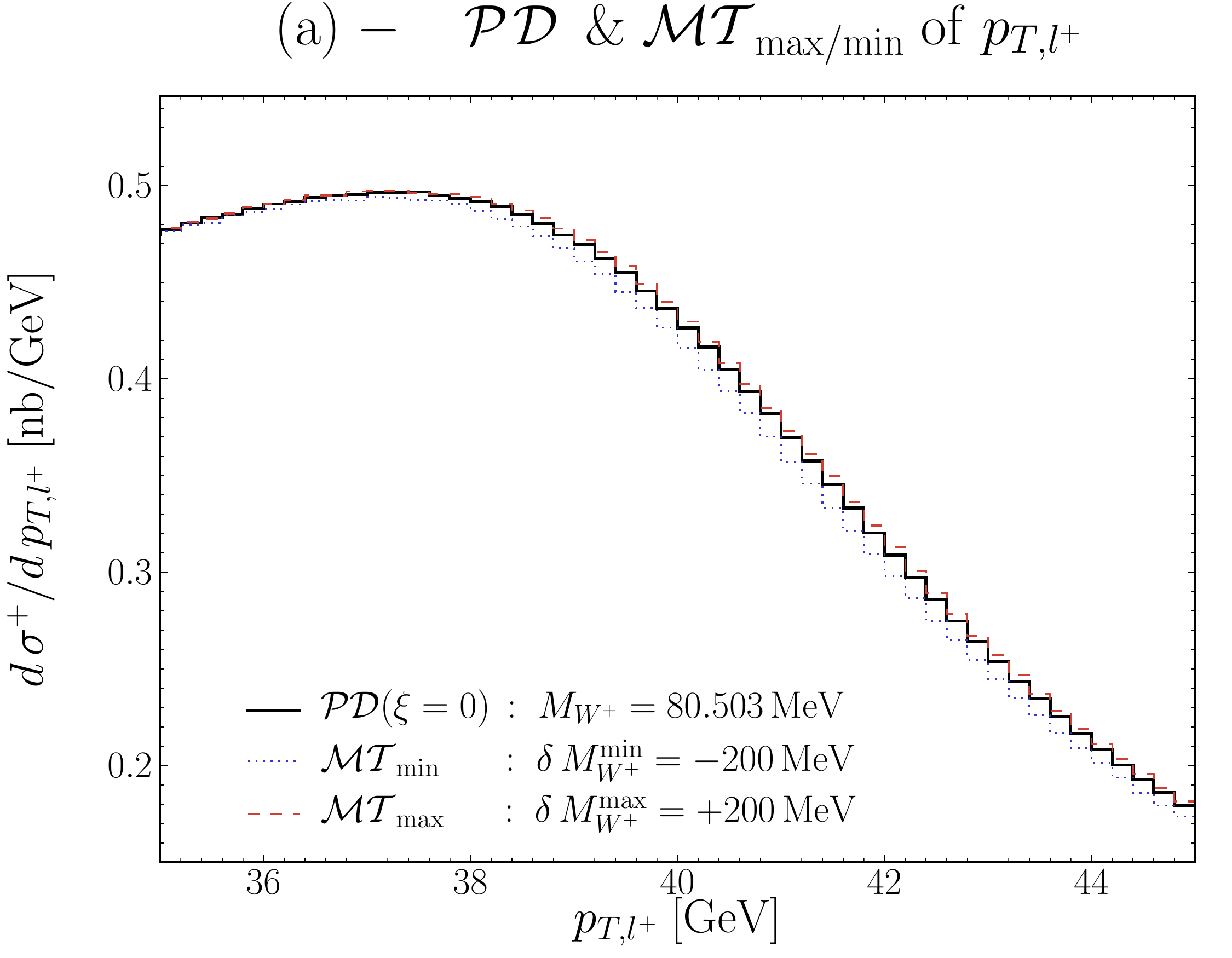}
    \hfill
    \includegraphics[width=0.495\tw]{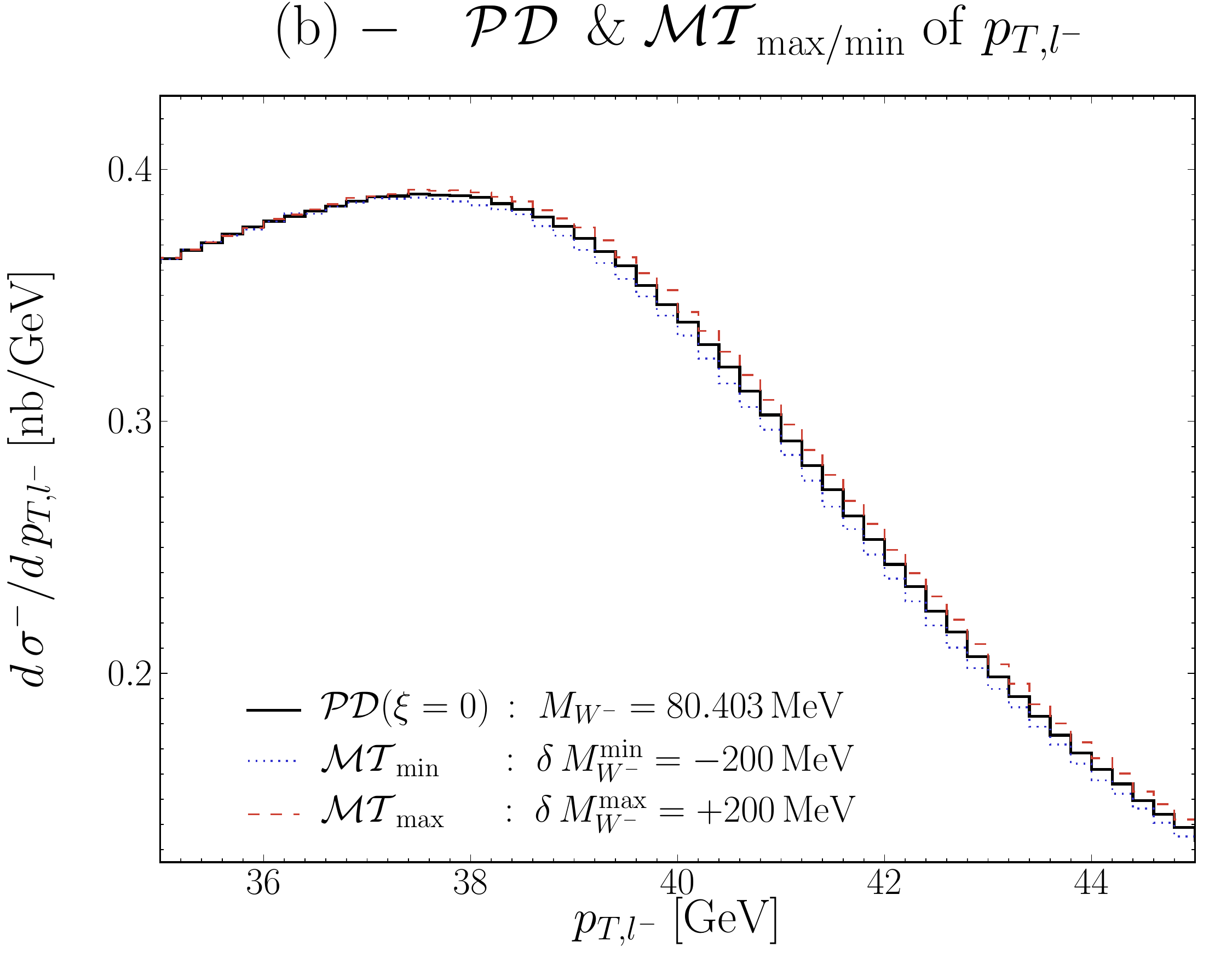}
    \vfill
    \includegraphics[width=0.495\tw]{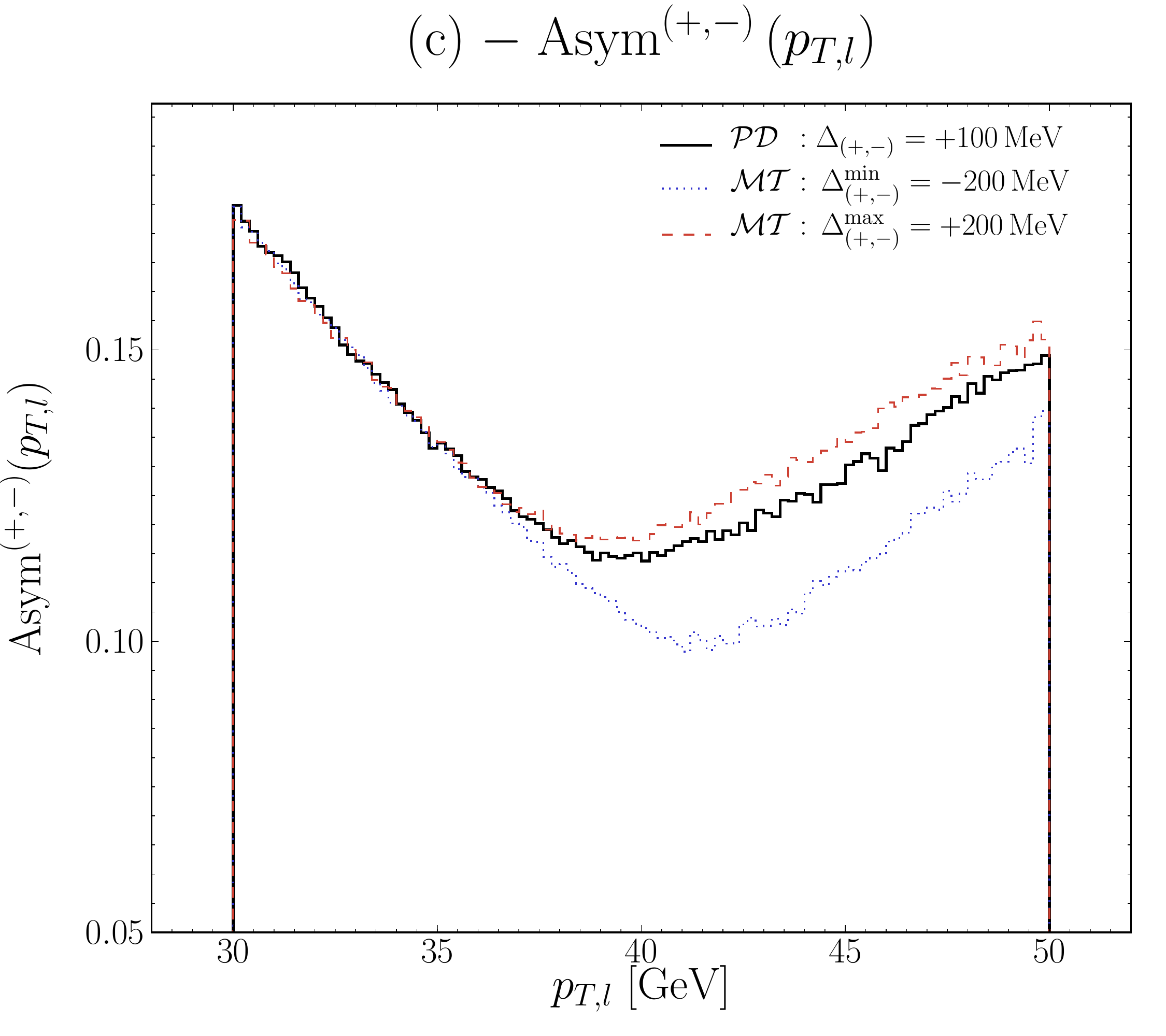}
    \hfill
    \includegraphics[width=0.495\tw]{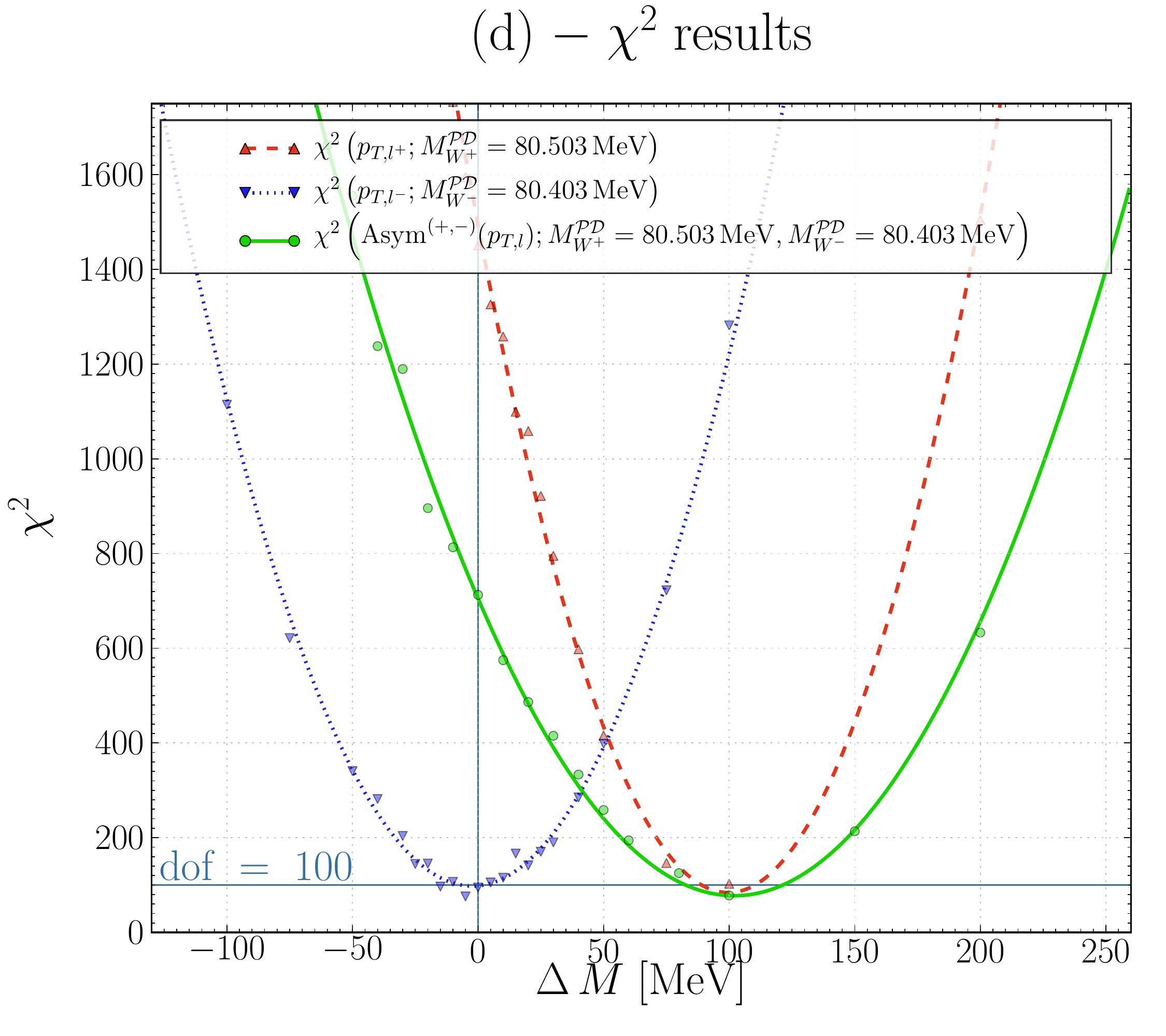}
    \caption[Histograms of the two extrema mass templates and particular unbiased
      pseudo-data $\PD$ having $\MWp=80.503\GeV$ and $\MWp=80.403\GeV$ for the bare 
      positive and negative $\pTl$ spectra and charge asymmetry of $\pTl$ along with the 
      corresponding $\chiD$ results]
            {\figtxt{Histograms of the two extrema mass templates and for the particular case of 
                unbiased pseudo-data $\PD$ having $\MWp=80.503\GeV$ and $\MWp=80.403\GeV$ for the 
                bare positive (a) and negative (b) $\pTl$ spectra 
                and charge asymmetry of $\pTl$ (c) along with
                the corresponding $\chiD$ results (d) where $\Delta\,M$ are defined according
                to Eqs.~(\ref{eq_DeltaM}).}}
            \label{fig_app_MWp_plus_100MeV}
  \end{center} 
\end{figure}

The Figure~\ref{fig_app_MWp_plus_100MeV} presents in the first three frames ((a), (b) and (c)) the 
$\PD$ histograms of $\pTlp$, $\pTlm$ and $\Asym{\pTl}$ for the case (1) of the previous item list 
where the mass of the $\Wp$ has been fixed to $80.503\MeV$ while the mass of the $\Wm$ has been kept
to $80.403\MeV$.\index{Charged lepton@Charged lepton from $W$ decay!Transverse momentum}
The two extrema $\MT$ samples, $\MT_\Min$ and $\MT_\Max$, have been drawn as well to enhance the 
deviation of the unbiased $\PD(\xi=0)$ with respect to the half way position between them.
Starting with Fig.~\ref{fig_app_MWp_plus_100MeV}.(a), the jacobian peak of the $\Wp$ $\PD$ is
slightly shifted to higher $\pT$ while the one for $\Wm$ in frame (b) is half-way between 
the two $\MT$.
In frame (c), the charge asymmetry of $\pTl$ is closer to the $\MT$ for which 
$\DeltaPM=+200\MeV$. This is understandable as in general beyond the jacobian peak 
(\ie{} $\pTl>40\GeV$ or $\rhoTl<0.025\,\mm{GeV}^{-1}$ in $\rhoTl$-space) the following equation holds
\begin{equation}
\DfDx{\sigma}{a}\Bigg|_{M_W^{(2)}} > \DfDx{\sigma}{a}\Bigg|_{M_W^{(1)}<M_W^{(2)}},
\end{equation}
where $a$ is to be replaced by $\pTl$ or $\rhoTl$.
Then, because of the form of the charge asymmetry (Eq.~(\ref{eq_def_charge_asym})), in this first
test, increasing the mass $\MWp$ increases the $\pTlp$ spectrum for each fixed $\pT$ value beyond the 
jacobian peak with respect to its former value eventually leading to an increase of both numerator 
and denominator. In the end, we observe that the kinematics at the LHC are such that the charge 
asymmetry of $\pTl$ or $\rhoTl$ are growing functions in the parameter $\DeltaPM$ as already hinted
by the behaviour of the $\MT$.

\index{Charged lepton@Charged lepton from $W$ decay!Transverse momentum}
Finally the $\chiD$ tests for the $\pTlp$, $\pTlm$ and $\Asym{\pTl}$ spectra are represented in
Fig.~\ref{fig_app_MWp_plus_100MeV}.(d). The meaning of $\Delta M$ depends on the parabola we look
at, we have\,:
\begin{eqnarray}
\Delta M &\equiv& M_\Wp^\PD-M_W^\mm{(ref.)}\qquad \mbox{if looking at } \chiD\left(\pTlp\right),
\nonumber\\
\Delta M &\equiv& M_\Wm^\PD-M_W^\mm{(ref.)}\qquad \mbox{if looking at } \chiD\left(\pTlm\right),
\label{eq_DeltaM}\\
\Delta M &\equiv& M_\Wp^\PD-M_\Wm^\PD   \,\;\qquad \mbox{if looking at } \chiD\left(\Asym{\pTl}\right).
\nonumber
\end{eqnarray}
Starting with the $\chiD$ associated to $\pTlp$ we find a shift of $100\MeV$ present in the 
pseudo-data with respect to the reference mass confirming that $M_\Wp^\PD=80.503\GeV$ while
looking at the negatively charged $\pTlm$ analysis we find $M_\Wp^\PD=80.403\GeV$.
Finally the case of the analysis based on the charge asymmetry gives direct access to the
charge difference between the masses of the pseudo-data $W$ bosons, that is 
$M_\Wp^\PD-M_\Wm^\PD=100\MeV$.
For the cases (2) and (3) of the previous item list the results where found to be relevant as well.

Let us note that in the case of the $\pTlp$ and $\Asym{\pTl}$ parabola the fact there are less
points in the vicinity of the parabola minimum does not decrease the degree of convergence of 
the calculus. For non trivial analysis where $\xi\neq 0$ we often observe shifts of such amplitude
that most of the points are localised on one side only of the fitted parabola. Hence a question
we can ask is at which point the small number of considered mass templates is sufficient 
to provide a relevant value. 
In the next Subsection we investigate this issue by using the present pseudo-data for unbiased 
$\chiD$ tests.

\subsection{Influence on the results of some input parameters to the analysis}\label{app_val_infl}
\subsubsection{Influence of the localisation and number of templates}\label{app_val3}
In this part we consider the problem where the systematic error is such that there are no 
simulated mass templates having masses of the order of the systematic error.
In other words we want to address the problem where we have to perform of parabola fit with the
points being far from the parabola minimum and localised in one side only of the minimum.
For that purpose we consider the same pseudo-data with $\xi=0$ than in the case (1) of the item list 
of the previous section, that is with $M_\Wp^\PD=M_W^\mm{(ref.)}+100\MeV$, $M_\Wm^\PD=M_W^\mm{(ref.)}$,
$M_W^\mm{(ref.)}=80.403\MeV$ and use the charge asymmetry method.
This emulates in a good way the cases where $\DeltaPM(\xi)\gg 0$ for $\xi\neq 0$.

To sees at which point the localisation of the templates matters we reduce the range covered by
the templates to see at which level the convergence and $\DeltaPM$ value changes.
The results are shown in Table.~\ref{table_app_localisation_MT}.
\begin{table}[]
\begin{center}
\renewcommand\arraystretch{1.2}
\begin{tabular}{c|cc}
  \hline
  $\MT$ range [MeV]  & $\chiDmin/\dof$ 
  & $\DeltaPM(\xi=0)$ [MeV] {\scriptsize{($M_\Wp^\PD-M_\Wm^\PD=100\MeV$})}\\
  \hline\hline
  $\pm\, 200$ & 0.77 & $102.0 $ \\
  $\pm\, 150$ & 0.84 & $103.7 $ \\
  $\pm\, 80$  & 1.20 & $\;\,95.4 $ \\
  $\pm\, 60$  & 1.69 & $\;\,89.3 $ \\
  $\pm\, 20$  & 2.34 & $\;\,73.9 $ \\
  \hline
\end{tabular}
\renewcommand\arraystretch{1.45}
\caption[Influence of the lack of mass templates in the vicinity of the $\chiD$ parabola
fit minimum looking at the trivial case where $\xi=0$ with $M_\Wp^\PD=M_\Wm^\PD+100\MeV$
and using the charge asymmetry of $\pTl$]
        {\figtxt{Influence of the lack of mass templates in the vicinity of the $\chiD$ parabola
fit minimum looking at the trivial case where $\xi=0$ with $M_\Wp^\PD=M_\Wm^\PD+100\MeV$
and using the $\Asym{\pTl}$ method.}}
\label{table_app_localisation_MT}
\end{center}
\end{table}
Then we come to the conclusion that as long as the systematic errors $\DeltaPM$ are such that
$|\DeltaPM(\xi)|<200\MeV$ if the convergence is poor ($\chiDmin/\dof\gg 1$) it has to be blamed 
purely on the distortion on the $\PD$ affected by the bias $\xi$.
On the other hand, for the cases where $|\DeltaPM(\xi)|>200\MeV$ a bad convergence will always be
observed because of the lack of mass templates which eventually will provide values of $\DeltaPM$
which will differ from the one to be observed if having generated much more templates.

Nonetheless, with the chosen biases $\xi$ values and the analysis techniques developed most of the
errors $\DeltaPM$ are within the $\pm\,200\MeV$ range which justify why the extension of the range
was not considered a vital problem to extract the essential physics from our analysis.

\subsubsection{Influence of the resolution of the histograms}\label{app_val4}
\index{Charged lepton@Charged lepton from $W$ decay!Transverse momentum}
In this part we considered changes to larger bins for the analysis with respect to the one used
for the generation to see at which level the bin width of the histograms is important. 
Let us remind the former binning corresponded to a resolution of $200\MeV$, which in the range
of $20\GeV<\pTl<60\GeV$ corresponded to $200$ bins. 

We present here the results for the classic and charge asymmetry methods in the case of
$\pp$ collisions for the standard cuts of $\pTl>20\GeV$ and $|\etal|<2.5$. The likelihood test
was carried for the case of unbiased pseudo-data ($\xi=0$). 
The procedure for the likelihood test between the pseudo-data $\PD(\xi=0)$ and the 
$n^{\mm{th.}}$-mass template $\MT^{(n)}$ was done in two steps\,:
(1) $\PD$ and $\MT^{(n)}$ histograms were re-binned to a lower resolution,
(2) the likelihood analysis was carried out, \ie{} the two histograms are cut to the standard 
analysis window of $30\GeV<\pTl<50\GeV$ and then the $\chiD$ test was done.

\begin{figure}[!ht] 
  \begin{center}
    \includegraphics[width=0.495\tw]{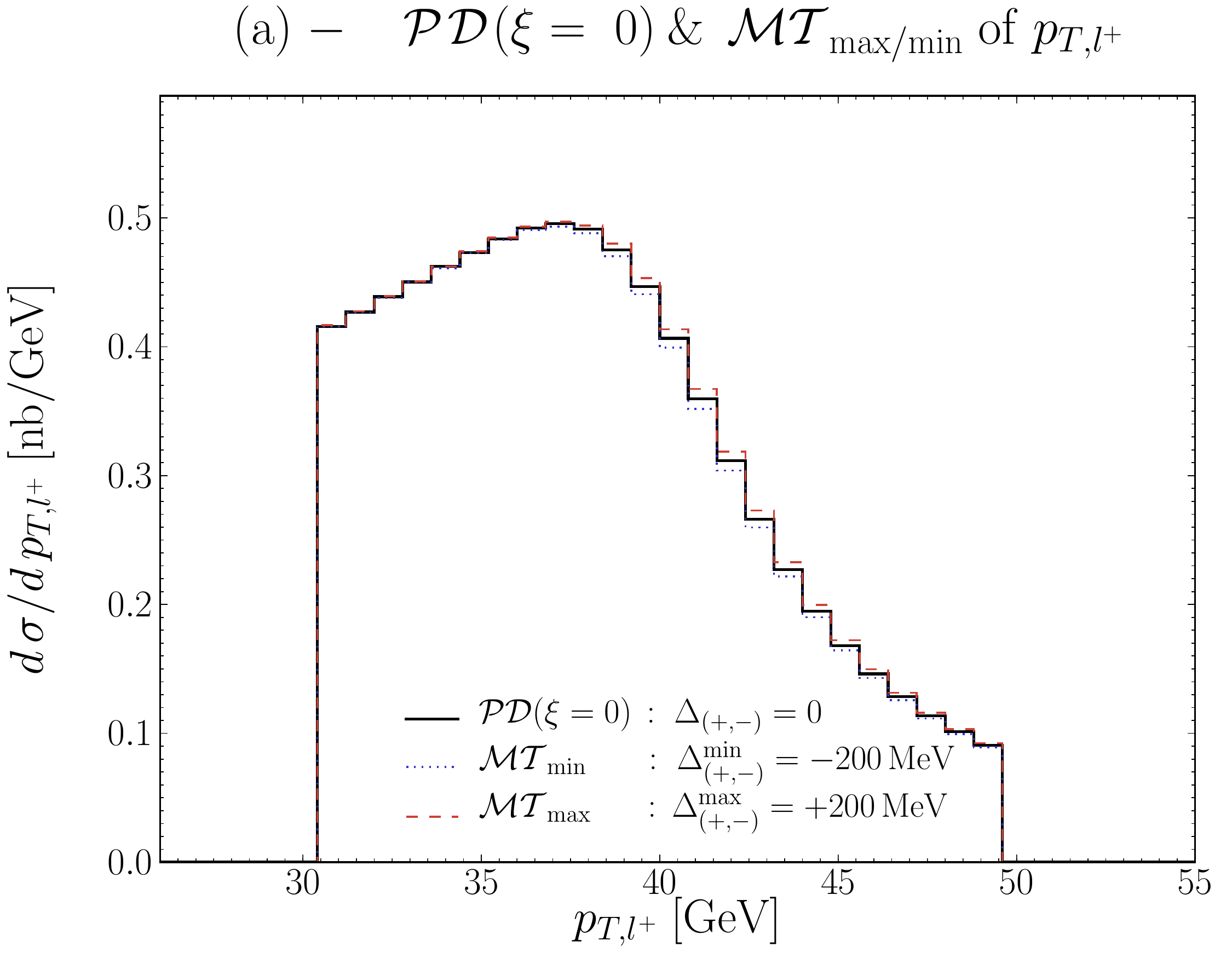}
    \hfill
    \includegraphics[width=0.495\tw]{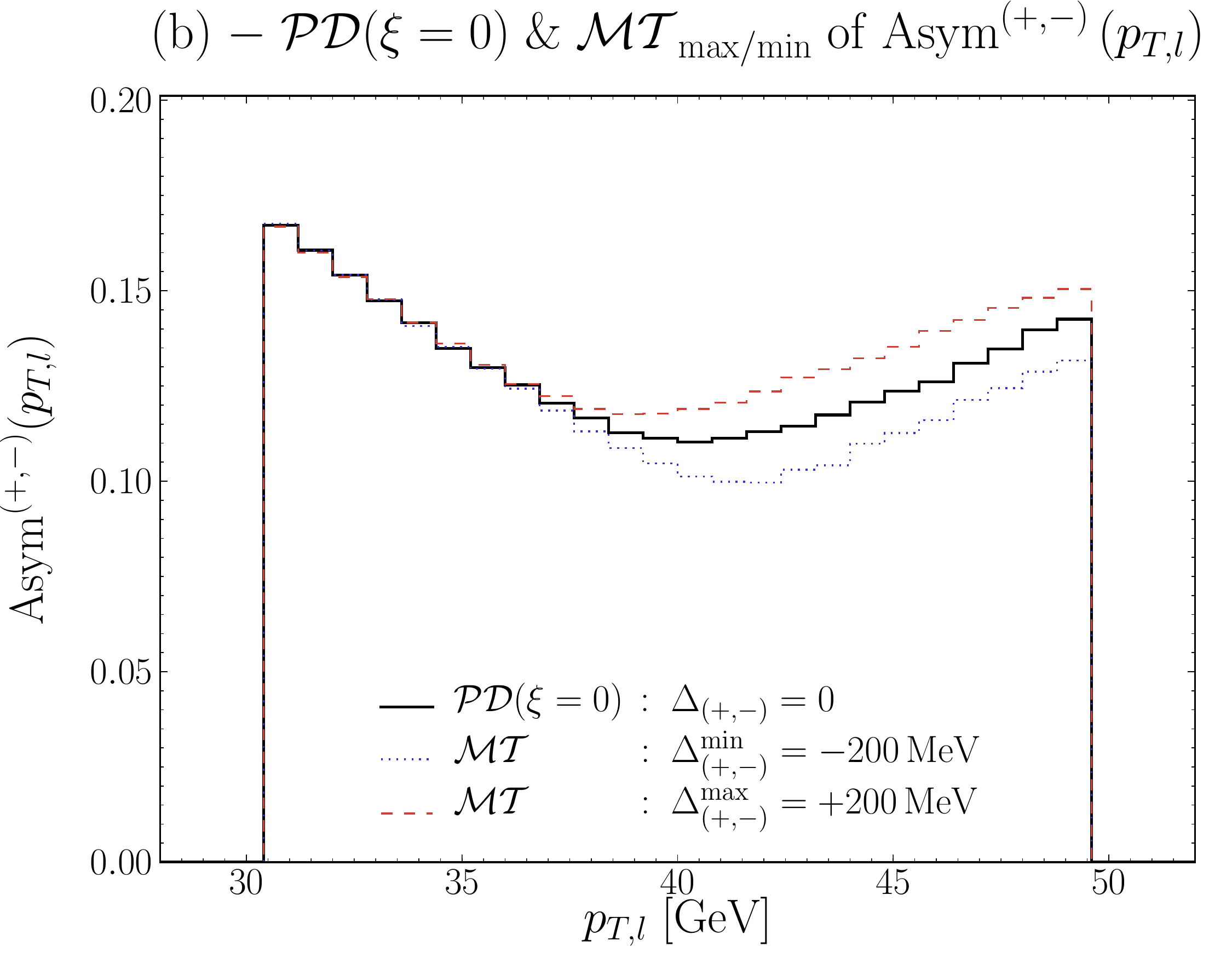}
    \caption[Pseudo-data and extrema mass templates for $\pTlp$ and $\Asym{\pTl}$ histograms for a  resolution
	of $\delta\pT=0.8\GeV$]
            {\figtxt{Pseudo-data	and extrema mass templates $\pTlp$ (a)
	and $\Asym{\pTl}$ (b) histograms for a resolution of $\delta\pT=0.8\GeV$.}		
            }
            \label{fig_app_binning}
  \end{center} 
\end{figure}

Table~\ref{table_app_binning} shows the influence of different $\delta\,\pT$ resolutions
for the charged lepton on the $\chiD$ results while
Figure~\ref{fig_app_binning} present $\pTlp$ and $\Asym{\pTl}$ spectra with a resolution of 
$\delta\,\pT\approx 0.8\GeV$.
\begin{table}[]
\begin{center}
\renewcommand\arraystretch{1.75}
\begin{tabular}{cccc}
  \hline
  Number of bins                        & $\delta\,\pT$ [GeV] & Classic $\Delta_{(+,-)} {}^{(\chi^2_{+,\Min}/\dof)}_{(\chi^2_{-,\Min}/\dof)}$ 
                                                                 & Charge Asymmetry $\Delta_{(+,-)} {}^{(\chi^2_{\Min}/\dof)}$ \\
  \hline\hline
  200                                    & 0.2                                     & $0.8 \,\pm\,5.8\CHIEXPPM{0.86}{0.98}$ &  $1.2 \,\pm\, 4.1\CHIEXP{0.82}$ \\
  100                                    & 0.4                                     & $0.9 \,\pm\,5.7\CHIEXPPM{0.94}{1.10}$ &  $1.3 \,\pm\, 4.1\CHIEXP{0.95}$ \\
   50                                    & 0.8                                     & $1.0 \,\pm\,5.8\CHIEXPPM{0.97}{1.06}$ &  $1.5 \,\pm\, 4.1\CHIEXP{0.90}$ \\
   25                                    & 1.6                                     & $1.2 \,\pm\,5.7\CHIEXPPM{1.02}{1.23}$ &  $1.9 \,\pm\, 4.1\CHIEXP{0.96}$ \\ 
  \hline
\end{tabular}

  \caption[Influence of the binning of the histograms for the likelihood analysis ]
          {\figtxt{Influence of the binning of the histograms for the likelihood analysis.
          These results corresponds to unbiased $(\xi=0)$, 
          the number of bins corresponds to the one for the range $20\GeV<\pTl<60\GeV$.}}
            \label{table_app_binning}
\renewcommand\arraystretch{1.45}
\end{center}
\end{table}

\subsubsection{Influence of the window in $\mathbf{\pTl}$}\label{app_val5}
\index{Charged lepton@Charged lepton from $W$ decay!Transverse momentum}
We look now at the influence of the window for the $\chiD$ analysis.
Enlarging the range at low and high $\pT$ would imply in reality to test
the sensibility of the charge asymmetry observable as a mean to extract the $W$ mass
respectively toward the modeling of the low $\pT$ background differences between 
the $\Wp$ and the $\Wm$ production and the influence of the $\GamW$ and/or 
$\GamWp\neq\GamWm$ in the high $\pT$ region.\index{W boson@$W$ boson!Width@Width $\GamW$}

In the first step the high $\pTl$ is maintained at $50\GeV$.
Then, taking the minimum value of $20\GeV$ for the low cut we found for trivial results
using the charge asymmetry of $\pTl$\,:
\begin{equation}
20\GeV<\pTl<50\GeV \qquad \Rightarrow \quad \DeltaPM(\xi=0)=1.3\,\pm\,4.1\;\;\;
(\mm{with\;} \chiDmin/\dof=0.84),
\end{equation}
which does not enhance the former result seen in Eq.~(\ref{eq_DeltaPM_asym_std}).
Now, considering the whole range used for the $\pTlp$ and $\pTlm$ histograms, still using the
charge asymmetry $\Asym{\pTl}$ we find\,:
\begin{equation}
20\GeV<\pTl<60\GeV \qquad \Rightarrow \quad \DeltaPM(\xi=0)=1.1\,\pm\,3.9\;\;\;
(\mm{with\;} \chiDmin/\dof=0.87),
\end{equation}
that is here we have a slightly better result.

Since the results do not change a lot this would eventually allow by first reducing the
lower cut on $\pT$ to see the influence of the different background contributions in the
positive and negative channels. Note that the lower cut should not be too close from the
trigger cut to avoid the possible charge dependency from the selection process.
On the other hand, maintaining the lower cut fixed and exploring higher $\pT$ cuts would
allow now to study the influence of the $W$ boson width on the results.

\subsubsection{Influence of $\mathbf{\slashiv{p}_{T,\nul}}$ cuts}\label{app_val6}
The influence of the $\ETmiss$ cut was investigated with the best emulation achievable in our 
framework, that is during the generation process in top of the requirements from 
Eq.~(\ref{pTl-etal-cuts}), the simulated events were selected if and only if the decaying 
neutrino displays $\pTnu>20\GeV$.
All $\chiD$ ($\xi=0$ and $\xi\neq 0$), up to non avoidable numerical discrepancies, were of the
same values than the data gathered without doing any cut on the neutrino.

Just to cite two numbers, the Monte Carlo truth test gave\,:
\begin{equation}
\DeltaPM(\xi=0)=-1.0\,\pm\, 3.3\MeV\quad \mm{with}\; \chiDmin/\dof=0.98,
\end{equation}
while the unbiased experimental test gave\,:
\begin{equation}
\DeltaPM(\xi=0)=1.0\,\pm\, 4\MeV\quad \mm{with}\; \chiDmin/\dof=0.81.
\end{equation}
Both of these values are very close to the one present in 
Table~\ref{table_exp_sys_classic_vs_casym_vs_dcasym}.
This justified for the rest of the study the non necessity to take care of such cuts.

\cleardoublepage
\section{Measurement of the W mass charge asymmetry\,: the How Not To} \label{app_how_not_to}
\setlength{\epigraphwidth}{0.375\tw}
\epigraph{
``That's not right. It's not even wrong.''
}%
{\textsc{Wolfgang Pauli}'s words of wisdom \\ (Kept alive nowadays by \textsc{M.W.K.})}

\index{Electroweak!VmA@$V-A$ coupling|(}
\index{Charged lepton@Charged lepton from $W$ decay!Transverse momentum}
We consider a particular context to extract $\MWp-\MWm$ that 
stresses the charge asymmetry between the $\pTlp$ and $\pTlm$ distributions at the LHC.
For that matter we propose to assume just for a moment that we are not aware of the $V-A$ coupling 
of $W$ bosons to fermions in the Standard Model and that the $\Wp$ and $\Wm$, which are particle 
and antiparticle of each other provide positively and negatively decaying charged leptons with the 
same kinematics.
Putting aside the fact that the detector measurement are different between positively and negatively 
charged leptons we consider the extraction of $\MWp-\MWm$ could be obtained directly by 
confronting the data from one charge to mass templates generated for the opposite charge.

To be more precise, the data is here represented by Monte Carlo pseudo-data simulations of the 
$\Wp$ confronted via $\chiD$ likelihood tests with Monte Carlo simulations of the $\Wm$ bosons. 
Unaware of the $V-A$ coupling and of all its consequences seen in  
Chapter~\ref{chap_w_pheno_in_drell-yan} one would expect that in this configuration
a trivial $\chiD$ test with unbiased $\Wp$ pseudo-data should give $\MWp-\MWm=0$ with a good 
convergence ($\chiDmin/\dof\approx 1$). \index{Electroweak!VmA@$V-A$ coupling|)}

\begin{figure}[!h] 
  \begin{center}
    \includegraphics[width=0.6\tw]{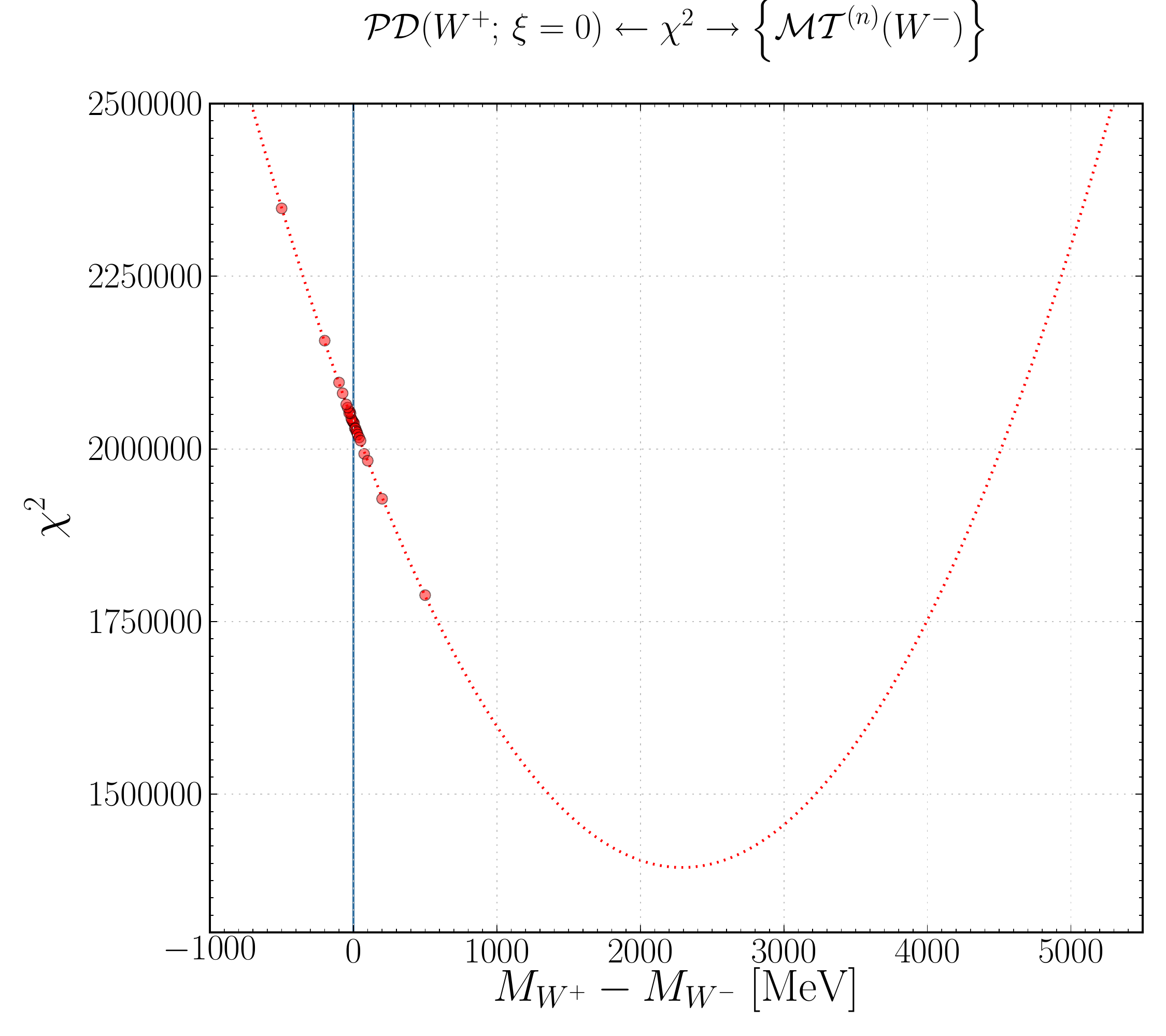}
    \caption[$\chi^2$ results when considering the unbiased pseudo-data built from the positive charged lepton
    data and the mass templates built from the negatively charged lepton data]
            {\figtxt{$\chi^2$ results when considering the unbiased ($\xi=0$) 
                pseudo-data built from the positive charged lepton
                data and the mass templates built from the negatively charged lepton data}
            }
            \label{fig_app_wp_vs_wm}
  \end{center} 
\end{figure}

Figure~\ref{fig_app_wp_vs_wm} shows the $\chiD$ test using the unbiased pseudo-data of the $\pTlp$
distribution and using $\pTlm$ for the mass templates (covering here the range $\pm\,500\MeV$).
The result obtained is $\chiDmin/\dof\approx 14,000$ and $\MWp-\MWm \approx 2.3\GeV$.
The poor quality of this result can be understood using the arguments from the previous 
Appendix\,: the number of $\MT$ is too small and far from the expected parabola minimum.
Nonetheless as we saw the order is still more or less correct and gives an idea of the size of the
error, hence here we can say that going to such a naive test imply an error of the order of the 
$\mm{GeV}$ which is completely ruled out by the actual measurements 
as seen in Table~\ref{table_cdf_mw_mwp_mwm}.

\cleardoublepage
\section{Step-by-step interpretations of systematic errors results}
\label{app_detailed_chi2_res}
\setlength{\epigraphwidth}{0.7\tw}
\epigraph{
Reggie\,:\;``Trucks come in and out of here all day long.
Truck comes in, you unload it.
Truck goes out, you load it.
Next truck comes in,
you unload it. Next truck\dots''\\
Randy (very confused)\,:\;``Wait a second. Go back to that first truck.''
}%
{\textit{My Name is Earl - Get a real job}\\}

In this Appendix more details are given on the results presented in the core of the Chapter.
For that purpose step-by-step comments of the tables are provided with the help of visual support.

The first Subsection introduces the notation implemented in the more detailed versions of the tables 
shown in the core of the Chapter.
Then in a second part the comments of these detailed tables are made based each time
on the relevant $\PD$ and $\MT$ histograms and $\chiD$. 
The influence of the following effects are addressed in that chronological order
\begin{itemize}
\item[\ref{ss_a1}] Energy scale of the charged lepton
\begin{itemize}
  \item[-] Coherent biases between the positively and negatively charged lepton tracks 
    (noted sometime for convenience ``coherent biases'')
  \item[-] Incoherent biases between the positively and negatively charged lepton tracks 
    (noted sometime for convenience ``incoherent biases'')
\end{itemize}
\item[\ref{ss_a2}] Resolution of the charged lepton track parameters 
\item[\ref{ss_a3}] Intrinsic $\kT$ of the partons
\item[\ref{ss_a5}] $u^\val-d^\val$ asymmetry
\item[\ref{ss_a6}] $s-c$ asymmetry
\end{itemize}

\subsection{Detailed tables}\label{app_detailed_tables}
The Tables ~\ref{table_app_exp_sys_classic_vs_casym_vs_dcasym} and 
\ref{table_app_charge_asym_results} in the following pages are just reproduction of 
Tables~\ref{table_exp_sys_classic_vs_casym_vs_dcasym} and \ref{table_charge_asym_results} 
from the core of the Chapter with some additive information.
More precisely, each systematic error is now given with a precision of $0.1\MeV$ along with the 
associated $\chiDmin/\dof$ value providing then further details on the degree of convergence of the 
calculus.

Besides, as explained previously, different random numbers sets were used for each event generation
to emulate as better as possible uncorrelated events.
For that purpose $100$ independent seeds set were produced and tagged from set number $00$ to set 
number $99$. 
Practically the generation of $\approx 10^8$ events needed to be achieved by splitting each generation
into several sub-generations.
Each sub-generation were produced so that all the $100\times n_\mm{s}$ seeds were 
different to ensure the best possible un-correlation between different event batches.
For the standard generation of one year of low luminosity, each $W$ generation were split in average to
$10$ sub-generations, hence we used here $n_\mm{s}=20$. Concerning the generation in the narrow 
lepton pseudo-rapidity or $W$ boson rapidity region this time each $W$ generation were split in average
into $50$ hence here $n_\mm{s}=100$.

The seeds tags used for producing each pseudo-data $\PD$ will be as well displayed in the
tables, the one for the $\Wp$ being labeled ``$(+)$'' each time written in top of the one used for 
$\Wm$ labeled ``$(-)$''.
Concerning the mass templates $\MT$, each $\MT^{(n)}$ template seed tag are indicated on the next page 
in Table~\ref{table_seed_set}.

\begin{table}[]
\begin{center}
\renewcommand\arraystretch{1}
\begin{tabular}{c|cc}
  \hline
  $\delta M_W^{(n)}$ [MeV] for $\MT^{(n)}$ & $\Wp$ seed tag & $\Wm$ seed tag \\
  \hline\hline
  0                       &                    10 &                    50 \\
  \arrayrulecolor{Greymin}
  \hline
  \arrayrulecolor{Black}
  \!\!-5                  &                    11 &                    51 \\
       5                  &                    12 &                    52 \\
   \arrayrulecolor{Greymin}
  \hline
  \arrayrulecolor{Black}
 \!\!-10                  &                    13 &                    53 \\
      10                  &                    14 &                    54 \\
  \arrayrulecolor{Greymin}
  \hline
  \arrayrulecolor{Black}
 \!\!-15                  &                    15 &                    55 \\
      15                  &                    16 &                    56 \\
  \arrayrulecolor{Greymin}
  \hline
  \arrayrulecolor{Black}
 \!\!-20                  &                    17 &                    57 \\
      20                  &                    18 &                    58 \\
\arrayrulecolor{Greymin}
  \hline
  \arrayrulecolor{Black}
 \!\!-25                  &                    19 &                    59 \\
      25                  &                    20 &                    60 \\
\arrayrulecolor{Greymin}
  \hline
  \arrayrulecolor{Black}
 \!\!-30                  &                    21 &                    61 \\
      30                  &                    22 &                    62 \\
\arrayrulecolor{Greymin}
  \hline
  \arrayrulecolor{Black}
 \!\!-40                  &                    23 &                    63 \\
      40                  &                    24 &                    64 \\
\arrayrulecolor{Greymin}
  \hline
  \arrayrulecolor{Black}
 \!\!-50                  &                    25 &                    65 \\
      50                  &                    26 &                    66 \\
\arrayrulecolor{Greymin}
  \hline
  \arrayrulecolor{Black}
 \!\!-75                  &                    27 &                    67 \\
      75                  &                    28 &                    68 \\
\arrayrulecolor{Greymin}
  \hline
  \arrayrulecolor{Black}
\!\!-100                  &                    29 &                    69 \\
     100                  &                    30 &                    70 \\
\arrayrulecolor{Greymin}
  \hline
  \arrayrulecolor{Black}
\!\!-200                  &                    31 &                    71 \\
     200                  &                    32 &                    72 \\
  \hline
\end{tabular}
\renewcommand\arraystretch{1.45}
\caption[Seed set associated to each mass templates generation]
{\figtxt{Seed set associated to each mass templates generation.}}
\label{table_seed_set}
\end{center}
\end{table}

\clearpage
\begin{table}[]
\begin{center}
\begin{TableSizeOne}
\renewcommand\arraystretch{1.7}
\begin{tabular}{|c|c|c||r@{.\kern\tabcolsep}>{\kern-\tabcolsep}l|r@{.\kern\tabcolsep}>{\kern-\tabcolsep}l|r@{.}l|}
  \cline{4-9}
  \multicolumn{3}{c|}{} & \multicolumn{6}{c|}   {$\MWp-\MWm^{(\chiDmin/\dof)}\quad[\mm{MeV}]$} \\ 
  \cline{2-9}
  \multicolumn{1}{c}{}  & \multicolumn{1}{|c|}  {Systematic $\xi$}
                        & \multicolumn{1}{c||}  {$^{\mm{Seed}}_{\mm{tag}}$}
                        & \multicolumn{2}{c|}   {Classic Method}
                        & \multicolumn{2}{c|}   {$\Asym{\pTl}$} 
                        & \multicolumn{2}{c|}   {$\DAsym{\rhoTl}$}  \\

  \hline
  \multicolumn{1}{|c}{MC truth}   & \multicolumn{1}{|c|}{$\xi=0$} & \multicolumn{1}{c||}{$^{01}_{04}$}
  & \multicolumn{2}{c|}{$\!\!\!-1.6 \,\pm\, 3.2\CHIEXPPM{1.05}{0.95}$ }
  & \multicolumn{2}{c|}{$\!\!\!\!-1.0 \,\pm\, 3.3\CHIEXP{0.98}$ }
  & \multicolumn{2}{c|}{$ 0.2     \,\pm\, 3.3\CHIEXP{0.85}$ } \\
  \multicolumn{1}{|c}{Cent. Exp.} & \multicolumn{1}{|c|}{$\xi=0$} & \multicolumn{1}{c||}{$^{01}_{04}$}
  & \multicolumn{2}{c|}{$\,0.8 \,\pm\, 4.1\CHIEXPPM{0.86}{0.98}$ } 
  & \multicolumn{2}{c|}{$1.2 \,\pm\, 4.1\CHIEXP{0.82}$ }
  & \multicolumn{2}{c|}{$ 0.4  \,\pm\, 4.1\CHIEXP{0.90}$} \\
  \hline\hline
  \multirow{8}{*}{ES [\%]}
  & $\es_\lp=+\es_\lm=+0.05\,\%$ & \multicolumn{1}{c||}{$^{01}_{04}$}
  &   \hspace*{0.5cm} $3$&$4\CHIEXPPM{1.35}{1.39}$
  &   \hspace*{0.6cm} $1$&$6\CHIEXP{0.99}$ 
  &   \multicolumn{2}{c|}{} \\
  & $\es_\lp=+\es_\lm=-0.05\,\%$ & \multicolumn{1}{c||}{$^{01}_{04}$}
  &   $-2$&$4\CHIEXPPM{1.19}{1.24}$
  &   $-0$&$3\CHIEXP{1.06}$
  &   \multicolumn{2}{c|}{ \multirow{2}{*}{$\huge\times$}} \\
  \arrayrulecolor{Greymin}
  \cline{2-7}
  \arrayrulecolor{Black}
  & $\es_\lp=+\es_\lm=+0.50\,\%$ & \multicolumn{1}{c||}{$^{01}_{04}$}
  &   $15$&$9\CHIEXPPM{30}{24}$
  &    $7$&$9\CHIEXP{1.53}$ 
  &   \multicolumn{2}{c|}{}  \\
  & $\es_\lp=+\es_\lm=-0.50\,\%$ & \multicolumn{1}{c||}{$^{01}_{04}$}
  &  $-35$&$8\CHIEXPPM{27}{21}$
  &   $-6$&$2\CHIEXP{1.59}$ 
  &   \multicolumn{2}{c|}{}  \\
  \arrayrulecolor{Greymax}
  \cline{2-9}
  \arrayrulecolor{Black}
  & $\es_\lp=-\es_\lm=+0.05\,\%$ & \multicolumn{1}{c||}{$^{01}_{04}$}
  &  $-56$&$1\CHIEXPPM{1.35}{1.24}$ 
  &  $-56$&$8\CHIEXP{1.48}$  
  &  \multicolumn{2}{c|}{\multirow{2}{*}{$0.9\CHIEXP{0.98}$}} \\
  & $\es_\lp=-\es_\lm=-0.05\,\%$ & \multicolumn{1}{c||}{$^{01}_{04}$}
  &   $57$&$1\CHIEXPPM{1.19}{1.39}$ 
  &   $56$&$9\CHIEXP{1.48}$ 
  &  \multicolumn{2}{c|}{} \\
  \arrayrulecolor{Greymin}
  \cline{2-9}
  \arrayrulecolor{Black}
  & $\es_\lp=-\es_\lm=+0.50\,\%$ & \multicolumn{1}{c||}{$^{01}_{04}$}
  & $-567$&$2\CHIEXPPM{30}{21}$ 
  & $-611$&$2\CHIEXP{30}$ 
  & \multicolumn{2}{c|}{\multirow{2}{*}{$-0.6\CHIEXP{1.13}$}} \\ 
  & $\es_\lp=-\es_\lm=-0.50\,\%$ & \multicolumn{1}{c||}{$^{01}_{04}$}
  &  $547$&$2\CHIEXPPM{27}{24}$ 
  &  $514$&$8\CHIEXP{59}$
  &  \multicolumn{2}{c|}{} \\ 
  \hline\hline
  \multirow{2}{*}{ERF}
  & $0.7$ & \multicolumn{1}{c||}{$^{01}_{04}$}
  &  $1$&$1\CHIEXPPM{16}{13}$ 
  & $-2$&$3\CHIEXP{0.95}$ 
  & \multicolumn{2}{c|}{ \multirow{2}{*}{$\huge\times$}} \\
  & $1.3$ & \multicolumn{1}{c||}{$^{01}_{04}$}
  & $-2$&$6\CHIEXPPM{23}{22}$ 
  & $2$&$5\CHIEXP{1.08}$ 
  & \multicolumn{2}{c|}{} \\
  \hline
\end{tabular}
\renewcommand\arraystretch{1.45}

\end{TableSizeOne}
  \caption[Experimental systematics errors for the classic method, the charge asymmetry and the 
    double charge asymmetry (Detailed)]
          {\figtxt{Experimental systematics errors for the classic method, the charge asymmetry and 
              the double charge asymmetry.
              The results in the case of the double charge asymmetry are the one for 
              $\es_\lp=-\es_\lm>0$ for the first six months and $\es_\lp=-\es_\lm<0$ for the last six 
              months.
            }}
          \label{table_app_exp_sys_classic_vs_casym_vs_dcasym}
          \index{Double charge asymmetry!Used for the extraction of MWpmMWm@
Used for the extraction of $\MWp-\MWm$}
\index{Chi2 Likelihood analysis@Chi-2 ($\chiD$) likelihood analysis!Results (detailed)}
\end{center}
\end{table}


\cleardoublepage
\begin{table}[]
\begin{center}
\begin{TableSizeSevenPt}
\renewcommand\arraystretch{1.7}
\begin{tabular}
{|c|l|c||r@{.\kern\tabcolsep}>{\kern-\tabcolsep}l|r@{.\kern\tabcolsep}>{\kern-\tabcolsep}l|r@{.\kern\tabcolsep}>{\kern-\tabcolsep}l|r@{.\kern\tabcolsep}>{\kern-\tabcolsep}l|}
  \cline{3-11}
  \multicolumn{2}{c|}{} & \multicolumn{1}{c||}  {\tiny{Seed}}
                        & \multicolumn{8}{c|}   {$\MWp-\MWm^{(\chiDmin/\dof)}\;[\mm{MeV}]$ 
                                                 using $\Asym{\pTl}$} \\ \cline{2-2}\cline{4-11}
  \multicolumn{1}{c|}{} & \multicolumn{1}{c|}   {Systematic $\xi$}
                        & \multicolumn{1}{c||}  {${}^{(+)}_{(-)}$}
                        & \multicolumn{2}{c|}   {$\pp$ - $|\etal|<2.5$}
                        & \multicolumn{2}{c|}   {$\pp$ - $|\etal|<0.3$  } 
                        & \multicolumn{2}{c|}   {$\pp$ - $|\yW|<0.3$}
                        & \multicolumn{2}{c|}   {$\dd$ - $|\etal|<2.5$}  \\

  \cline{2-11}\hline
  \multicolumn{1}{|c|}{MC truth}   
  & \multicolumn{1}{c|}{$\xi=0$} & \multicolumn{1}{c||}{$^{01}_{04}$}
  & \multicolumn{2}{c|}{$\!\!\!\!-1.0 \,\pm\, 3.3\CHIEXP{0.98}$ }
  & \multicolumn{2}{c|}{$-0.2 \,\pm\, 1.2\CHIEXP{0.82}$ }
  & \multicolumn{2}{c|}{$-0.1 \,\pm\, 1.2\CHIEXP{0.83}$ }
  & \multicolumn{2}{c|}{$\!\!\!\!-0.2 \,\pm\, 4.5\CHIEXP{1.01}$ } \\
  \multicolumn{1}{|c|}{Cent. Exp.} 
  & \multicolumn{1}{c|}{$\xi=0$} & \multicolumn{1}{c||}{$^{01}_{04}$}
  & \multicolumn{2}{c|}{$1.2  \,\pm\, 4.1\CHIEXP{0.82}$ }
  & \multicolumn{2}{c|}{$-0.2 \,\pm\, 4.1\CHIEXP{1.19}$   }
  & \multicolumn{2}{c|}{$-1.1 \,\pm\, 4.1\CHIEXP{1.06}$   }
  & \multicolumn{2}{c|}{$ 4.2 \,\pm\, 5.2\CHIEXP{0.94}$   } \\
  \hline\hline
  \multirow{5}{*}{$\Mean{\kT}$ [GeV]}
  & \multicolumn{1}{c|}{$2$} & $^{90}_{93}$
  & \hspace*{.5cm} $7$&$7\CHIEXP{1.02}$
  & \hspace*{.5cm} \cellcolor{hl}{$0$}&\cellcolor{hl}{$0\CHIEXP{1.26}$} 
  & \hspace*{.6cm} \cellcolor{hl}{$1$}&\cellcolor{hl}{$8\CHIEXP{0.98}$} 
  & \hspace*{.4cm} $27$&$9\CHIEXP{1.28}$ \\
  & \multicolumn{1}{c|}{$3$} & $^{00}_{03}$
  &  $7$&$0\CHIEXP{1.04}$
  &  \cellcolor{hl}{$2$}&\cellcolor{hl}{$8\CHIEXP{1.10}$}
  & \cellcolor{hl}{$-1$}&\cellcolor{hl}{$7\CHIEXP{1.01}$} 
  & $19$&$9\CHIEXP{1.10}$ \\
  & \multicolumn{1}{c|}{$5$} & $^{02}_{05}$
  & $-3$&$8\CHIEXP{0.89}$ 
  & \cellcolor{hl}{$-3$}&\cellcolor{hl}{$4\CHIEXP{1.12}$}
  & \cellcolor{hl}{$-6$}&\cellcolor{hl}{$4\CHIEXP{1.15}$} 
  &$-14$&$7\CHIEXP{1.07}$ \\
  & \multicolumn{1}{c|}{$6$} & $^{91}_{94}$
  & $-7$&$6\CHIEXP{1.14}$
  &  \cellcolor{hl}{$2$}&\cellcolor{hl}{$3\CHIEXP{0.98}$}
  & \cellcolor{hl}{$-5$}&\cellcolor{hl}{$1\CHIEXP{1.11}$} 
  &$-34$&$5\CHIEXP{1.65}$  \\
  & \multicolumn{1}{c|}{$7$} & $^{92}_{95}$
  & $-15$&$8\CHIEXP{1.36}$ 
  &   \cellcolor{hl}{$1$}&\cellcolor{hl}{$8\CHIEXP{1.24}$}
  &  \cellcolor{hl}{$-8$}&\cellcolor{hl}{$1\CHIEXP{1.00}$} 
  & $-48$&$5\CHIEXP{2.53}$ \\
  \hline\hline
  \multirow{2}{*}{PDF$^{(\ast)}$}
  & \multicolumn{1}{c|}{Min.} & $^{01}_{04}$
  & $-4$&$1\CHIEXP{0.85}$ 
  &  $5$&$6\CHIEXP{1.23}$
  & $-0$&$4\CHIEXP{1.13}$ 
  & $-2$&$5\CHIEXP{0.94}$ \\
  & \multicolumn{1}{c|}{Max.} & $^{01}_{04}$
  &  $4$&$2\CHIEXP{0.81}$
  & $-8$&$4\CHIEXP{1.20}$
  &  $4$&$6\CHIEXP{1.11}$ 
  &  $7$&$5\CHIEXP{0.92}$ \\
  \hline\hline
  \multirow{8}{*}{$u^\val,\,d^\val{}^{(\ast)}$}
  & \twolinebox{${u^\val_\Max=1.05\,u^\val}$}{${d^\val_\Min\,=d^\val-0.05\,u^\val}$} & $^{76}_{77}$
  & $114$&$5\CHIEXP{7.52}$
  &  $68$&$7\CHIEXP{3.21}$
  & $-38$&$1\CHIEXP{1.44}$
  &  \cellcolor{hl}{$2$}&\cellcolor{hl}{$9\CHIEXP{0.95}$} \\
  \arrayrulecolor{Greymin}
  \cline{2-11}
  \arrayrulecolor{Black}
  & \twolinebox{${u^\val_\Min\,=0.95\,u^\val}$}{${d^\val_\Max=d^\val+0.05\,u^\val}$} & $^{78}_{79}$
  & $-138$&$5\CHIEXP{7.93}$
  &  $-87$&$2\CHIEXP{3.41}$
  &   $59$&$8\CHIEXP{1.44}$
  &  \cellcolor{hl}{$4$}&\cellcolor{hl}{$5\CHIEXP{0.94}$} \\
  \arrayrulecolor{Greymax}
  \cline{2-11}
  \arrayrulecolor{Black}
  & \twolinebox{${u^\val_\Max=1.02\,u^\val}$}{${d^\val_\Min\,=0.92\,d^\val}$} & $^{86}_{87}$
  &  $83$&$7\CHIEXP{3.92}$ 
  &  $52$&$6\CHIEXP{2.09}$ 
  & $-30$&$5\CHIEXP{1.27}$
  & \cellcolor{hl}{$1$}&\cellcolor{hl}{$0\CHIEXP{0.99}$} \\
  \arrayrulecolor{Greymin}
  \cline{2-11}
  \arrayrulecolor{Black}
  & \twolinebox{${u^\val_\Min\,=0.98\,u^\val}$}{${d^\val_\Max=1.08\,d^\val}$} & $^{88}_{89}$
  & $-88$&$5\CHIEXP{4.30}$
  & $-56$&$7\CHIEXP{2.26}$ 
  &  $44$&$4\CHIEXP{1.36}$ 
  &  \cellcolor{hl}{$5$}&\cellcolor{hl}{$7\CHIEXP{1.13}$} \\
  \hline\hline
  \multirow{6}{*}{$s,\,c^{(\ast)}$}
  & \twolinebox{${c_\Min\,=0.9\,c},$}{${s_\Max=s+0.1\,c}$} & $^{78}_{79}$
  & $17$&$1\CHIEXP{0.99}$
  &  $9$&$9\CHIEXP{1.12}$
  &  \cellcolor{hl}{$7$}&\cellcolor{hl}{$3\CHIEXP{1.01}$}
  & $19$&$7\CHIEXP{0.99}$ \\
  \arrayrulecolor{Greymin}
  \cline{2-11}
  \arrayrulecolor{Black}
  & \twolinebox{${c_\Max=1.1\,c},$}{${s_\Min\,=s-0.1\,c}$} & $^{80}_{81}$
  & $-10$&$8\CHIEXP{1.22}$ 
  & $-10$&$3\CHIEXP{1.24}$ 
  &  \cellcolor{hl}{$-0$}&\cellcolor{hl}{$3\CHIEXP{1.16}$} 
  & $-15$&$7\CHIEXP{1.17}$ \\
  \arrayrulecolor{Greymax}
  \cline{2-11}
  \arrayrulecolor{Black}
  & \twolinebox{${c_\Min\,=0.8\,c},$}{${s_\Max=s+0.2\,c}$} & $^{76}_{77}$
  & $38$&$8\CHIEXP{1.38}$ 
  & $24$&$7\CHIEXP{1.31}$ 
  &  \cellcolor{hl}{$6$}&\cellcolor{hl}{$1\CHIEXP{1.10}$} 
  & $38$&$0\CHIEXP{1.71}$ \\
  \arrayrulecolor{Greymin}
  \cline{2-11}
  \arrayrulecolor{Black}
  & \twolinebox{${c_\Max=1.2\,c},$}{${s_\Min\,=s-0.2\,c}$} & $^{82}_{83}$
  & $-29$&$0\CHIEXP{1.42}$ 
  & $-23$&$7\CHIEXP{1.29}$ 
  &   \cellcolor{hl}{$1$}&\cellcolor{hl}{$0\CHIEXP{1.27}$}
  & $-33$&$8\CHIEXP{2.00}$ \\
\hline
\end{tabular}
\renewcommand\arraystretch{1.45}

\end{TableSizeSevenPt}
  \caption[The shifts of the $W$-mass charge asymmetry
    corresponding to various modeling effects using the charge asymmetry of $\pTl$ for the
    analysis (Detailed)]
          {\figtxt{The shifts of the $W$-mass charge asymmetry
                   corresponding to various modeling effects.
                   The systematic labeled $\ast$ are obtained using the scaling trick mentioned in 
                   \S\,\ref{ss_scaling_trick}.}}
\label{table_app_charge_asym_results}
\index{Quarks!smc@$s-c$ asymmetry}
\index{Quarks!uvmdv@$u^\val-d^\val$ asymmetry}
\index{Quarks!Valence quarks}
\index{Quarks!Intrinsic transverse momenta}
\index{Chi2 Likelihood analysis@Chi-2 ($\chiD$) likelihood analysis!Results (detailed)}
\end{center}
\end{table}

\clearpage
\subsection{Detailed comments and graphics}
\subsubsection{Energy scale of the charged lepton}\label{ss_a1}
\begin{figure}[!ht] 
  \begin{center}
    \includegraphics[width=0.495\tw]{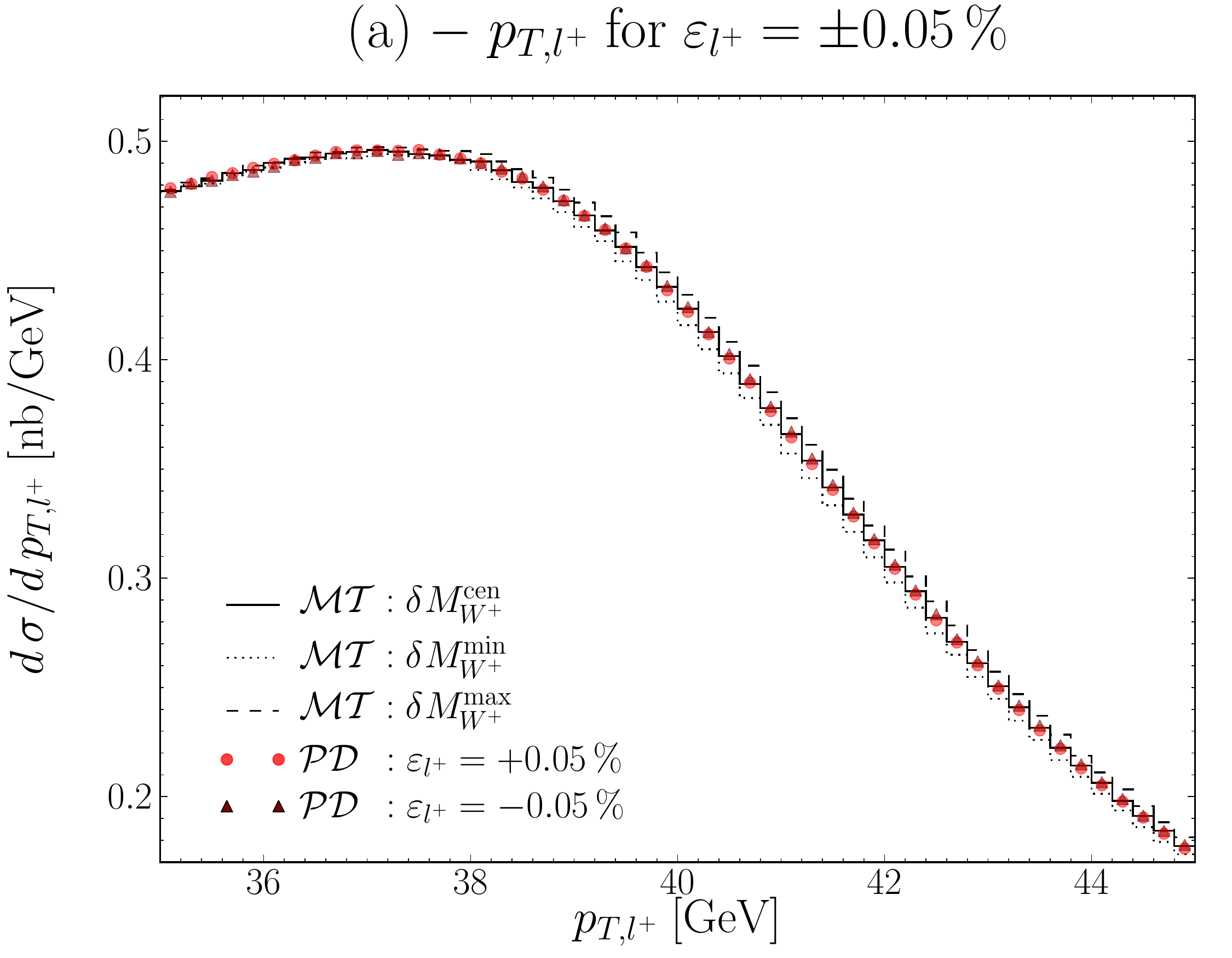}
    \hfill
    \includegraphics[width=0.495\tw]{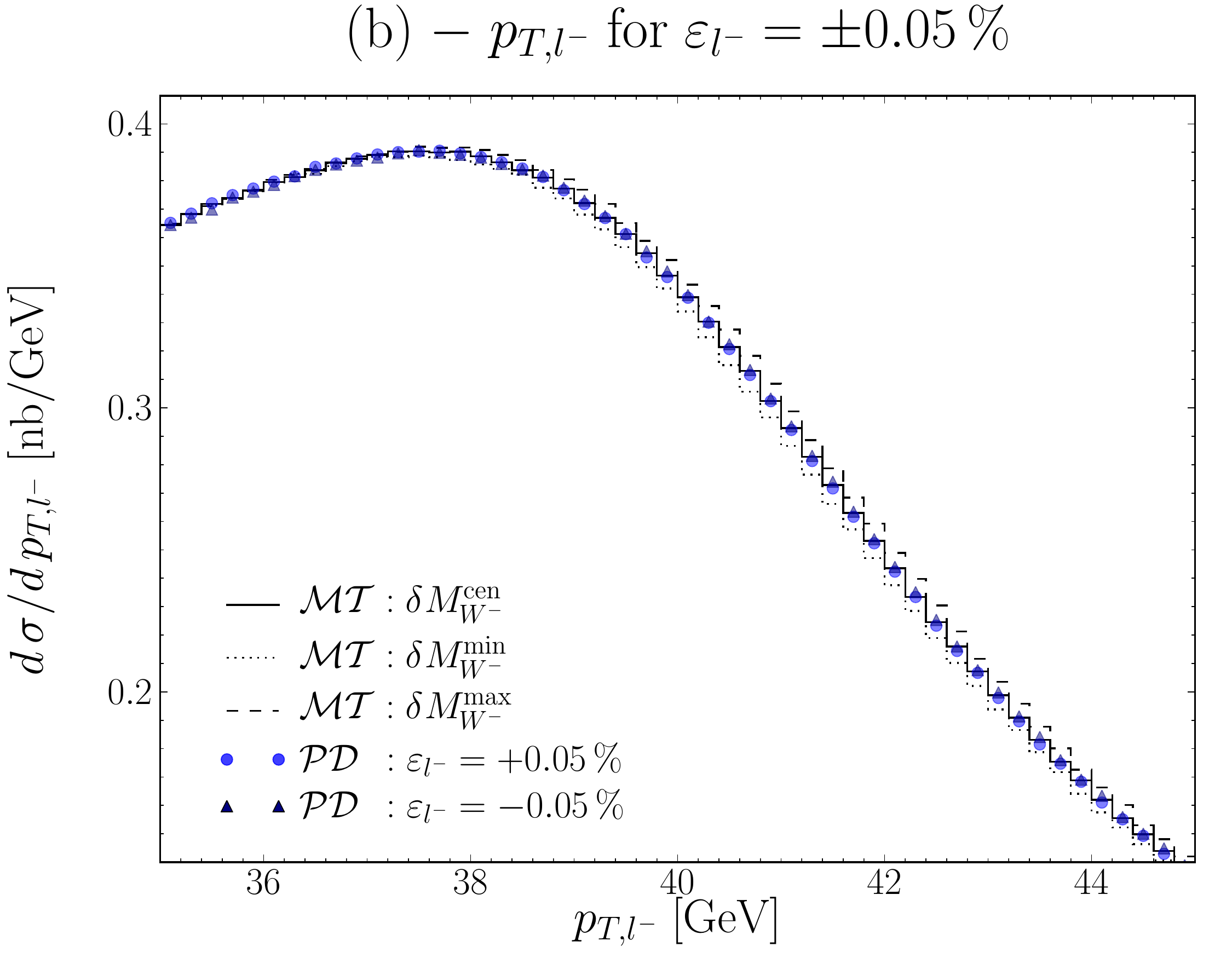}
    \vfill
    \includegraphics[width=0.495\tw]{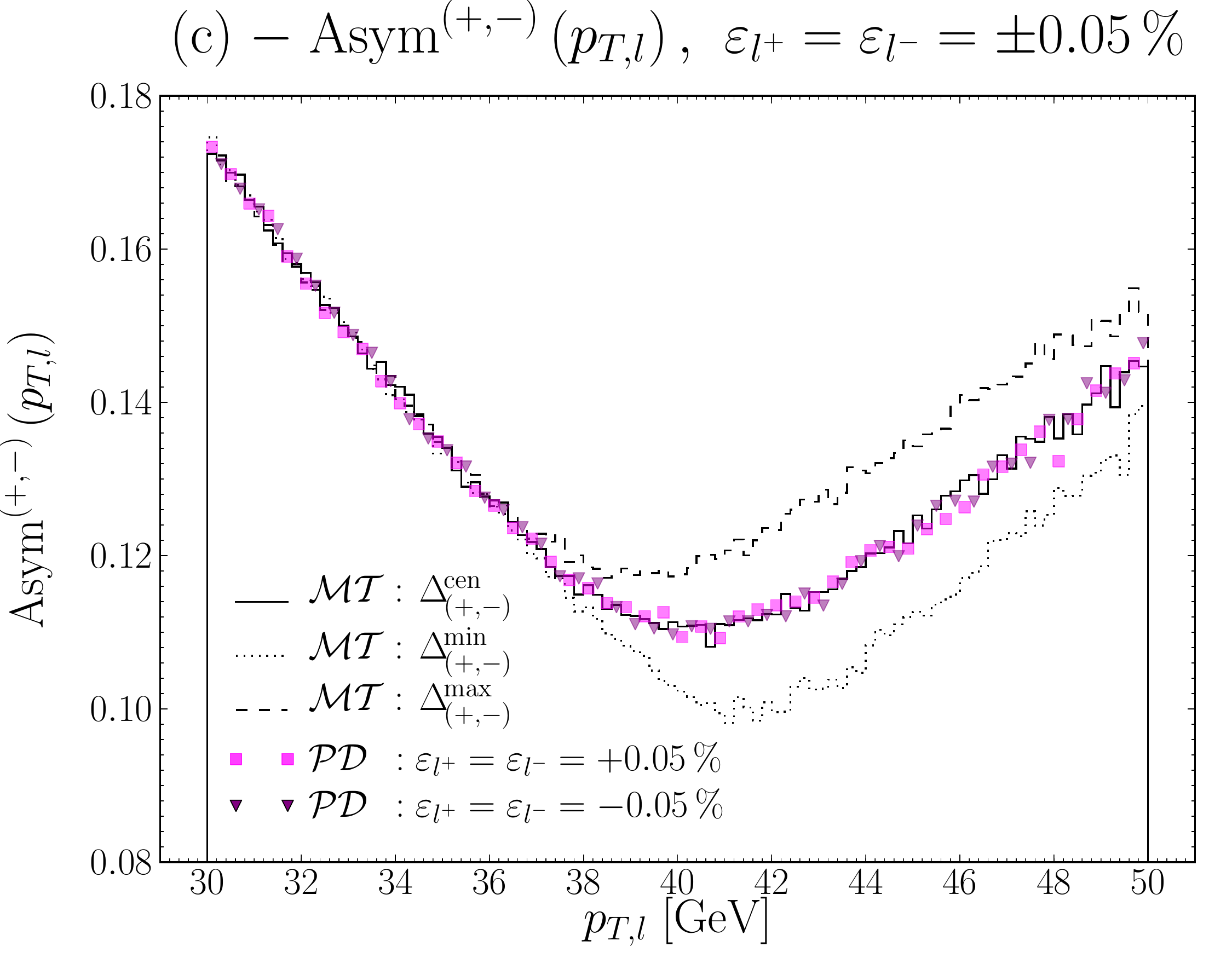}
    \hfill
    \includegraphics[width=0.495\tw]{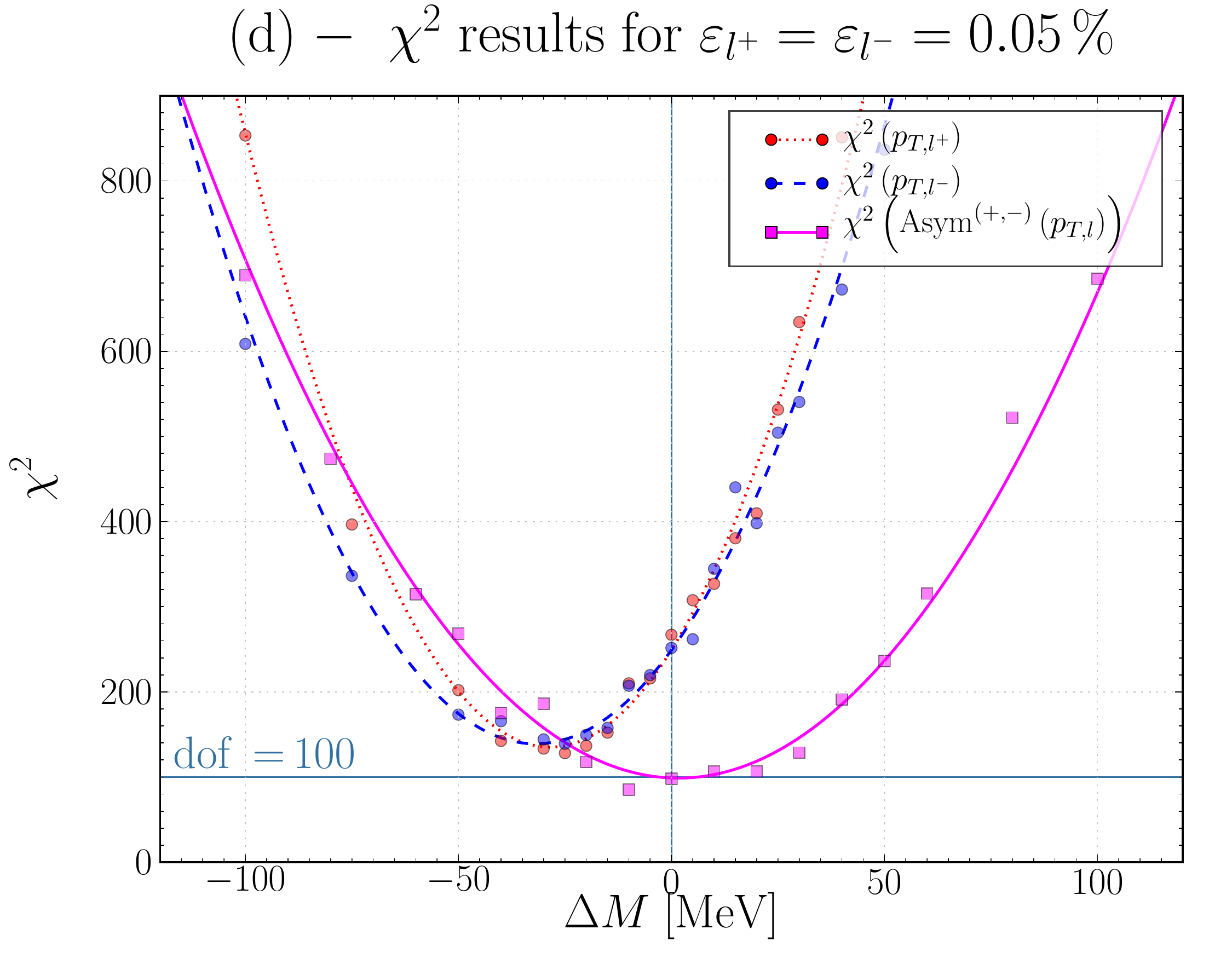}
    \caption[Systematic error on $\MWp-\MWm$ due to coherent energy scale biases 
      ($\es_l=0.05\percent$) between the positively and negatively charged lepton reconstructed 
      transverse momenta and for both classic and charge asymmetry methods]
            {\figtxt{Systematic error on $\MWp-\MWm$ due to coherent energy scale biases 
                ($\es_l=0.05\percent$) between the positively and negatively charged lepton 
                reconstructed transverse momenta and for both classic and charge asymmetry methods.
                Frames (a) and (b) display respectively the jacobian peaks of the $\pTlp$ and $\pTlm$ 
                spectra for the minimum, central and maximum $\MT$ and for the two biased $\PD$ 
                distributions corresponding to $\es_l=\pm 0.05\percent$ while
                frame (c) displays the same data but for $\Asym{\pTl}$.
                Frame (d) presents the $\chiD$ results for the classic and charge asymmetry methods 
                for the case of $\es_\lp=\es_\lm=0.05\percent$, $\Delta M\equiv M_\Wpm^{\PD}-M_W^{(\mm{ref.})}$
                for the classic method results and $\Delta M\equiv \MWp-\MWm$ for the charge asymmetry result.
}}
              \label{fig_a1}
  \end{center} 
\end{figure}
We start our detailed review on the energy scale systematics looking at the coherent biases.
The plots above in Fig.~\ref{fig_a1} illustrate the consequences of coherent biases between the 
positively and negatively charged lepton reconstructed tracks.
Fig.~\ref{fig_a1}.(a) represents several distributions of the transverse momentum of the $\lp$ 
lepton.
For the sake of clarity, only three mass templates $\MT$ from the 
entire collection have been drawn, the minimum template ($\delta\,M_\Wp=-200\MeV$), 
the central one ($\delta\,M_\Wp=0\MeV$) and the maximum one  ($\delta\,M_\Wp=+200\MeV$). 
Although the histograms are considered for the analysis in the range of $30\GeV<\pTl<50\GeV$, a zoom was 
made on the jacobian peaks to make it possible to resolve by the eye the differences between the 
three $\MT$.
In top of these histograms the $\PD$ histograms corresponding to $\es_\lp=-0.05\percent$ and 
$\es_\lp=\pm 0.05\percent$ are shown. 
For such values the deviation from $\Delta_{(+,-)}^\Cen$ is hardly decipherable.
Fig.~\ref{fig_a1}.(b) shows the exact same thing than in (a) but this time for the negative lepton. 
Fig.~\ref{fig_a1}.(c) represents the charge asymmetry of $\pTl$ for its central $\MT$
($\delta M_\Wp^\Cen\equiv 0\MeV$), minimum $\MT$ ($\Delta_{(+,-)}^\Min\equiv -200\MeV$) and 
maximum $\MT$ ($\Delta_{(+,-)}^\Max\equiv +200\MeV$).
Are also present the two $\PD$ for which $\es_\lp=\es_\lm=\pm 0.05\percent$.
Note here that the difference between the $\MT$ is much more magnified which is directly associated 
with the fact that the charge asymmetry, by its form, get rid of the common features from both 
$\pTlp$ and $\pTlm$ histograms and as a consequence emphasise the discrepancies among them.
Note also that to decipher between the points of the two different $\PD$, in each distribution a point 
is being skipped.
Fig.~\ref{fig_a1}.(d) presents the $\chiD$ results for both classic and charge asymmetry 
methods for the case of $\es_\lp=\es_\lm=0.05\percent$.
As seen in Table~\ref{table_app_exp_sys_classic_vs_casym_vs_dcasym} both errors are quite steady with
respect to the considered value of $|\es_l|=0.05\percent$, nonetheless the convergence for the charge 
asymmetry method is slightly better than for the classic method.

\begin{figure}[!ht] 
  \begin{center}
    \includegraphics[width=0.495\tw]{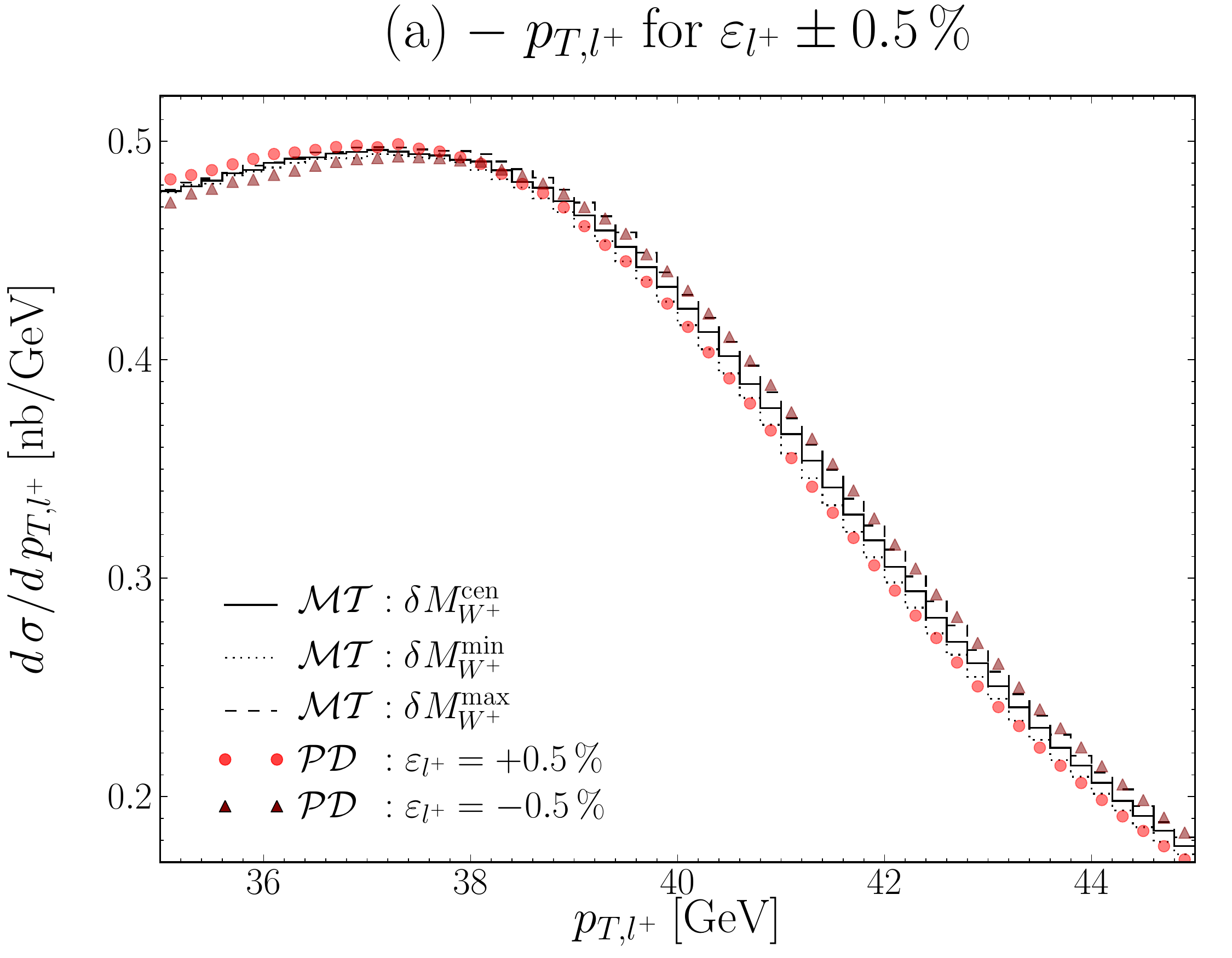}
    \hfill
    \includegraphics[width=0.495\tw]{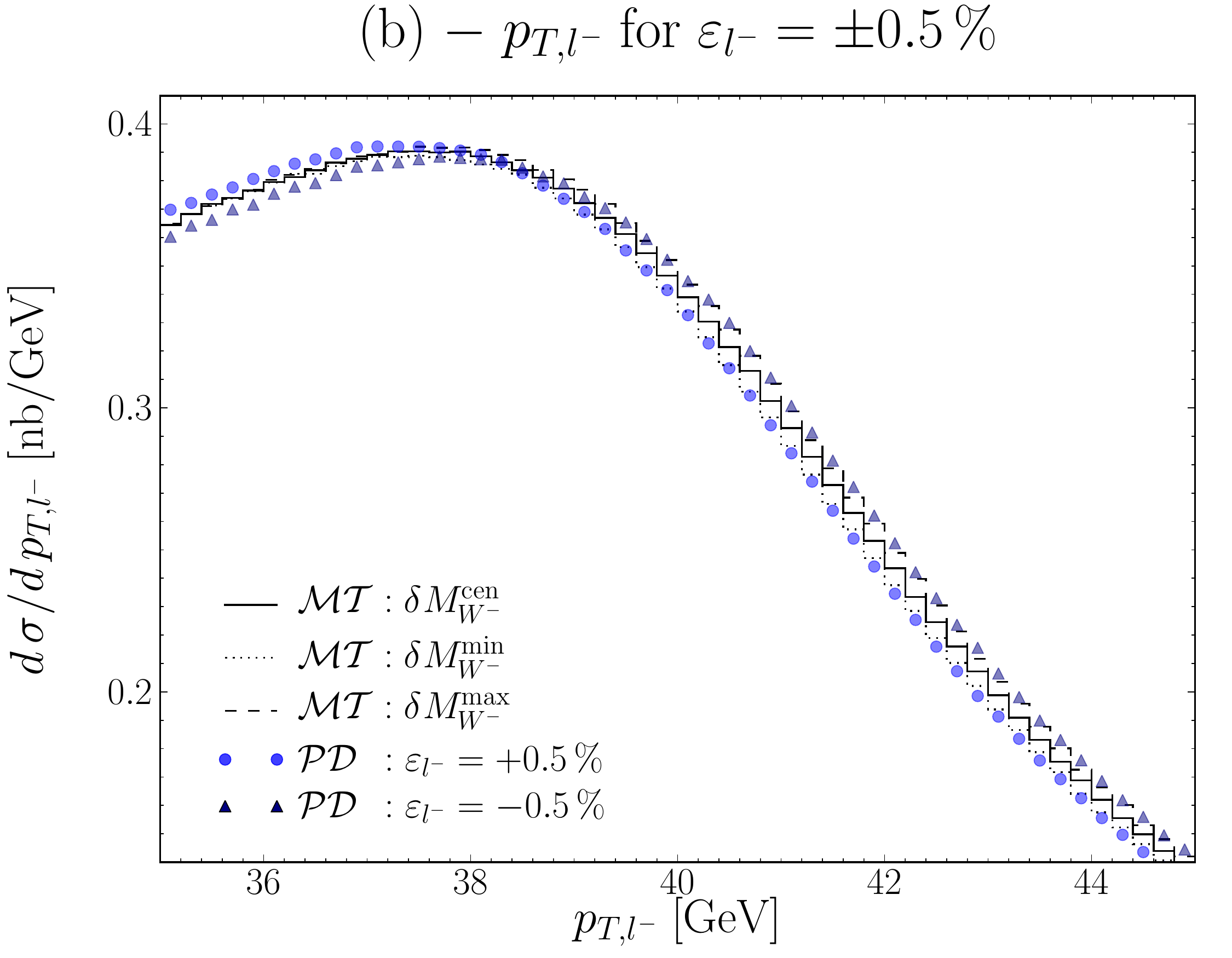}
    \vfill
    \includegraphics[width=0.495\tw]{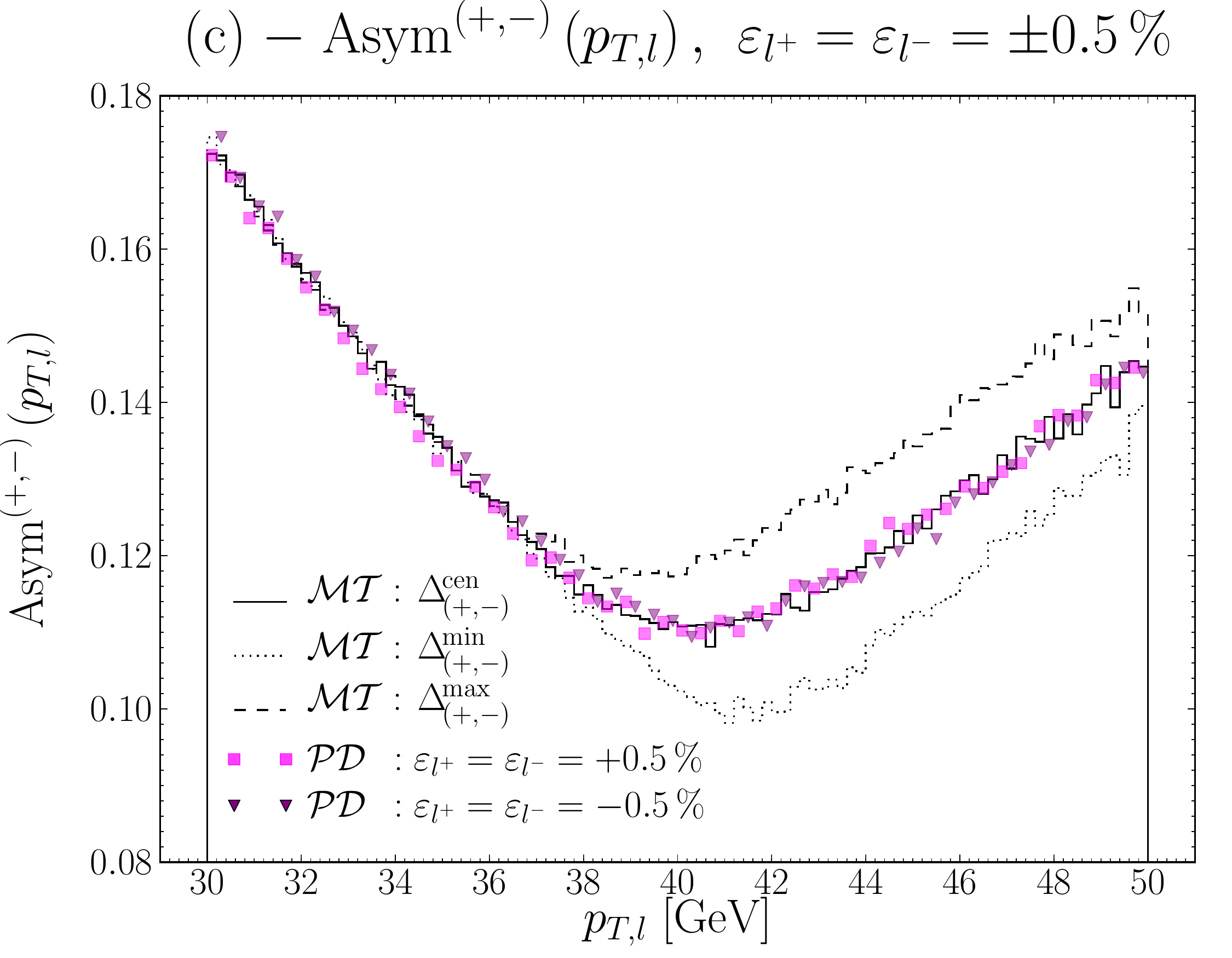}
    \hfill
    \includegraphics[width=0.495\tw]{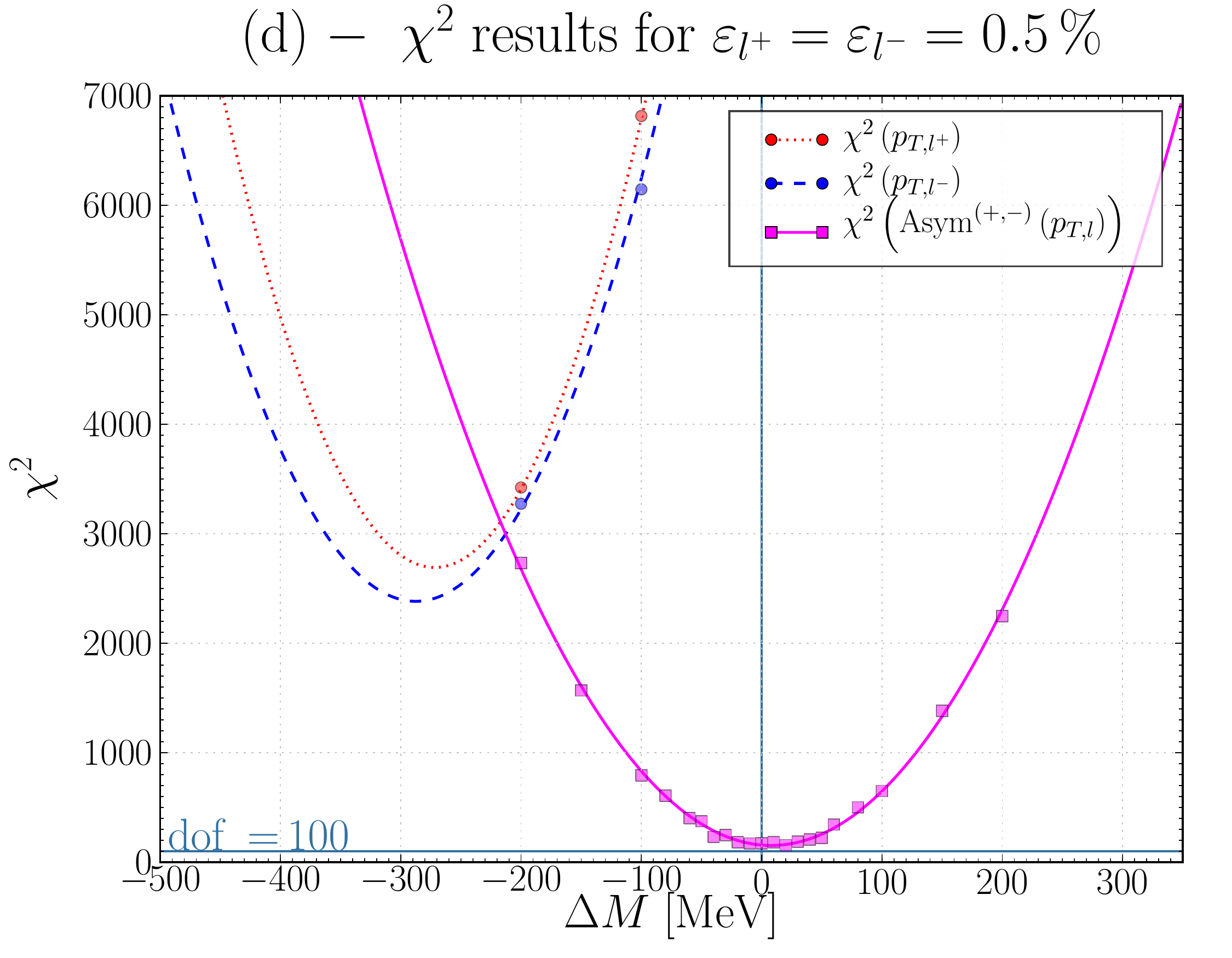}
    \caption[Systematic error on $\MWp-\MWm$ due to coherent energy scale biases ($|\es_l|=0.5\percent$) 
      between the positively and negatively charged lepton reconstructed transverse momenta and
      for both classic and charge asymmetry methods]
            {\figtxt{Systematic error on $\MWp-\MWm$ due to coherent energy scale biases 
                ($\es_l=\pm 0.5\percent$) between the positively and negatively charged lepton 
                reconstructed transverse momenta and for both classic and charge asymmetry methods.
                Frames (a) and (b) display respectively the jacobian peaks of the $\pTlp$ and $\pTlm$ 
                spectra for the minimum, central and maximum $\MT$ and for the two biased $\PD$ 
                distributions corresponding to $\es_l=\pm 0.5\percent$ while frame (c) displays the 
                latter distributions for $\Asym{\pTl}$.
                Frame (d) presents the $\chiD$ results for the classic and charge asymmetry methods 
                for the case of $\es_\lp=\es_\lm=0.5\percent$, $\Delta M\equiv M_\Wpm^{\PD}-M_W^{(\mm{ref.})}$
                for the classic method results and $\Delta M\equiv \MWp-\MWm$ for the charge asymmetry result.
            }}
            \label{fig_a2}
  \end{center} 
\end{figure}
Now Fig.~\ref{fig_a2} represents the same histograms than in Fig.~\ref{fig_a1} but this time with 
$\es_l=\pm 0.5\percent$.
Now the bias is large enough so we can see some differences. Indeed, starting with 
Fig.~\ref{fig_a2}.(a) the jacobian peaks of the $\PD$ are slightly shifted to higher $\pT$ for 
$\es_l=-0.5\percent$ and to lower $\pT$ for $\es_l=0.5\percent$.
This can seem quite non intuitive but let us remind the biases are applied to $\rhoTl$, the inverse 
of $\pTl$, hence as long as $\es_l\ll 1$, the bias on $p_{T,l}^\rec$ can be deduced at first order
from the expression of $\rho_{T,l}^\rec$\,:
\begin{equation}
\rho_{T,l}^\rec=\rho_{T,l}^\true\left(1+\es_l\right)   
\quad\Rightarrow\quad p_{T,l}^\rec=p_{T,l}^\true\left(1-\es_l\right),
\end{equation}
justifying the inverse behaviour between $\rhoTl$ and $\pTl$ biases. 
Then, considering for example the case of $\es_l=-0.5\percent$, 
in both Figs.~\ref{fig_a2}.(a) and (b) the jacobian peaks of the $\PD$ 
being shifted to lower $\pT$ both $M_\Wp^{\PD}$ and $M_\Wm^{\PD}$ are underestimated by 
$\approx 300\MeV$ as seen in frame (c). Nonetheless, since these biases are coherent the impact on
$\MWp-\MWm$ is of the order of $16\MeV$ only.
The low convergences $\chiDmin/\dof\approx 30$ is due to the lack of enough mass templates.
In Fig.~\ref{fig_a2}.(c) the charge asymmetry of $\pTl$ the two $\PD$ show more
steadiness than the bare $\pTl$ spectra.
Again for the sake of clarity in each $\PD$ distribution one point is being skipped each time.
As expected the stability of the charge asymmetry with respect to biases of $|\es_l|=0.05\percent$ is 
such that the $\chiD$ shows good convergence and no particular deviation from the central 
$\MWp-\MWm=0$.

\index{Double charge asymmetry!Used for the extraction of MWpmMWm@
Used for the extraction of $\MWp-\MWm$|(}
\begin{figure}[!ht] 
  \begin{center}
    \includegraphics[width=0.495\tw]{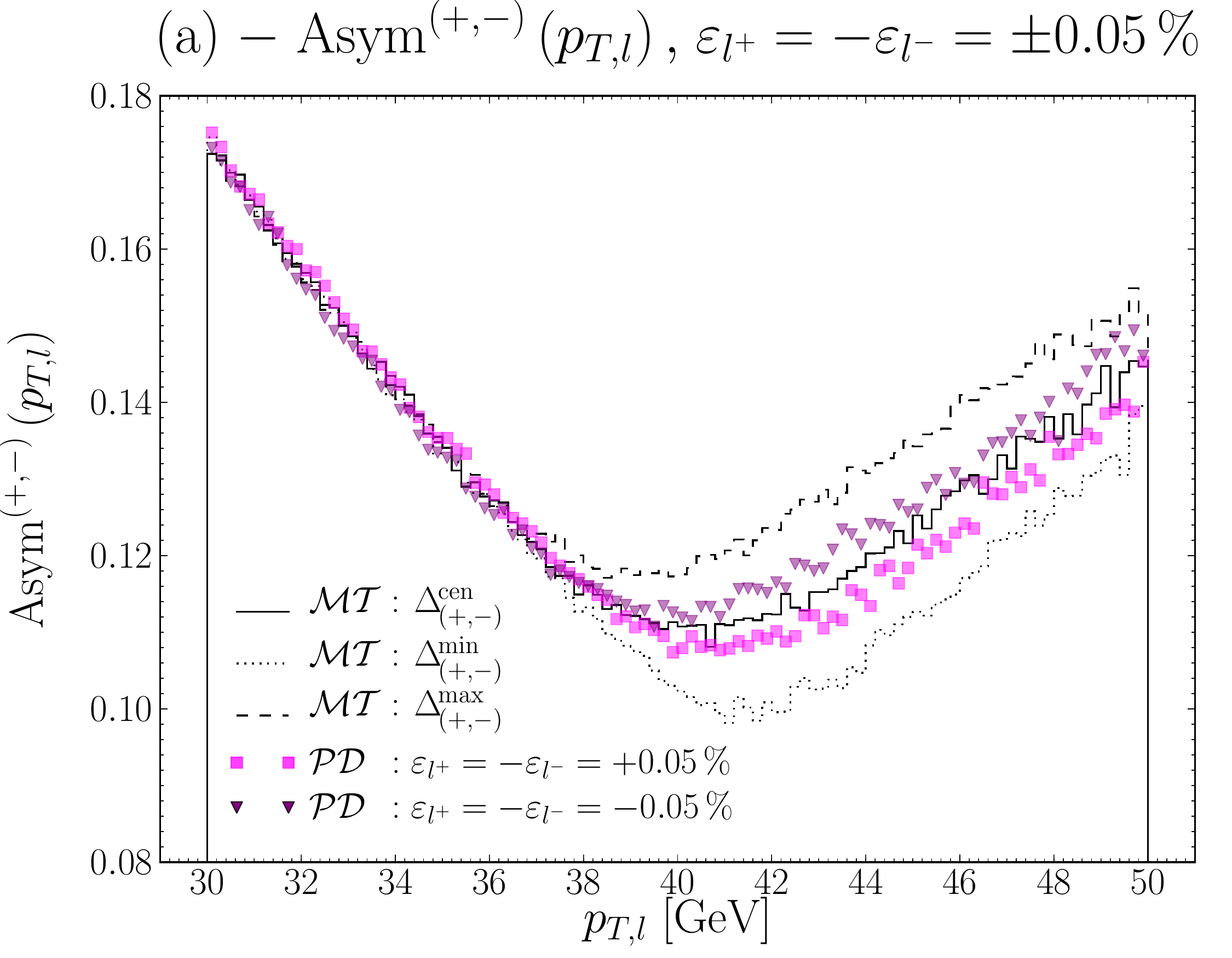}
    \hfill
    \includegraphics[width=0.495\tw]{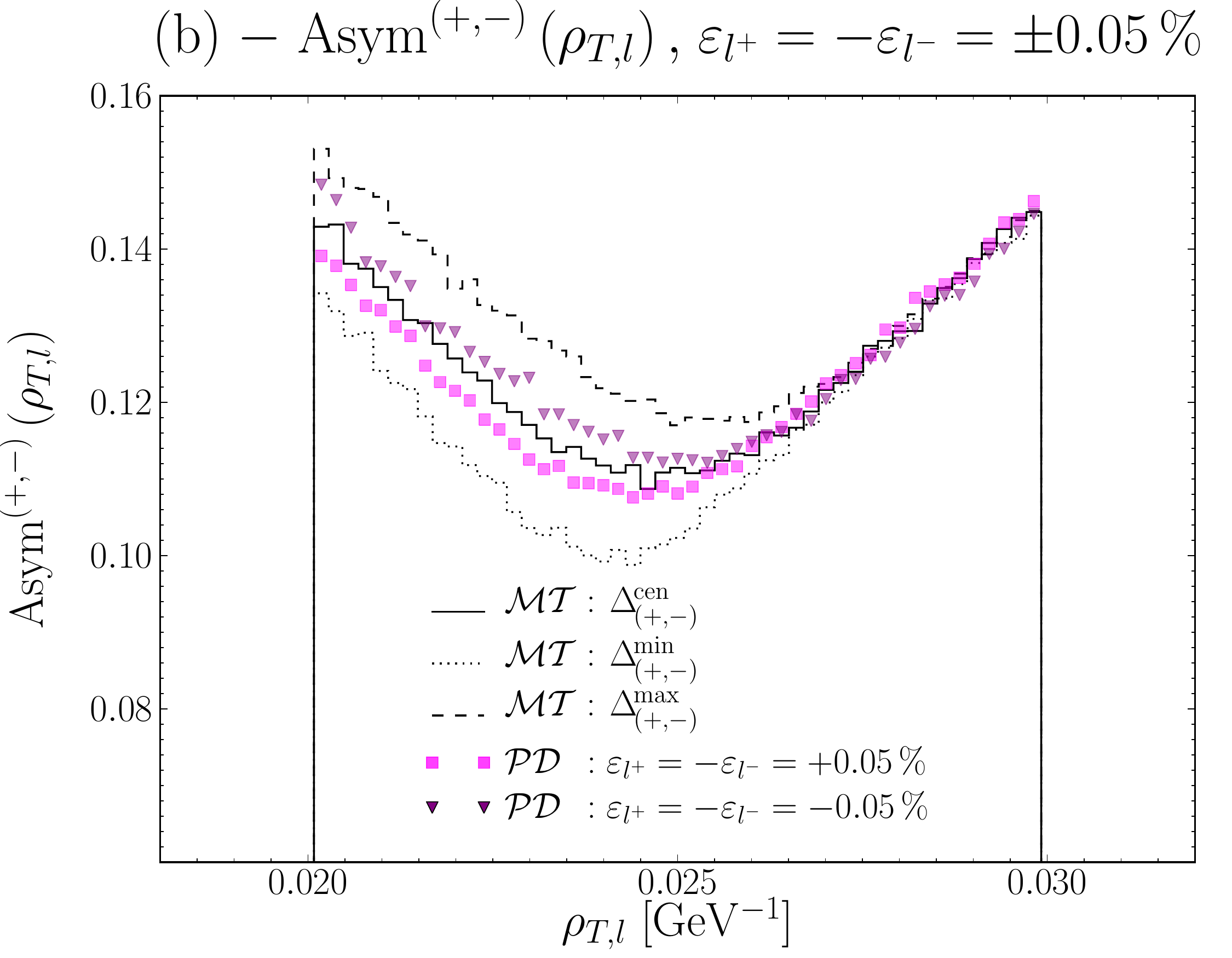}
    \vfill
    \includegraphics[width=0.495\tw]{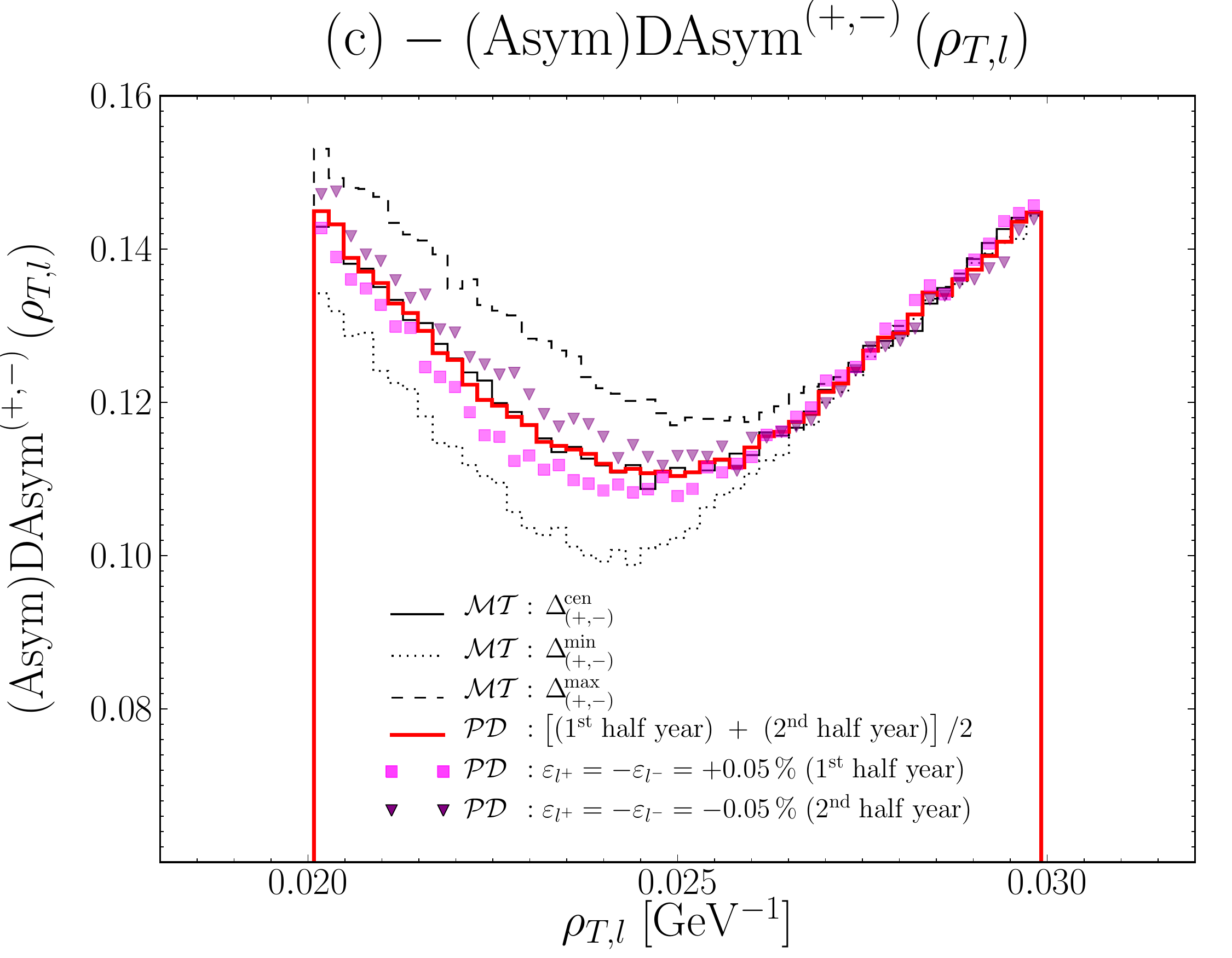}
    \hfill
    \includegraphics[width=0.495\tw]{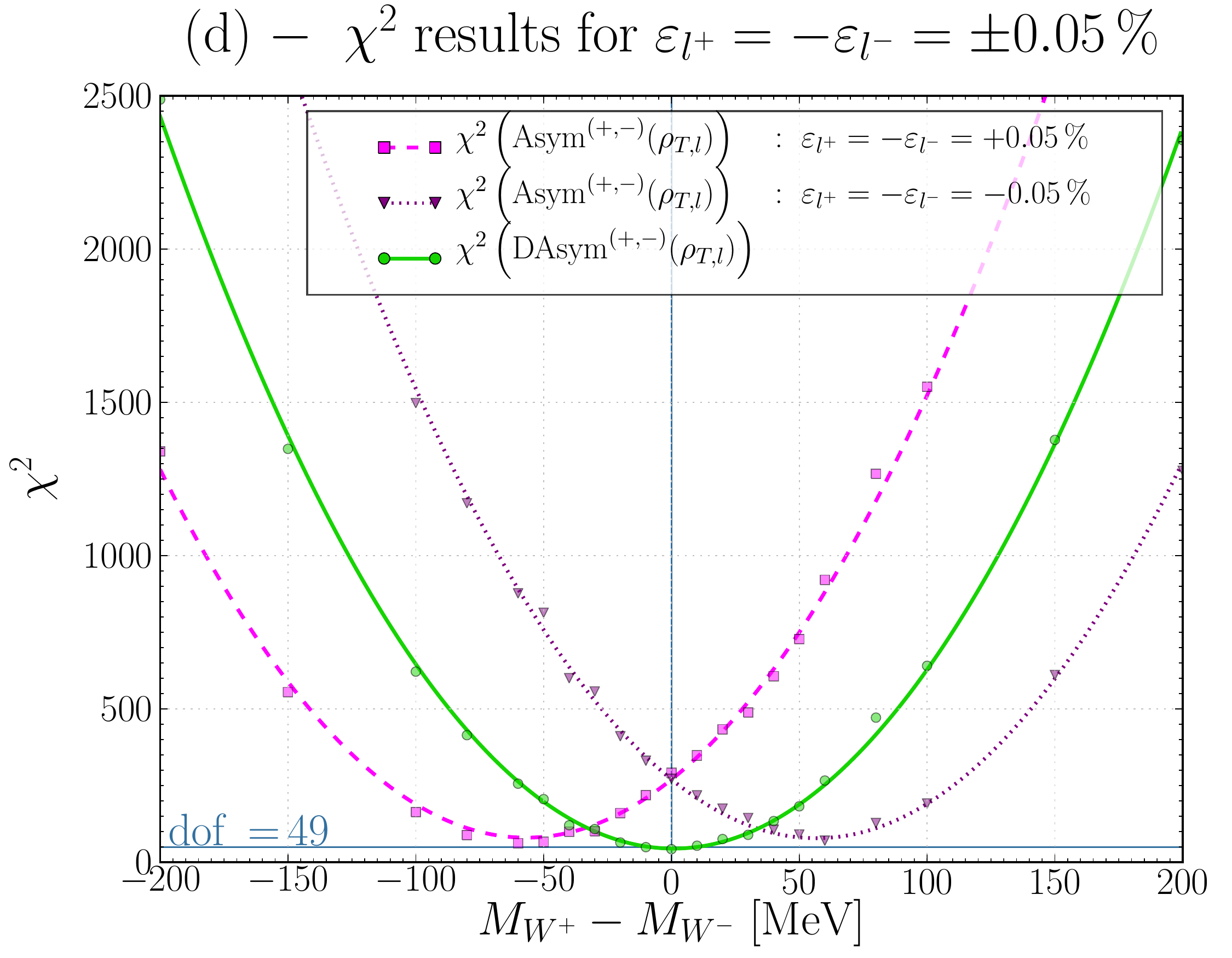}
    \caption[Systematic error on $\MWp-\MWm$ due to incoherent energy scale biases ($\es_l=\pm 0.05\percent$) 
      for both charge and double charge asymmetry methods]
            {\figtxt{Systematic error on $\MWp-\MWm$ due to incoherent energy scale biases 
                ($\es_l=\pm 0.05\percent$) for both charge and double charge asymmetry methods.
                Frames (a) and (b) display respectively the charge asymmetry of $\pTl$ and $\rhoTl$ 
                for the minimum, central and maximum $\MT$ along with the $\PD$ corresponding to
                incoherent biases of size $\es_l=\pm 0.05\percent$.
                Frame (c) displays the latter distributions for $\DAsym{\rhoTl}$.
                Frame (d) presents the $\chiD$ results for both charge and double charge asymmetry 
                methods for the case of $\es_\lp=\es_\lm=0.05\percent$.}
            }
            \label{fig_af3}
  \end{center} 
\end{figure}
The incoherent biases for the energy scale between the positive and negative charged lepton is now 
addressed starting with Fig.~\ref{fig_af3} that considers the case of 
$\es_\lp=-\es_\lm=\pm 0.05\percent$. This time the classic method is no longer treated since 
the $\chiD$ are too much out of charts. Instead the charge and double charge asymmetry are confronted. 
First, Fig.~\ref{fig_af3}.(a) shows the charge asymmetry of $\pTl$ for the usual central and 
extrema $\MT$ and the two $\PD$ biased by energy scales of $\es_\lp=-\es_\lm=\pm 0.05\percent$.
In Fig.~\ref{fig_af3}.(b) the same data is drawn in function of $\rhoTl$ since it is in that space
the $\chiD$ is performed.
Unlike the ``coherent biases'' here the $\PD$ deviation from the central $\MT$ are such that in (d) 
the systematic error are quite important, more precisely of the order of $\pm 50\MeV$.
Now in Fig.~\ref{fig_af3}.(c) the case of the double charge asymmetry is considered in the $\rhoTl$ 
space. The $\MT$ are exactly the same than the one used in (b), and the pseudo-data shows the cases
where $\es_\lp=-\es_\lm=0.05\percent$ during the first year of the data collection 
and $\es_\lp=-\es_\lm=-0.05\percent$ for the second half of the year due to 
the reversing of the solenoidal magnetic field of the tracker.
Averaging the two data batches from the two six months period gives eventually a data collection 
which is localised near the central $\MT$. The consequence can be seen in  
Fig.~\ref{fig_af3}.(d), the double charge asymmetry is robust against such incoherent biases.

\begin{figure}[!ht] 
  \begin{center}
    \includegraphics[width=0.495\tw]{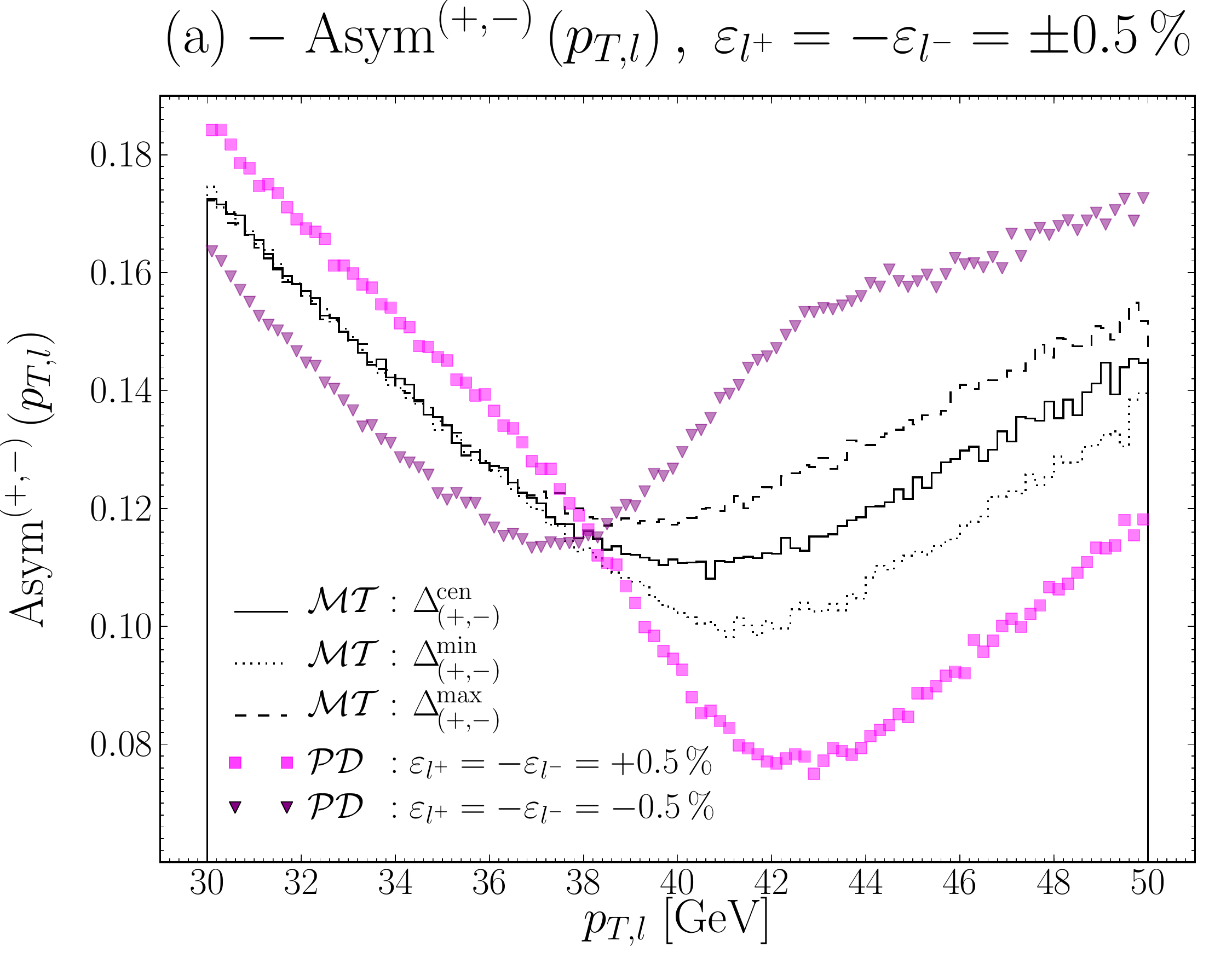}
    \hfill
    \includegraphics[width=0.495\tw]{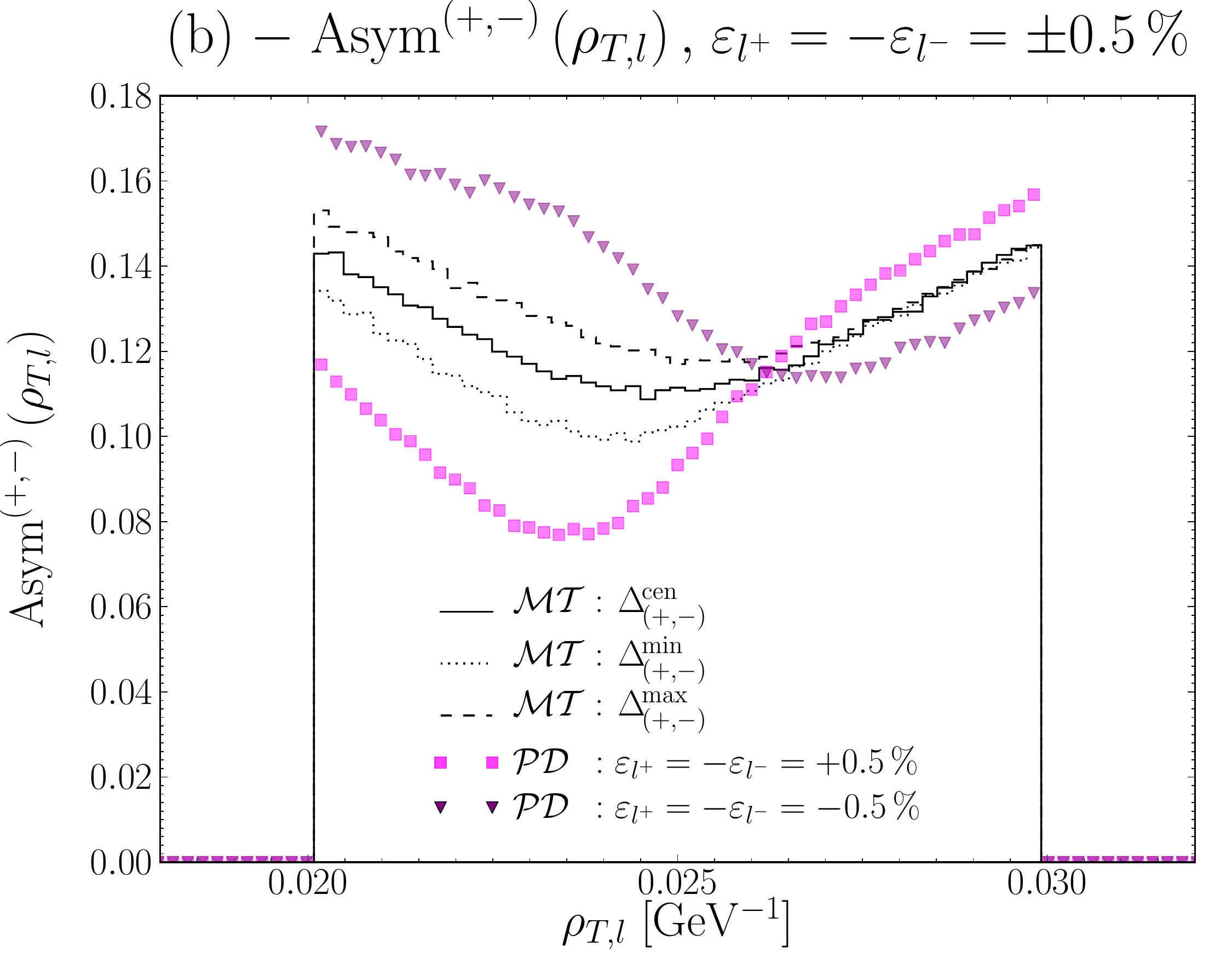}
    \vfill
    \includegraphics[width=0.495\tw]{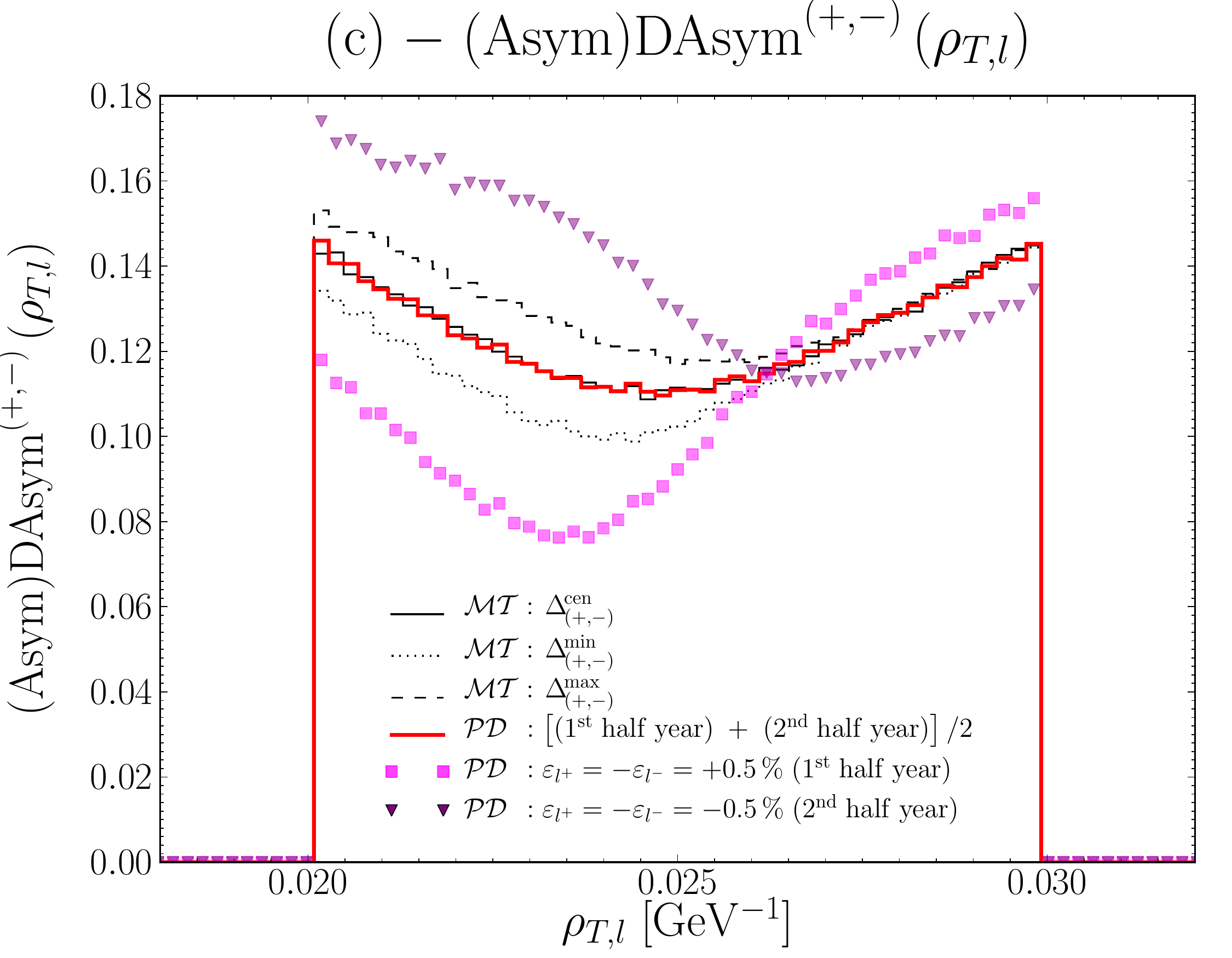}
    \hfill
    \includegraphics[width=0.495\tw]{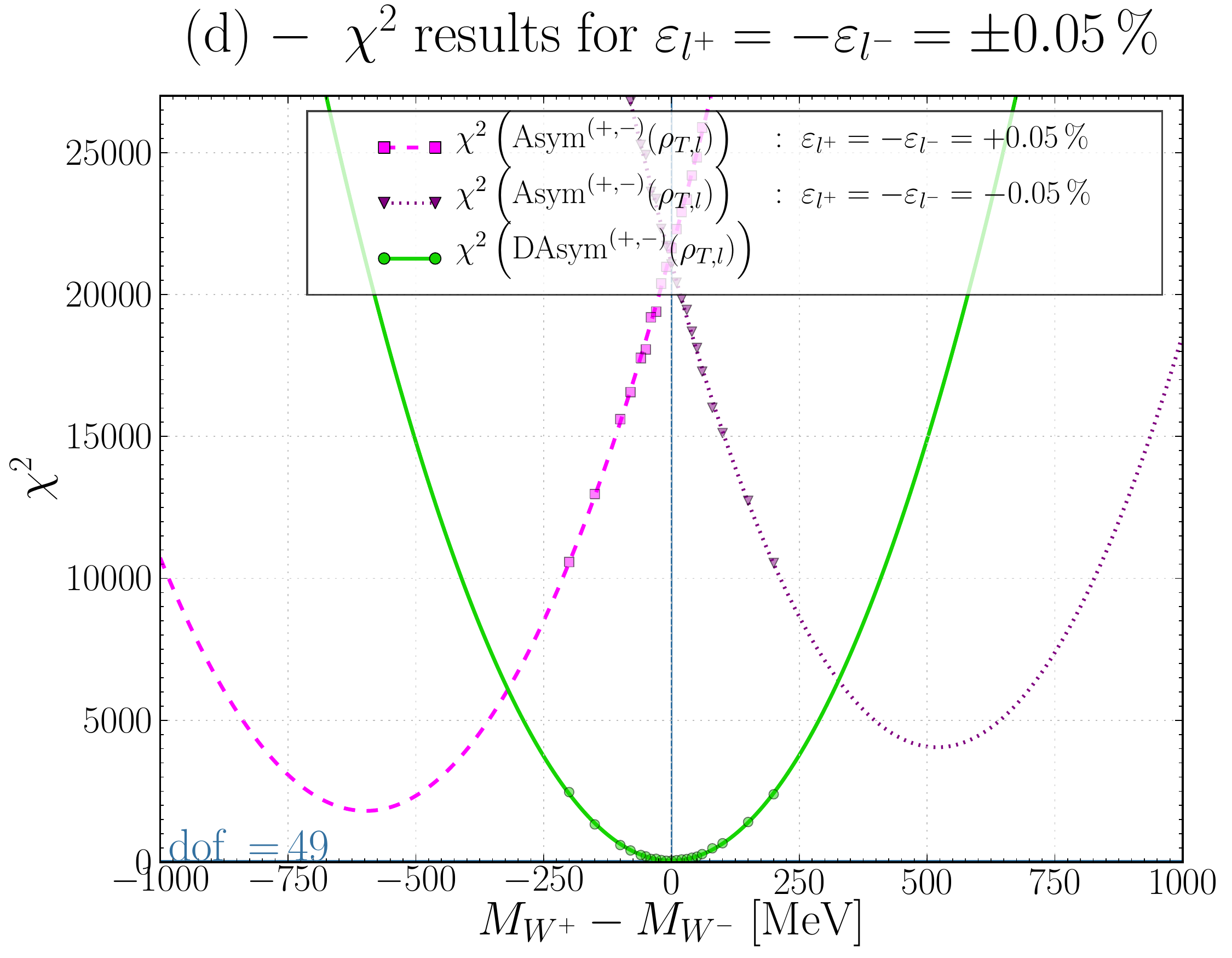}
    \caption[Systematic error on $\MWp-\MWm$ due to incoherent energy scale biases ($\es_l=\pm 0.5\percent$) for both charge and double charge asymmetry methods]
            {\figtxt{Systematic error on $\MWp-\MWm$ due to incoherent energy scale biases 
                ($\es_l=\pm 0.5\percent$) for both charge and double charge asymmetry methods.
                Frames (a) and (b) display respectively the charge asymmetry of $\pTl$ and $\rhoTl$ 
                for the minimum, central and maximum $\MT$ along with the $\PD$ corresponding to
                incoherent biases of size $\es_l=\pm 0.5\percent$.
                Frame (c) displays the latter distributions for $\DAsym{\rhoTl}$.
                Frame (d) presents the $\chiD$ results for both charge and double charge asymmetry 
                methods for the case of $\es_\lp=\es_\lm=0.5\percent$.}
            }
            \label{fig_af4}
  \end{center} 
\end{figure}

Finally, the ``incoherent biases'' are considered for $\es_\lp=-\es_\lm=\pm 0.5\percent$ as shown in 
Fig.~\ref{fig_af4}. Here we can see the biased data are completely different from the templates. 
Nonetheless, for the double charge asymmetry the average of the two ``six-months-data'' gives 
again a total pseudo-data batch centered on the central $\MT$ distribution. Then, once again, the
systematic error is still negligible for the double charge asymmetry while the charge asymmetry is 
not sufficient to draw any conclusion. 
\index{Double charge asymmetry!Used for the extraction of MWpmMWm@
Used for the extraction of $\MWp-\MWm$|)}

\clearpage
\subsubsection{Resolution of the charged lepton track parameters}\label{ss_a2}
\begin{figure}[!ht] 
  \begin{center}
    \includegraphics[width=0.495\tw]{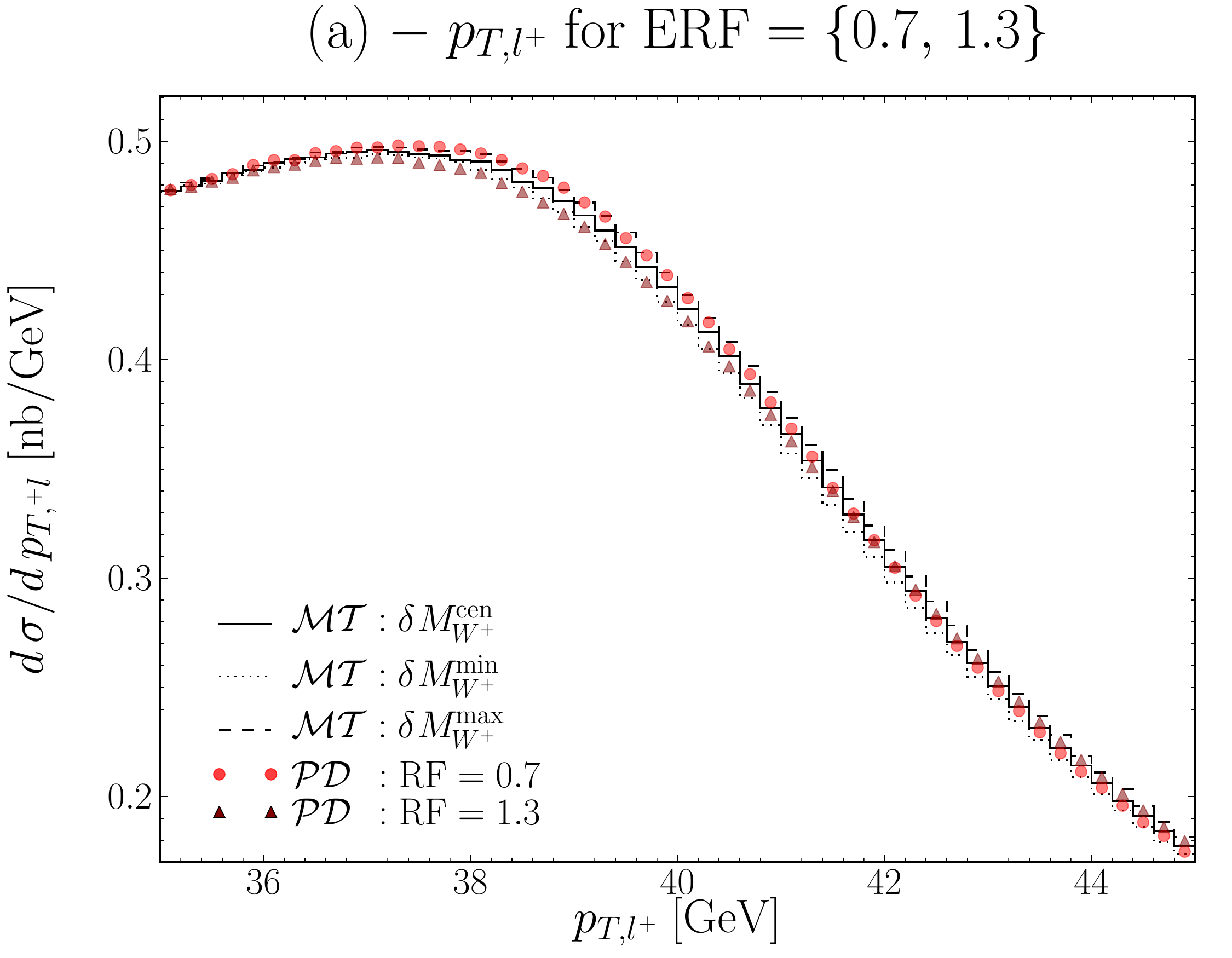}
    \hfill
    \includegraphics[width=0.495\tw]{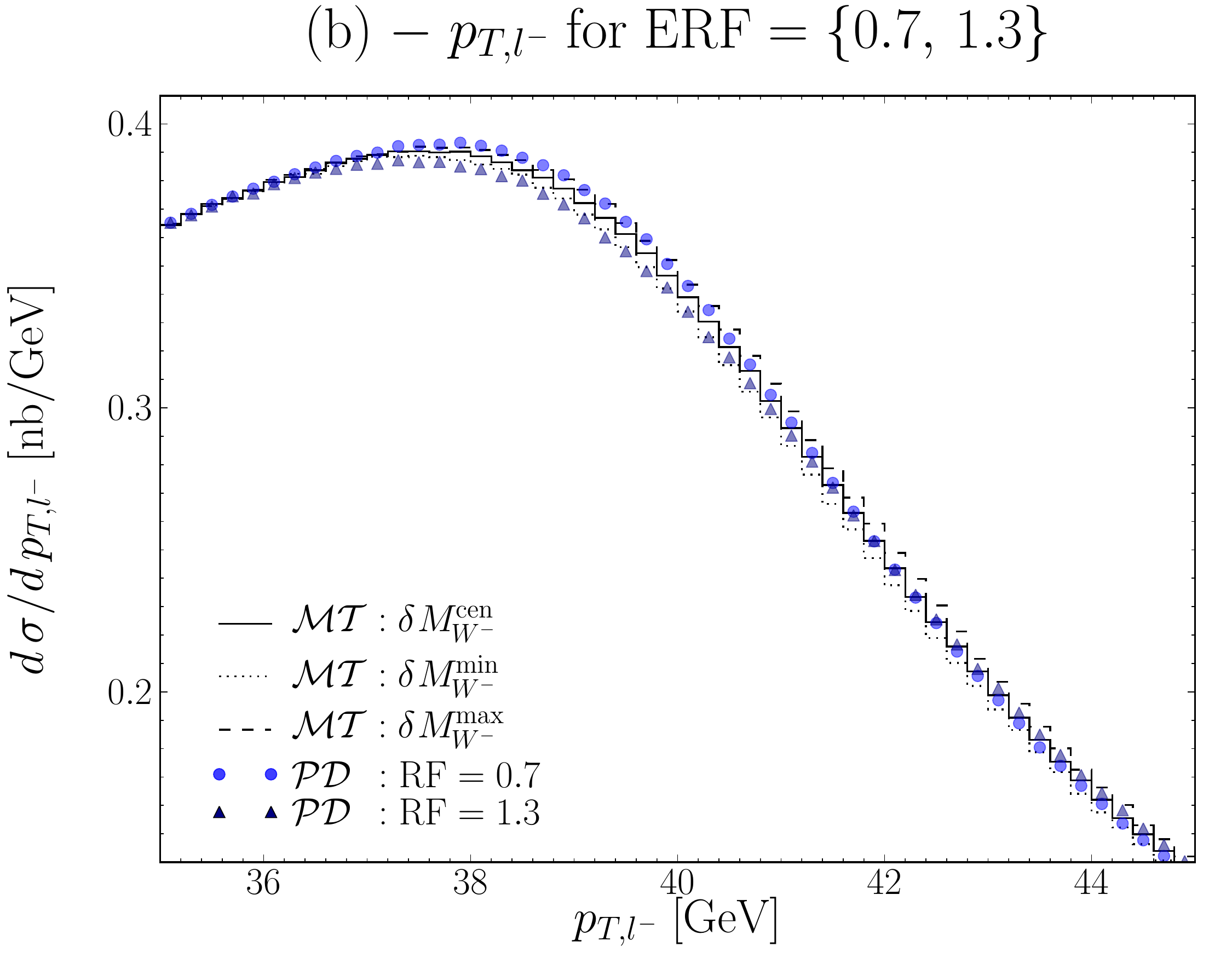}
    \vfill
    \includegraphics[width=0.495\tw]{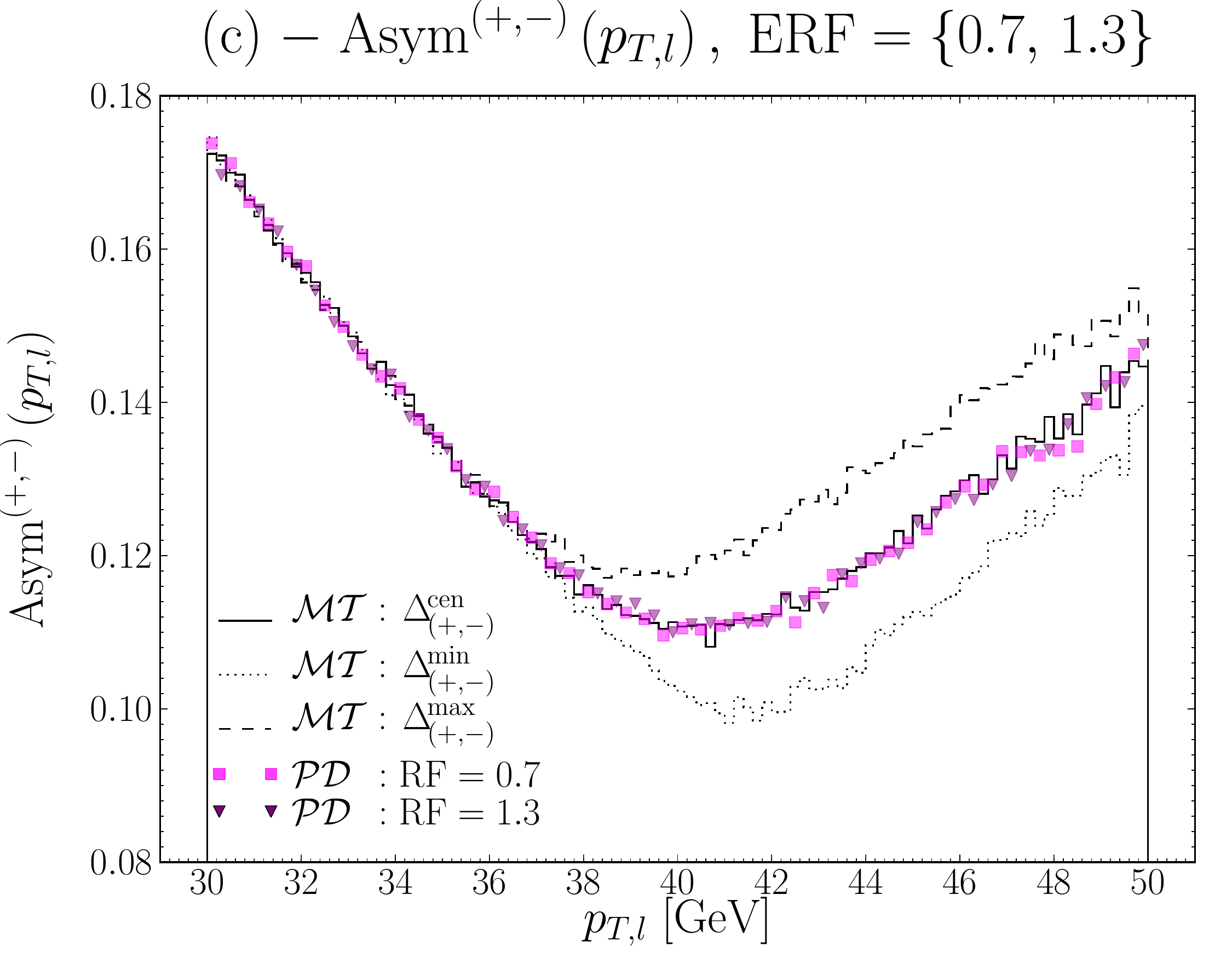}
    \hfill
    \includegraphics[width=0.495\tw]{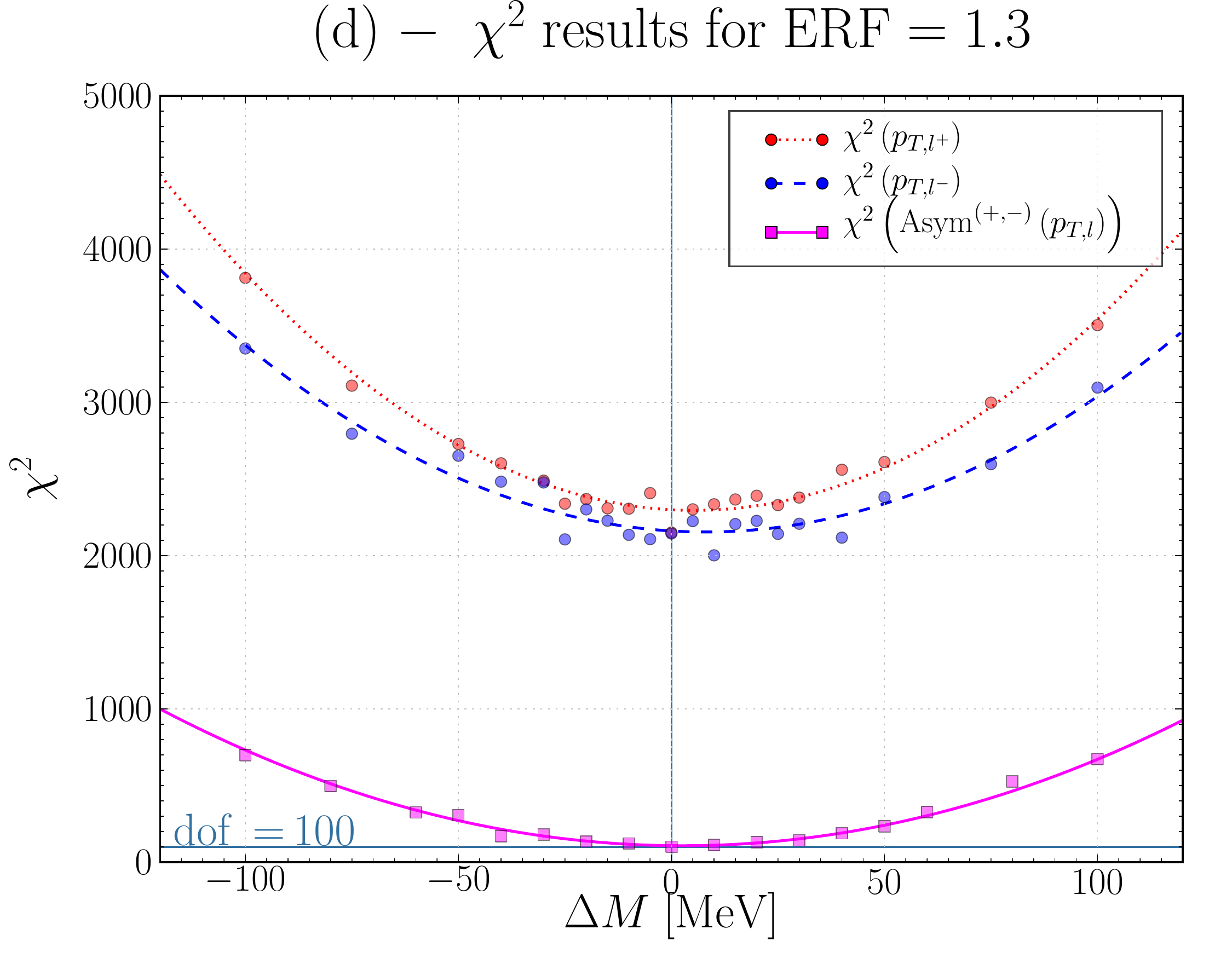}
    \caption[Systematic error on $\MWp-\MWm$ due to the resolution on the charged 
      lepton track for both classic and charge asymmetry method]
            {\figtxt{Systematic error on $\MWp-\MWm$ due to the resolution 
                on the charged lepton track for $\pTlp$ (a) and $\pTlm$ (b) and 
                charge asymmetry (c) spectra. In each frame the central and extrema $\MT$ are 
                drawn along with the two $\mm{ERF}=\{0.7,\,1.3\}$ $\PD$.
                In frame (d) the corresponding $\chiD$ for $\mm{ERF}=1.3$.
              }
            }
            \label{fig_af5}
  \end{center} 
\end{figure}

The influence of the resolution of the charged lepton track are displayed in Fig.~\ref{fig_af5}
for both $p_{T,\lp}$ (a,b) and charge asymmetry of $\pTl$ (c).
The value $\mm{ERF}=0.7$ tends to narrow the bare $\pTl$ distribution while having $\mm{ERF}=1.3$, 
by widening this time the width of the Gaussian response, tends to smear the sharpness of the distributions.
Starting with the classic method we see that no matter which value is used here the bias on the
intrinsic determination of $\MWp$ and $\MWm$, and as a consequence on $\MWp-\MWm$, 
is not very strong.
We can note though that here in top of shifting the unbiased bare spectra to different $\pT$ the
widening/narrowing change locally the normalisation. The consequence is that without any other
technical refinement the $\PD$ can hardly apparent itself to any of the histogram. Also, even
though the impact is negligible on the mass determination the convergence is very low, \ie{}
$\chiDmin/\dof\approx 20$.
In the frame (c) the charge asymmetry shows, no matter the used ERF values good steadiness and 
stays again in the vicinity of the central $\MT$. Then, the result for the likelihood are good
from both physics and convergence point of view.

\clearpage
\subsubsection{Systematic due to the intrinsic $\mbf{\kT}$ of the partons}\label{ss_a3}
\begin{figure}[!ht] 
  \begin{center}
    \includegraphics[width=0.495\tw]{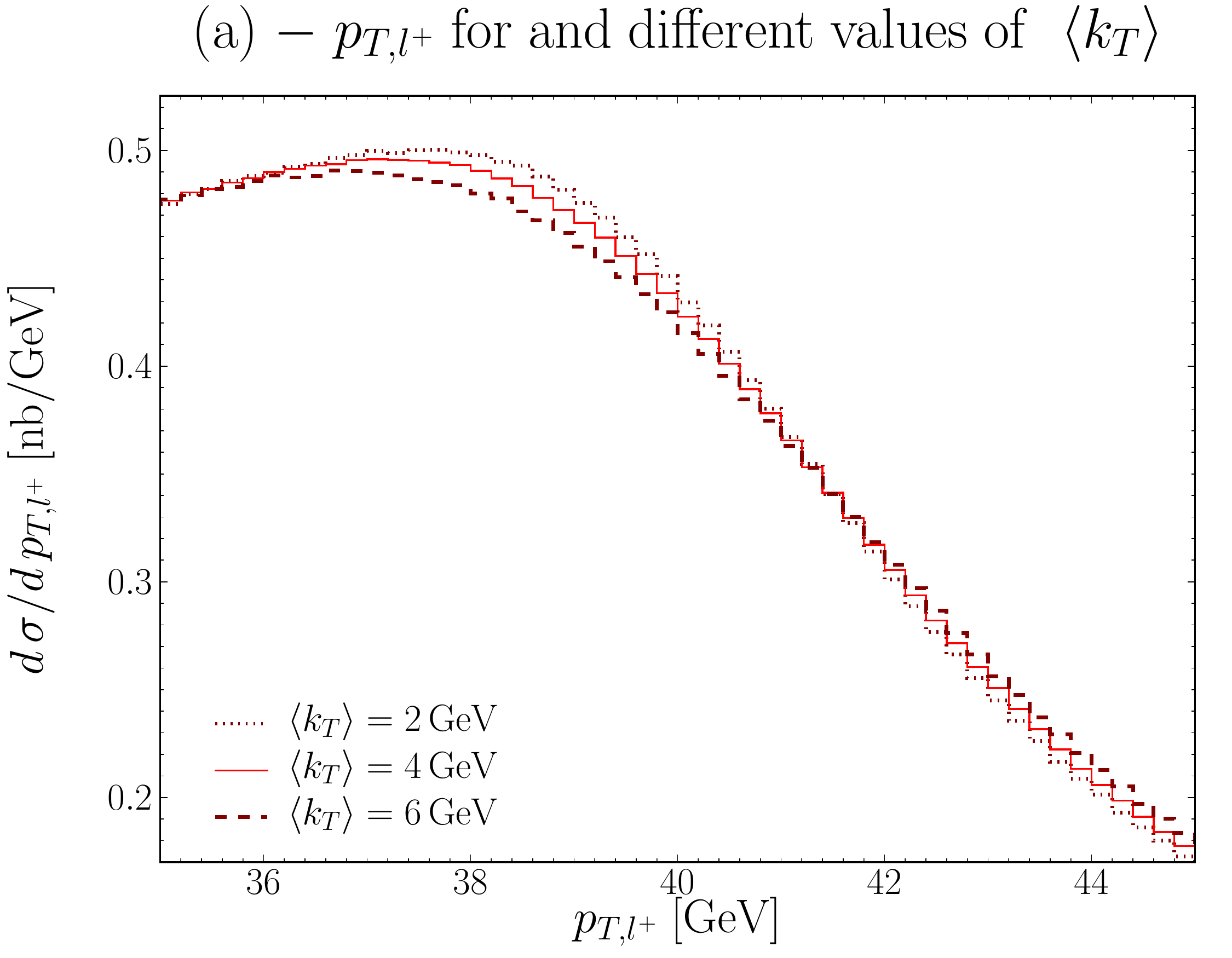}
    \hfill
    \includegraphics[width=0.495\tw]{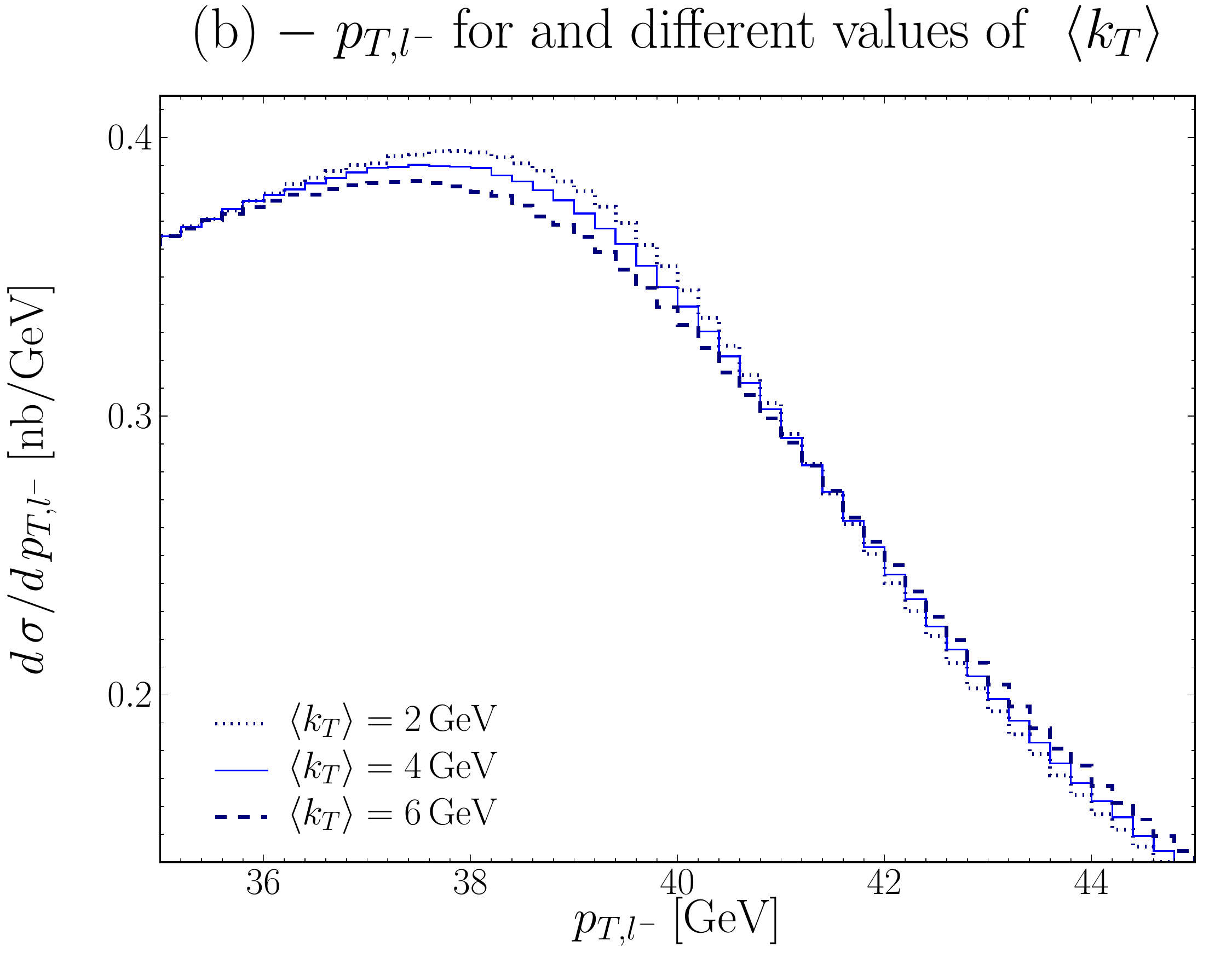}
    \caption[Transverse momentum distribution for several values of the intrinsic $\Mean{\kT}$ of the partons]
            {\figtxt{Transverse momentum distribution for several values of the intrinsic $\Mean{\kT}=\{2,\,4,\,6\}\GeV$ 
                of the partons for positively charge lepton (a) and the negatively charged lepton (b),}
            }
            \label{fig_af6_1}
  \end{center} 
\end{figure}

\index{Quarks!Intrinsic transverse momenta|(}
The impact of the intrinsic $\kT$ of partons is studied with the charge asymmetry method.
First, Fig.~\ref{fig_af6_1} shows the impact of the change of $\Mean{\kT}$ on the jacobian 
peaks of the $\pTlp$ and $\pTlm$ spectra. As expected when the more important is the average
intrinsic $\kT$ the more the jacobian peaks are shifted to higher $\pTl$ values.

\begin{figure}[!ht] 
  \begin{center}
    \includegraphics[width=0.495\tw]{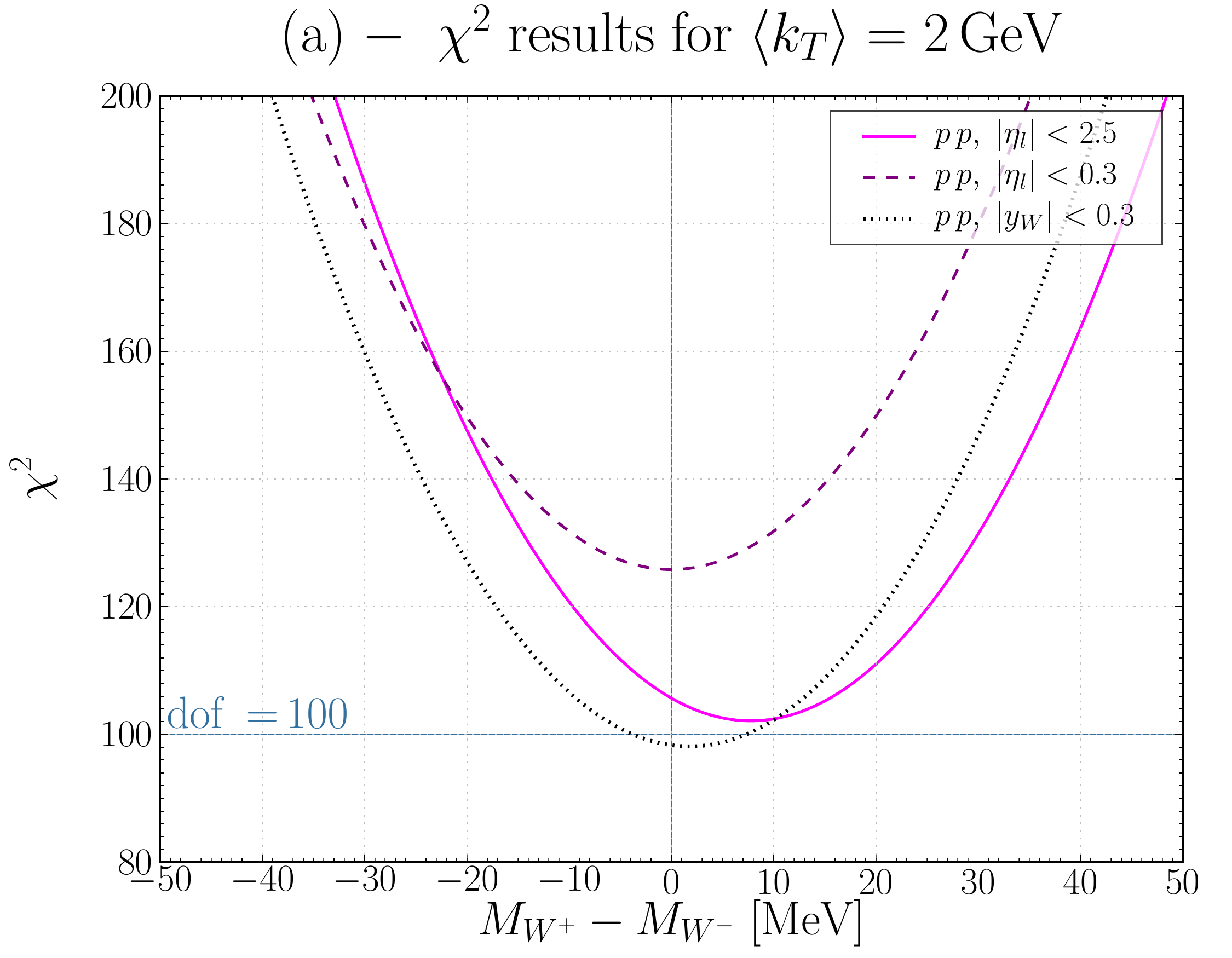}
    \hfill
    \includegraphics[width=0.495\tw]{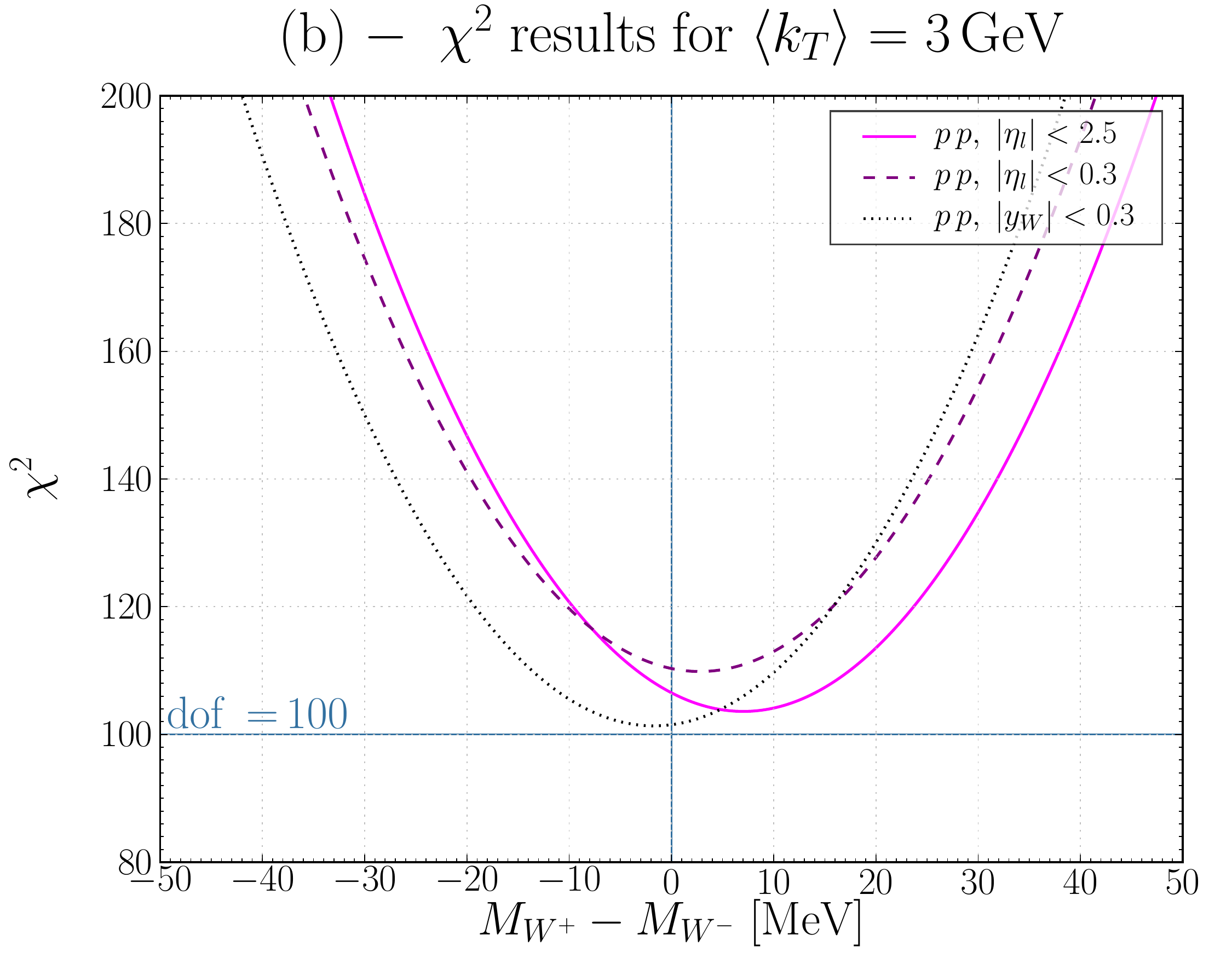}
    \vfill
    \includegraphics[width=0.495\tw]{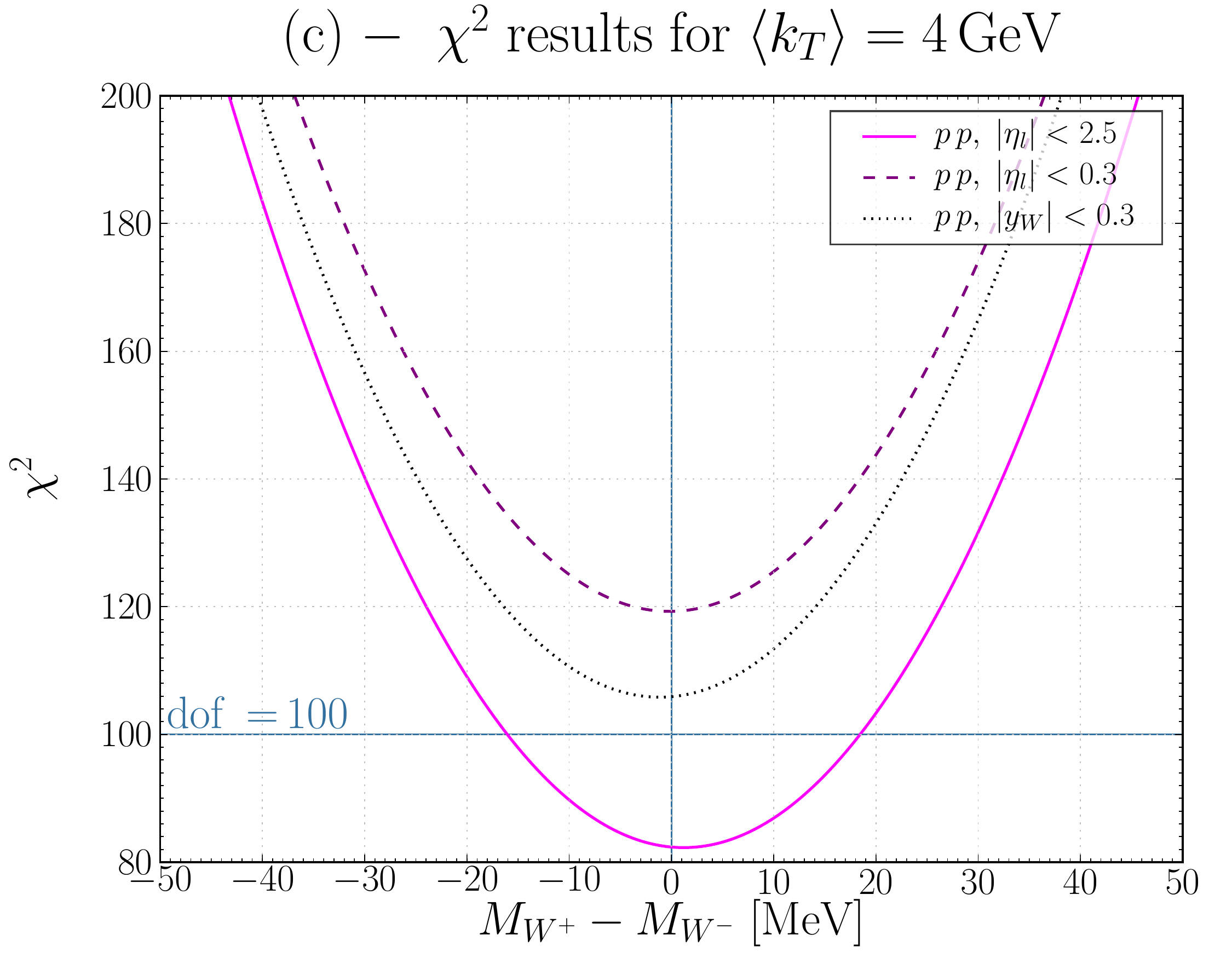}
    \hfill
    \includegraphics[width=0.495\tw]{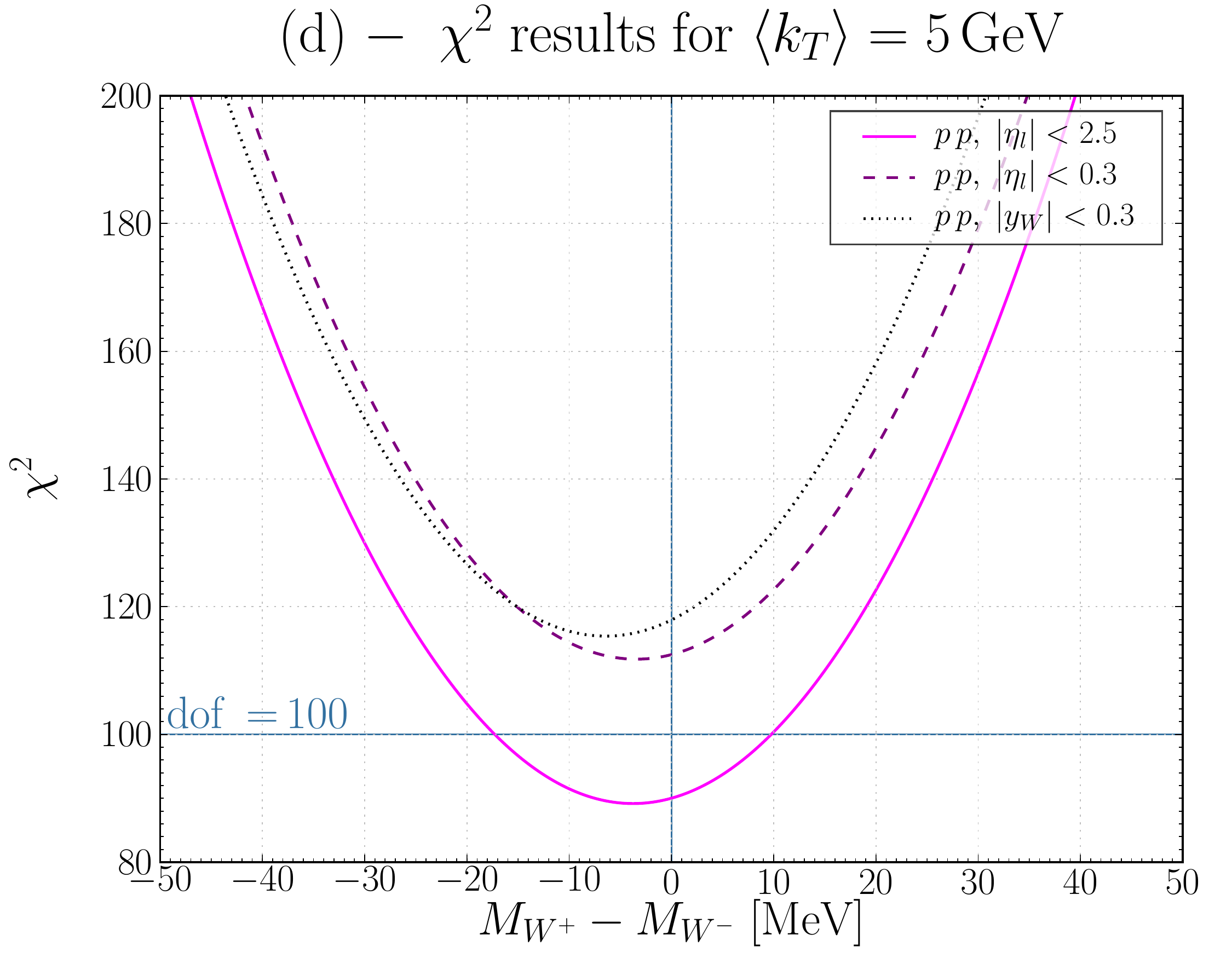}
    \vfill
    \includegraphics[width=0.495\tw]{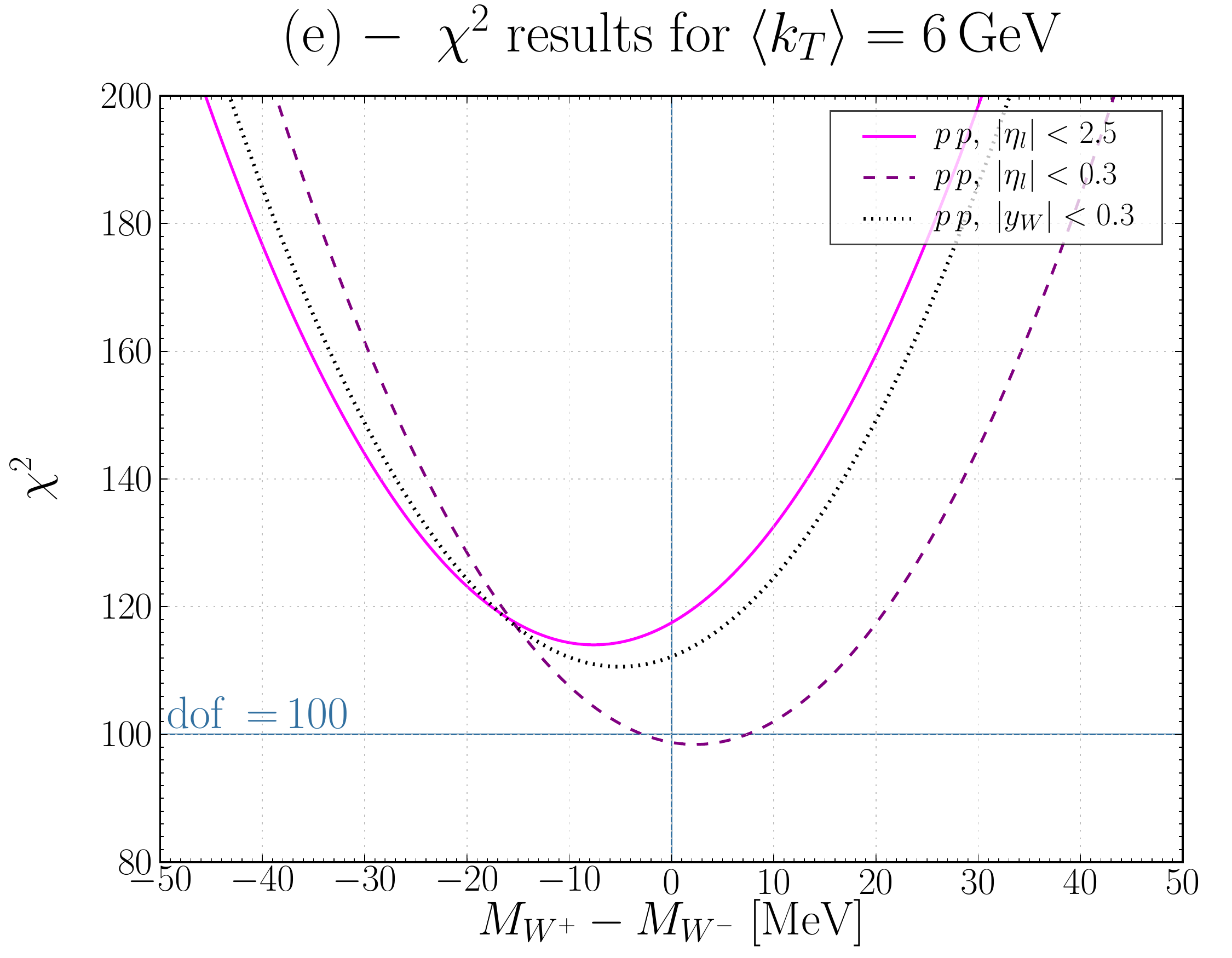}
    \hfill
    \includegraphics[width=0.495\tw]{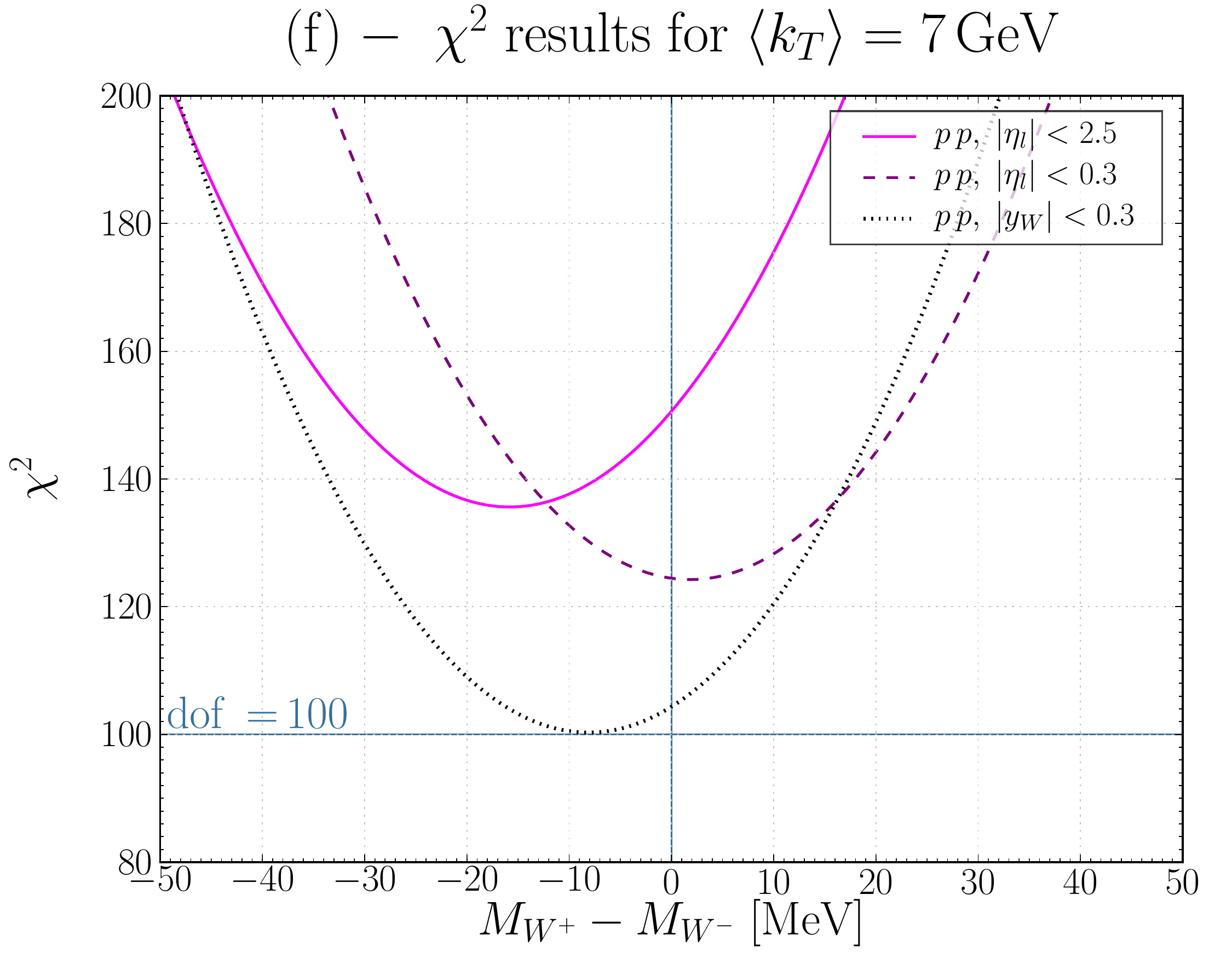}
    \caption[$\chiD$ results for the asymmetry method depending on the intrinsic $\Mean{\kT}$ and for the cuts $\pTl>20\GeV$
    and respectively $|\etal|<2.5$, $|\etal|<0.3$ and $|\yW|<0.3$]
            {\figtxt{$\chiD$ results for the asymmetry method depending on the intrinsic $\Mean{\kT}$ and for the cuts $\pTl>20\GeV$
                and respectively $|\etal|<2.5$, $|\etal|<0.3$ and $|\yW|<0.3$.}
            }
            \label{fig_af6}
  \end{center} 
\end{figure}

Fig.~\ref{fig_af6} on the next page presents in each frame the $\chiD$ results for $\pp$ collisions with $\pTl>20\GeV$ 
and respectively for the three following acceptance cuts\,:
\begin{itemize}
\item[] $|\etal|<2.5$,
\item[] $|\etal|<0.3$ and
\item[] $|\yW|<0.3$.
\end{itemize}
The frames in Fig.~\ref{fig_af6} represent respectively the $\chiD$ results obtained for these
three cases and for the values of $\Mean{\kT}$ of $2$, $3$, $4$, $5$, $6$ and $7\GeV$. The value
$\Mean{\kT}=4\GeV$ is the central one but it was repeated to ensure the continuity in the pattern
the $\chiD$ follow as $\Mean{\kT}$ increases.
The cuts made using $|\etal|<0.3$, up to a lower convergence accuracy, displays a good
steadiness with respect to the uncertainty on $\Mean{\kT}$.
\index{Quarks!Intrinsic transverse momenta|)}

\clearpage
\subsubsection{$\mathbf{u^\val-d^\val}$ asymmetry}\label{ss_a5}
\begin{figure}[!ht] 
  \begin{center}
    \includegraphics[width=0.495\tw]{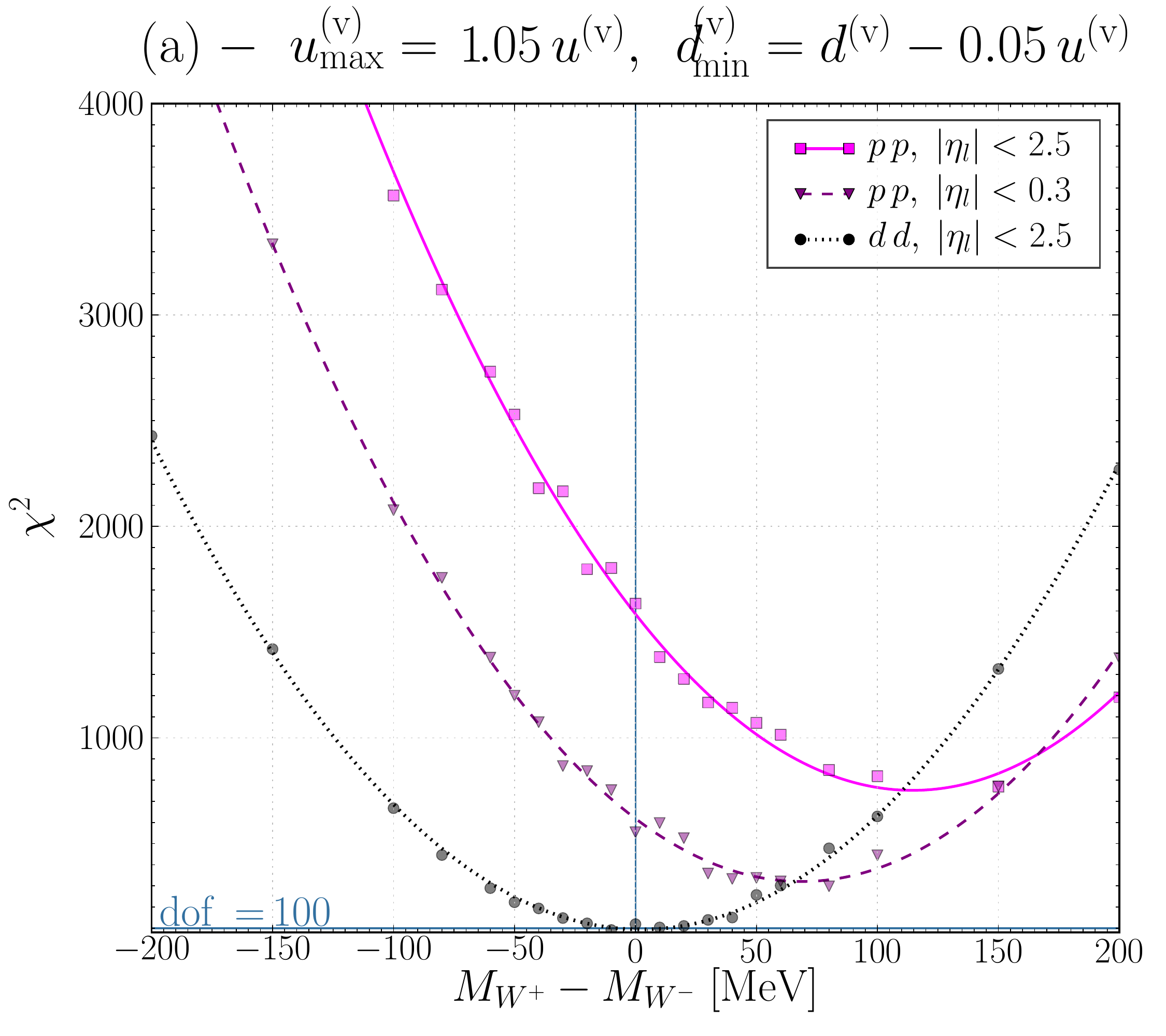}
    \hfill
    \includegraphics[width=0.495\tw]{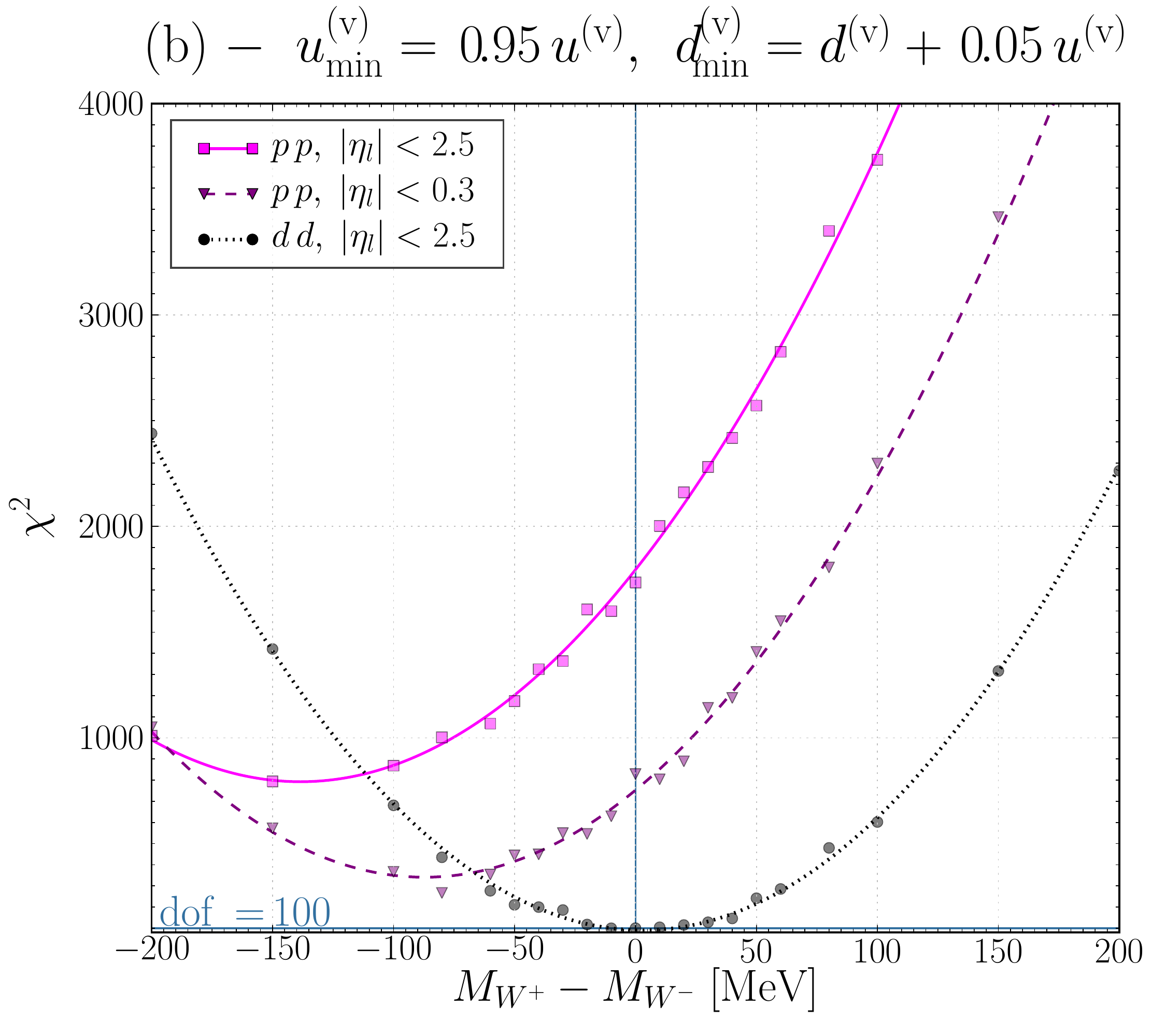}
    \caption[$\chiD$ results for $u^\val_\Max=1.05\,u^\val$ and $d^\val_\Min=d^\val-0.05\,u^\val$
      and $u^\val_\Max=0.95\,u^\val$ and $d^\val_\Min=d^\val+0.05\,u^\val$ cases using $\pp$ and $\dd$
      collisions]
            {\figtxt{$\chiD$ results for $u^\val_\Max=1.05\,u^\val$ and $d^\val_\Min=d^\val-0.05\,u^\val$ (a) 
                and $u^\val_\Max=0.95\,u^\val$ and $d^\val_\Min=d^\val+0.05\,u^\val$ (b)
                for $\pp$ and $\dd$ collisions.}
            }
            \label{fig_af7}
  \end{center} 
\end{figure}

\index{Quarks!Valence quarks}
\index{Quarks!uvmdv@$u^\val-d^\val$ asymmetry|(}
\index{Quarks!Valence quarks}
Figure~\ref{fig_af7} presents the $\chiD$ results obtained for the study of the $u^\val-d^\val$
asymmetry in the cases where $u^\val_\Max=1.05\,u^\val$ and $d^\val_\Min=d^\val-0.05\,u^\val$ in (a) 
and where $u^\val_\Max=0.95\,u^\val$ and $d^\val_\Min=d^\val+0.05\,u^\val$ in (b).
The narrow selection on $\etal$ shows only a slight enhancement while the use of 
$\dd$ collisions, even with an acceptance of $|\etal|<2.5$, remove the ambiguities on the
valence PDF errors.

This can be understood quite easily looking at the impact of the modeled systematic error on the
production of a $\Wp$ and a $\Wm$ for both $\pp$ and $\dd$ collisions.
In the first case, for $\pp$ collision we consider the $\Wp$ and $\Wm$ are respectively produced
via $u^\val\,\dbar\to\Wp$ and $d^\val\,\ubar\to\Wm$. Making explicit the expressions of the 
biased valence distributions in functions of the LHAPDF (LHA)\index{Parton Distribution Functions (PDFs)!LHAPDF}
predictions gives
\begin{eqnarray}
\Wp\,:\quad u^{p,\val}_\mm{biased}\,\dbar &=& \left(u^{p,\val}_\mm{L}+ \delta\right)\,\dbar, \\
\Wm\,:\quad d^{p,\val}_\mm{biased}\,\ubar &=& \left(d^{p,\val}_\mm{L}- \delta\right)\,\ubar,
\end{eqnarray}
where $\delta$ is the error on the PDF we've been introducing by hand.
In this configuration we observe incoherent shift between the production of $\Wp$ and $\Wm$.
Now, for $\dd$ collisions we explicit this time the biased valence 
$u^{d,\val}$ and $d^{d,\val}$ expressions having in mind that in a neutron $n$ we 
have $u^n=d^p$ and $d^n=u^p$. This leads to cancel incoherent biases between the $u^\val$ and $d^\val$\,:
\begin{eqnarray}
\Wp\,:\quad u^{d,\val}_\mm{biased}\,\dbar &=& \left(u^{p,\val}_\mm{LHA}+ \delta + d^{p,\val}_\mm{LHA}- \delta\right)\,\dbar, \\
                                   &=& \left(u^{p,\val}_\mm{LHA} + d^{p,\val}_\mm{LHA}\right)\,\dbar, \\
                                   &=& u_\mm{LHA}^{d,\val}\,\dbar, \\
\Wm\,:\quad d^{d,\val}_\mm{biased}\,\dbar &=&  \left(d^{p,\val}_\mm{LHA}- \delta + u^{p,\val}_\mm{LHA}+ \delta\right)\,\ubar,  \\
                                   &=&  \left(d^{p,\val}_\mm{LHA} + u^{p,\val}_\mm{LHA}\right)\,\ubar,  \\
                                   &=& d_\mm{LHA}^{d,\val}\,\ubar,
\end{eqnarray}
\index{Quarks!uvmdv@$u^\val-d^\val$ asymmetry|)}

\clearpage
\subsubsection{$\mathbf{s-c}$ asymmetry}\label{ss_a6}
\begin{figure}[!ht] 
  \begin{center}
    \includegraphics[width=0.495\tw]{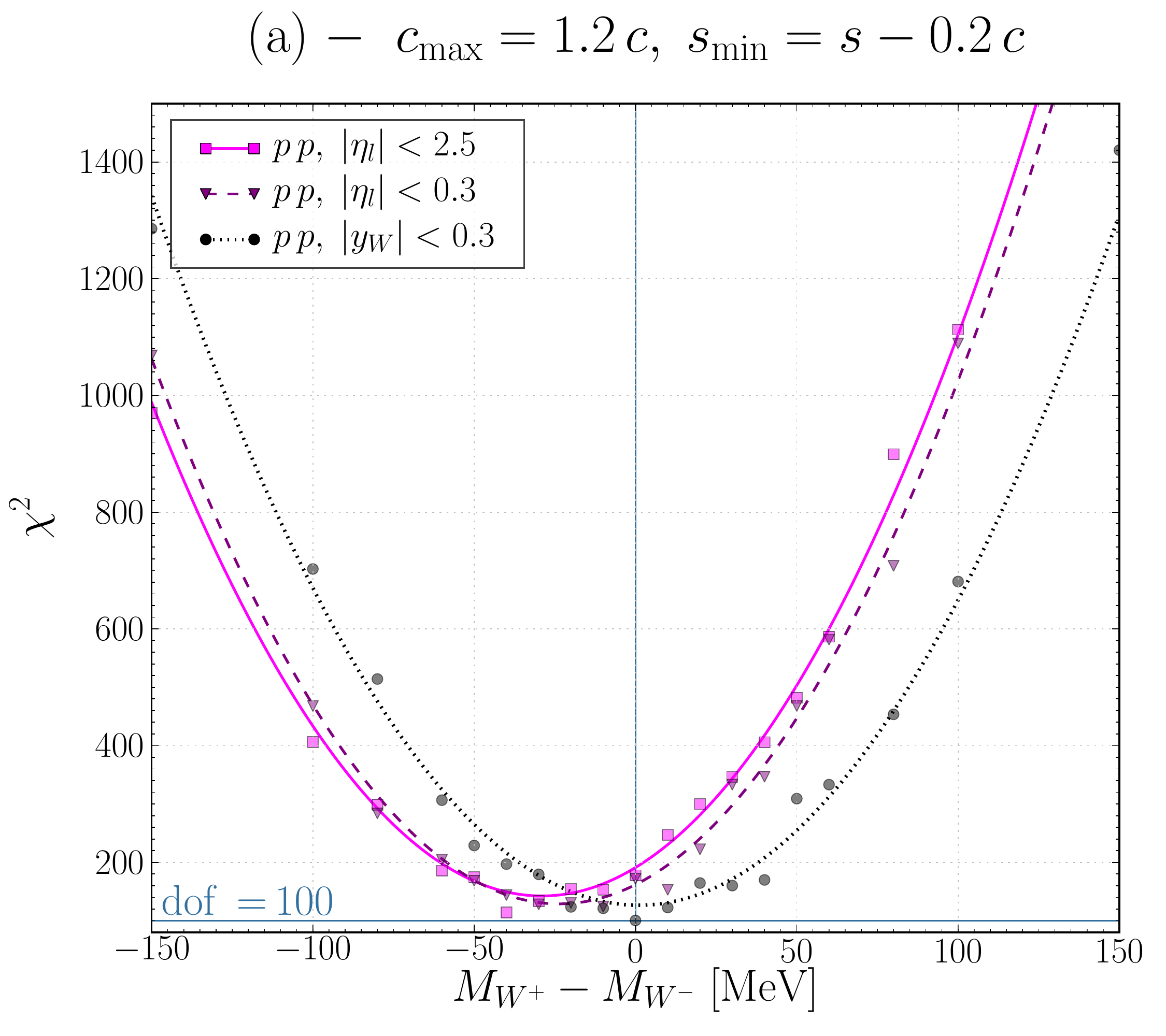}
    \hfill
    \includegraphics[width=0.495\tw]{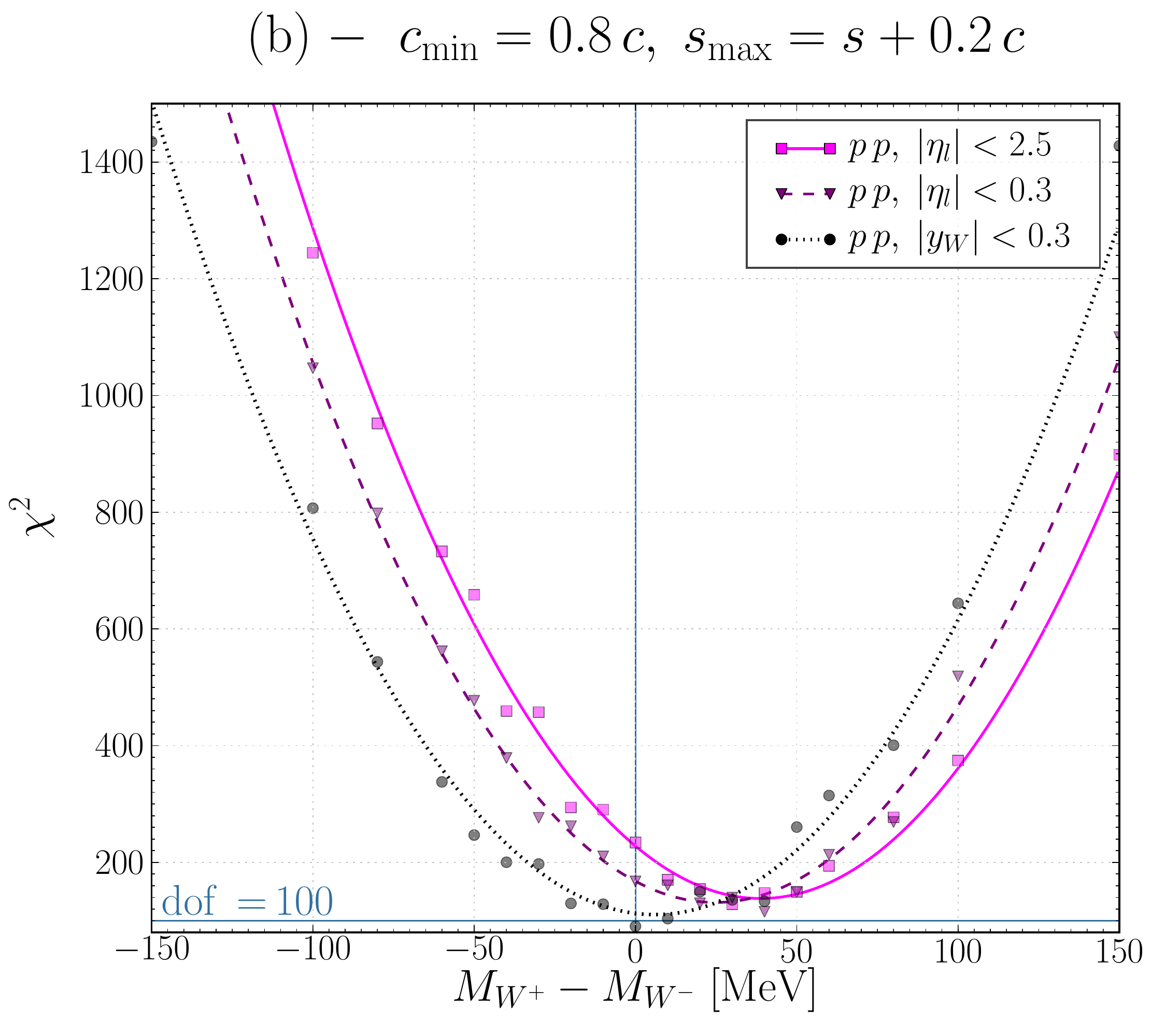}
    \caption[$\chiD$ results for $c_\Max=1.2\,c$ and $s_\Min=s-0.2\,c$
                and $c_\Min=0.8\,c$ and $s_\Max=s+0.2\,c$ for different acceptance criteria for the
                decaying charged lepton]
            {\figtxt{$\chiD$ results for $c_\Max=1.2\,c$ and $s_\Min=s-0.2\,c$ (a) 
                and $c_\Min=0.8\,c$ and $s_\Max=s+0.2\,c$ (b) for different acceptance criteria for the
                decaying charged lepton.}
            }
            \label{fig_af8}
  \end{center} 
\end{figure}

\index{Quarks!smc@$s-c$ asymmetry}
Figure.~\ref{fig_af8} shows the $\chiD$ results for the study of $c_\Max=1.2\,c$ and $s_\Min=s-0.2\,c$
in (a) and $c_\Min=0.8\,c$ and $s_\Max=s+0.2\,c$ in (b). In that case we see that having access to a narrow
cut in the rapidity would improve a lot the independence from these quark flavours asymmetry in our measure.

\end{subappendices}
\cleardoublepage
\ClearShipoutPicture

\backmatter
\chapter{Conclusion}\label{chap_conclusion}
\setlength{\epigraphwidth}{0.3\tw}
\epigraph{
Dobranoc, id{\c e} spa{\'c},\\
\dots
}%
{\textit{Piosenka dla Edka}\\ \textsc{Maria Peszek}}

The work that has been presented in this document corresponds to the second stage in the 
optimisation of measurement strategies of the Standard Model parameters at the LHC.  
It presented a brand new dedicated strategy for the precision measurement of the charge asymmetry 
of the $W$ boson mass specific to the LHC collider physics.

This measurement must, in our view, precede the measurement of the charge averaged mass of the 
$W$ boson and the measurement of $\sin\theta_W$, in order to diminish the risk of false absorption 
of variety of unknown Beyond Standard Model effects within the Standard Model parameter space. 
This measurement is of particular importance for the following two reasons. 
Firstly, at the LHC  -- contrary to the Tevatron $\ppbar$ collider -- we have seen the measurement 
of the averaged mass of the $W$ boson cannot be dissociated from the measurement of the masses of 
its charge states. Secondly, the precision of verification of the charge universality of the Fermi 
coupling constant $\GF$, measured via the charge asymmetry of the muon life time, must be matched 
by the precision of  verification of the charge universality of the $SU(2)$ coupling strength $g_W$.
This can be achieved only if the mass difference $\MWp - \MWm$ can be determined with the precision 
of a few MeV, \ie{} a factor of $\sim 20$ better than the best present measurement.   

\index{Tevatron collider|(}
The Tevatron $\ppbar$ collision scheme, as far as the systematic and modeling errors are concerned,
is better suited for the precision measurement of the $W$ boson charge asymmetry.
However, even if dedicated strategies, as \eg{} those proposed in this document are used, the 
measurement will be limited by the statistical precision, affecting both the $W$ boson samples and, 
more importantly, the $Z$ boson calibration sample. 

At the LHC the requisite statistical precision can be achieved already for the integrated luminosity
of $10\,{\rm fb}^{-1}$, \ie{} in the first year of the LHC operation at the ``low'', 
${\cal O}(10^{33}\,{\rm cm}^{-2}\,{\rm s}^{-1})$, luminosity.\index{Luminosity!At the LHC (expected)!Low}
However, in order to achieve a comparable systematic precision in an analysis based on the 
calibration and measurement strategies developed at the Tevatron, the charge dependent biases in 
the energy (momentum) scale of positive (negative) leptons must be controlled to the precision 
of $0.005\percent$. As it was argued, it will be extremely hard, if not impossible, 
to achieve such a precision using the calibration methods developed at the Tevatron. 
Moreover, we have identified the LHC-specific sources of errors, related to the uncertainties in the
present knowledge of the flavour composition of the WBpB which limit, at present, the measurement 
precision to ${\cal O}(100\MeV)$.\index{Tevatron collider|)}
Certainly, this uncertainty will be reduced once the high statistic $W$ boson and $Z$ boson samples 
are collected. Nevertheless, it will be hard, if not impossible, to improve by a factor of 10 or 
more the precision of the $u^\val$--$d^\val$ and $s$--$c$ quark-momentum distribution asymmetries, 
as they are hardly  detectable in the $Z$-boson production processes.  
Whether or not the requisite precision target will be reached using the standard measurement 
strategies remains to be seen. In our view, it will be indispensable to use the dedicated 
LHC-specific measurement strategies.

The strategy proposed and discussed here makes a full use of the flexibility of the machine
and detector configurations which, we hope, will be exploited in the mature phase of the LHC 
experimental program. It requires\,: (1) running for a fraction of time the inverted polarity 
current in the detector solenoid, (2) the dedicated trigger and data-acquisition  configuration 
in the ``high'', ${\cal O}(10^{34}\,{\rm cm}^{-2}\,{\rm s}^{-1})$, luminosity LHC operation mode,
and (3) replacing the LHC proton beams by light isoscalar ion beams.\index{Luminosity!At the LHC (expected)!Nominal/High}  

The underlying principle of the proposed dedicated strategy is that, instead of diminishing the 
systematic measurement and modeling uncertainties, it  minimizes their influence on the measured 
value of $\MWp -\MWm$. We have demonstrated that already for the modest (easy to fulfill) 
measurement and modeling precision requirements,  the resulting uncertainty of $\MWp -\MWm$ can be 
kept at the level comparable to the statistical measurement uncertainty, \ie{} at the level of 
${\cal O}(5\MeV)$.

It remains to be demonstrated that the remaining  systematic measurement errors, of secondary 
importance at the Tevatron and not discussed here, could be neglected at this level of the 
measurement precision. This can, however, be proved only when the real data are collected and 
analysed, both for the standard and for the dedicated measurement strategies.

\cleardoublepage
\listoffigures{}
\cleardoublepage
\listoftables{}

\cleardoublepage
\providecommand{\href}[2]{#2}\begingroup\raggedright\endgroup

\cleardoublepage
\printindex

\end{document}